\documentclass[12pt,a4paper]{article}
\usepackage[utf8]{inputenc}
\usepackage[english]{babel}
\usepackage{amsmath}
\usepackage{amsfonts}
\usepackage{amssymb}
\usepackage{graphicx}
\usepackage[left= 1.3cm,right=1.3cm,top=2cm,bottom=2cm]{geometry}
\usepackage[none]{hyphenat}
\usepackage{authblk}

\begin{document}
\title{\bf Mapping Multivariate Phenotypes in the Presence of Missing Observations for Family-Based Data}
  \author[1]{Soumya Sahu}
  \author[2]{Saurabh Ghosh}
  \affil[1]{Biostatistics,  School of Public Health, University of Illinois Chicago, USA}
  \affil[2]{Human Genetics Unit, Indian Statistical Institute, Kolkata}
  \date{}
  \maketitle

\begin{abstract}
Clinical end-point traits are often characterized by quantitative or qualitative precursors and it has been argued that it may be statistically a more powerful strategy to analyze these precursor traits to decipher the genetic architecture of the underlying complex end-point trait. While association methods for both quantitative and qualitative traits have been extensively developed to analyze population level data, development of such methods are of current research interest for family-level data that pose additional challenges of incorporation of correlation of trait values within a family. Haldar and Ghosh ( 2015 ) developed a test which is Statistical equivalent of the classical TDT for quantitative traits and multivariate phenotypes. The model does not require a priori assumptions on the probability distributions of the phenotypes. However, it may often arise in practice that data on the phenotype of interest may not be available for all offspring in a nuclear family. In this study, we explore methodologies to estimate missing phenotypes conditioned on the available ones and carry out the transmission-based test for association on the “complete” data. We consider three types of phenotypes: continuous, count and categorical. For a missing continuous phenotype, the trait value is estimated using a conditional normal model. For a missing count phenotypes, the trait value is estimated using a conditional Poisson model. For a missing categorical phenotype, the risk of the phenotype status is estimated using a conditional logistic model. We shall carry out simulations under a wide spectrum of genetic models and assess the effect of the proposed imputation strategy on the power of the association test vis-à-vis the ideal situation with no missing data.
\end{abstract}

\section{Introduction and Background}

A clinical trait or disease with a complex end-point is generally characterized by multiple quantitative precursors. In such situations, one can improve statistical power by considering a multivariate phenotype. Galesloot et al. ( 2014 ) showed that the use of multivariate association analysis can yield higher power compared with univariate analysis even when genetic correlations between traits are weak or when a genetic variant is associated with only one of the phenotypes ( i.e. when there is no multivariate association ). For example, cardiovascular disease ( CVD ) is characterized by high levels of both low density serum lipoprotein levels ( LDL ) and systolic blood pressure ( SBP ) [ Newman et al., 1986 ]. In order to identify genomic regions associated with CVD, it may be intuitively more appealing and statistically more powerful to carry out association tests with a bivariate phenotype vector comprising LDL and SBP instead of the binary CVD phenotype.\\

In this aspect there are strong methodologies available in literature to map multivariate phenotypes for population level data but problem of such methods are that they are affected severely when there is population stratification. So, one of the current research interest to find the methodologies which can do the same for family level data. Family-based genetic association tests are based on the transmission-disequilibrium test (TDT). Families are recruited based on a disease phenotype in the offspring. If a particular disease locus
is uninvolved in the disease, one would expect the parents' allele at that locus to be transmitted
randomly. On the other hand, if the allele is actually associated with disease, and the offspring was
selected based on the disease, there will be an apparent over-transmission of the allele. The classical
transmission disequilibrium test ( TDT ) for binary traits based on a trio design is a popular family-based alternative to population-based case-control studies as it tests for association in the presence of
linkage, and hence is protected against population stratification.\\

Haldar and Ghosh ( 2015 ) developed a test which is Statistical equivalent of the classical TDT for quantitative traits and multivariate phenotypes. They used Generalized Linear Model and the method is free from any priori assumptions on the distributions of the phenotypes.\\

In practical it is very much likely that data on some phenotypes is missing and important thing is that the phenotypes of the offspring in a particular family is correlated. In this paper we discuss the different type of missingness. We suggest methodologies to use this correlation to estimate those missing data. After estimation we perform a power comparison between three cases - non-missing, estimated missing (by our proposed methods) and omitted missing.

\section{Data Description and Model}

We assume that vector of QT, $\textbf{Y} = (Y_1, Y_2,..., Y_k)^T$ is controlled by a biallelic QT locus (QTL) with alleles $D_1$ and $D_2$. We consider a biallelic marker locus with alleles $M_1$ and $M_2$ such that the recombination fraction between the QTL and the marker locus is $ \theta $ and the coefficient of LD between the loci is measured by $ \delta = P(D_2M_2) - P(D_2)P(M_2) $, suppose that the allele frequencies of $D_2$ and $M_2$ are $d$ and $m$, respectively. We assume that the probability density of Y conditioned on the genotype at the QTL is $f_1$, if the genotype is $D_1D_1$, $f_2$, if it is $D_1D_2$ and $f_3$, if it
is $D_2D_2$. The data comprise marker genotypes of trios (two parents and an offspring) such that at least one of the parents is heterozygous and QT values of the offspring.

We define $Z$ to be indicator variable corresponding to the transmission of the allele $M_2$ by the heterozygous parent. If the both parents and the offspring are heterozygous at marker locus, we assign $Z$ randomly as 0 or 1.

\subsection{Model : ( by Halder \& Ghosh [ 2015 ] )}

They consider a logistic regression model-

$$logit(P(Z=1|y_1, y_2,..., y_k)) = \sum \limits_{i=1}^k \beta_i(y_i - c_i) \quad ,$$

where, $logit(x) = ln(\frac{x}{1-x})$ , $c_i$ is some central summary measure of the QT $Y_i$ in the population.\\

The test for no linkage or no association versus the presence of both linkage and association is equivalent to testing $H_0: \beta_i = 0 \quad \forall \quad i = 1, 2, . . . , k$ vs $H_1: \beta_i \neq 0$ for at least one i. We use suitable central tendency measure ( like mean , median ) in place of $c_i$ in the logistic model. The standard log-likelihood ratio test statistic is distributed as chi-squares with k degrees of freedom under the null hypothesis.

We should note that these two models are discussed for Trio's only but in this paper we shall handle the data for sib-pairs. As we have family level data we determine the transmissions at marker locus based on the parental genotypes, so given parental information transmissions of two sibs should be independent under null hypothesis i.e. there is no association. So, we shall treat this data as $2n$ independent transmissions for $n$ families but we shall lose some power as under alternative hypothesis transmissions within a family are not independent.

\section{Type of Missing Data}

We shall consider two types of data and discuss possible missing types for each of them. Here we consider traits can be continuous type or count type or categorical type.
\begin{itemize}
\item \textbf{Type 1:}  we consider one trait and for missing data, for each family either we have trait values for one of the offspring or we have trait values for both of the offspring.

\item \textbf{Type 2:} we consider two traits so for each family we have for trait values. let trait values for 1st offspring are $x_1, y_1$ and those for 2nd offspring are $x_2, y_2.$ For missing data, we can have following types of missing-

\begin{itemize}
\item \textbf{Type 2.1:} any one of $x_1, y_1, x_2, y_2$ is missing.
\item \textbf{Type 2.2:} any one of $x_1, y_1$ and any one of $x_2, y_2$ are missing.

\item \textbf{Type 2.3:} either both of $x_1, y_1$ are missing or both of $x_2, y_2$ are missing.

\item \textbf{Type 2.4:} any three of $x_1, y_1, x_2, y_2$ is missing.

\end{itemize} 
\end{itemize}

In this report we shall discuss the imputation method for missing types - 1, 2.1, 2.3, 2.4.

\section{Missing Data Estimation Method}

Suppose the k-dimensional phenotype vector $\mathbf{Y}$ is partitioned into three subsets as:

\begin{itemize}
\item SN: the set of continuous phenotypes.
\item SP: the set of count phenotypes.
\item SB: the set of binary phenotypes.
\end{itemize}

For a particular sampling unit, suppose $Y_i$ ; $i= 1,2,...,k$ denotes one of those phenotypes corresponding to which ,trait values are missing. For the same sampling unit, let $SN^\star, SP^\star$ and $SB^\star$ respectively denote the sets of continuous, count and binary phenotypes, whose trait values are available.
Suppose, $SN^\star \cup SP^\star \cup SB^\star = S^\star$ and $s^\star$ denotes a set of observed values for $S^\star.$ Then, we consider the following assumptions:

\begin{itemize}
\item If $Y_i \in SN,$  then $Y_i |S^\star = s^\star$ is normally distributed with mean $\theta^Ty^{\star\star}$ and constant variance, $\sigma^2$.
\item If $Y_i \in SP,$  then $Y_i |S^\star = s^\star \sim Poisson(e^{\theta^Ty^{\star\star}} )$
\item If $Y_i \in SB,$  then $Y_i |S^\star = s^\star \sim Bernoulli( \frac{e^{\theta^Ty^{\star\star}}}{ 1 + e^{\theta^Ty^{\star\star}} }).$
\end{itemize}

Here, $\theta$ is a vector of unknown constants, $y^{\star\star}$ is the vector of known trait values and we shall discuss different choices of $y^{\star\star}$ for different types of missing.\\

{\large \textbf{Imputation:}}

If $Y_i$ is missing, we impute $Y_i$ by $E(Y_i| S^\star = s^\star),$ where we estimate $\theta$ by using regression-

\begin{enumerate}
\item if the trait values are continuous we use OLS regression.
\item if the trait values are count type we use Poisson regression.
\item if the trait values are binary type we use logistic regression.

\end{enumerate}

\textbf{{\large Choice of $y^{\star\star}:$}}

\begin{itemize}
\item In the missing type 1,  $y^{\star\star}$ is the value of the trait of other offspring which is available.

\item In the missing type 2.1, we have used following choices of  $y^{\star\star}$ - 
\begin{itemize}
\item \textbf{Strategy 1:}  estimate the missing trait using the value of the other trait of that offspring. For example if $x_1$ is missing, $y^{\star\star}$ will be $y_1$.
\item \textbf{Strategy 2:} estimate the missing trait using the value of same trait of the other offspring \hspace{.2cm}( sib ) of that family. For example if $x_1$ is missing, $y^{\star\star}$ will be $x_2$.
\item \textbf{Strategy 3:} estimate the missing trait using both of the trait values mentioned in above two cases. For example if $x_1$ is missing, $y^{\star\star}$ will be $( y_1, x_2 )^T$.

We varied two types of correlations - $\rho_1$ is correlation between two traits of same offspring (like $x_1, y_1$) and $\rho_2$ is correlation between same trait of two offspring (like $x_1, x_2$). By simulation study we have found out which of the above choices of $y^{\star\star}$ yields best power for different choices of $\rho_1$ and $\rho_2.$ 

\item In the missing type 2.3, we have estimated the missing trait using the value of same trait of the other offspring( sib ). For example if $x_1, y_1$ are missing, to estimate $x_1$, $y^{\star\star}$ is $x_2$ and to estimate $y_1$, $y^{\star\star}$ is $y_2.$

\item In the missing type 2.4, there is only one trait available for each family, so we have estimated other three missing trait using the available one.

\end{itemize}

\end{itemize}

Here we have talked about families with sib pairs but in reality there may be some families who have one sib, in those cases if we are observing only one trait and that is missing, we have to delete the family from data or if we are observing more than one trait and any trait is missing we can estimate them using other trait values using our proposed methodology.
 
\section{Method of Simulation}

\subsection{simulating n families with sib pairs and obtaining $Z_1$ and $Z_2$ }

At first we shall fix $n$(sample size), $d, m, \delta^\star$(this can take value between 0 and 1) and 

$\delta = \delta^\star \times min\{d(1-m), m(1-d)\}$. We generate haplotypes from the following discrete distribution:

\begin{center}
\begin{tabular}{|c|c|c|c|c|}
 \hline 
 haplotype & $M_2D_2$ & $M_1D_2$ & $M_2D_1$ & $M_1D_1$ \\ 
 \hline 
 prob & $md+\delta$ & $(1-m)d - \delta$ & $m(1-d) - \delta$ & $(1-m)(1-d) + \delta$ \\ 
 \hline 
 \end{tabular}  
\end{center}

we generate $4n$ observations from this distribution and joining(pairwise) them randomly we get $2n$ parents.So, we have parents like $M_2D_2M_1D_1$, $M_2D_2M_2D_1$ etc. We randomly made $n$ pairs from these $2n$ many and each pair mate to create offspring. But many of the parents-pair may be non informative in our purpose. So we delete those pair of parents where both of them are homozygous in the marker locus.\\

We fix recombination factor, $\theta$. For a particular parent we generate haplotype in the following method-\\

if the parent is homozygous it will generate haplotype with alleles in its 1st and 2nd position w.p. 1, if the parent is heterozygous in exactly one loci then  it will generate haplotype with alleles-

\vspace{.2cm}
\begin{tabular}{|c|c|c|}
\hline 
position & 1st \& 2nd & 3rd \& 4th \\ 
\hline 
prob & .5 & .5 \\ 
\hline 
\end{tabular}   
\vspace{.4cm}

if the parent is heterozygous is both loci then it will generate haplotype with alleles-

\vspace{.4cm}
\begin{tabular}{|c|c|c|c|c|}
\hline 
position & 1st \& 2nd & 3rd \& 4th & 1st \& 4th & 2nd \& 3rd \\ 
\hline 
prob & (1-$\theta$)/2 & (1-$\theta$)/2 & $\theta$/2 & $\theta$/2 \\ 
\hline 
\end{tabular} 
\vspace{.4cm}

In this way we generate two offspring from each pair of parents. Now we divide genotypes of each offspring into marker genotype and genotype of QTL, 1st one will be used( this is the part we can observe in real life data ) to find the value of $Z$ and 2nd one will be used to find trait values. We find $Z$ in the following way-

If offspring has $M_1M_1$ then $Z = 0$ if offspring has $M_2M_2$ then $Z = 1$ If offspring has $M_1M_2$ then

\vspace{.4cm}

 \begin{tabular}{|c|c|}
\hline 
parents & $Z$\\ 
\hline 
$M_1M_2$, $M_1M_2$ & rand\\ 
\hline 
$M_1M_2$, $M_2M_2$ & 0\\ 
\hline 
$M_1M_2$, $M_1M_1$ & 1\\ 
\hline 
\end{tabular} 
, where rand is a random sample from bernoulli(.5).

\subsection{simulating trait values}

As defined in section 3, $$f_1 = P(Y|x=0), f_2 = P(Y|x=1), f_3 = P(Y|x=2)$$,

where, $Y$ is the value of the trait, $x$ denotes no of allele $D_2$ at QTL and assuming HWE,
$$x \sim binomial(2, d).$$

Let, QTL explains $p^\star$ proportion of the total variance of trait values, $p^\star = \frac{V(E(Y|x))}{V(Y)}.$

\vspace{.4cm}

We have considered four types of trait values - 

\begin{enumerate}
\item \textbf{continuous trait from a symmetric distribution}: we have done for normal, we generate trait from the model:

 $f_1:N(\alpha, \sigma^2)$, $f_2: N(\alpha + \beta, \sigma^2)$, $f_3:N(\alpha + 2\beta, \sigma^2).$
 
\item \textbf{continuous trait from a skewed distribution}: we have done for Chi Squares, we generate trait from the model:

 $f_1: \alpha + \chi^2_{df} $, $f_2: \alpha + \beta + \chi^2_{df}$, $f_3:\alpha + 2\beta + \chi^2_{df}.$

\item \textbf{discrete trait of count type}: we have done for Poisson, we generate trait from the model:

 $f_1: \alpha + P(\lambda) $, $f_2: \alpha + \beta + P(\lambda)$, $f_3:\alpha + 2\beta + P(\lambda).$
 
\item \textbf{discrete trait of binary type}: we generate trait from the model:
(We generated $z$ from normal distribution and we have assigned 1 if $z>c$ and 0 otherwise)

$f_1: bernoulli(\Phi(\frac{c - \alpha}{\sigma}))$, $f_2: bernoulli(\Phi(\frac{c - \alpha - \beta}{\sigma}))$, 
$f_3: bernoulli(\Phi(\frac{c - \alpha - 2\beta}{\sigma})).$

\end{enumerate}

Let, $miss_p$ denote proportion missing observations. We randomly choose $miss_p$ proportion families and for each them, omit trait values depending on the type of missing. So, we create three datasets- without missing, estimated missing, deleted missing and compare the power of the test in Model-1 for these three datasets. We draw the power curves of these datasets by using colours green, blue and red respectively.

\section{Simulation and Results}

We fix $\theta = .01$, sample size(number of families) $=$ 500, number of replication for each power computation $=$ 1000. We fix the proportion of missing and vary other parameters such that $p^\star$ remains constant. we fix the size of the test is .05 and in each power curve we calculate powers for five choices of $\delta^\star,$ 0(to check if size is being attained), .33, .67, 1.

\subsection{For Missing Type 1}

We have done simulation for 20\% and 30\% missing for each of the four types of traits.

\subsubsection{When Distribution of Trait is Continuous \& Symmetric ( Normal )}

Here we vary $\beta$, $d$ and $var$ (=$\sigma^2$) such a way that $p^\star$ is close to .1. We take two values of $m$, .1 and .5 and for each $m$ we vary $d$ and $var$ in the following way-
\begin{center}
\begin{tabular}{|c|c|c|c|}
\hline 
$d$ & .1 & .2 & .5 \\ 
\hline 
$var$ & 1.62 & 2.88 & 4.5 \\ 
\hline 
\end{tabular} \hspace{2cm} {\large $\beta = 1$}

\end{center}

\begin{center}

 $m = .1 \quad miss_p = .2$

\includegraphics[width = 2.3in, height = 1.5in]{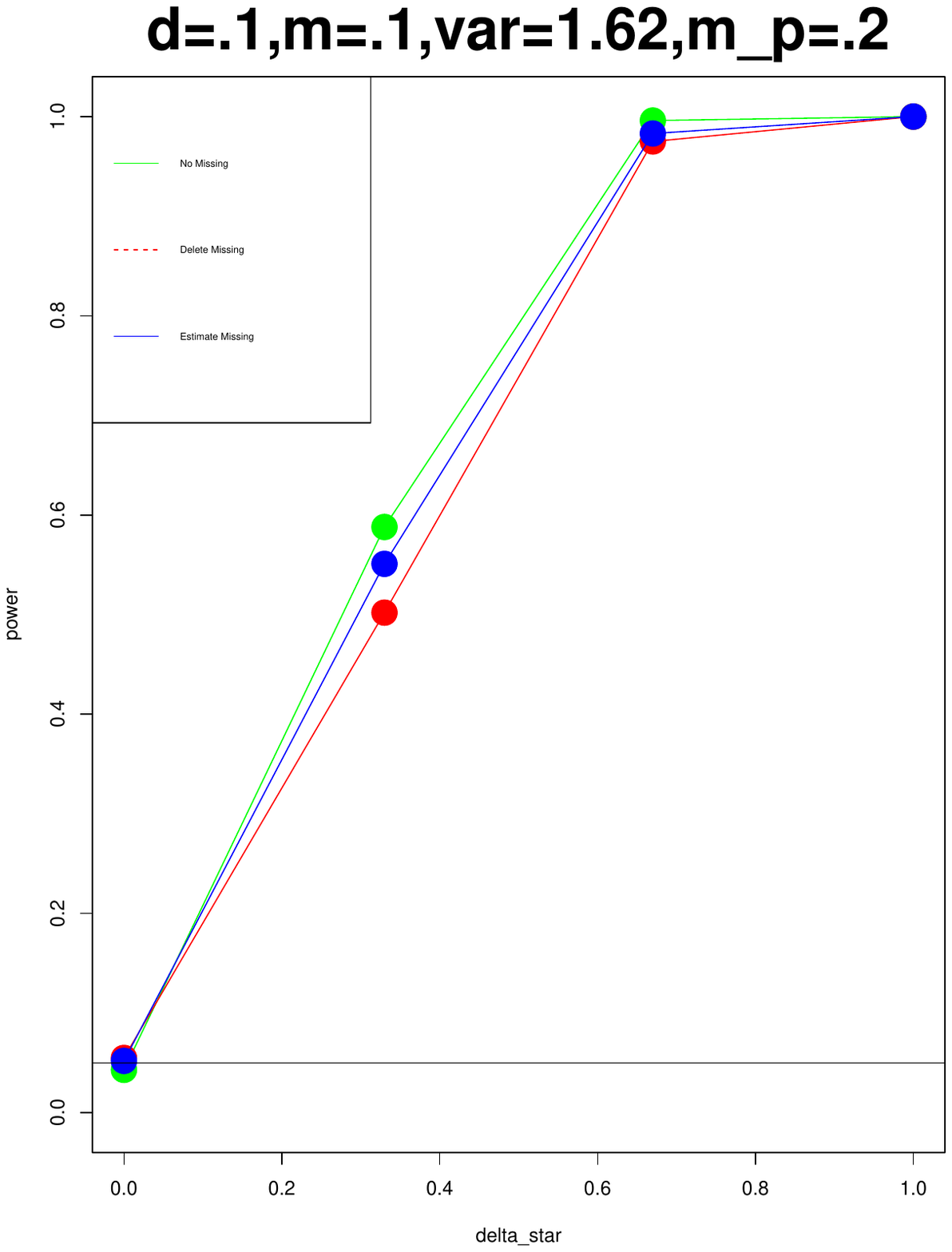}
\includegraphics[width = 2.3in, height = 1.5in]{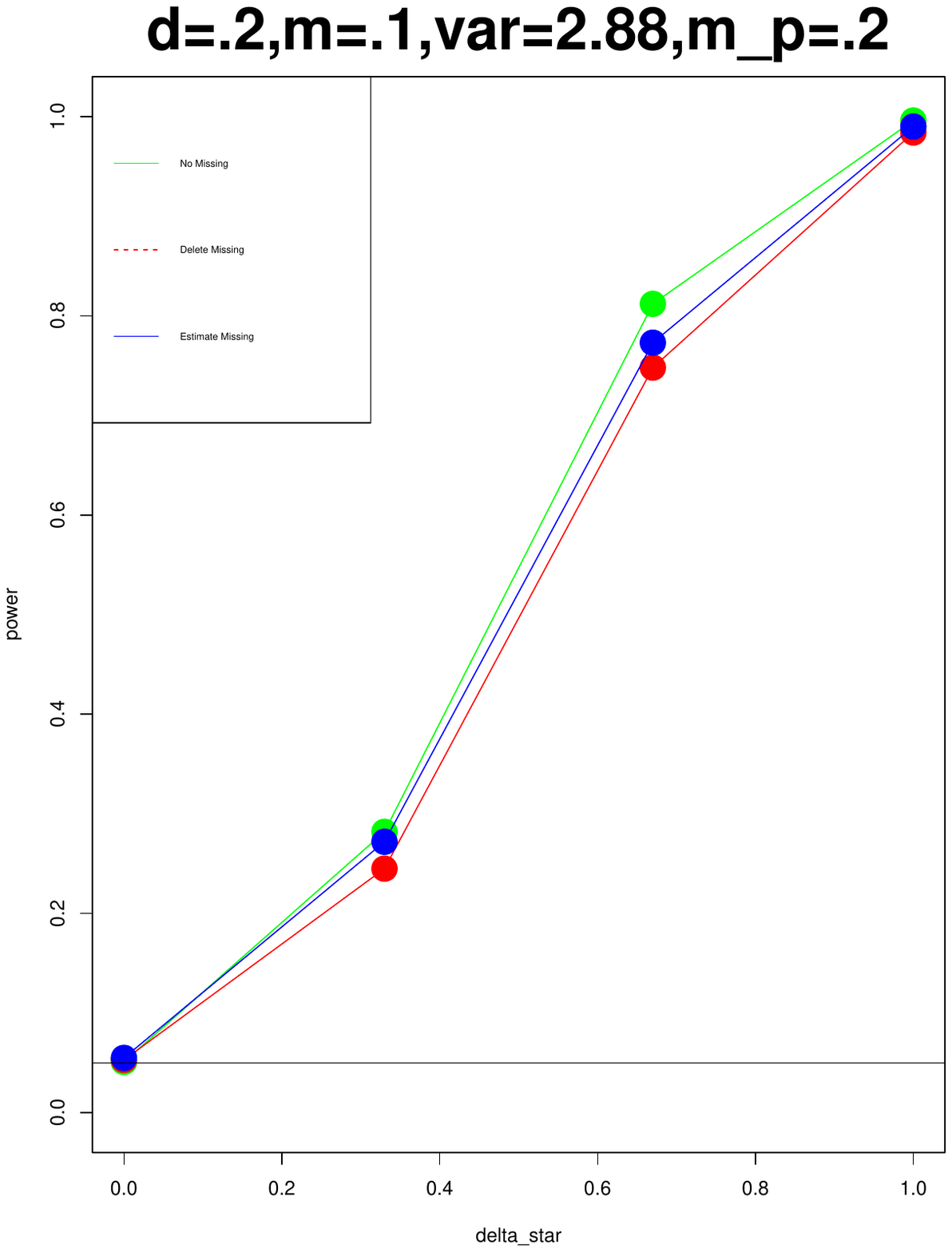}
\includegraphics[width = 2.3in, height = 1.5in]{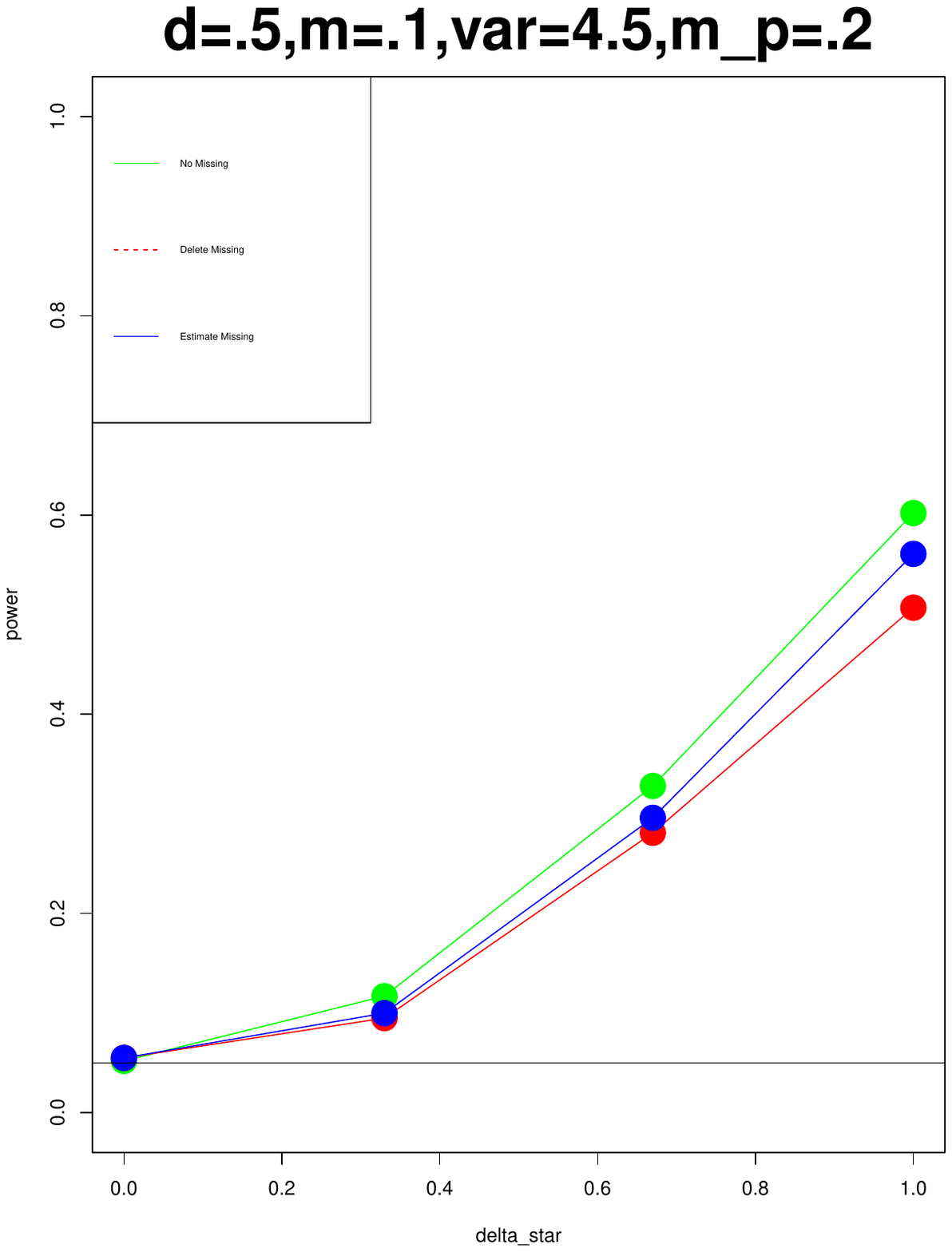}

 $m = .5 \quad miss_p = .2$

\includegraphics[width = 2.3in, height = 1.5in]{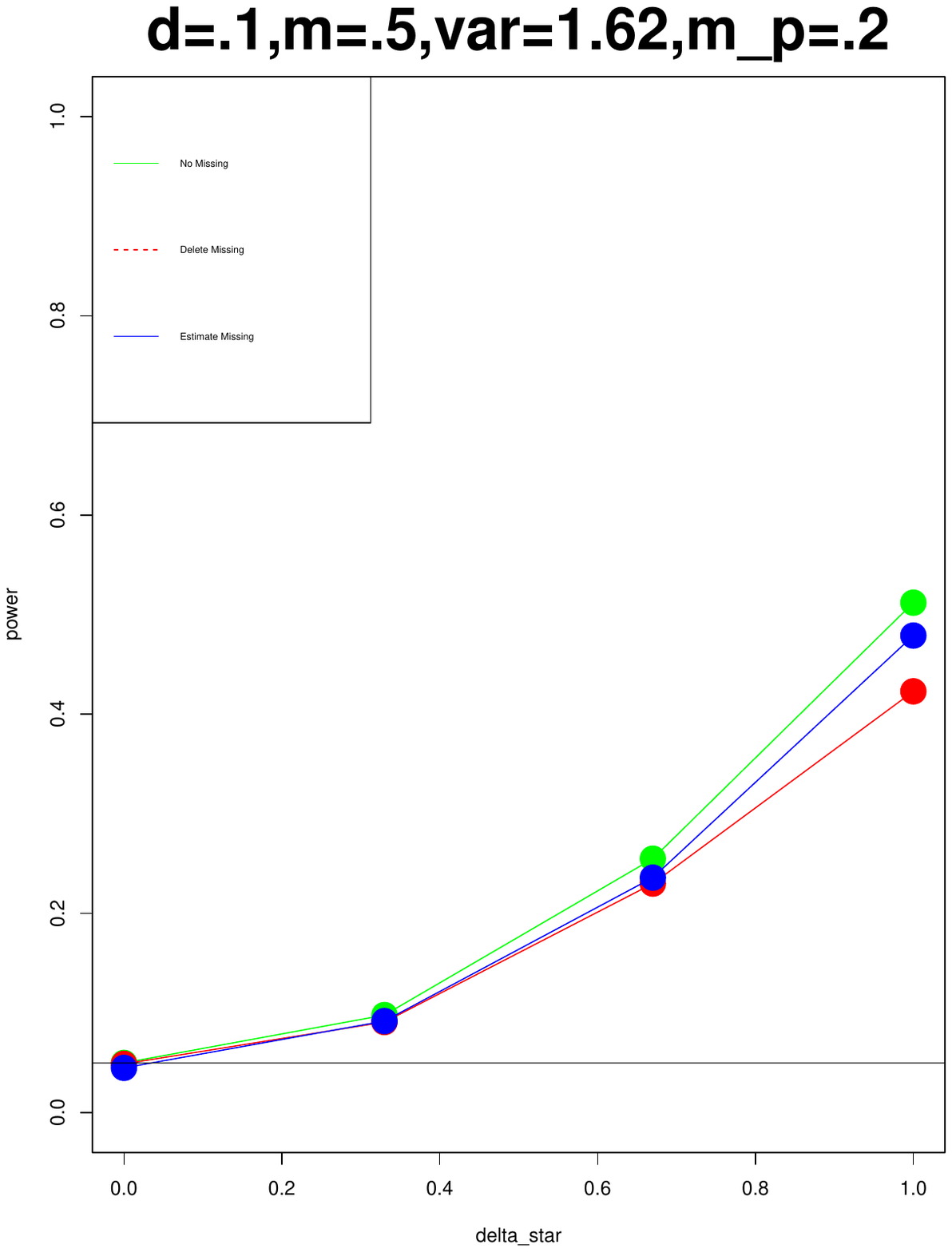}
\includegraphics[width = 2.3in, height = 1.5in]{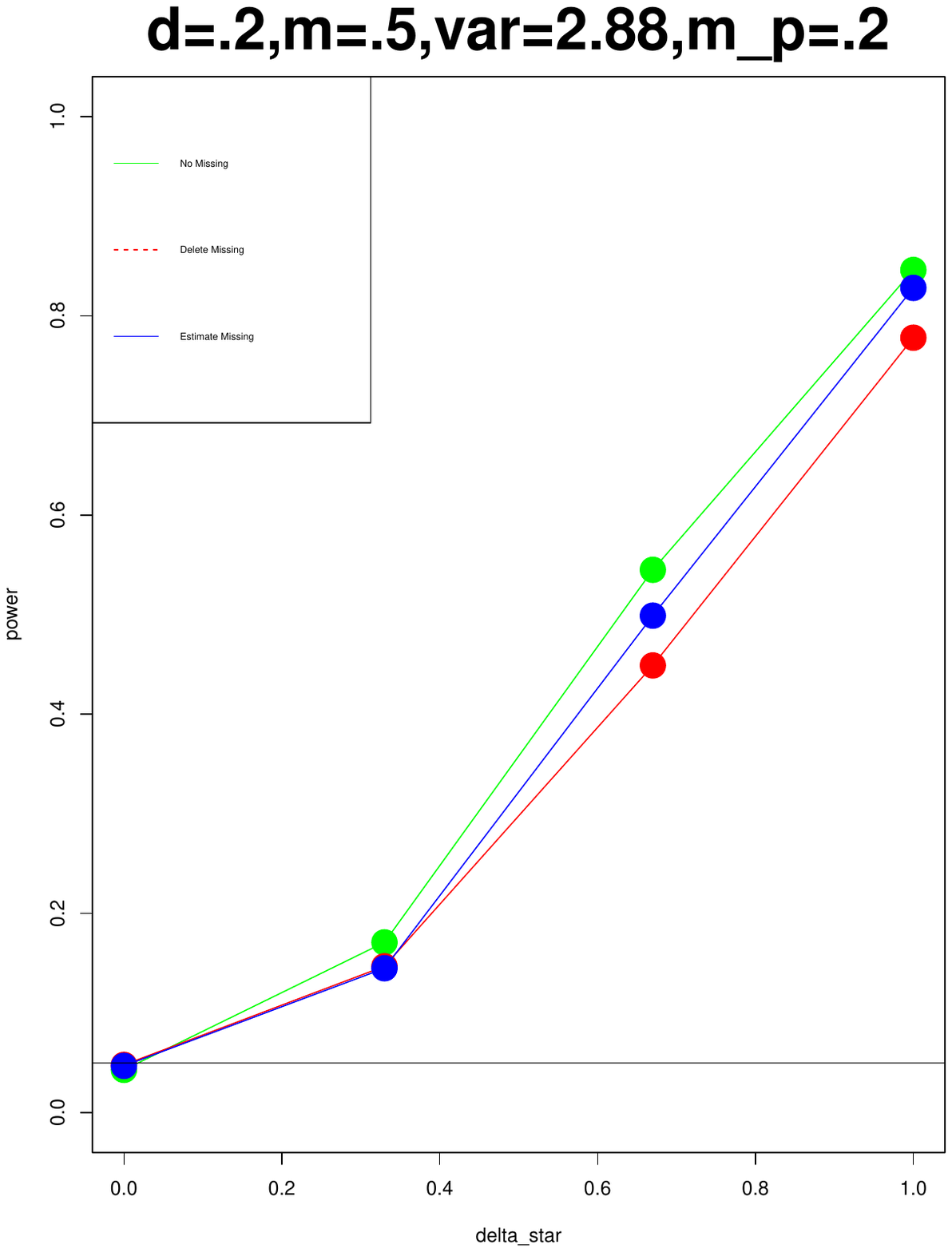}
\includegraphics[width = 2.3in, height = 1.5in]{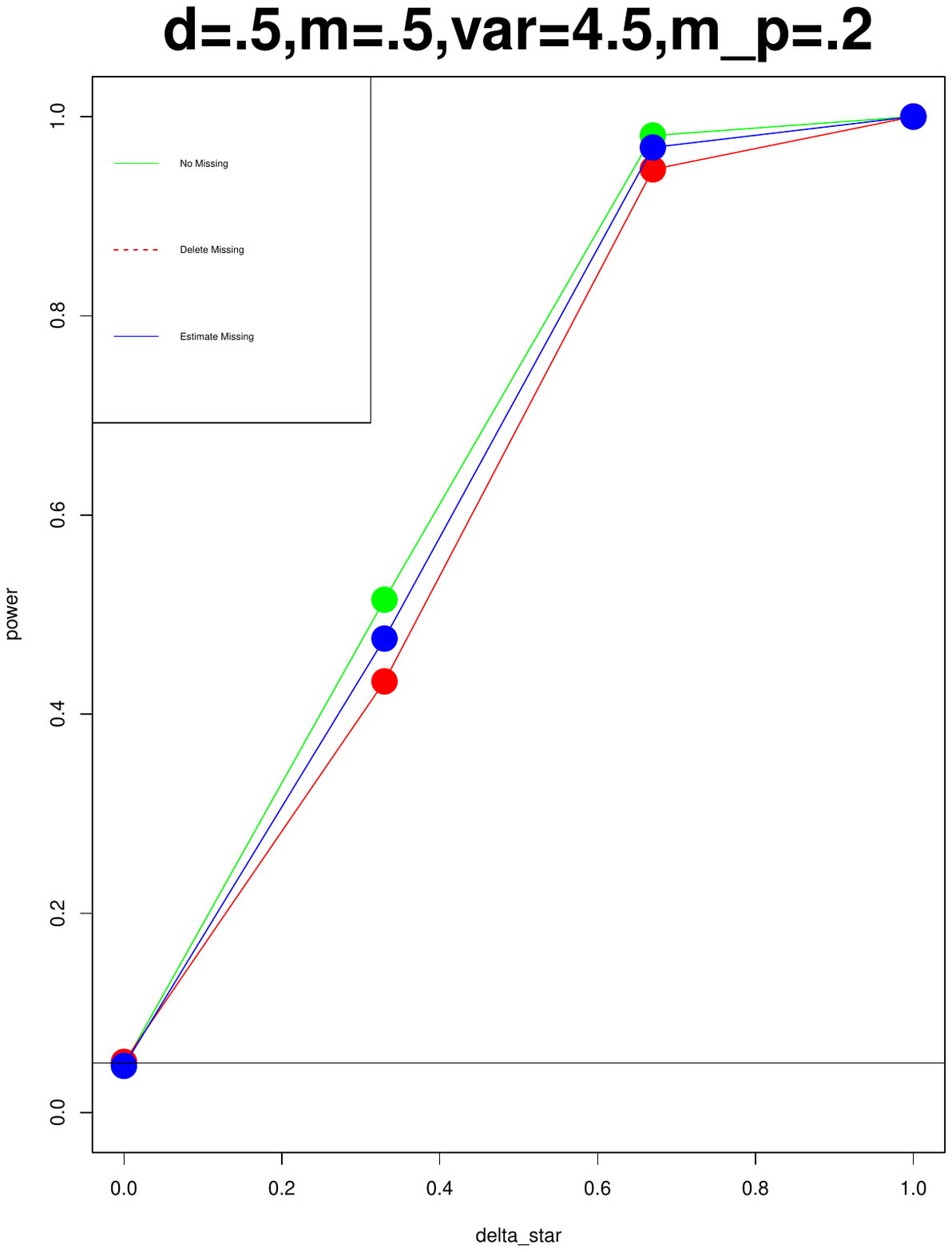}

 $m = .1 \quad miss_p = .3$

\includegraphics[width = 2.3in, height = 1.5in]{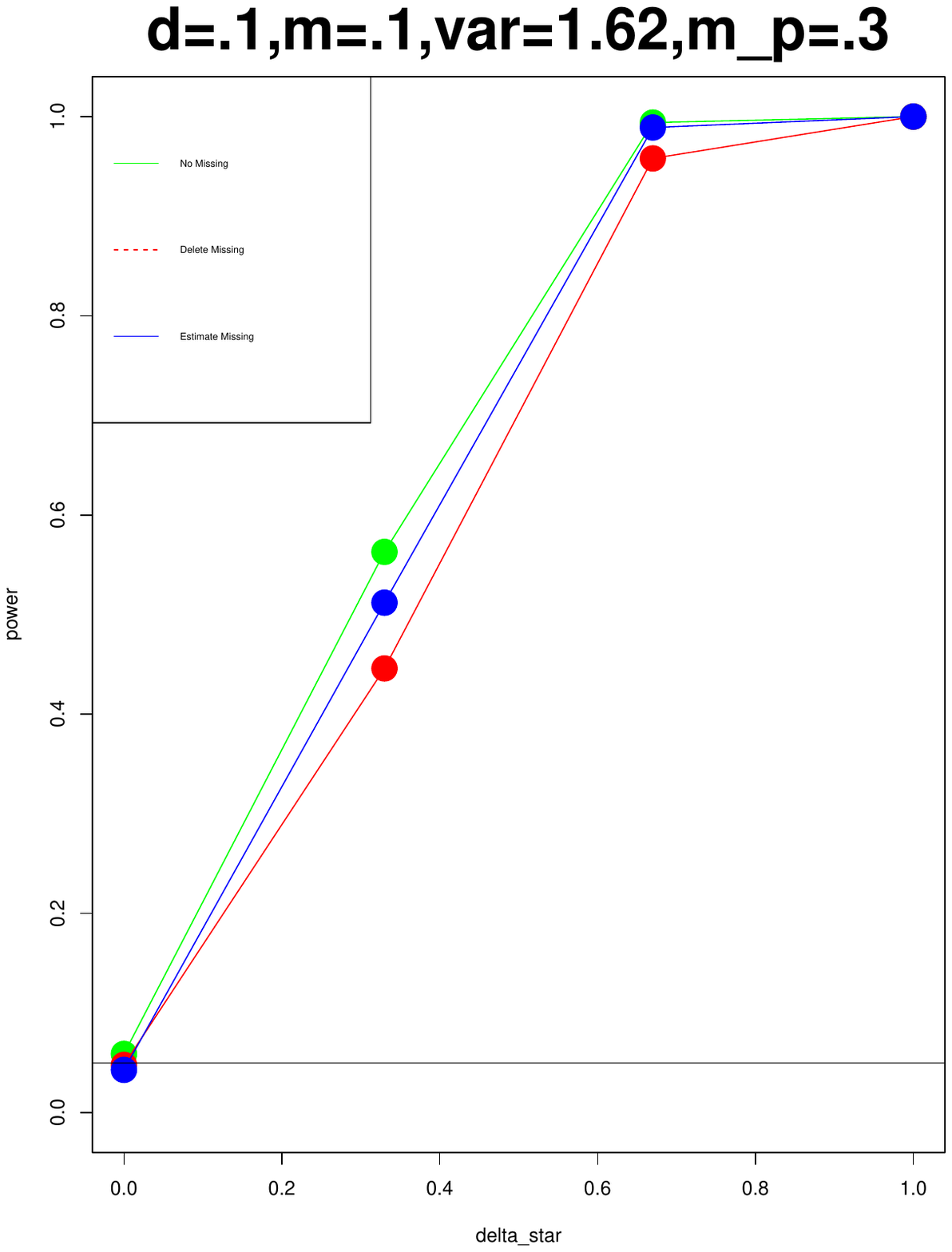}
\includegraphics[width = 2.3in, height = 1.5in]{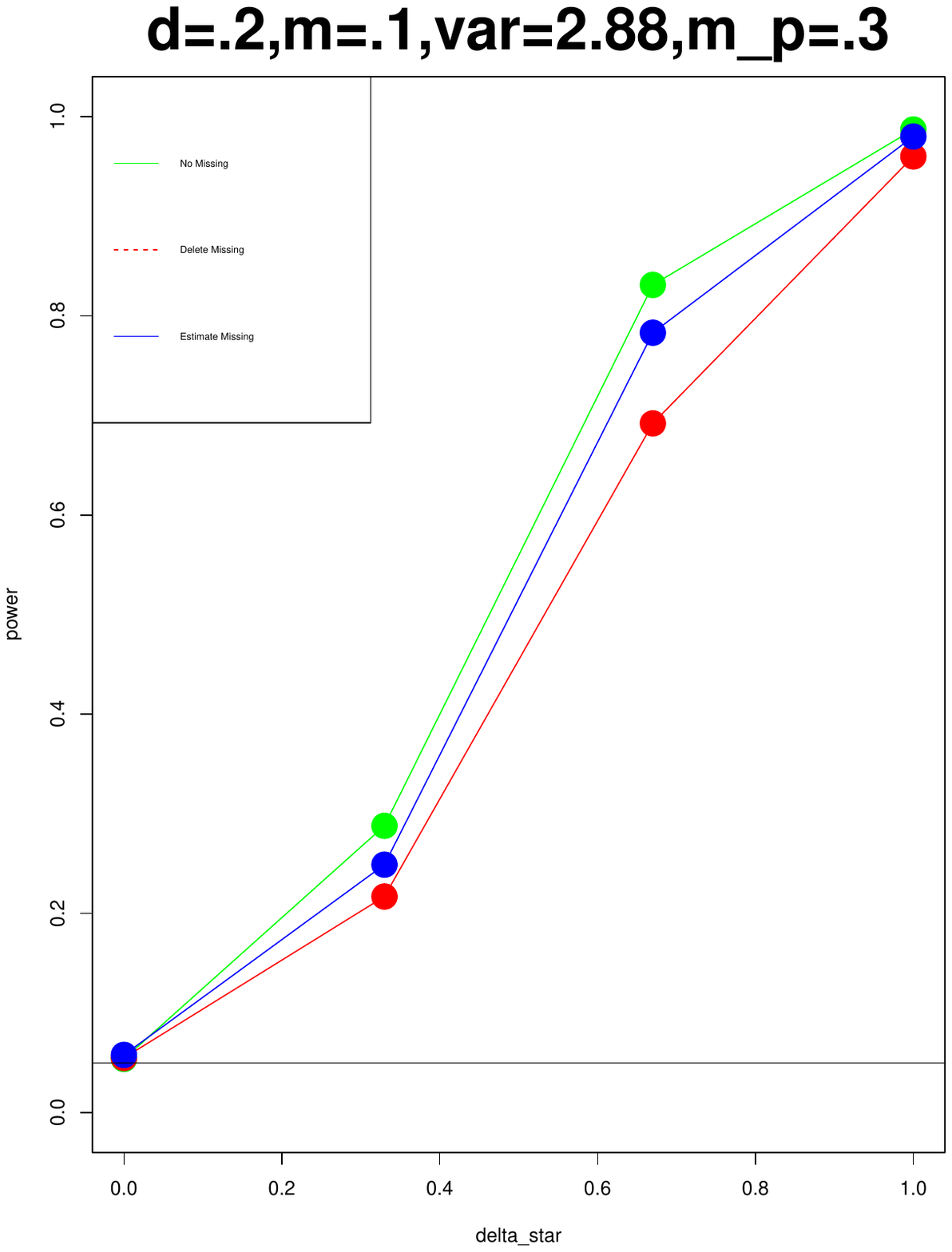}
\includegraphics[width = 2.3in, height = 1.5in]{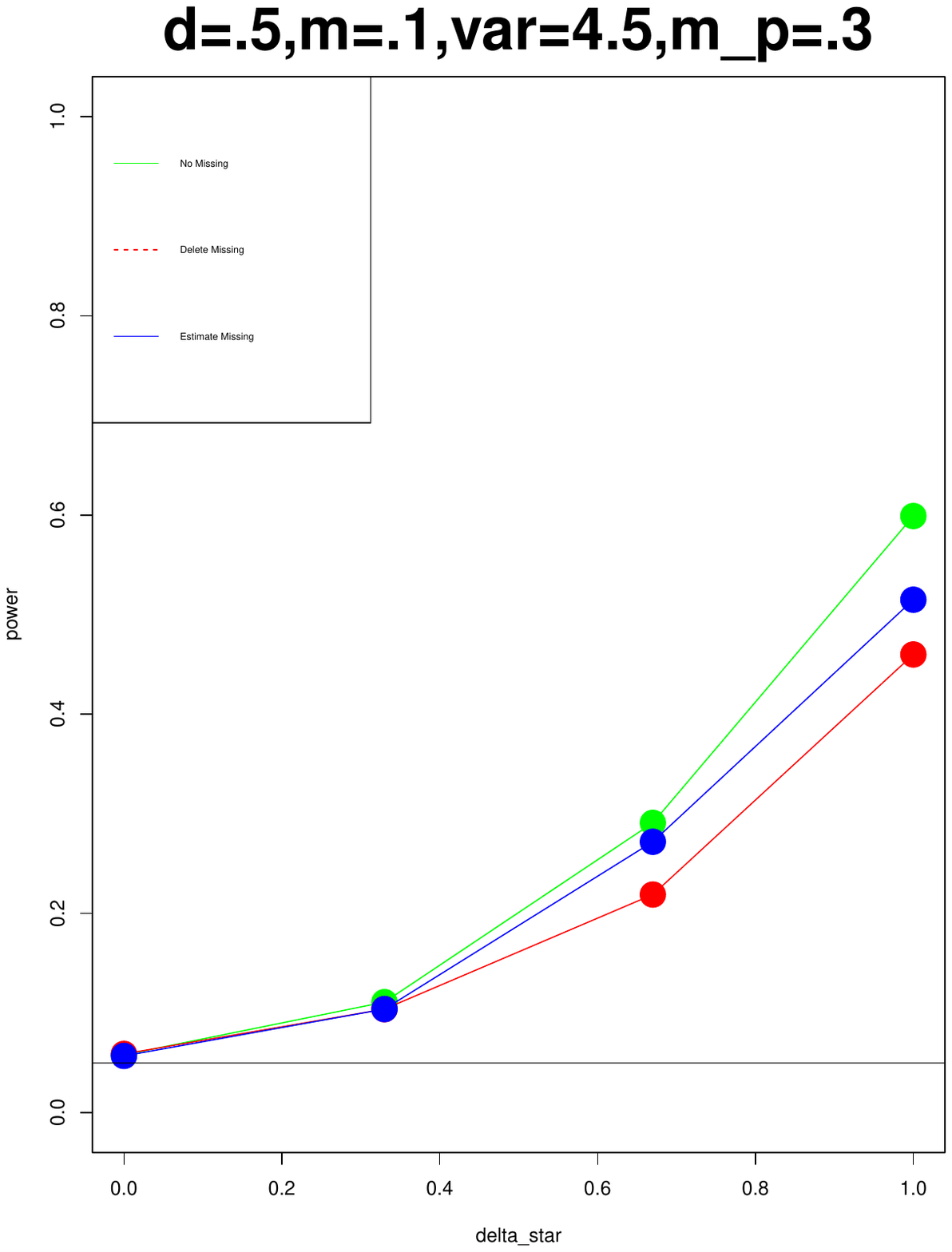}

 $m = .5 \quad miss_p = .3$

\includegraphics[width = 2.3in, height = 1.5in]{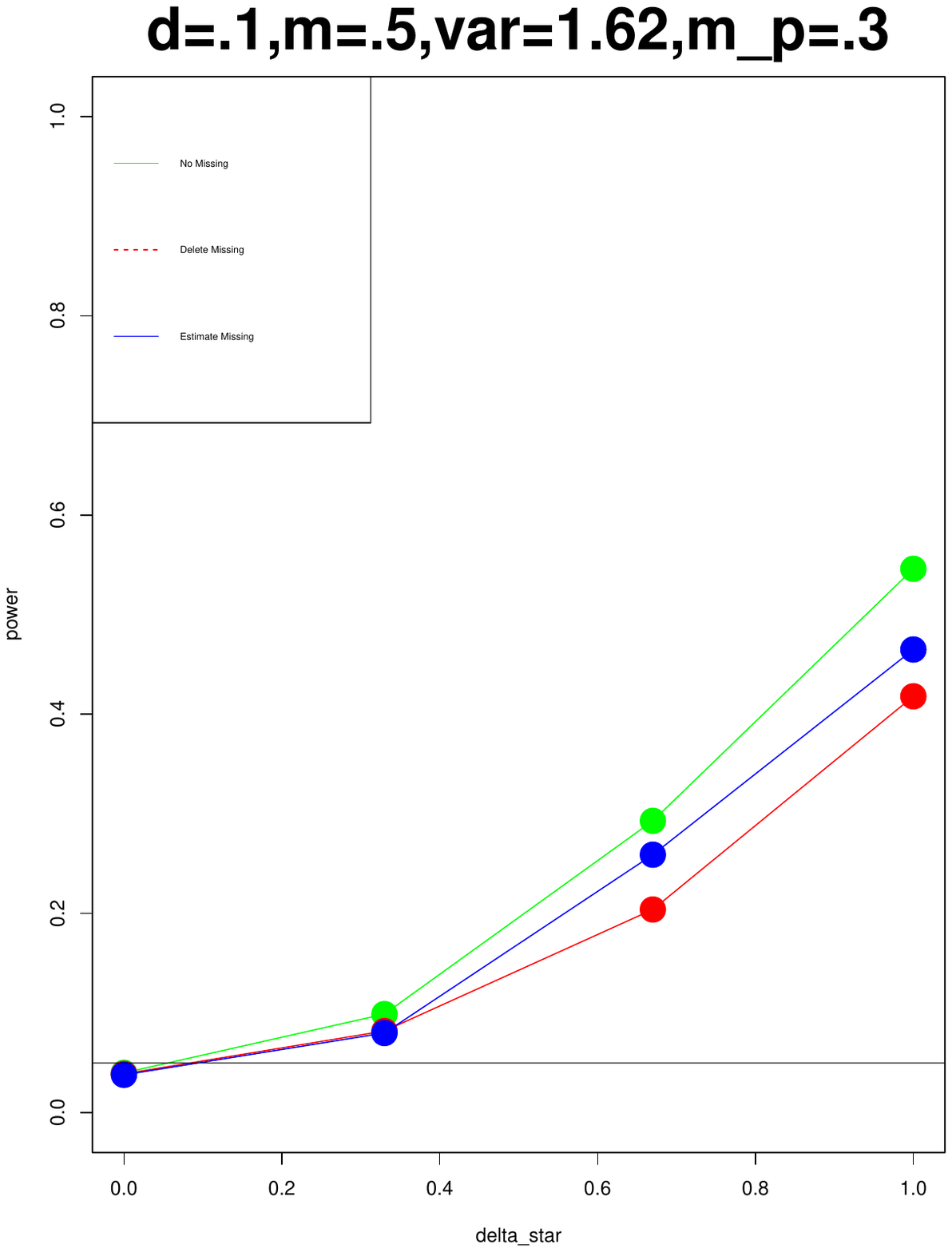}
\includegraphics[width = 2.3in, height = 1.5in]{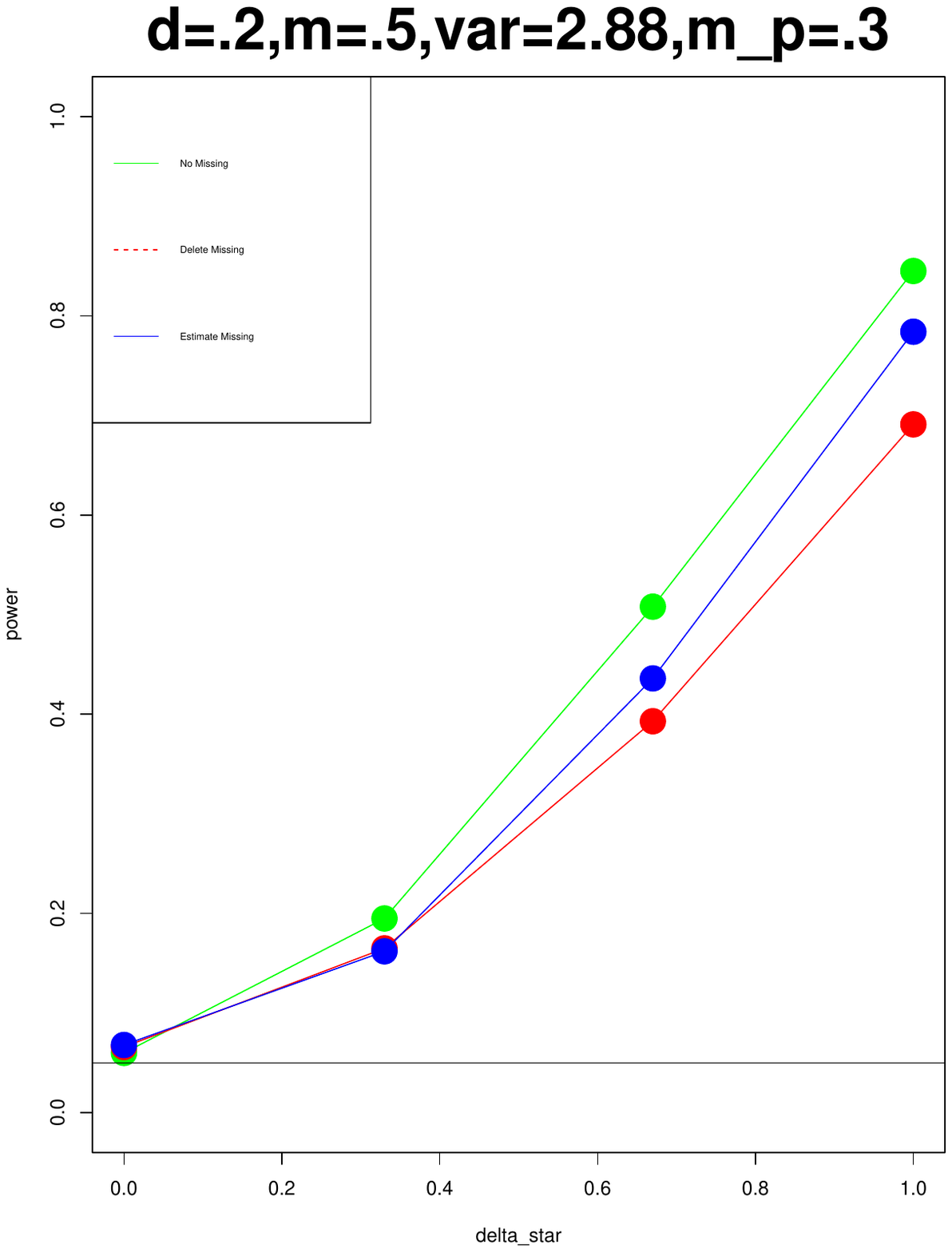}
\includegraphics[width = 2.3in, height = 1.5in]{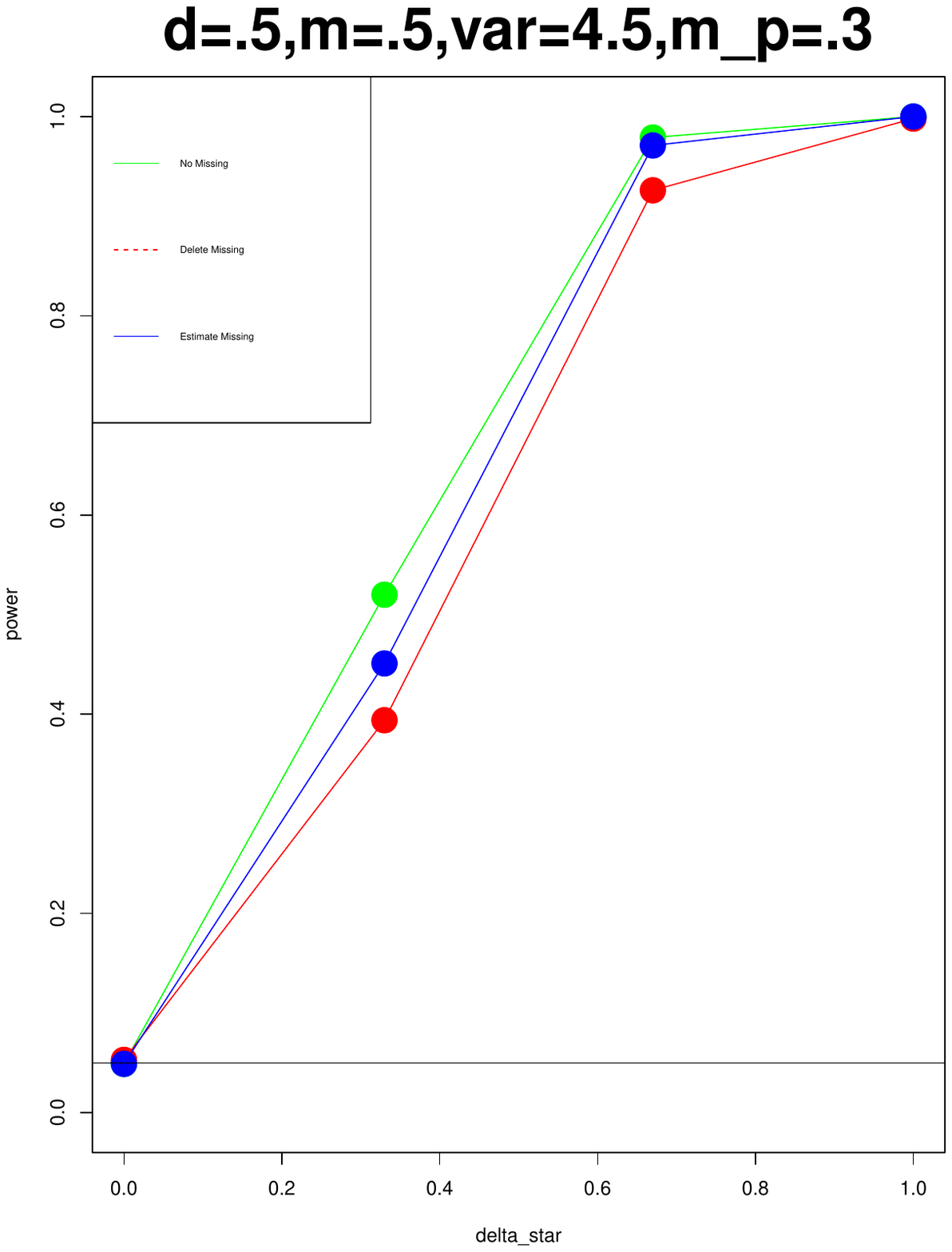}
\\

\end{center}

Now we shall change $\beta$ and adjust others such that $p^\star$ is close to .1. 

\begin{center}
\begin{tabular}{|c|c|c|c|}
\hline 
$d$ & .1 & .2 & .5 \\ 
\hline 
$var$ & 6.48 & 11.5 & 18 \\ 
\hline 
\end{tabular} \hspace{2cm} {\large $\beta = 2$}
\end{center}

\begin{center}

 $m = .1 \quad miss_p = .2$

\includegraphics[width = 2.3in, height = 1.5in]{n7}
\includegraphics[width = 2.3in, height = 1.5in]{n11}
\includegraphics[width = 2.3in, height = 1.5in]{n9}

 $m = .5 \quad miss_p = .2$

\includegraphics[width = 2.3in, height = 1.5in]{n8}
\includegraphics[width = 2.3in, height = 1.5in]{n12}
\includegraphics[width = 2.3in, height = 1.5in]{n10}

 $m = .1 \quad miss_p = .3$

\includegraphics[width = 2.3in, height = 1.5in]{n13}
\includegraphics[width = 2.3in, height = 1.5in]{n17}
\includegraphics[width = 2.3in, height = 1.5in]{n15}

 $m = .5 \quad miss_p = .3$

\includegraphics[width = 2.3in, height = 1.5in]{n14}
\includegraphics[width = 2.3in, height = 1.5in]{n18}
\includegraphics[width = 2.3in, height = 1.5in]{n16}
\\

\end{center}

\subsubsection{When Distribution of Trait is Continuous \& Skewed ( Chi Square )}

Here traits comes from the skewed distribution so we have done a bit of modification in our imputation method. We use logarithm transformation for all the traits and we can think transformed traits comes from a symmetric distribution and use the same methods to estimate the missing ones and we perform exponential transformation on these estimated trait to get the actual estimate.\\

Here we vary $d$ and $df$ such a way that $p^\star$ is close to .1. Notice that m has no effect on $p^\star$, so we take two values of $m$, .1 and .5 and for each $m$ we vary $d$ and $df$ in the following way-

\begin{center}
\begin{tabular}{|c|c|c|c|}
\hline 
$d$ & .1 & .2 & .3 \\ 
\hline 
$df$ & .81 & 1.44 & 2.25 \\ 
\hline 
\end{tabular}

\end{center}

\begin{center}

 $m = .1 \quad miss_p = .2$

\includegraphics[width = 2.3in, height = 1.5in]{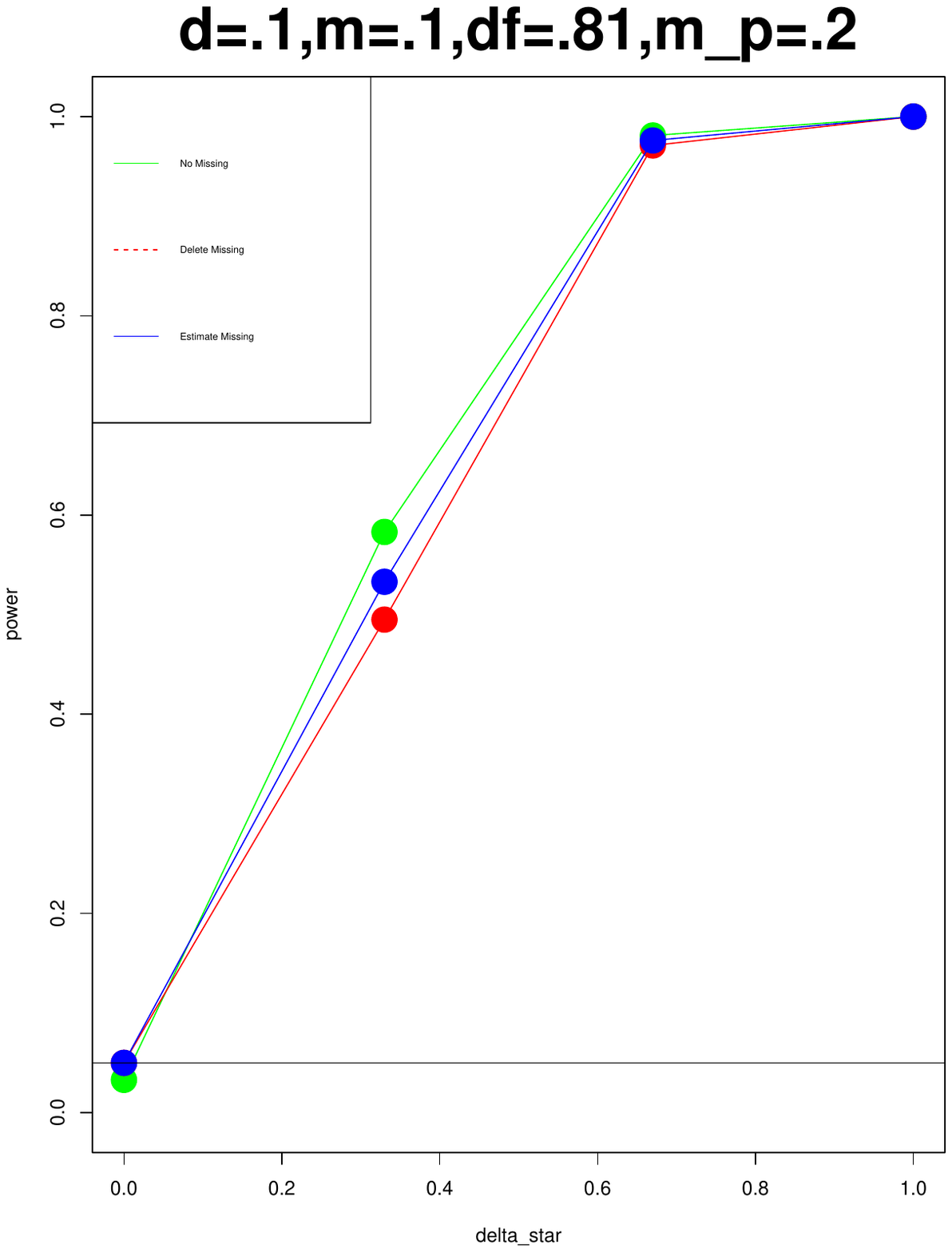}
\includegraphics[width = 2.3in, height = 1.5in]{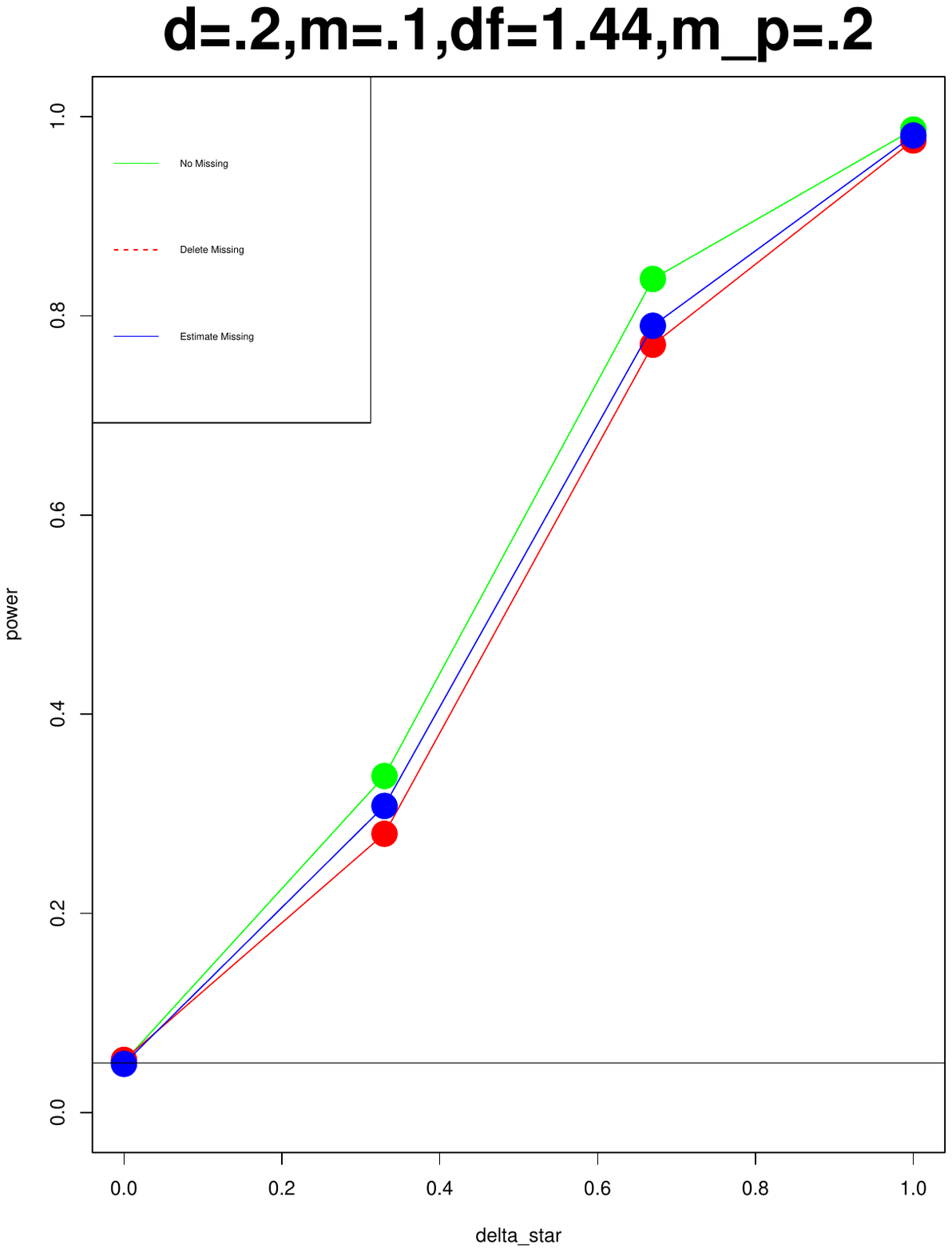}
\includegraphics[width = 2.3in, height = 1.5in]{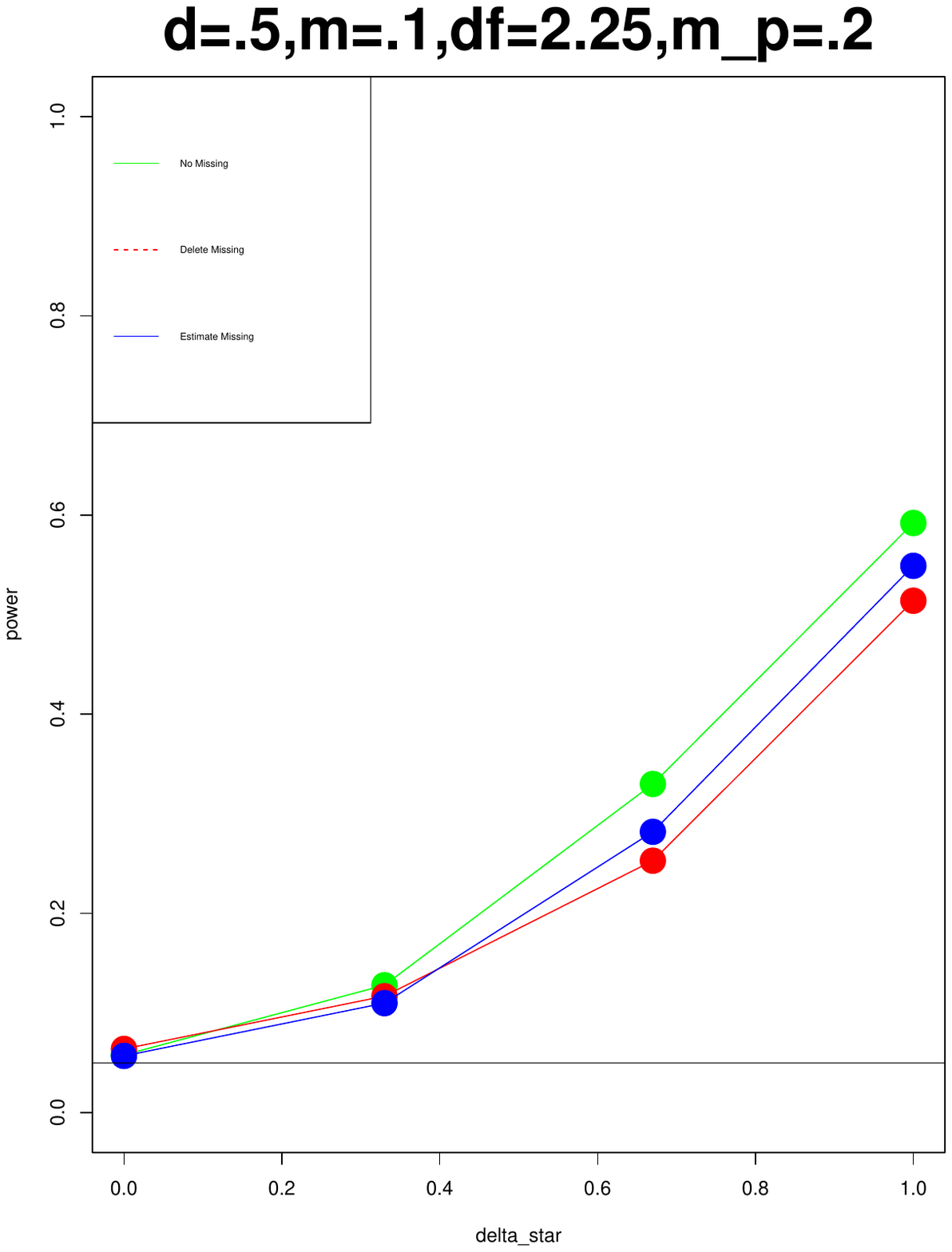}

 $m = .5 \quad miss_p = .2$

\includegraphics[width = 2.3in, height = 1.5in]{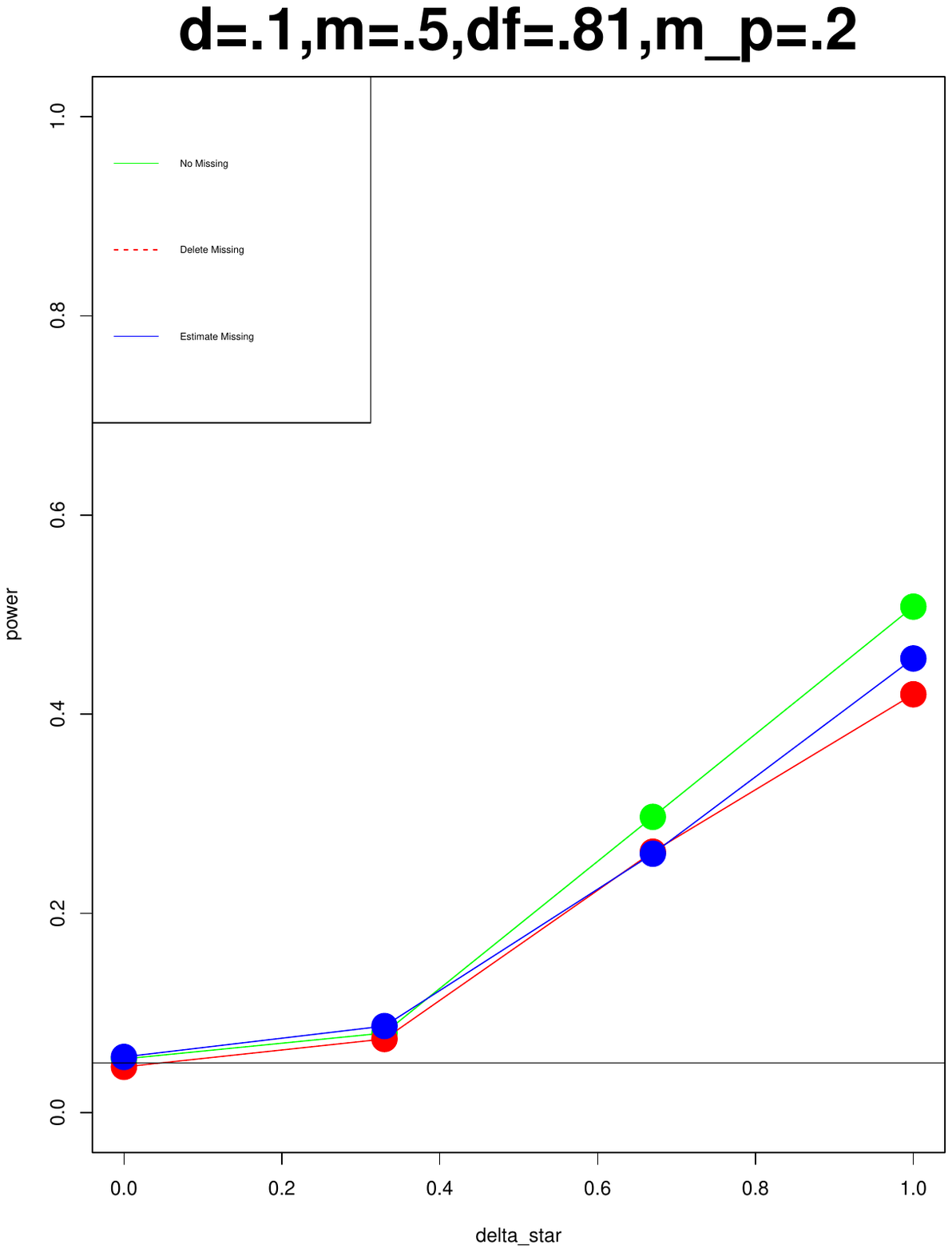}
\includegraphics[width = 2.3in, height = 1.5in]{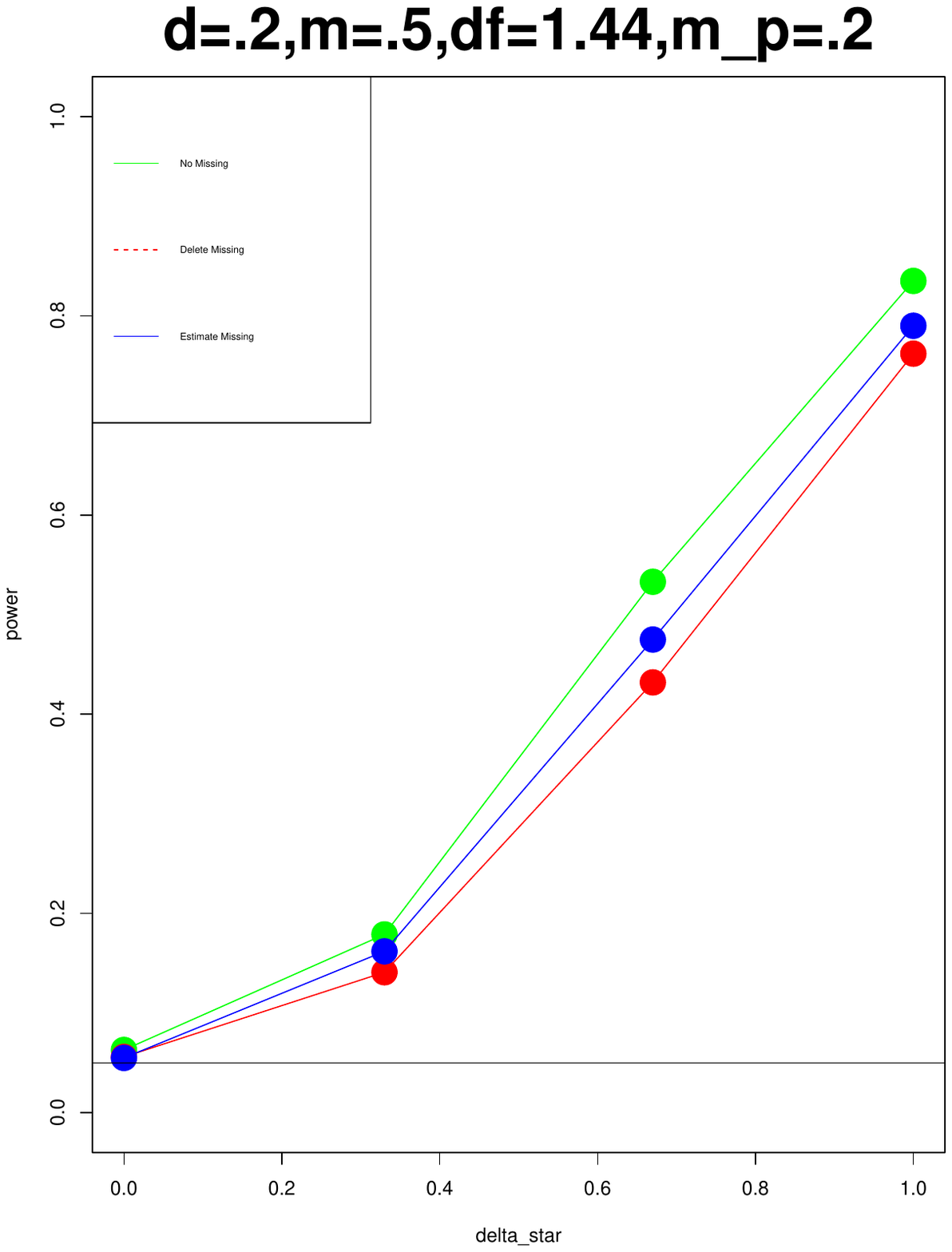}
\includegraphics[width = 2.3in, height = 1.5in]{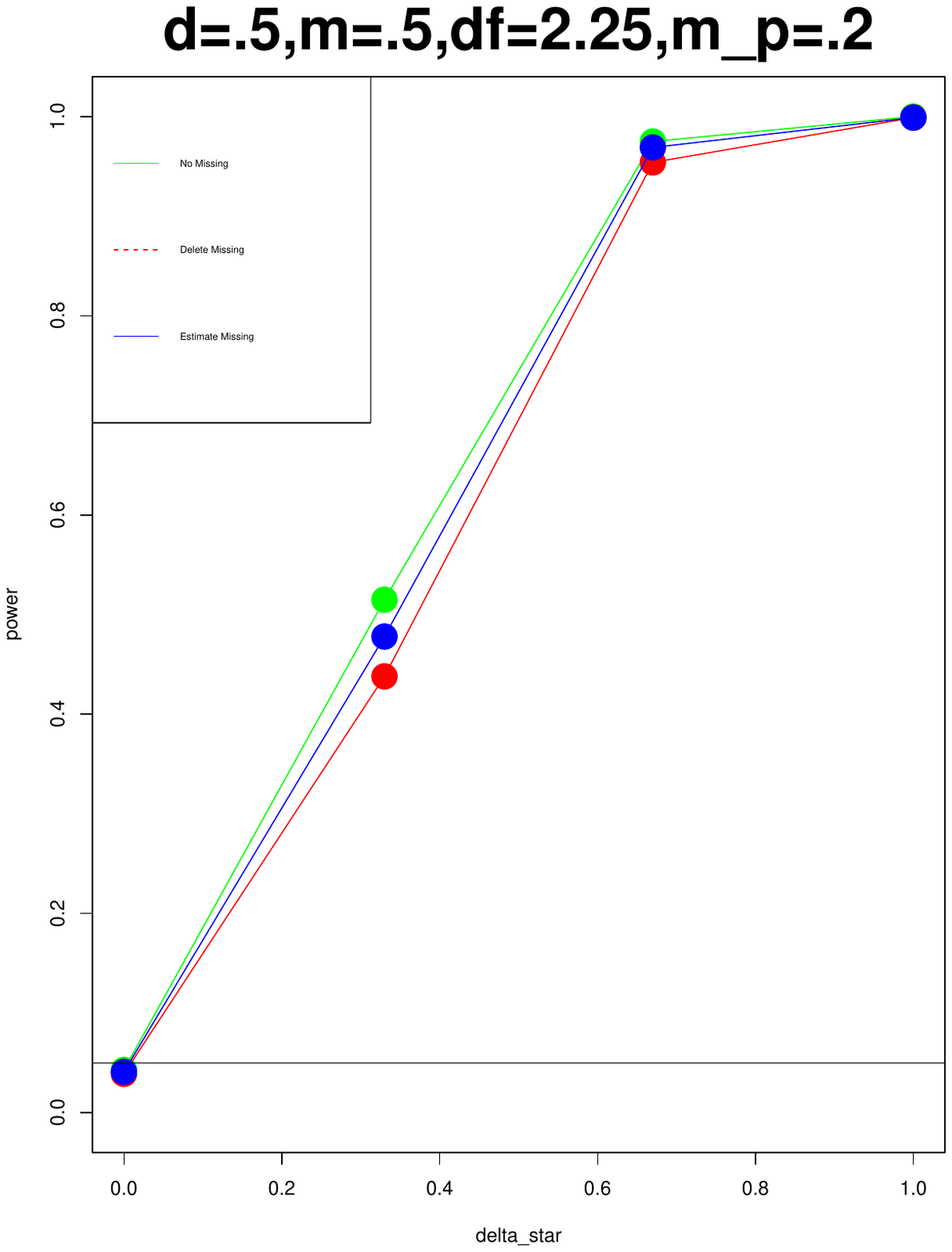}

 $m = .1 \quad miss_p = .3$

\includegraphics[width = 2.3in, height = 1.5in]{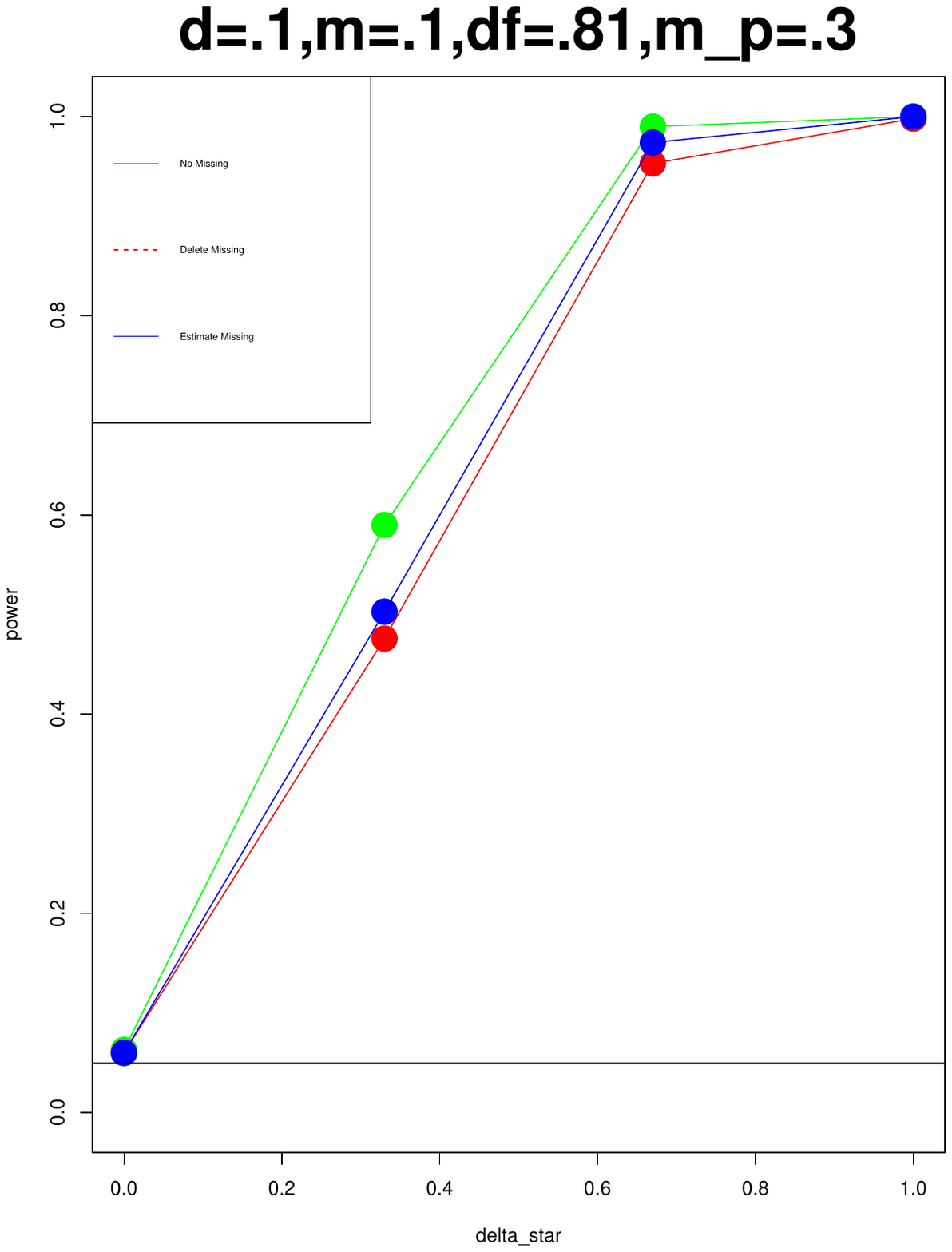}
\includegraphics[width = 2.3in, height = 1.5in]{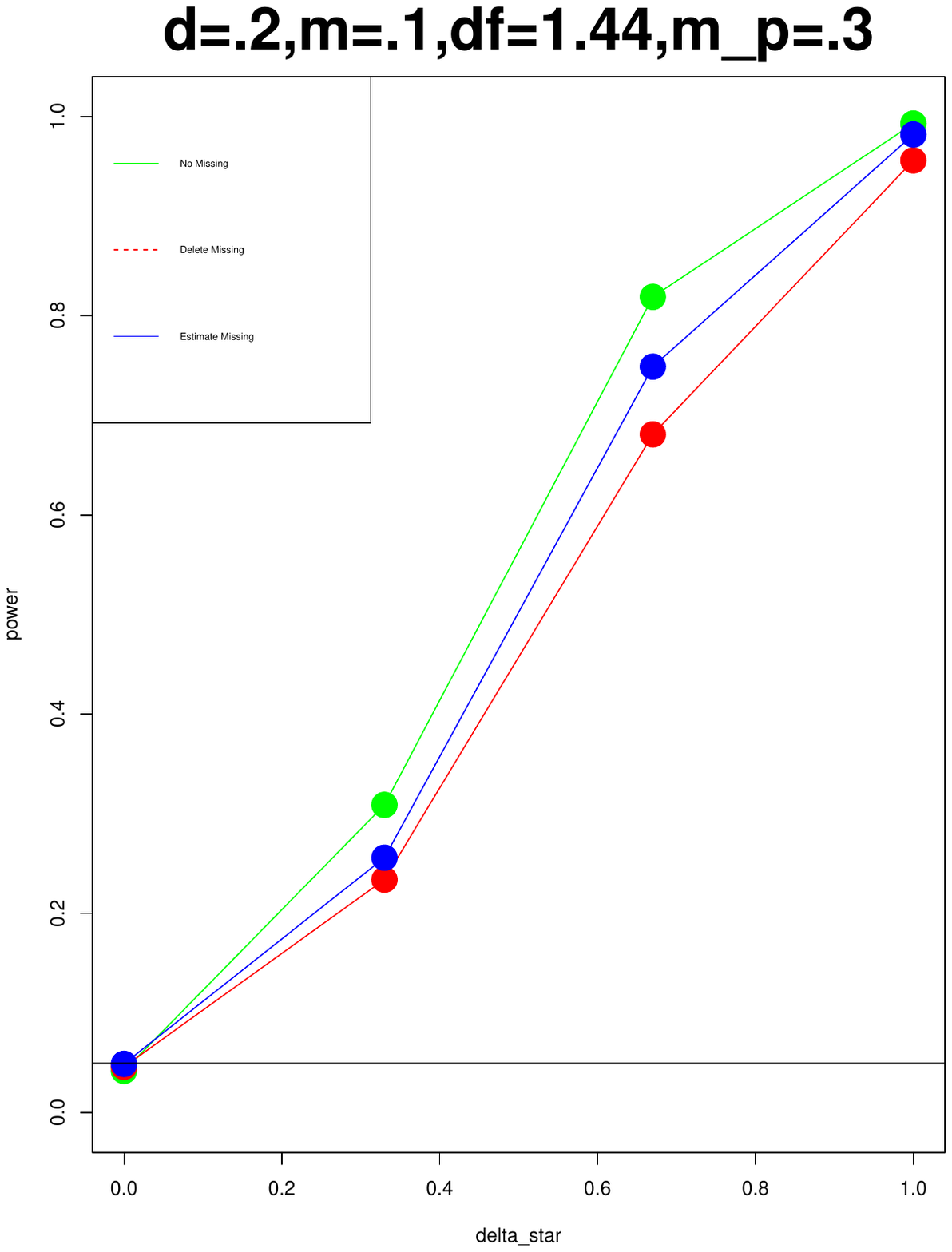}
\includegraphics[width = 2.3in, height = 1.5in]{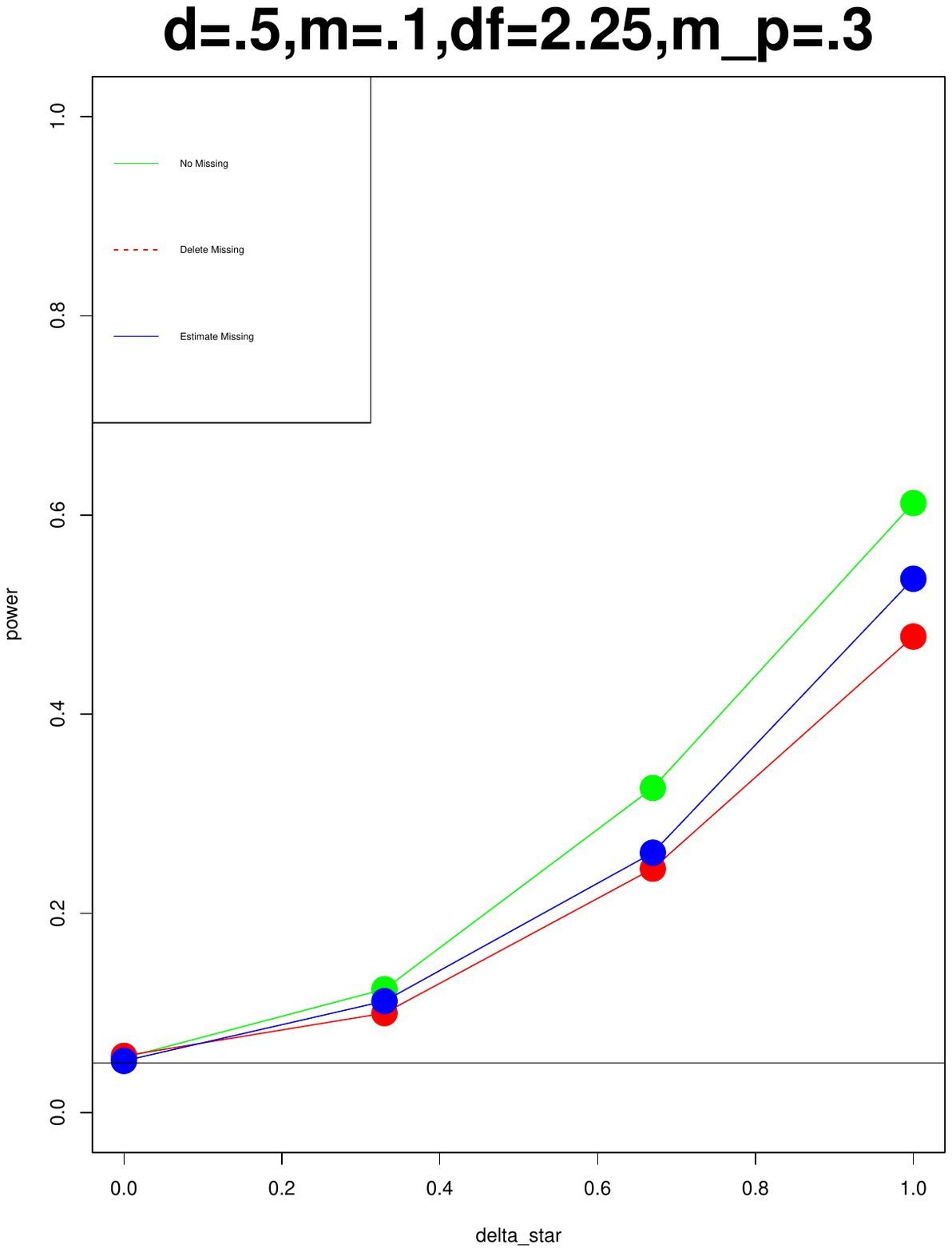}

 $m = .5 \quad miss_p = .3$

\includegraphics[width = 2.3in, height = 1.5in]{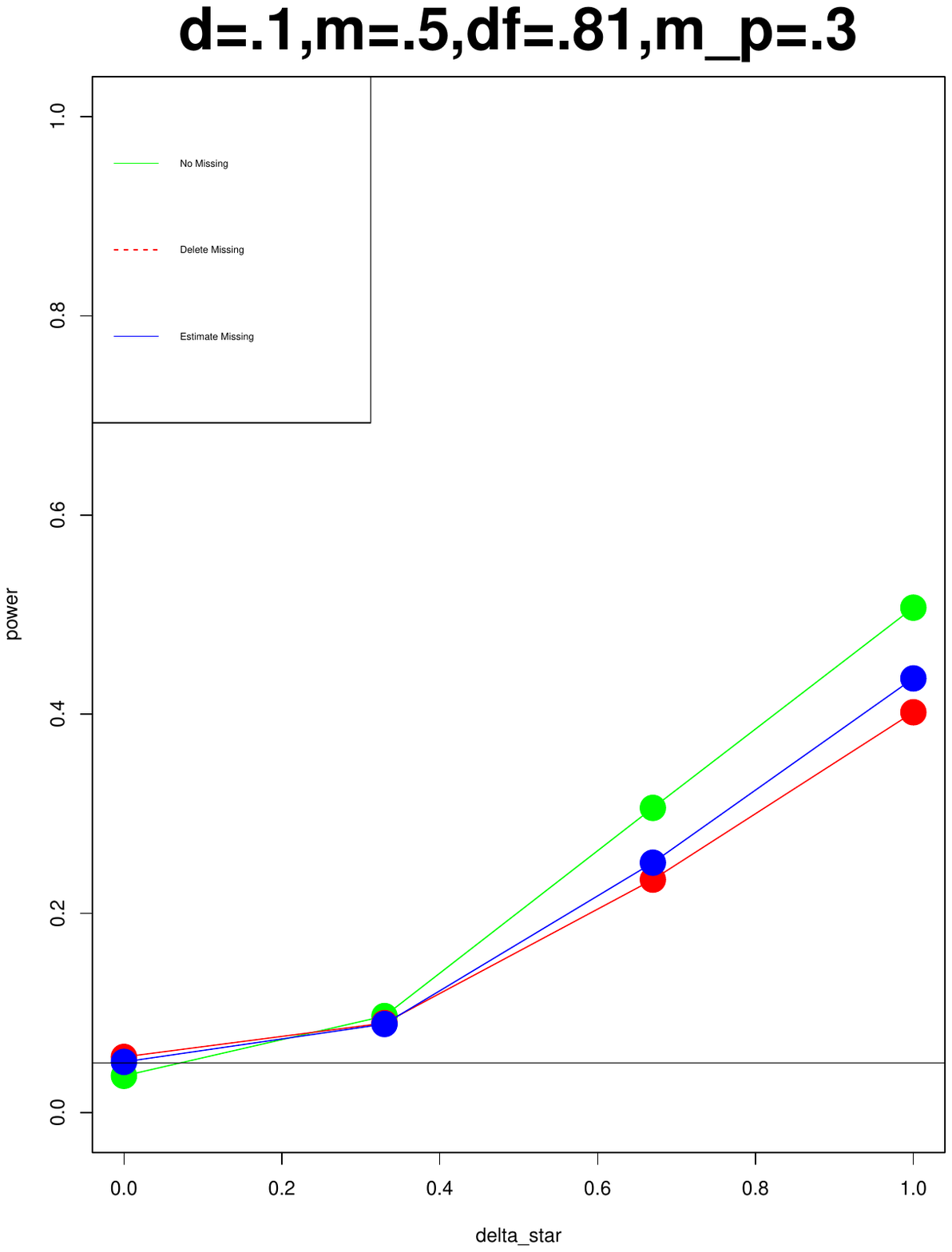}
\includegraphics[width = 2.3in, height = 1.5in]{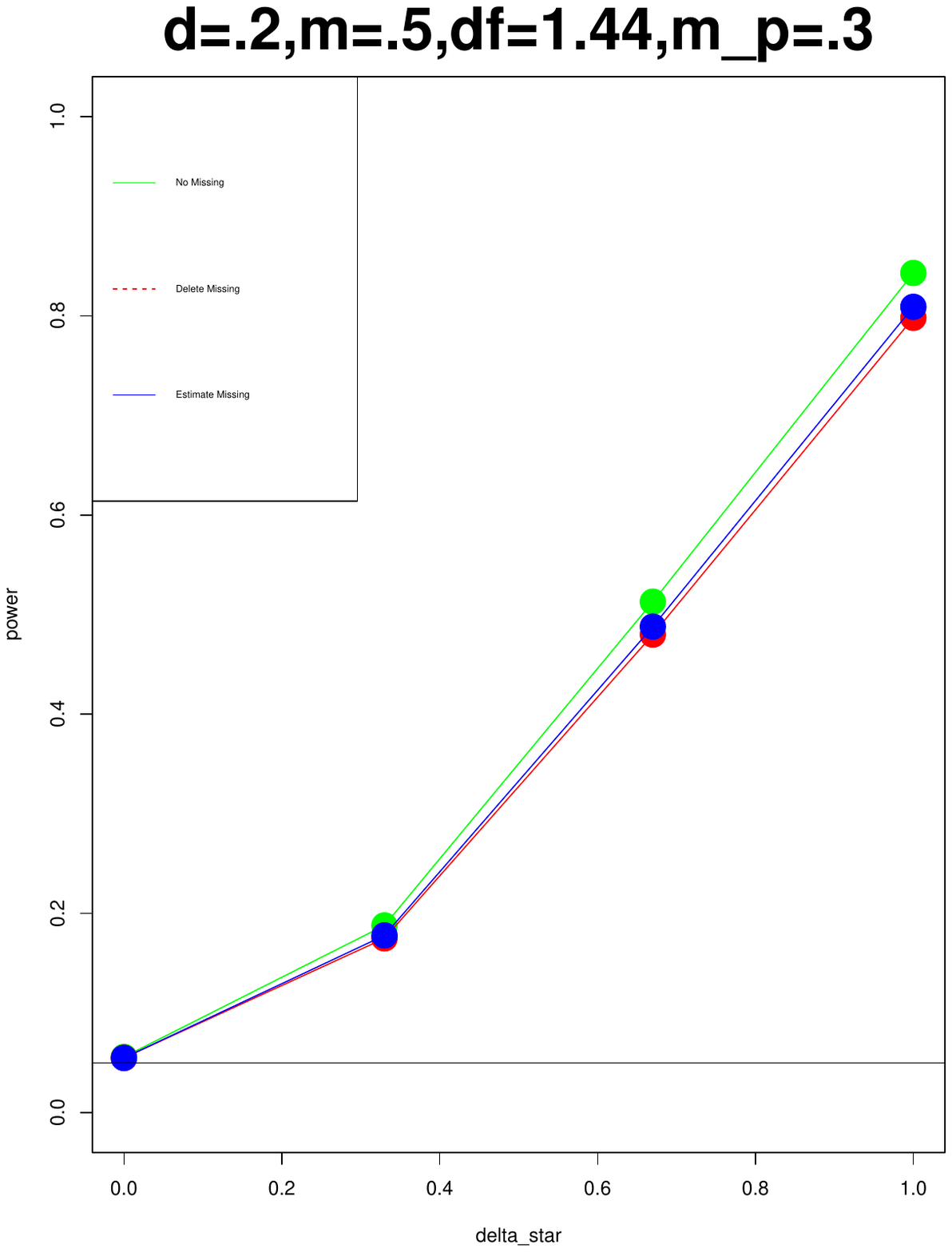}
\includegraphics[width = 2.3in, height = 1.5in]{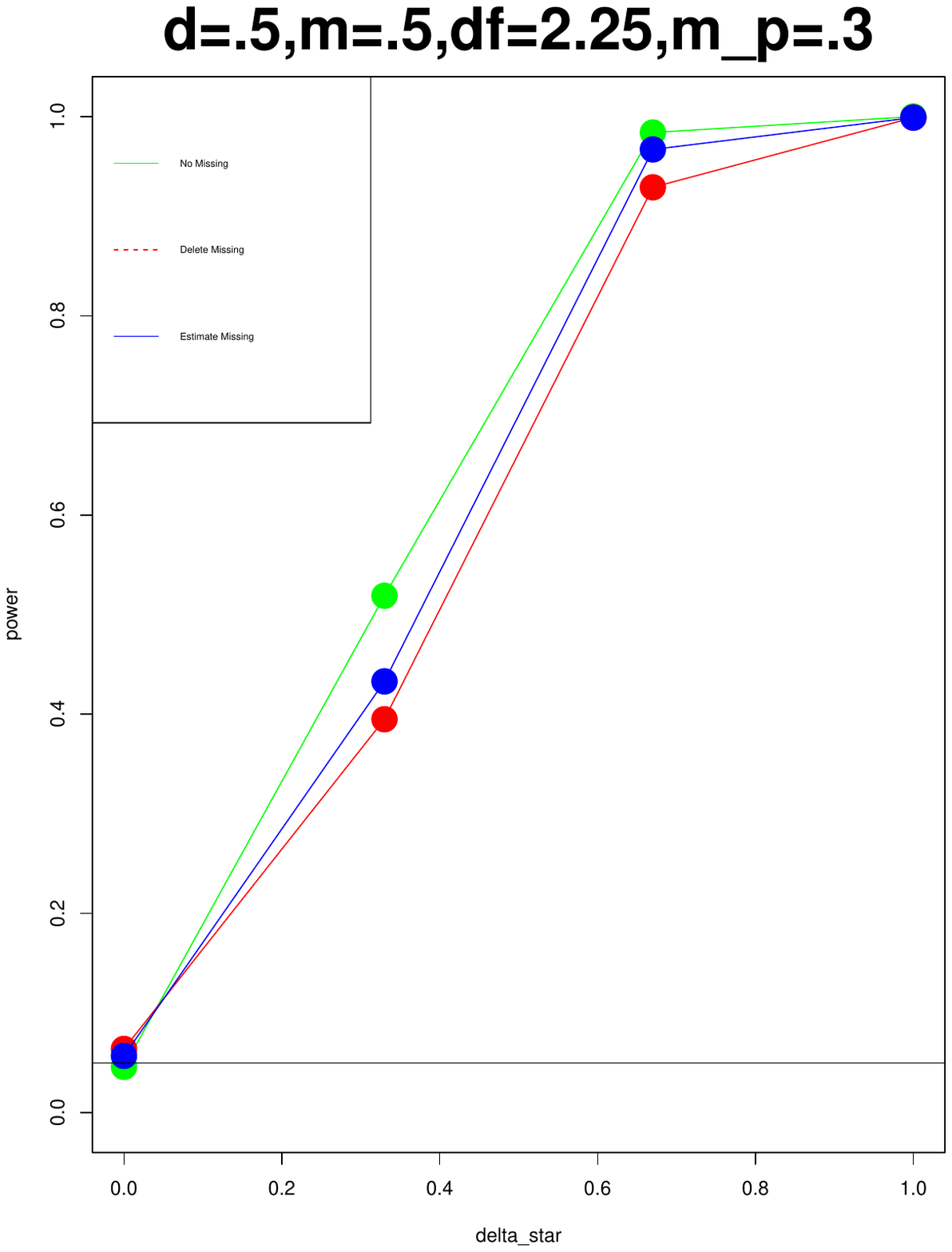}

\end{center}

\subsubsection{When Distribution of Trait is Count Type }

Here we vary $d$ and $\lambda$ such a way that $p^\star$ is close to .1. Notice that m has no effect on $p^\star$, so we take two values of $m$, .1 and .5 and for each $m$ we vary $d$ and $\lambda$ in the following way-

\begin{center}
\begin{tabular}{|c|c|c|c|}
\hline 
$d$ & .1 & .2 & .5 \\ 
\hline 
$\lambda$ & 1.62 & 2.88 & 4.5 \\ 
\hline 
\end{tabular}

\end{center}

\begin{center}

 $m = .1 \quad miss_p = .2$

\includegraphics[width = 2.3in, height = 1.5in]{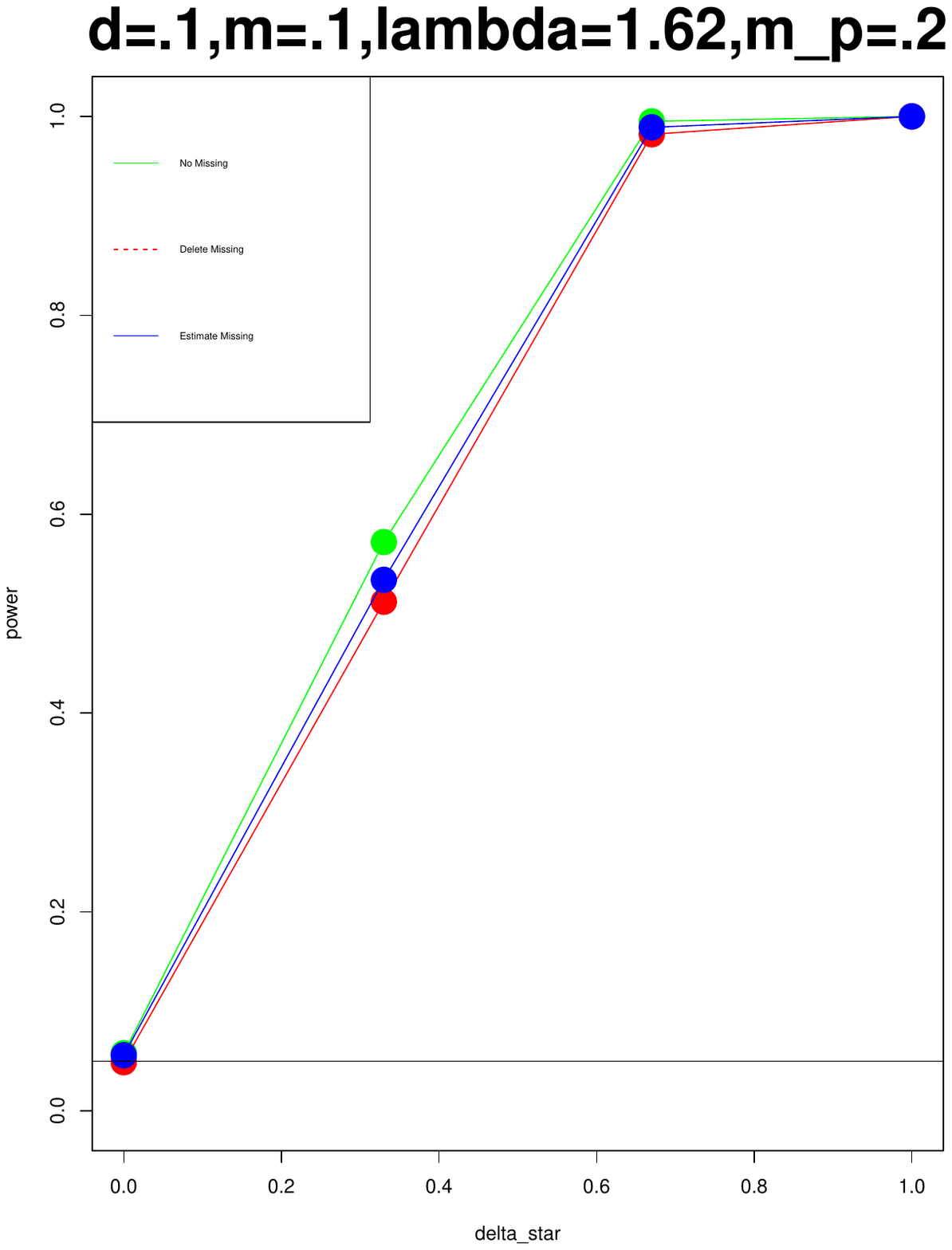}
\includegraphics[width = 2.3in, height = 1.5in]{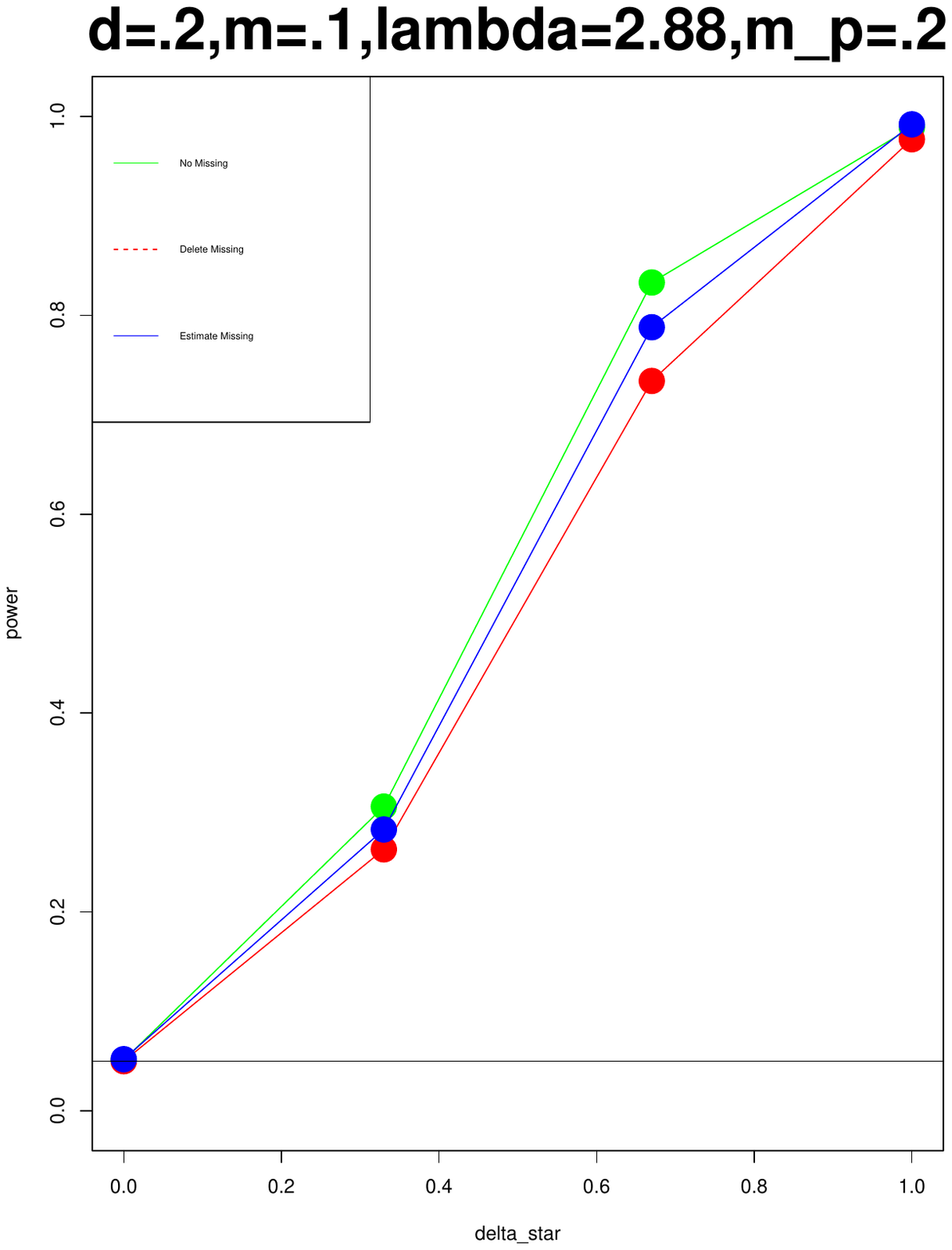}
\includegraphics[width = 2.3in, height = 1.5in]{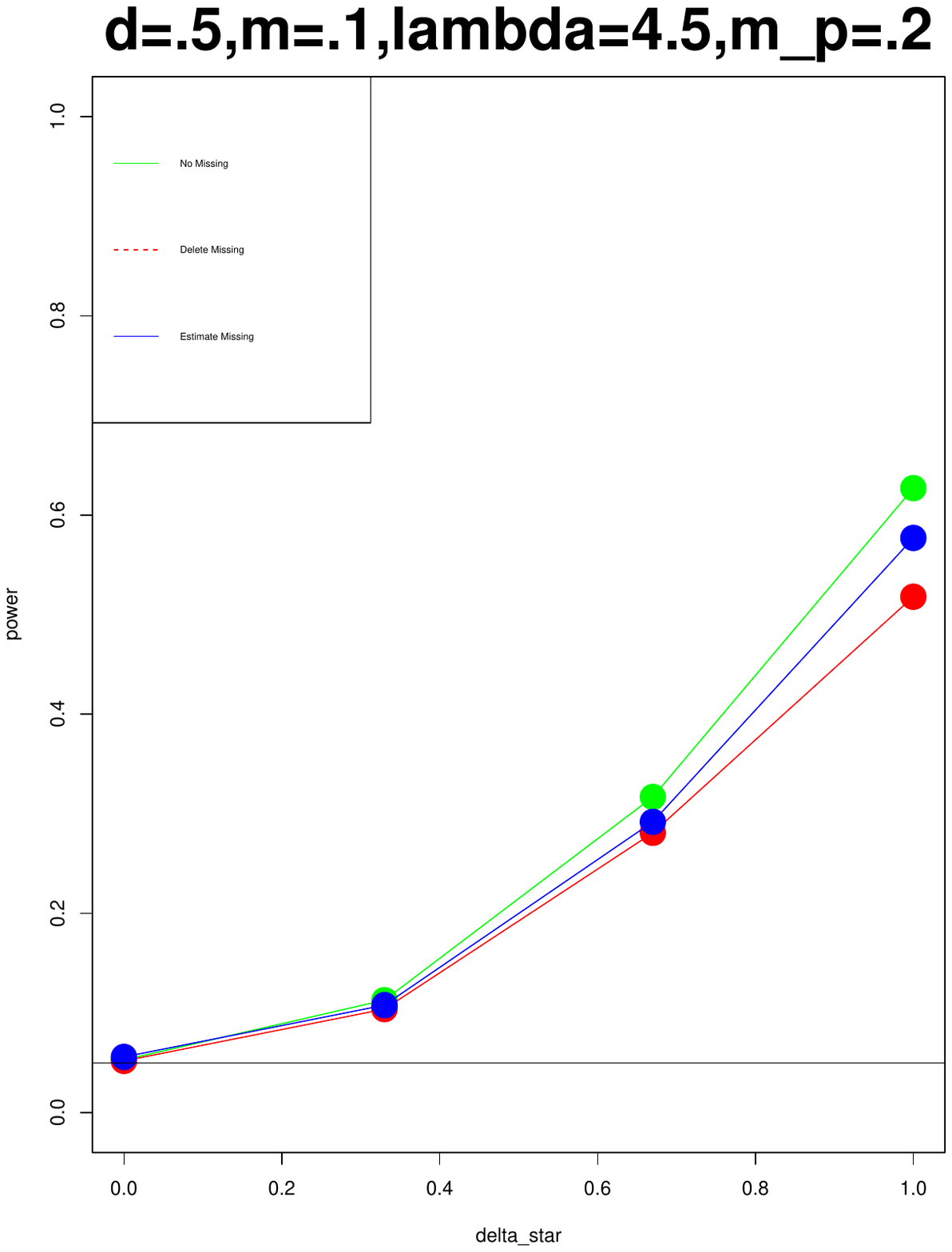}

 $m = .5 \quad miss_p = .2$

\includegraphics[width = 2.3in, height = 1.5in]{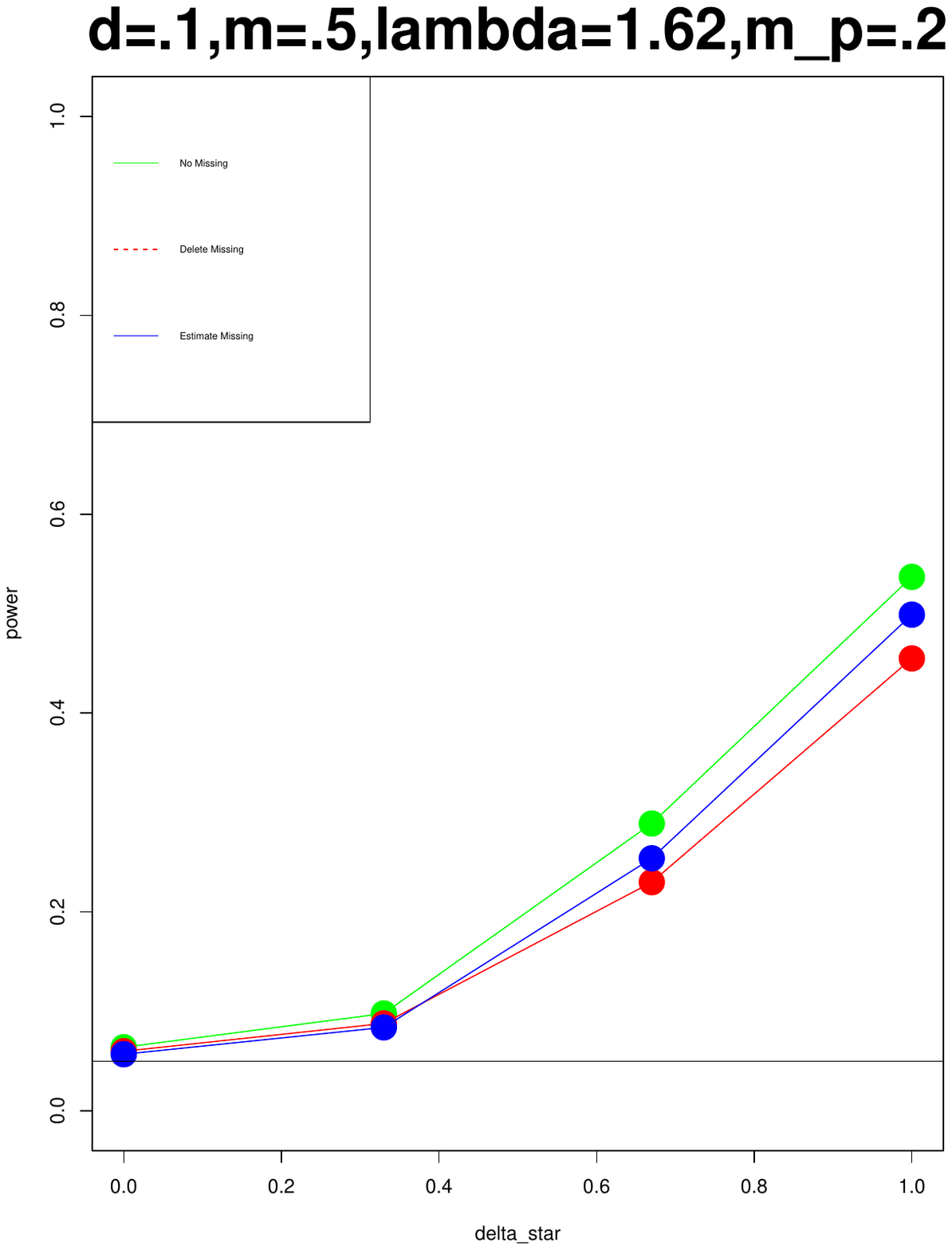}
\includegraphics[width = 2.3in, height = 1.5in]{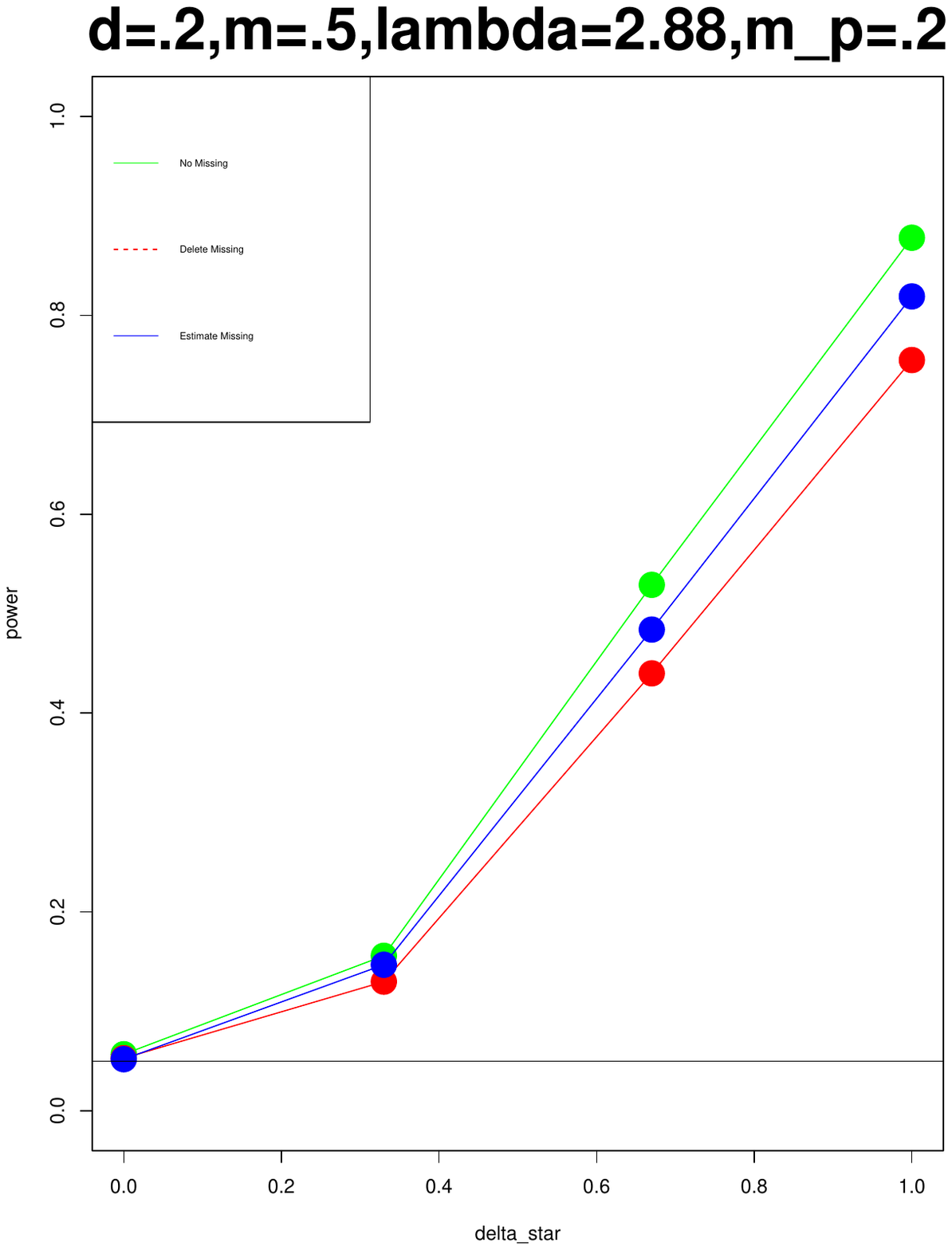}
\includegraphics[width = 2.3in, height = 1.5in]{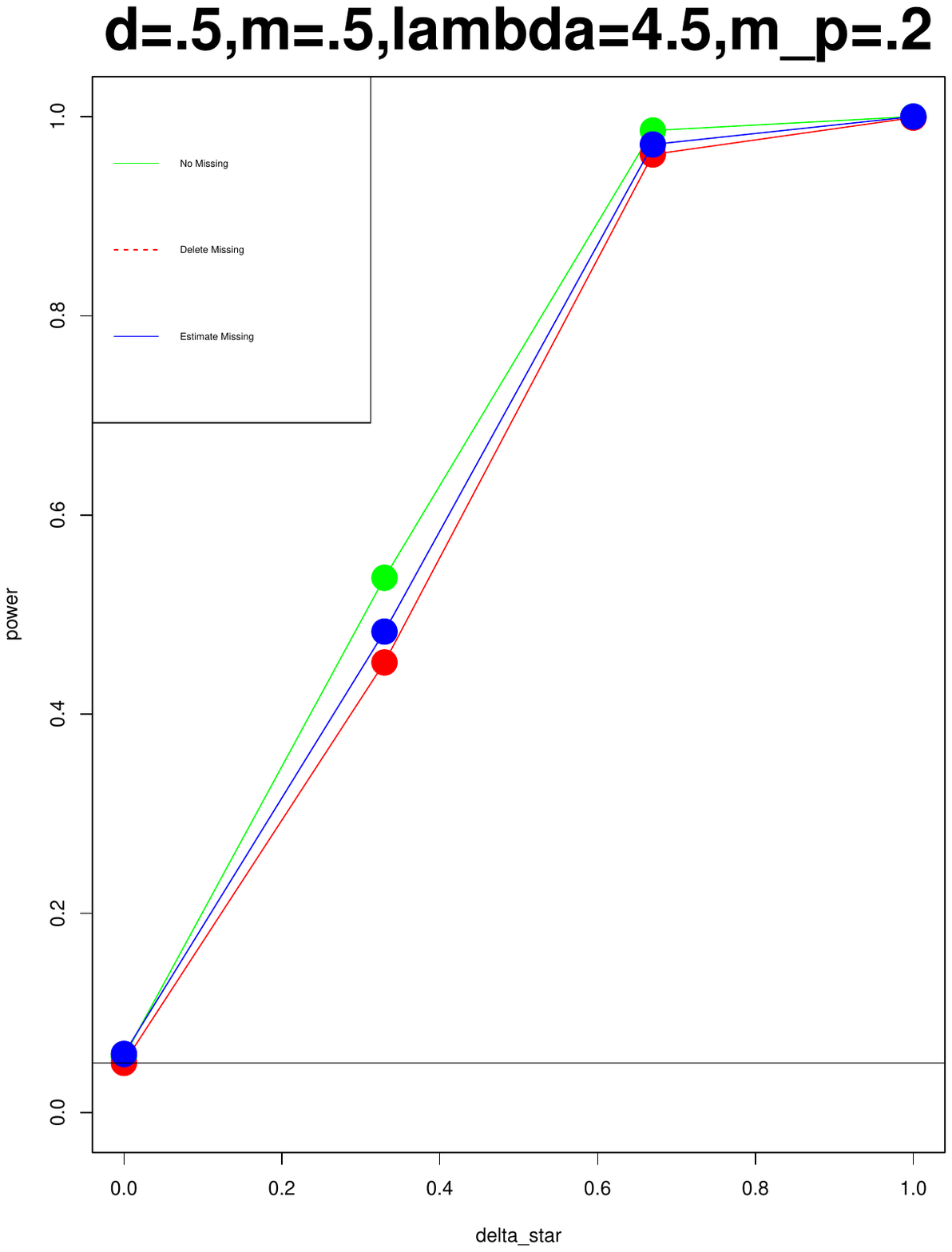}

 $m = .1 \quad miss_p = .3$

\includegraphics[width = 2.3in, height = 1.5in]{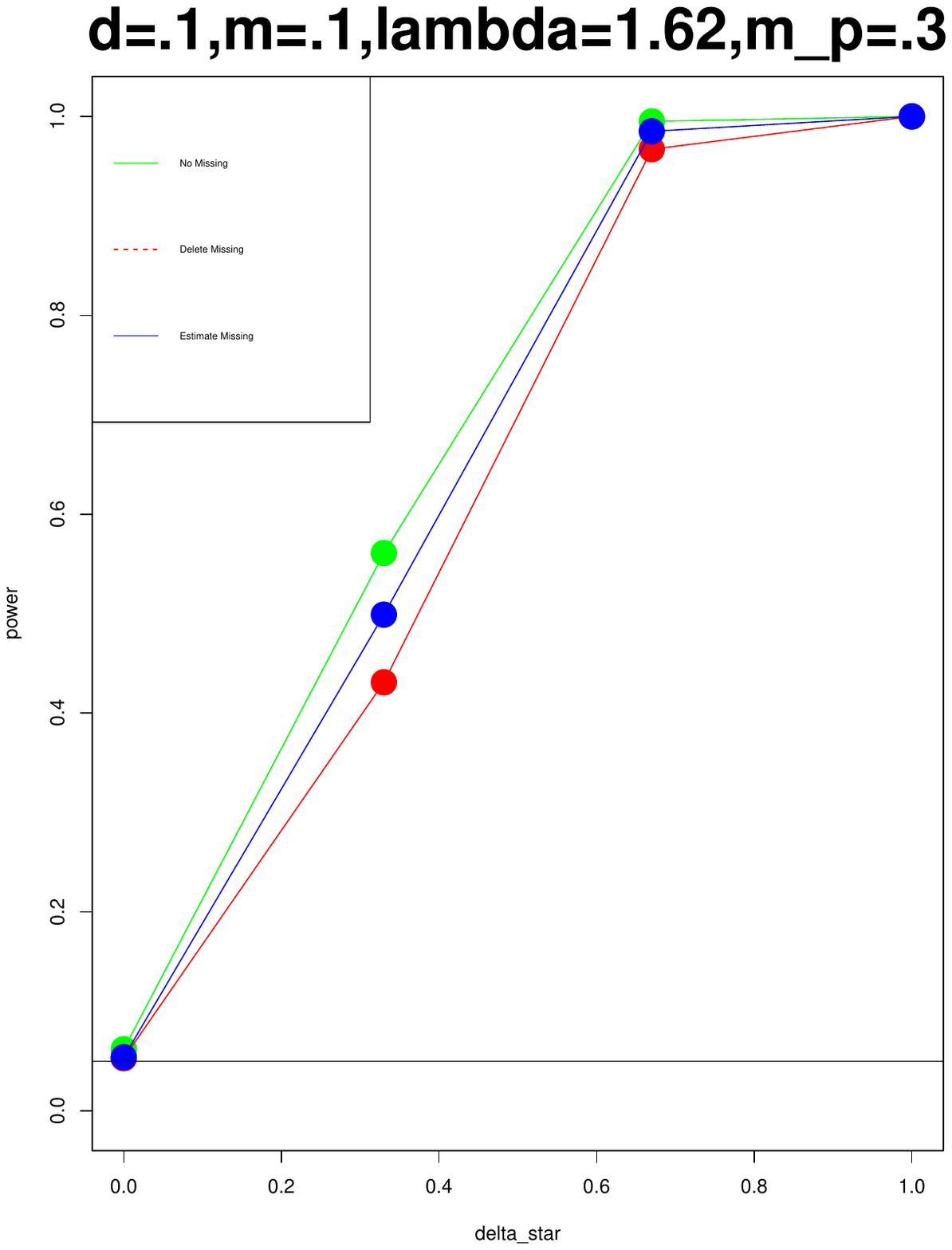}
\includegraphics[width = 2.3in, height = 1.5in]{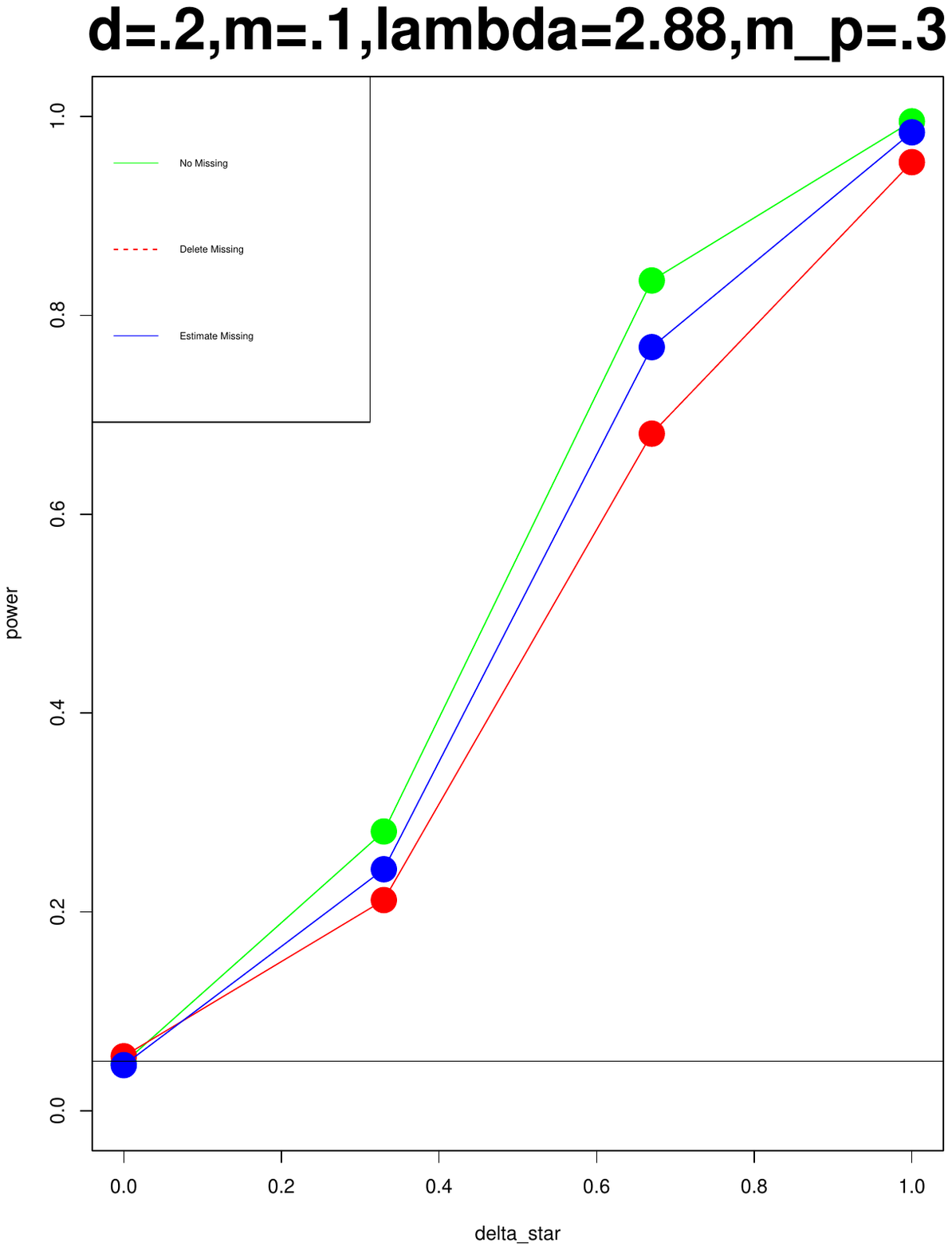}
\includegraphics[width = 2.3in, height = 1.5in]{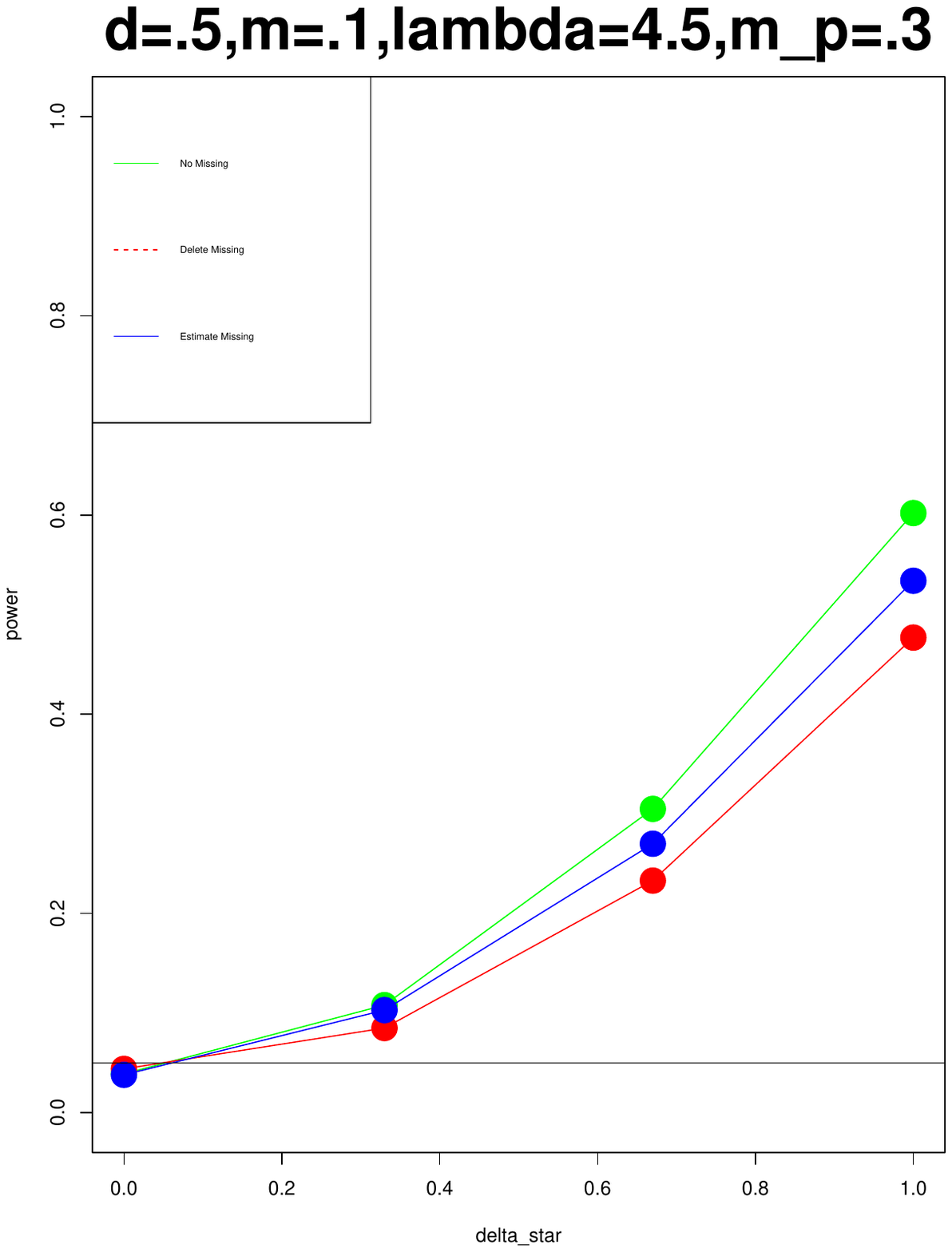}

 $m = .5 \quad miss_p = .3$

\includegraphics[width = 2.3in, height = 1.5in]{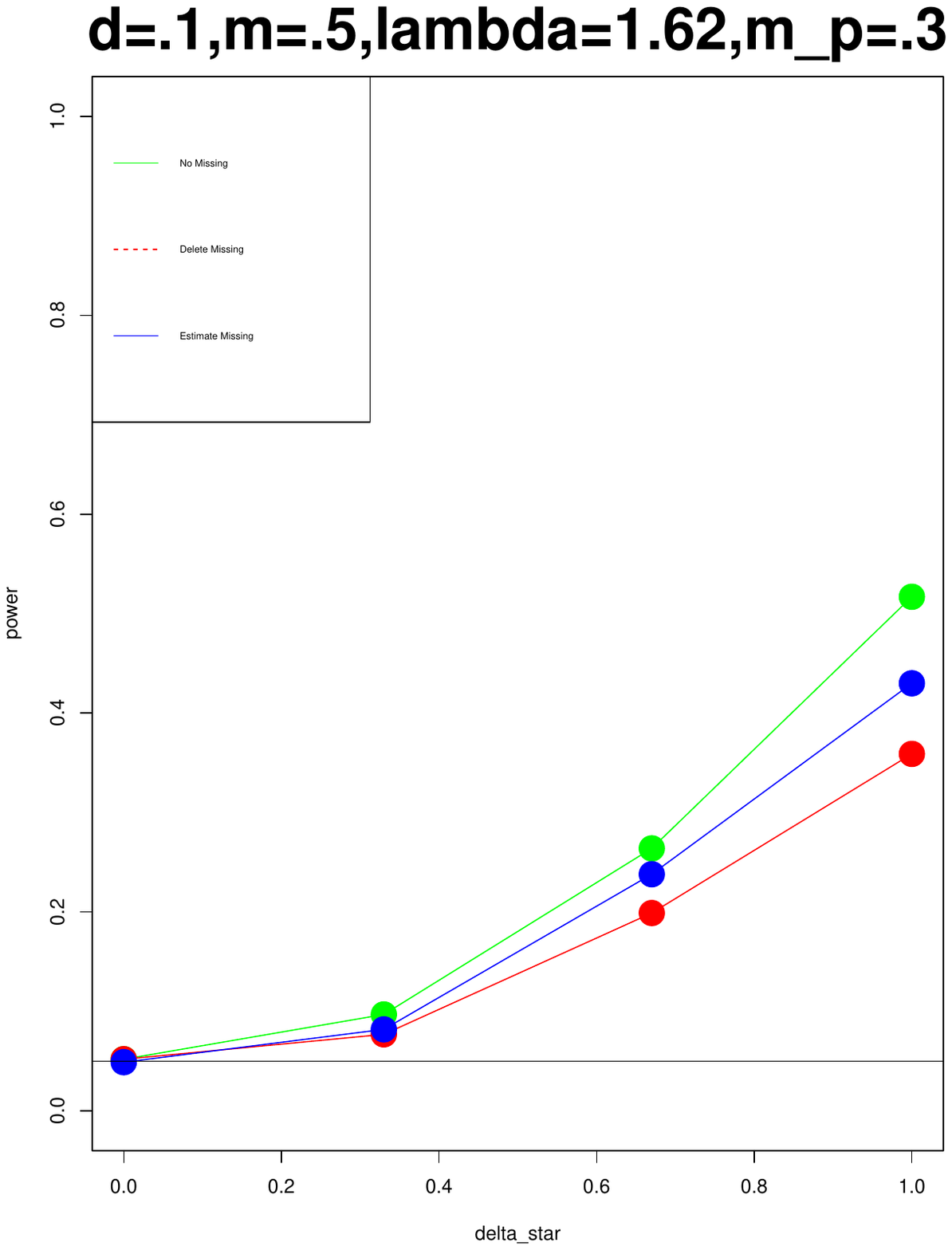}
\includegraphics[width = 2.3in, height = 1.5in]{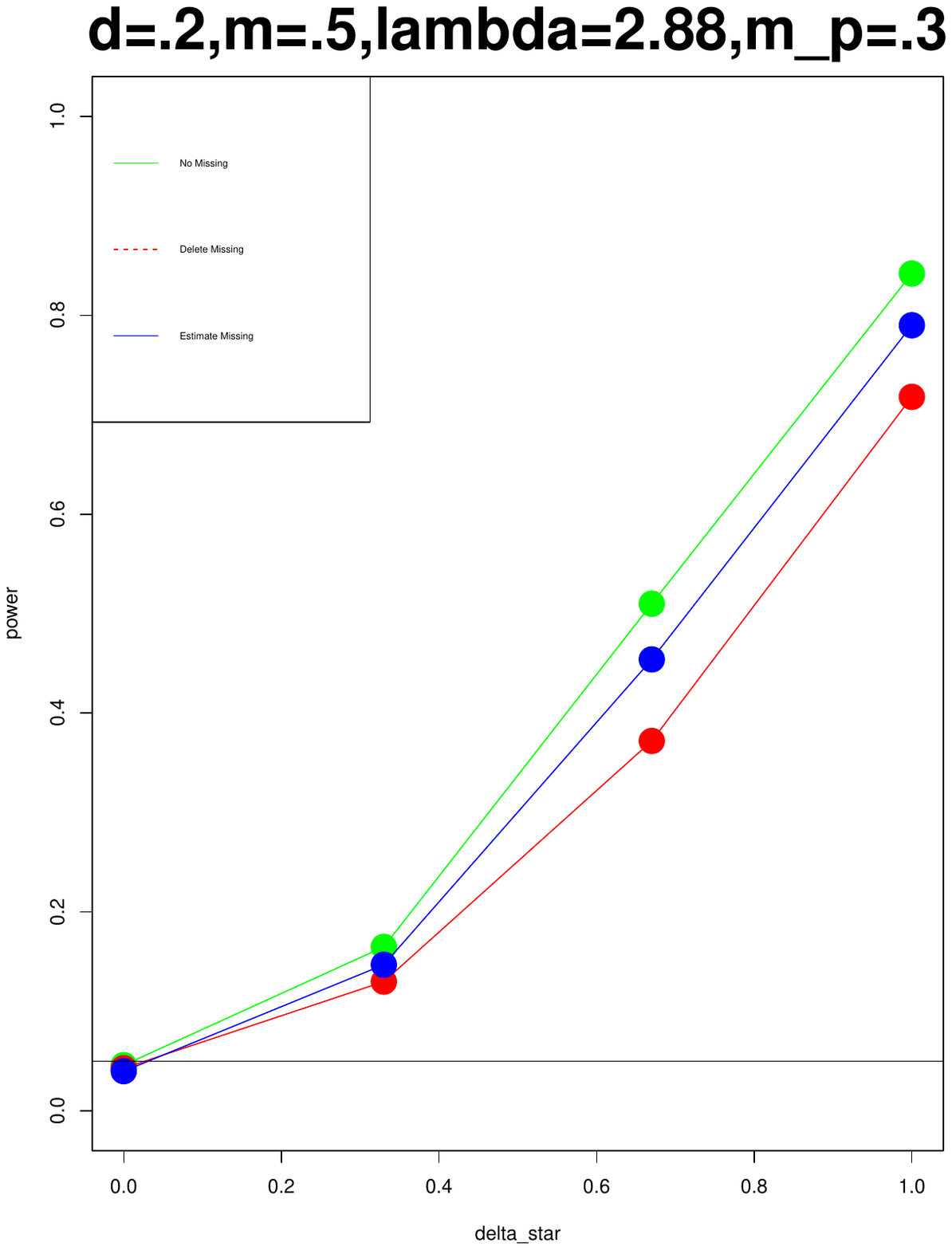}
\includegraphics[width = 2.3in, height = 1.5in]{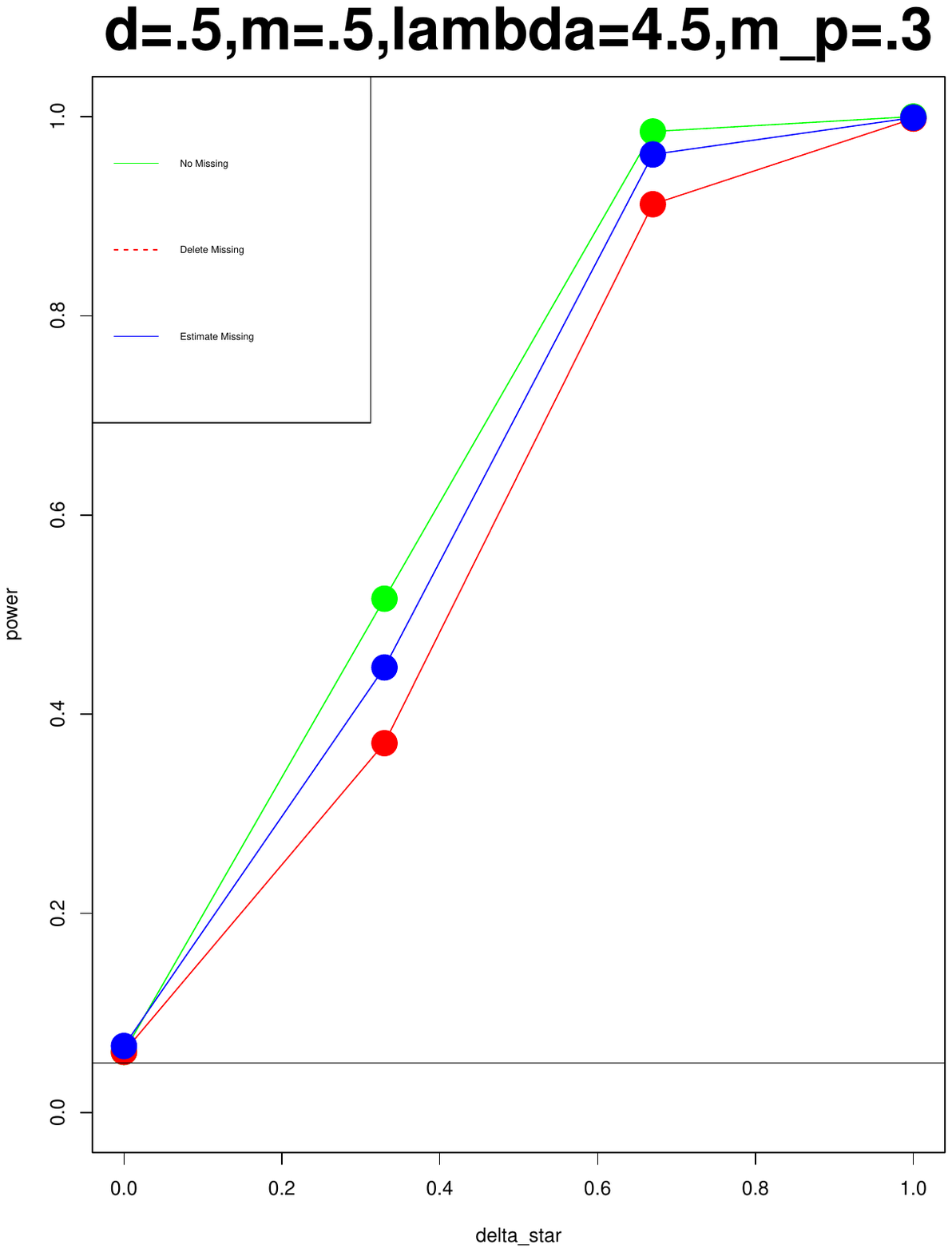}
\\

\end{center}

\subsubsection{When Distribution of Trait is Binary Type}

Let, $p_{inf}$ be the proportion of 1 the population. In this case $p^{\star}$ depends of $d, \alpha, \beta, \sigma, c$ and also $p_{inf}.$ In the simulation study we have used three choices of $p_{inf}$ and varied $d, \alpha, \beta, \sigma, c$ such that $p^{\star} = .1$

\begin{center}

\begin{tabular}{|c|c|c|c|c|c|c|}
\hline 
No. & $p_{inf}$ & $d$ & $c$ & $\sigma$ & $\alpha$ & $\beta$ \\ 
\hline 
1 & .3 & .1 & 0.186 & 0.6 & 0 & -2 \\ 
\hline 
2 & .1 & .3 & 0.53 & 0.6 & 0 & -2 \\ 
\hline 
3 & .05 & .4 & 0.41 & 0.4 & 0 & -2 \\ 
\hline 
\end{tabular} 

\end{center}

We repeat this for two values of $m,$ .1 and .5.

\begin{center}

 $m = .1 \quad miss_p = .2$

\includegraphics[width = 2.3in, height = 1.5in]{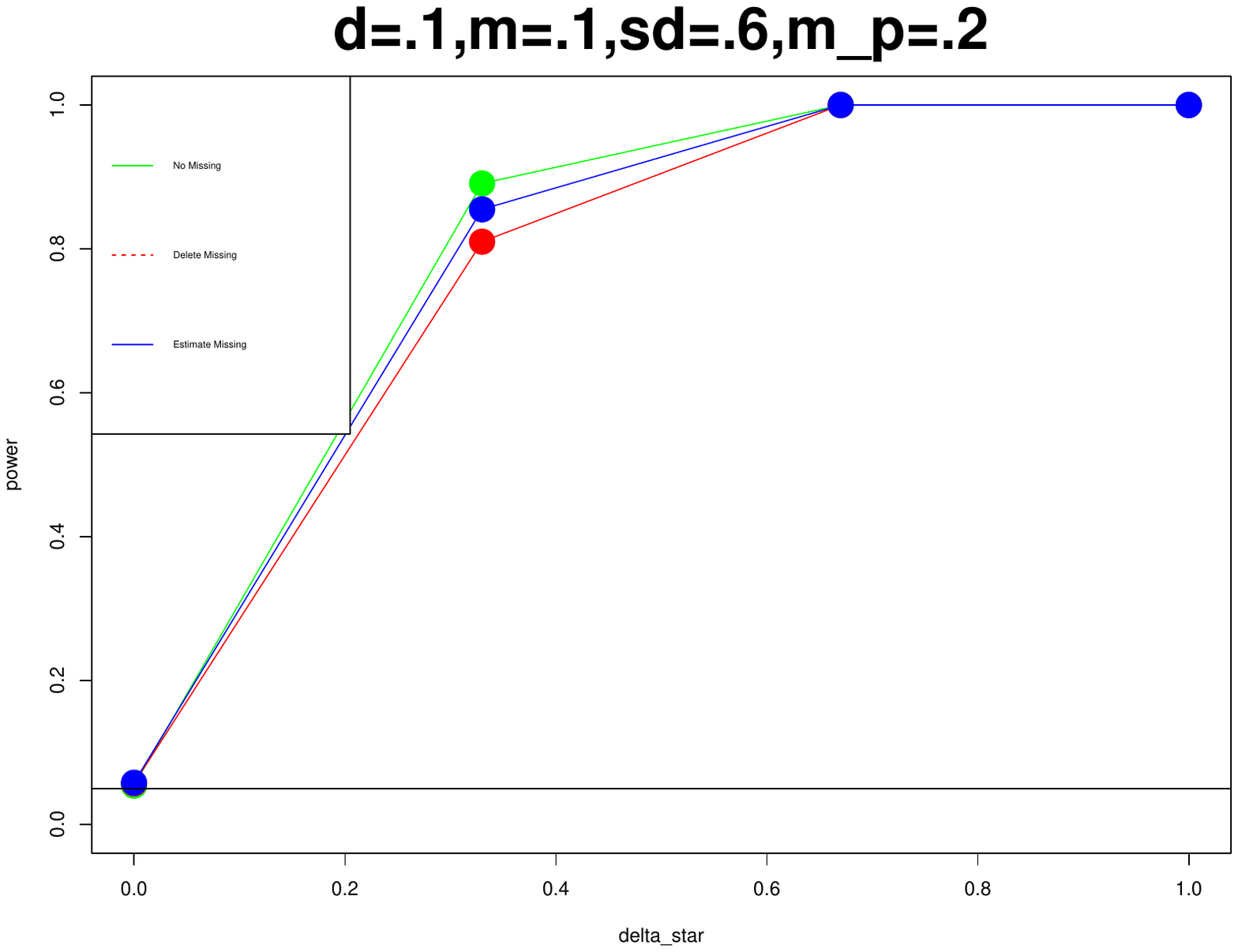}
\includegraphics[width = 2.3in, height = 1.5in]{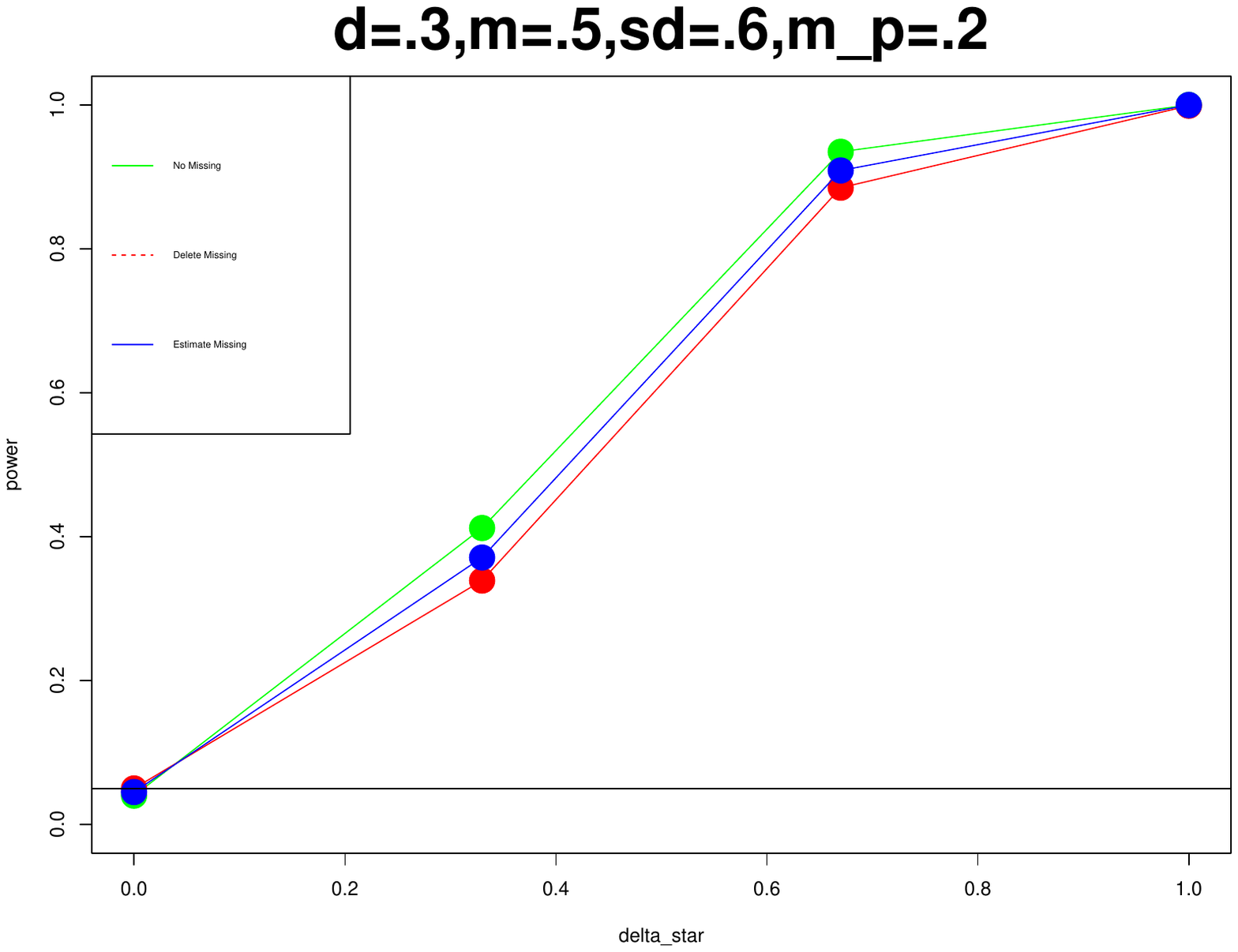}
\includegraphics[width = 2.3in, height = 1.5in]{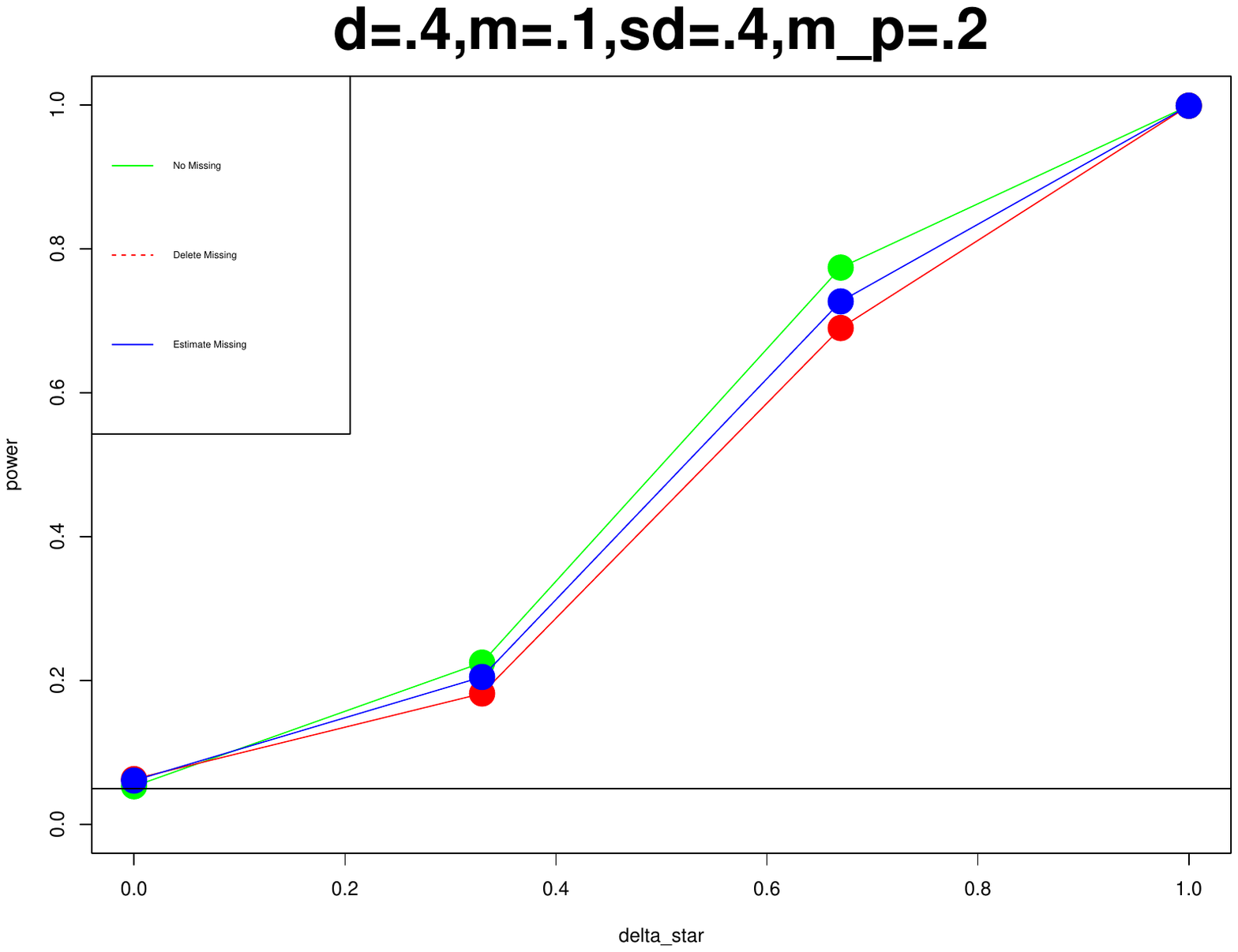}

 $m = .5 \quad miss_p = .2$

\includegraphics[width = 2.3in, height = 1.5in]{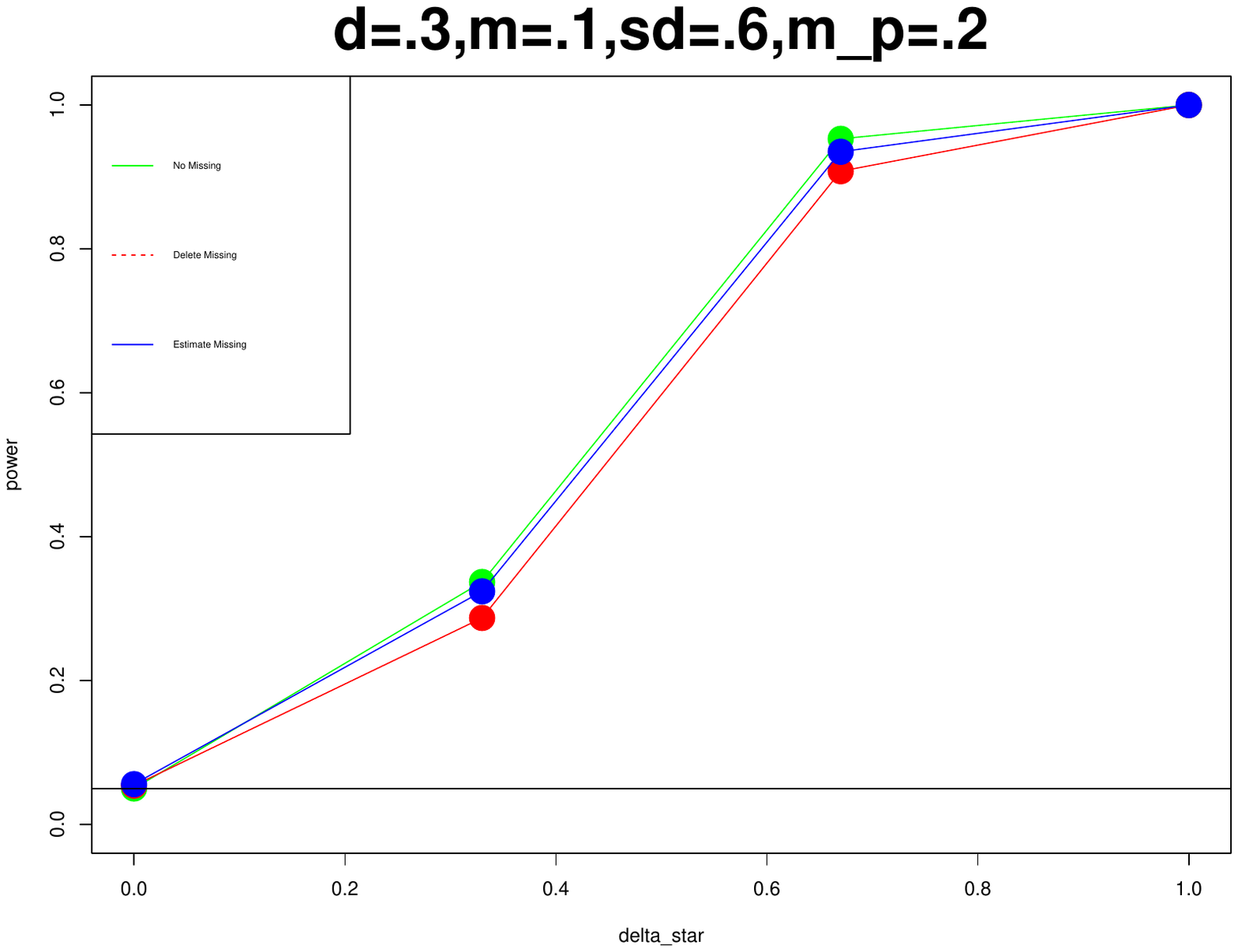}
\includegraphics[width = 2.3in, height = 1.5in]{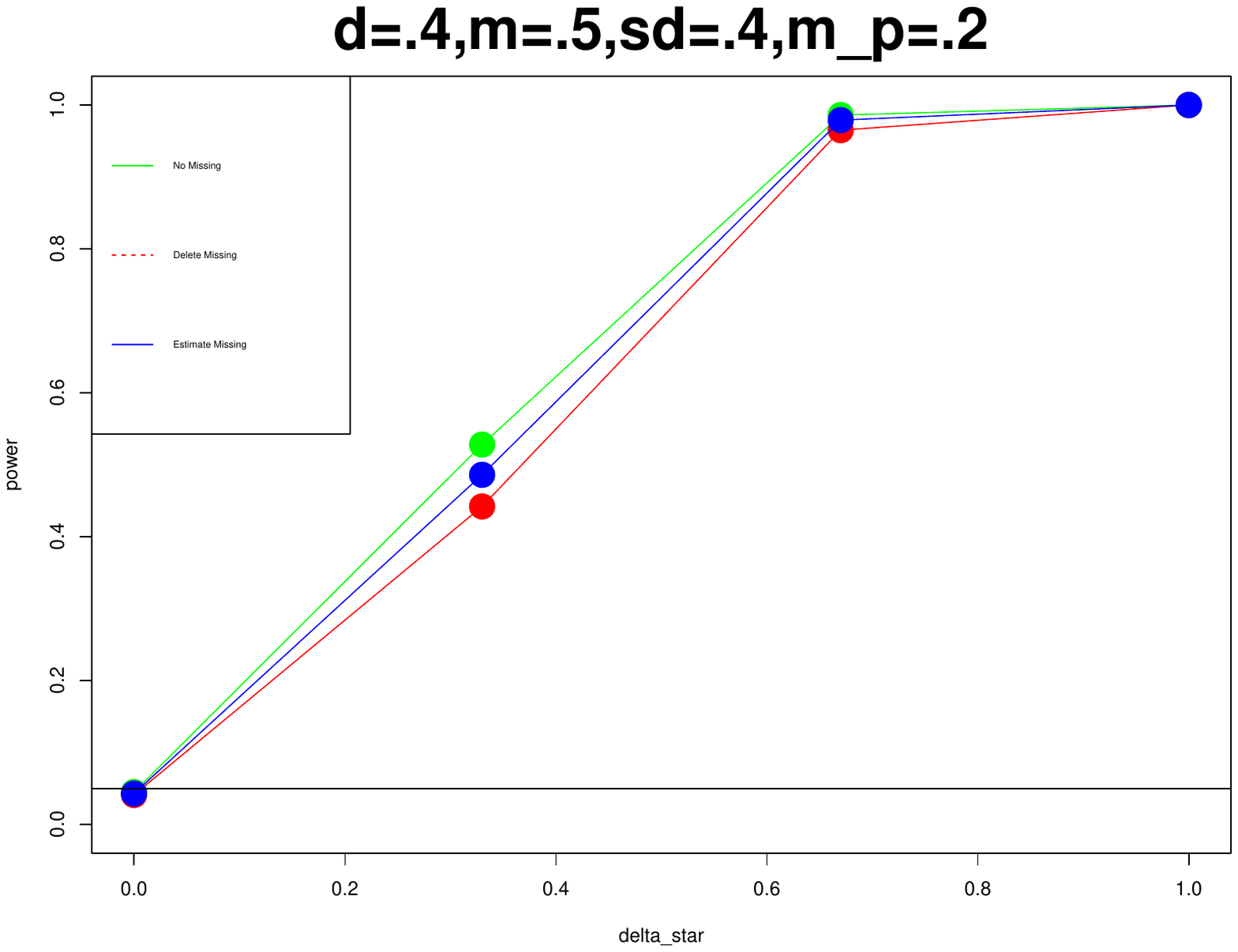}
\includegraphics[width = 2.3in, height = 1.5in]{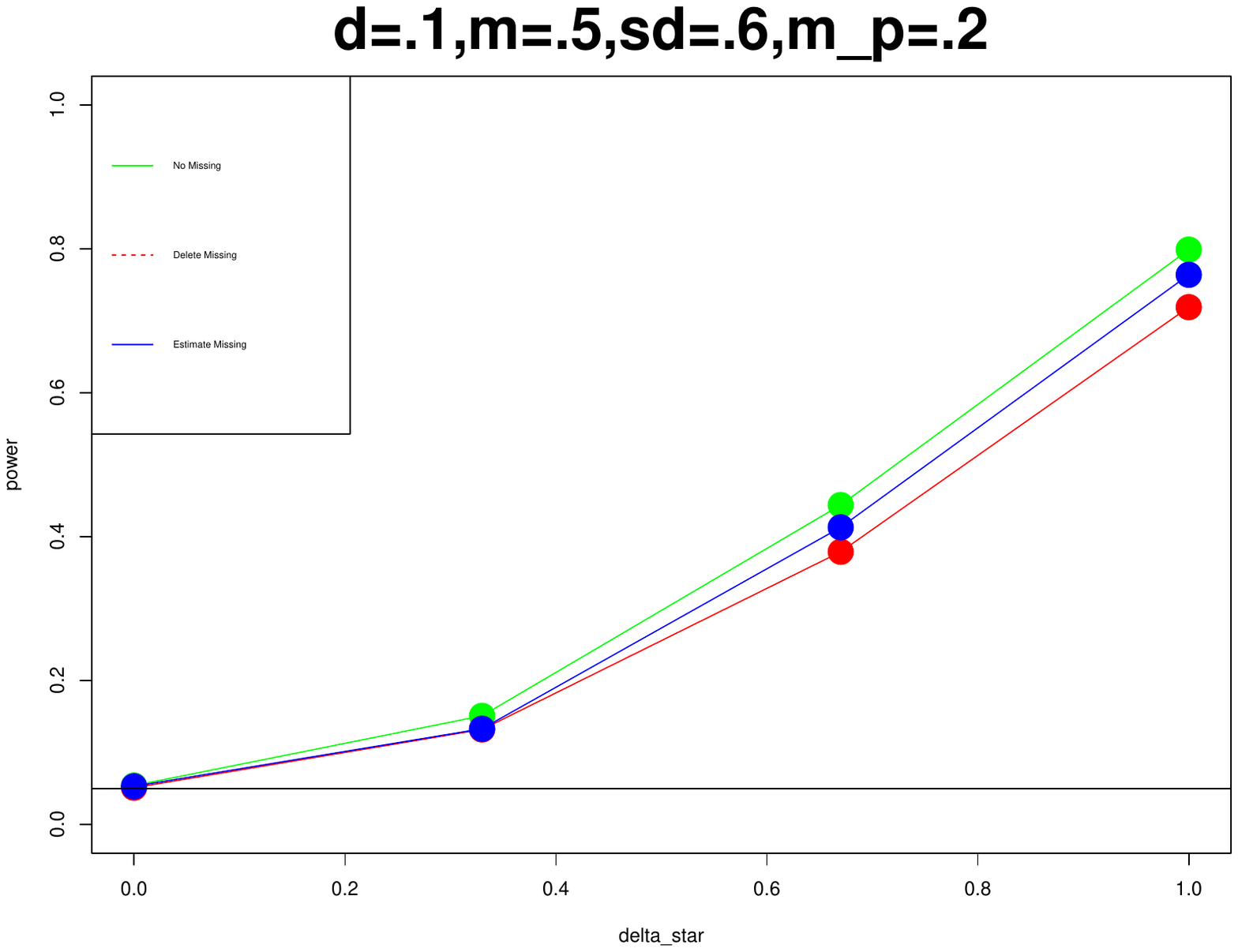}

 $m = .1 \quad miss_p = .3$

\includegraphics[width = 2.3in, height = 1.5in]{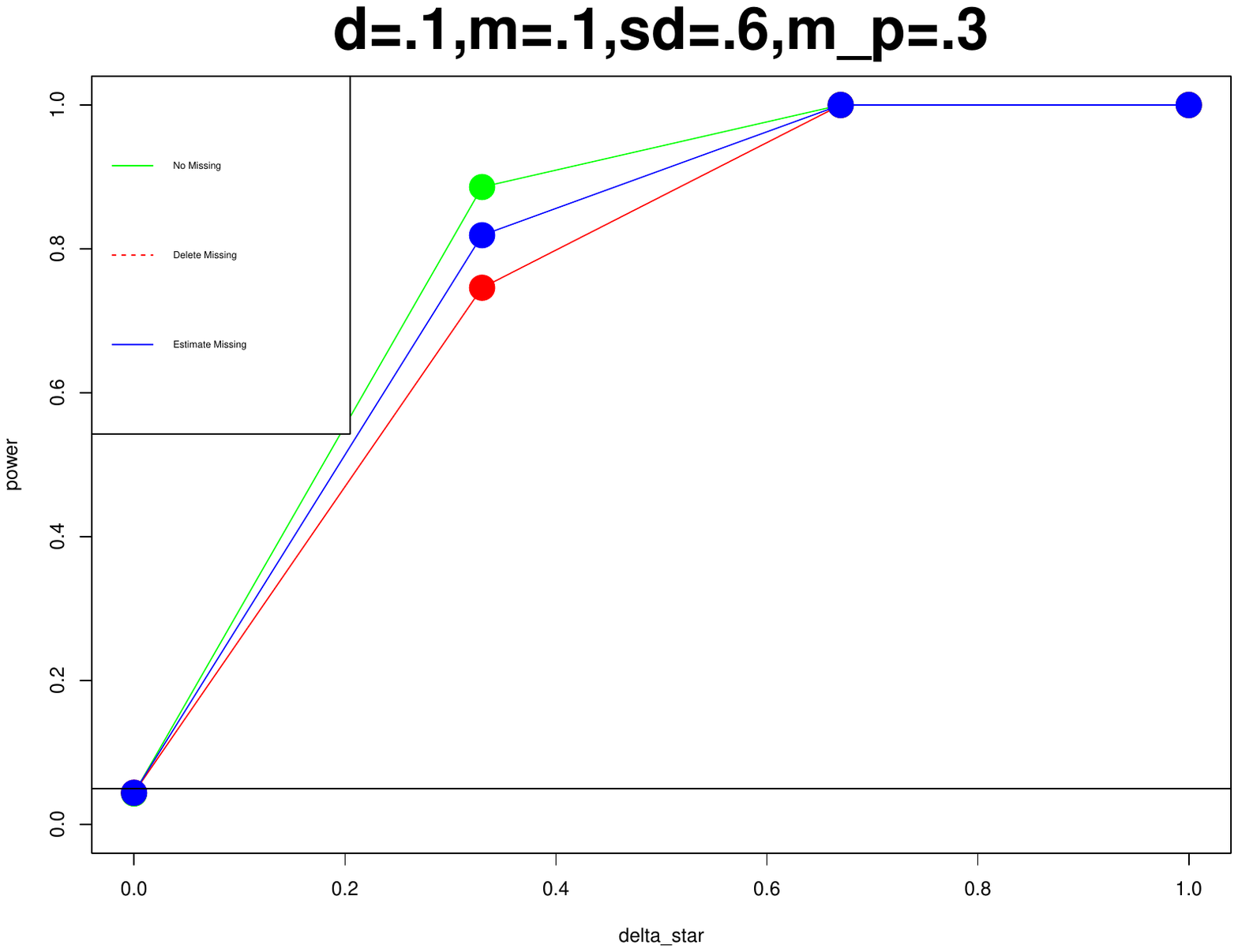}
\includegraphics[width = 2.3in, height = 1.5in]{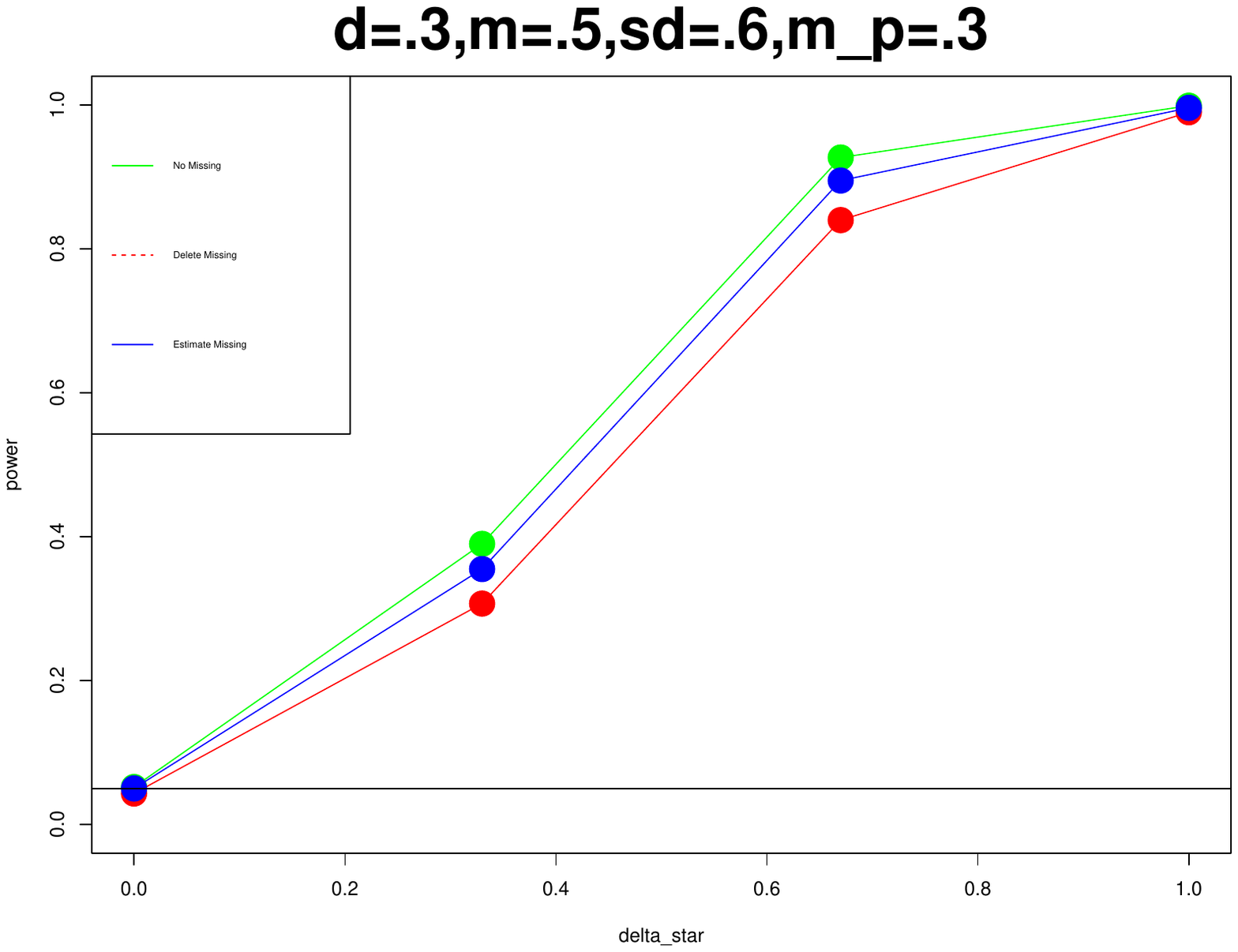}
\includegraphics[width = 2.3in, height = 1.5in]{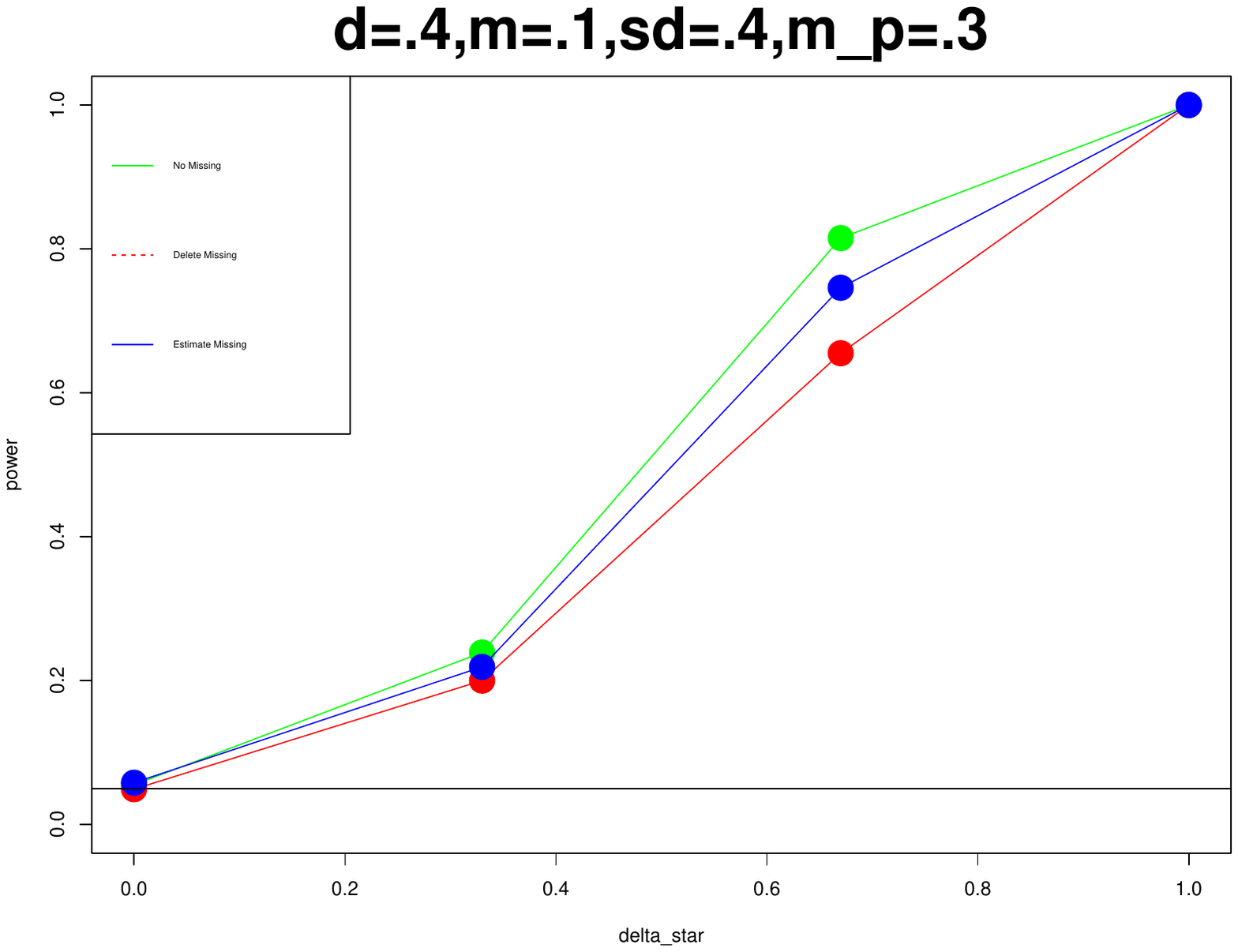}

 $m = .5 \quad miss_p = .3$

\includegraphics[width = 2.3in, height = 1.5in]{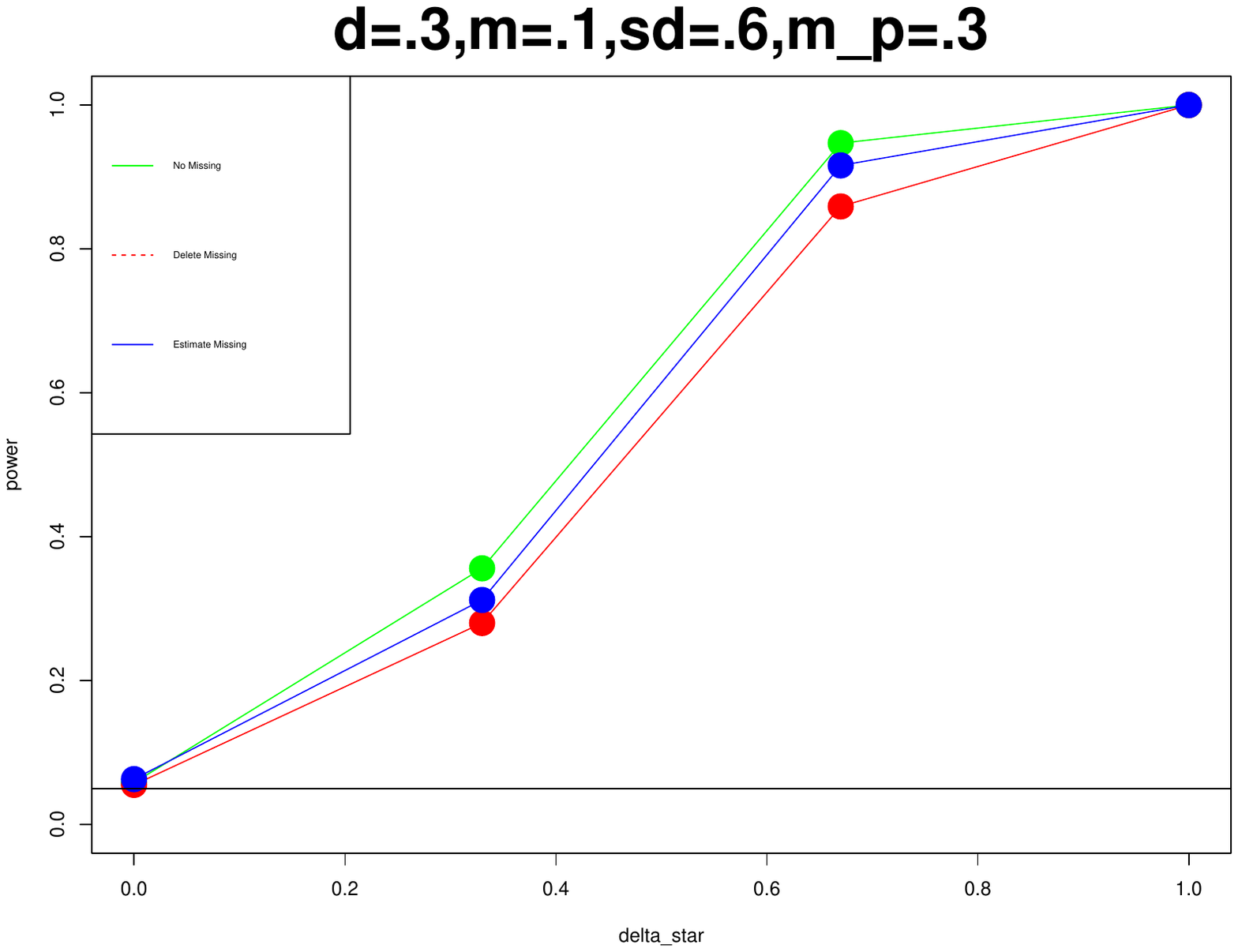}
\includegraphics[width = 2.3in, height = 1.5in]{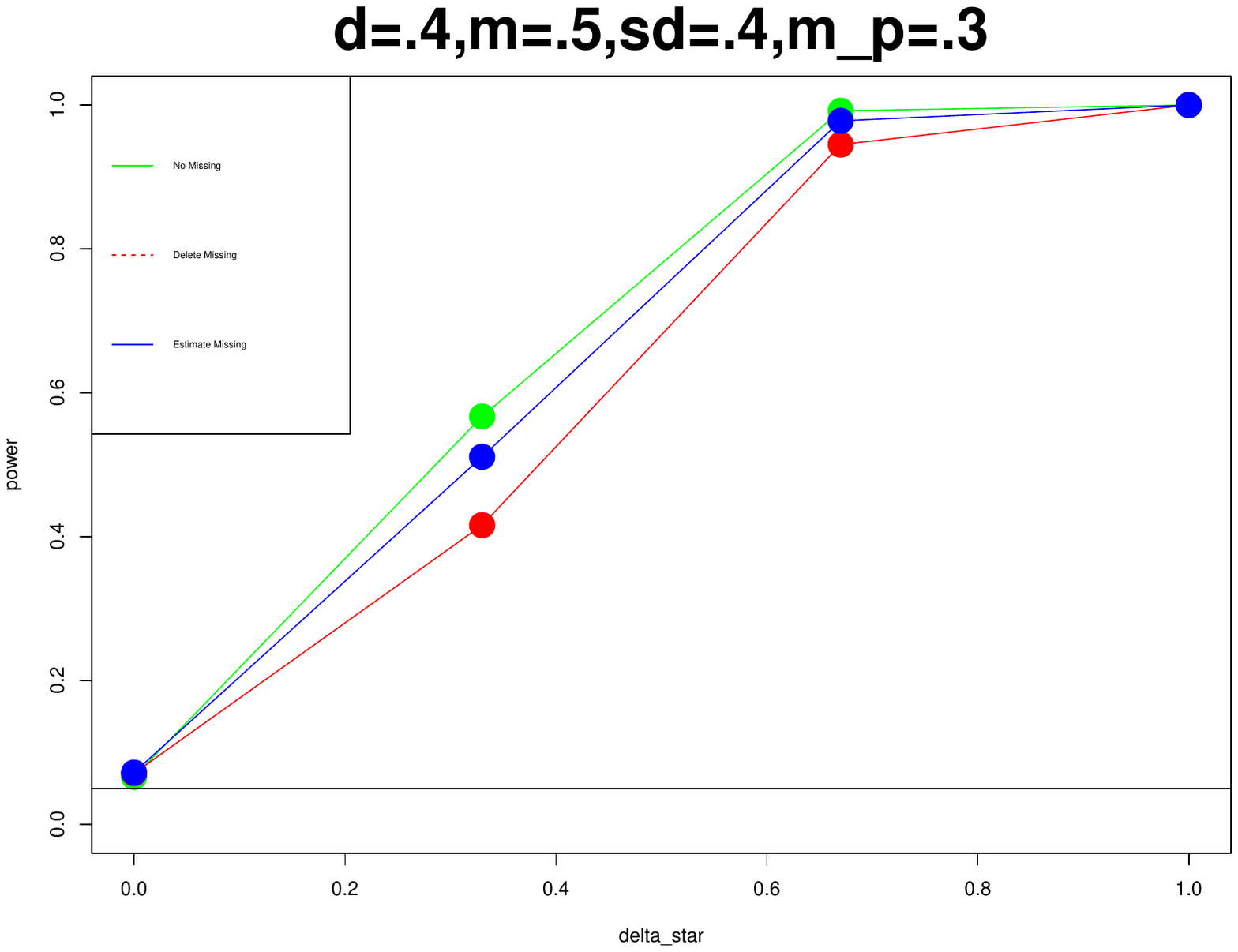}
\includegraphics[width = 2.3in, height = 1.5in]{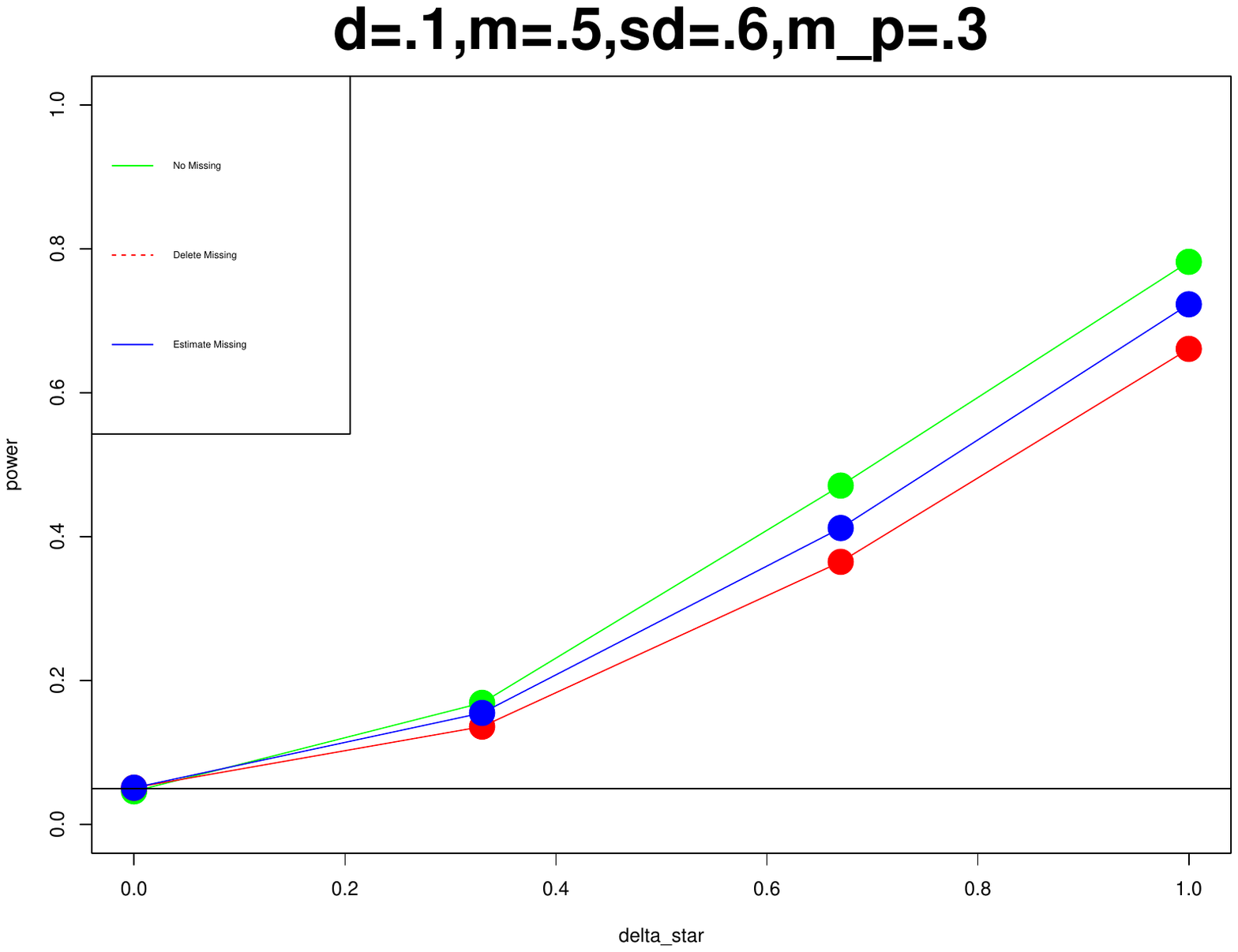}

\end{center}

\subsection{For Missing Type 2.1}

We have stated three strategies of imputation in this type of missing. Let us call strategy 1, 2 and 3 as \textbf{use other}, \textbf{use same} and \textbf{use both} respectively.

We have defined two types of correlations $\rho_1$ and $\rho_2$. Now, $\rho_2$ can take two values - correlation between 1st trait of two offspring and correlation between 2nd trait of two offspring. We have considered maximum of those two correlation as $\rho_2$.

In each of the trait combinations we shall consider three situations $\rho_1>\rho_2$, $\rho_1<\rho_2$ and $\rho_1=\rho_2$. We adjusted other parameters in a suitable way and considered three choices of $\rho_2$ as .15, .3, .45 while $\rho_1$ is fixed at .3. We have found out which strategy is better in each of three cases.

(If in some case we see \textbf{use both} is the best strategy and any of \textbf{use other} or \textbf{use same} is almost as good as that, then we shall choose that instead of choosing \textbf{use both} as the best strategy.)

\vspace{.2cm}

We have done simulation for 30\% missing for all the following cases. 

\subsubsection{When both Traits have Normal Distribution}

The following tables represents powers for the three strategies evaluated at $\delta = 0, .33, .67$ and 1.

\textbf{ For $\rho_1 > \rho_2$}

\vspace{.4cm}

\hspace{3.5cm}$m=.1$
\hspace{8.5cm}$m=.5$

\vspace{.2cm}

\begin{tabular}{|c|c|c|c|c|}
\hline 
Strategy & $\delta = 0$ & $\delta = .33$ & $\delta = .67$ & $\delta = 1$ \\ 
\hline 
use same & 0.051 & 0.448 & 0.965 & 1 \\ 
\hline 
use other & 0.053 & 0.456 & 0.965 & 1 \\ 
\hline 
use both & 0.052 & 0.452 & 0.964 & 1 \\ 
\hline 
\end{tabular} \hspace{1.5cm}
\begin{tabular}{|c|c|c|c|c|}
\hline 
Strategy & $\delta = 0$ & $\delta = .33$ & $\delta = .67$ & $\delta = 1$ \\ 
\hline 
use same & 0.050 & 0.754 & 1 & 1 \\ 
\hline 
use other & 0.051 & 0.762 & 1 & 1 \\ 
\hline 
use both & 0.052 & 0.760 & 1 & 1 \\ 
\hline 
\end{tabular} 

\hspace{.4cm}

\textbf{ For $\rho_1 < \rho_2$}

\vspace{.4cm}

\hspace{3.5cm}$m=.1$
\hspace{8.5cm}$m=.5$

\vspace{.2cm}

\begin{tabular}{|c|c|c|c|c|}
\hline 
Strategy & $\delta = 0$ & $\delta = .33$ & $\delta = .67$ & $\delta = 1$ \\ 
\hline 
use same & 0.048 & 0.365 & 0.932 & 1 \\ 
\hline 
use other & 0.052 & 0.368 & 0.937 & 1 \\ 
\hline 
use both & 0.047 & 0.359 & 0.934 & 1 \\ 
\hline 
\end{tabular} \hspace{1.5cm}
\begin{tabular}{|c|c|c|c|c|}
\hline 
Strategy & $\delta = 0$ & $\delta = .33$ & $\delta = .67$ & $\delta = 1$ \\ 
\hline 
use same & 0.048 & 0.901 & 1 & 1 \\ 
\hline 
use other & 0.051 & 0.908 & 1 & 1 \\ 
\hline 
use both & 0.049 & 0.898 & 1 & 1 \\ 
\hline 
\end{tabular}

\hspace{.4cm}

\textbf{ For $\rho_1 = \rho_2$}

\vspace{.4cm}

\hspace{3.5cm}$m=.1$
\hspace{8.5cm}$m=.5$

\vspace{.2cm}

\begin{tabular}{|c|c|c|c|c|}
\hline 
Strategy & $\delta = 0$ & $\delta = .33$ & $\delta = .67$ & $\delta = 1$ \\ 
\hline 
use same & 0.052 & 0.512 & 0.993 & 1 \\ 
\hline 
use other & 0.054 & 0.518 & 0.994 & 1 \\ 
\hline 
use both & 0.053 & 0.515 & 0.993 & 1 \\ 
\hline 
\end{tabular} \hspace{1.5cm}
\begin{tabular}{|c|c|c|c|c|}
\hline 
Strategy & $\delta = 0$ & $\delta = .33$ & $\delta = .67$ & $\delta = 1$ \\ 
\hline 
use same & 0.053 & 0.830 & 1 & 1 \\ 
\hline 
use other & 0.049 & 0.836 & 1 & 1 \\ 
\hline 
use both & 0.046 & 0.829 & 1 & 1 \\ 
\hline 
\end{tabular} 

\vspace{1cm}

According to the above results we can conclude-

\begin{center}
\begin{tabular}{|c|c|}
\hline 
Case & Best Strategy \\ 
\hline 
$\rho_1 > \rho_2$ & use other \\ 
\hline 
$\rho_1 < \rho_2$ & use other \\ 
\hline 
$\rho_1 = \rho_2$ & use other \\ 
\hline 
\end{tabular} 

\end{center}

Now we go for power comparison among no missing, estimated missing and deleted missing.

We generate 1st trait from normal distribution (section 6.2) with the parameters $\alpha = \alpha_1, \beta = \beta_1, \sigma = \sigma1$ keeping $p^\star = p^\star_1$ and we generate 2nd trait from normal distribution (section 6.2) with the parameters $\alpha = \alpha_2, \beta = \beta_2, \sigma = \sigma_2$ keeping $p^\star = p^\star_2$.

We have done simulation for three choices of $d$ as .1, .2, .3 and for each $d$ we take $(p^\star_1, p^\star_2)$ as (.1, .2), (.2, .2).

We take $\alpha_1 = 5, \alpha_2 = 10, \beta_1 = 1, and \beta_2 = 2$ and varied $\sigma_1$ and $\sigma_2$ in the following way,

\vspace{.4cm}

\begin{tabular}{|c|c|c|}
\hline 
d & $(p^\star_1, p^\star_2)$ & ($\sigma_1^2, \sigma_2^2$) \\ 
\hline 
.1 & (.1, .2) & (1.62, 2.88) \\ 
\hline 
.1 & (.2, .2) & (.72, 2.88) \\ 
\hline 

\end{tabular} \hspace{1.5cm}
\begin{tabular}{|c|c|c|}
\hline 
d & $(p^\star_1, p^\star_2)$ & ($\sigma_1^2, \sigma_2^2$) \\ 
\hline 
.2 & (.1, .2) & (2.88, 5.12) \\ 
\hline 
.2 & (.2, .2) & (1.28, 5.12) \\ 
\hline 

\end{tabular} \hspace{1.5cm}
\begin{tabular}{|c|c|c|}
\hline 
d & $(p^\star_1, p^\star_2)$ & ($\sigma_1^2, \sigma_2^2$) \\ 
\hline 
.3 & (.1, .2) & (3.78, 6.72) \\ 
\hline 
.3 & (.2, .2) & (1.68, 6.72) \\ 
\hline 
\end{tabular} 

\vspace{.4cm}

We replicate these for $m =$ .1 and .2.

Note that as we have seen in all 3 cases \textbf{use other} is the best strategy, we have done all the following imputations using this strategy.

\vspace{.4cm}

\hspace{1.5cm}
$d=.1 \quad m=.5$
\hspace{3cm}
$d=.2 \quad m=.5$
\hspace{3cm}
$d=.3 \quad m=.5$

\includegraphics[width = 2.3in, height = 1.5in]{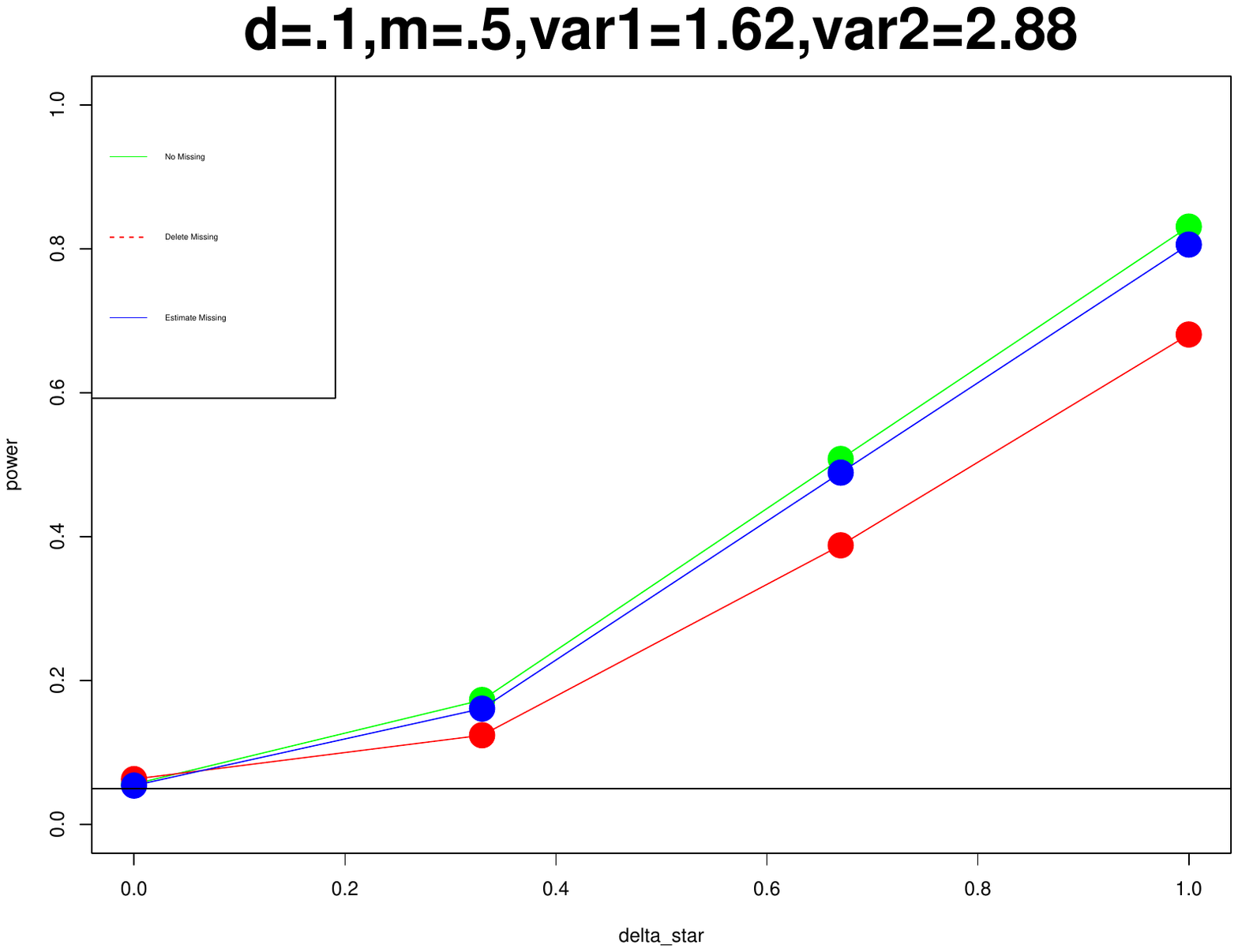}
\includegraphics[width = 2.3in, height = 1.5in]{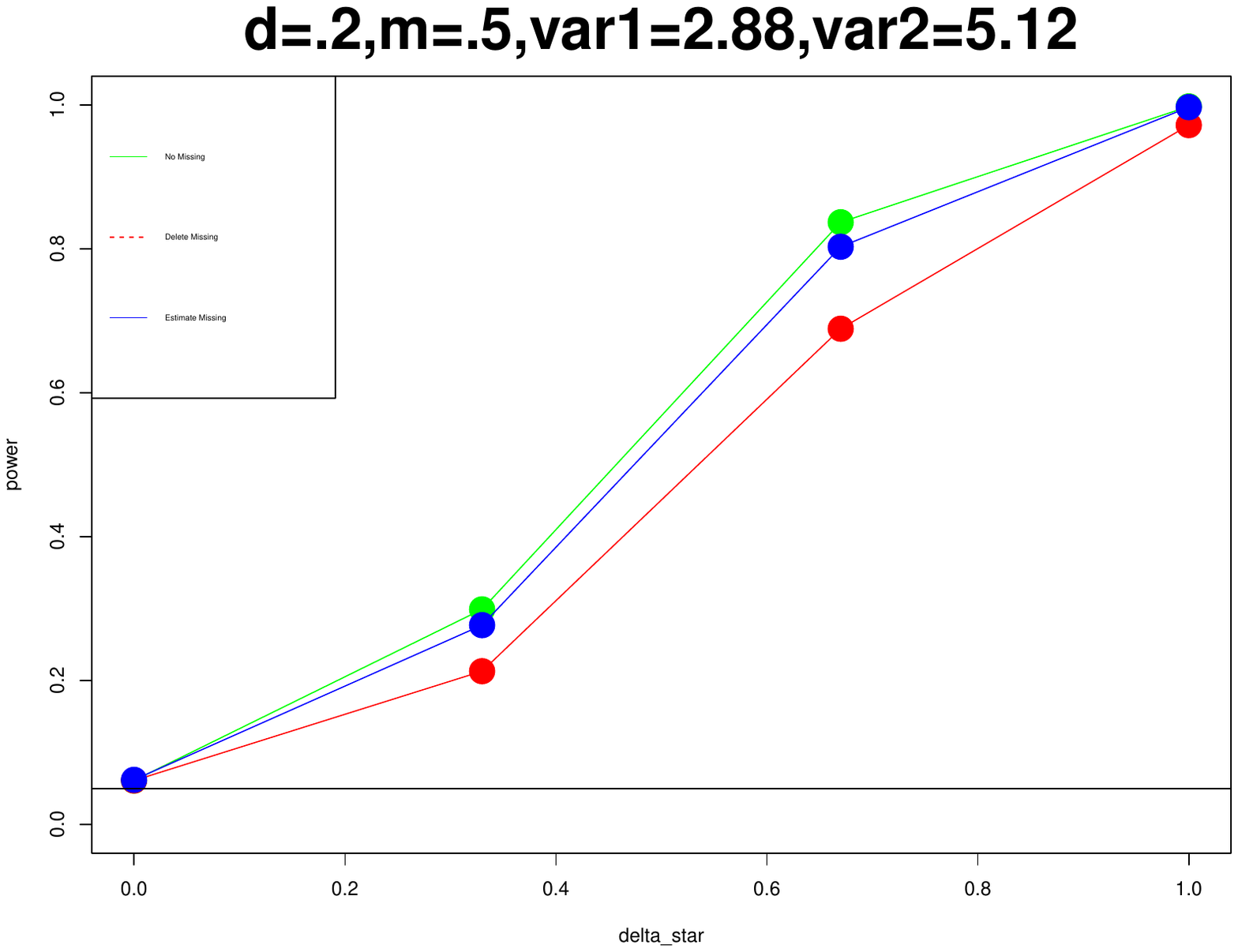}
\includegraphics[width = 2.3in, height = 1.5in]{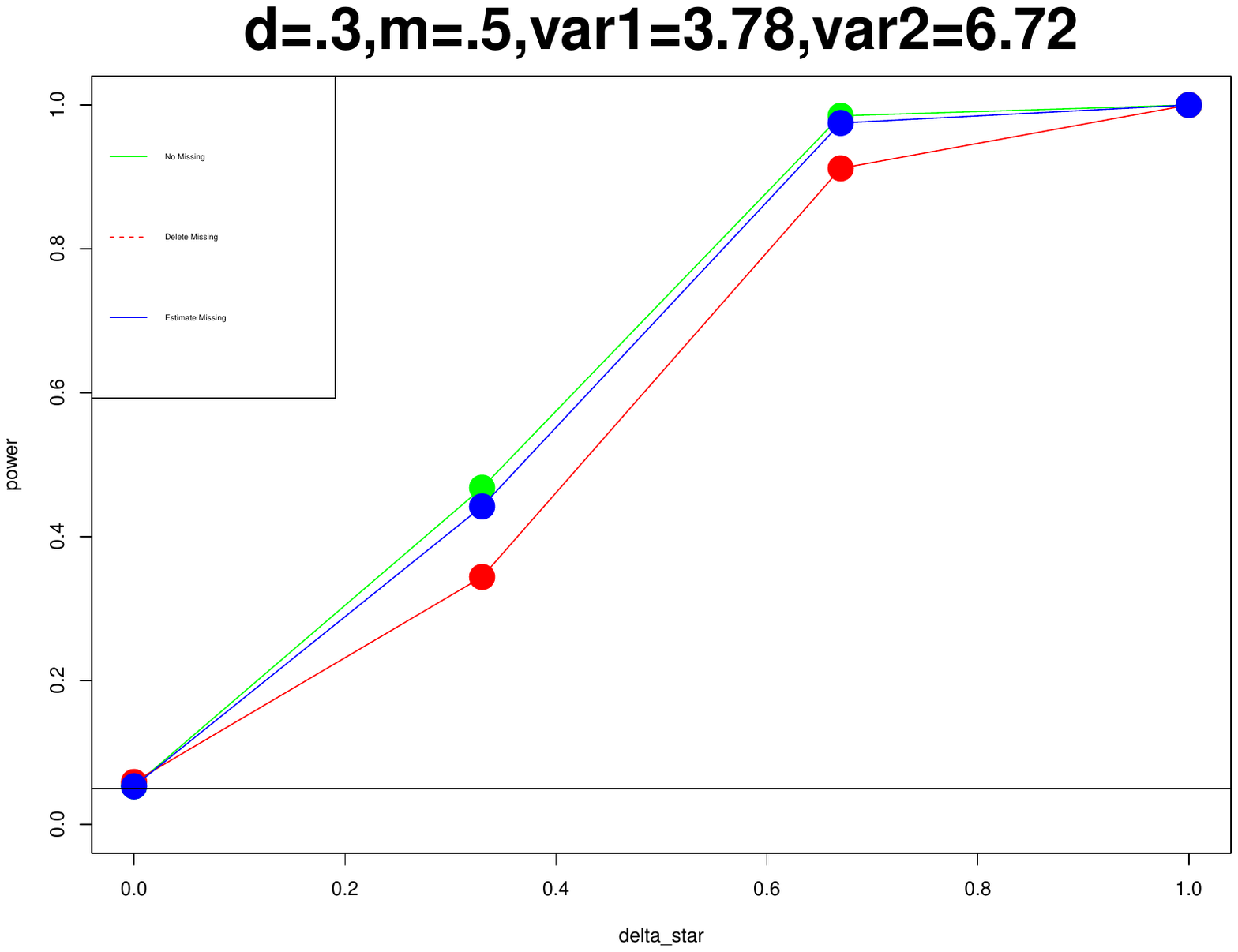}

\hspace{1.5cm}
$d=.1 \quad m=.1$
\hspace{3cm}
$d=.2 \quad m=.1$
\hspace{3cm}
$d=.3 \quad m=.1$

\includegraphics[width = 2.3in, height = 1.5in]{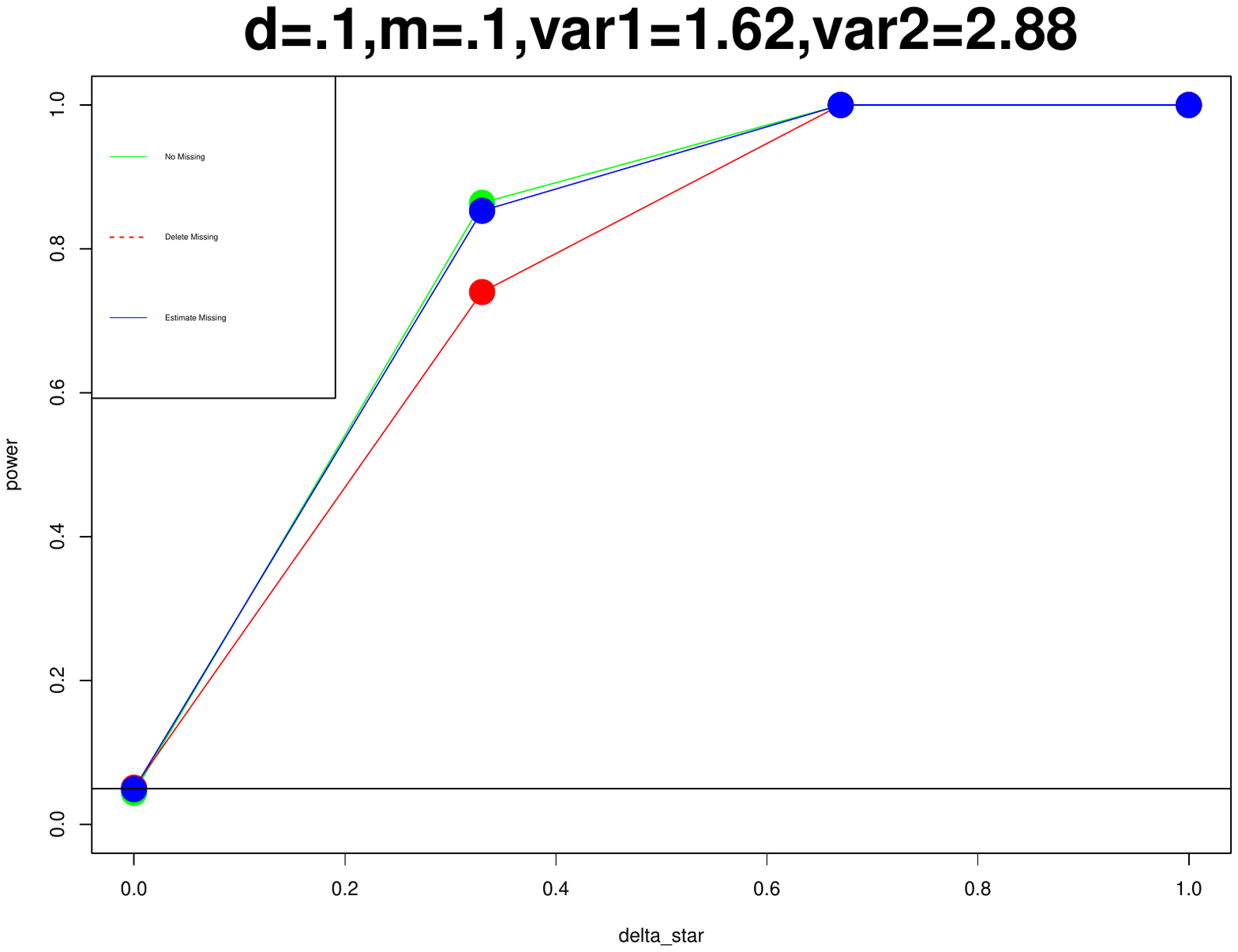}
\includegraphics[width = 2.3in, height = 1.5in]{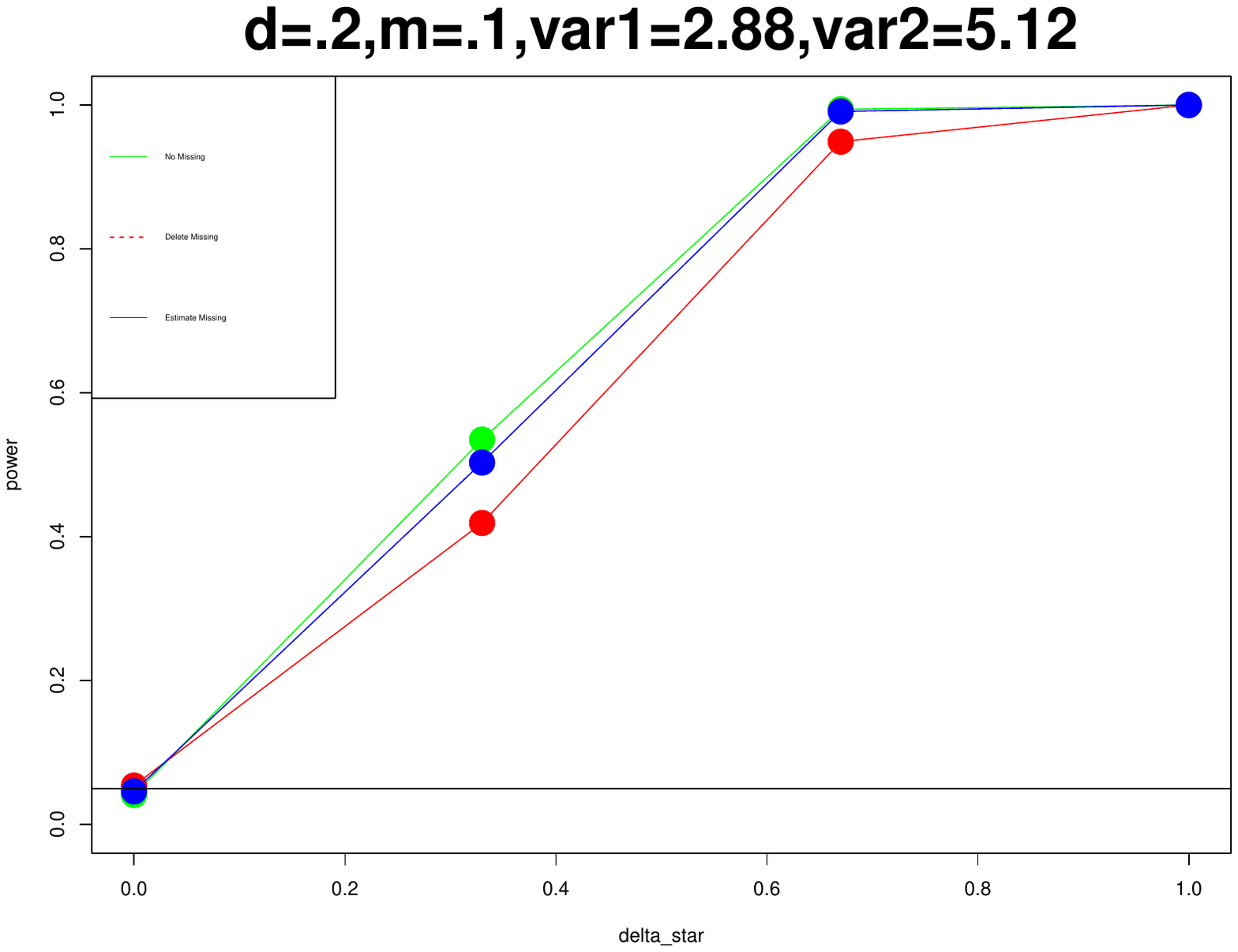}
\includegraphics[width = 2.3in, height = 1.5in]{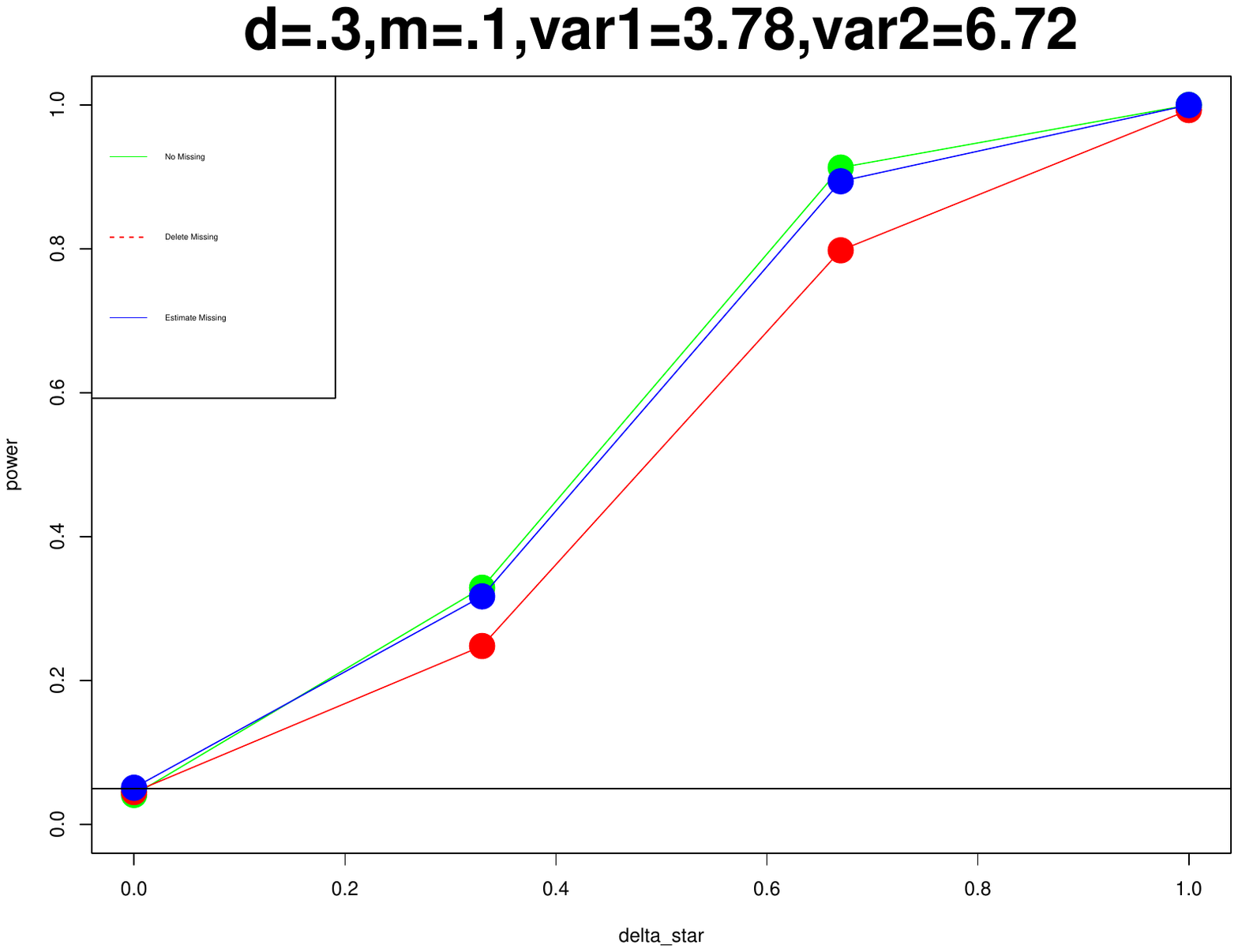}

\hspace{1.5cm}
$d=.1 \quad m=.5$
\hspace{3cm}
$d=.2 \quad m=.5$
\hspace{3cm}
$d=.3 \quad m=.5$

\includegraphics[width = 2.3in, height = 1.5in]{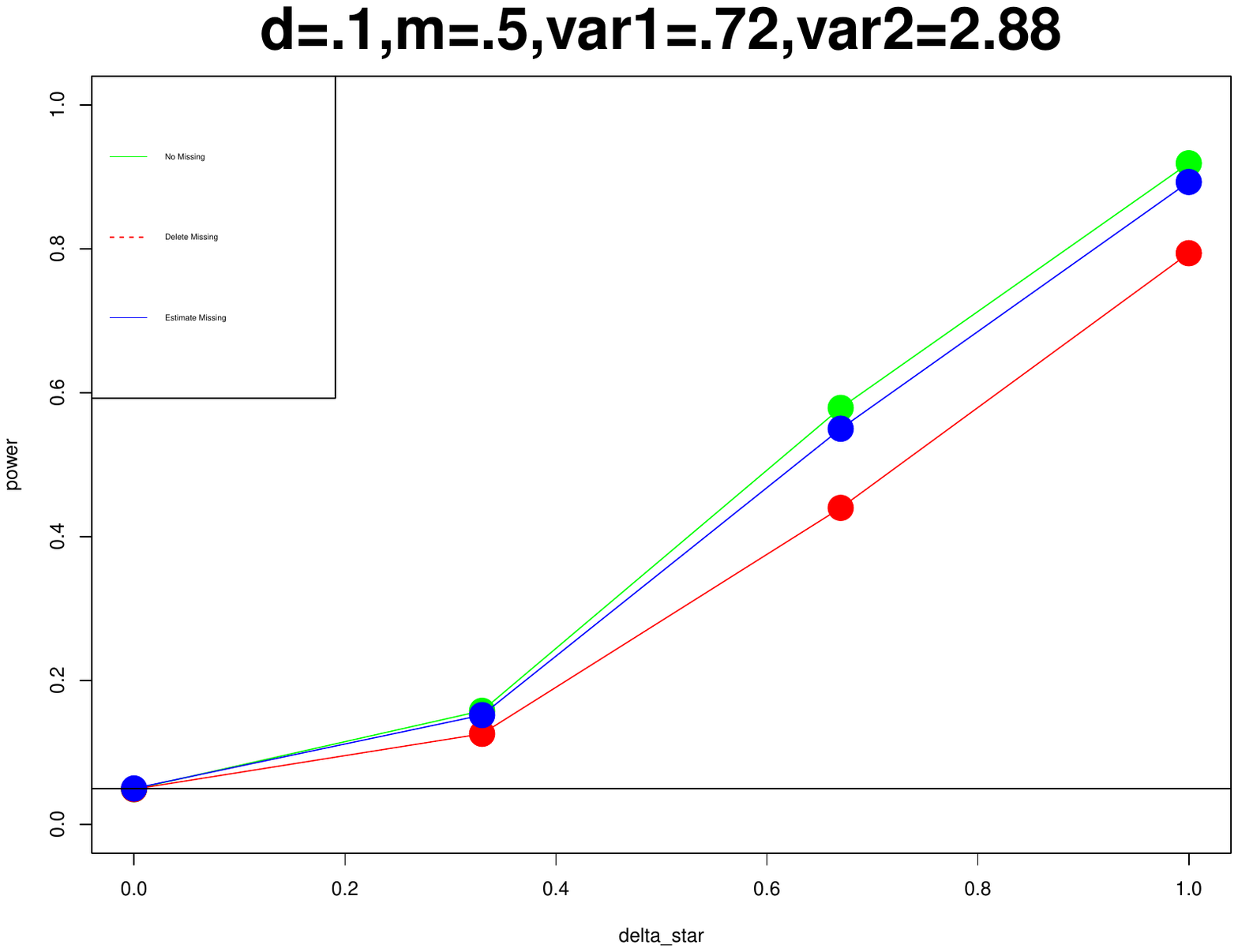}
\includegraphics[width = 2.3in, height = 1.5in]{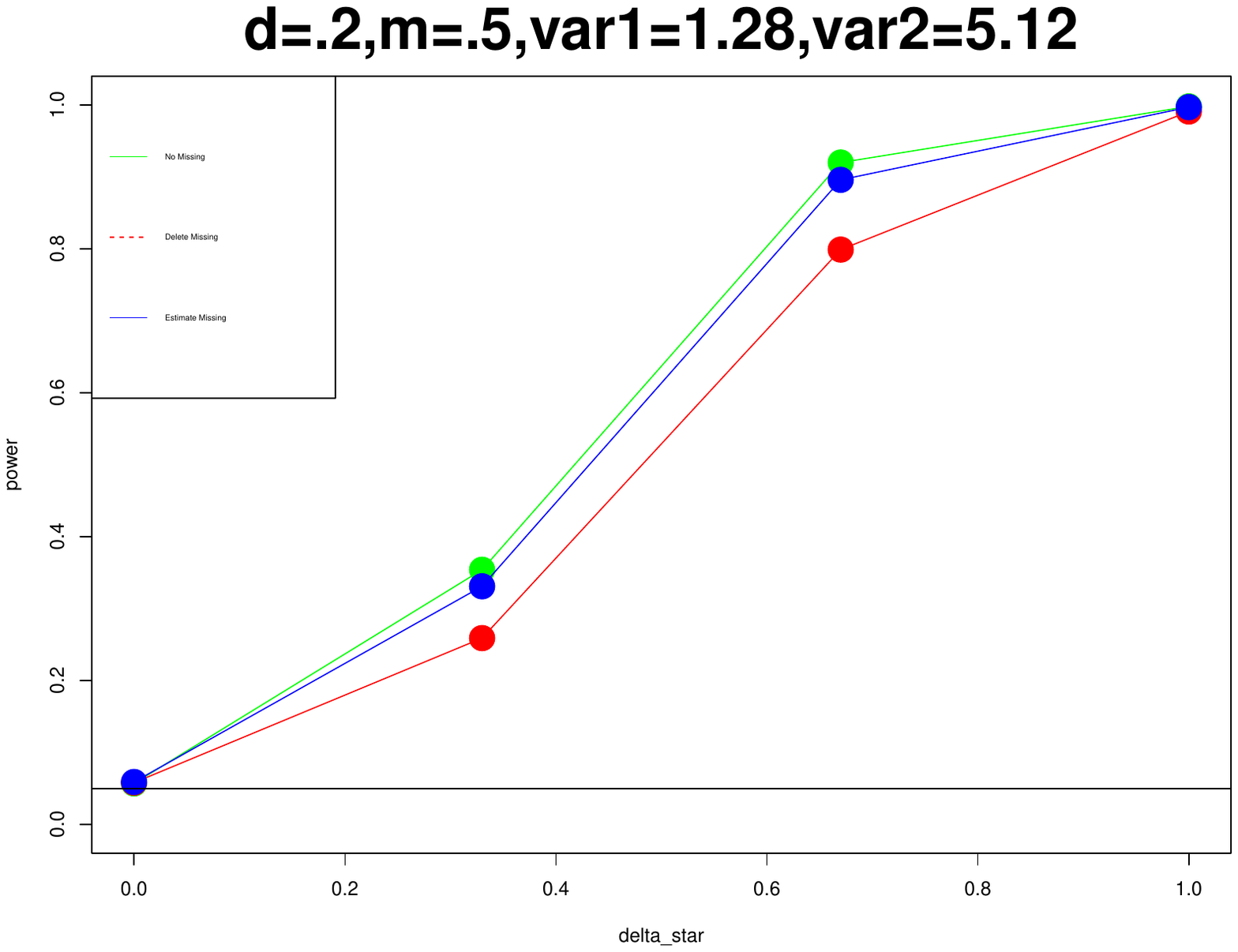}
\includegraphics[width = 2.3in, height = 1.5in]{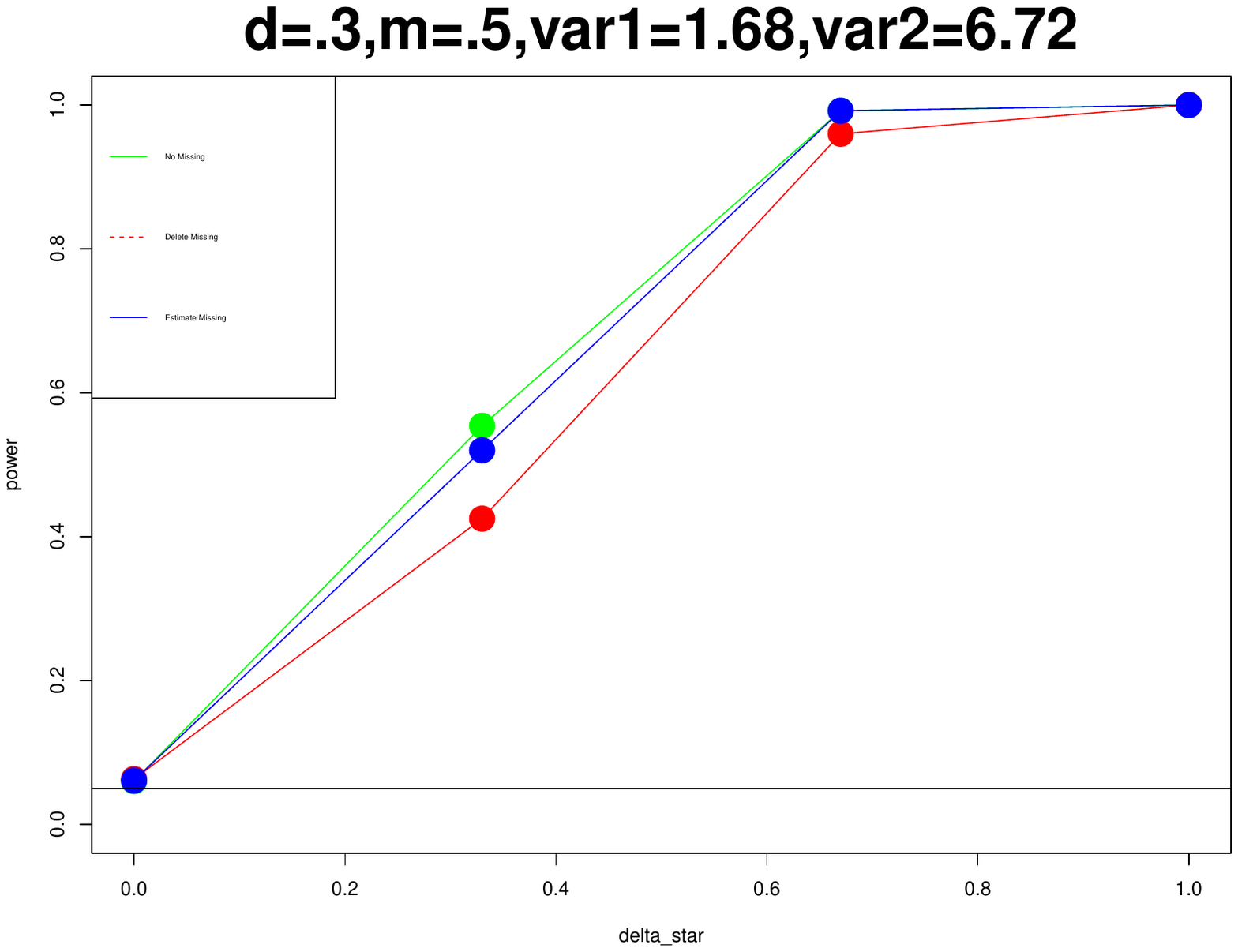}

\hspace{1.5cm}
$d=.1 \quad m=.1$
\hspace{3cm}
$d=.2 \quad m=.1$
\hspace{3cm}
$d=.3 \quad m=.1$

\includegraphics[width = 2.3in, height = 1.5in]{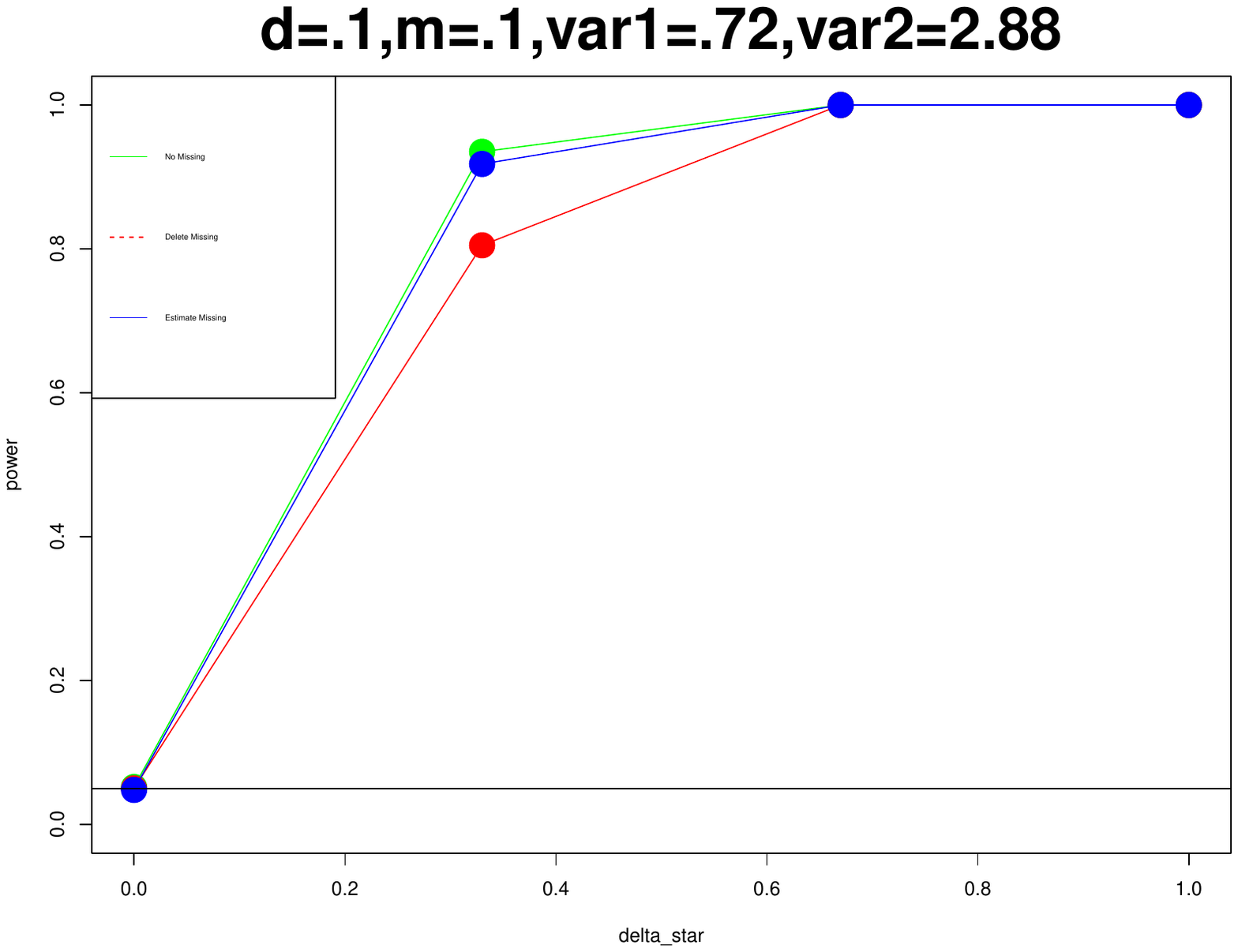}
\includegraphics[width = 2.3in, height = 1.5in]{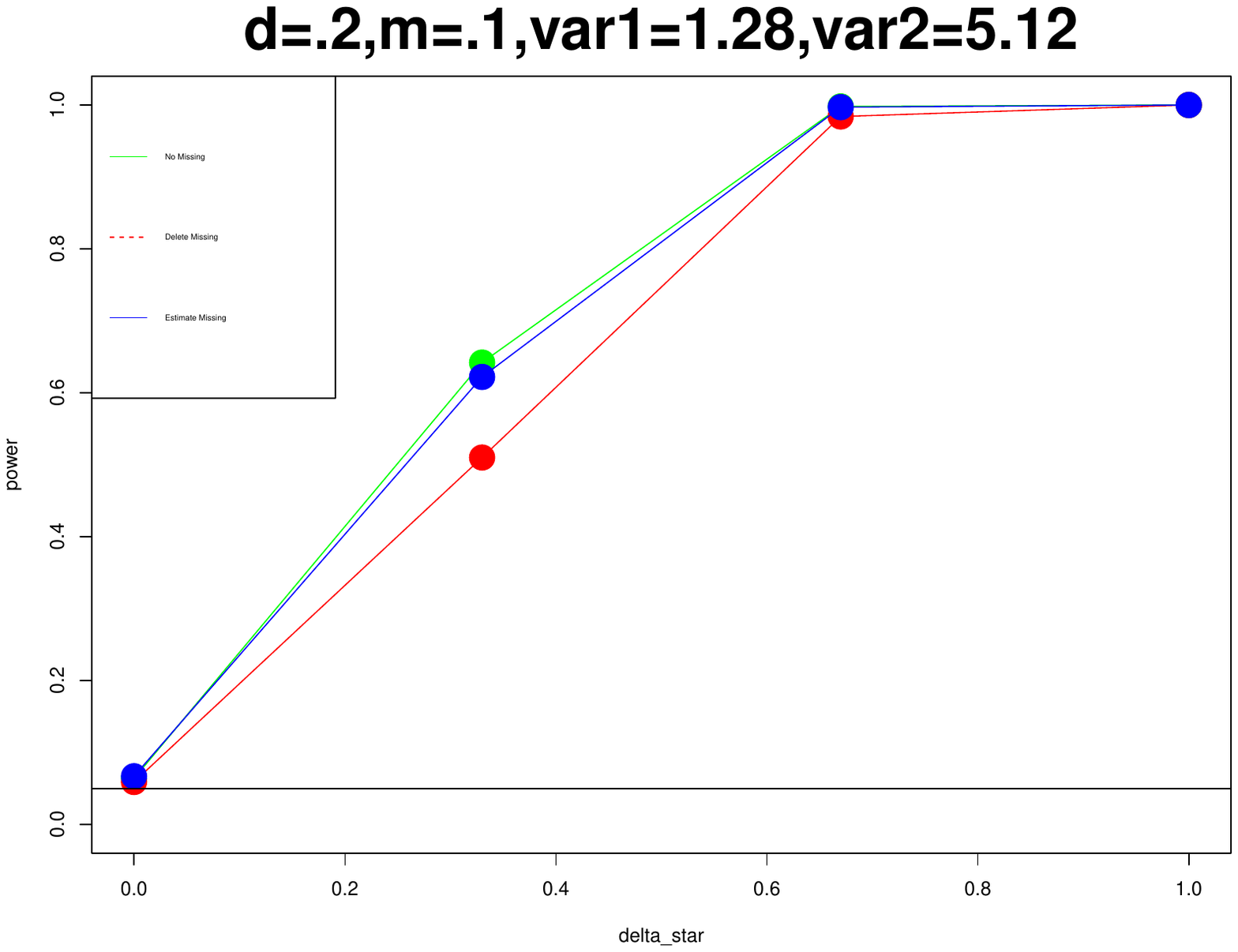}
\includegraphics[width = 2.3in, height = 1.5in]{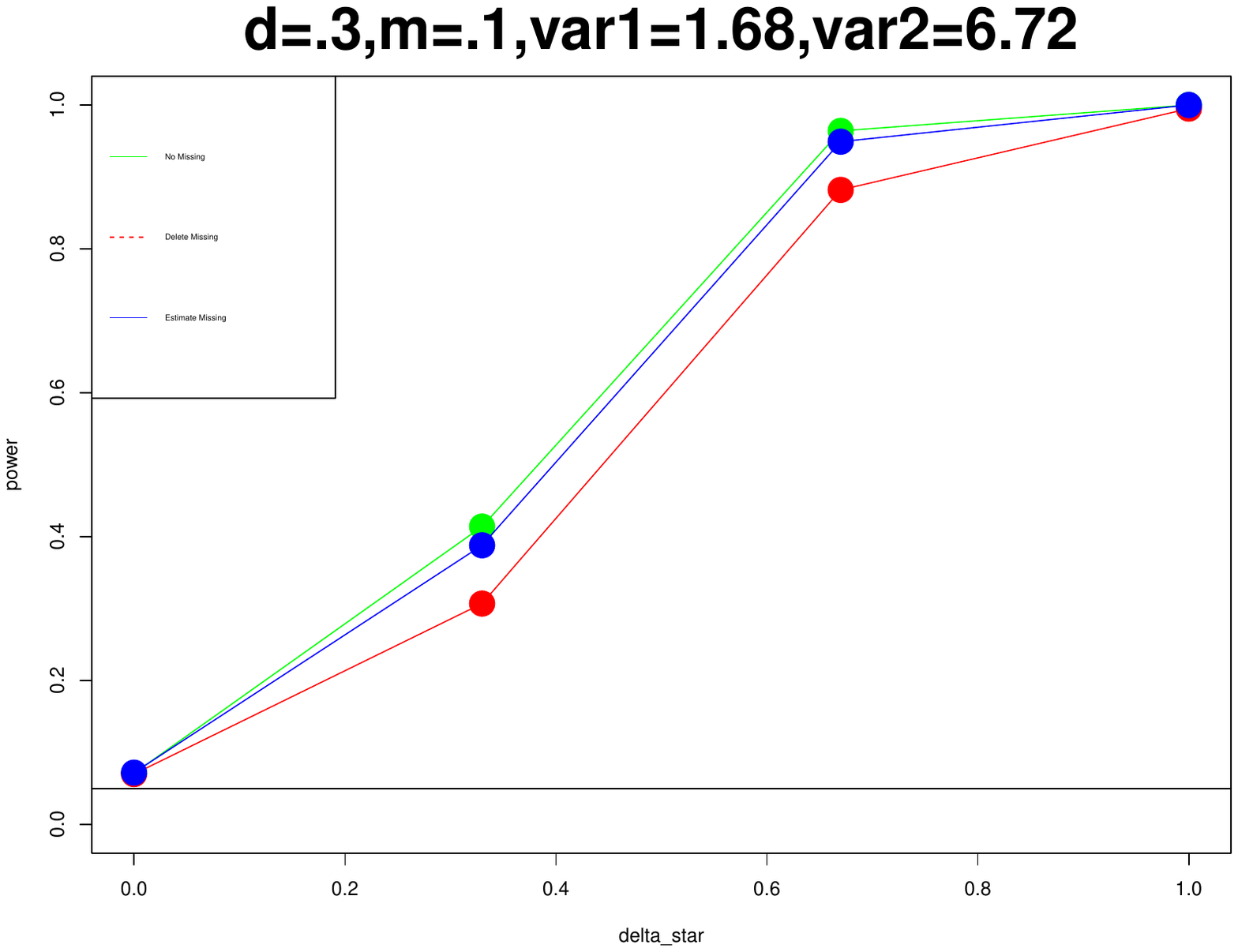}

\subsubsection{When one Trait has Normal Distribution and other Trait has Poisson Distribution}

The following tables represents powers for the three strategies evaluated at $\delta = 0, .33, .67$ and 1.

\textbf{ For $\rho_1 > \rho_2$}

\vspace{.4cm}

\hspace{3.5cm}$m=.1$
\hspace{8.5cm}$m=.5$

\vspace{.2cm}

\begin{tabular}{|c|c|c|c|c|}
\hline 
Strategy & $\delta = 0$ & $\delta = .33$ & $\delta = .67$ & $\delta = 1$ \\ 
\hline 
use same & 0.046 & 0.464 & 0.975 & 1 \\ 
\hline 
use other & 0.047 & 0.471 & 0.978 & 1 \\ 
\hline 
use both & 0.046 & 0.470 & 0.978 & 1 \\ 
\hline 
\end{tabular} \hspace{1.5cm}
\begin{tabular}{|c|c|c|c|c|}
\hline 
Strategy & $\delta = 0$ & $\delta = .33$ & $\delta = .67$ & $\delta = 1$ \\ 
\hline 
use same & 0.051 & 0.742 & 1 & 1 \\ 
\hline 
use other & 0.056 & 0.743 & 1 & 1 \\ 
\hline 
use both & 0.053 & 0.740 & 1 & 1 \\ 
\hline 
\end{tabular} 

\hspace{.6cm}

\textbf{ For $\rho_1 < \rho_2$}

\pagebreak

\hspace{3.5cm}$m=.1$
\hspace{8.5cm}$m=.5$

\vspace{.2cm}

\begin{tabular}{|c|c|c|c|c|}
\hline 
Strategy & $\delta = 0$ & $\delta = .33$ & $\delta = .67$ & $\delta = 1$ \\ 
\hline 
use same & 0.051 & 0.394 & 0.935 & 1 \\ 
\hline 
use other & 0.052 & 0.385 & 0.933 & 1 \\ 
\hline 
use both & 0.046 & 0.388 & 0.936 & 1 \\ 
\hline 
\end{tabular} \hspace{1.5cm}
\begin{tabular}{|c|c|c|c|c|}
\hline 
Strategy & $\delta = 0$ & $\delta = .33$ & $\delta = .67$ & $\delta = 1$ \\ 
\hline 
use same & 0.048 & 0.900 & 1 & 1 \\ 
\hline 
use other & 0.050 & 0.903 & 1 & 1 \\ 
\hline 
use both & 0.051 & 0.900 & 1 & 1 \\ 
\hline 
\end{tabular}

\hspace{.4cm}

\textbf{ For $\rho_1 = \rho_2$}

\vspace{.4cm}

\hspace{3.5cm}$m=.1$
\hspace{8.5cm}$m=.5$

\vspace{.2cm}

\begin{tabular}{|c|c|c|c|c|}
\hline 
Strategy & $\delta = 0$ & $\delta = .33$ & $\delta = .67$ & $\delta = 1$ \\ 
\hline 
use same & 0.050 & 0.514 & 0.988 & 1 \\ 
\hline 
use other & 0.054 & 0.519 & 0.989 & 1 \\ 
\hline 
use both & 0.048 & 0.513 & 0.987 & 1 \\ 
\hline 
\end{tabular} \hspace{1.5cm}
\begin{tabular}{|c|c|c|c|c|}
\hline 
Strategy & $\delta = 0$ & $\delta = .33$ & $\delta = .67$ & $\delta = 1$ \\ 
\hline 
use same & 0.049 & 0.815 & 1 & 1 \\ 
\hline 
use other & 0.048 & 0.830 & 1 & 1 \\ 
\hline 
use both & 0.051 & 0.824 & 1 & 1 \\ 
\hline 
\end{tabular} 

\vspace{1cm}

According to the above results we can conclude-

\begin{center}
\begin{tabular}{|c|c|}
\hline 
Case & Best Strategy \\ 
\hline 
$\rho_1 > \rho_2$ & use other \\ 
\hline 
$\rho_1 < \rho_2$ & use same \\ 
\hline 
$\rho_1 = \rho_2$ & use other \\ 
\hline 
\end{tabular} 

\end{center}

Now we go for power comparison among no missing, estimated missing and deleted missing.

We generate 1st trait from normal distribution (section 6.2) with the parameters $\alpha = \alpha_1, \beta = \beta_1, \sigma = \sigma_1$ keeping $p^\star = p^\star_1$ and we generate 2nd trait from poisson distribution (section 6.2) with the parameters $\alpha = \alpha_2, \beta = \beta_2, \lambda=\lambda_2$ keeping $p^\star = p^\star_2$.

We have done simulation for three choices of $d$ as .1, .2, .3 and for each $d$ we take $(p^\star_1, p^\star_2)$ as (.1, .2), (.2, .2).

We take $\alpha_1 = 5, \alpha_2 = 10, \beta_1 = 1, and \beta_2 = 2$ and varied $\sigma_1$ and $\lambda_2$ in the following way,

\vspace{.4cm}

\begin{tabular}{|c|c|c|}
\hline 
d & $(p^\star_1, p^\star_2)$ & ($\sigma_1^2, \lambda_2$) \\ 
\hline 
.1 & (.1, .2) & (1.62, 2.88) \\ 
\hline 
.1 & (.2, .2) & (.72, 2.88) \\ 
\hline
.1 & (.2, .1) & (.72, 6.48) \\ 
\hline 

\end{tabular} \hspace{1.5cm}
\begin{tabular}{|c|c|c|}
\hline 
d & $(p^\star_1, p^\star_2)$ & ($\sigma_1^2, \lambda_2$) \\ 
\hline 
.2 & (.1, .2) & (2.88, 5.12) \\ 
\hline 
.2 & (.2, .2) & (1.28, 5.12) \\ 
\hline
.2 & (.2, .1) & (1.28, 11.52) \\ 
\hline 

\end{tabular} \hspace{1.5cm}
\begin{tabular}{|c|c|c|}
\hline 
d & $(p^\star_1, p^\star_2)$ & ($\sigma_1^2, \lambda_2$) \\ 
\hline 
.3 & (.1, .2) & (3.78, 6.72) \\ 
\hline 
.3 & (.2, .2) & (1.68, 6.72) \\ 
\hline
.3 & (.2, .1) & (1.68, 15.12) \\ 
\hline 
\end{tabular} 

\vspace{.4cm}

We replicate these for $m =$ .1 and .2.

Note that for each of the following simulations we have calculated $\rho_1$ and $\rho_2$ and accordingly we have used the best imputation strategy.

\vspace{.4cm}

\hspace{1.5cm}
$d=.1 \quad m=.5$
\hspace{3cm}
$d=.2 \quad m=.5$
\hspace{3cm}
$d=.3 \quad m=.5$

\includegraphics[width = 2.3in, height = 1.5in]{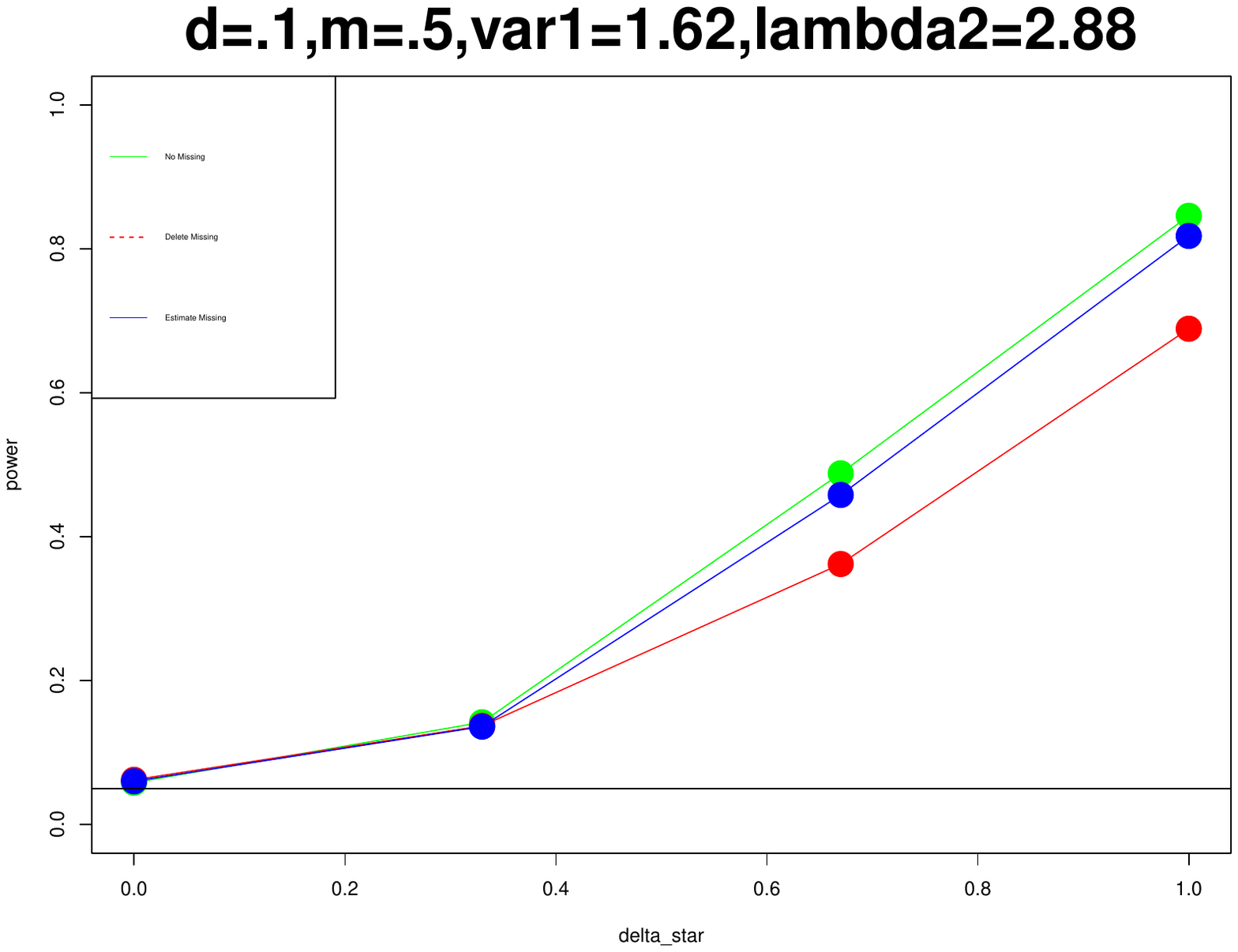}
\includegraphics[width = 2.3in, height = 1.5in]{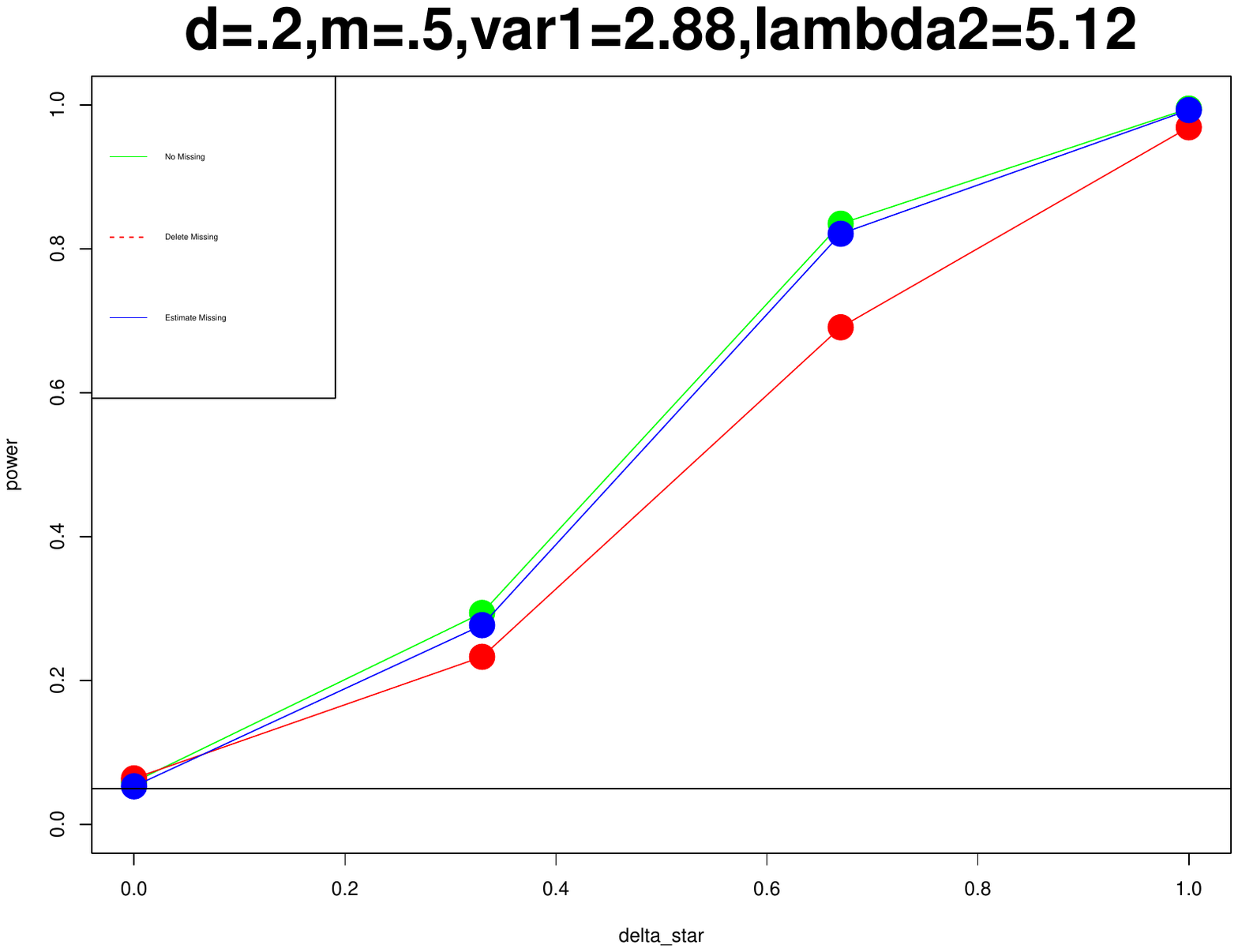}
\includegraphics[width = 2.3in, height = 1.5in]{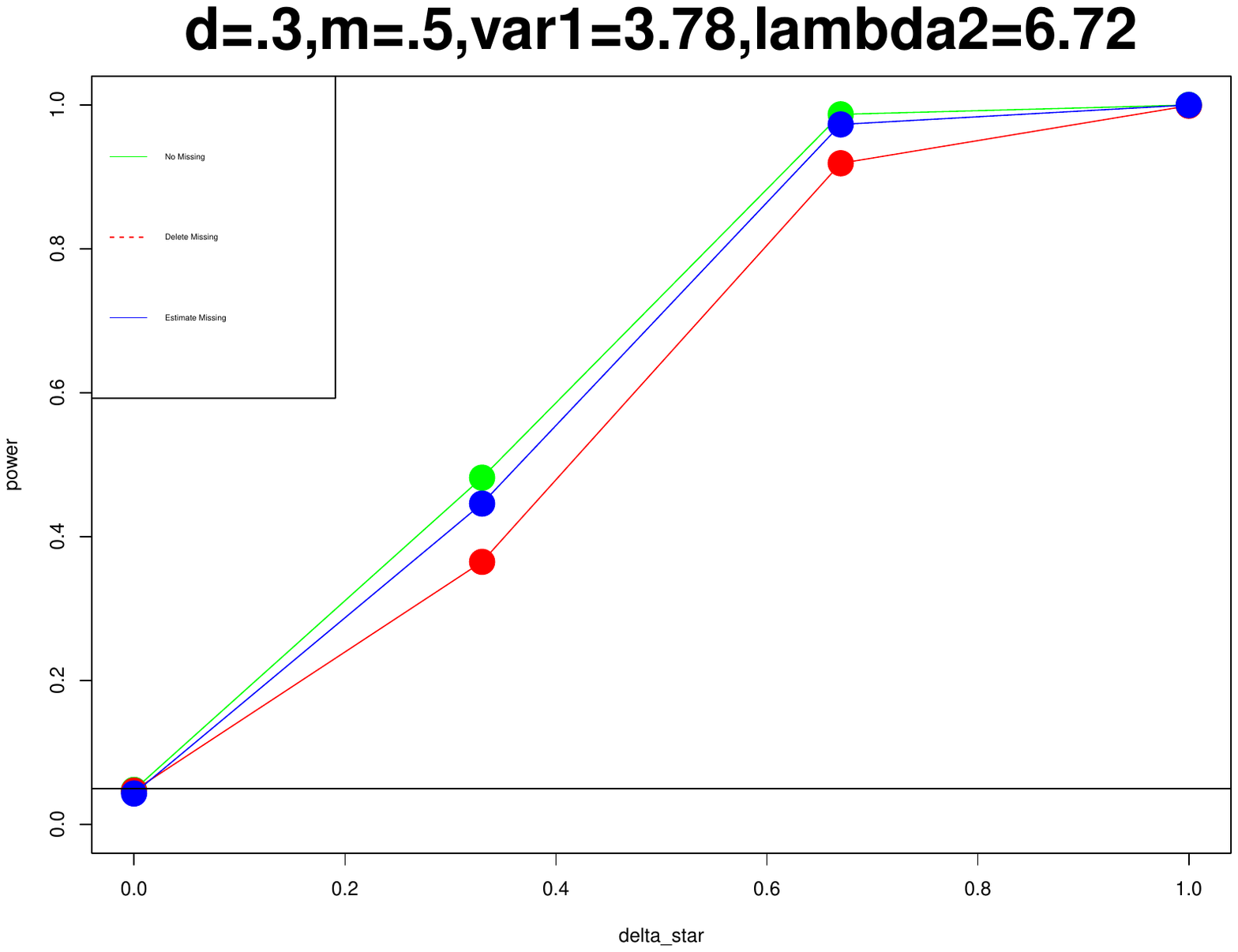}

\hspace{1.5cm}
$d=.1 \quad m=.1$
\hspace{3cm}
$d=.2 \quad m=.1$
\hspace{3cm}
$d=.3 \quad m=.1$

\includegraphics[width = 2.3in, height = 1.5in]{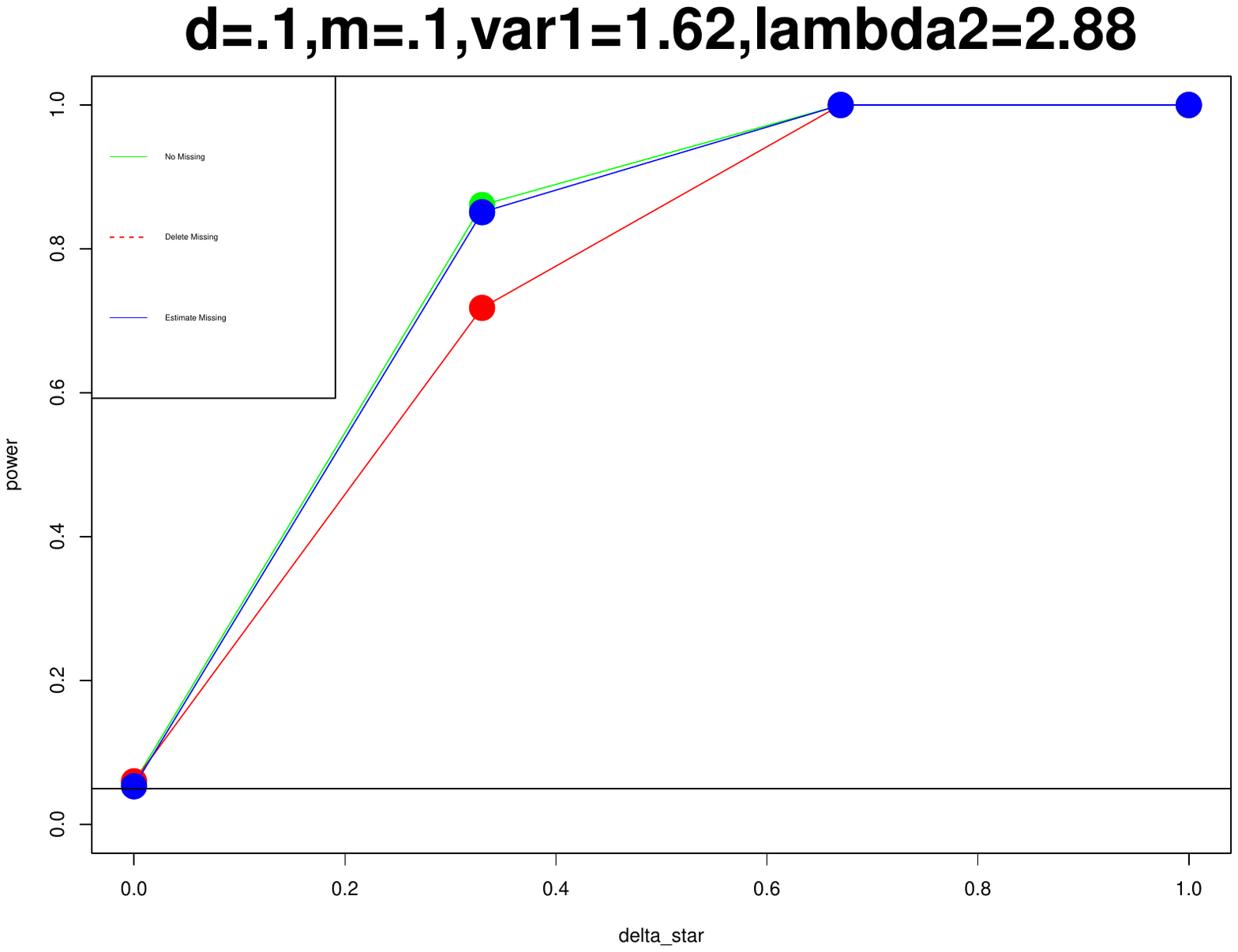}
\includegraphics[width = 2.3in, height = 1.5in]{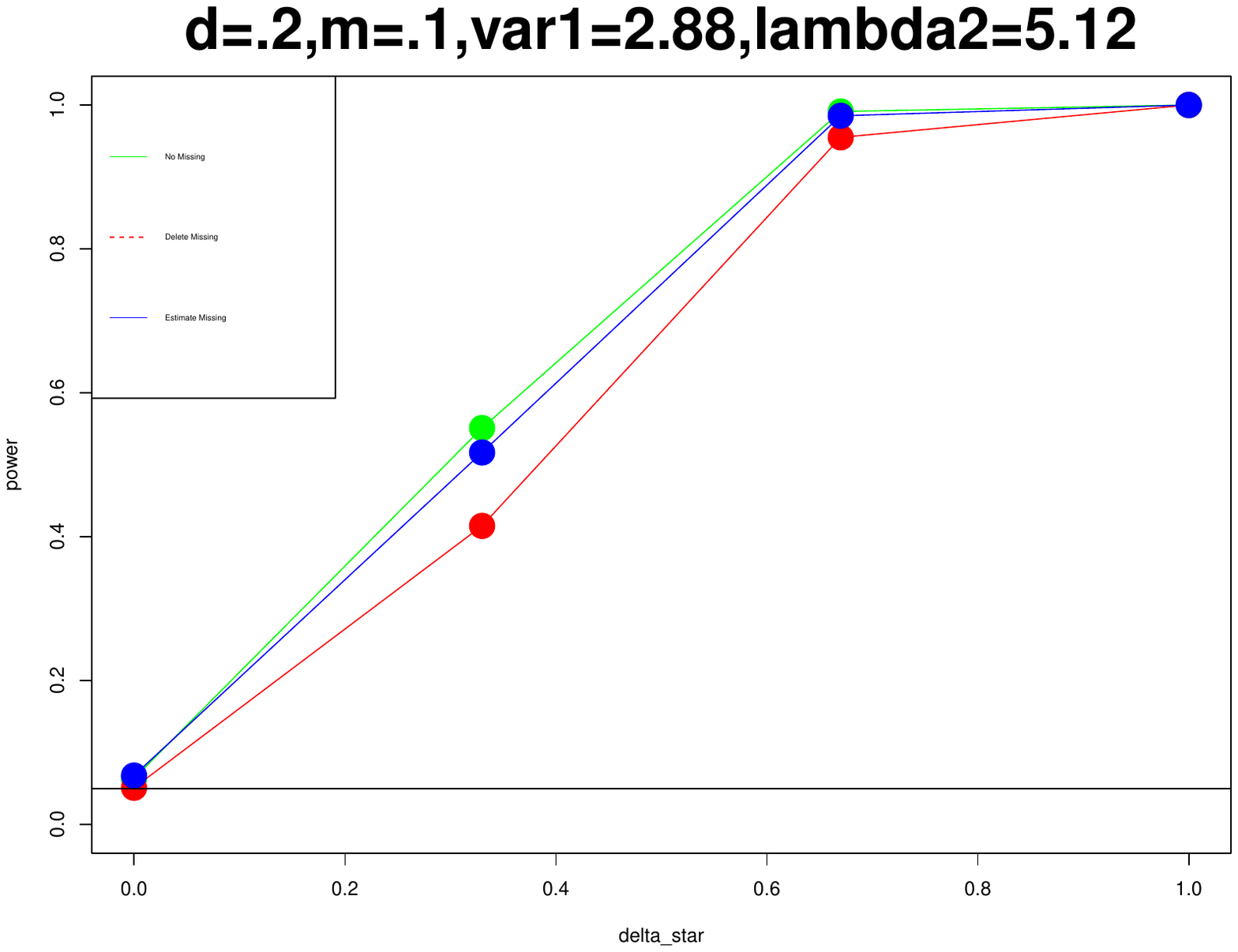}
\includegraphics[width = 2.3in, height = 1.5in]{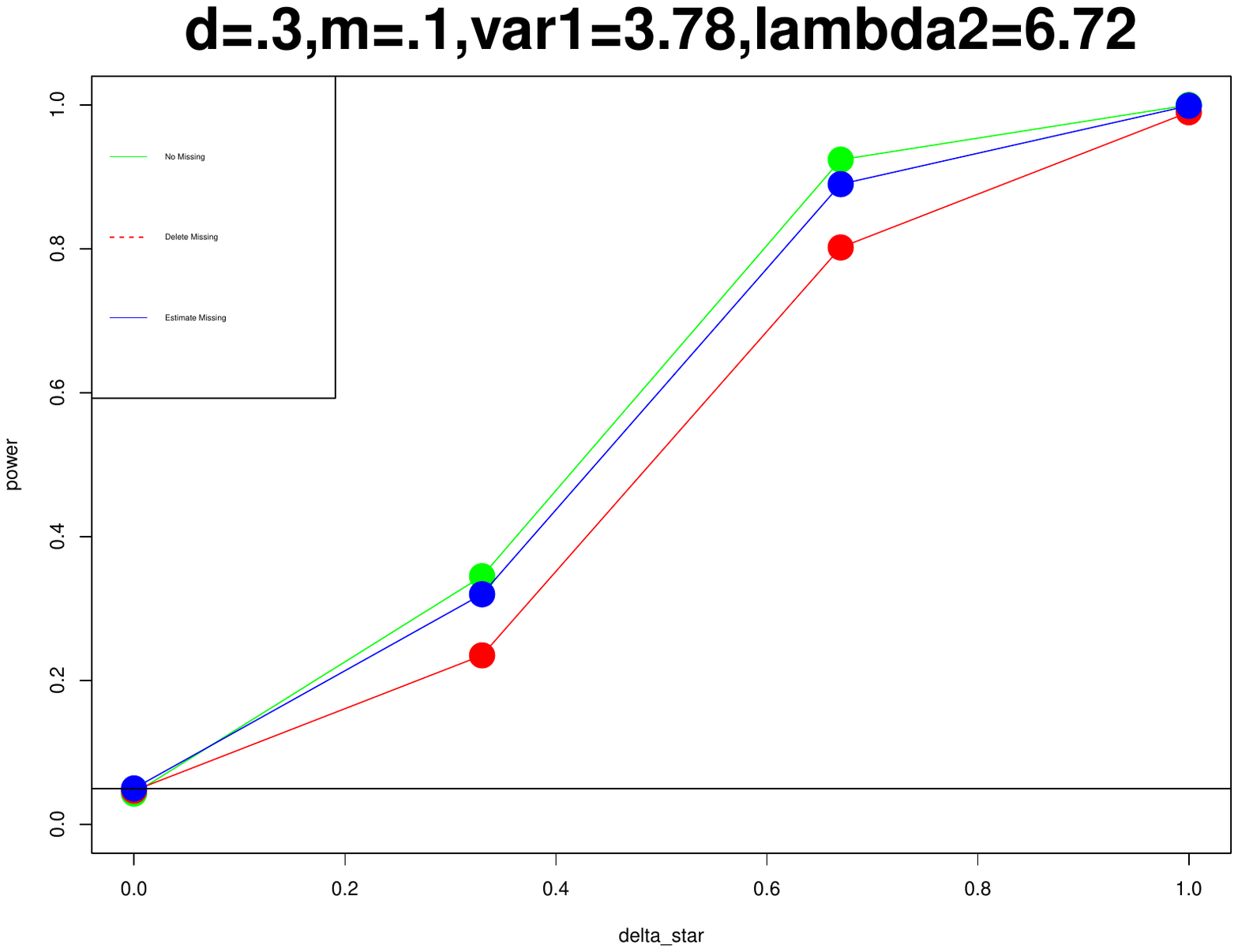}

\hspace{1.5cm}
$d=.1 \quad m=.5$
\hspace{3cm}
$d=.2 \quad m=.5$
\hspace{3cm}
$d=.3 \quad m=.5$

\includegraphics[width = 2.3in, height = 1.5in]{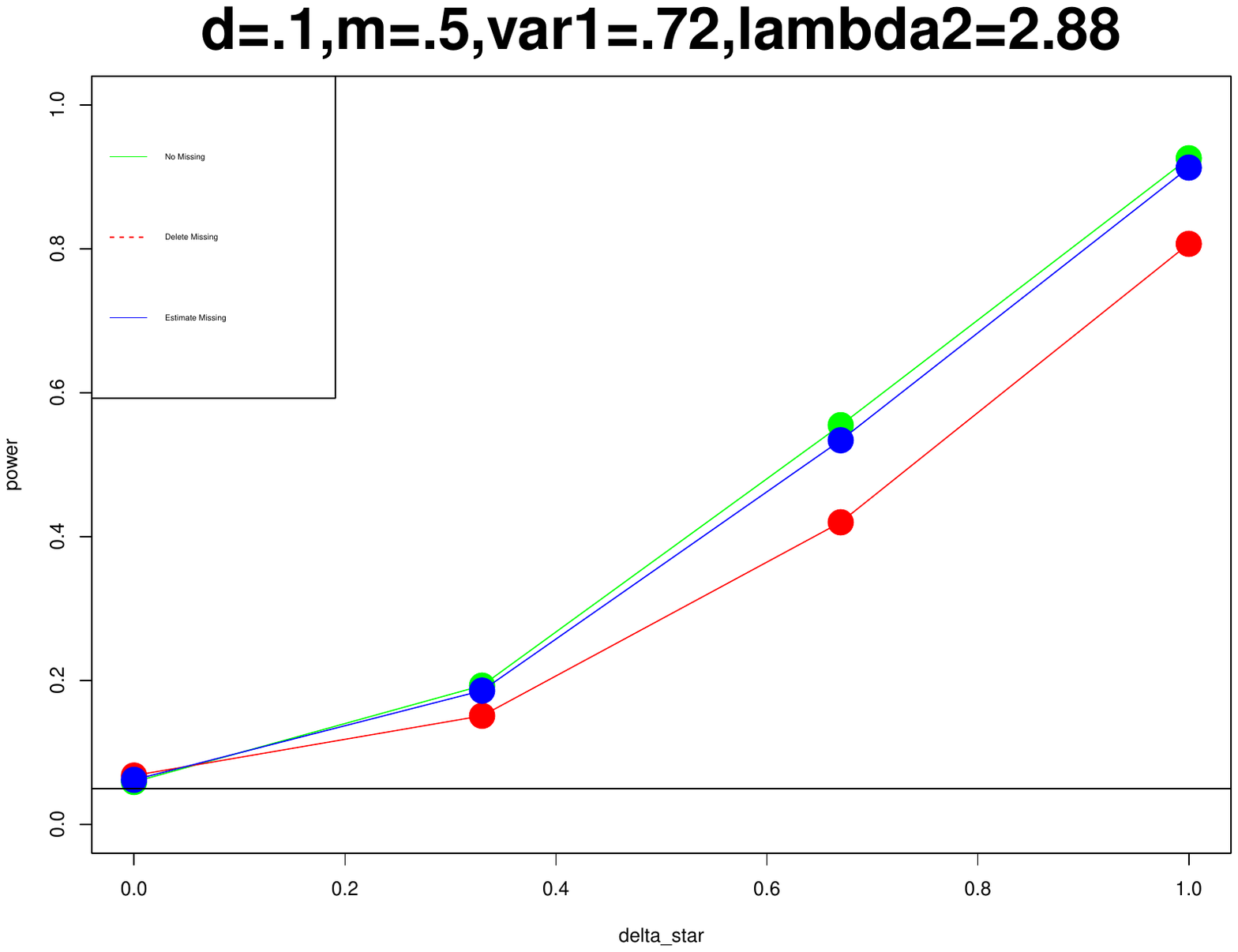}
\includegraphics[width = 2.3in, height = 1.5in]{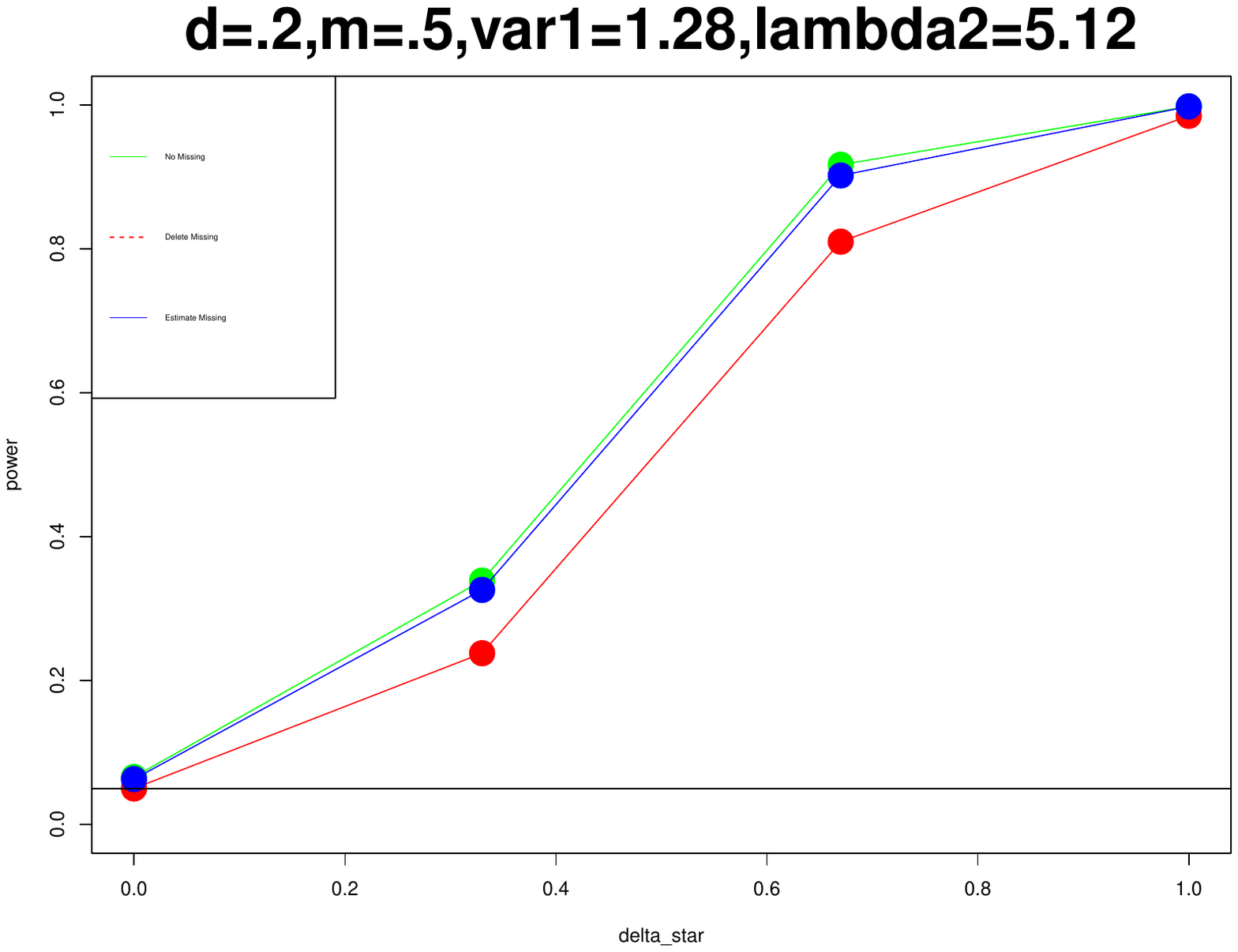}
\includegraphics[width = 2.3in, height = 1.5in]{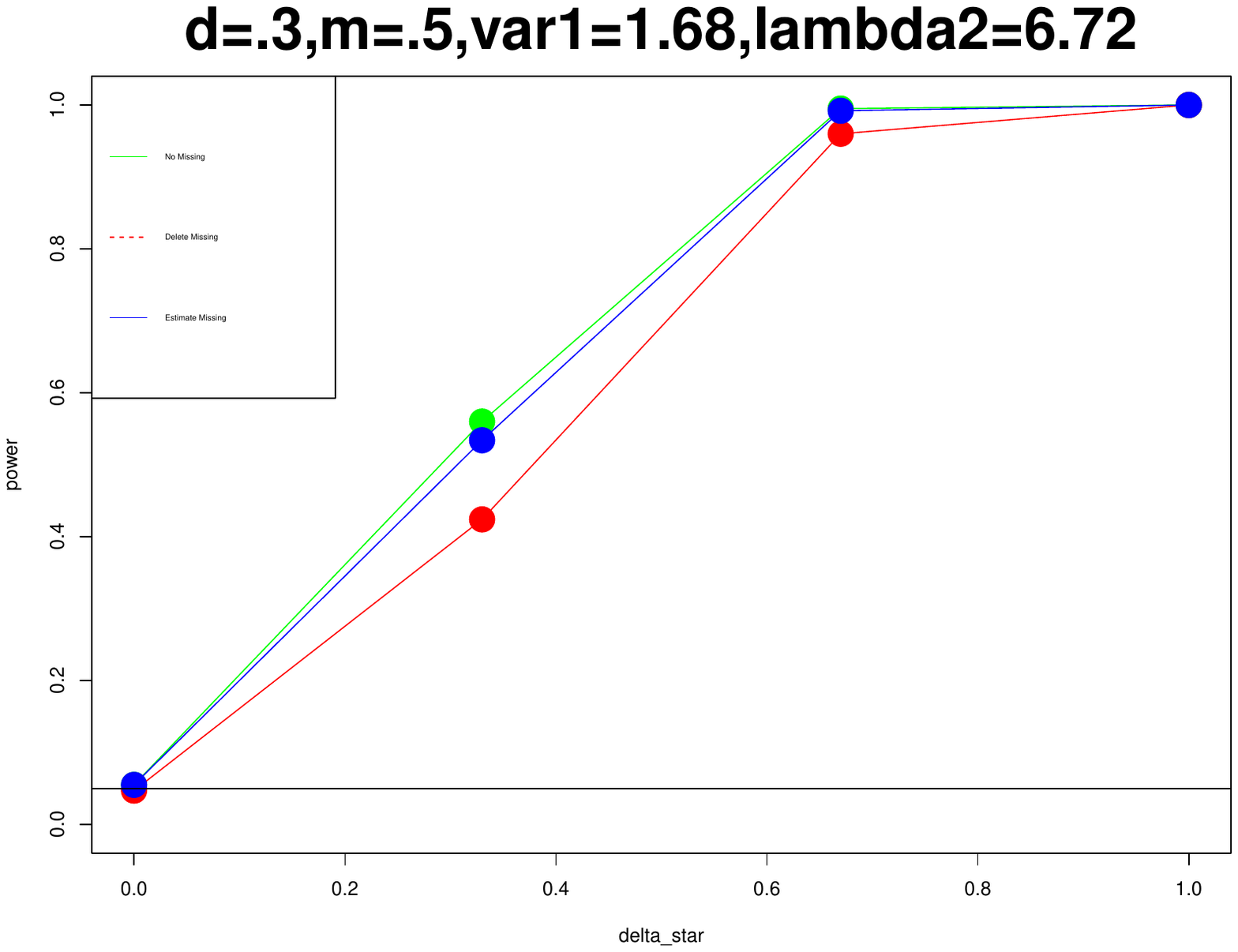}

\hspace{1.5cm}
$d=.1 \quad m=.1$
\hspace{3cm}
$d=.2 \quad m=.1$
\hspace{3cm}
$d=.3 \quad m=.1$

\includegraphics[width = 2.3in, height = 1.5in]{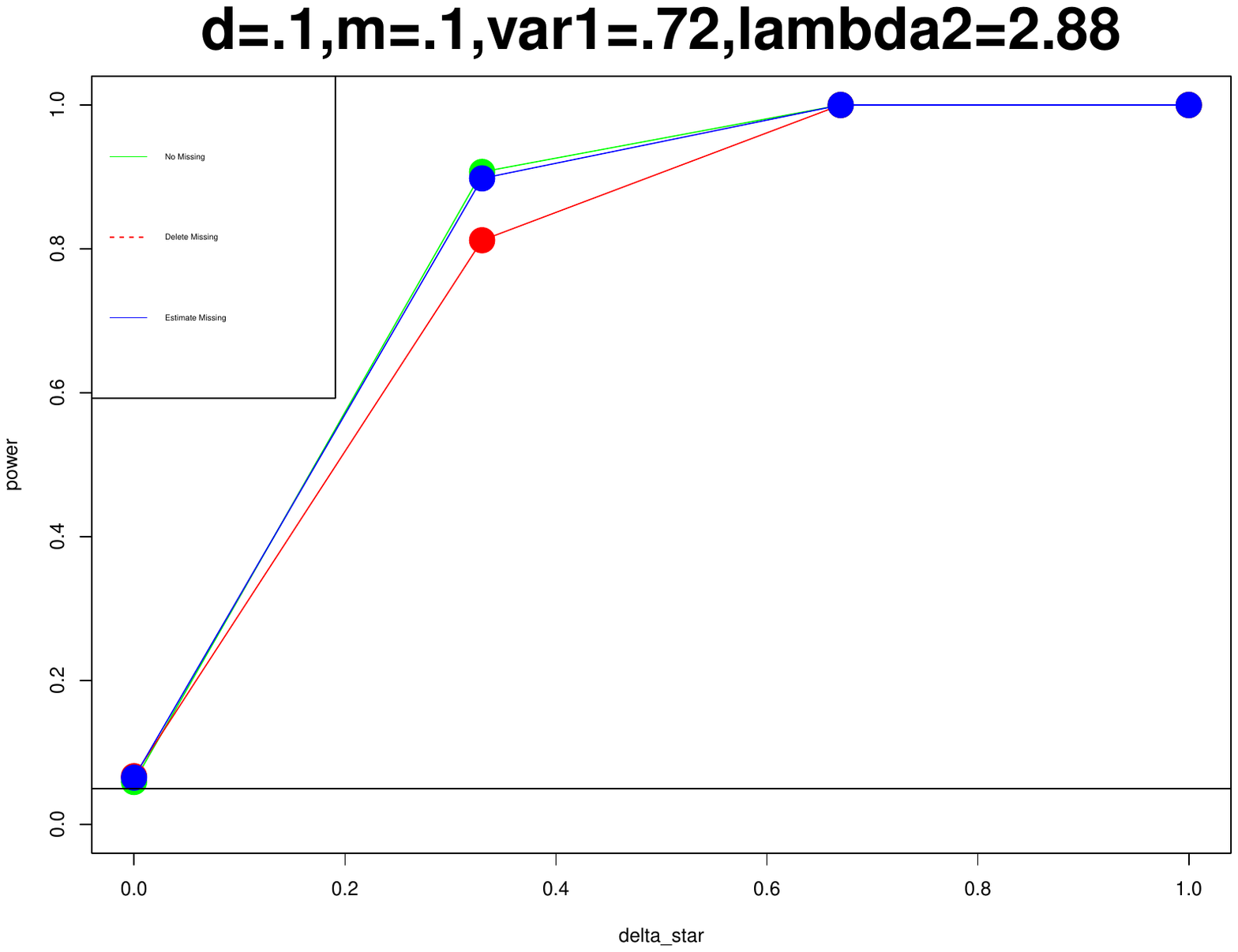}
\includegraphics[width = 2.3in, height = 1.5in]{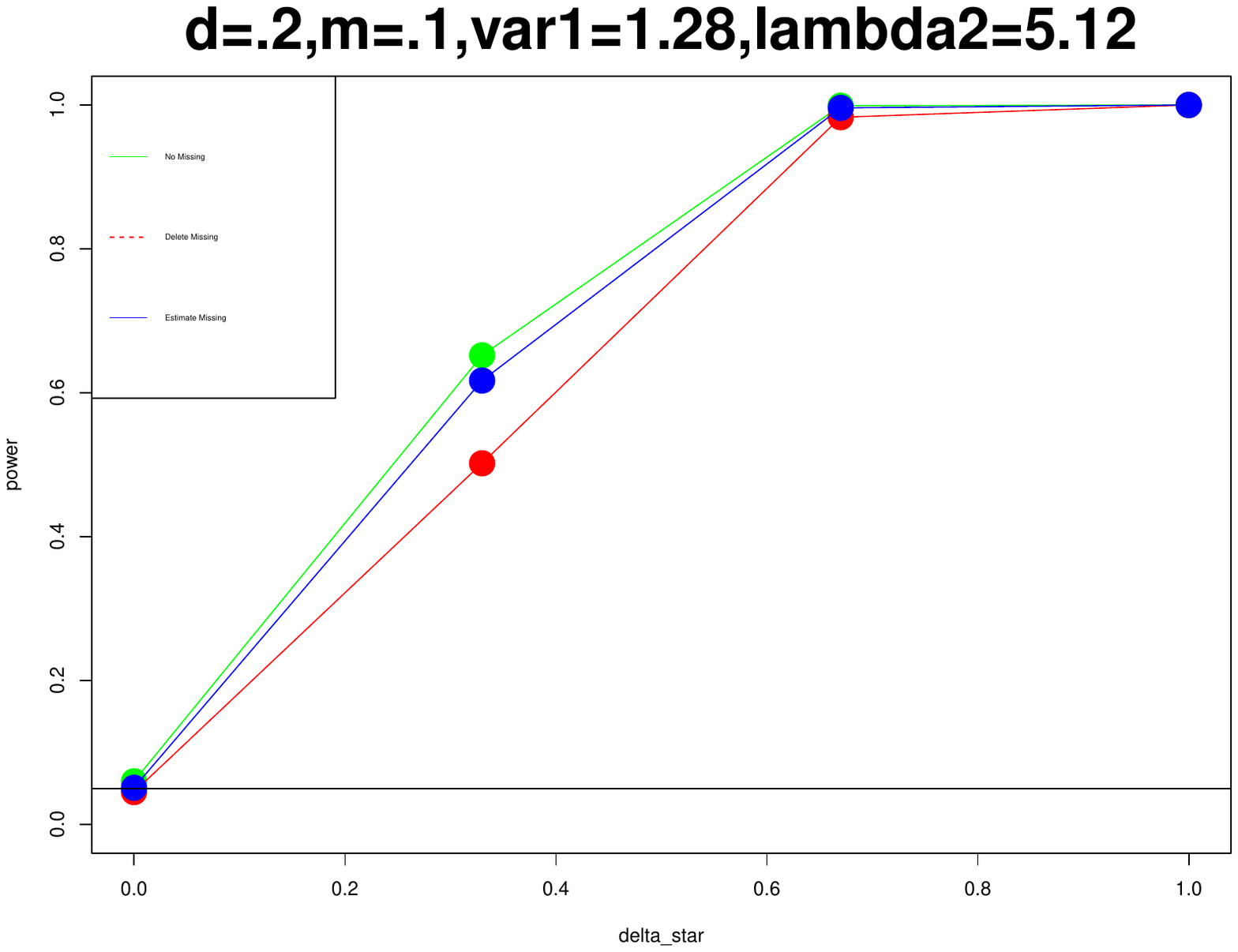}
\includegraphics[width = 2.3in, height = 1.5in]{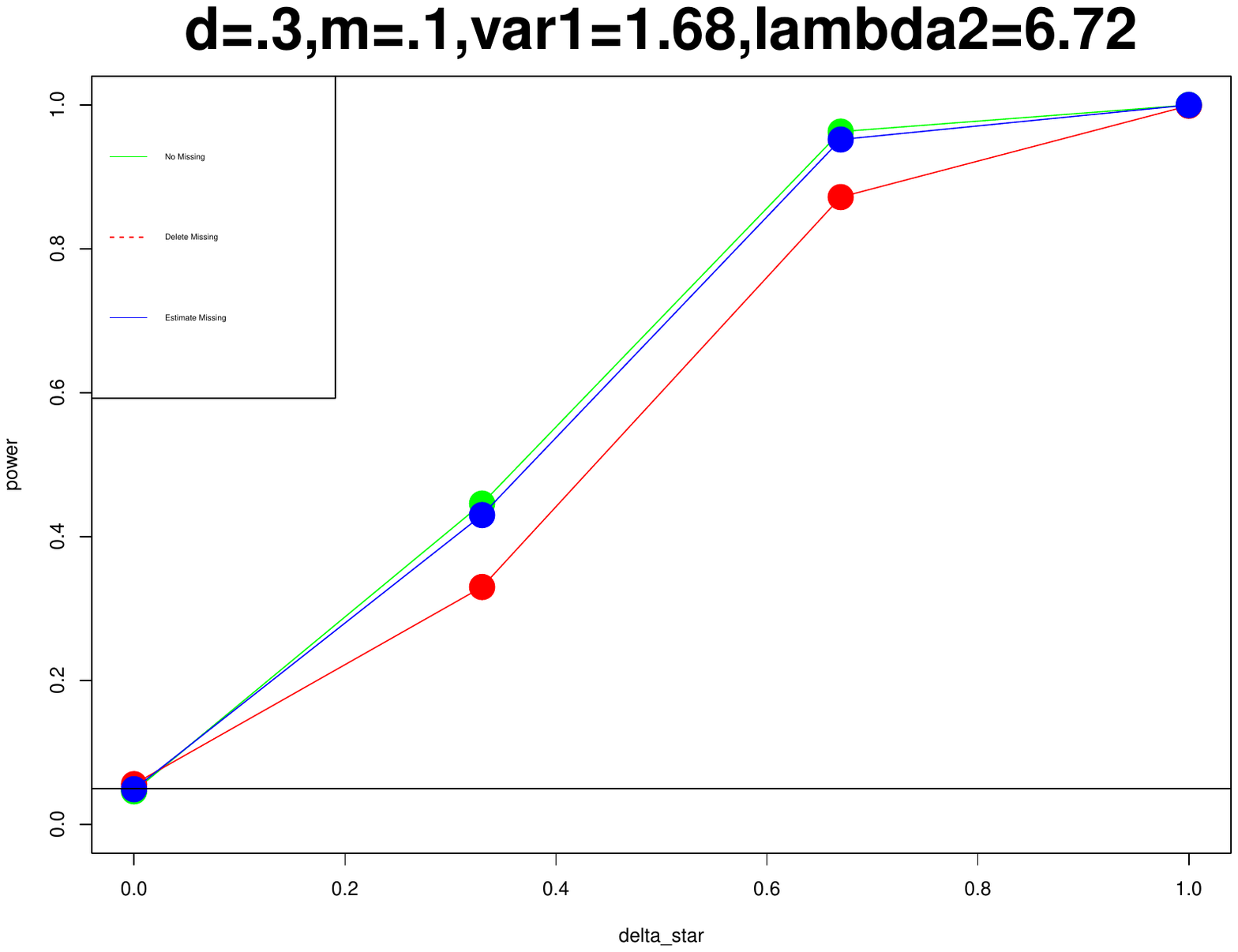}

\hspace{1.5cm}
$d=.1 \quad m=.5$
\hspace{3cm}
$d=.2 \quad m=.5$
\hspace{3cm}
$d=.3 \quad m=.5$

\includegraphics[width = 2.3in, height = 1.5in]{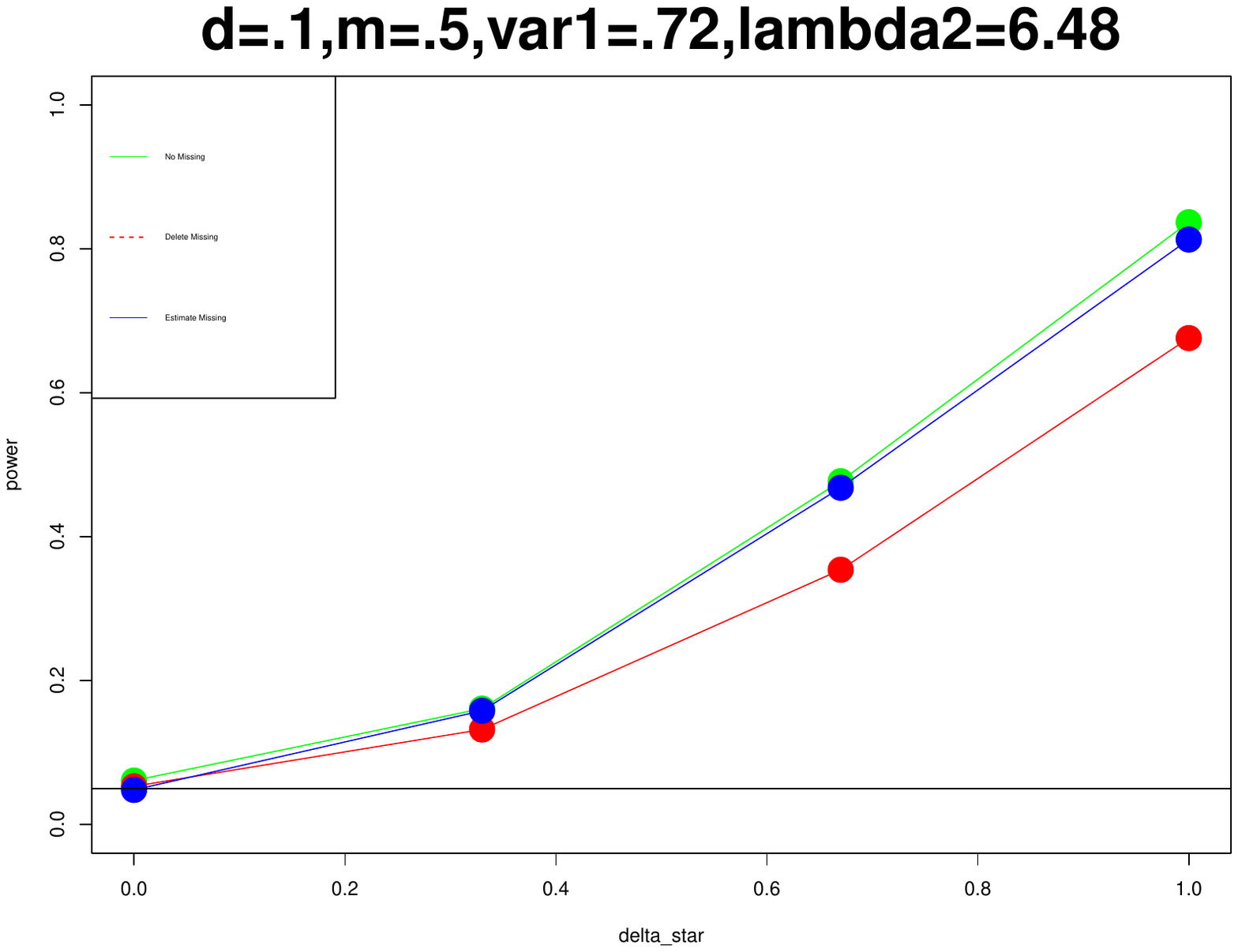}
\includegraphics[width = 2.3in, height = 1.5in]{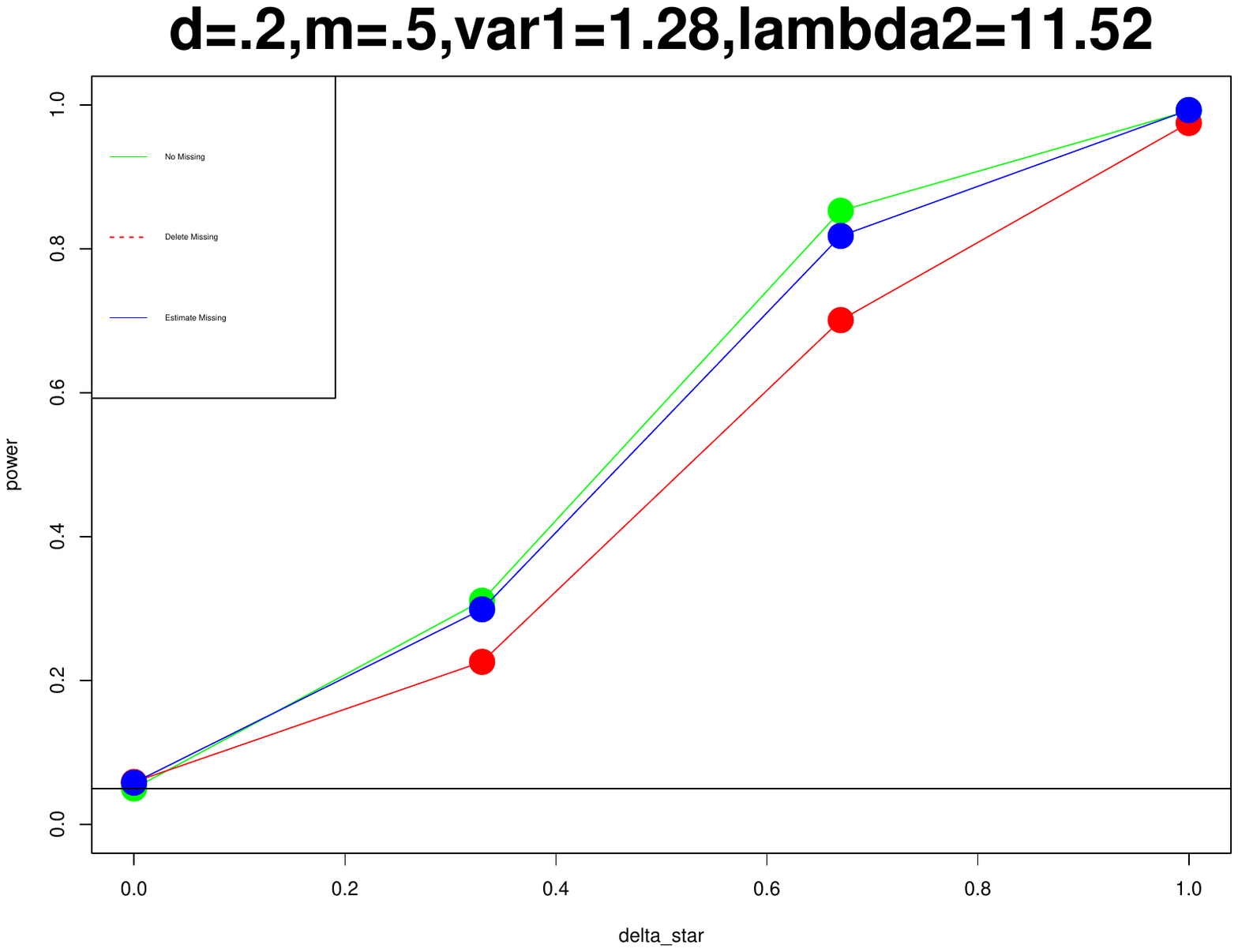}
\includegraphics[width = 2.3in, height = 1.5in]{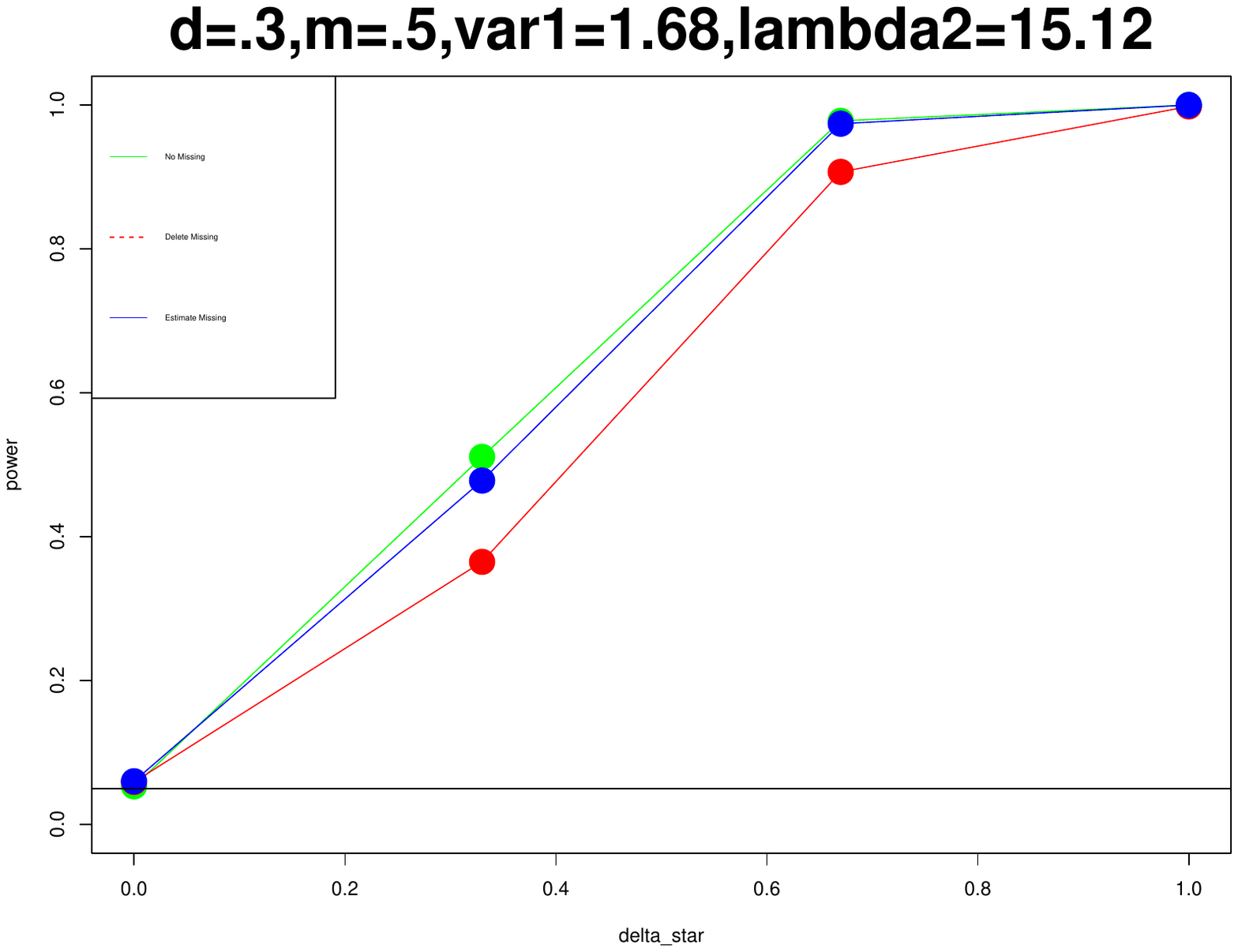}

\hspace{1.5cm}
$d=.1 \quad m=.1$
\hspace{3cm}
$d=.2 \quad m=.1$
\hspace{3cm}
$d=.3 \quad m=.1$

\includegraphics[width = 2.3in, height = 1.5in]{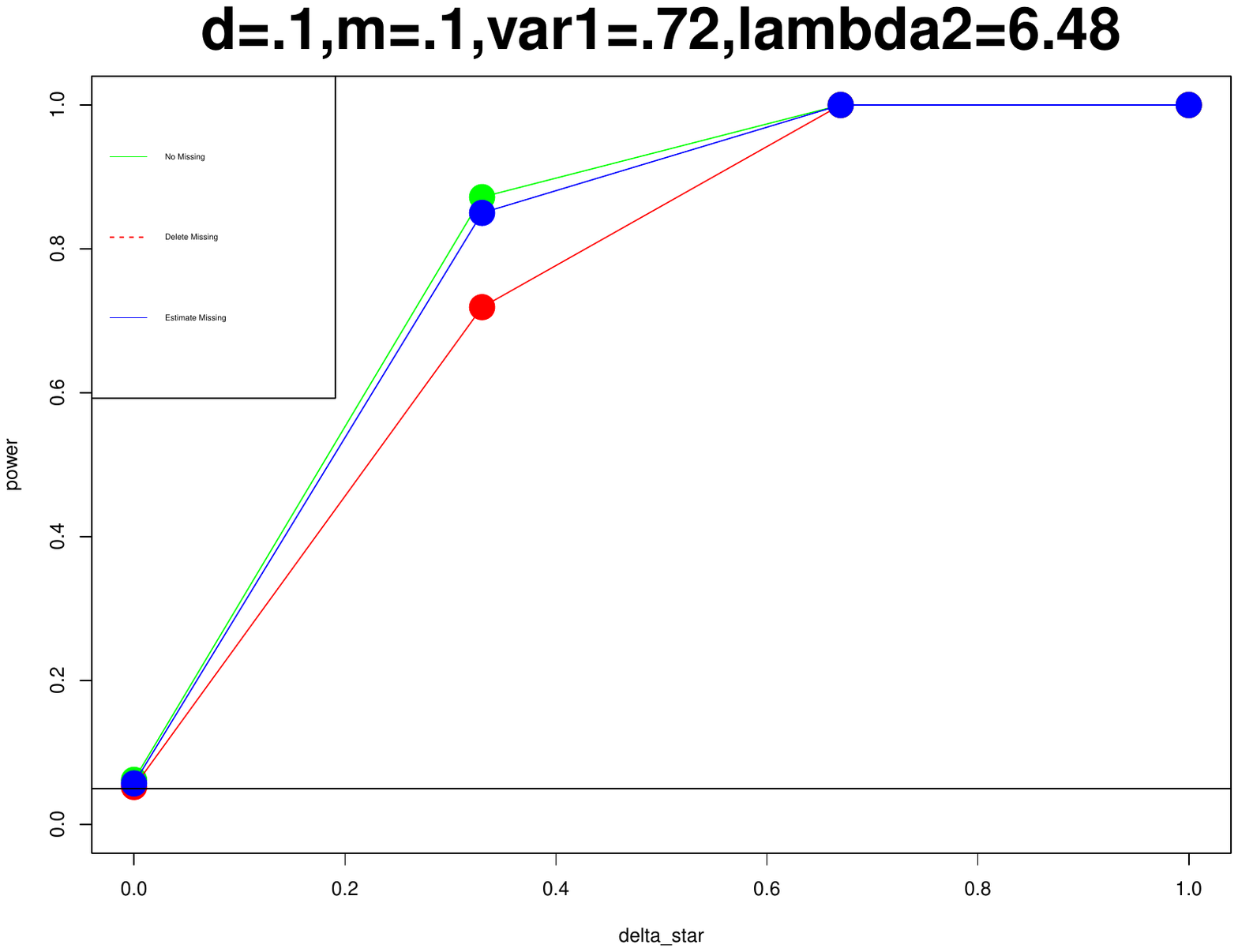}
\includegraphics[width = 2.3in, height = 1.5in]{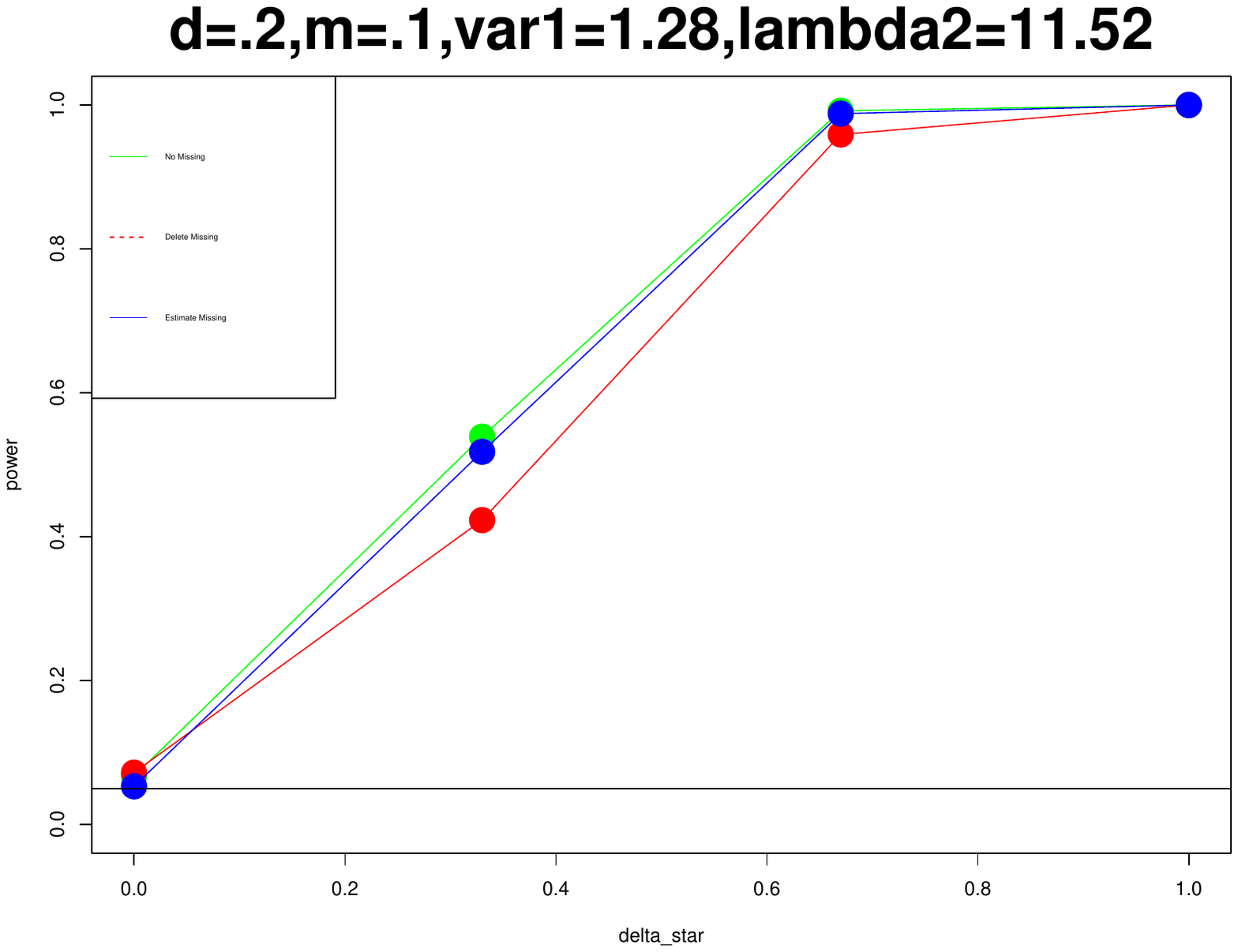}
\includegraphics[width = 2.3in, height = 1.5in]{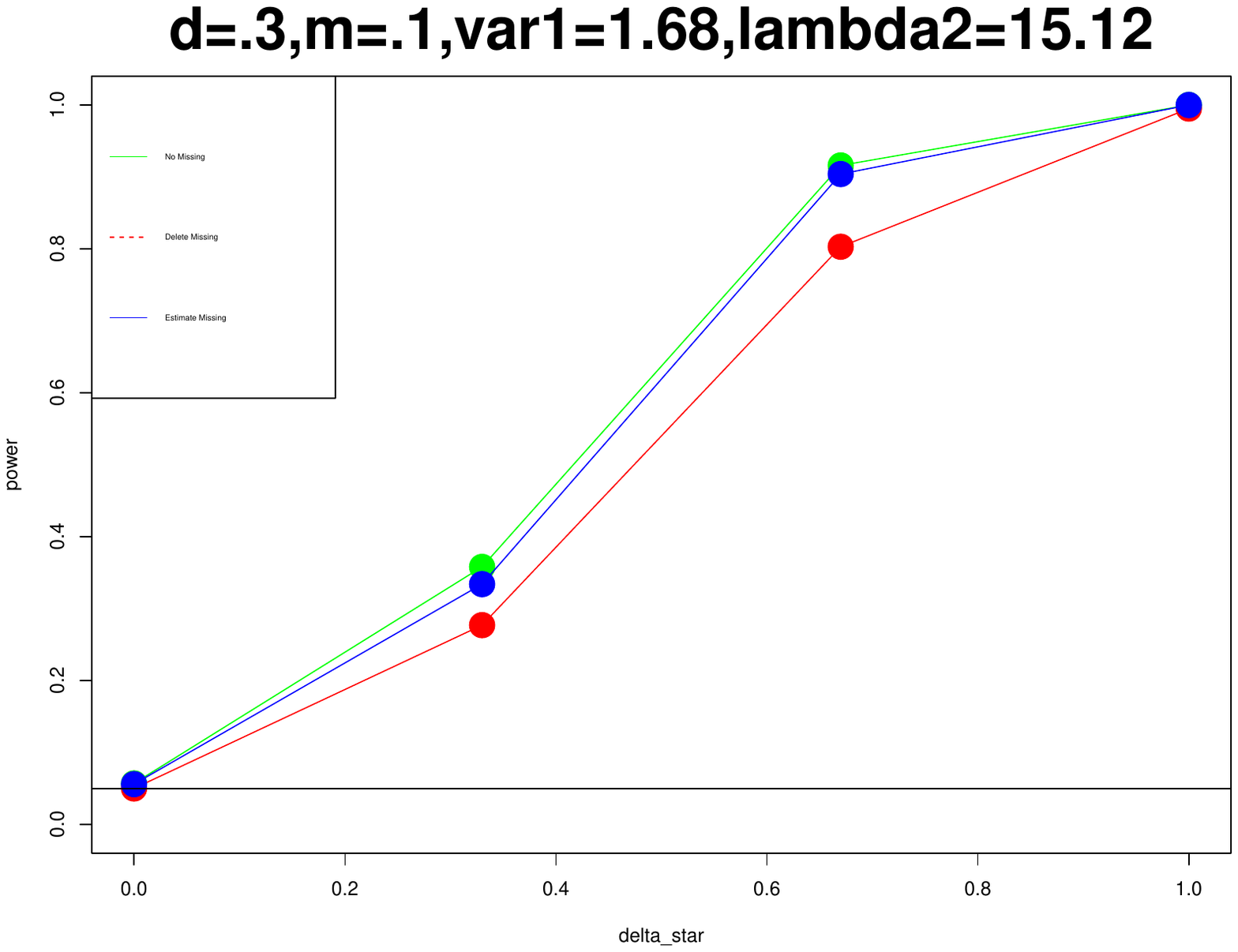}

\subsubsection{When both Traits have Poisson Distribution}

The following tables represents powers for the three strategies evaluated at $\delta = 0, .33, .67$ and 1.

\vspace{.2cm}

\textbf{ For $\rho_1 > \rho_2$}

\pagebreak

\hspace{3.5cm}$m=.1$
\hspace{8.5cm}$m=.5$

\vspace{.2cm}

\begin{tabular}{|c|c|c|c|c|}
\hline 
Strategy & $\delta = 0$ & $\delta = .33$ & $\delta = .67$ & $\delta = 1$ \\ 
\hline 
use same & 0.051 & 0.456 & 0.969 & 1 \\ 
\hline 
use other & 0.050 & 0.456 & 0.970 & 1 \\ 
\hline 
use both & 0.053 & 0.450 & 0.973 & 1 \\ 
\hline 
\end{tabular} \hspace{1.5cm}
\begin{tabular}{|c|c|c|c|c|}
\hline 
Strategy & $\delta = 0$ & $\delta = .33$ & $\delta = .67$ & $\delta = 1$ \\ 
\hline 
use same & 0.051 & 0.760 & 1 & 1 \\ 
\hline 
use other & 0.053 & 0.775 & 1 & 1 \\ 
\hline 
use both & 0.050 & 0.769 & 1 & 1 \\ 
\hline 
\end{tabular} 

\hspace{.4cm}

\textbf{ For $\rho_1 < \rho_2$}

\vspace{.4cm}

\hspace{3.5cm}$m=.1$
\hspace{8.5cm}$m=.5$

\vspace{.2cm}

\begin{tabular}{|c|c|c|c|c|}
\hline 
Strategy & $\delta = 0$ & $\delta = .33$ & $\delta = .67$ & $\delta = 1$ \\ 
\hline 
use same & 0.049 & 0.388 & 0.943 & 1 \\ 
\hline 
use other & 0.052 & 0.402 & 0.949 & 1 \\ 
\hline 
use both & 0.051 & 0.393 & 0.946 & 1 \\ 
\hline 
\end{tabular} \hspace{1.5cm}
\begin{tabular}{|c|c|c|c|c|}
\hline 
Strategy & $\delta = 0$ & $\delta = .33$ & $\delta = .67$ & $\delta = 1$ \\ 
\hline 
use same & 0.052 & 0.899 & 1 & 1 \\ 
\hline 
use other & 0.051 & 0.902 & 1 & 1 \\ 
\hline 
use both & 0.047 & 0.902 & 1 & 1 \\ 
\hline 
\end{tabular}

\hspace{.4cm}

\textbf{ For $\rho_1 = \rho_2$}

\vspace{.4cm}

\hspace{3.5cm}$m=.1$
\hspace{8.5cm}$m=.5$

\vspace{.2cm}

\begin{tabular}{|c|c|c|c|c|}
\hline 
Strategy & $\delta = 0$ & $\delta = .33$ & $\delta = .67$ & $\delta = 1$ \\ 
\hline 
use same & 0.050 & 0.523 & 0.992 & 1 \\ 
\hline 
use other & 0.050 & 0.524 & 0.992 & 1 \\ 
\hline 
use both & 0.051 & 0.530 & 0.992 & 1 \\ 
\hline 
\end{tabular} \hspace{1.5cm}
\begin{tabular}{|c|c|c|c|c|}
\hline 
Strategy & $\delta = 0$ & $\delta = .33$ & $\delta = .67$ & $\delta = 1$ \\ 
\hline 
use same & 0.048 & 0.836 & 1 & 1 \\ 
\hline 
use other & 0.049 & 0.839 & 1 & 1 \\ 
\hline 
use both & 0.051 & 0.833 & 1 & 1 \\ 
\hline 
\end{tabular} 

\vspace{.4cm}

According to the above results we can conclude-

\begin{center}
\begin{tabular}{|c|c|}
\hline 
Case & Best Strategy \\ 
\hline 
$\rho_1 > \rho_2$ & use other \\ 
\hline 
$\rho_1 < \rho_2$ & use other \\ 
\hline 
$\rho_1 = \rho_2$ & use other \\ 
\hline 
\end{tabular} 

\end{center}

Now we go for power comparison among no missing, estimated missing and deleted missing.

We generate 1st trait from poisson distribution (section 6.2) with the parameters $\alpha = \alpha_1, \beta = \beta_1, \lambda = \lambda1$ keeping $p^\star = p^\star_1$ and we generate 2nd trait from poisson distribution (section 6.2) with the parameters $\alpha = \alpha_2, \beta = \beta_2, \lambda = \lambda_2$ keeping $p^\star = p^\star_2$.

We have done simulation for three choices of $d$ as .1, .2, .3 and for each $d$ we take $(p^\star_1, p^\star_2)$ as (.1, .2), (.2, .2).

We take $\alpha_1 = 5, \alpha_2 = 10, \beta_1 = 1, and \beta_2 = 2$ and varied $\lambda_1$ and $\lambda_2$ in the following way,

\vspace{.2cm}

\begin{tabular}{|c|c|c|}
\hline 
d & $(p^\star_1, p^\star_2)$ & ($\lambda_1, \lambda_2$) \\ 
\hline 
.1 & (.1, .2) & (1.62, 2.88) \\ 
\hline 
.1 & (.2, .2) & (.72, 2.88) \\ 
\hline 

\end{tabular} \hspace{1.5cm}
\begin{tabular}{|c|c|c|}
\hline 
d & $(p^\star_1, p^\star_2)$ & ($\lambda_1, \lambda_2$) \\ 
\hline 
.2 & (.1, .2) & (2.88, 5.12) \\ 
\hline 
.2 & (.2, .2) & (1.28, 5.12) \\ 
\hline 

\end{tabular} \hspace{1.5cm}
\begin{tabular}{|c|c|c|}
\hline 
d & $(p^\star_1, p^\star_2)$ & ($\lambda_1, \lambda_2$) \\ 
\hline 
.3 & (.1, .2) & (3.78, 6.72) \\ 
\hline 
.3 & (.2, .2) & (1.68, 6.72) \\ 
\hline 
\end{tabular} 

\vspace{.4cm}

We replicate these for $m =$ .1 and .2.

Note that as we have seen in all 3 cases \textbf{use other} is the best strategy, we have done all the following imputations using this strategy.

\vspace{.4cm}

\hspace{1.5cm}
$d=.1 \quad m=.5$
\hspace{3cm}
$d=.2 \quad m=.5$
\hspace{3cm}
$d=.3 \quad m=.5$

\includegraphics[width = 2.3in, height = 1.5in]{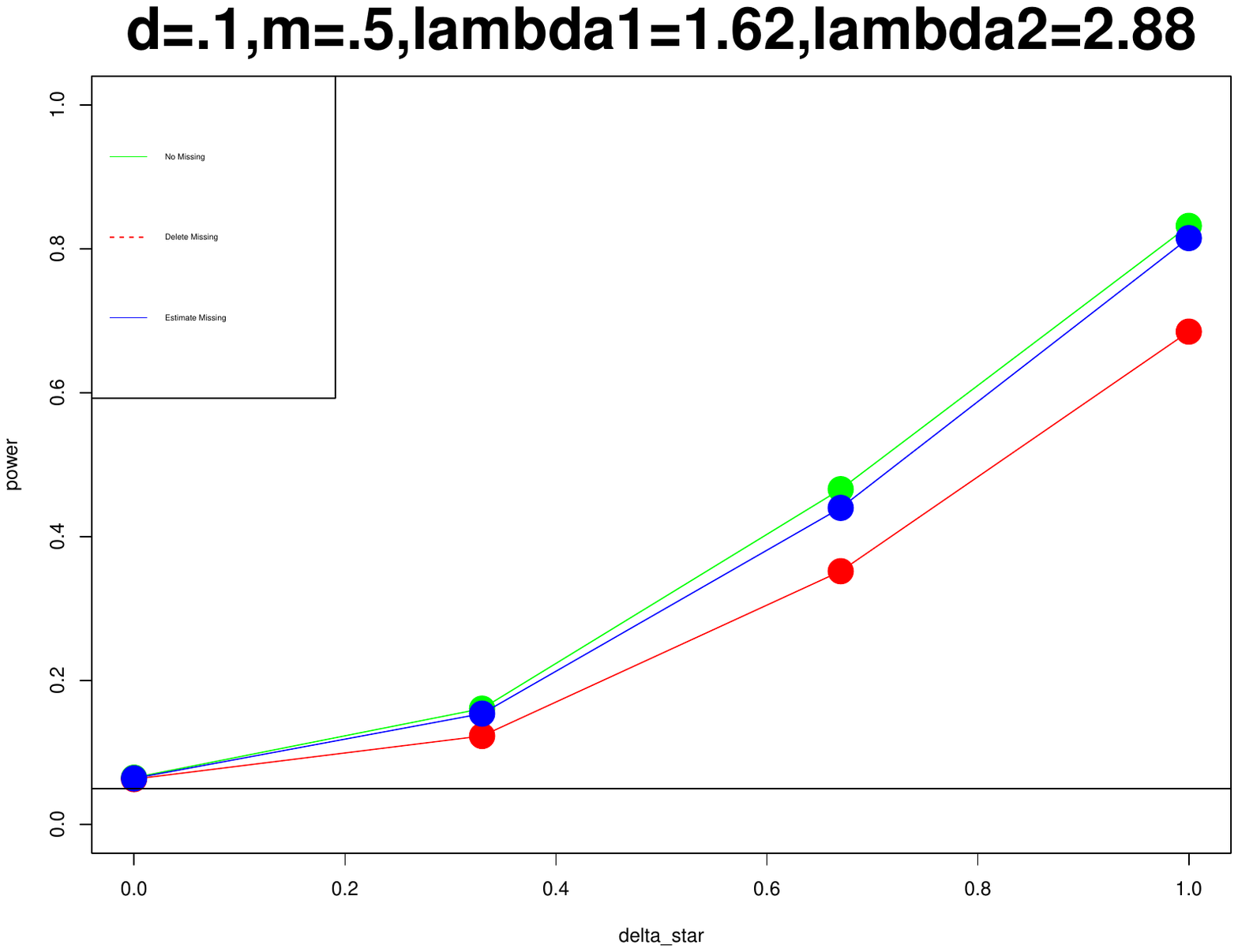}
\includegraphics[width = 2.3in, height = 1.5in]{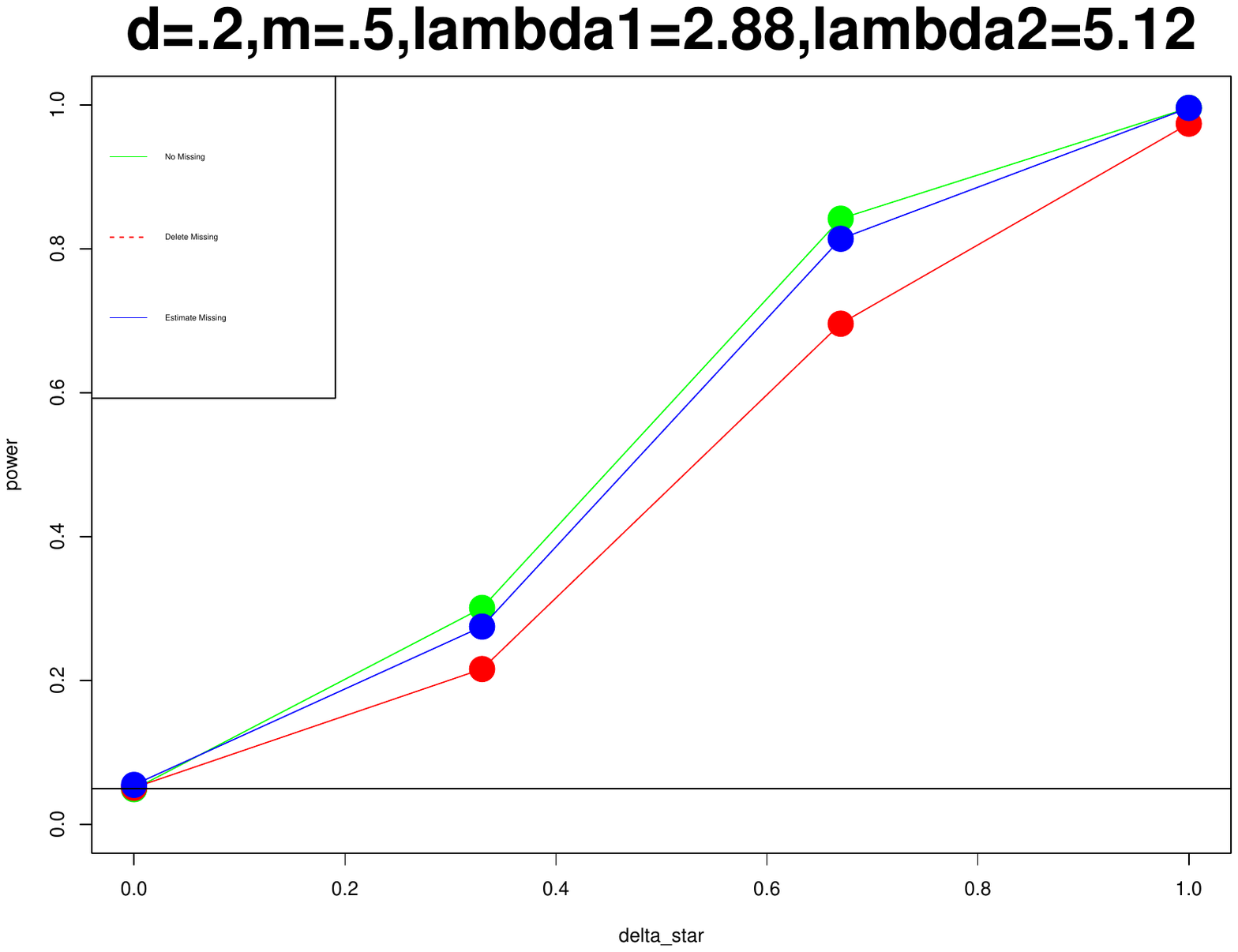}
\includegraphics[width = 2.3in, height = 1.5in]{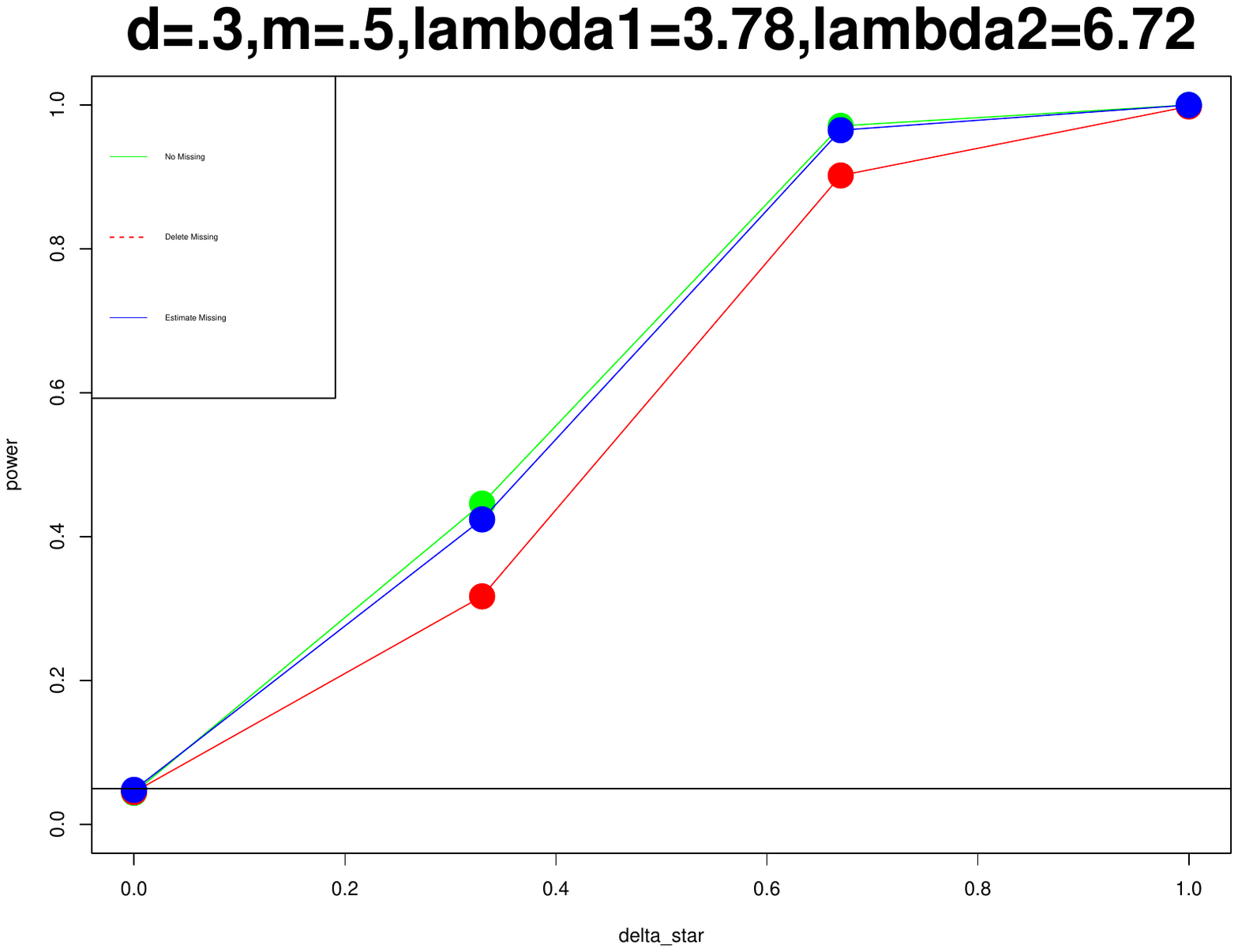}

\hspace{1.5cm}
$d=.1 \quad m=.1$
\hspace{3cm}
$d=.2 \quad m=.1$
\hspace{3cm}
$d=.3 \quad m=.1$

\includegraphics[width = 2.3in, height = 1.5in]{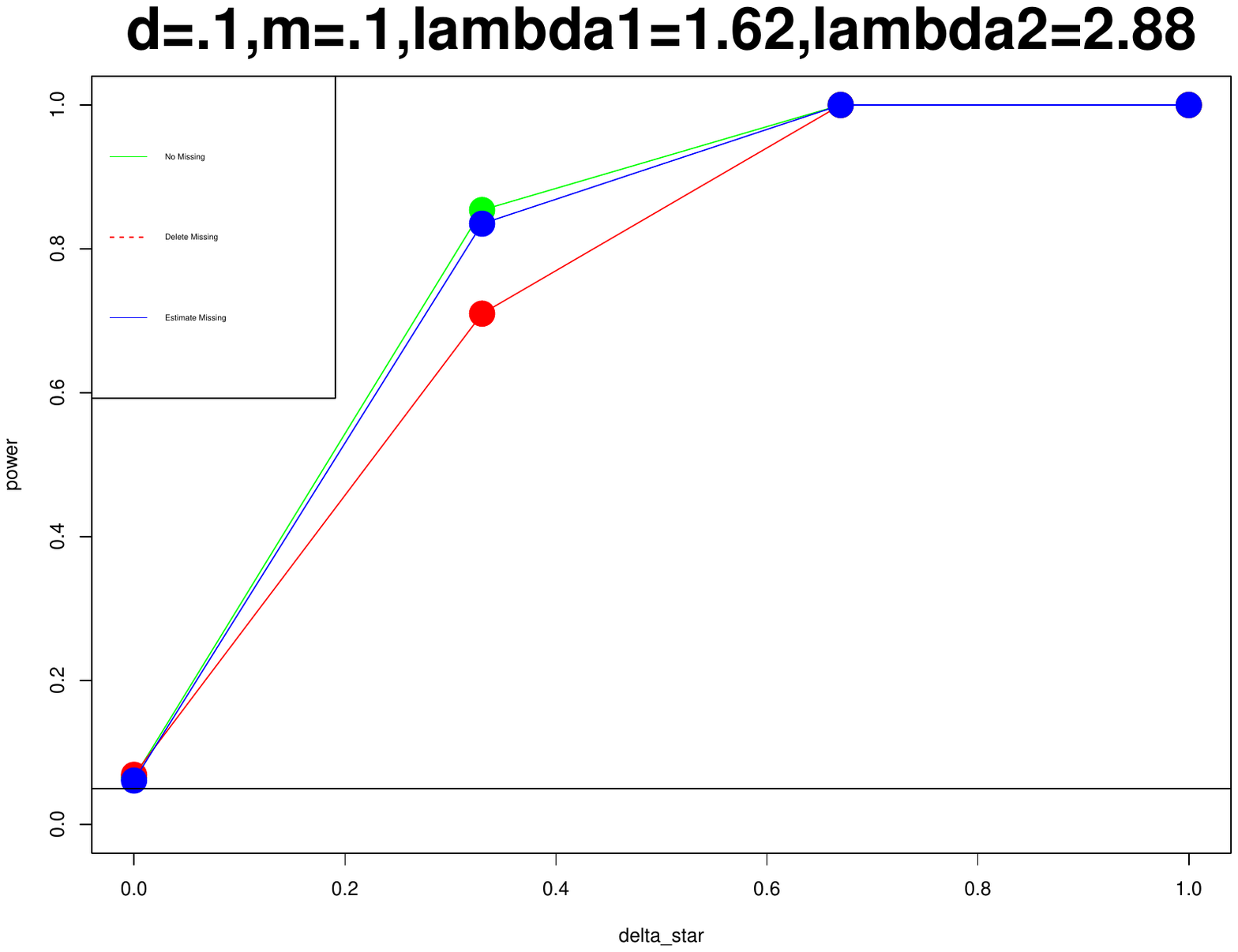}
\includegraphics[width = 2.3in, height = 1.5in]{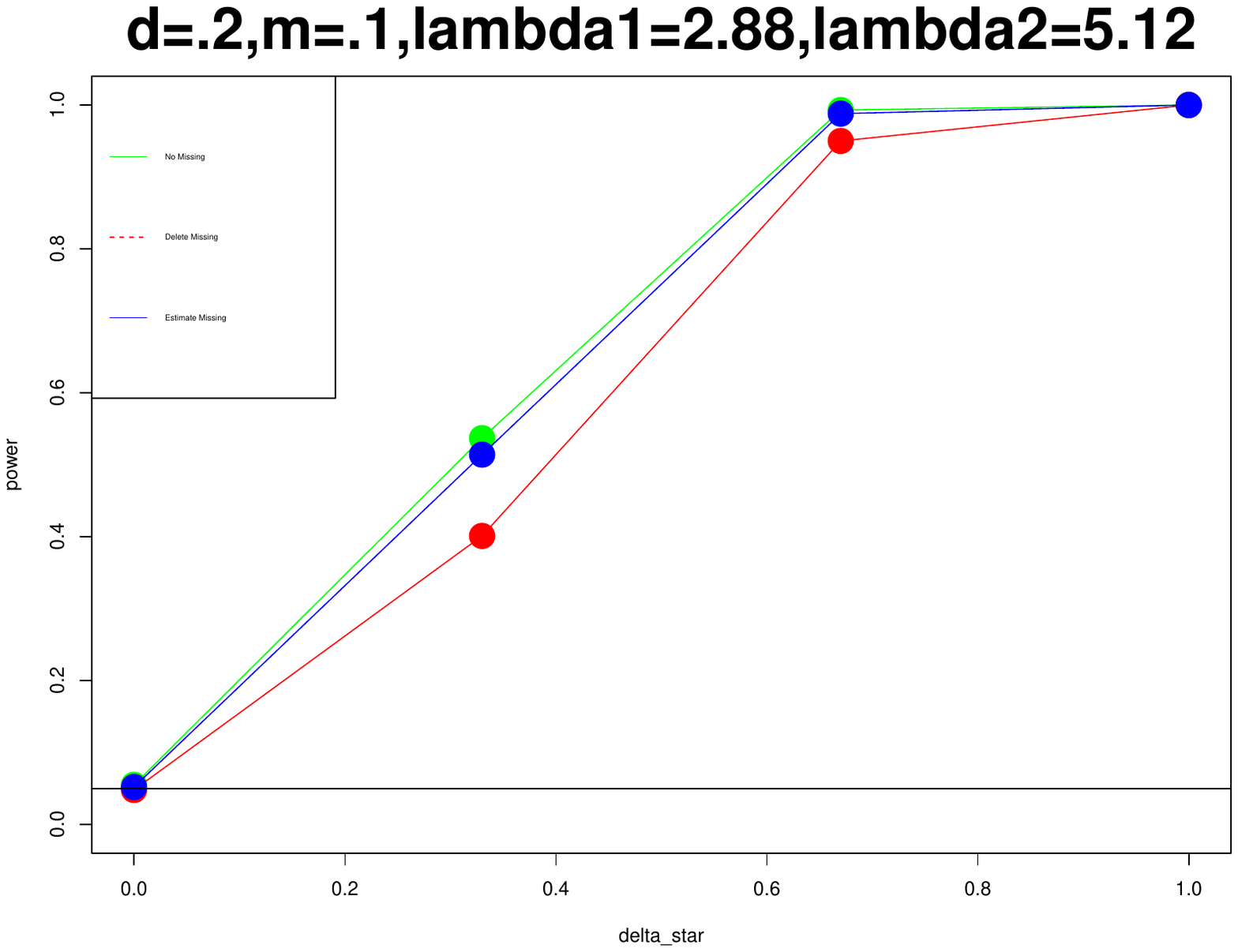}
\includegraphics[width = 2.3in, height = 1.5in]{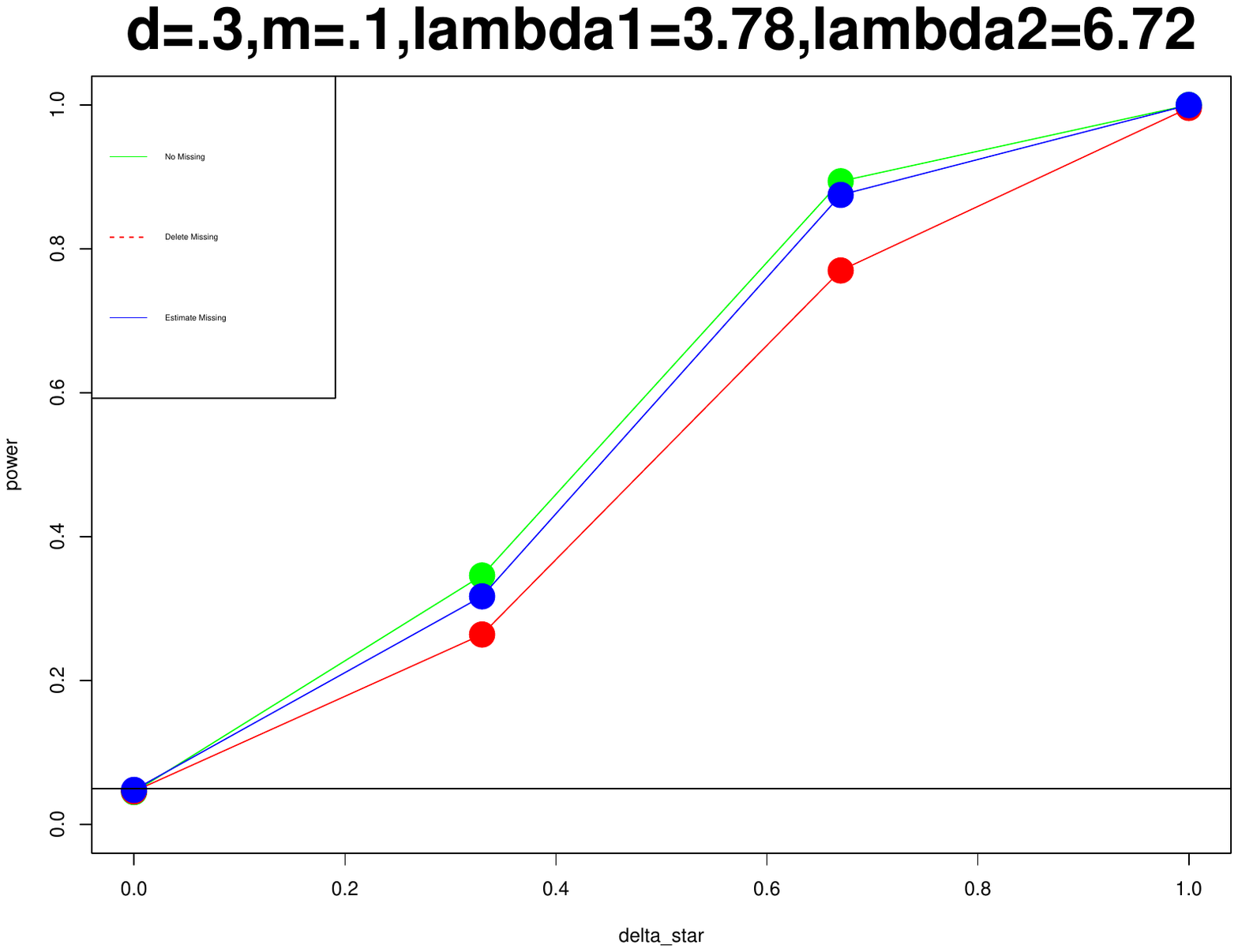}

\hspace{1.5cm}
$d=.1 \quad m=.1$
\hspace{3cm}
$d=.2 \quad m=.1$
\hspace{3cm}
$d=.3 \quad m=.1$

\includegraphics[width = 2.3in, height = 1.5in]{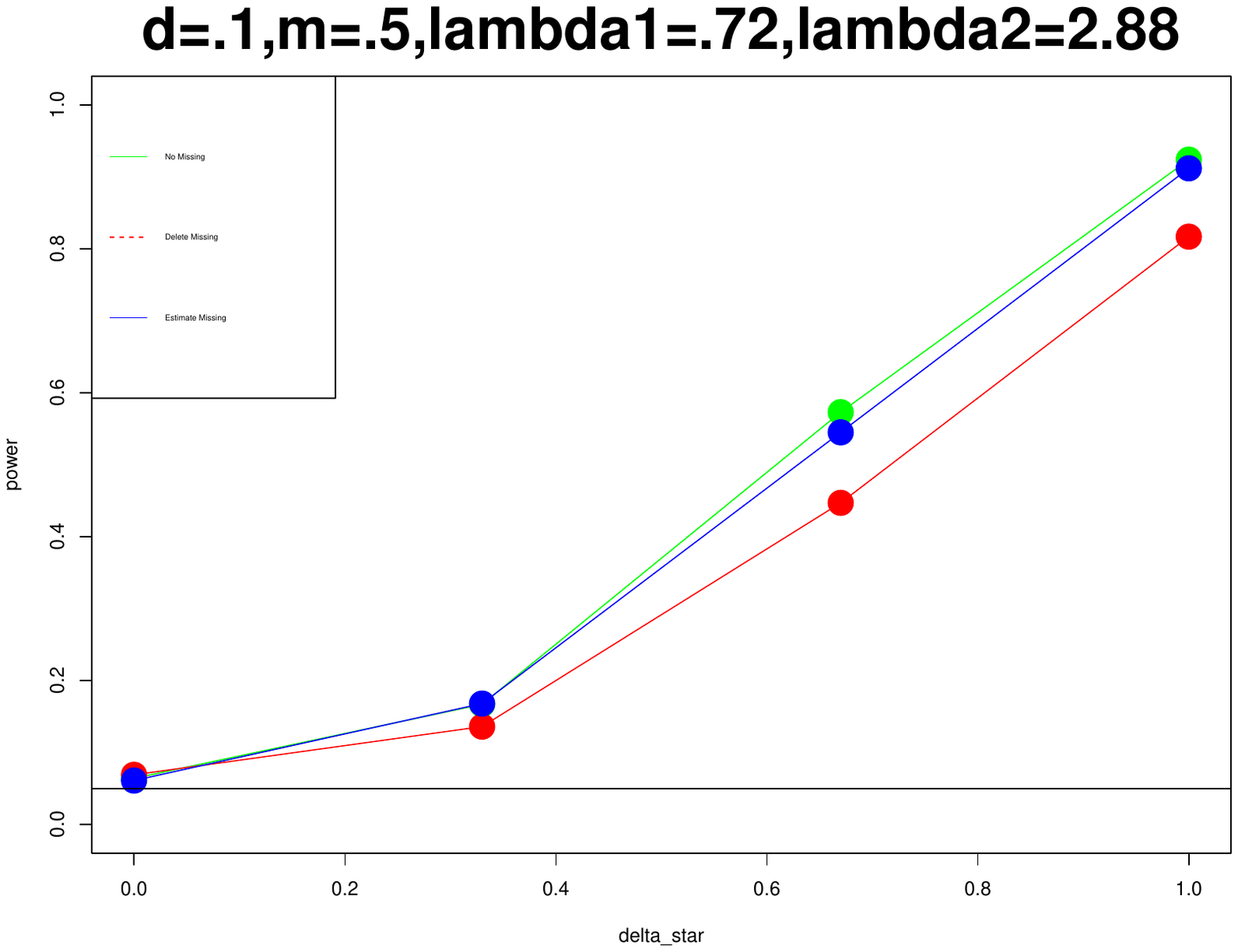}
\includegraphics[width = 2.3in, height = 1.5in]{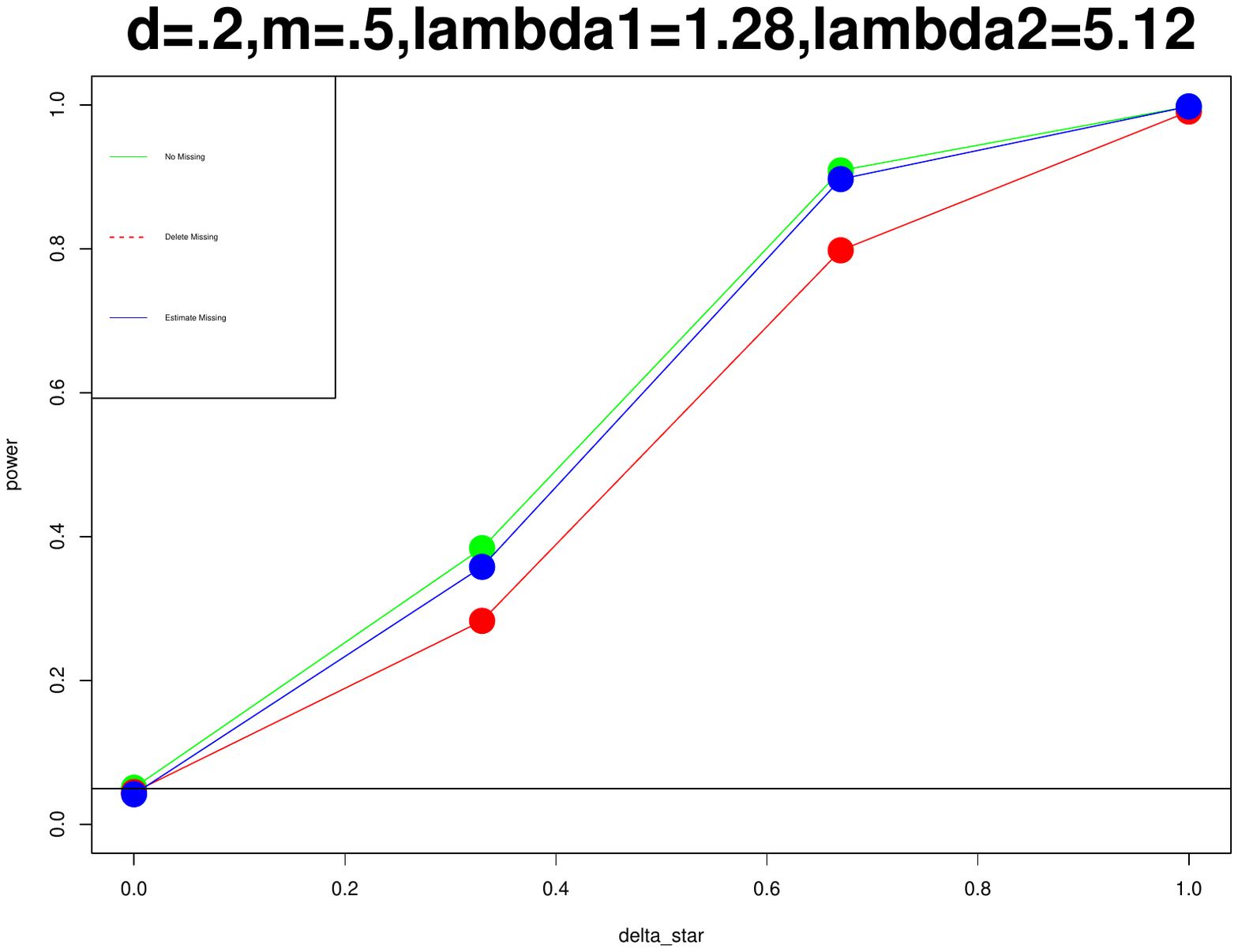}
\includegraphics[width = 2.3in, height = 1.5in]{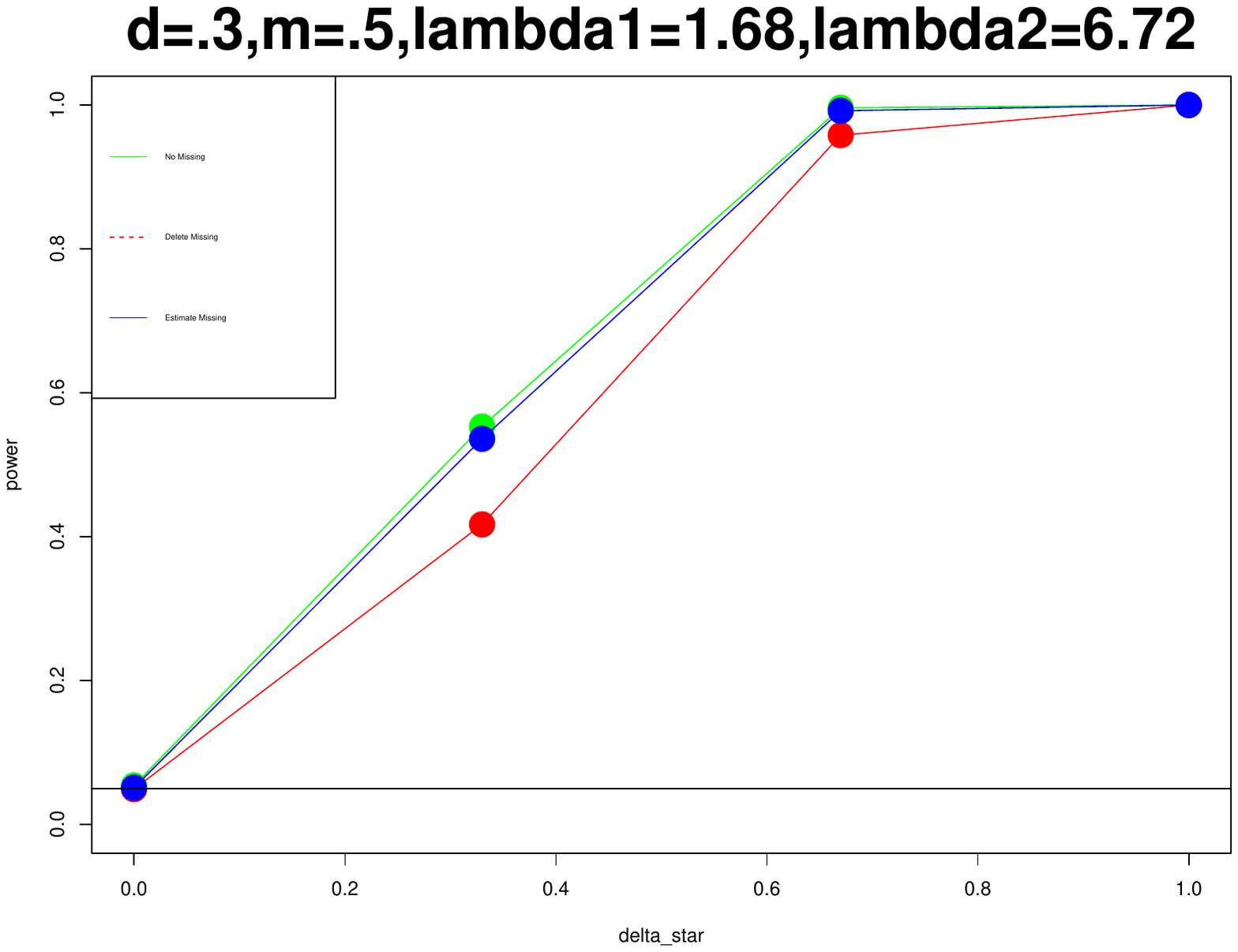}

\hspace{1.5cm}
$d=.1 \quad m=.5$
\hspace{3cm}
$d=.2 \quad m=.5$
\hspace{3cm}
$d=.3 \quad m=.5$

\includegraphics[width = 2.3in, height = 1.5in]{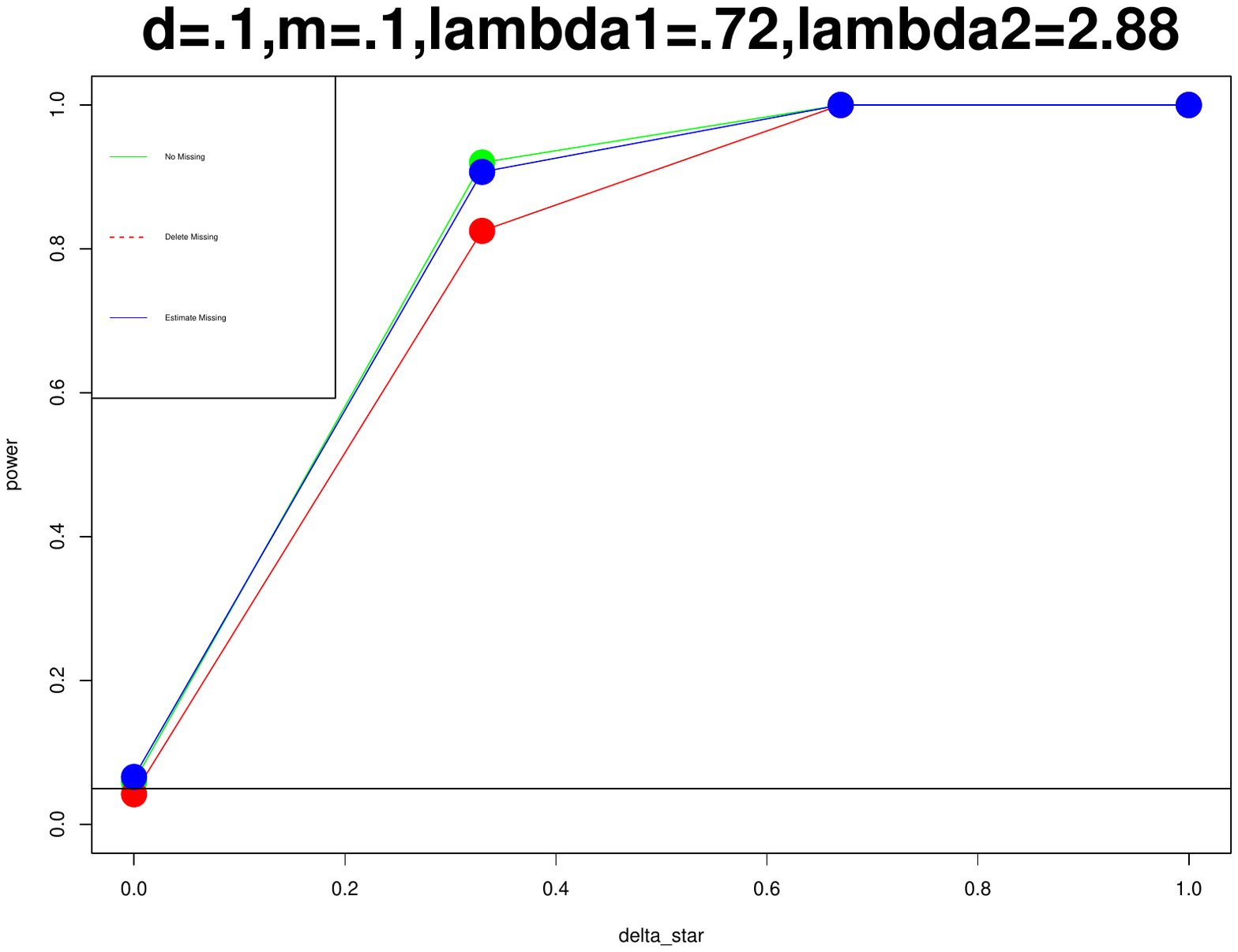}
\includegraphics[width = 2.3in, height = 1.5in]{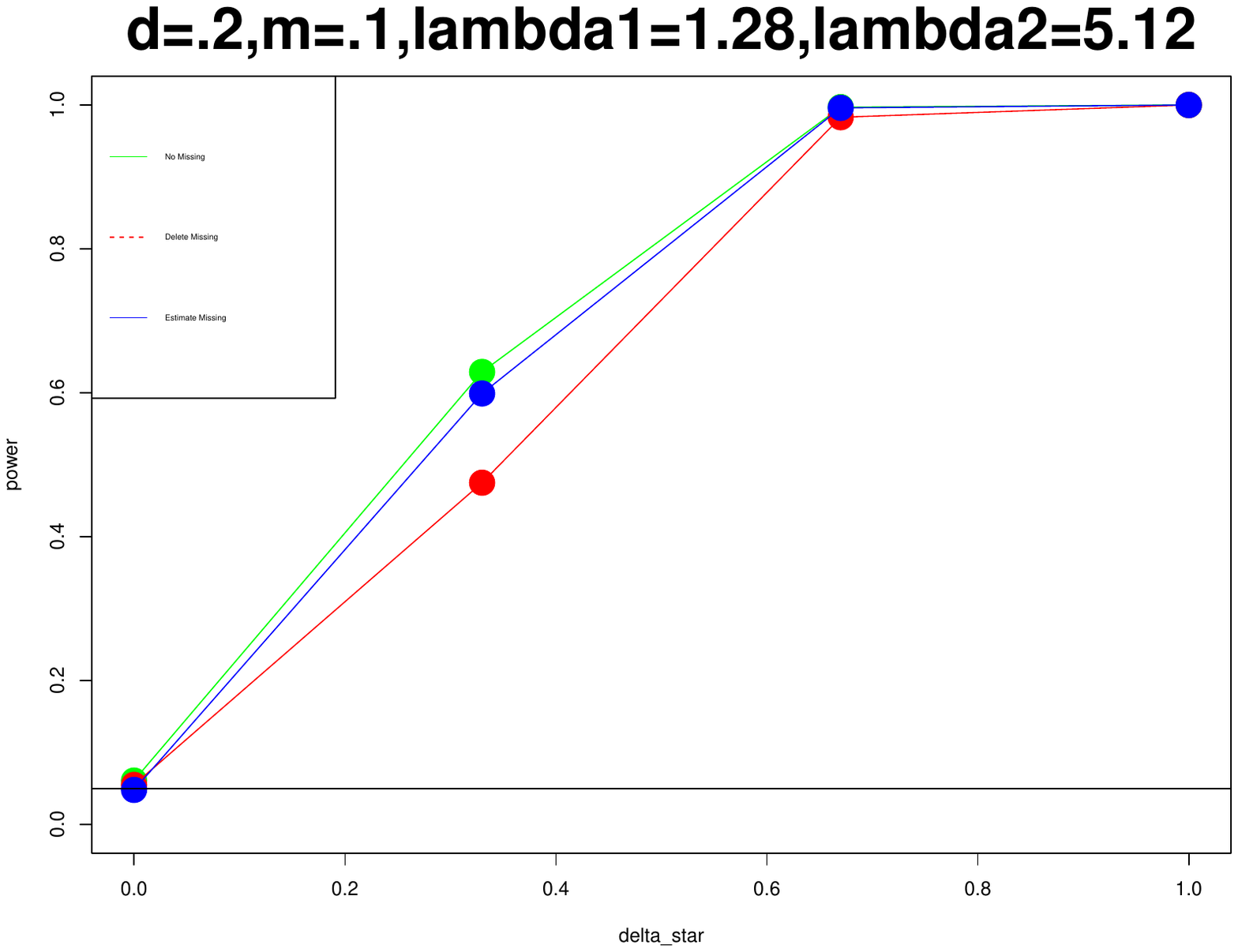}
\includegraphics[width = 2.3in, height = 1.5in]{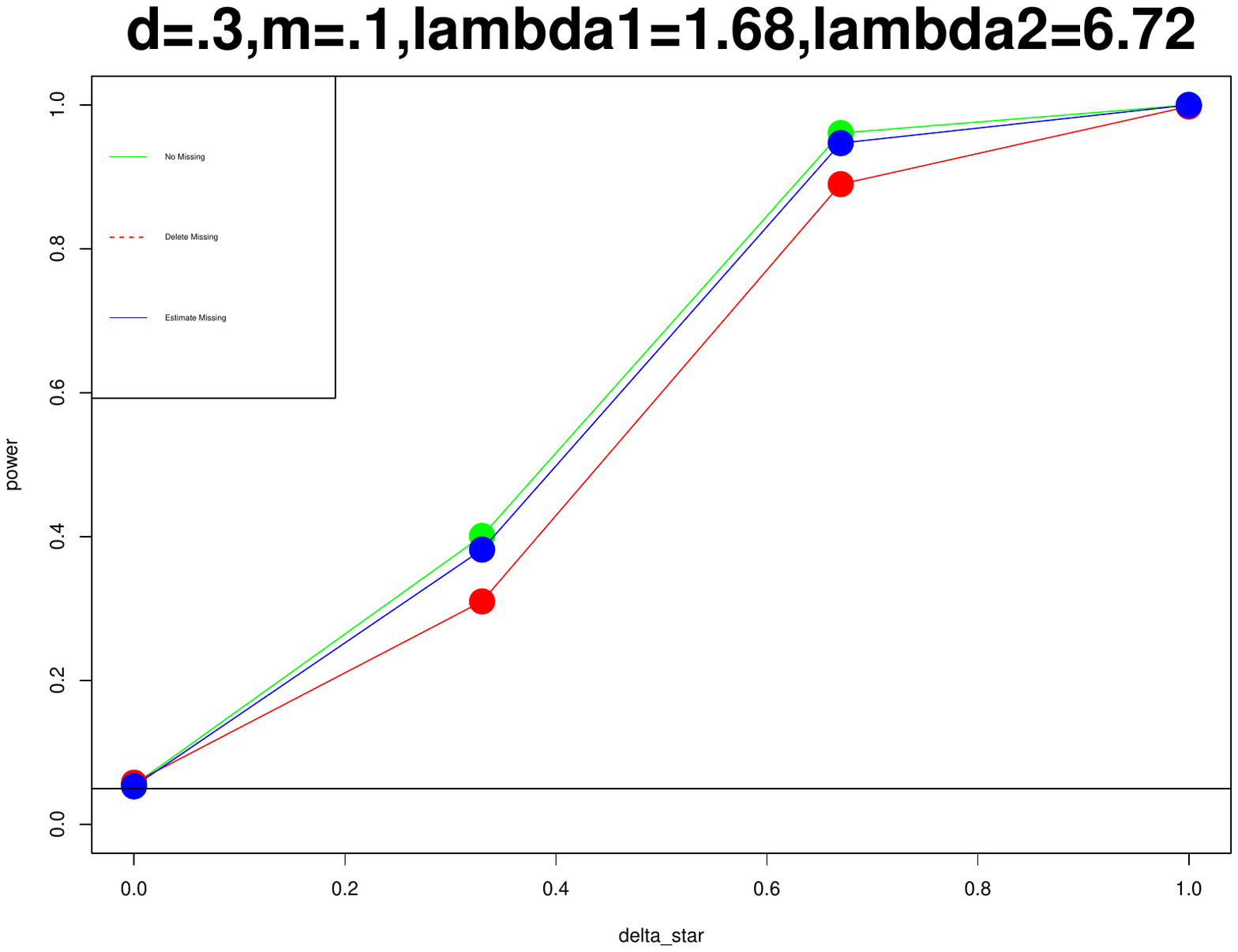}

\subsubsection{When one Trait has Normal Distribution and other Trait has Chi Squares Distribution}

The following tables represents powers for the three strategies evaluated at $\delta = 0, .33, .67$ and 1.

\textbf{ For $\rho_1 > \rho_2$}

\vspace{.4cm}

\hspace{3.5cm}$m=.1$
\hspace{8.5cm}$m=.5$

\vspace{.2cm}

\begin{tabular}{|c|c|c|c|c|}
\hline 
Strategy & $\delta = 0$ & $\delta = .33$ & $\delta = .67$ & $\delta = 1$ \\ 
\hline 
use same & 0.050 & 0.389 & 0.920 & 0.999 \\ 
\hline 
use other & 0.047 & 0.321 & 0.842 & 0.996 \\ 
\hline 
use both & 0.051 & 0.318 & 0.838 & 0.996 \\ 
\hline 
\end{tabular} \hspace{1.5cm}
\begin{tabular}{|c|c|c|c|c|}
\hline 
Strategy & $\delta = 0$ & $\delta = .33$ & $\delta = .67$ & $\delta = 1$ \\ 
\hline 
use same & 0.050 & 0.698 & 1 & 1 \\ 
\hline 
use other & 0.048 & 0.559 & 0.992 & 1 \\ 
\hline 
use both & 0.051 & 0.535 & 0.988 & 1 \\ 
\hline 
\end{tabular} 

\hspace{.4cm}

\textbf{ For $\rho_1 < \rho_2$}

\vspace{.4cm}

\hspace{3.5cm}$m=.1$
\hspace{8.5cm}$m=.5$

\vspace{.2cm}

\begin{tabular}{|c|c|c|c|c|}
\hline 
Strategy & $\delta = 0$ & $\delta = .33$ & $\delta = .67$ & $\delta = 1$ \\ 
\hline 
use same & 0.051 & 0.313 & 0.898 & 0.998 \\ 
\hline 
use other & 0.050 & 0.367 & 0.935 & 1 \\ 
\hline 
use both & 0.048 & 0.316 & 0.902 & 0.998 \\ 
\hline 
\end{tabular} \hspace{1.5cm}
\begin{tabular}{|c|c|c|c|c|}
\hline 
Strategy & $\delta = 0$ & $\delta = .33$ & $\delta = .67$ & $\delta = 1$ \\ 
\hline 
use same & 0.047 & 0.871 & 1 & 1 \\ 
\hline 
use other & 0.053 & 0.915 & 1 & 1 \\ 
\hline 
use both & 0.052 & 0.880 & 1 & 1 \\ 
\hline 
\end{tabular}

\hspace{.4cm}

\textbf{ For $\rho_1 = \rho_2$}

\pagebreak

\hspace{3.5cm}$m=.1$
\hspace{8.5cm}$m=.5$

\vspace{.2cm}

\begin{tabular}{|c|c|c|c|c|}
\hline 
Strategy & $\delta = 0$ & $\delta = .33$ & $\delta = .67$ & $\delta = 1$ \\ 
\hline 
use same & 0.051 & 0.375 & 0.935 & 1 \\ 
\hline 
use other & 0.053 & 0.415 & 0.955 & 1 \\ 
\hline 
use both & 0.049 & 0.367 & 0.930 & 1 \\ 
\hline 
\end{tabular} \hspace{1.5cm}
\begin{tabular}{|c|c|c|c|c|}
\hline 
Strategy & $\delta = 0$ & $\delta = .33$ & $\delta = .67$ & $\delta = 1$ \\ 
\hline 
use same & 0.052 & 0.669 & 0.996 & 1 \\ 
\hline 
use other & 0.053 & 0.712 & 1 & 1 \\ 
\hline 
use both & 0.048 & 0.643 & 0.993 & 1 \\ 
\hline 
\end{tabular} 

\vspace{1cm}

According to the above results we can conclude-

\begin{center}
\begin{tabular}{|c|c|}
\hline 
Case & Best Strategy \\ 
\hline 
$\rho_1 > \rho_2$ & use same \\ 
\hline 
$\rho_1 < \rho_2$ & use other \\ 
\hline 
$\rho_1 = \rho_2$ & use other \\ 
\hline 
\end{tabular} 

\end{center}

Now we go for power comparison among no missing, estimated missing and deleted missing.

We generate 1st trait from normal distribution (section 6.2) with the parameters $\alpha = \alpha_1, \beta = \beta_1, \sigma = \sigma_1$ keeping $p^\star = p^\star_1$ and we generate 2nd trait from chi squares distribution (section 6.2) with the parameters $\alpha = \alpha_2, \beta = \beta_2, df=df_2$ keeping $p^\star = p^\star_2$.

We have done simulation for three choices of $d$ as .1, .2, .3 and for each $d$ we take $(p^\star_1, p^\star_2)$ as (.1, .2), (.2, .2).

We take $\alpha_1 = 5, \alpha_2 = 10, \beta_1 = 1, and \beta_2 = 2$ and varied $\sigma_1$ and $df_2$ in the following way,

\vspace{.4cm}

\begin{tabular}{|c|c|c|}
\hline 
d & $(p^\star_1, p^\star_2)$ & ($\sigma_1^2, df_2$) \\ 
\hline 
.1 & (.1, .2) & (1.62, 1.44) \\ 
\hline 
.1 & (.2, .2) & (.72, 1.44) \\ 
\hline
.1 & (.2, .1) & (.72, 3.24) \\ 
\hline 

\end{tabular} \hspace{1.5cm}
\begin{tabular}{|c|c|c|}
\hline 
d & $(p^\star_1, p^\star_2)$ & ($\sigma_1^2, df_2$) \\ 
\hline 
.2 & (.1, .2) & (2.88, 2.56) \\ 
\hline 
.2 & (.2, .2) & (1.28, 2.56) \\ 
\hline
.2 & (.2, .1) & (1.28, 5.76) \\ 
\hline 

\end{tabular} \hspace{1.5cm}
\begin{tabular}{|c|c|c|}
\hline 
d & $(p^\star_1, p^\star_2)$ & ($\sigma_1^2, df_2$) \\ 
\hline 
.3 & (.1, .2) & (3.78, 3.36) \\ 
\hline 
.3 & (.2, .2) & (1.68, 3.36) \\ 
\hline
.3 & (.2, .1) & (1.68, 7.56) \\ 
\hline 
\end{tabular} 

\vspace{.4cm}

We replicate these for $m =$ .1 and .2.

Note that for each of the following simulations we have calculated $\rho_1$ and $\rho_2$ and accordingly we have used the best imputation strategy.

\vspace{.4cm}

\hspace{1.5cm}
$d=.1 \quad m=.5$
\hspace{3cm}
$d=.2 \quad m=.5$
\hspace{3cm}
$d=.3 \quad m=.5$

\includegraphics[width = 2.3in, height = 1.5in]{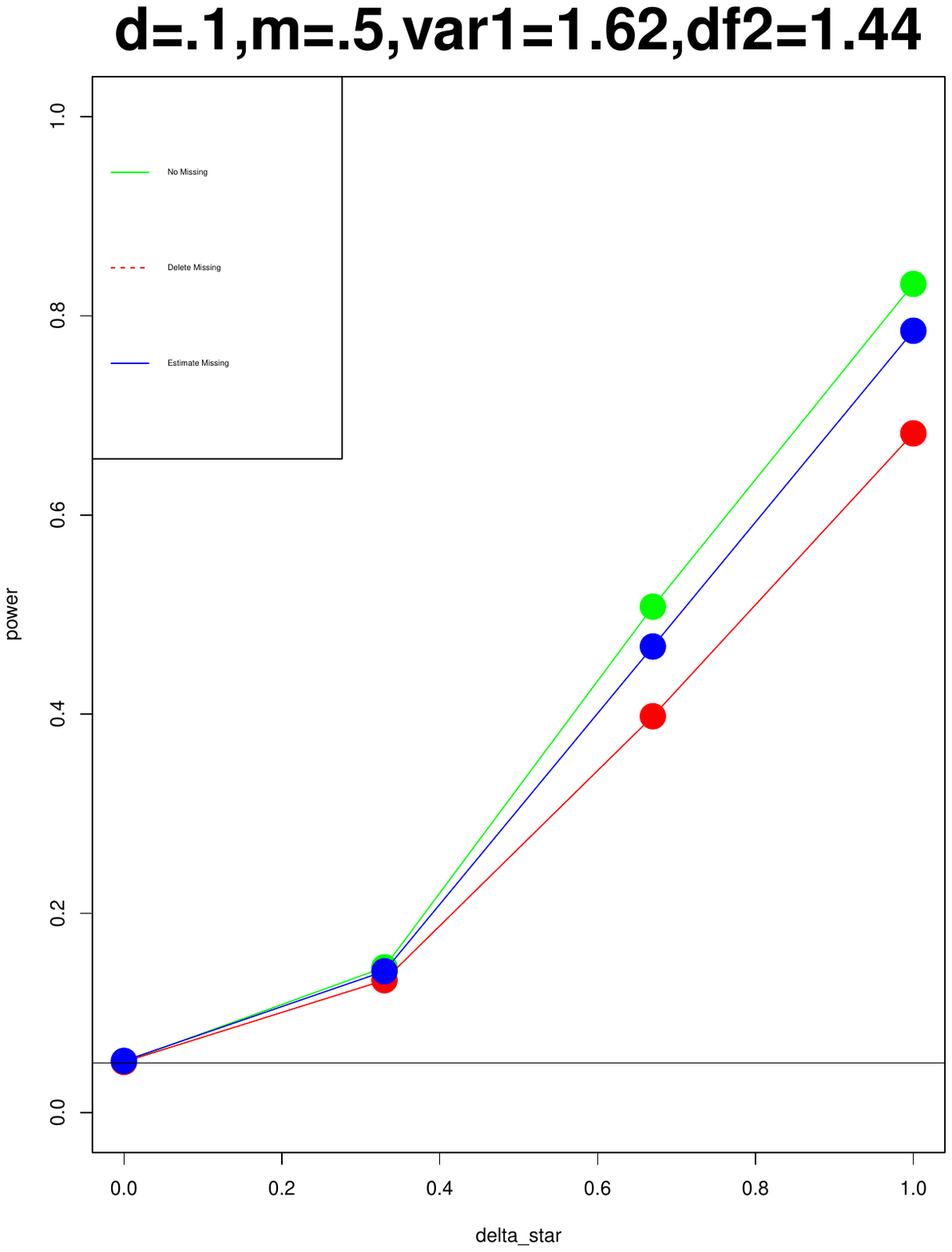}
\includegraphics[width = 2.3in, height = 1.5in]{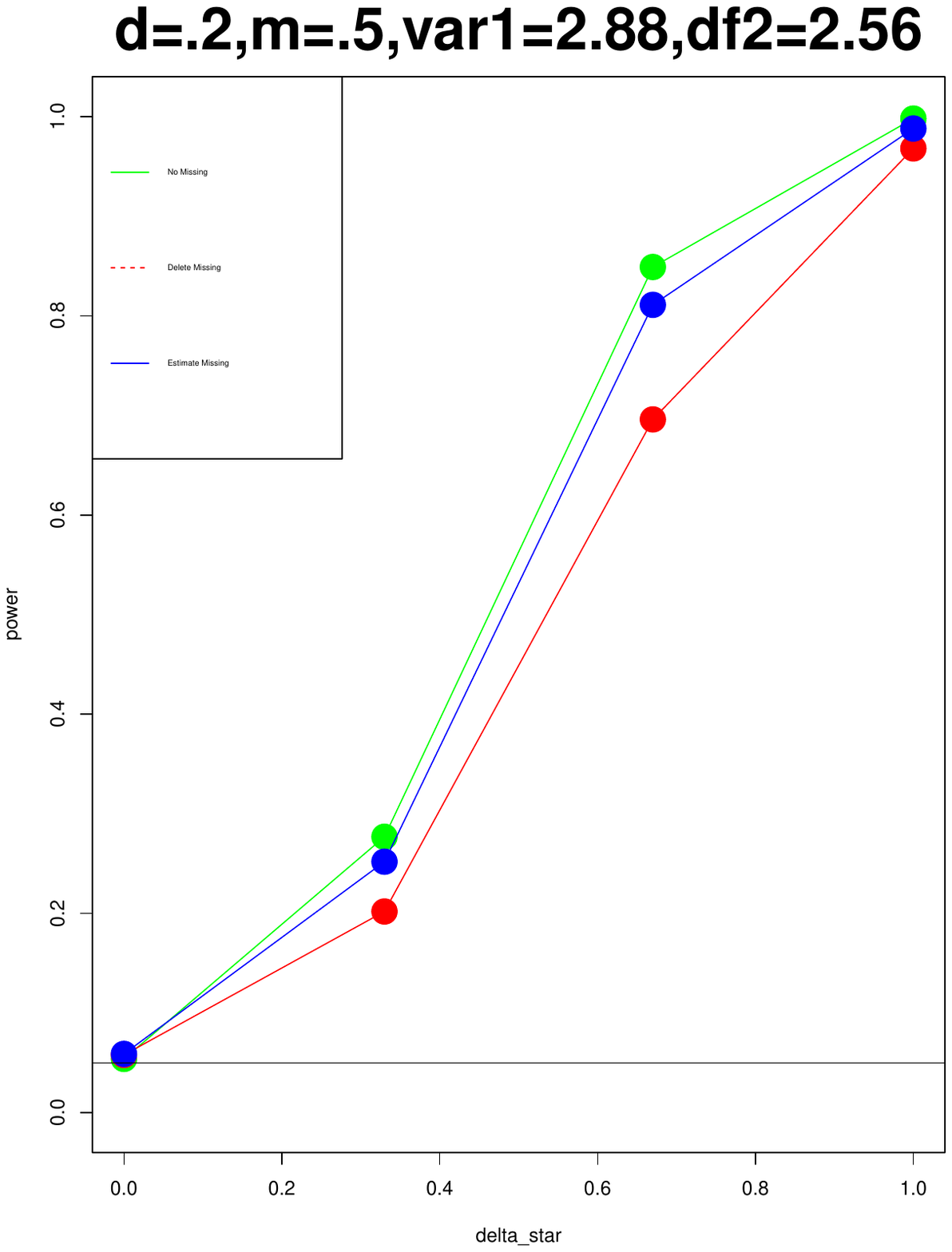}
\includegraphics[width = 2.3in, height = 1.5in]{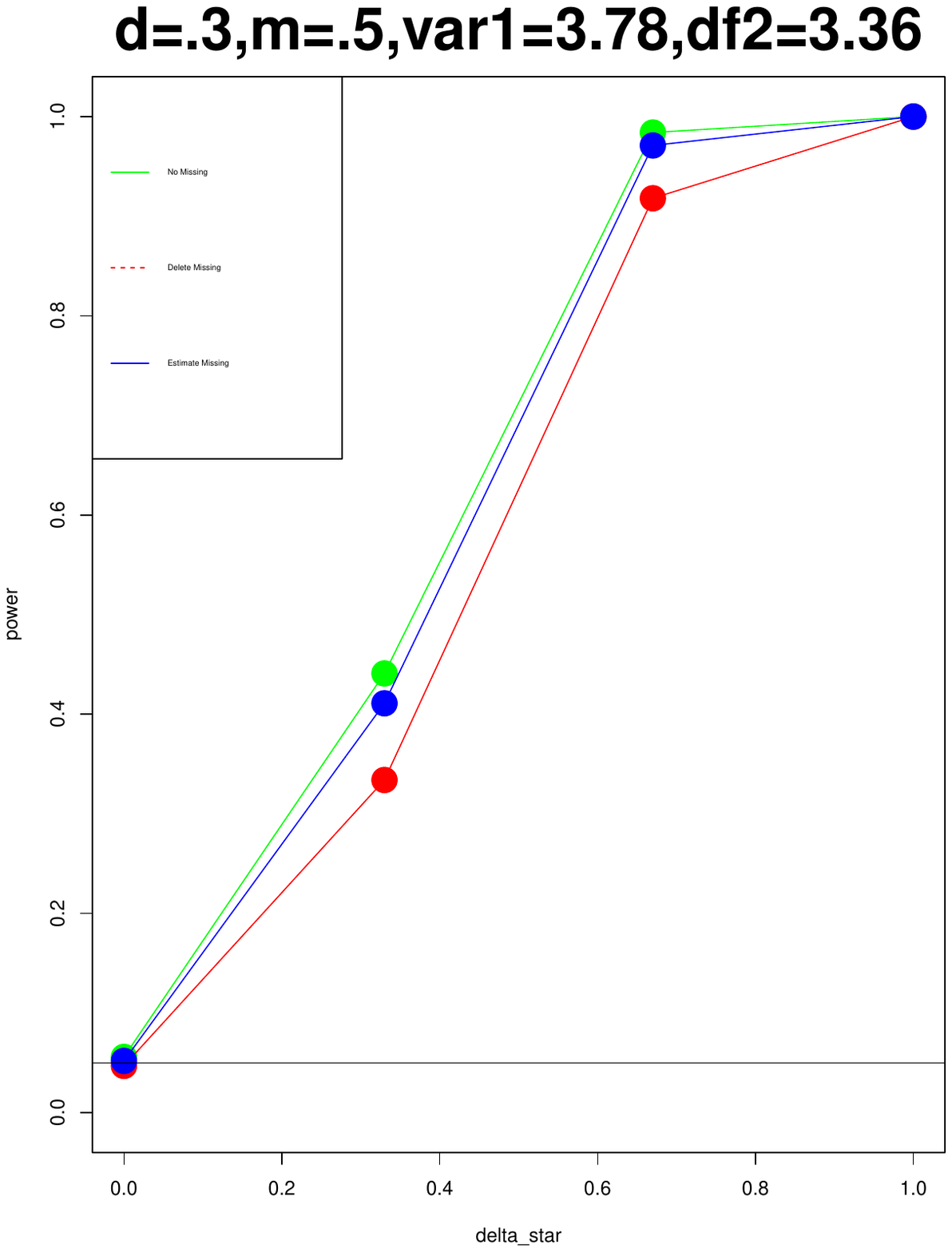}

\hspace{1.5cm}
$d=.1 \quad m=.1$
\hspace{3cm}
$d=.2 \quad m=.1$
\hspace{3cm}
$d=.3 \quad m=.1$

\includegraphics[width = 2.3in, height = 1.5in]{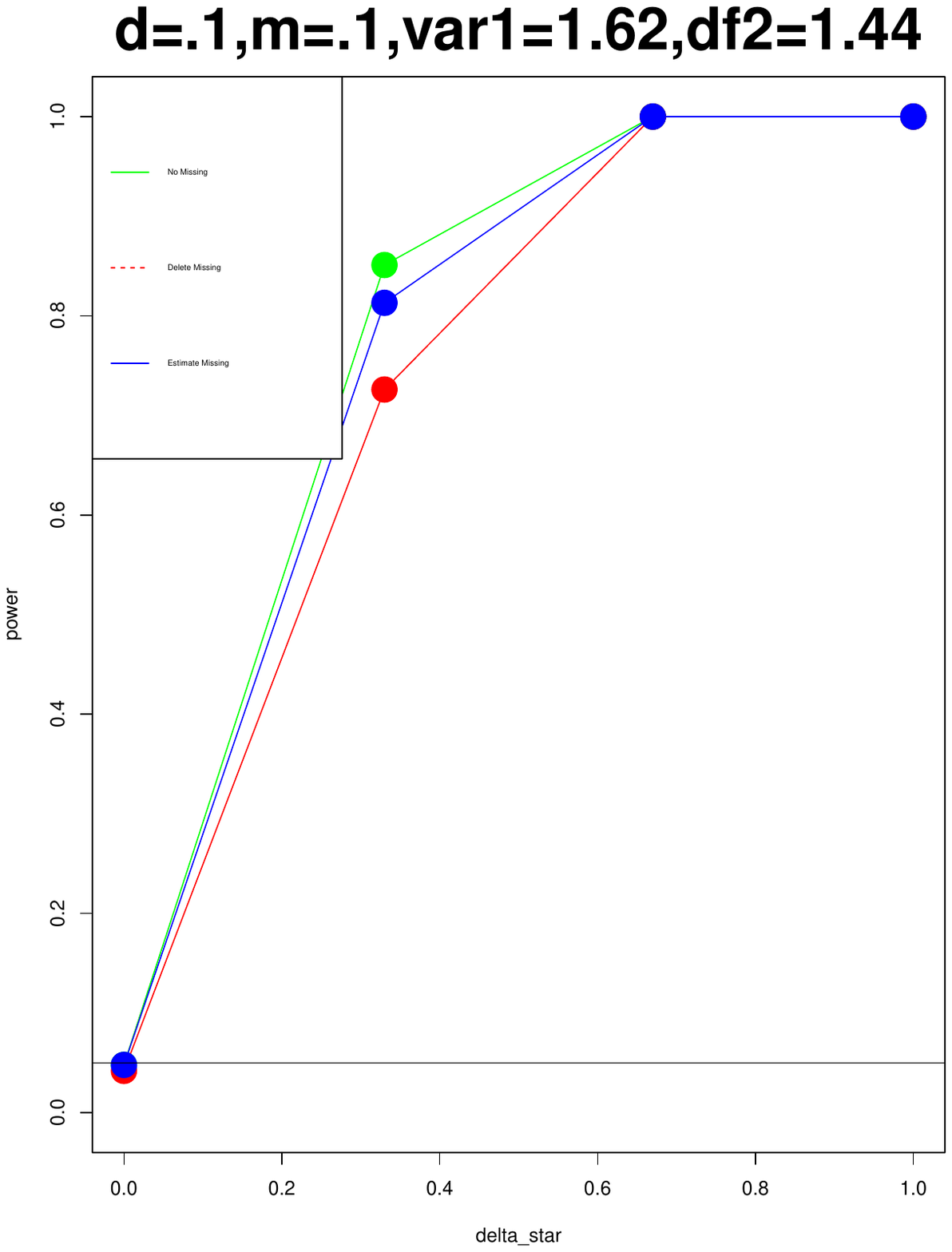}
\includegraphics[width = 2.3in, height = 1.5in]{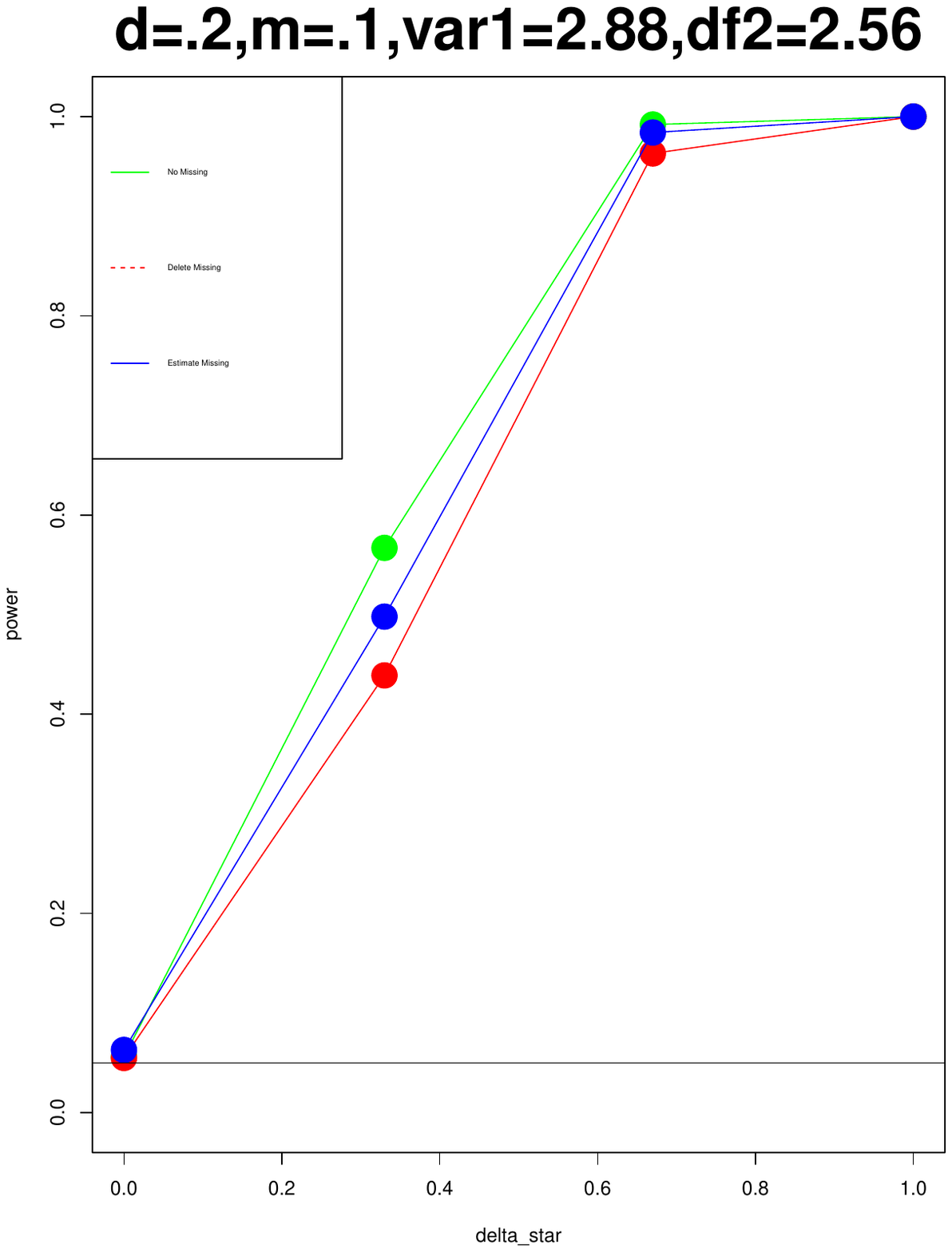}
\includegraphics[width = 2.3in, height = 1.5in]{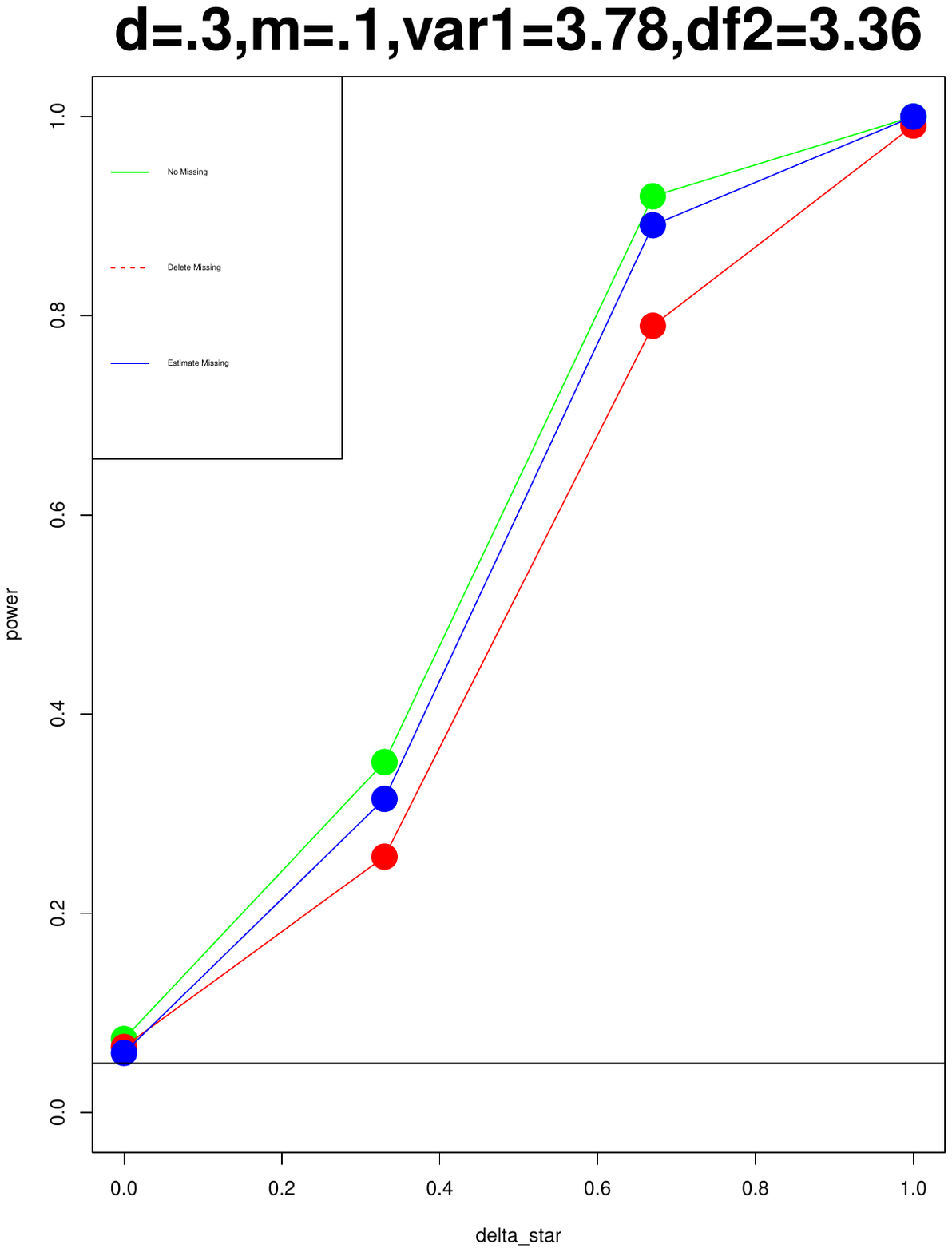}

\pagebreak

\hspace{1.5cm}
$d=.1 \quad m=.5$
\hspace{3cm}
$d=.2 \quad m=.5$
\hspace{3cm}
$d=.3 \quad m=.5$

\includegraphics[width = 2.3in, height = 1.5in]{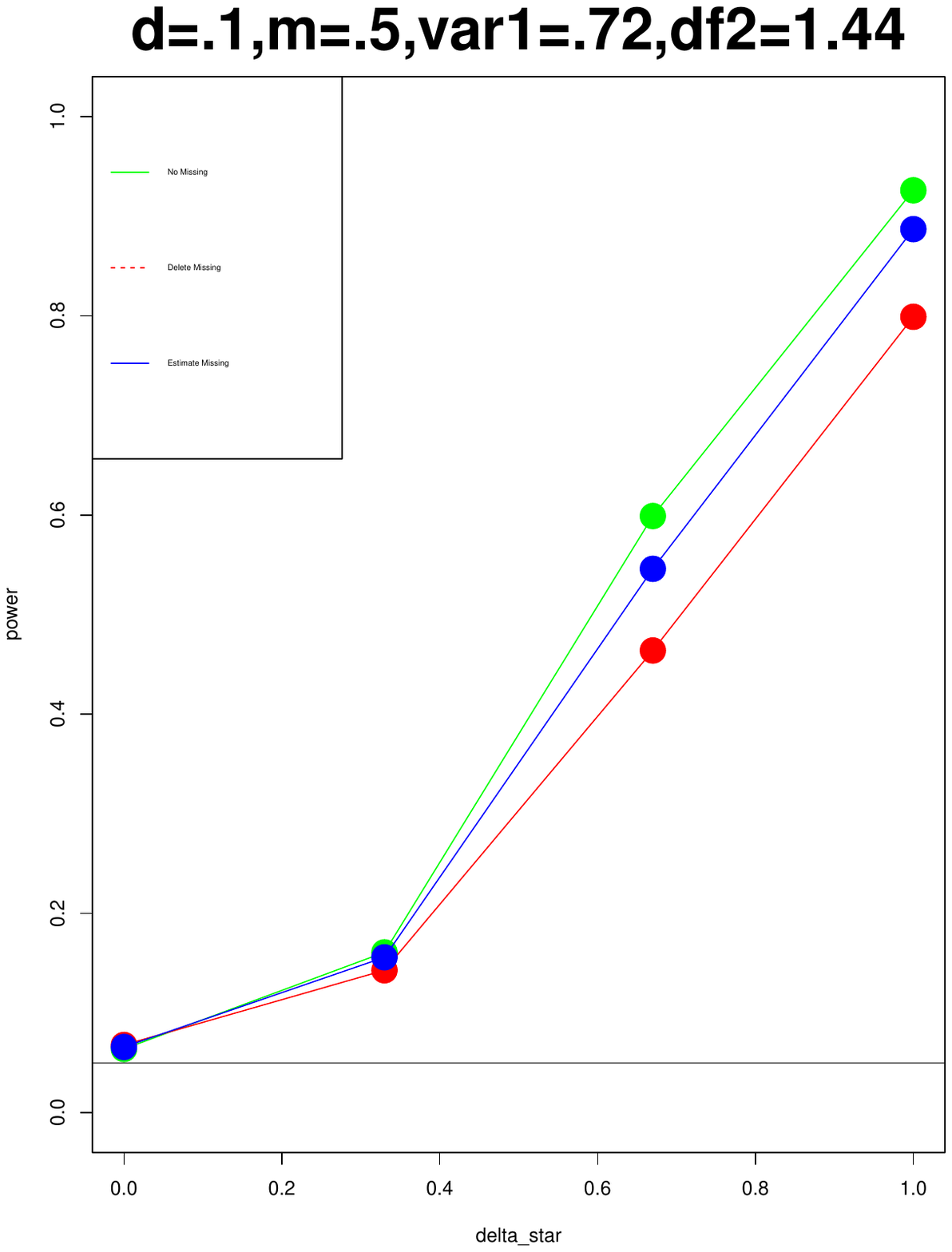}
\includegraphics[width = 2.3in, height = 1.5in]{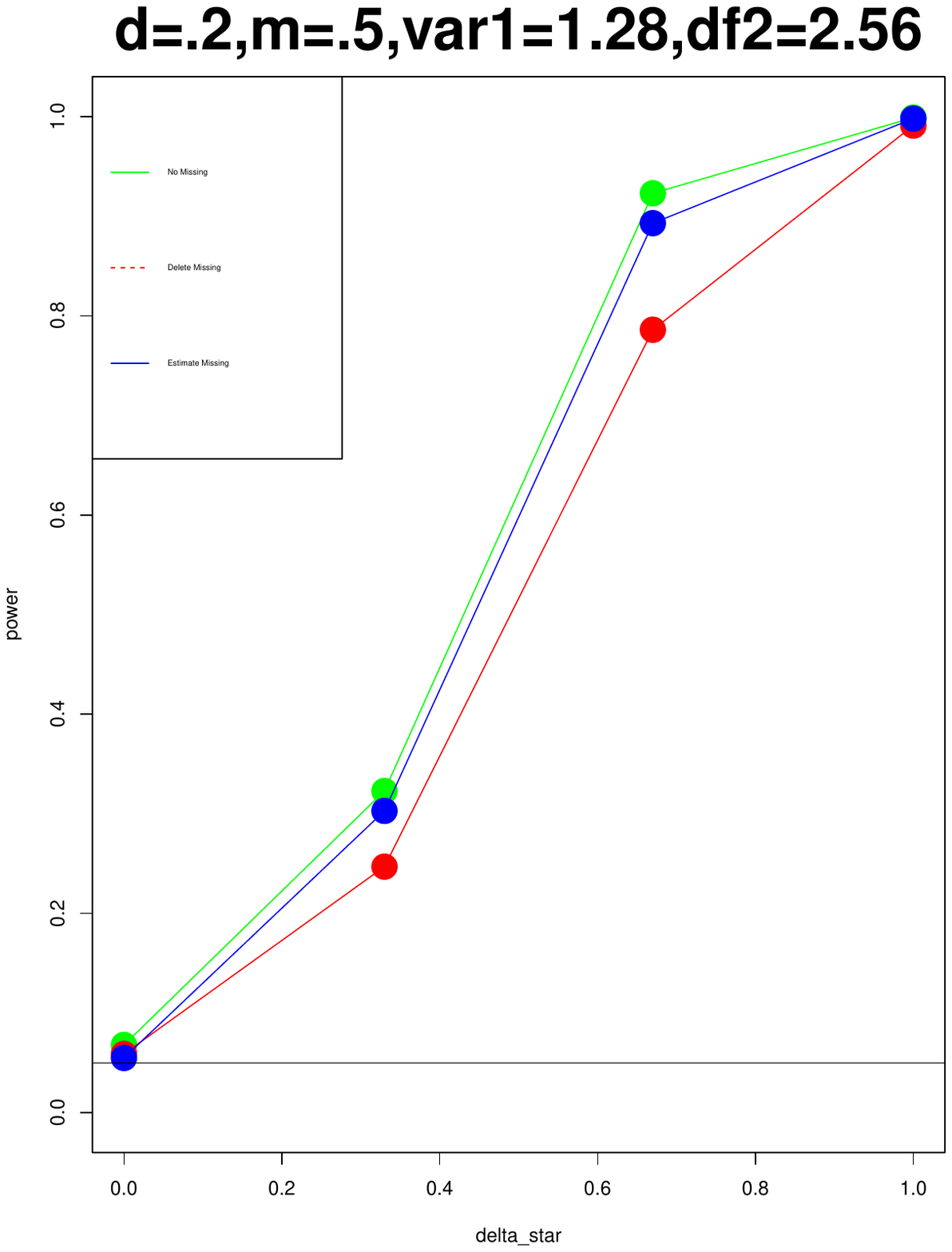}
\includegraphics[width = 2.3in, height = 1.5in]{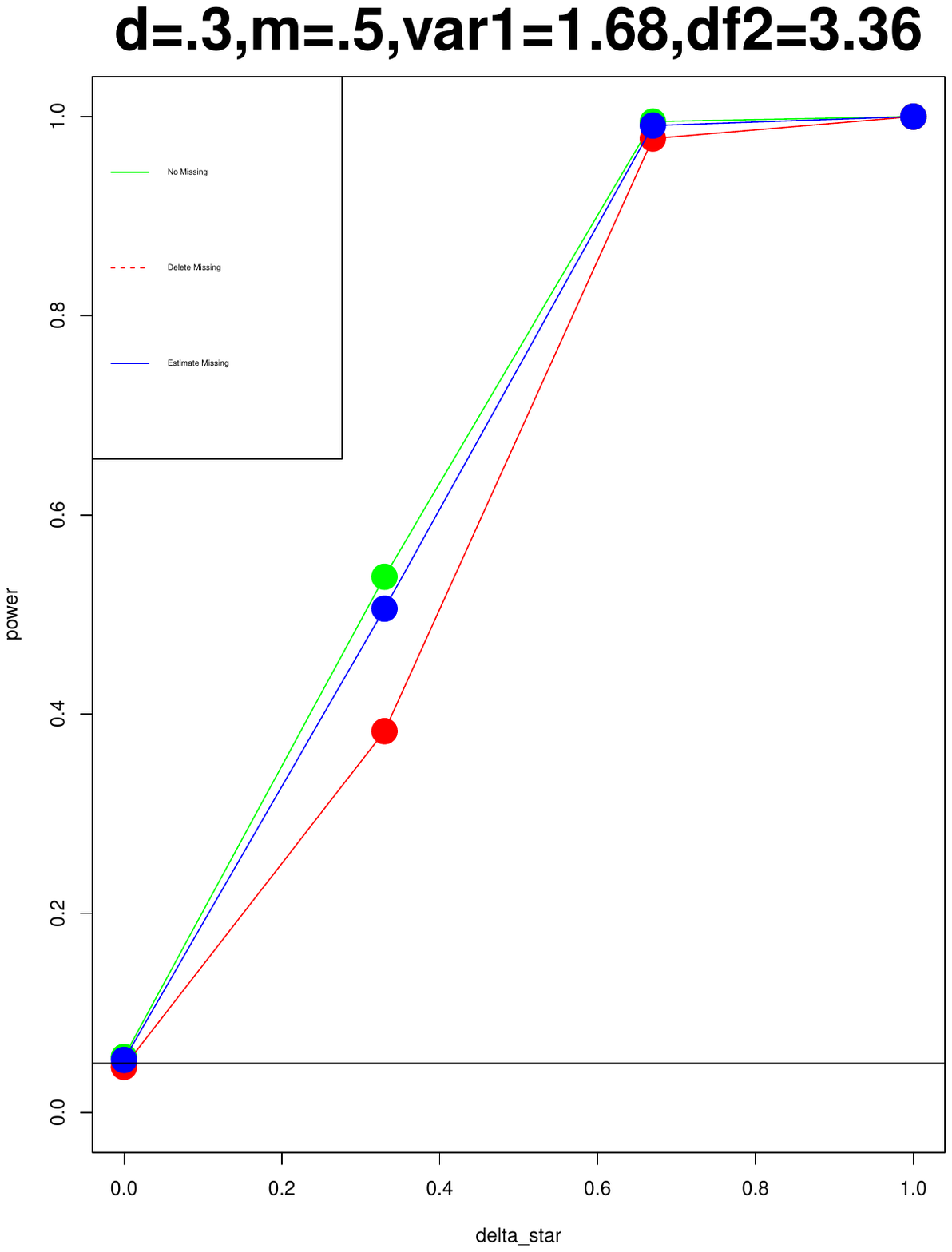}

\hspace{1.5cm}
$d=.1 \quad m=.1$
\hspace{3cm}
$d=.2 \quad m=.1$
\hspace{3cm}
$d=.3 \quad m=.1$

\includegraphics[width = 2.3in, height = 1.5in]{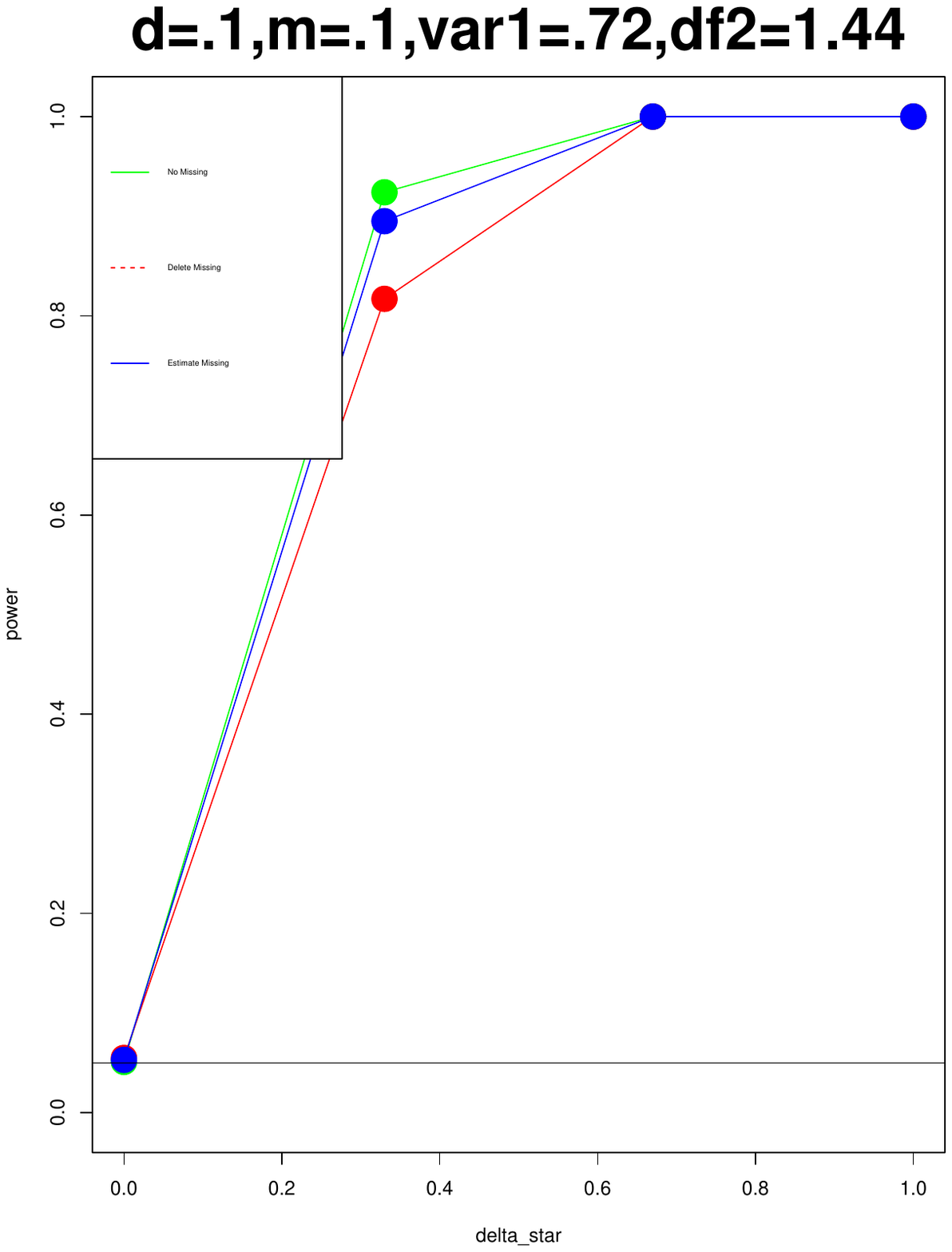}
\includegraphics[width = 2.3in, height = 1.5in]{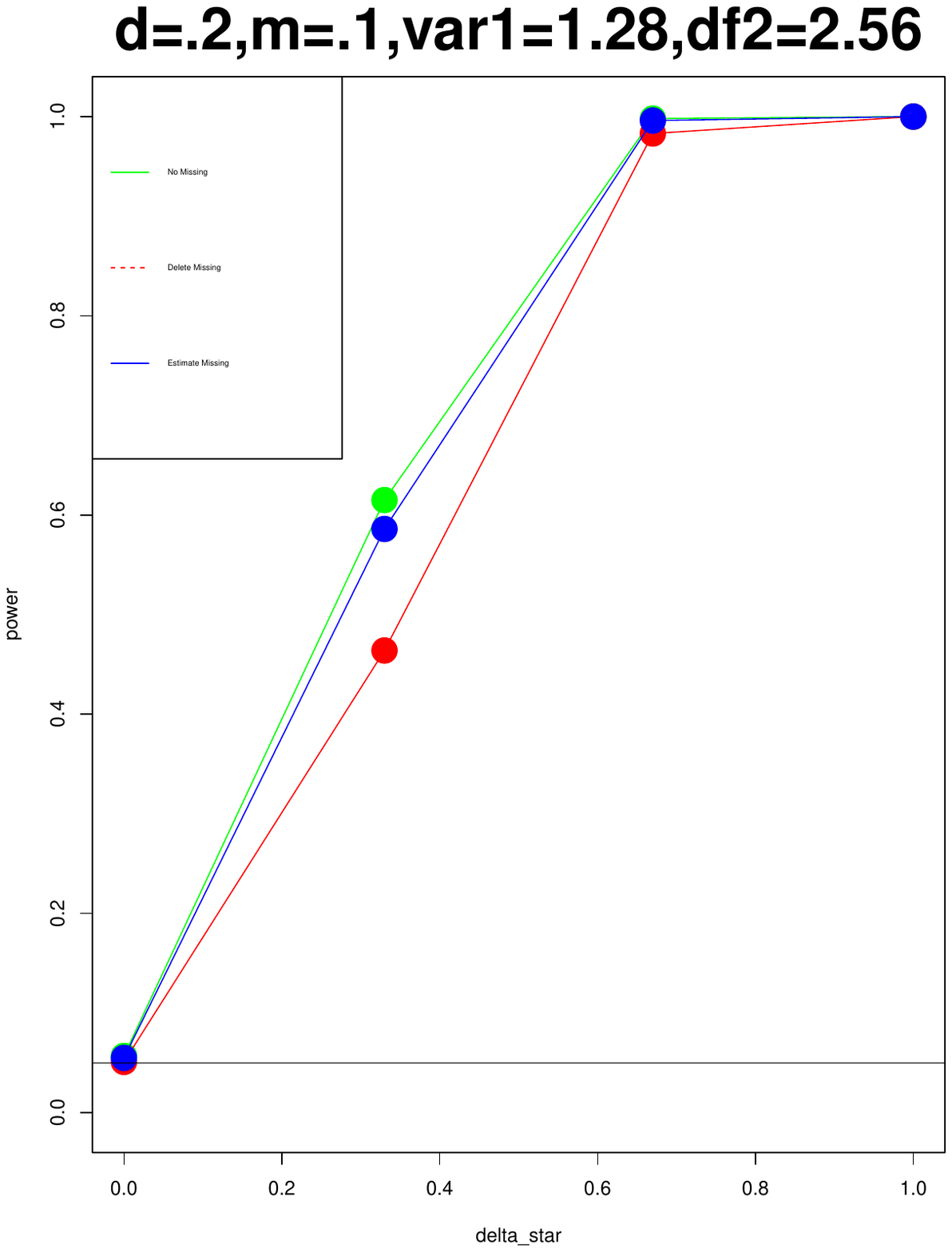}
\includegraphics[width = 2.3in, height = 1.5in]{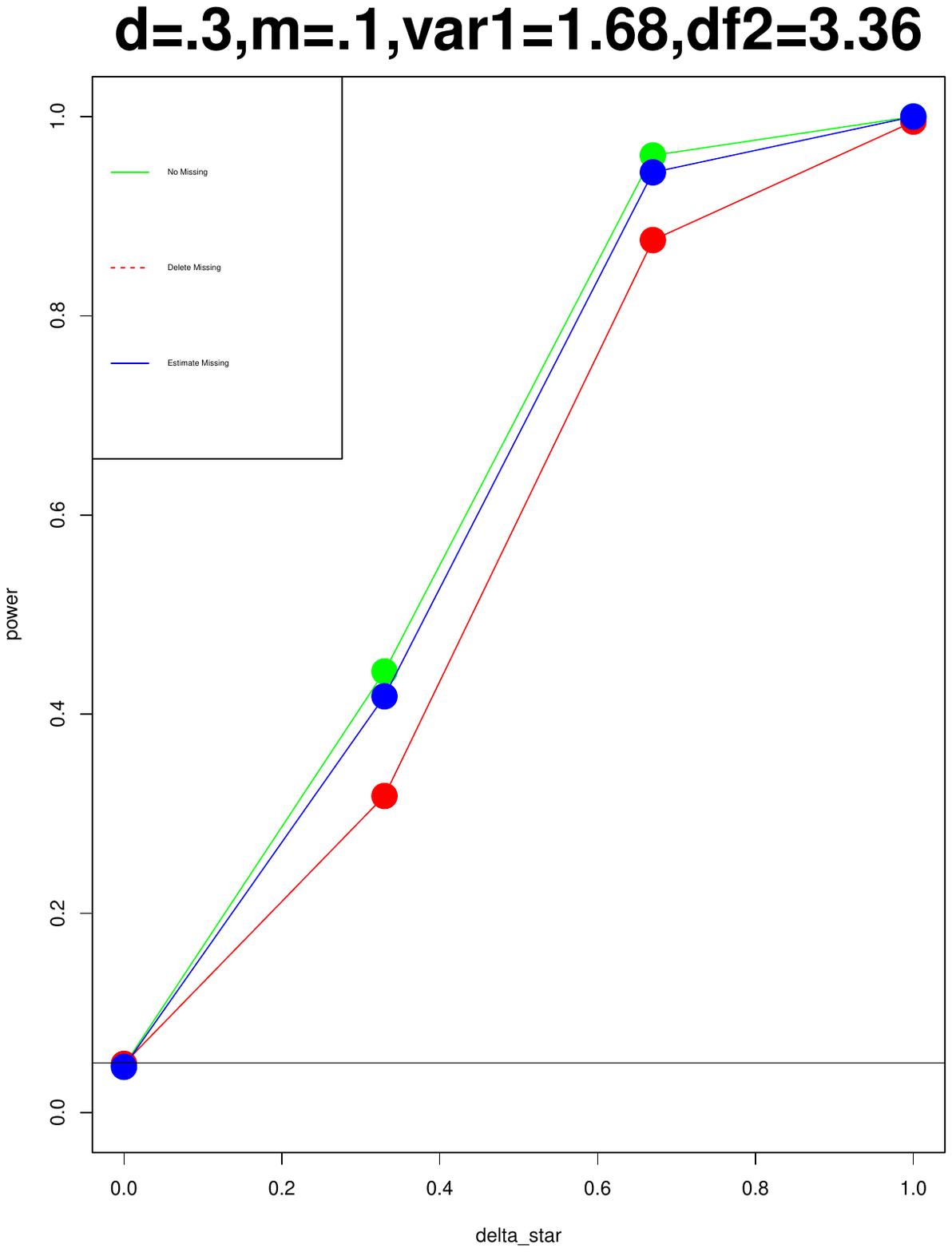}

\hspace{1.5cm}
$d=.1 \quad m=.5$
\hspace{3cm}
$d=.2 \quad m=.5$
\hspace{3cm}
$d=.3 \quad m=.5$

\includegraphics[width = 2.3in, height = 1.5in]{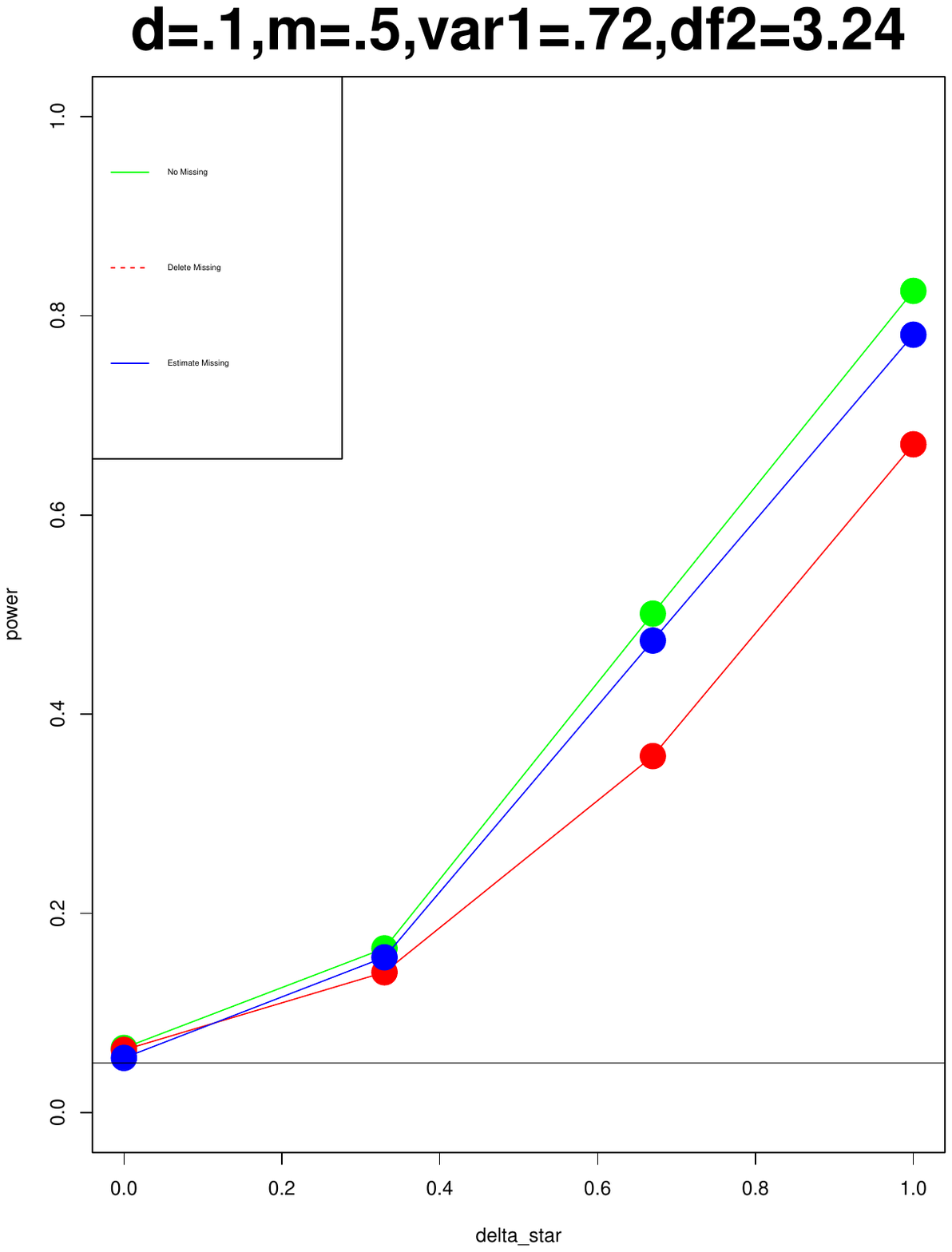}
\includegraphics[width = 2.3in, height = 1.5in]{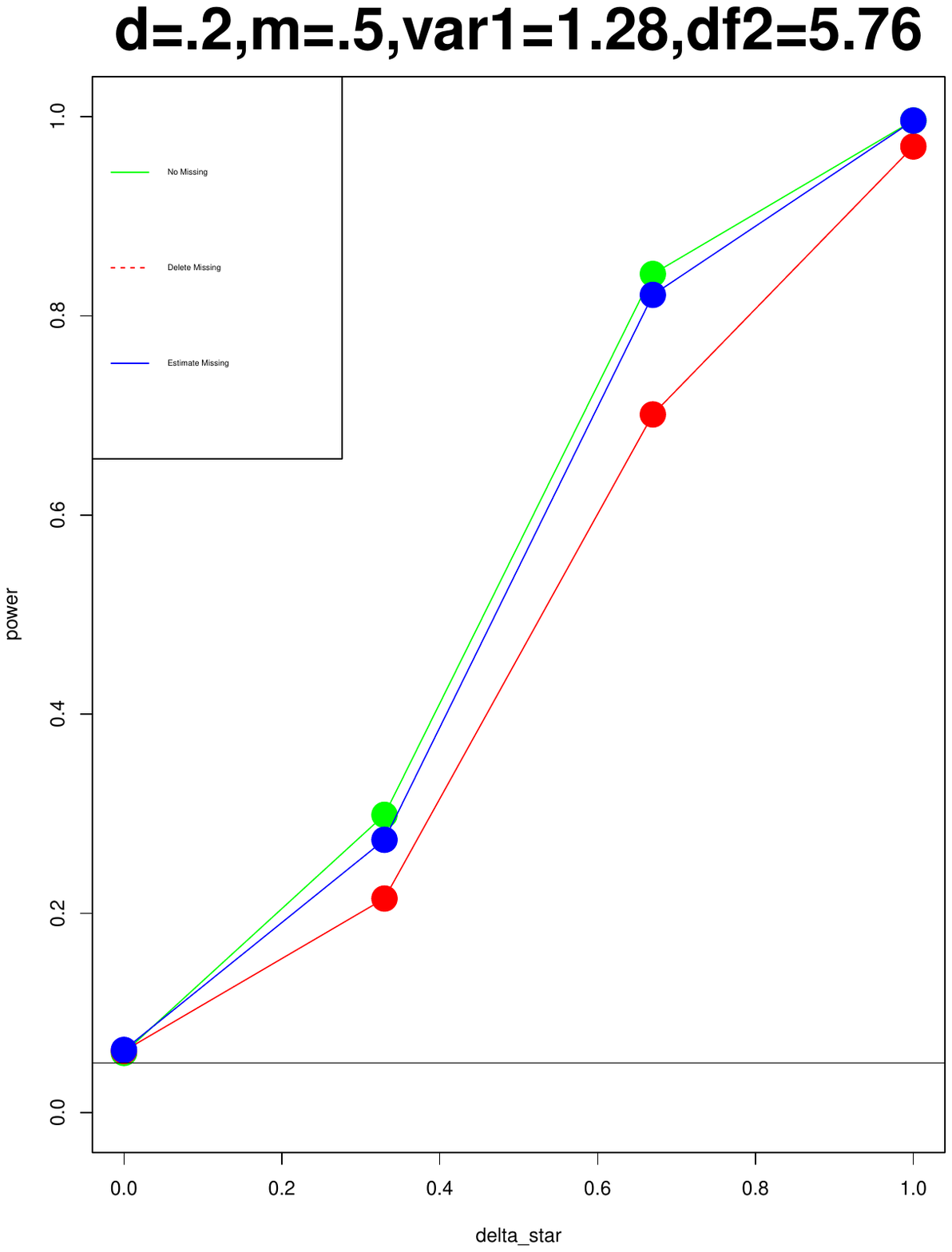}
\includegraphics[width = 2.3in, height = 1.5in]{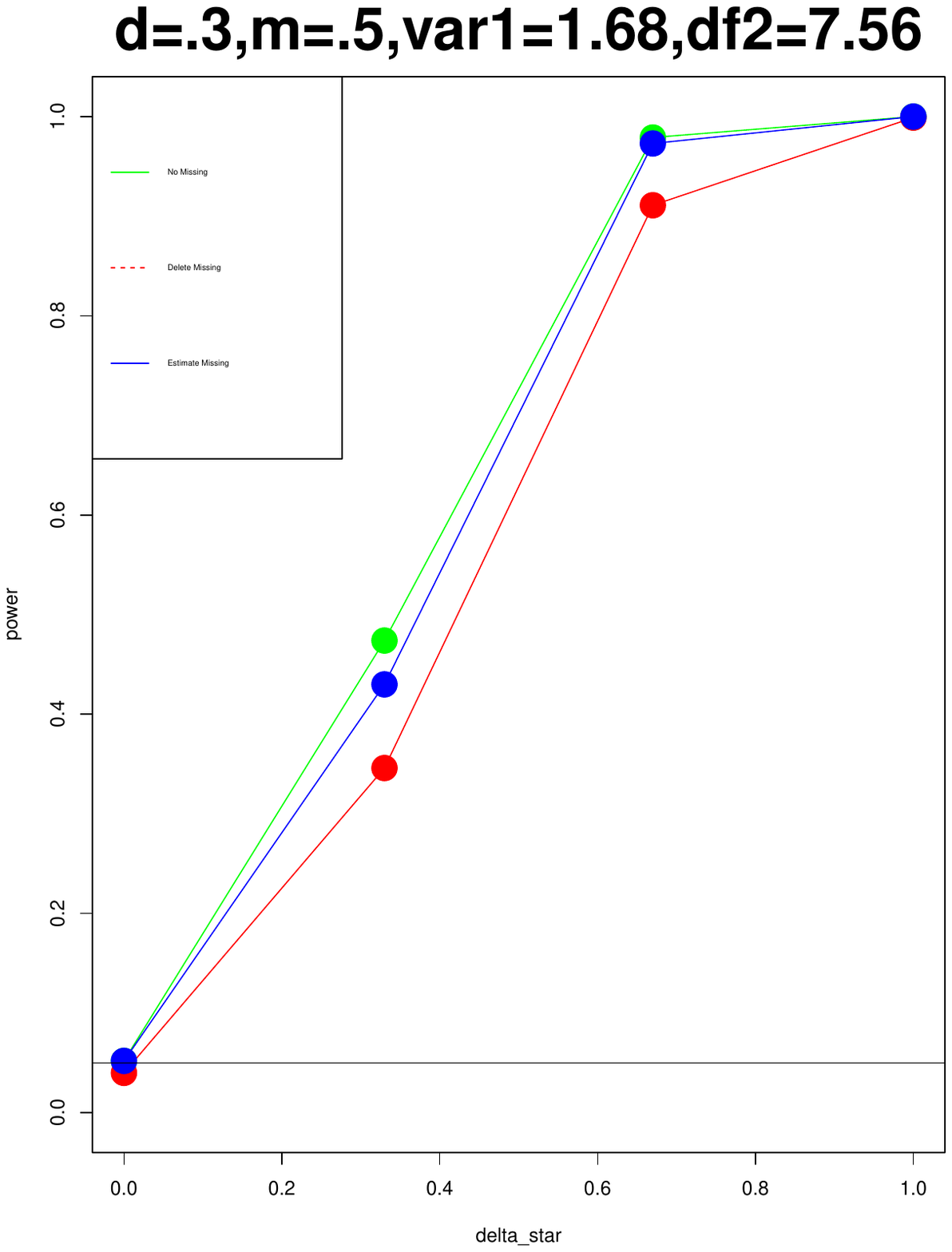}

\hspace{1.5cm}
$d=.1 \quad m=.1$
\hspace{3cm}
$d=.2 \quad m=.1$
\hspace{3cm}
$d=.3 \quad m=.1$

\includegraphics[width = 2.3in, height = 1.5in]{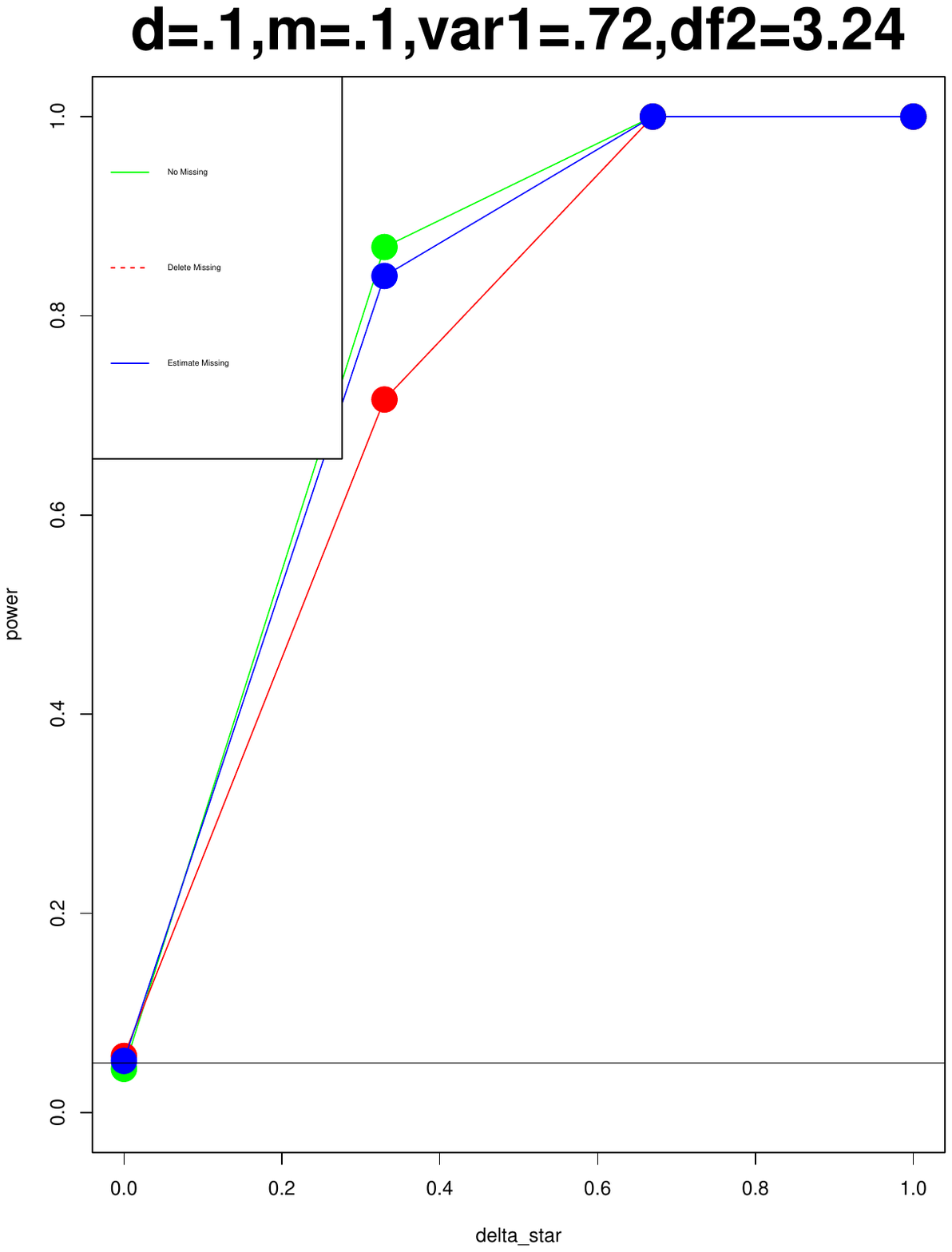}
\includegraphics[width = 2.3in, height = 1.5in]{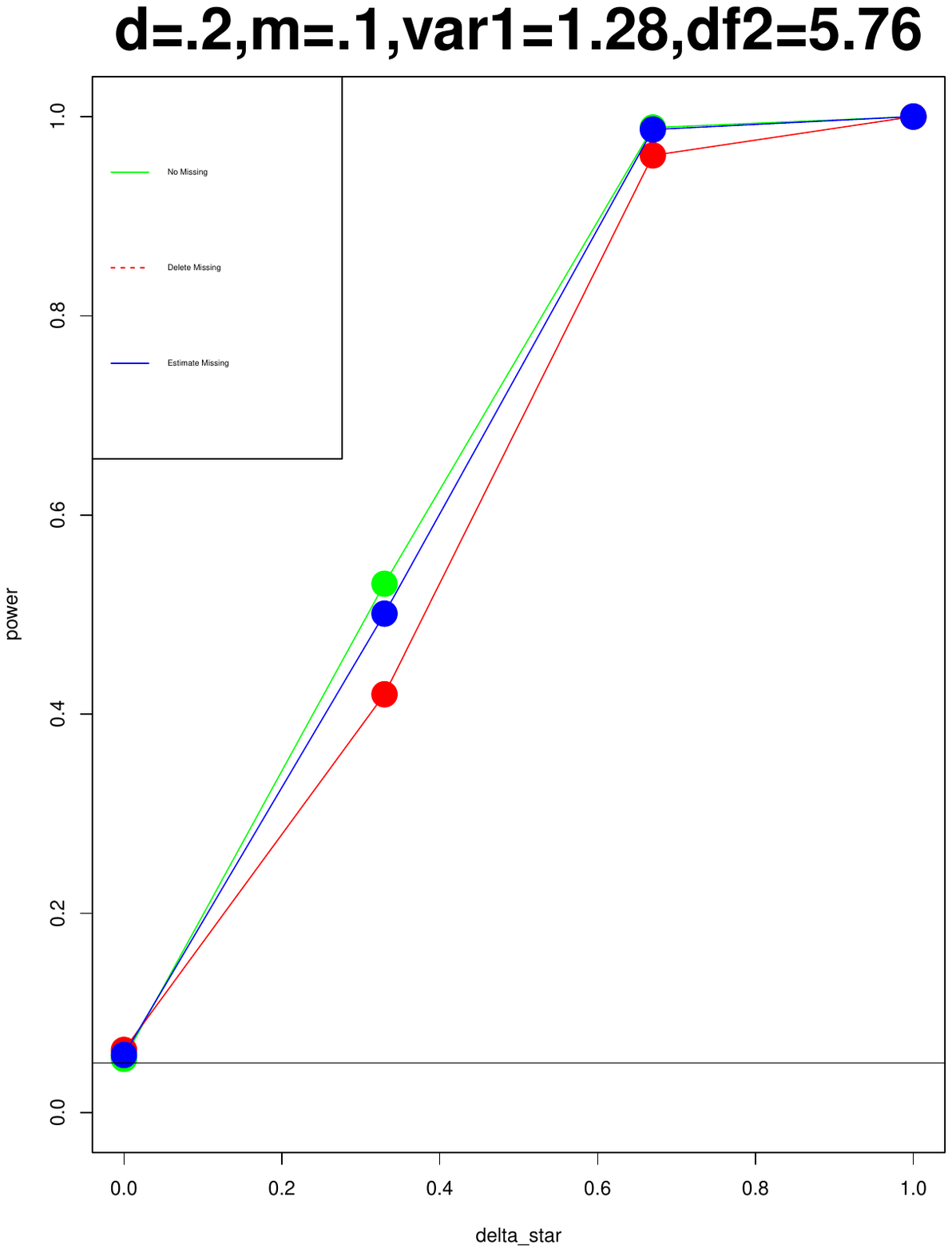}
\includegraphics[width = 2.3in, height = 1.5in]{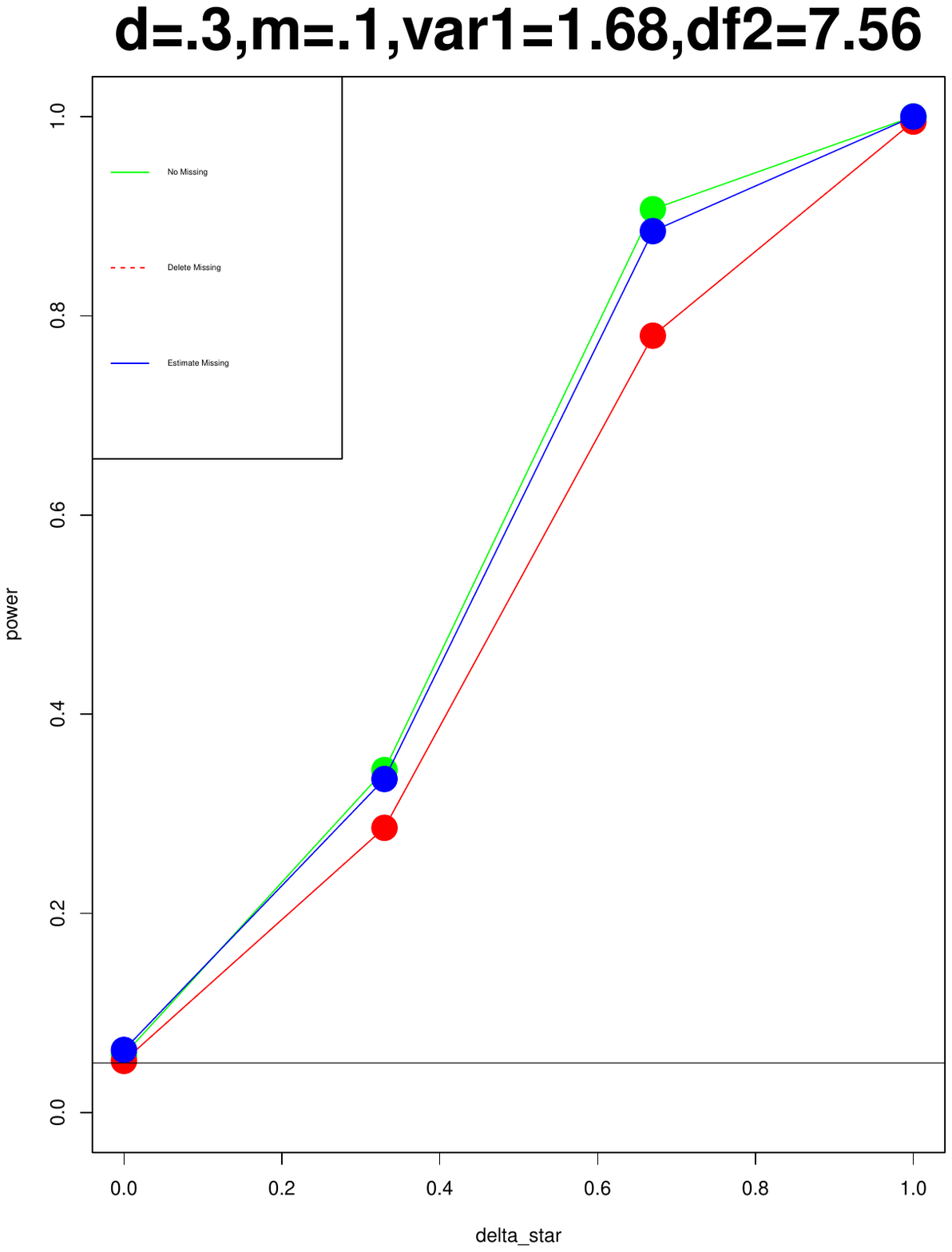}

\subsubsection{When one Trait has Poisson Distribution and other Trait has Chi Squares Distribution}

The following tables represents powers for the three strategies evaluated at $\delta = 0, .33, .67$ and 1.

\textbf{ For $\rho_1 > \rho_2$}

\vspace{.4cm}

\hspace{3.5cm}$m=.1$
\hspace{8.5cm}$m=.5$

\vspace{.2cm}

\begin{tabular}{|c|c|c|c|c|}
\hline 
Strategy & $\delta = 0$ & $\delta = .33$ & $\delta = .67$ & $\delta = 1$ \\ 
\hline 
use same & 0.051 & 0.390 & 0.945 & 0.999 \\ 
\hline 
use other & 0.046 & 0.368 & 0.924 & 1 \\ 
\hline 
use both & 0.050 & 0.401 & 0.948 & 0.999 \\ 
\hline 
\end{tabular} \hspace{1.5cm}
\begin{tabular}{|c|c|c|c|c|}
\hline 
Strategy & $\delta = 0$ & $\delta = .33$ & $\delta = .67$ & $\delta = 1$ \\ 
\hline 
use same & 0.047 & 0.682 & 1 & 1 \\ 
\hline 
use other & 0.049 & 0.642 & 1 & 1 \\ 
\hline 
use both & 0.047 & 0.686 & 1 & 1 \\ 
\hline 
\end{tabular} 

\hspace{.4cm}

\textbf{ For $\rho_1 < \rho_2$}

\pagebreak

\hspace{3.5cm}$m=.1$
\hspace{8.5cm}$m=.5$

\vspace{.2cm}

\begin{tabular}{|c|c|c|c|c|}
\hline 
Strategy & $\delta = 0$ & $\delta = .33$ & $\delta = .67$ & $\delta = 1$ \\ 
\hline 
use same & 0.054 & 0.354 & 0.943 & 1 \\ 
\hline 
use other & 0.053 & 0.366 & 0.956 & 1 \\ 
\hline 
use both & 0.051 & 0.358 & 0.945 & 1 \\ 
\hline 
\end{tabular} \hspace{1.5cm}
\begin{tabular}{|c|c|c|c|c|}
\hline 
Strategy & $\delta = 0$ & $\delta = .33$ & $\delta = .67$ & $\delta = 1$ \\ 
\hline 
use same & 0.049 & 0.904 & 1 & 1 \\ 
\hline 
use other & 0.053 & 0.916 & 1 & 1 \\ 
\hline 
use both & 0.054 & 0.904 & 1 & 1 \\ 
\hline 
\end{tabular}

\hspace{.4cm}

\textbf{ For $\rho_1 = \rho_2$}

\vspace{.4cm}

\hspace{3.5cm}$m=.1$
\hspace{8.5cm}$m=.5$

\vspace{.2cm}

\begin{tabular}{|c|c|c|c|c|}
\hline 
Strategy & $\delta = 0$ & $\delta = .33$ & $\delta = .67$ & $\delta = 1$ \\ 
\hline 
use same & 0.051 & 0.545 & 0.984 & 1 \\ 
\hline 
use other & 0.053 & 0.544 & 0.988 & 1 \\ 
\hline 
use both & 0.050 & 0.545 & 0.986 & 1 \\ 
\hline 
\end{tabular} \hspace{1.5cm}
\begin{tabular}{|c|c|c|c|c|}
\hline 
Strategy & $\delta = 0$ & $\delta = .33$ & $\delta = .67$ & $\delta = 1$ \\ 
\hline 
use same & 0.050 & 0.816 & 1 & 1 \\ 
\hline 
use other & 0.053 & 0.829 & 1 & 1 \\ 
\hline 
use both & 0.047 & 0.818 & 1 & 1 \\ 
\hline 
\end{tabular} 

\vspace{1cm}

According to the above results we can conclude-

\begin{center}
\begin{tabular}{|c|c|}
\hline 
Case & Best Strategy \\ 
\hline 
$\rho_1 > \rho_2$ & use same \\ 
\hline 
$\rho_1 < \rho_2$ & use other \\ 
\hline 
$\rho_1 = \rho_2$ & use other \\ 
\hline 
\end{tabular} 

\end{center}

Now we go for power comparison among no missing, estimated missing and deleted missing.

We generate 1st trait from chi squares distribution (section 6.2) with the parameters $\alpha = \alpha_1, \beta = \beta_1, df = df_1$ keeping $p^\star = p^\star_1$ and we generate 2nd trait from poisson distribution (section 6.2) with the parameters $\alpha = \alpha_2, \beta = \beta_2, \lambda = \lambda_2$ keeping $p^\star = p^\star_2$.

We have done simulation for three choices of $d$ as .1, .2, .3 and for each $d$ we take $(p^\star_1, p^\star_2)$ as (.1, .2), (.2, .2).

We take $\alpha_1 = 5, \alpha_2 = 10, \beta_1 = 1, and \beta_2 = 2$ and varied $df_1$ and $\lambda_2$ in the following way,

\vspace{.4cm}

\begin{tabular}{|c|c|c|}
\hline 
d & $(p^\star_1, p^\star_2)$ & ($df_1, \lambda_2$) \\ 
\hline 
.1 & (.1, .2) & (.81, 2.88) \\ 
\hline 
.1 & (.2, .2) & (.36, 2.88) \\ 
\hline
.1 & (.2, .1) & (.36, 6.48) \\ 
\hline 

\end{tabular} \hspace{1.5cm}
\begin{tabular}{|c|c|c|}
\hline 
d & $(p^\star_1, p^\star_2)$ & ($df_1, \lambda_2$) \\ 
\hline 
.2 & (.1, .2) & (1.44, 5.12) \\ 
\hline 
.2 & (.2, .2) & (.64, 5.12) \\ 
\hline
.2 & (.2, .1) & (.64, 11.52) \\ 
\hline 

\end{tabular} \hspace{1.5cm}
\begin{tabular}{|c|c|c|}
\hline 
d & $(p^\star_1, p^\star_2)$ & ($df_1, \lambda_2$) \\ 
\hline 
.3 & (.1, .2) & (1.89, 6.72) \\ 
\hline 
.3 & (.2, .2) & (.84, 6.72) \\ 
\hline
.3 & (.2, .1) & (.84, 15.12) \\ 
\hline 
\end{tabular} 

\vspace{.4cm}

We replicate these for $m =$ .1 and .2.

Note that for each of the following simulations we have calculated $\rho_1$ and $\rho_2$ and accordingly we have used the best imputation strategy.

\vspace{.4cm}

\hspace{1.5cm}
$d=.1 \quad m=.5$
\hspace{3cm}
$d=.2 \quad m=.5$
\hspace{3cm}
$d=.3 \quad m=.5$

\includegraphics[width = 2.3in, height = 1.5in]{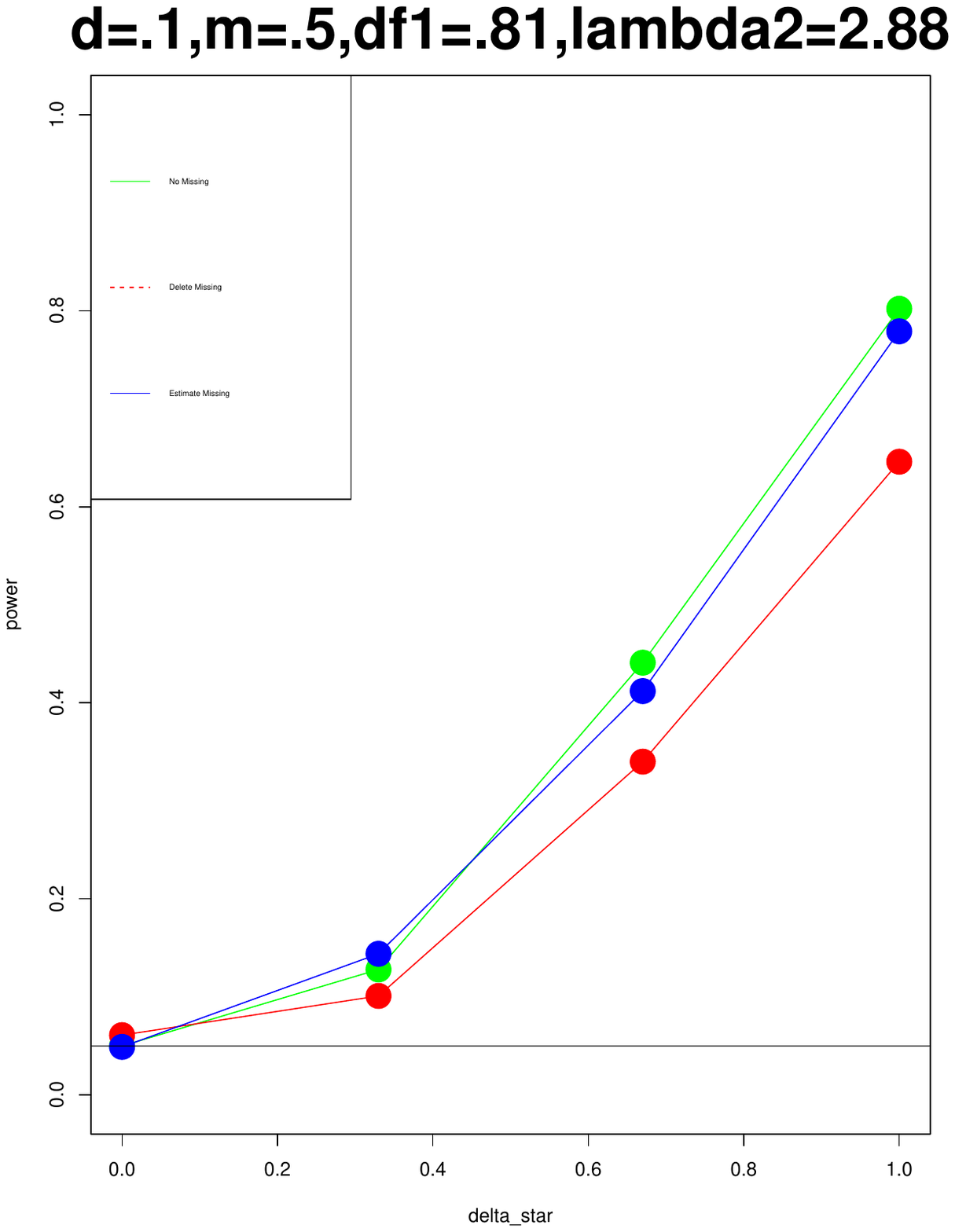}
\includegraphics[width = 2.3in, height = 1.5in]{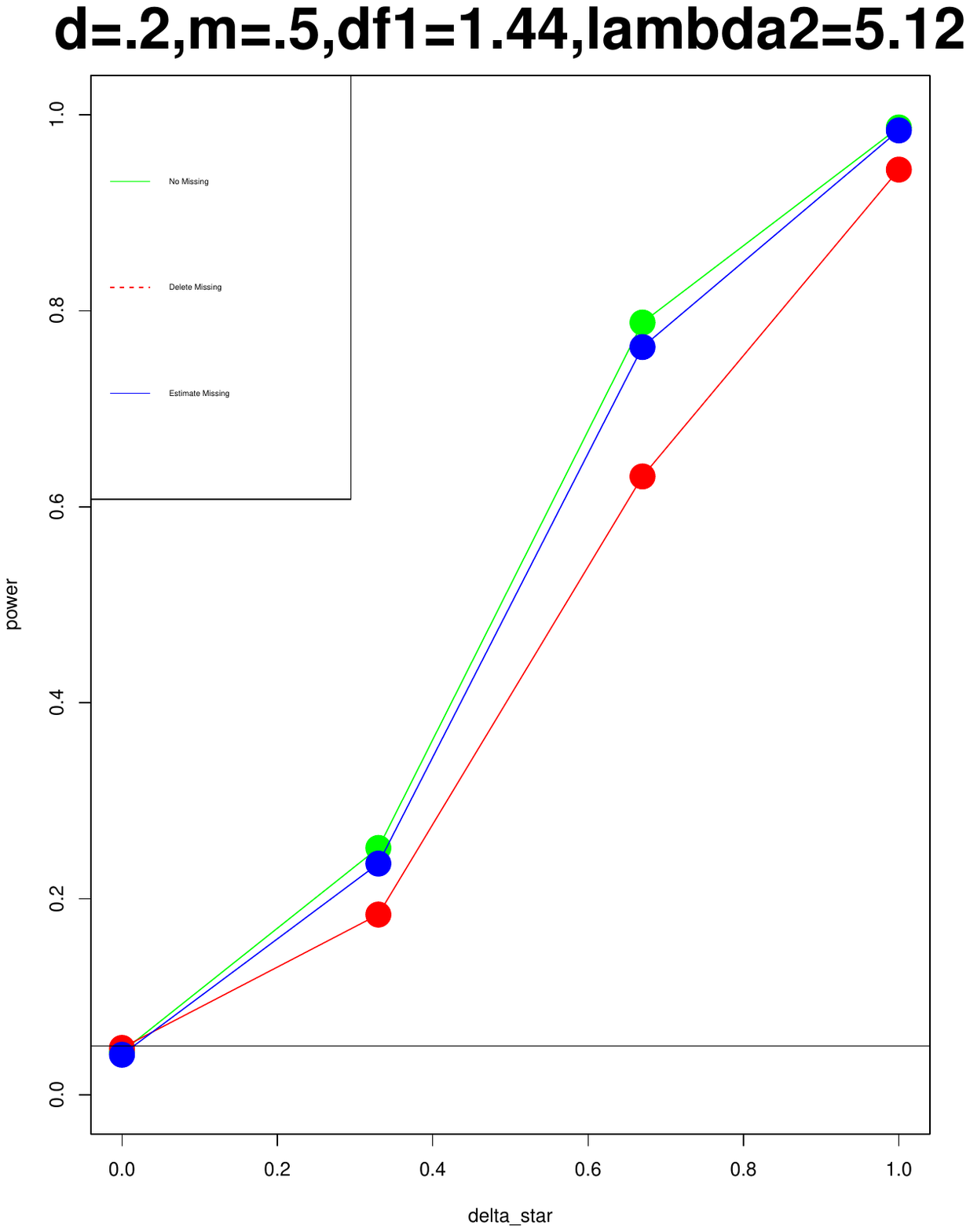}
\includegraphics[width = 2.3in, height = 1.5in]{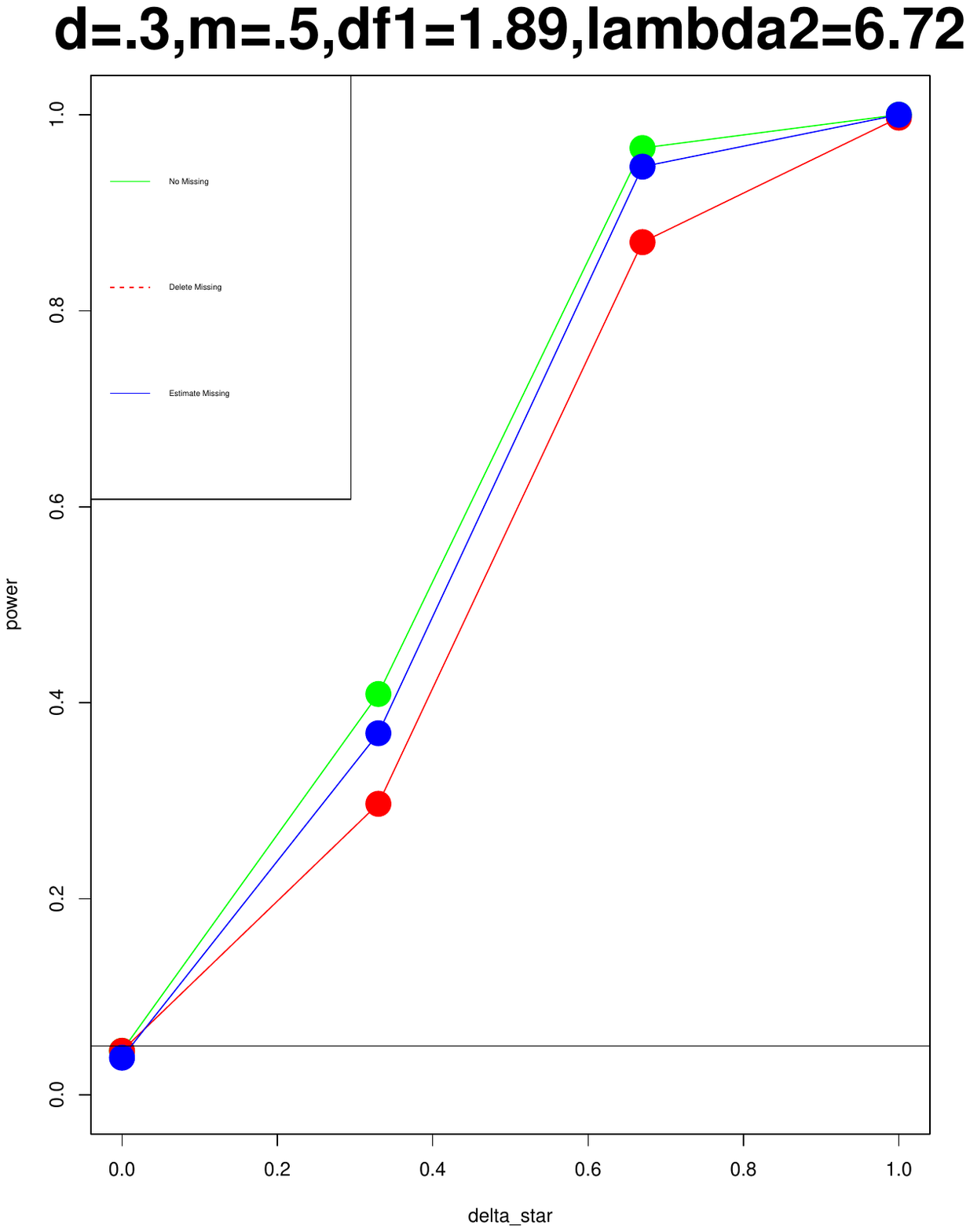}

\pagebreak

\hspace{1.5cm}
$d=.1 \quad m=.1$
\hspace{3cm}
$d=.2 \quad m=.1$
\hspace{3cm}
$d=.3 \quad m=.1$

\includegraphics[width = 2.3in, height = 1.5in]{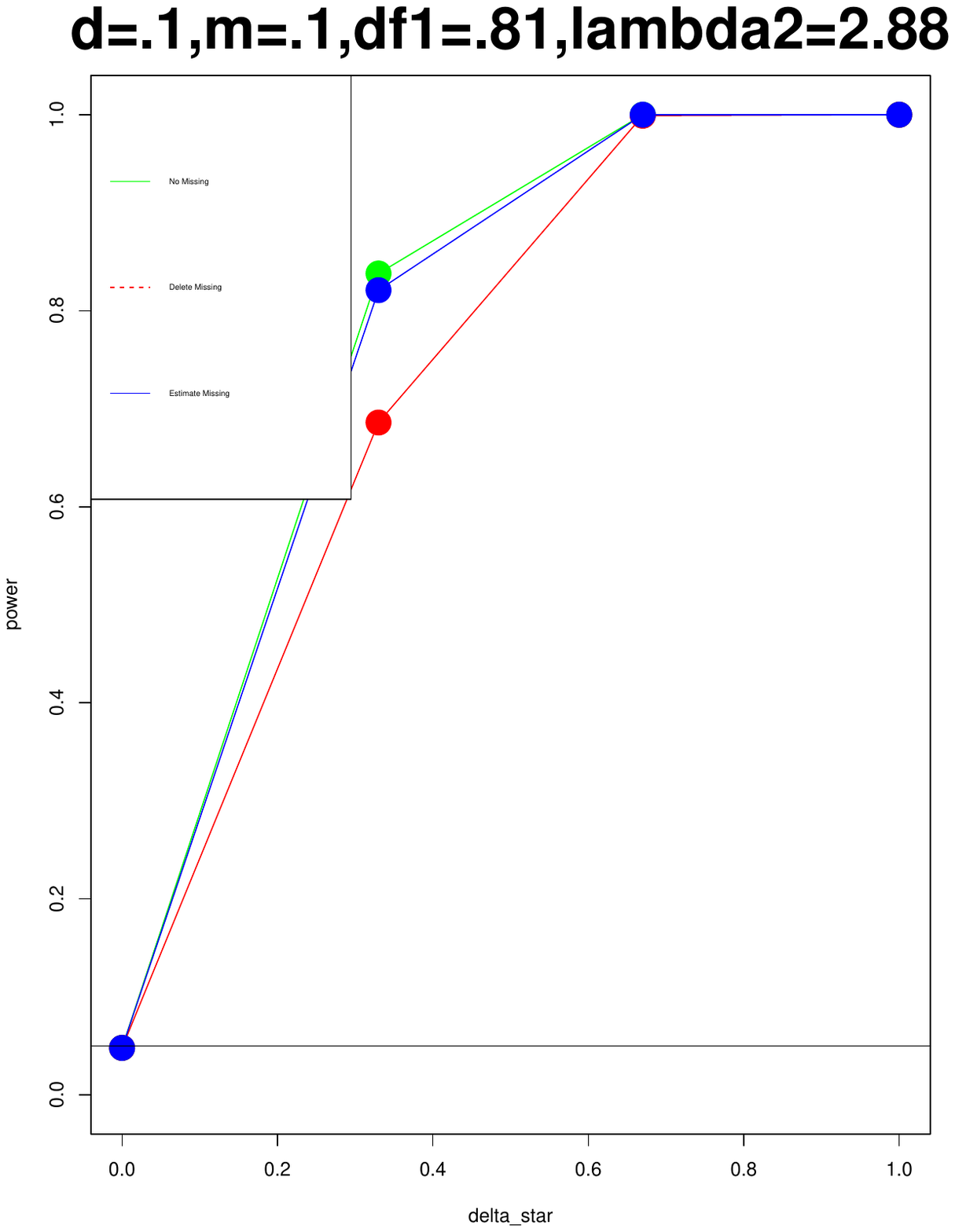}
\includegraphics[width = 2.3in, height = 1.5in]{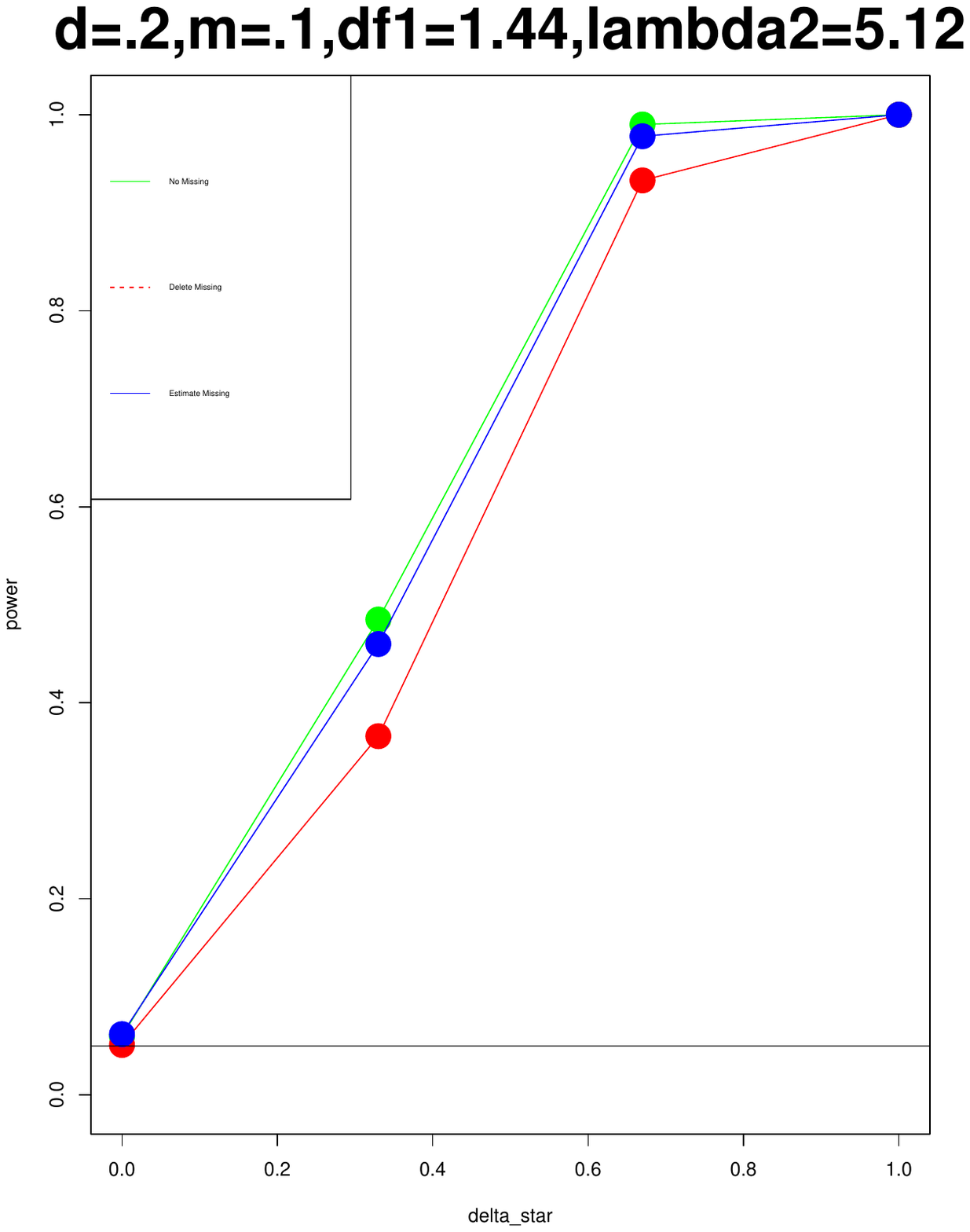}
\includegraphics[width = 2.3in, height = 1.5in]{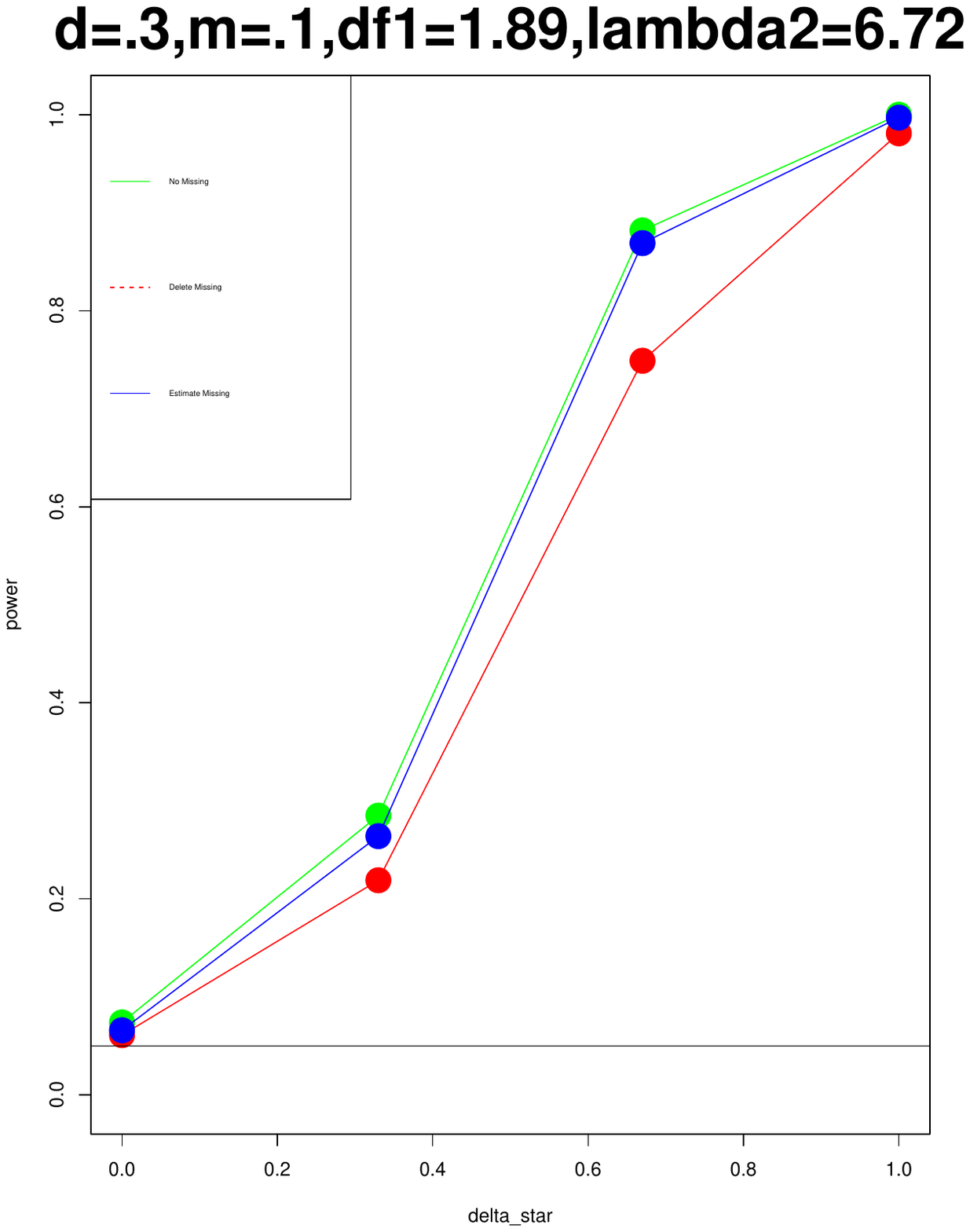}

\hspace{1.5cm}
$d=.1 \quad m=.5$
\hspace{3cm}
$d=.2 \quad m=.5$
\hspace{3cm}
$d=.3 \quad m=.5$

\includegraphics[width = 2.3in, height = 1.5in]{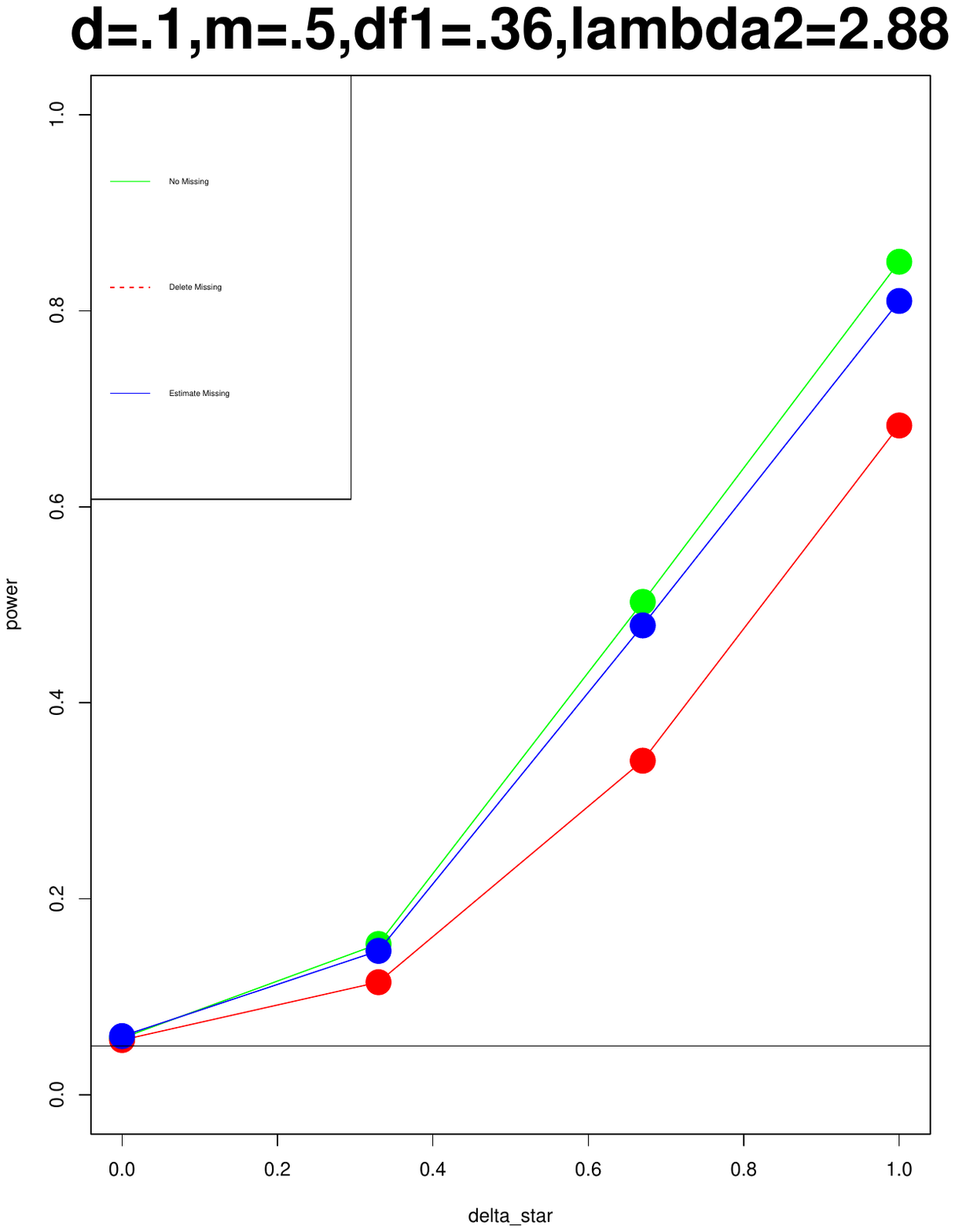}
\includegraphics[width = 2.3in, height = 1.5in]{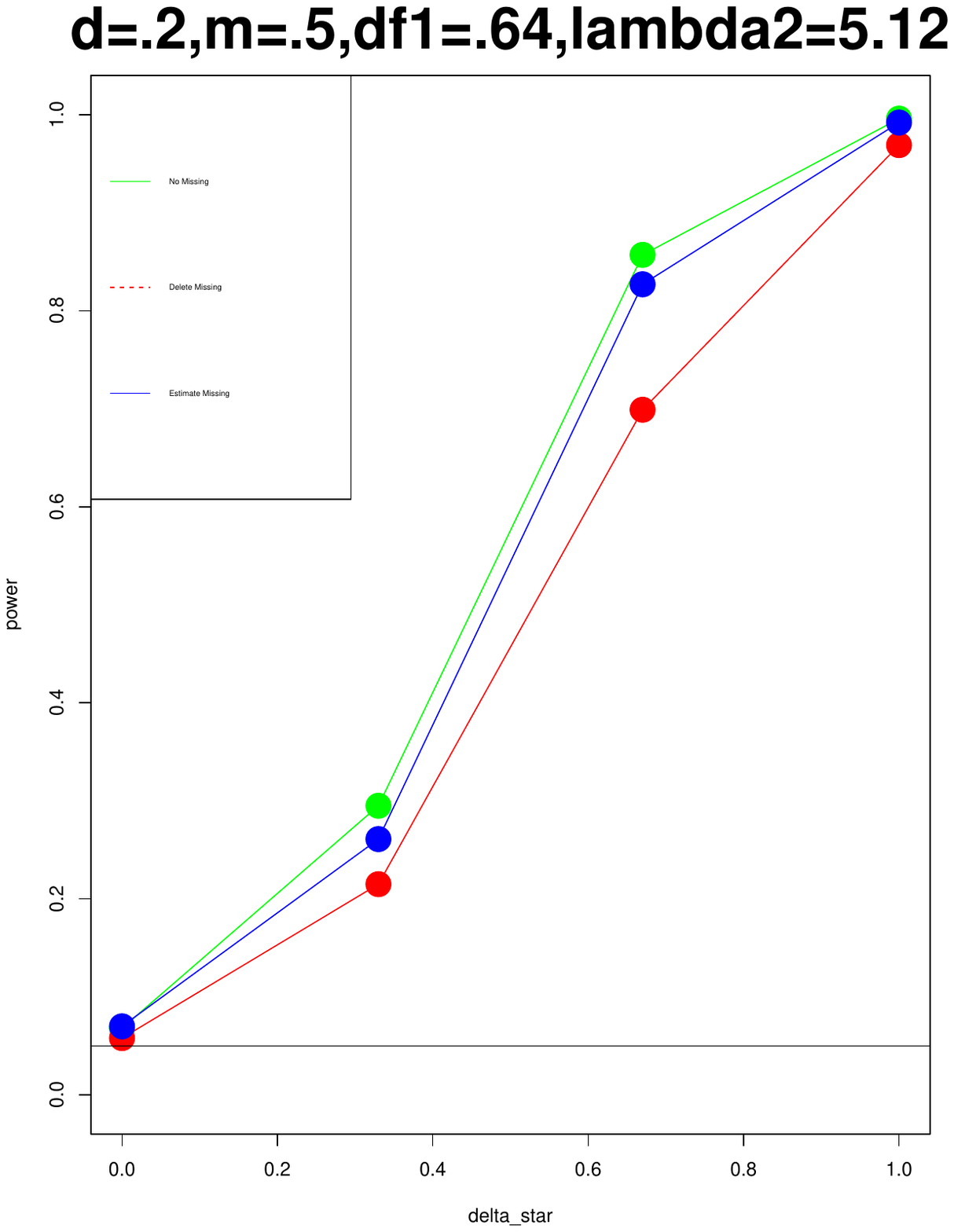}
\includegraphics[width = 2.3in, height = 1.5in]{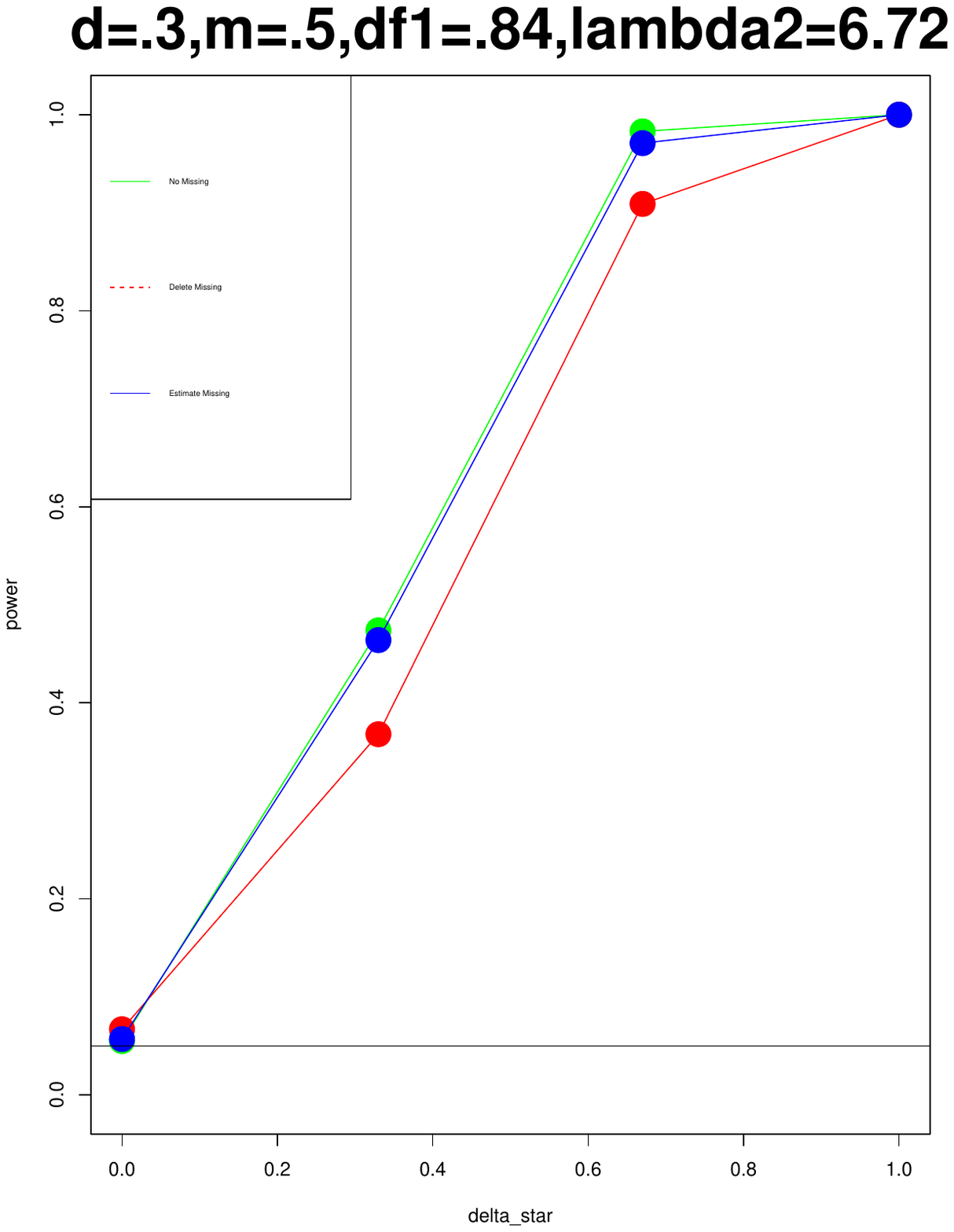}

\hspace{1.5cm}
$d=.1 \quad m=.1$
\hspace{3cm}
$d=.2 \quad m=.1$
\hspace{3cm}
$d=.3 \quad m=.1$

\includegraphics[width = 2.3in, height = 1.5in]{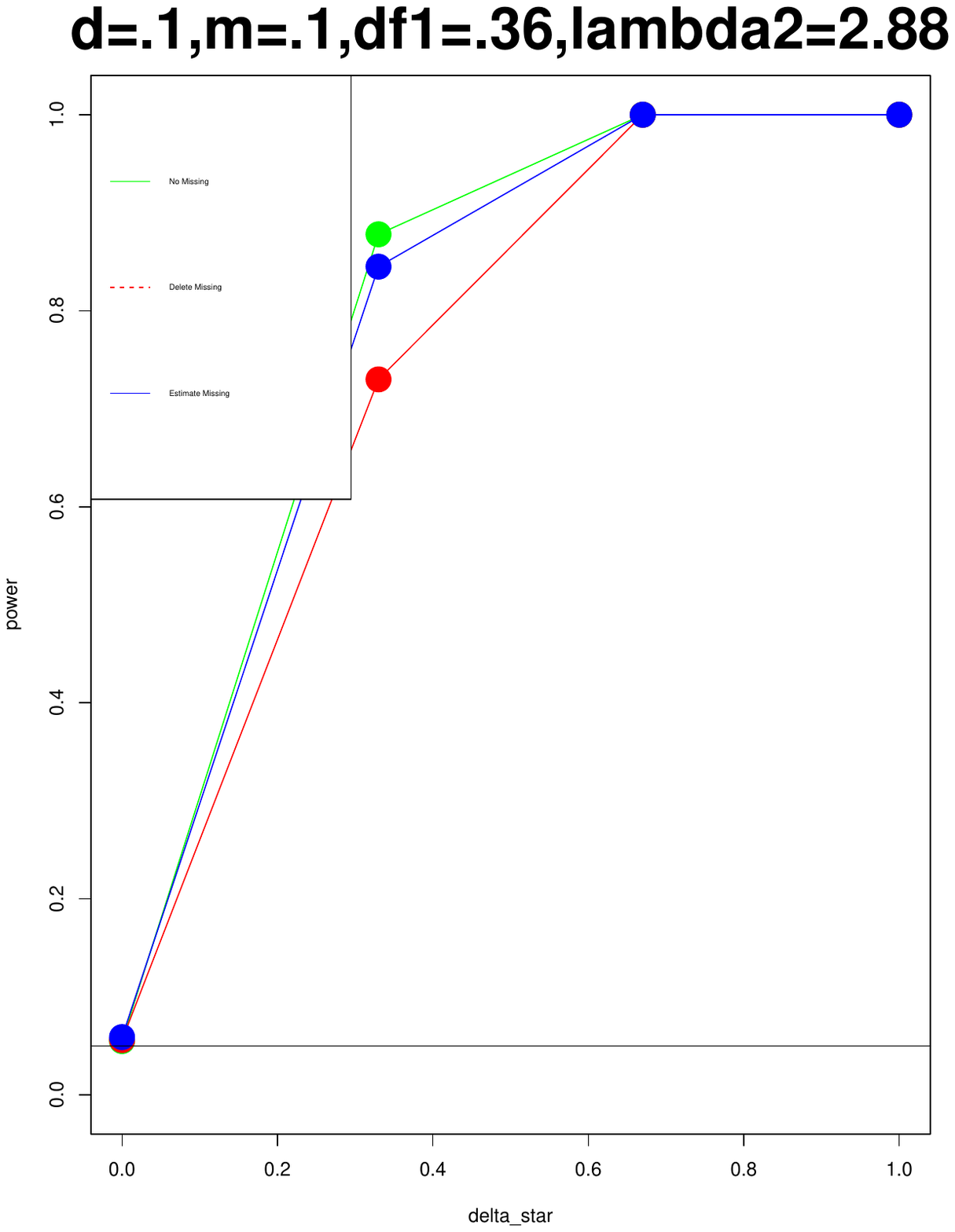}
\includegraphics[width = 2.3in, height = 1.5in]{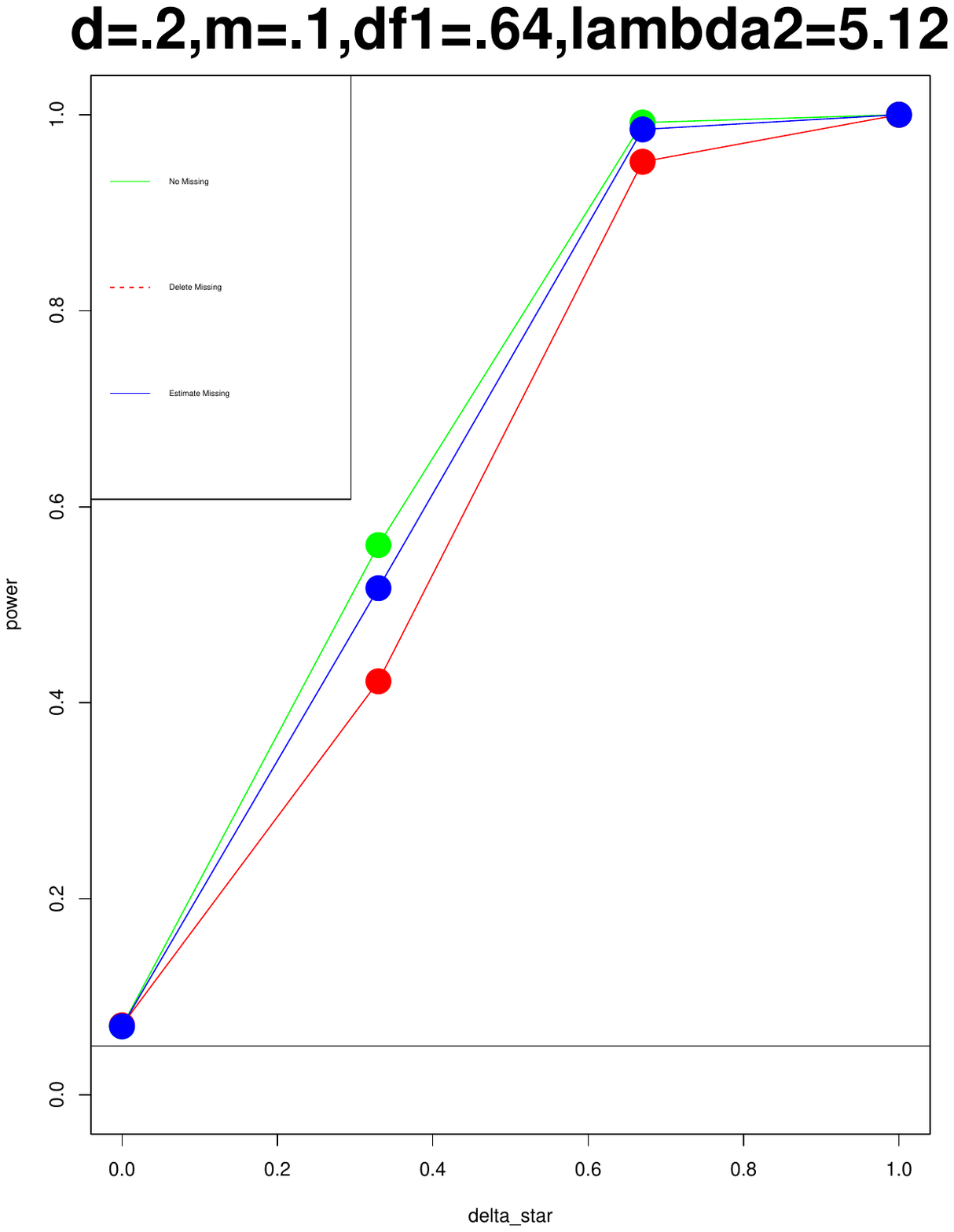}
\includegraphics[width = 2.3in, height = 1.5in]{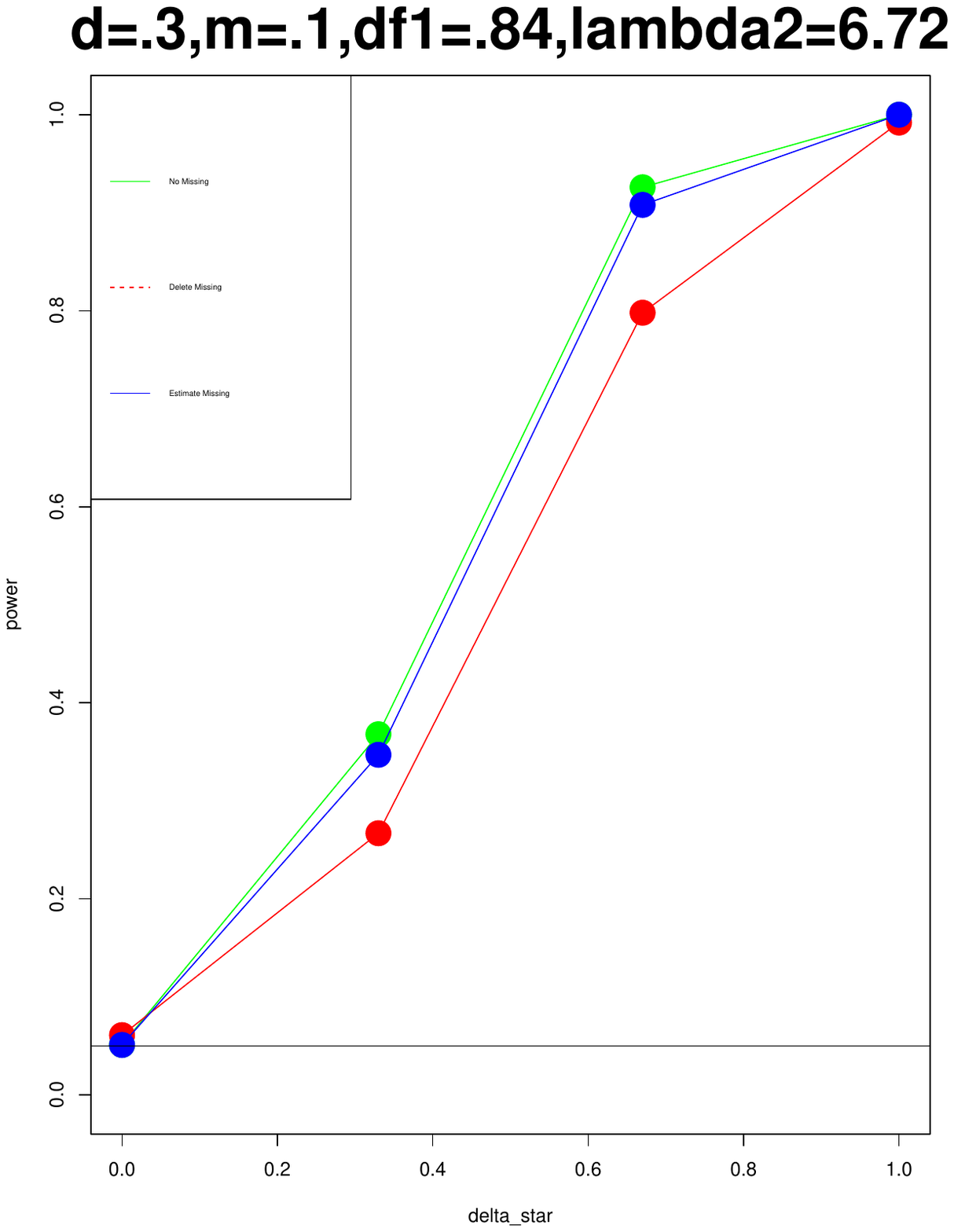}

\hspace{1.5cm}
$d=.1 \quad m=.5$
\hspace{3cm}
$d=.2 \quad m=.5$
\hspace{3cm}
$d=.3 \quad m=.5$

\includegraphics[width = 2.3in, height = 1.5in]{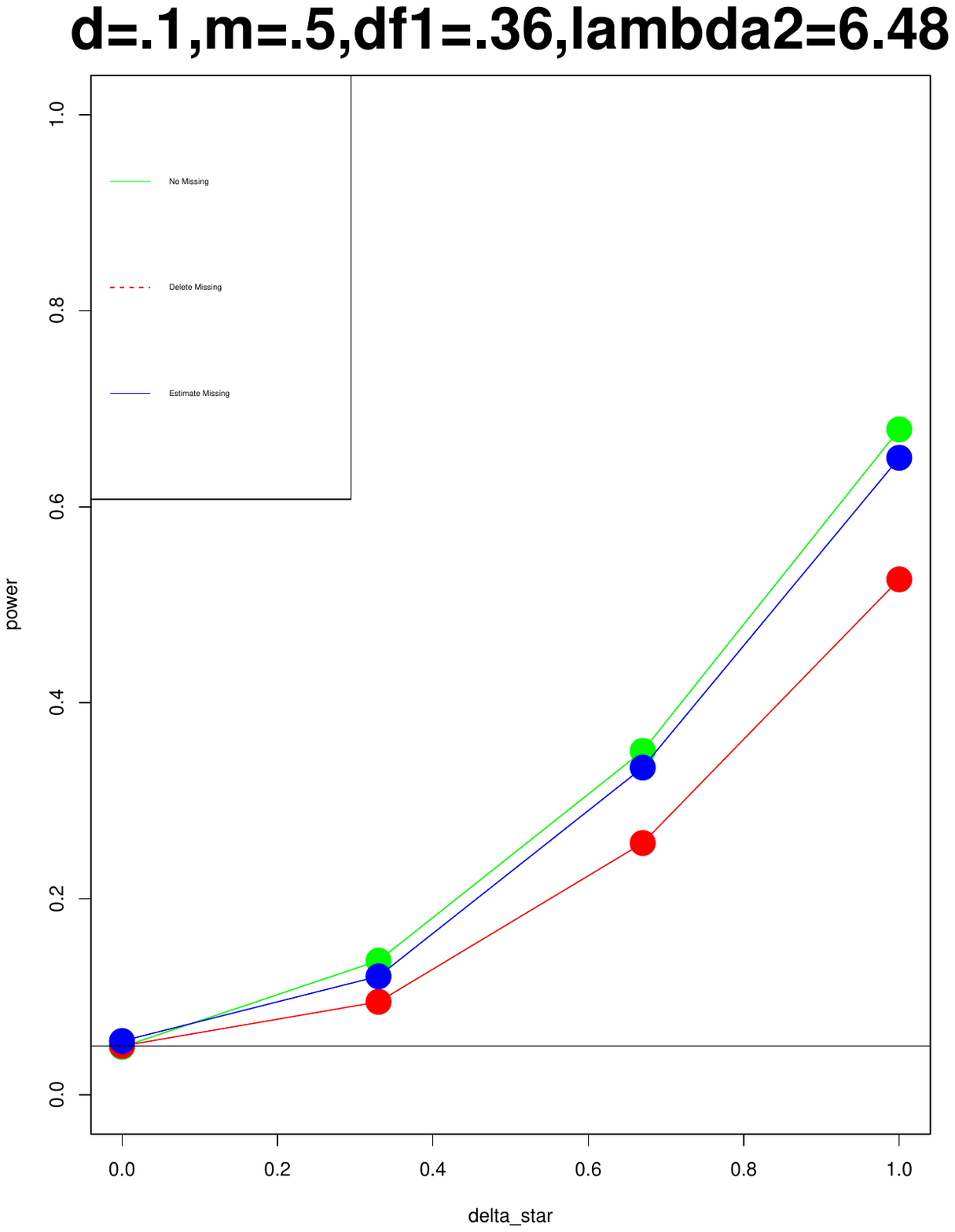}
\includegraphics[width = 2.3in, height = 1.5in]{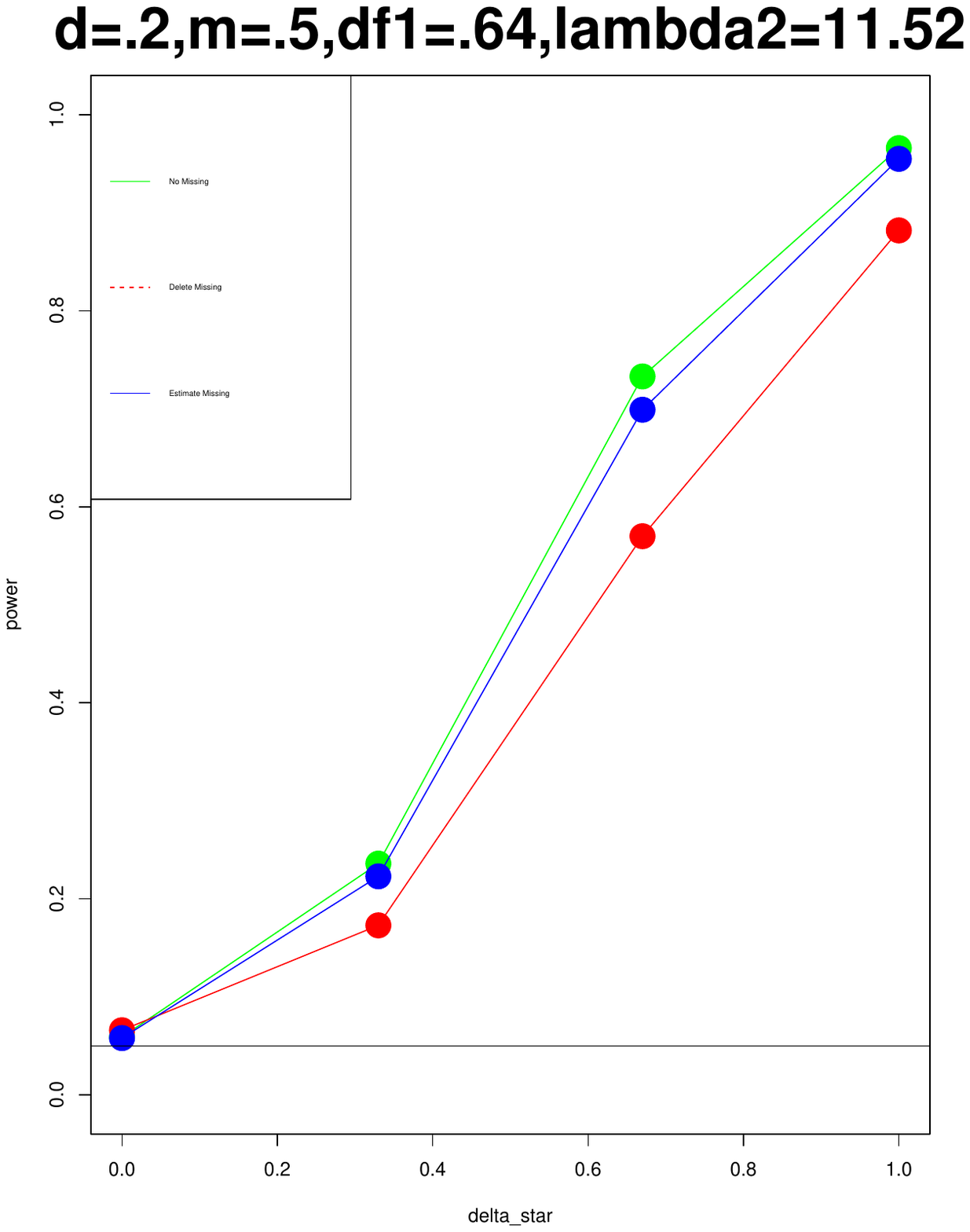}
\includegraphics[width = 2.3in, height = 1.5in]{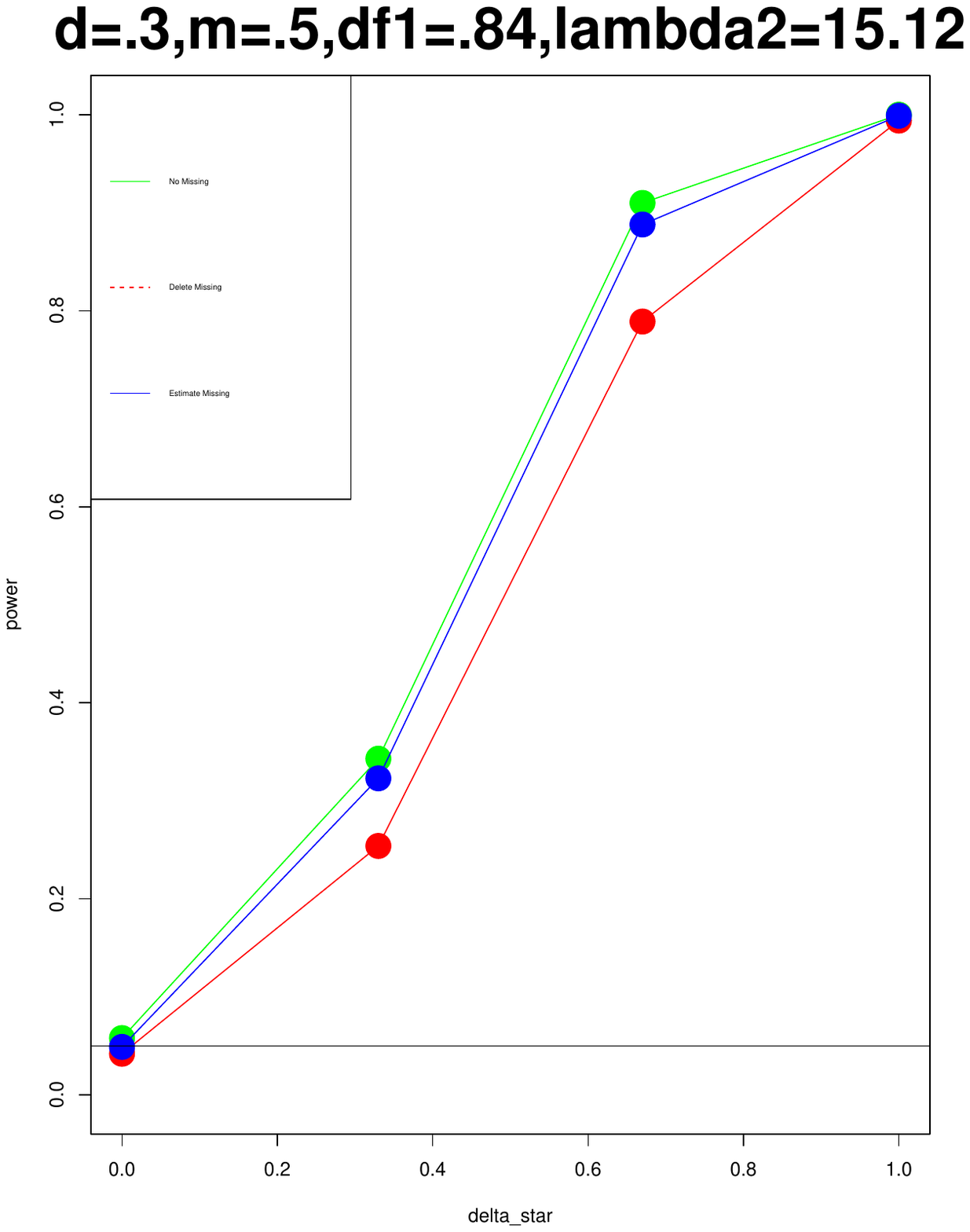}

\hspace{1.5cm}
$d=.1 \quad m=.1$
\hspace{3cm}
$d=.2 \quad m=.1$
\hspace{3cm}
$d=.3 \quad m=.1$

\includegraphics[width = 2.3in, height = 1.5in]{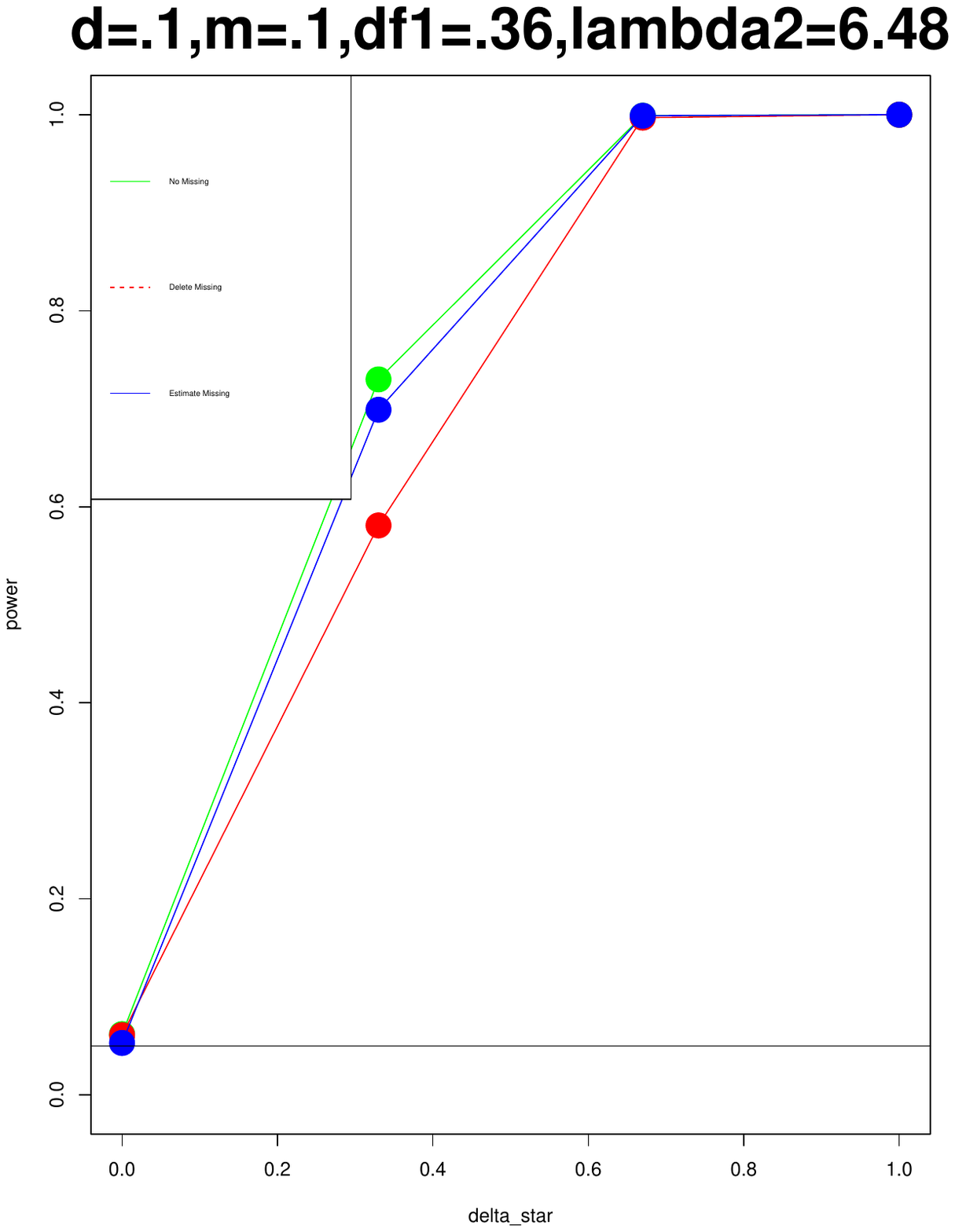}
\includegraphics[width = 2.3in, height = 1.5in]{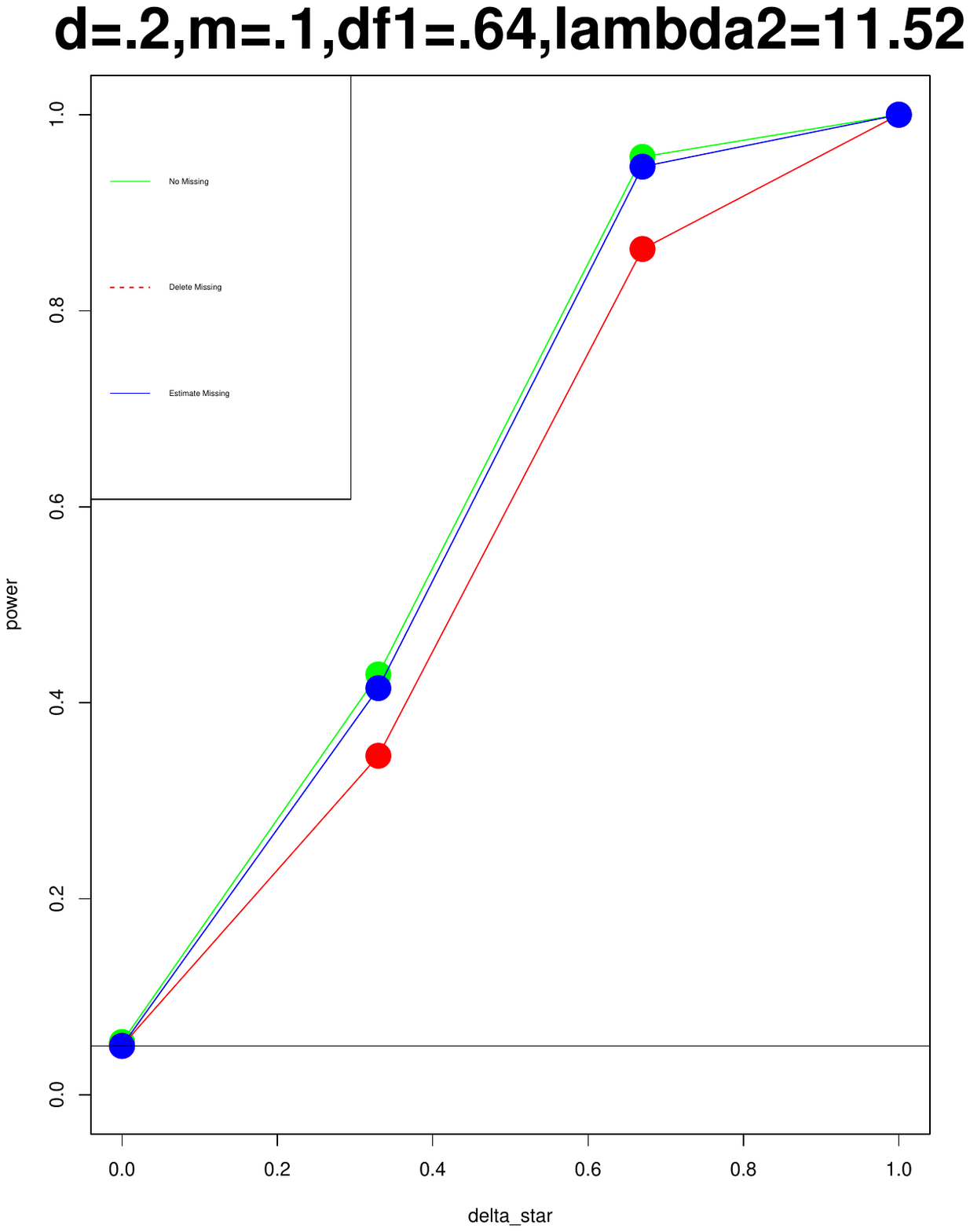}
\includegraphics[width = 2.3in, height = 1.5in]{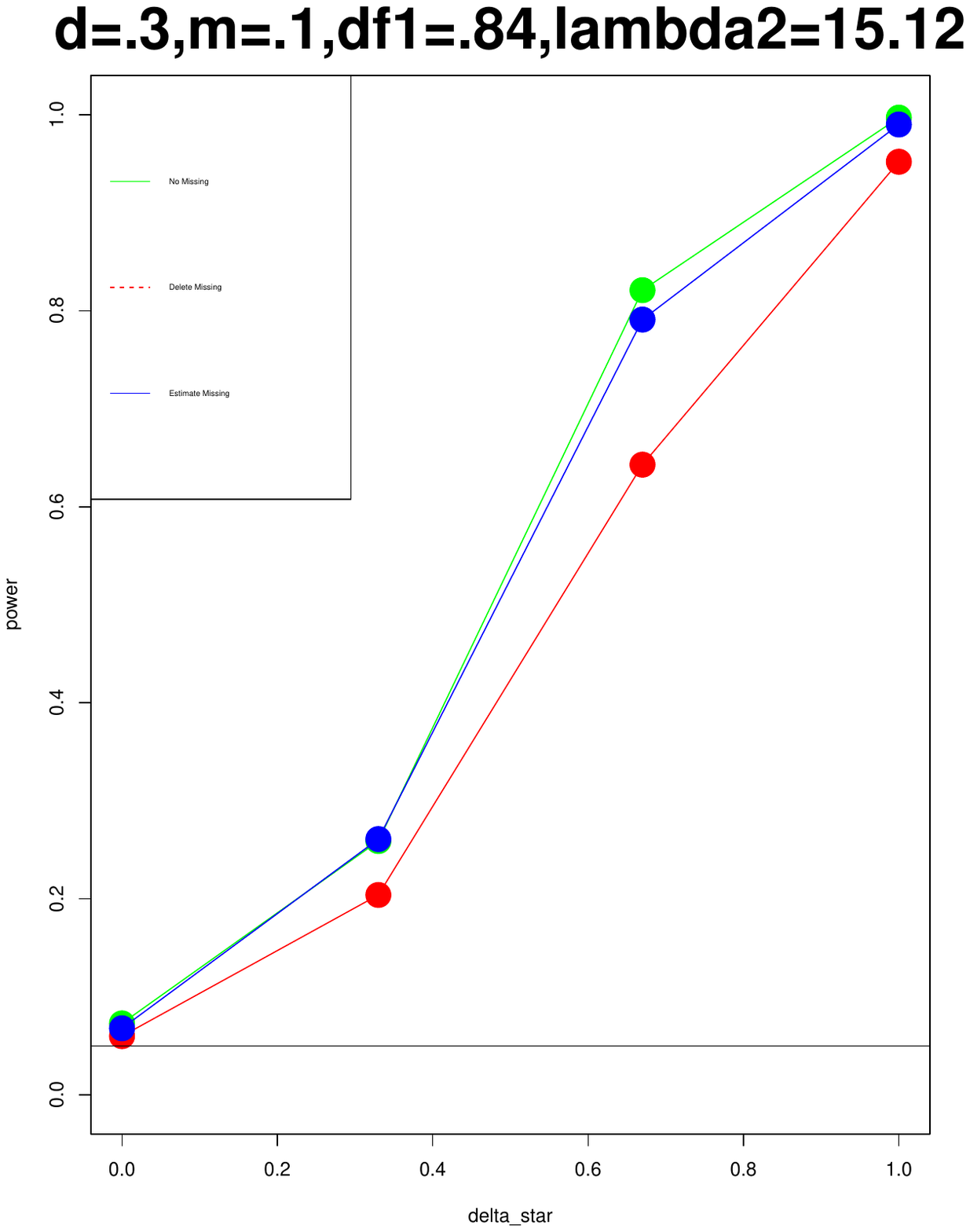}

\subsubsection{When both Traits have Chi Squares Distribution}

The following tables represents powers for the three strategies evaluated at $\delta = 0, .33, .67$ and 1.

\textbf{ For $\rho_1 > \rho_2$}

\pagebreak

\hspace{3.5cm}$m=.1$
\hspace{8.5cm}$m=.5$

\vspace{.2cm}

\begin{tabular}{|c|c|c|c|c|}
\hline 
Strategy & $\delta = 0$ & $\delta = .33$ & $\delta = .67$ & $\delta = 1$ \\ 
\hline 
use same & 0.052 & 0.282 & 0.850 & 1 \\ 
\hline 
use other & 0.051 & 0.267 & 0.823 & 0.997 \\ 
\hline 
use both & 0.053 & 0.254 & 0.790 & 0.992 \\ 
\hline 
\end{tabular} \hspace{1.5cm}
\begin{tabular}{|c|c|c|c|c|}
\hline 
Strategy & $\delta = 0$ & $\delta = .33$ & $\delta = .67$ & $\delta = 1$ \\ 
\hline 
use same & 0.050 & 0.540 & 0.988 & 1 \\ 
\hline 
use other & 0.048 & 0.519 & 0.984 & 1 \\ 
\hline 
use both & 0.051 & 0.484 & 0.977 & 1 \\ 
\hline 
\end{tabular} 

\hspace{.4cm}

\textbf{ For $\rho_1 < \rho_2$}

\vspace{.4cm}

\hspace{3.5cm}$m=.1$
\hspace{8.5cm}$m=.5$

\vspace{.2cm}

\begin{tabular}{|c|c|c|c|c|}
\hline 
Strategy & $\delta = 0$ & $\delta = .33$ & $\delta = .67$ & $\delta = 1$ \\ 
\hline 
use same & 0.048 & 0.322 & 0.875 & 0.998 \\ 
\hline 
use other & 0.050 & 0.337 & 0.899 & 0.998 \\ 
\hline 
use both & 0.051 & 0.309 & 0.854 & 0.997 \\ 
\hline 
\end{tabular} \hspace{1.5cm}
\begin{tabular}{|c|c|c|c|c|}
\hline 
Strategy & $\delta = 0$ & $\delta = .33$ & $\delta = .67$ & $\delta = 1$ \\ 
\hline 
use same & 0.054 & 0.863 & 1 & 1 \\ 
\hline 
use other & 0.052 & 0.883 & 1 & 1 \\ 
\hline 
use both & 0.047 & 0.836 & 1 & 1 \\ 
\hline 
\end{tabular}

\hspace{.4cm}

\textbf{ For $\rho_1 = \rho_2$}

\vspace{.4cm}

\hspace{3.5cm}$m=.1$
\hspace{8.5cm}$m=.5$

\vspace{.2cm}

\begin{tabular}{|c|c|c|c|c|}
\hline 
Strategy & $\delta = 0$ & $\delta = .33$ & $\delta = .67$ & $\delta = 1$ \\ 
\hline 
use same & 0.053 & 0.339 & 0.913 & 0.998 \\ 
\hline 
use other & 0.051 & 0.376 & 0.944 & 0.999 \\ 
\hline 
use both & 0.050 & 0.314 & 0.885 & 0.997 \\ 
\hline 
\end{tabular} \hspace{1.5cm}
\begin{tabular}{|c|c|c|c|c|}
\hline 
Strategy & $\delta = 0$ & $\delta = .33$ & $\delta = .67$ & $\delta = 1$ \\ 
\hline 
use same & 0.051 & 0.615 & 0.991 & 1 \\ 
\hline 
use other & 0.048 & 0.670 & 0.998 & 1 \\ 
\hline 
use both & 0.050 & 0.569 & 0.984 & 1 \\ 
\hline 
\end{tabular} 

\vspace{.3cm}

According to the above results we can conclude-

\begin{center}
\begin{tabular}{|c|c|}
\hline 
Case & Best Strategy \\ 
\hline 
$\rho_1 > \rho_2$ & use same \\ 
\hline 
$\rho_1 < \rho_2$ & use other \\ 
\hline 
$\rho_1 = \rho_2$ & use other \\ 
\hline 
\end{tabular} 

\end{center}

Now we go for power comparison among no missing, estimated missing and deleted missing.

We generate 1st trait from chi squares distribution (section 6.2) with the parameters $\alpha = \alpha_1, \beta = \beta_1, df = df_1$ keeping $p^\star = p^\star_1$ and we generate 2nd trait from chi squares distribution (section 6.2) with the parameters $\alpha = \alpha_2, \beta = \beta_2, df = df_2$ keeping $p^\star = p^\star_2$.

We have done simulation for three choices of $d$ as .1, .2, .3 and for each $d$ we take $(p^\star_1, p^\star_2)$ as (.1, .2), (.2, .2).

We take $\alpha_1 = 5, \alpha_2 = 10, \beta_1 = 1, and \beta_2 = 2$ and varied $df_1$ and $df_2$ in the following way,

\vspace{.4cm}

\begin{tabular}{|c|c|c|}
\hline 
d & $(p^\star_1, p^\star_2)$ & ($df_1, df_2$) \\ 
\hline 
.1 & (.1, .2) & (.81, 1.44) \\ 
\hline 
.1 & (.2, .2) & (.36, 1.44) \\ 
\hline 

\end{tabular} \hspace{1.5cm}
\begin{tabular}{|c|c|c|}
\hline 
d & $(p^\star_1, p^\star_2)$ & ($df_1, df_2$) \\ 
\hline 
.2 & (.1, .2) & (1.44, 2.56) \\ 
\hline 
.2 & (.2, .2) & (.64, 2.56) \\ 
\hline 

\end{tabular} \hspace{1.5cm}
\begin{tabular}{|c|c|c|}
\hline 
d & $(p^\star_1, p^\star_2)$ & ($df_1, df_2$) \\ 
\hline 
.3 & (.1, .2) & (1.89, 3.36) \\ 
\hline 
.3 & (.2, .2) & (.84, 3.36) \\ 
\hline 
\end{tabular} 

\vspace{.4cm}

We replicate these for $m =$ .1 and .2.

Note that for each of the following simulations we have calculated $\rho_1$ and $\rho_2$ and accordingly we have used the best imputation strategy.

\vspace{.4cm}

\hspace{1.5cm}
$d=.1 \quad m=.5$
\hspace{3cm}
$d=.2 \quad m=.5$
\hspace{3cm}
$d=.3 \quad m=.5$

\includegraphics[width = 2.3in, height = 1.5in]{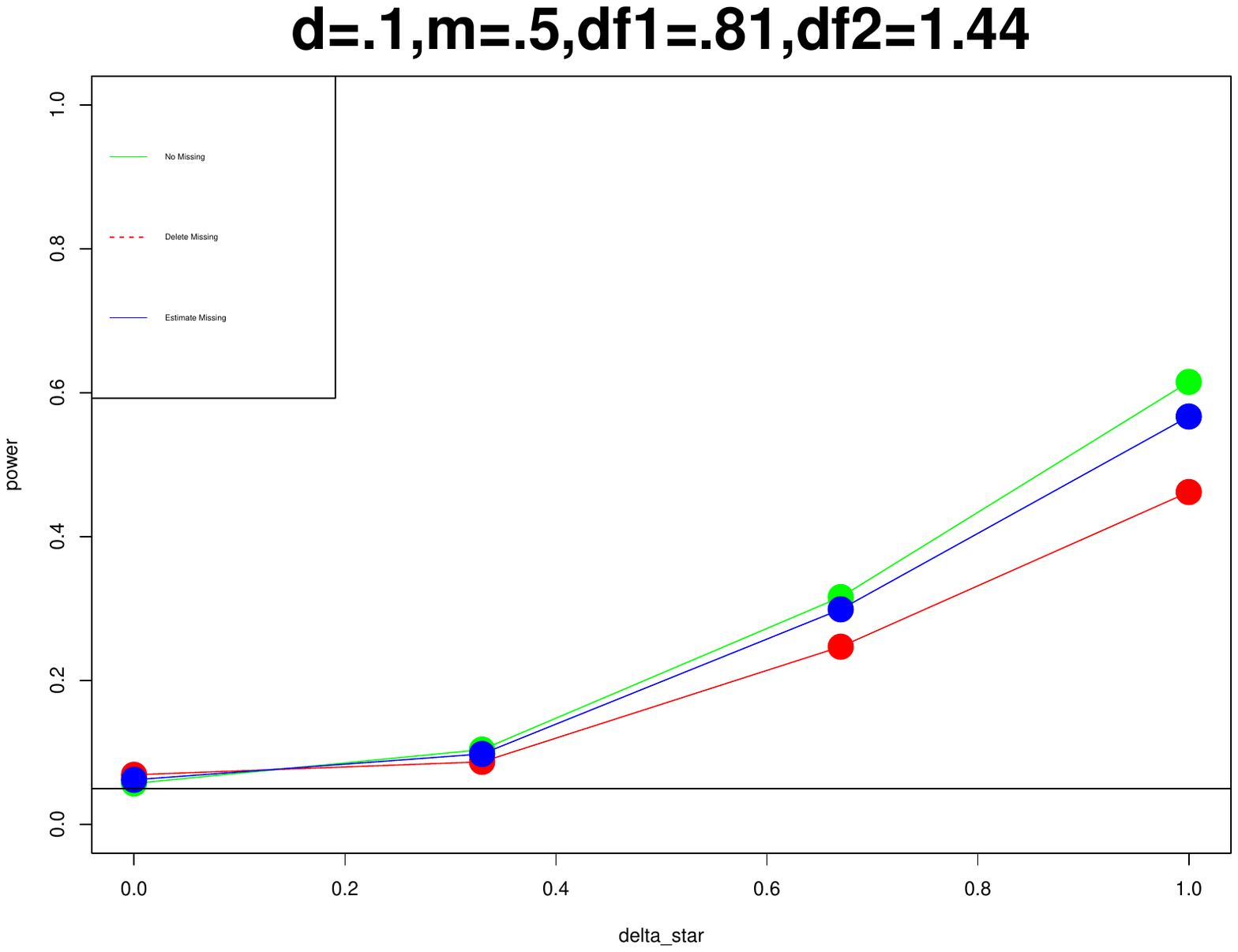}
\includegraphics[width = 2.3in, height = 1.5in]{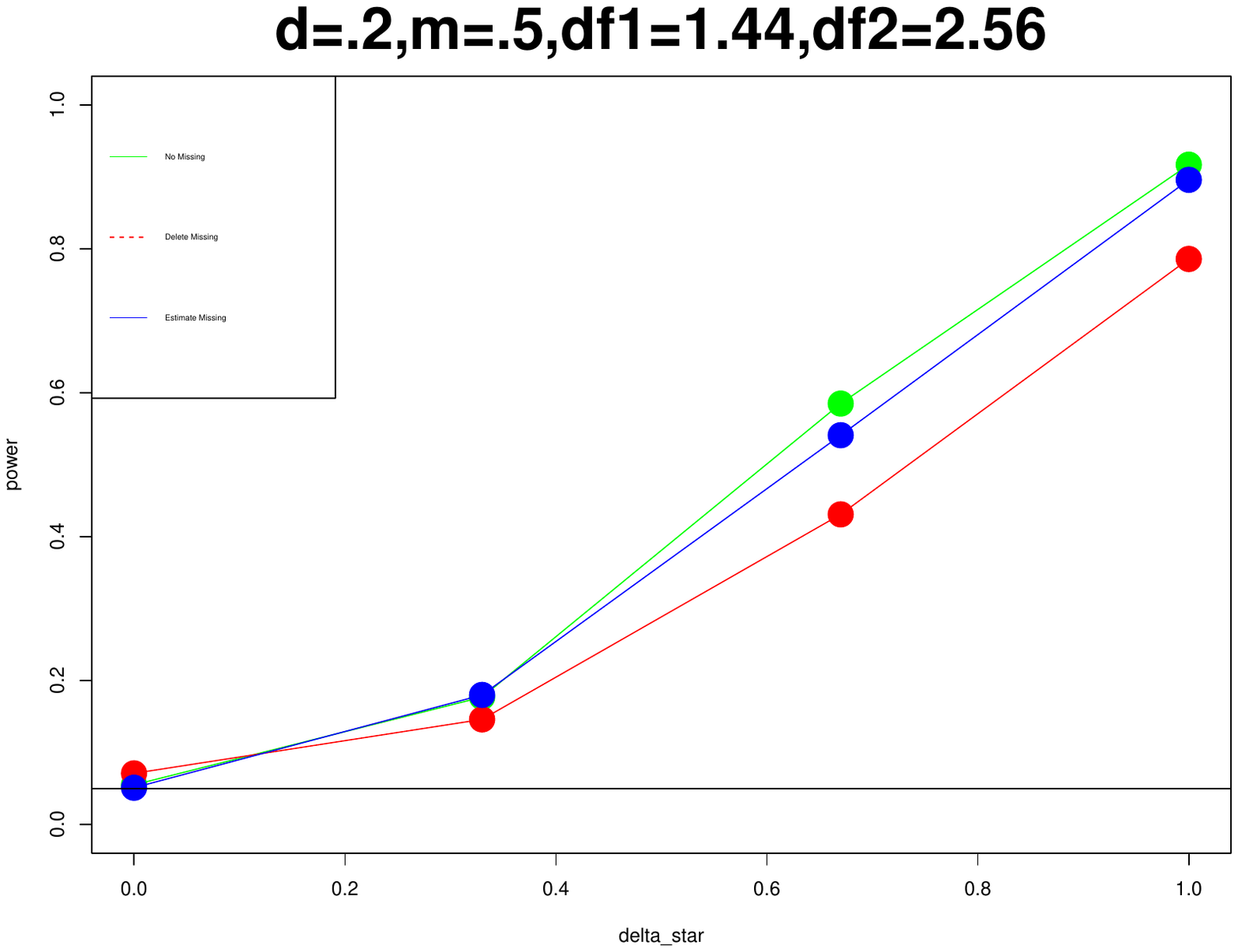}
\includegraphics[width = 2.3in, height = 1.5in]{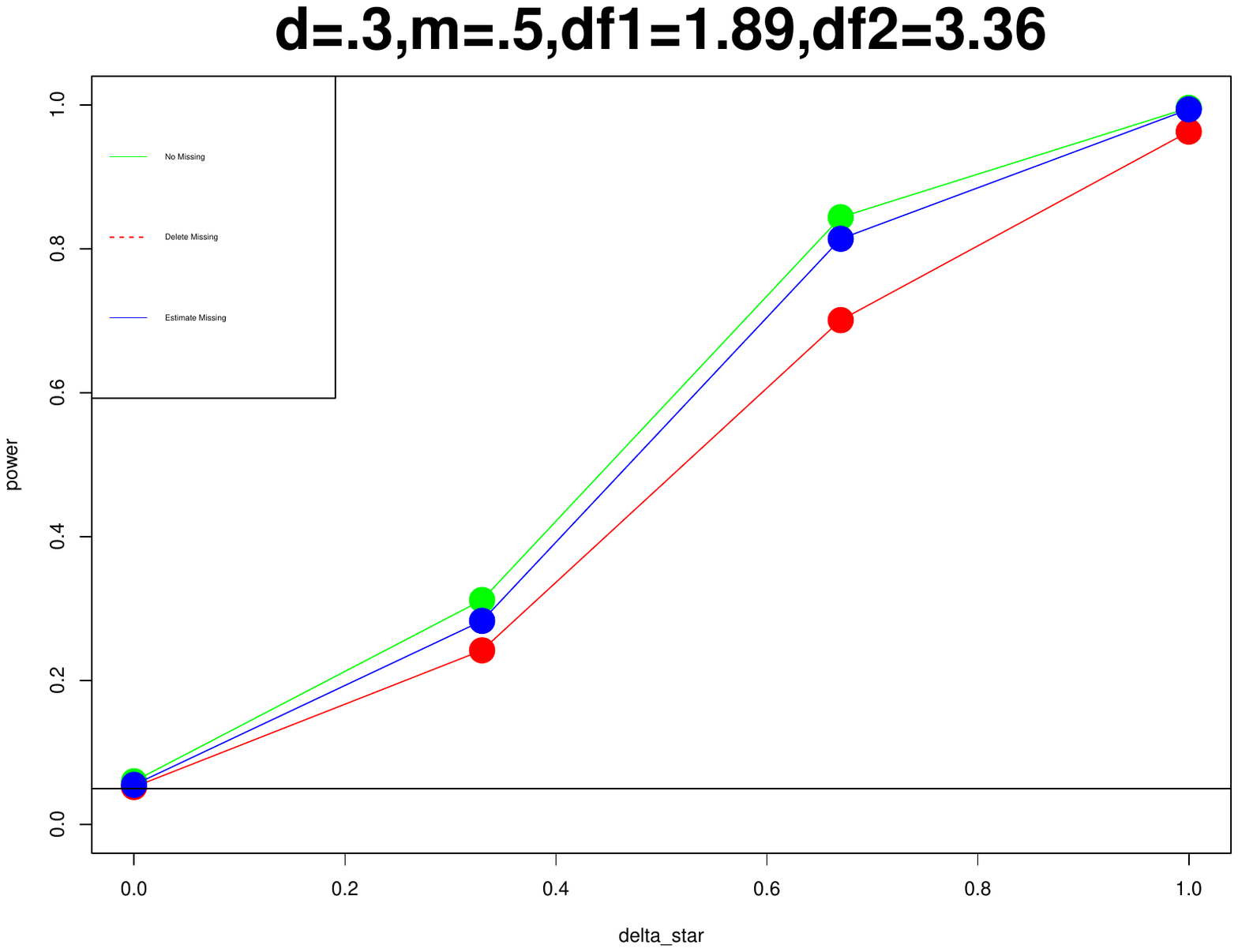}

\hspace{1.5cm}
$d=.1 \quad m=.1$
\hspace{3cm}
$d=.2 \quad m=.1$
\hspace{3cm}
$d=.3 \quad m=.1$

\includegraphics[width = 2.3in, height = 1.5in]{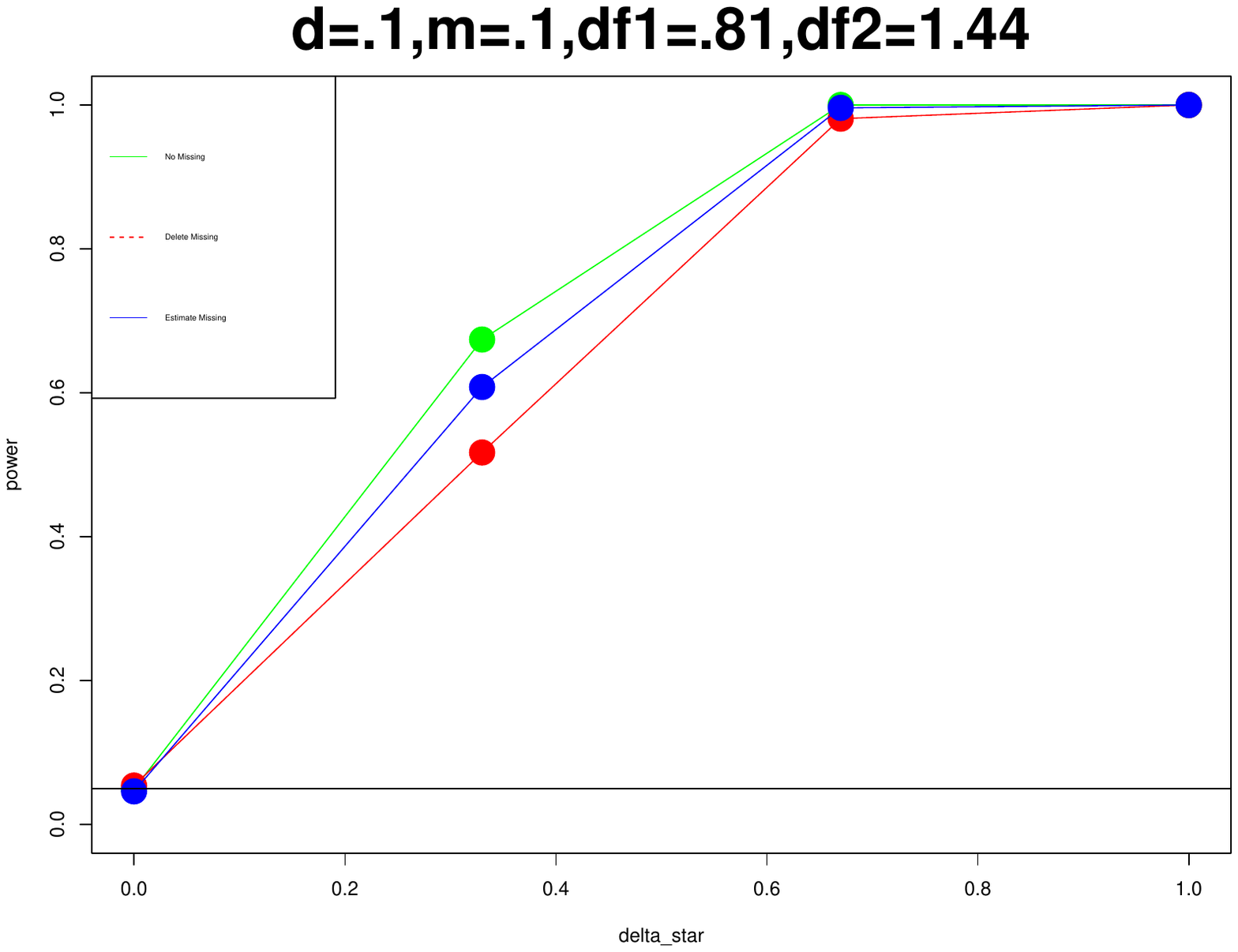}
\includegraphics[width = 2.3in, height = 1.5in]{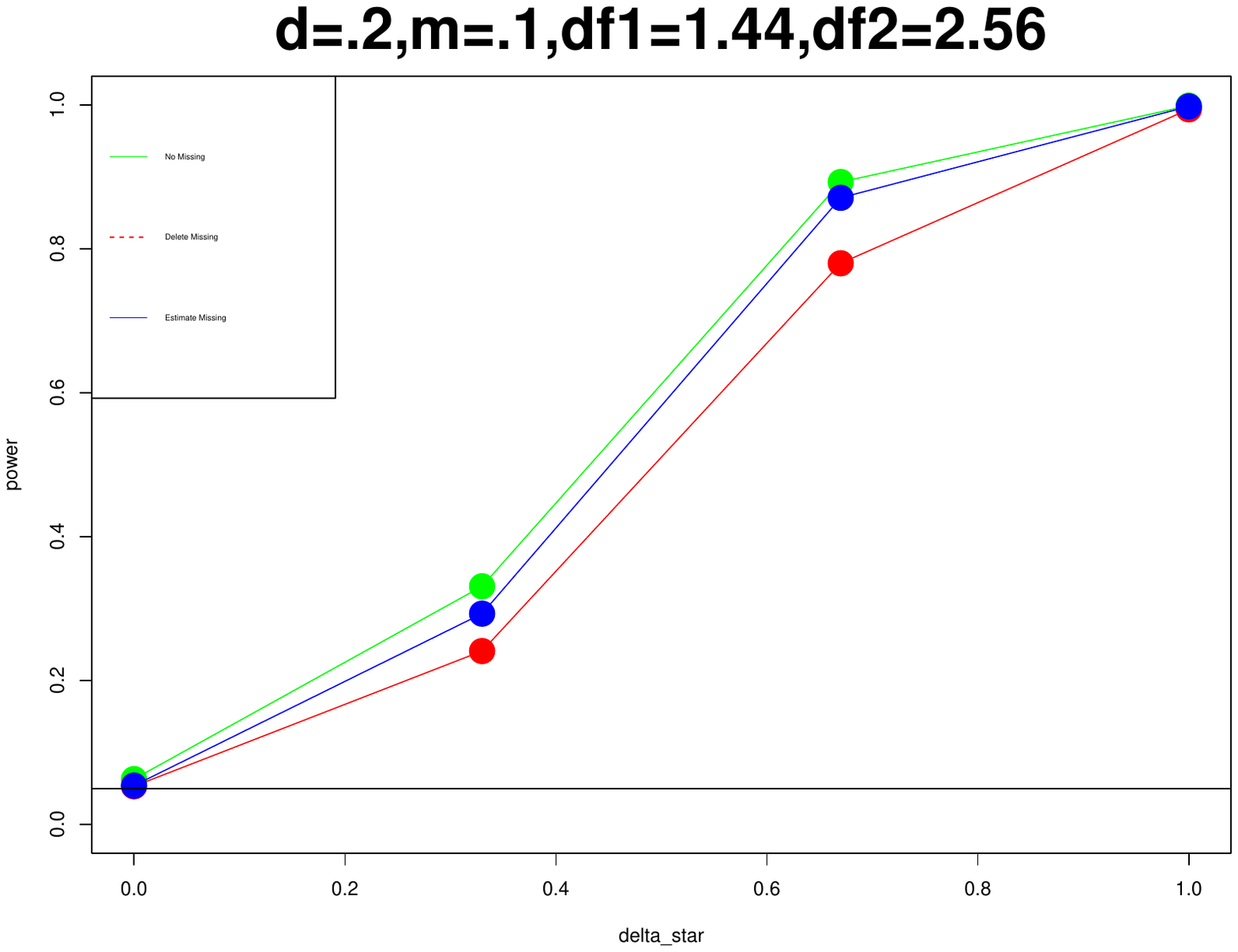}
\includegraphics[width = 2.3in, height = 1.5in]{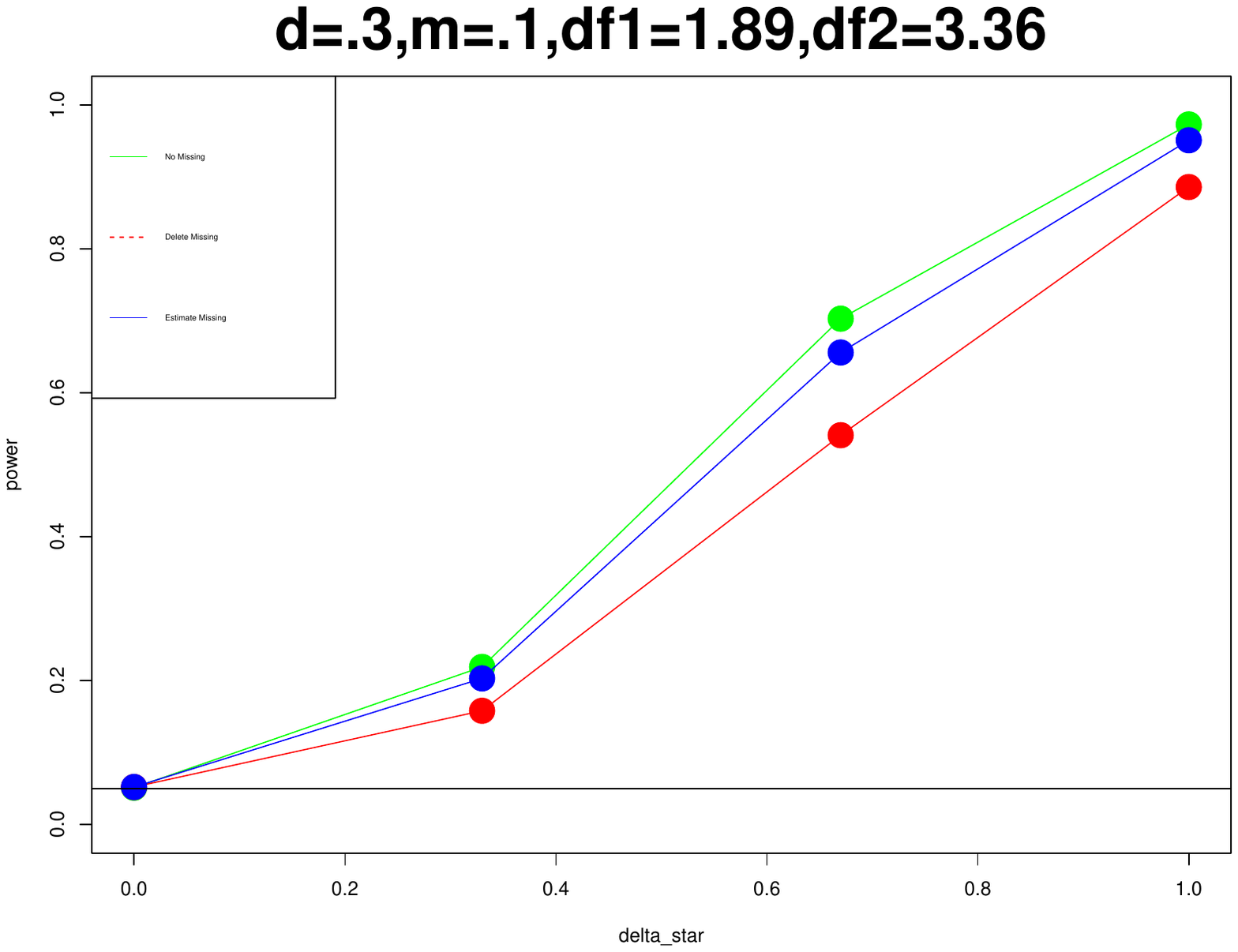}

\hspace{1.5cm}
$d=.1 \quad m=.1$
\hspace{3cm}
$d=.2 \quad m=.1$
\hspace{3cm}
$d=.3 \quad m=.1$

\includegraphics[width = 2.3in, height = 1.5in]{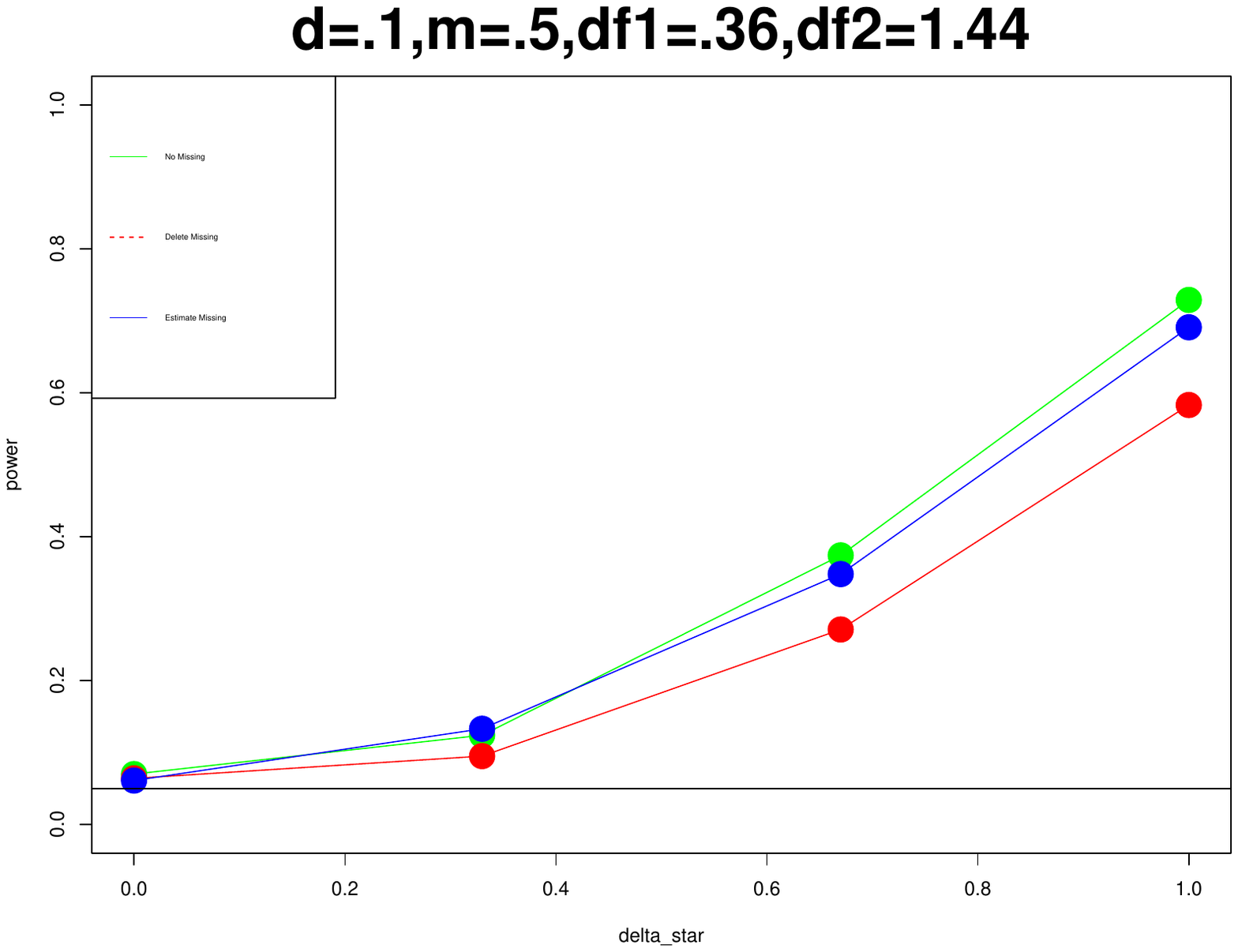}
\includegraphics[width = 2.3in, height = 1.5in]{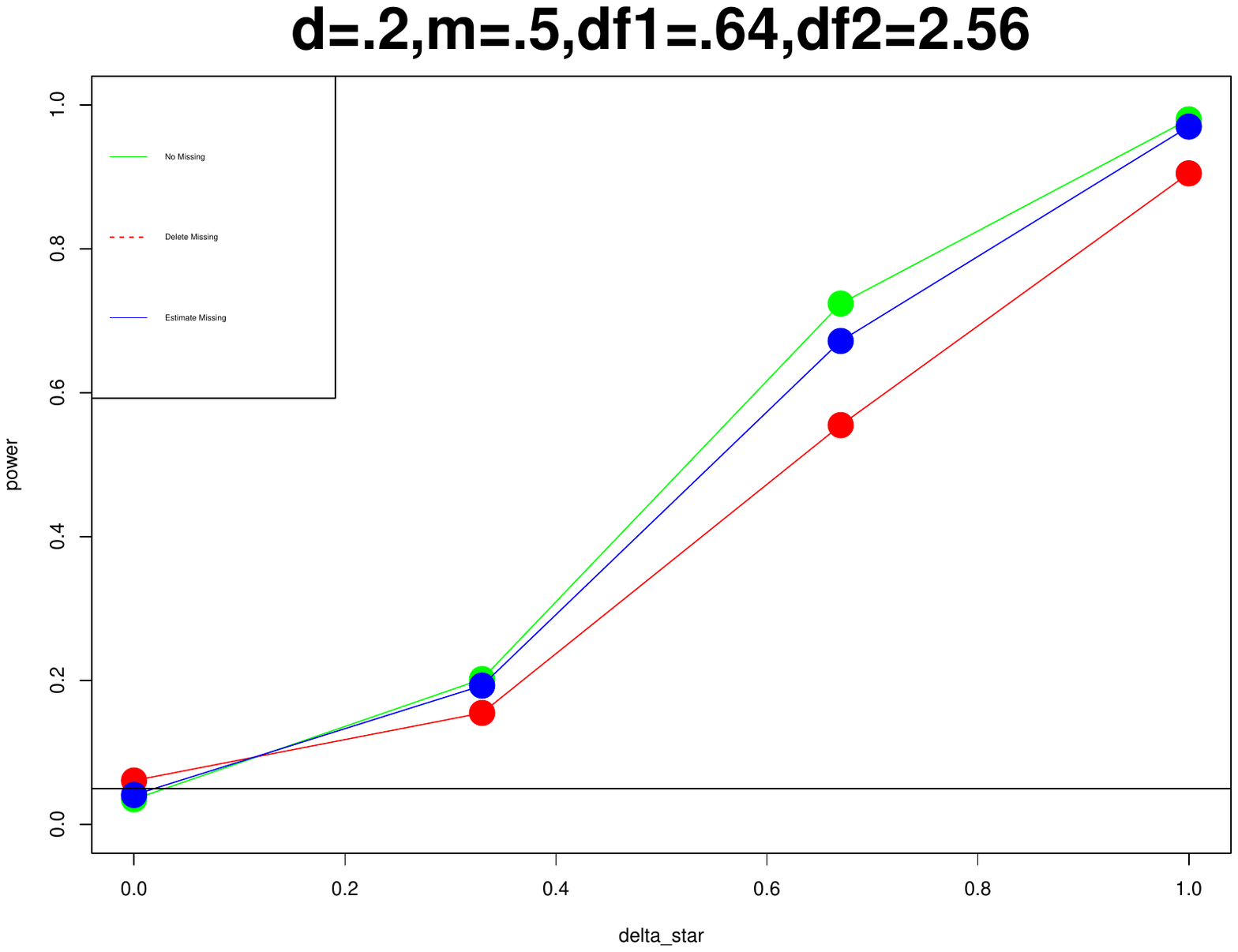}
\includegraphics[width = 2.3in, height = 1.5in]{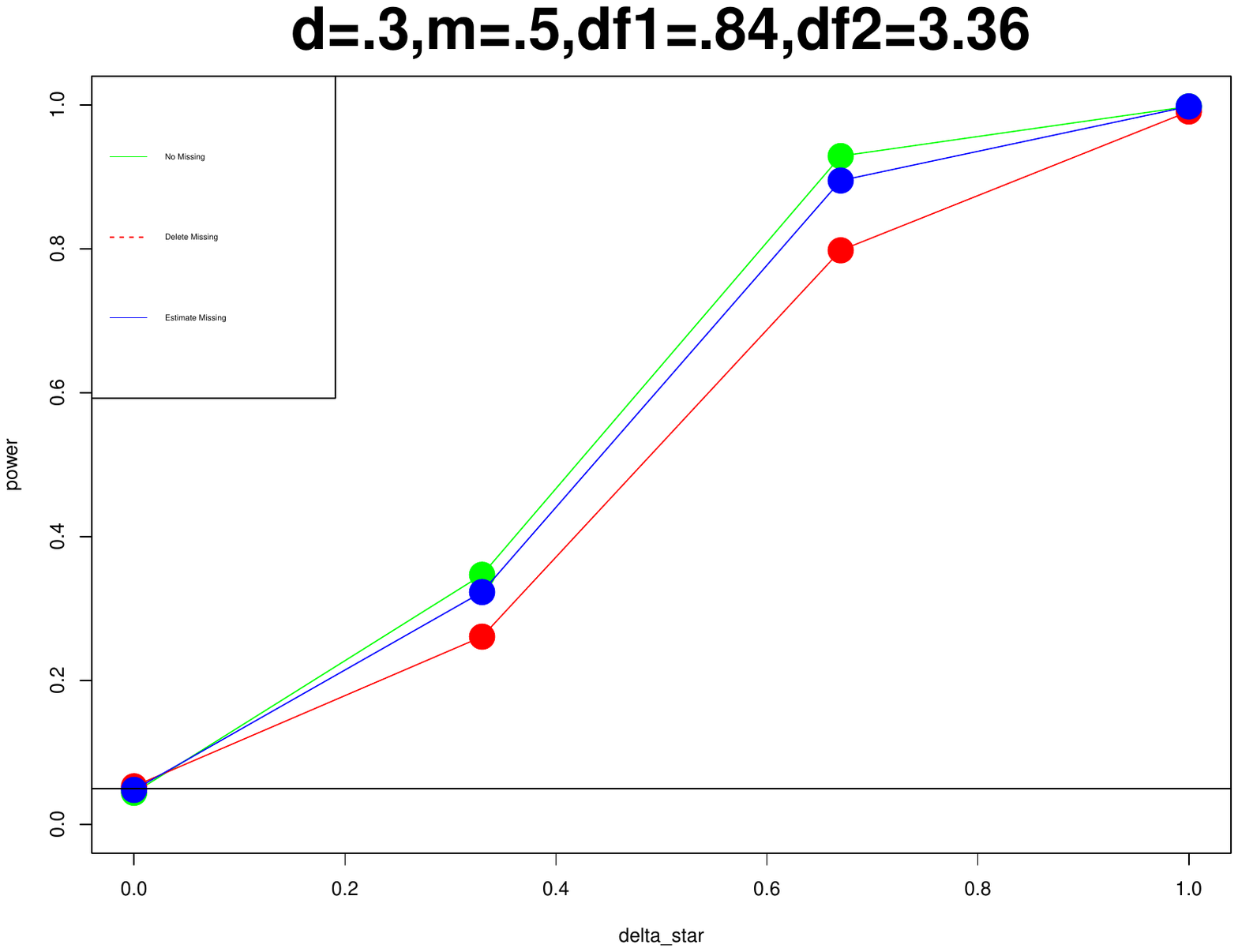}

\hspace{1.5cm}
$d=.1 \quad m=.5$
\hspace{3cm}
$d=.2 \quad m=.5$
\hspace{3cm}
$d=.3 \quad m=.5$

\includegraphics[width = 2.3in, height = 1.5in]{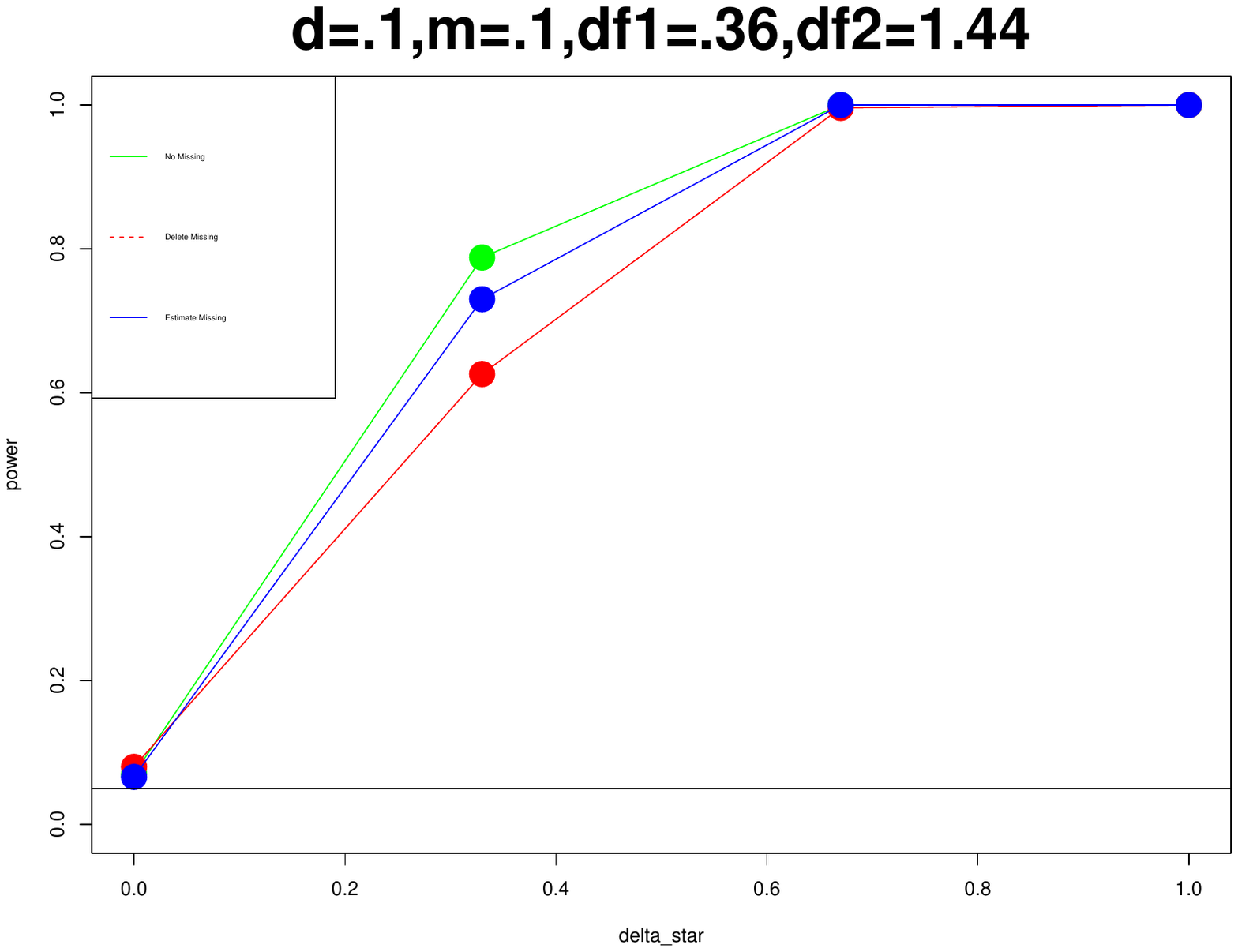}
\includegraphics[width = 2.3in, height = 1.5in]{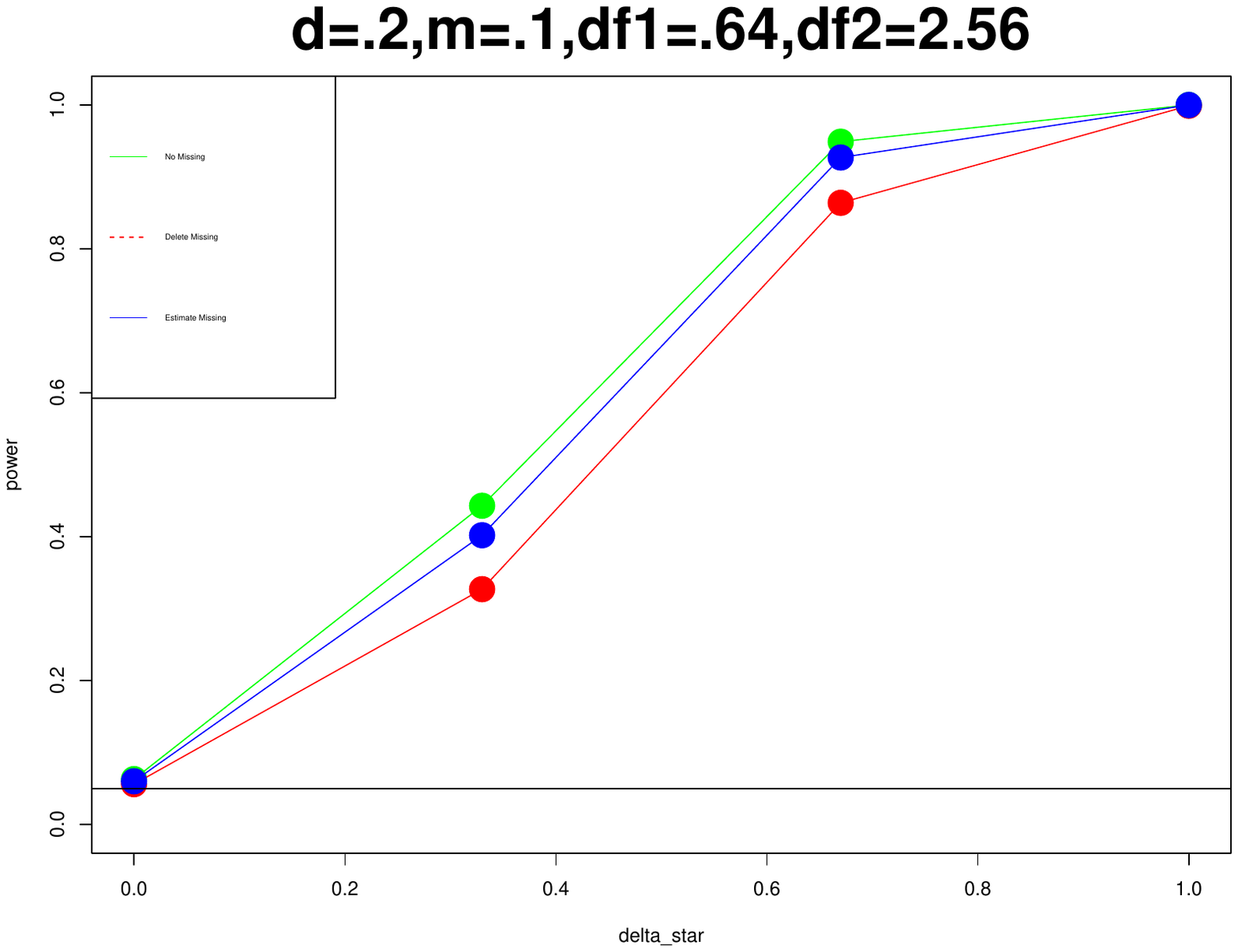}
\includegraphics[width = 2.3in, height = 1.5in]{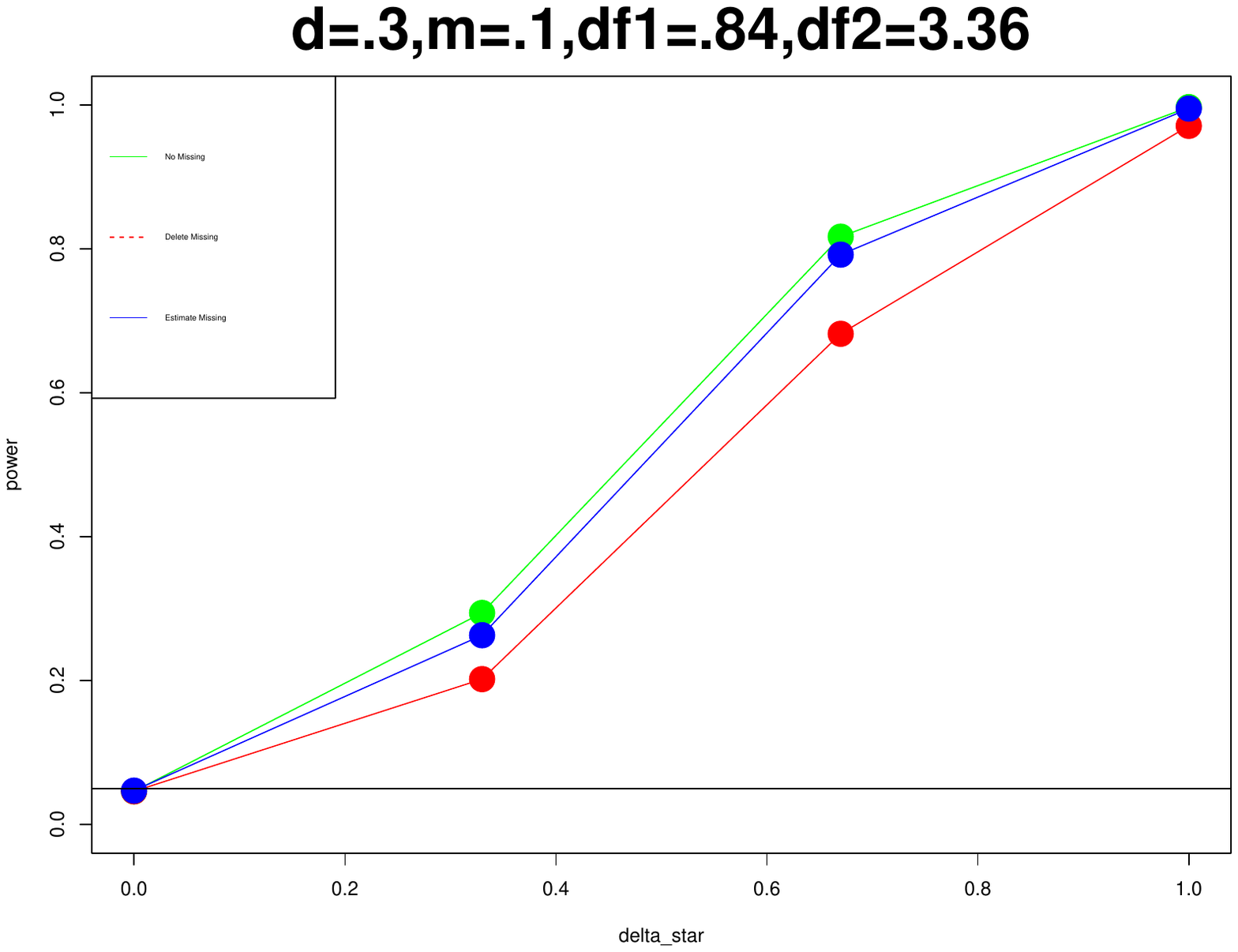}

\subsection{For Missing Type 2.3}

As stated in section 5, we have estimated the missing trait using the value of same trait of the other offspring ( sib ).

In each of the cases simulation details are exactly same as the simulation details in the section 7.2.

\subsubsection{When both traits have Normal Distribution}

\hspace{1.5cm}
$d=.1 \quad m=.5$
\hspace{3cm}
$d=.2 \quad m=.5$
\hspace{3cm}
$d=.3 \quad m=.5$

\includegraphics[width = 2.3in, height = 1.4in]{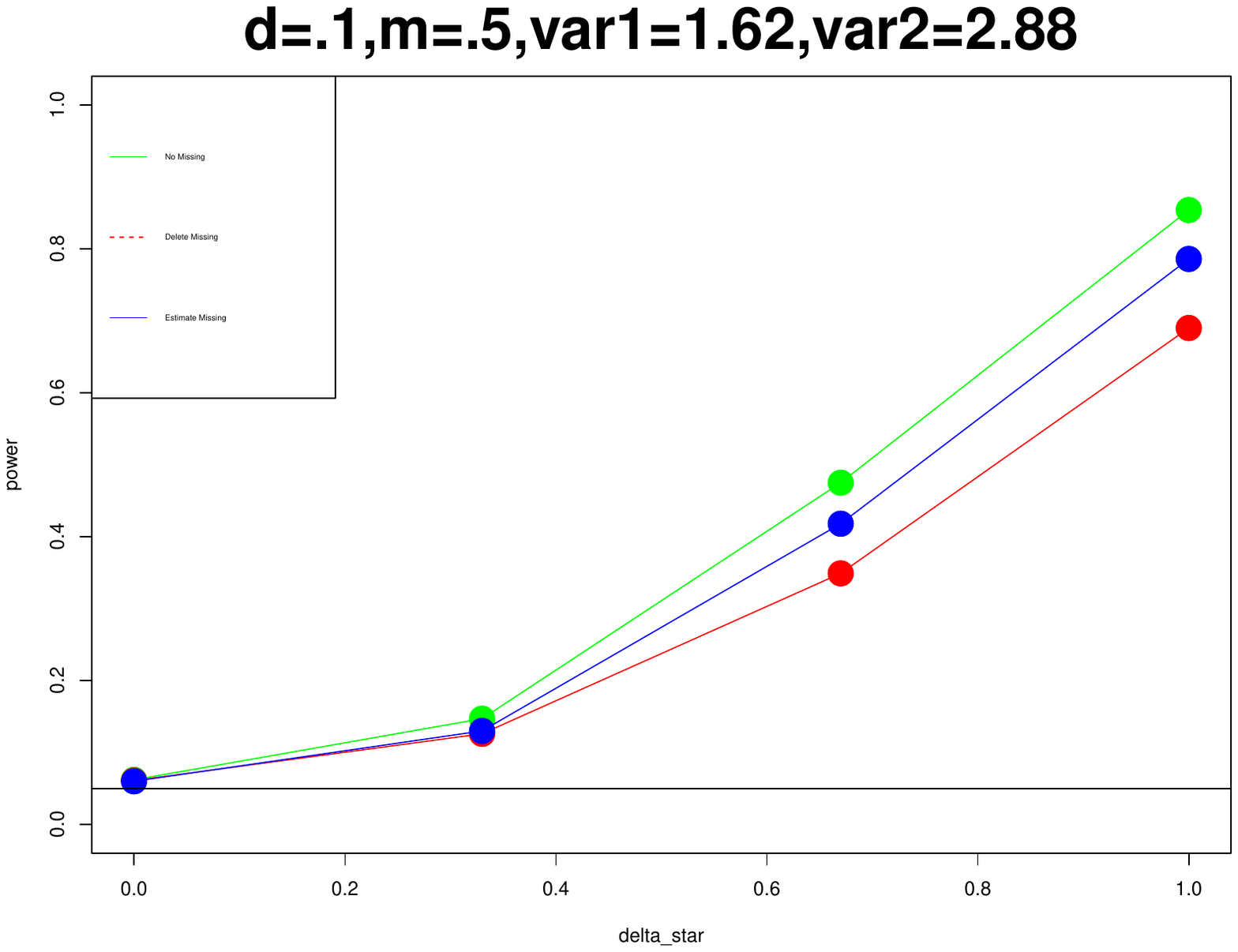}
\includegraphics[width = 2.3in, height = 1.4in]{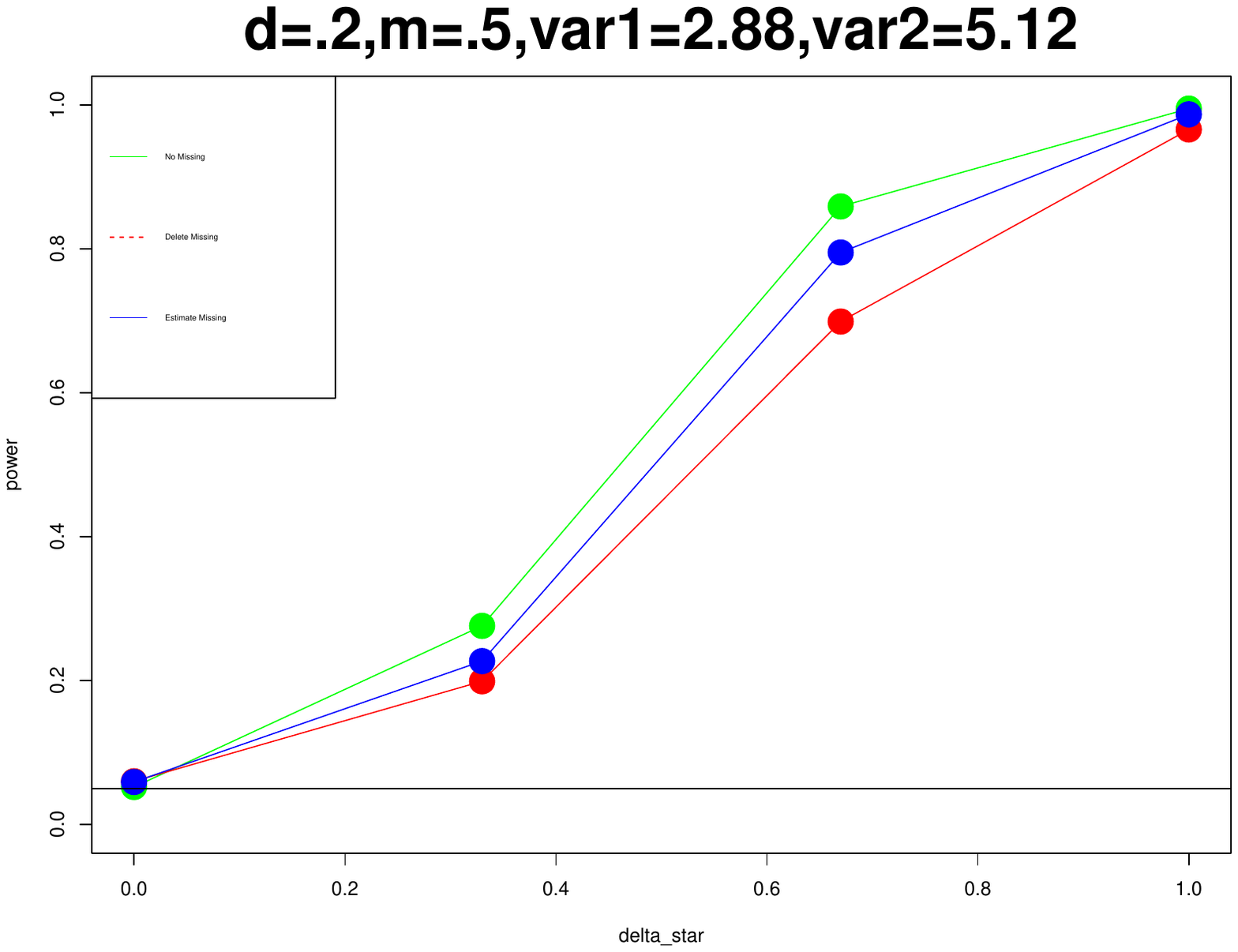}
\includegraphics[width = 2.3in, height = 1.4in]{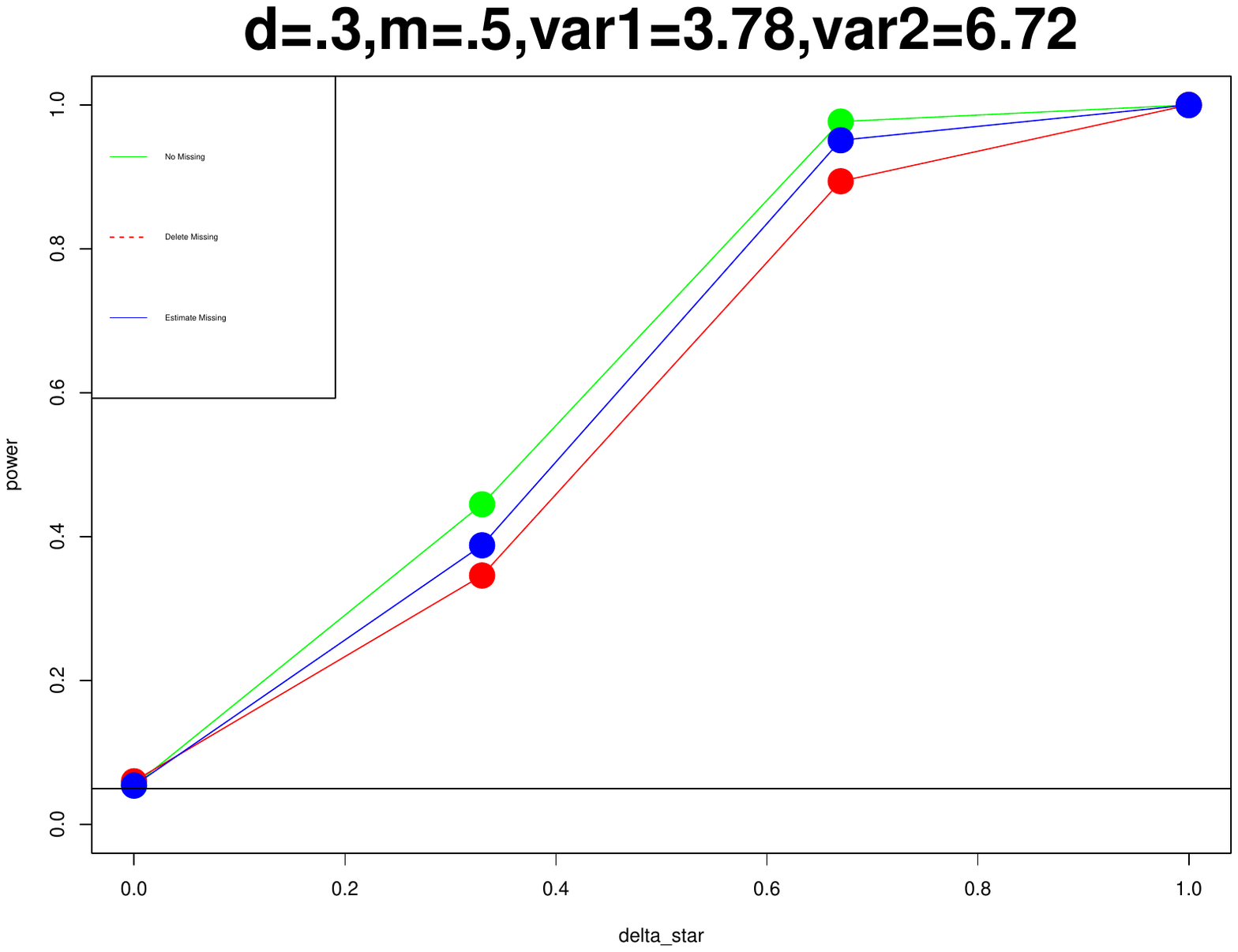}

\hspace{1.5cm}
$d=.1 \quad m=.1$
\hspace{3cm}
$d=.2 \quad m=.1$
\hspace{3cm}
$d=.3 \quad m=.1$

\includegraphics[width = 2.3in, height = 1.4in]{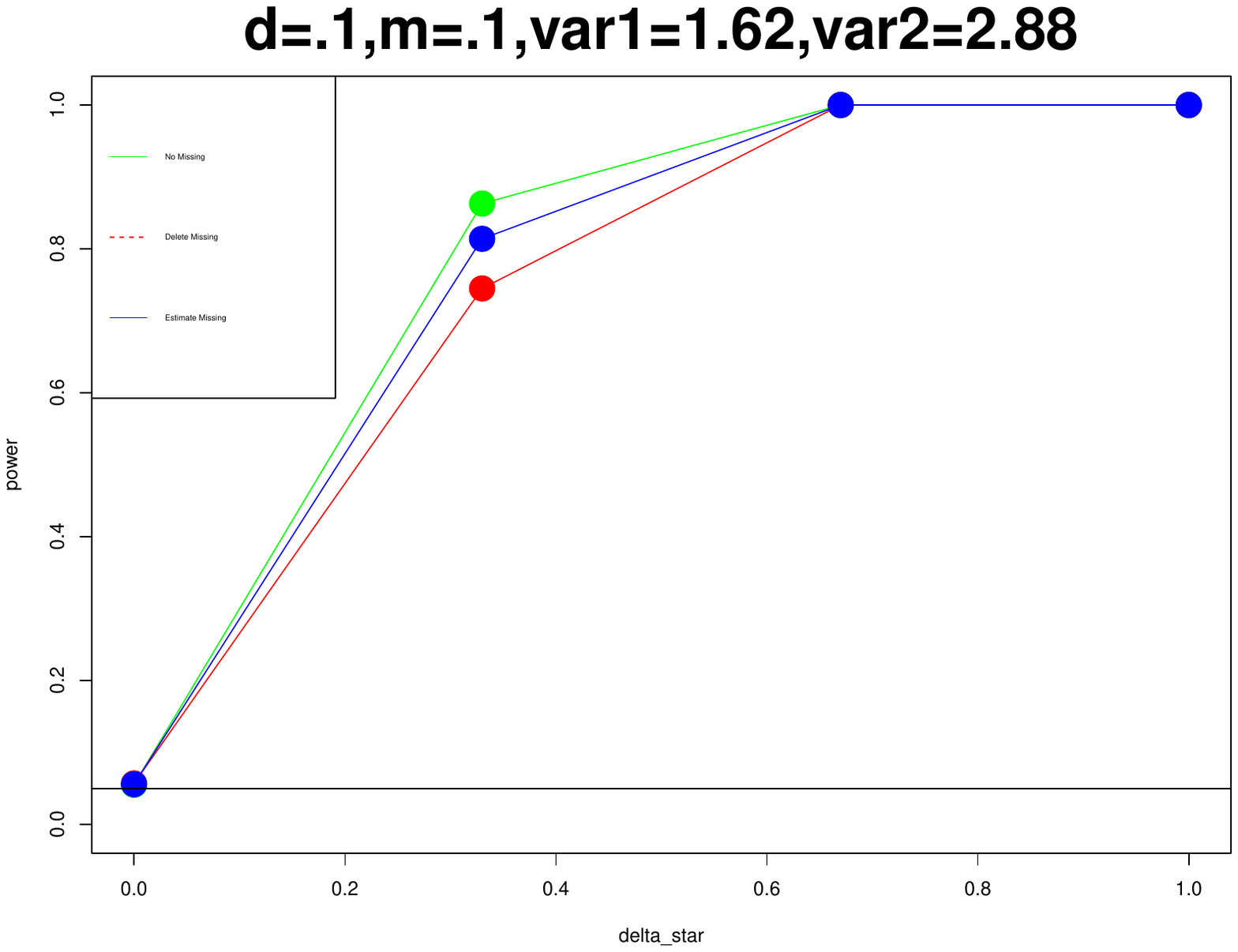}
\includegraphics[width = 2.3in, height = 1.4in]{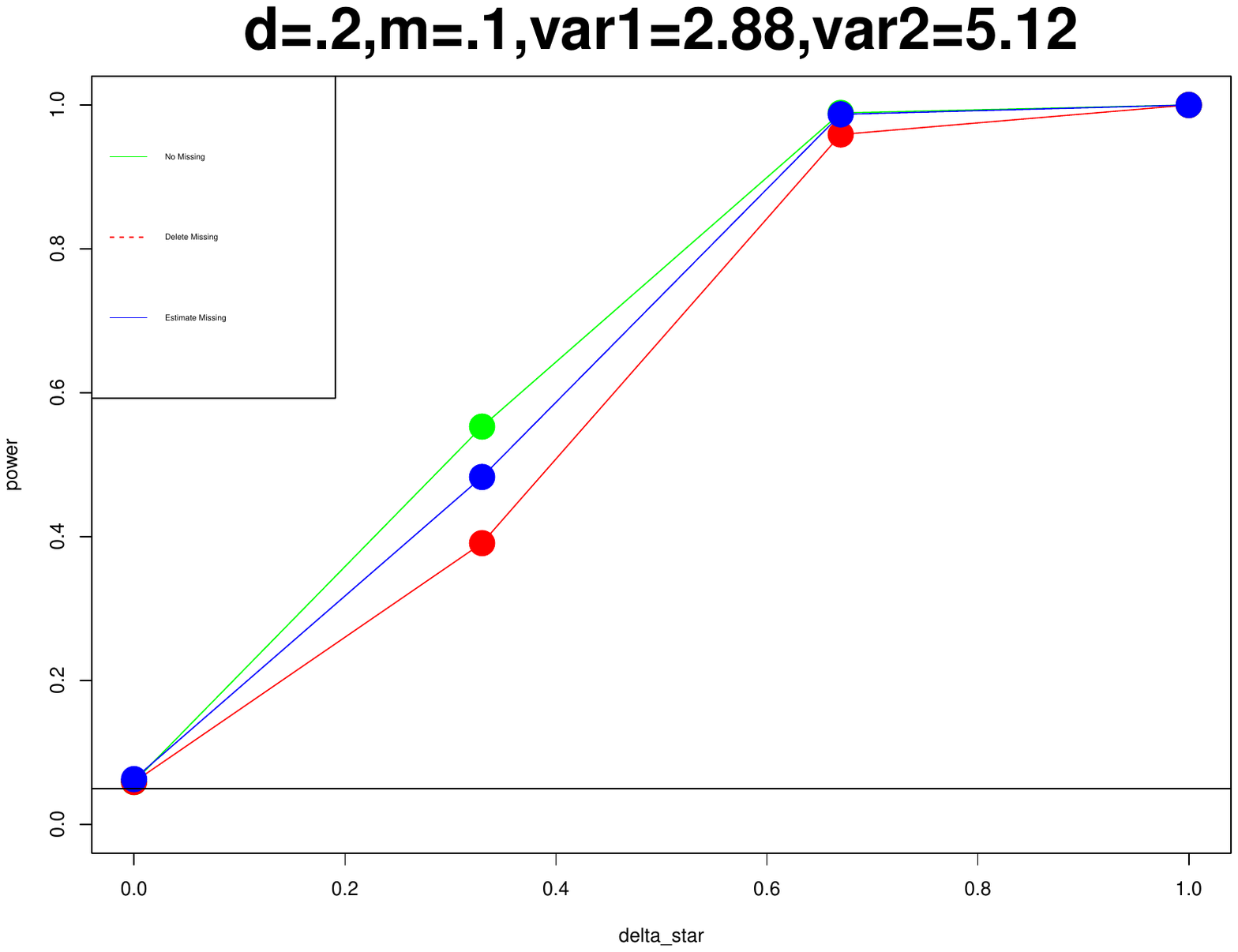}
\includegraphics[width = 2.3in, height = 1.4in]{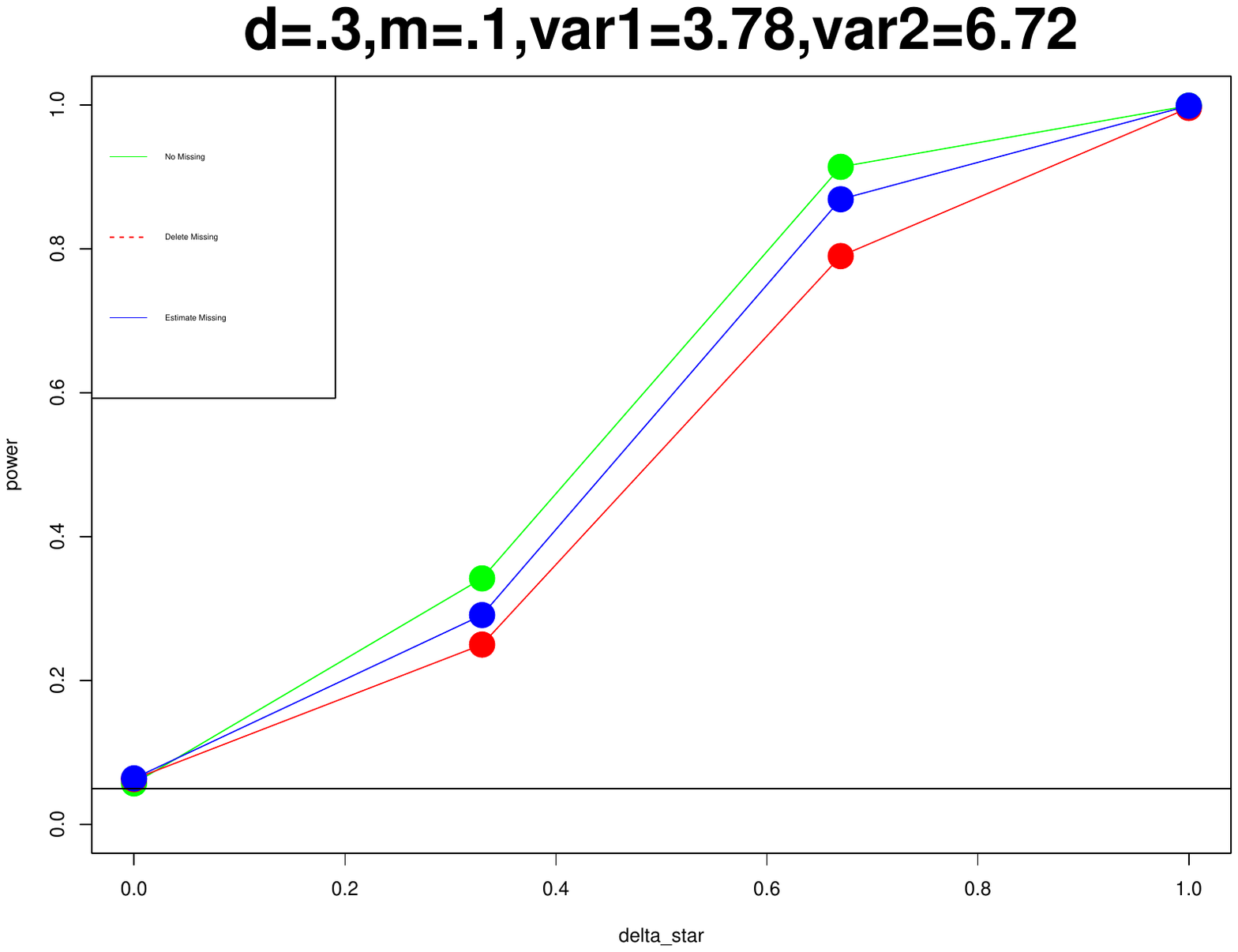}

\hspace{1.5cm}
$d=.1 \quad m=.1$
\hspace{3cm}
$d=.2 \quad m=.1$
\hspace{3cm}
$d=.3 \quad m=.1$

\includegraphics[width = 2.3in, height = 1.5in]{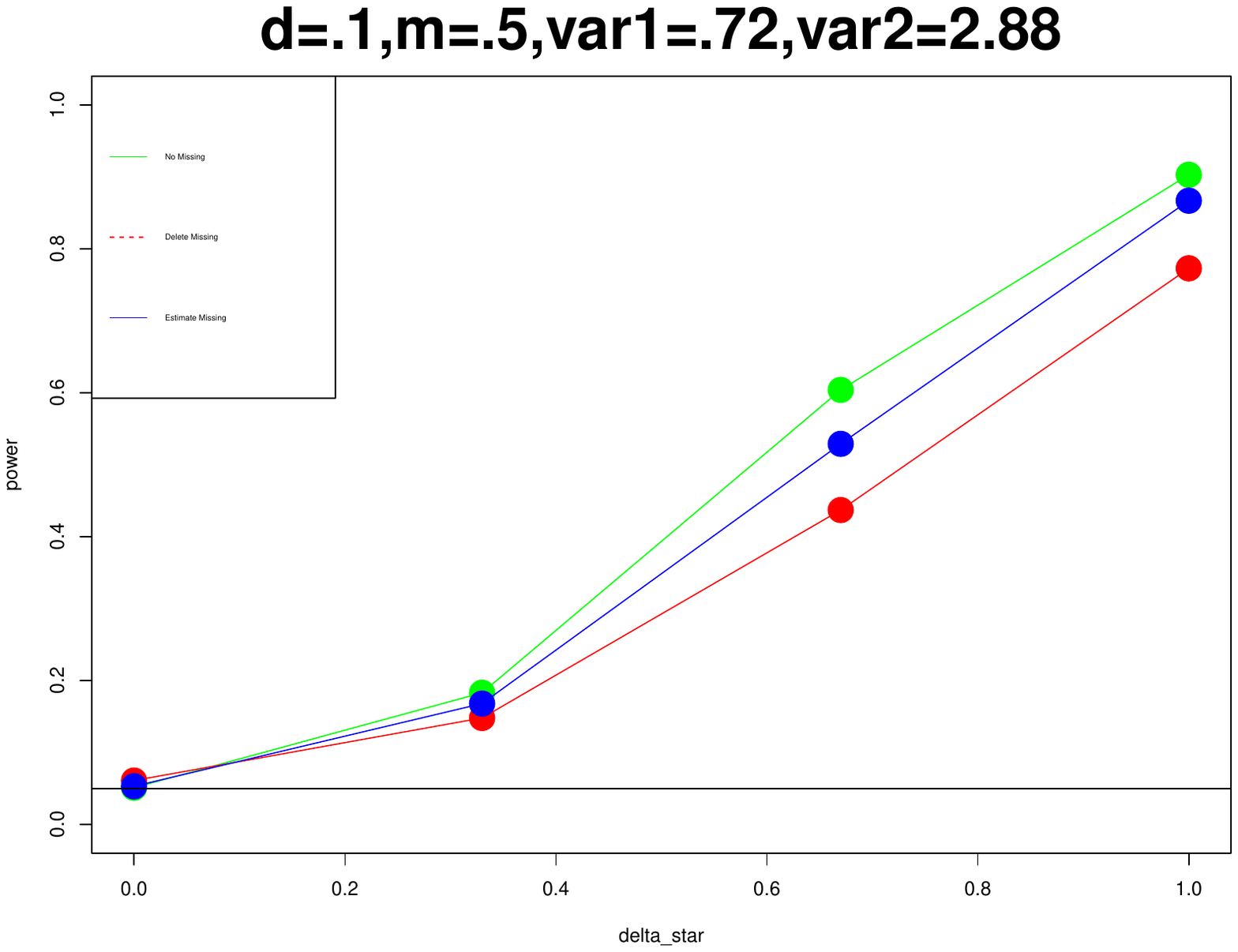}
\includegraphics[width = 2.3in, height = 1.5in]{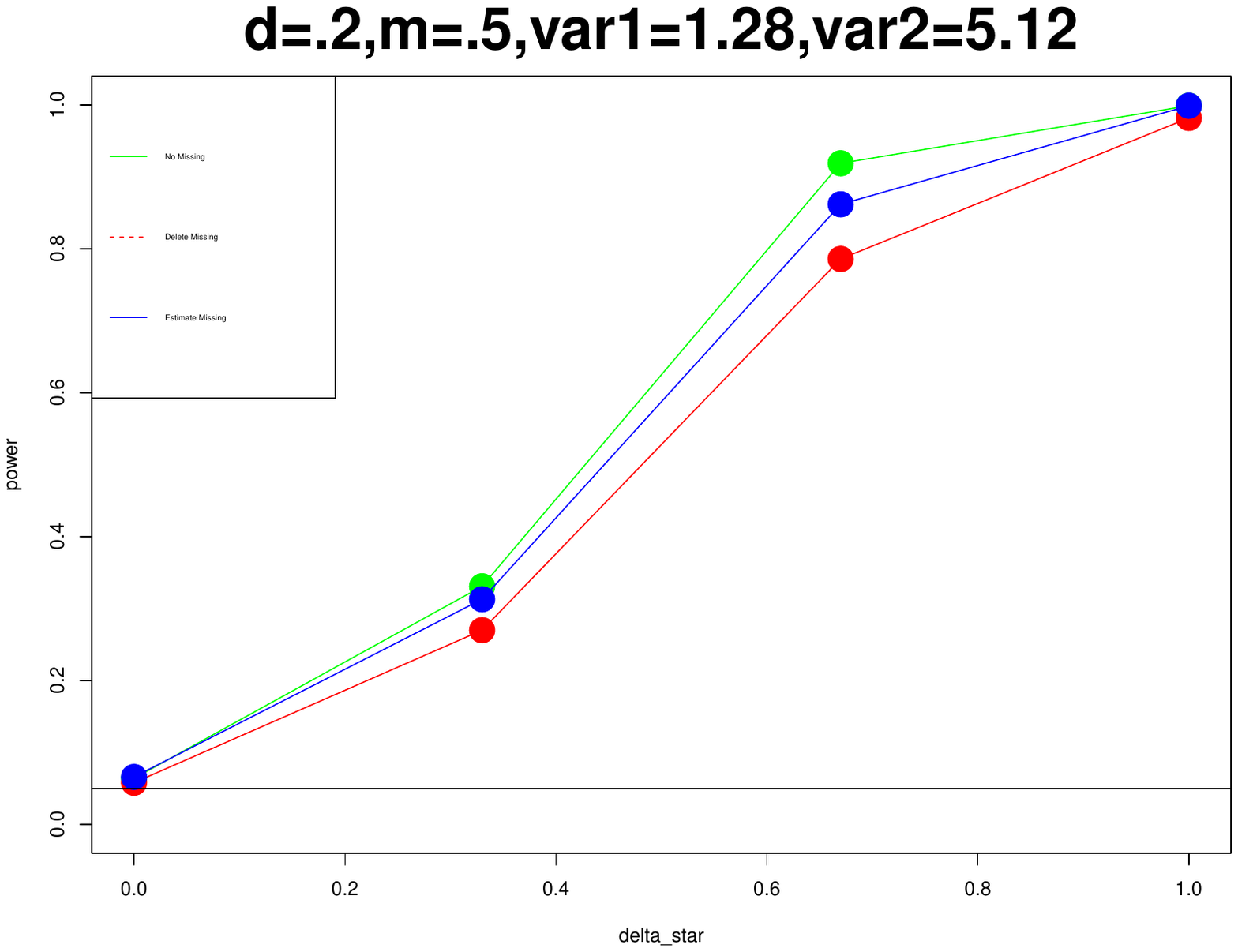}
\includegraphics[width = 2.3in, height = 1.5in]{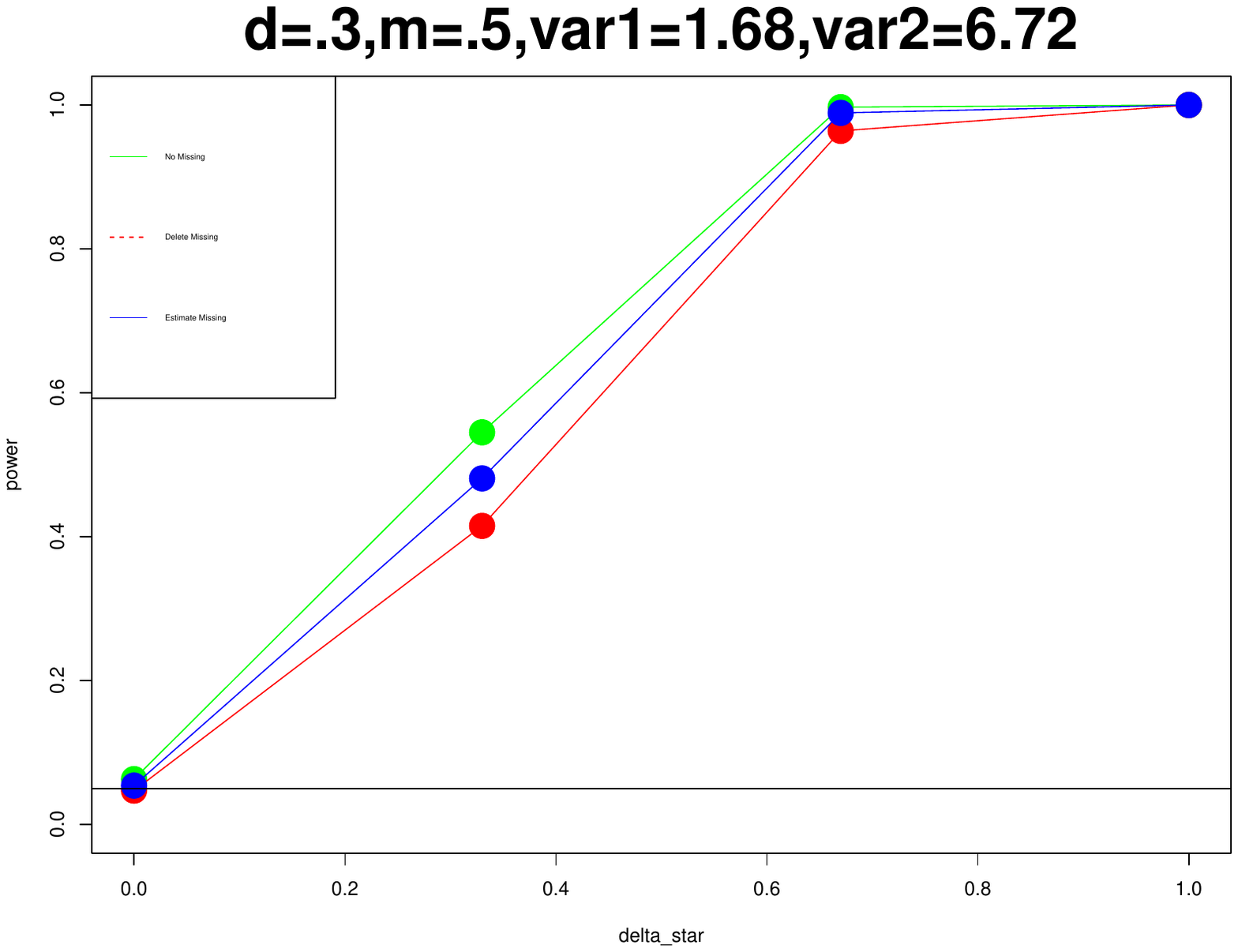}

\hspace{1.5cm}
$d=.1 \quad m=.5$
\hspace{3cm}
$d=.2 \quad m=.5$
\hspace{3cm}
$d=.3 \quad m=.5$

\includegraphics[width = 2.3in, height = 1.5in]{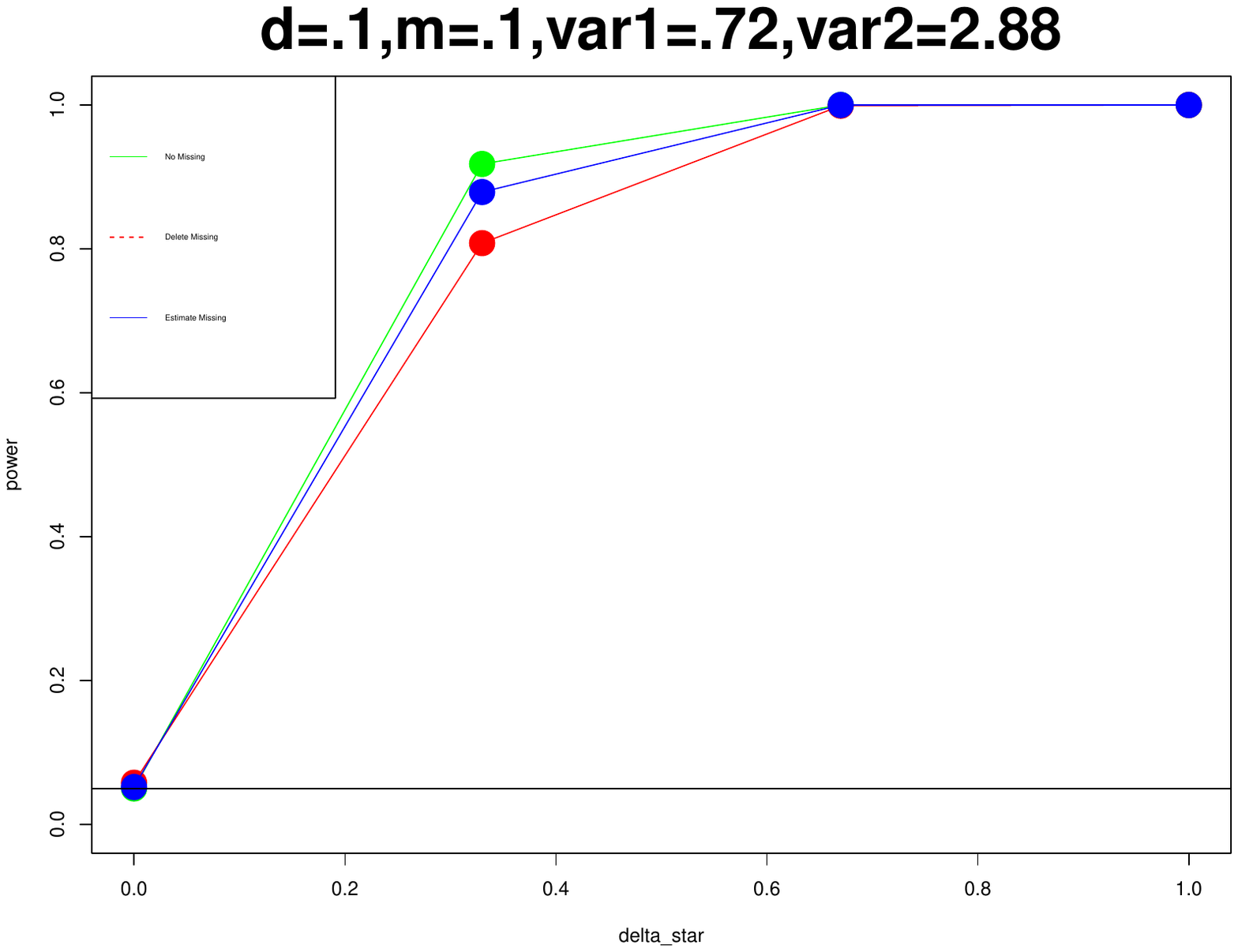}
\includegraphics[width = 2.3in, height = 1.5in]{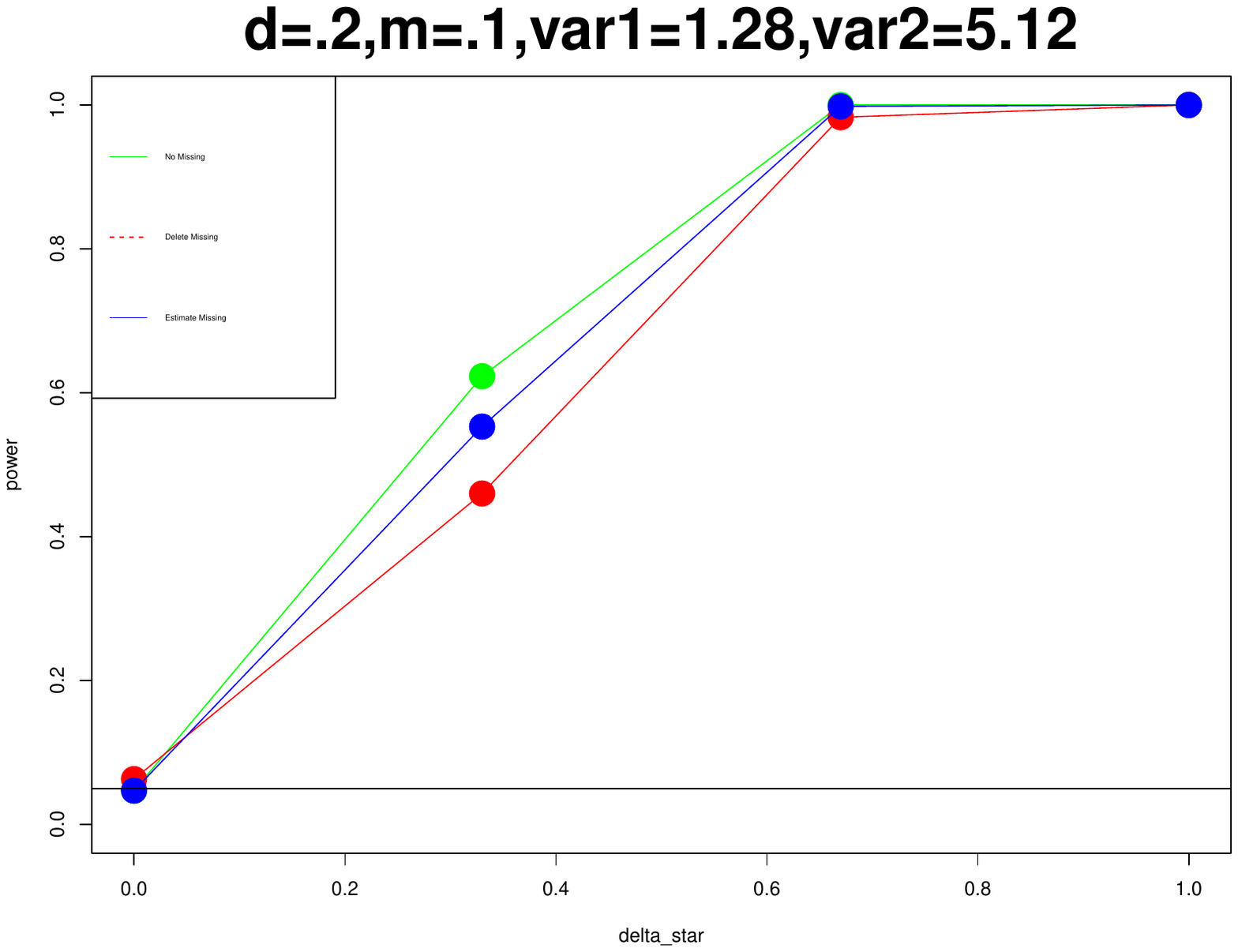}
\includegraphics[width = 2.3in, height = 1.5in]{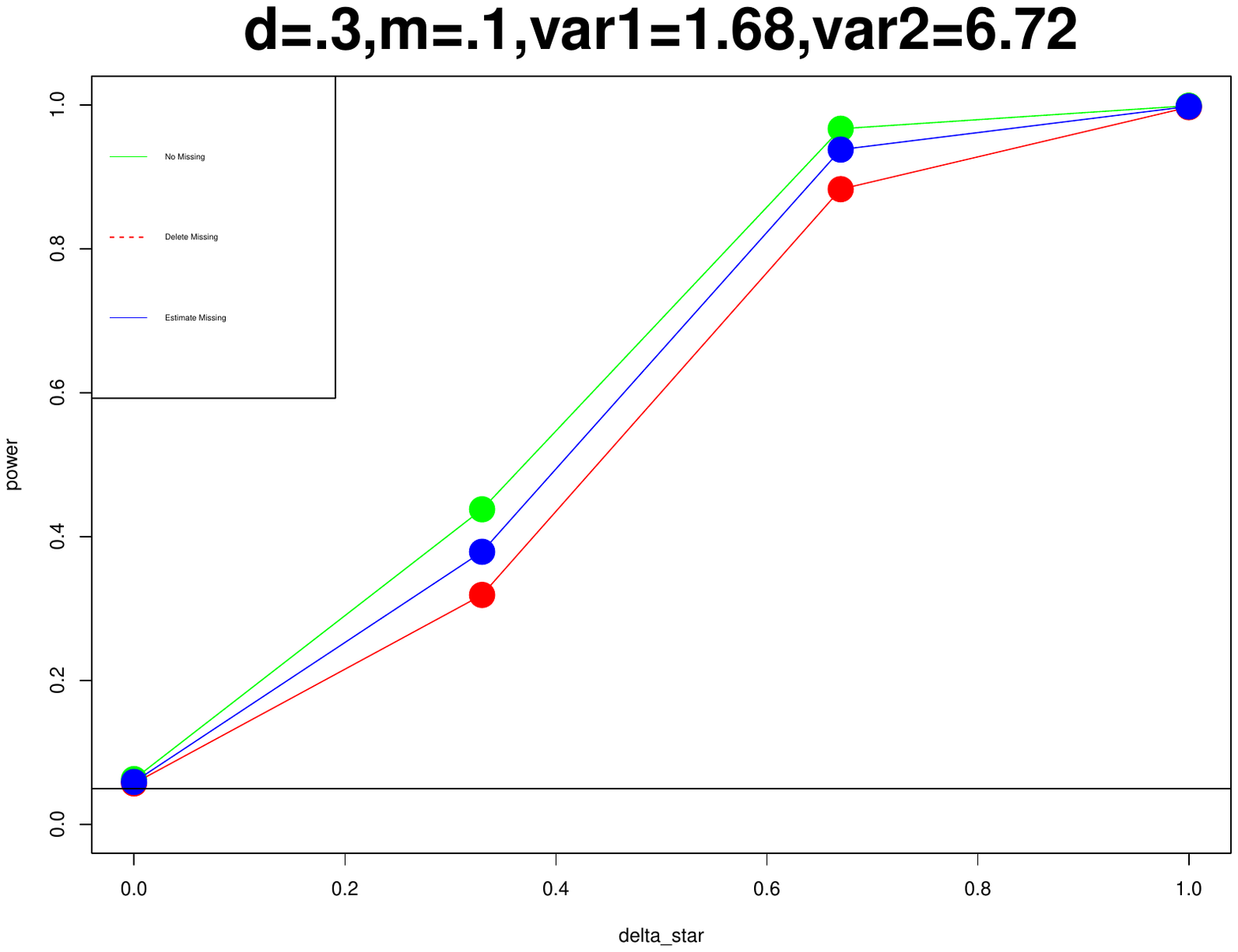}

\subsubsection{When one Trait has Normal Distribution and other Trait has Poisson Distribution}

\hspace{1.5cm}
$d=.1 \quad m=.5$
\hspace{3cm}
$d=.2 \quad m=.5$
\hspace{3cm}
$d=.3 \quad m=.5$

\includegraphics[width = 2.3in, height = 1.3in]{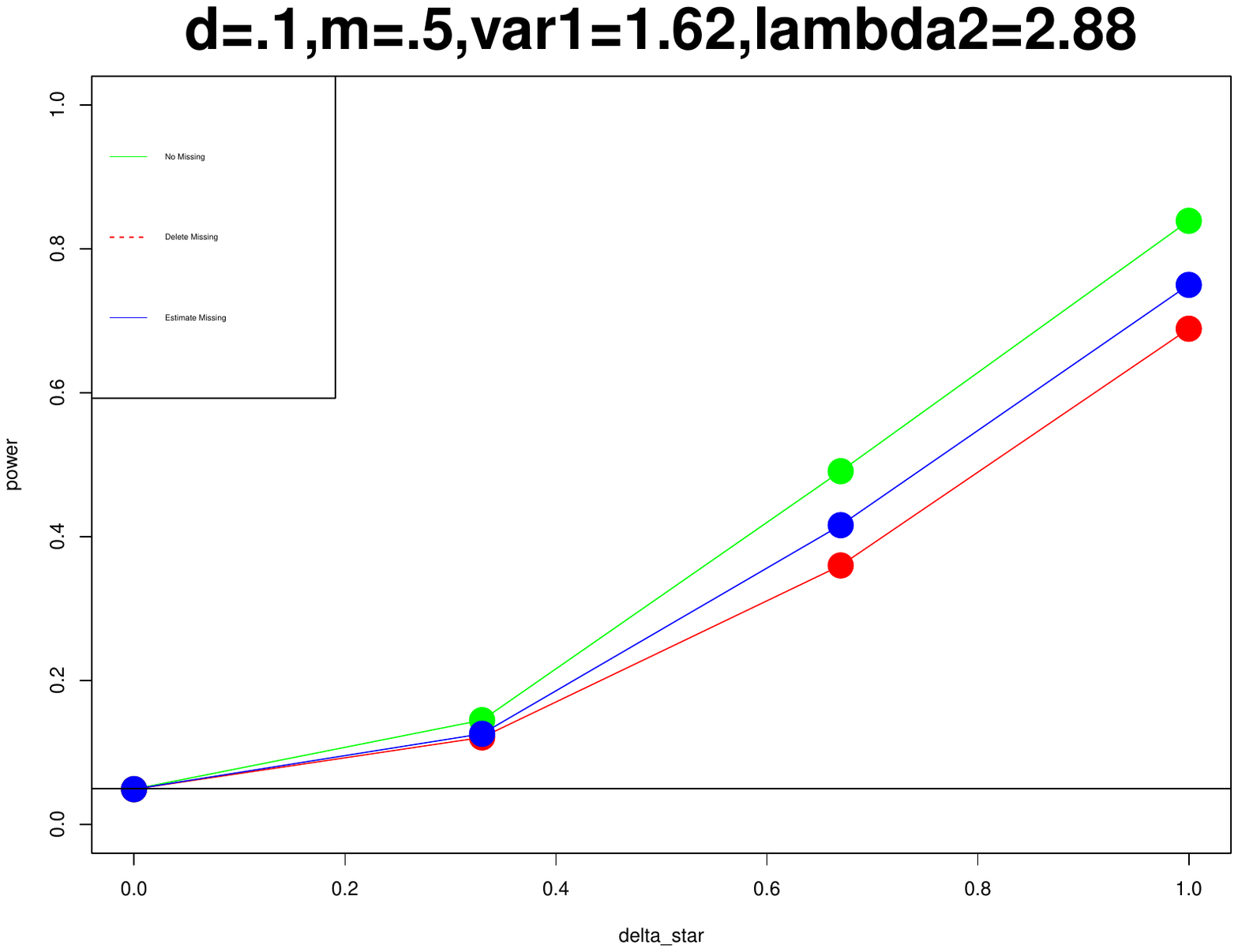}
\includegraphics[width = 2.3in, height = 1.3in]{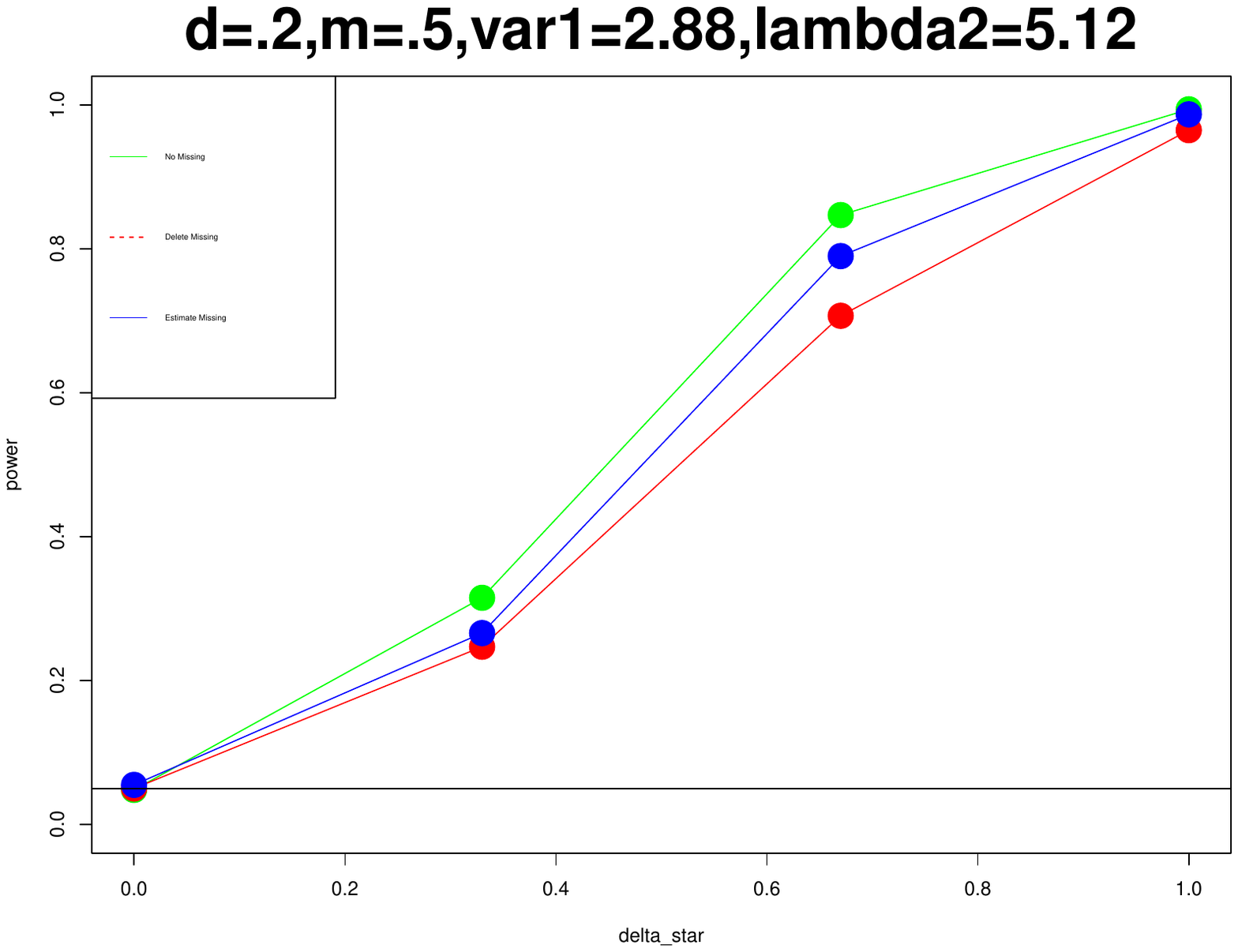}
\includegraphics[width = 2.3in, height = 1.3in]{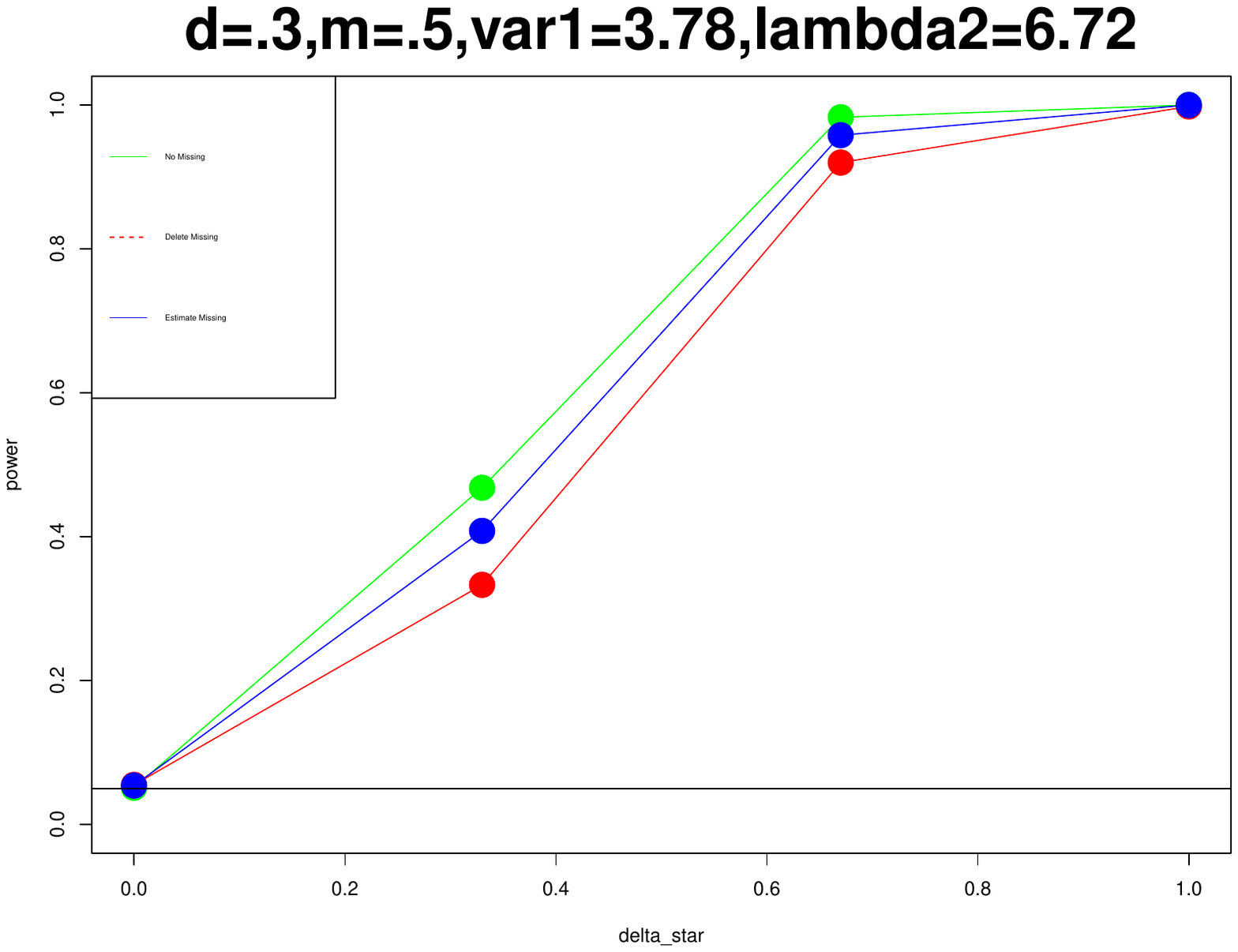}

\hspace{1.5cm}
$d=.1 \quad m=.1$
\hspace{3cm}
$d=.2 \quad m=.1$
\hspace{3cm}
$d=.3 \quad m=.1$

\includegraphics[width = 2.3in, height = 1.3in]{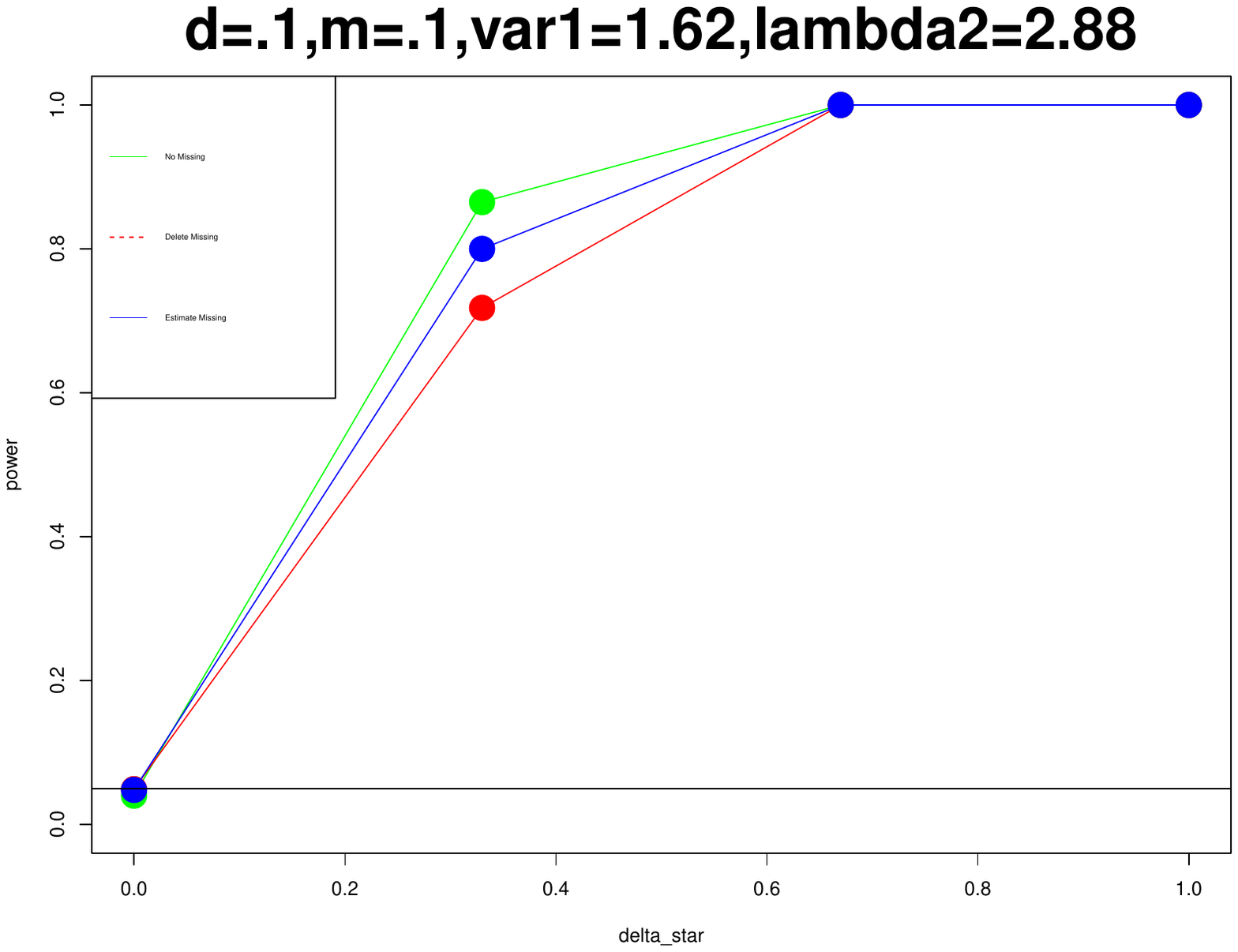}
\includegraphics[width = 2.3in, height = 1.3in]{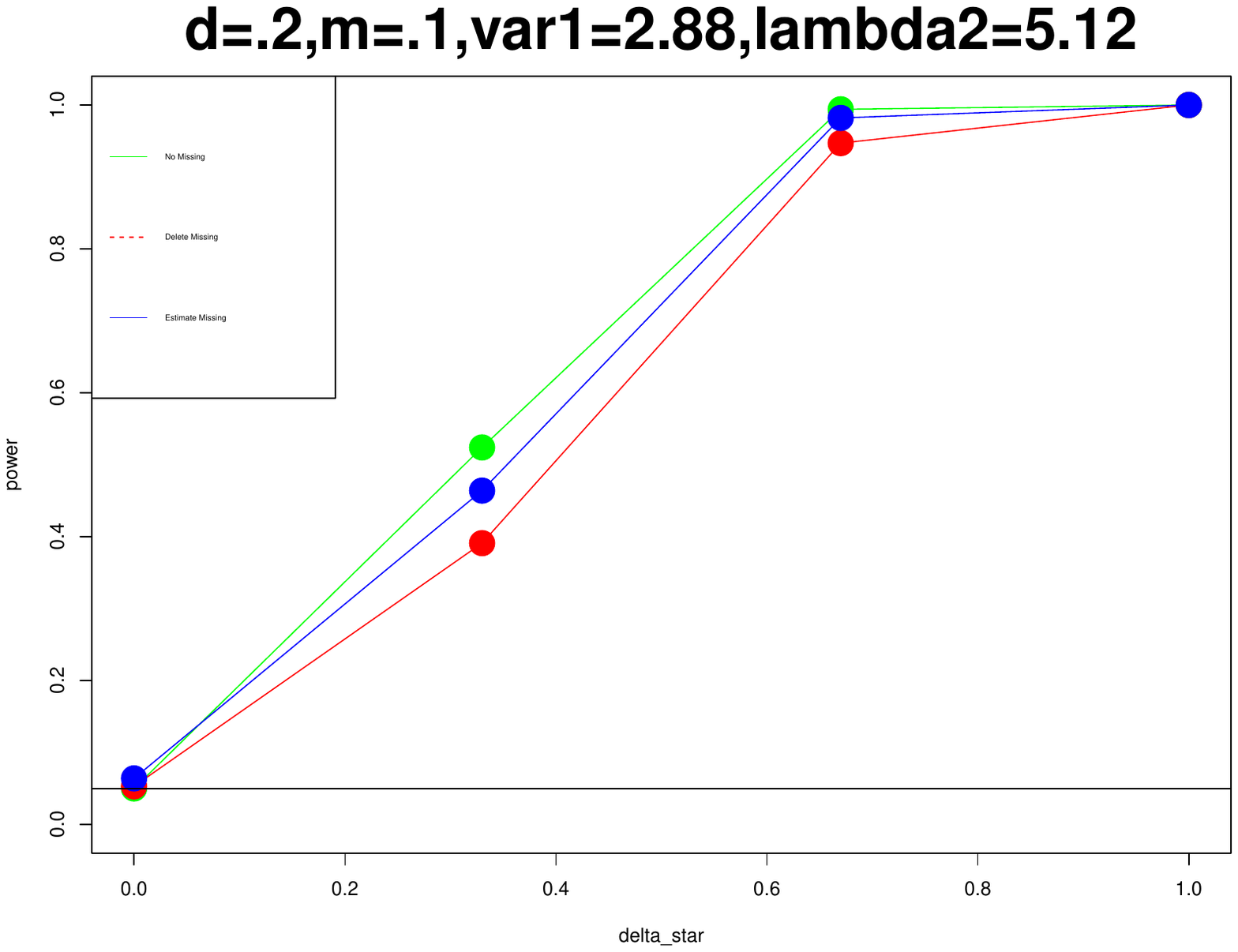}
\includegraphics[width = 2.3in, height = 1.3in]{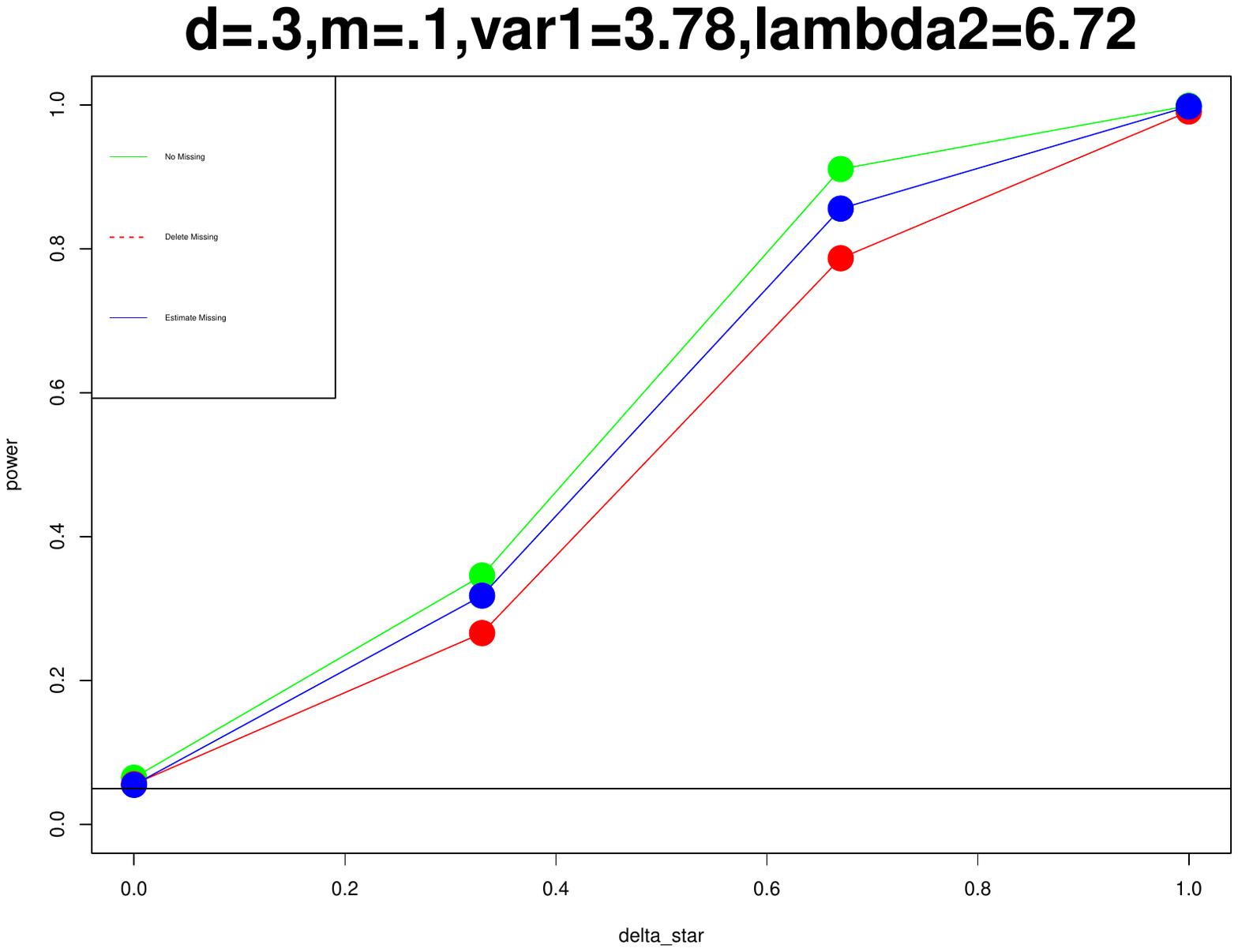}

\hspace{1.5cm}
$d=.1 \quad m=.1$
\hspace{3cm}
$d=.2 \quad m=.1$
\hspace{3cm}
$d=.3 \quad m=.1$

\includegraphics[width = 2.3in, height = 1.3in]{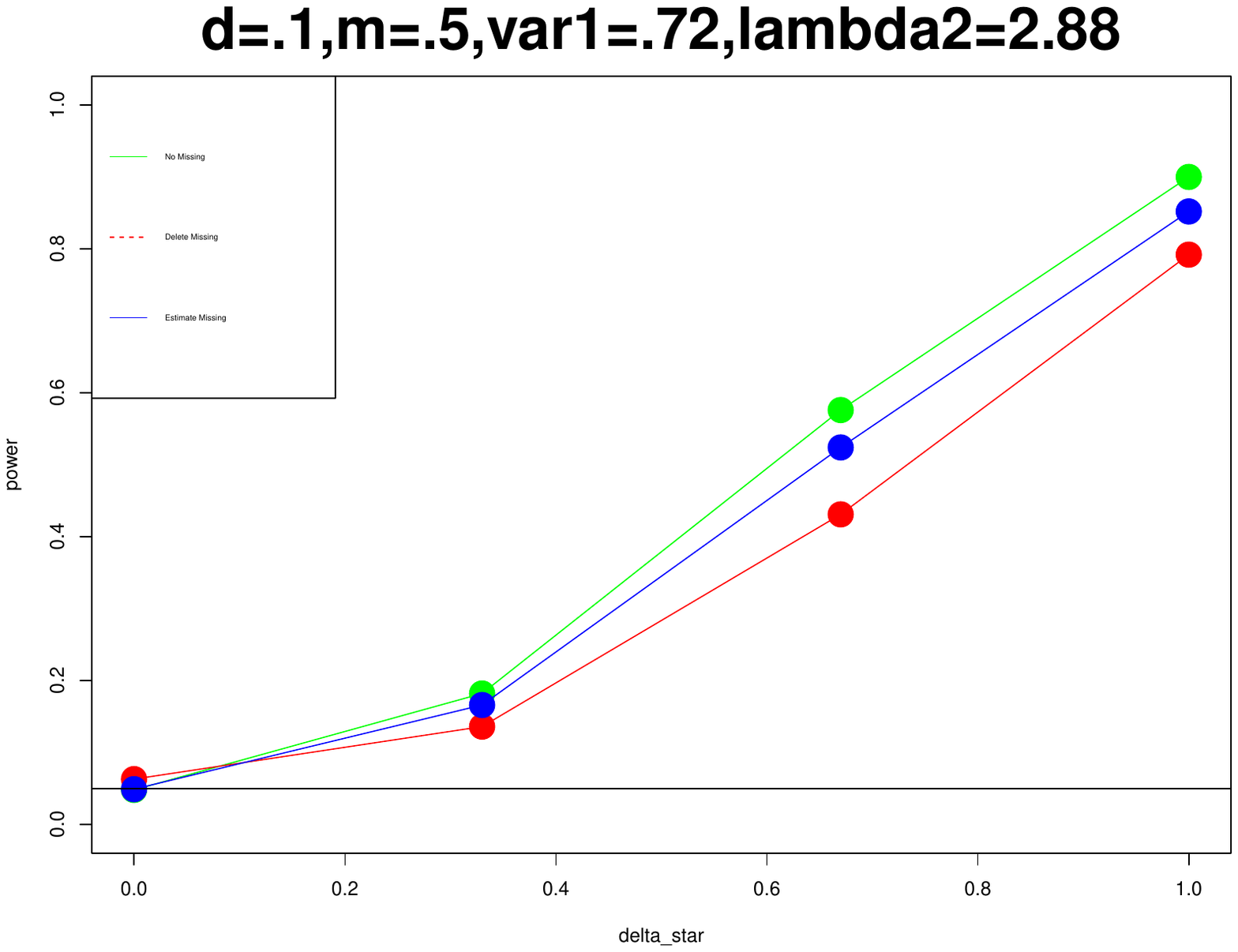}
\includegraphics[width = 2.3in, height = 1.3in]{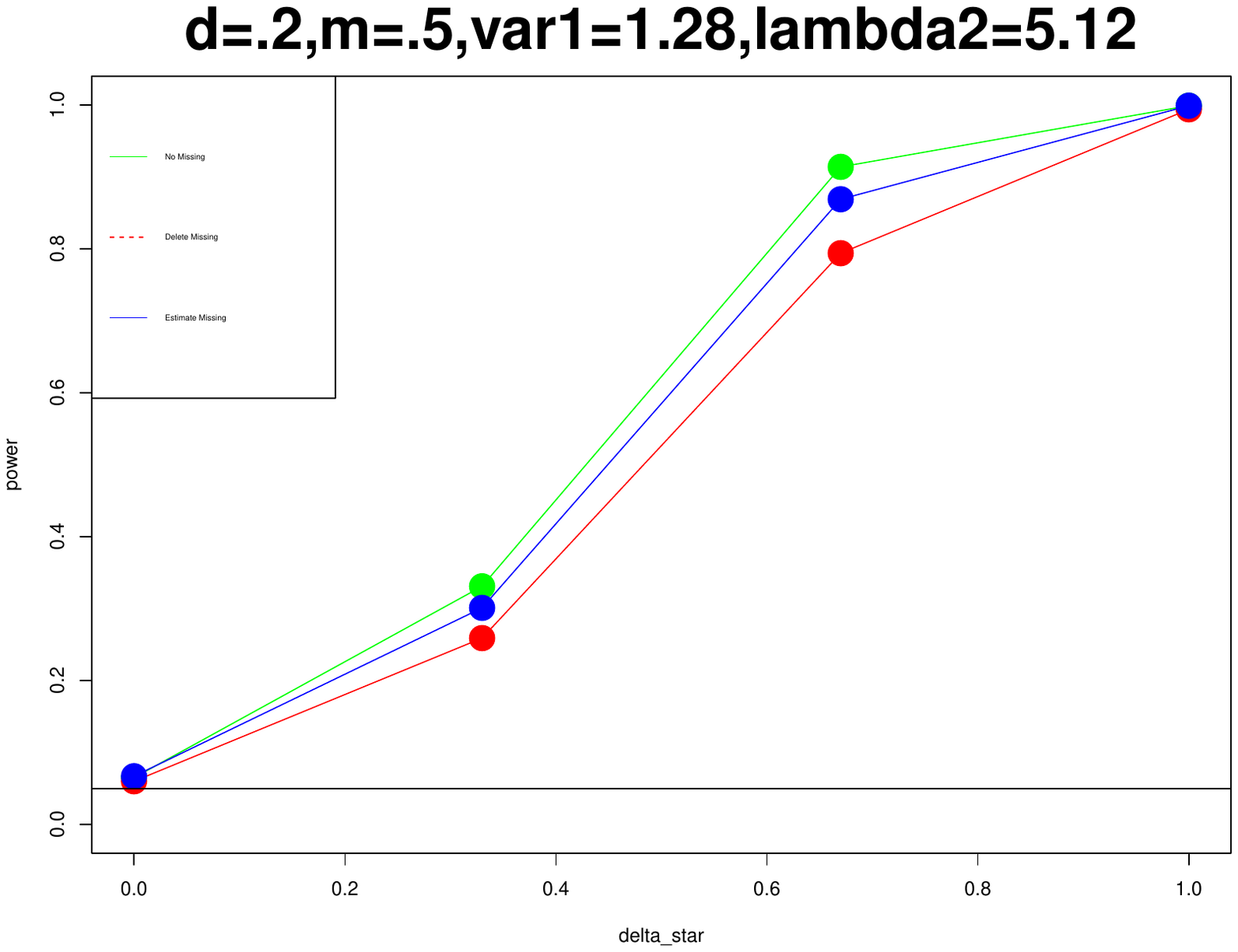}
\includegraphics[width = 2.3in, height = 1.3in]{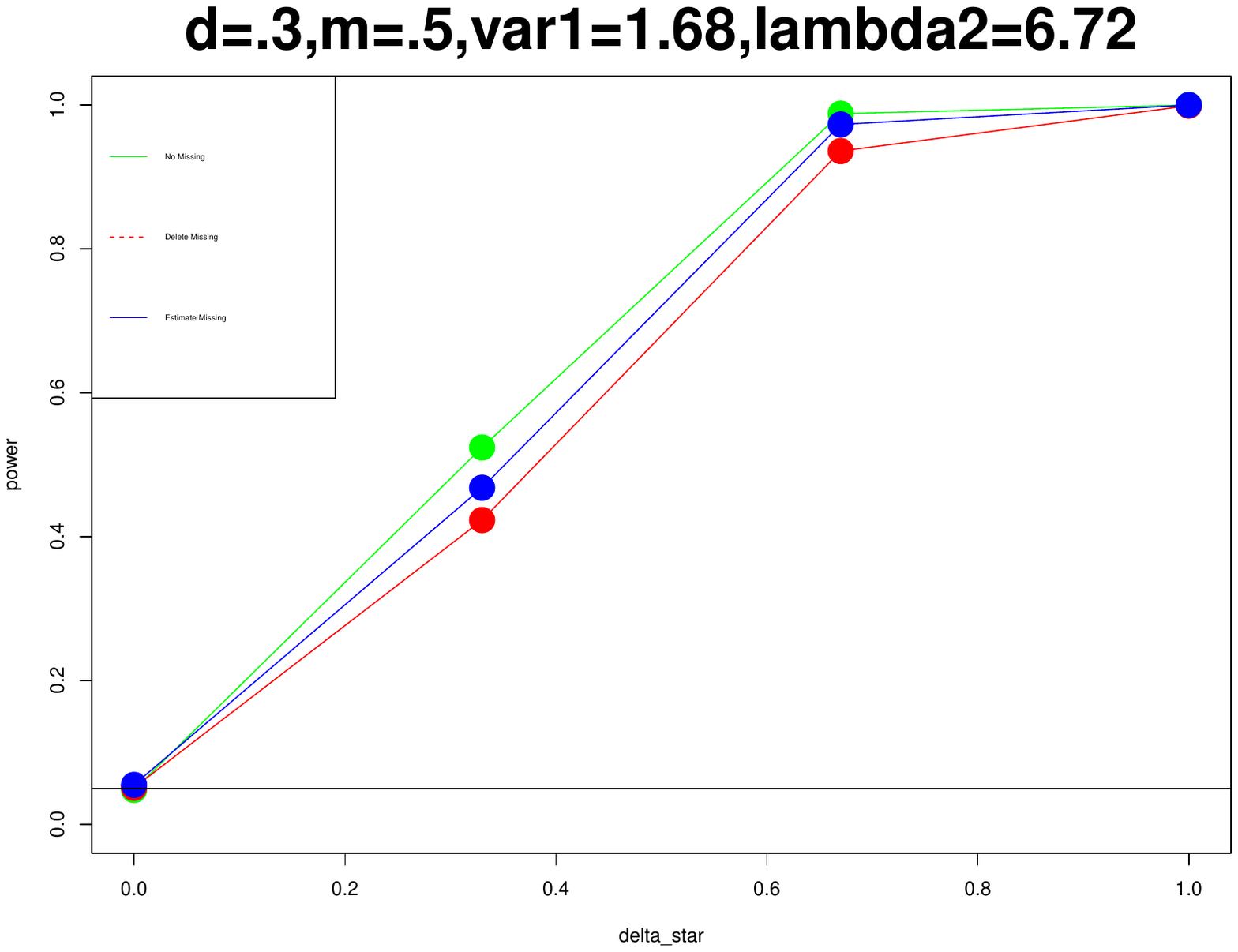}

\hspace{1.5cm}
$d=.1 \quad m=.5$
\hspace{3cm}
$d=.2 \quad m=.5$
\hspace{3cm}
$d=.3 \quad m=.5$

\includegraphics[width = 2.3in, height = 1.3in]{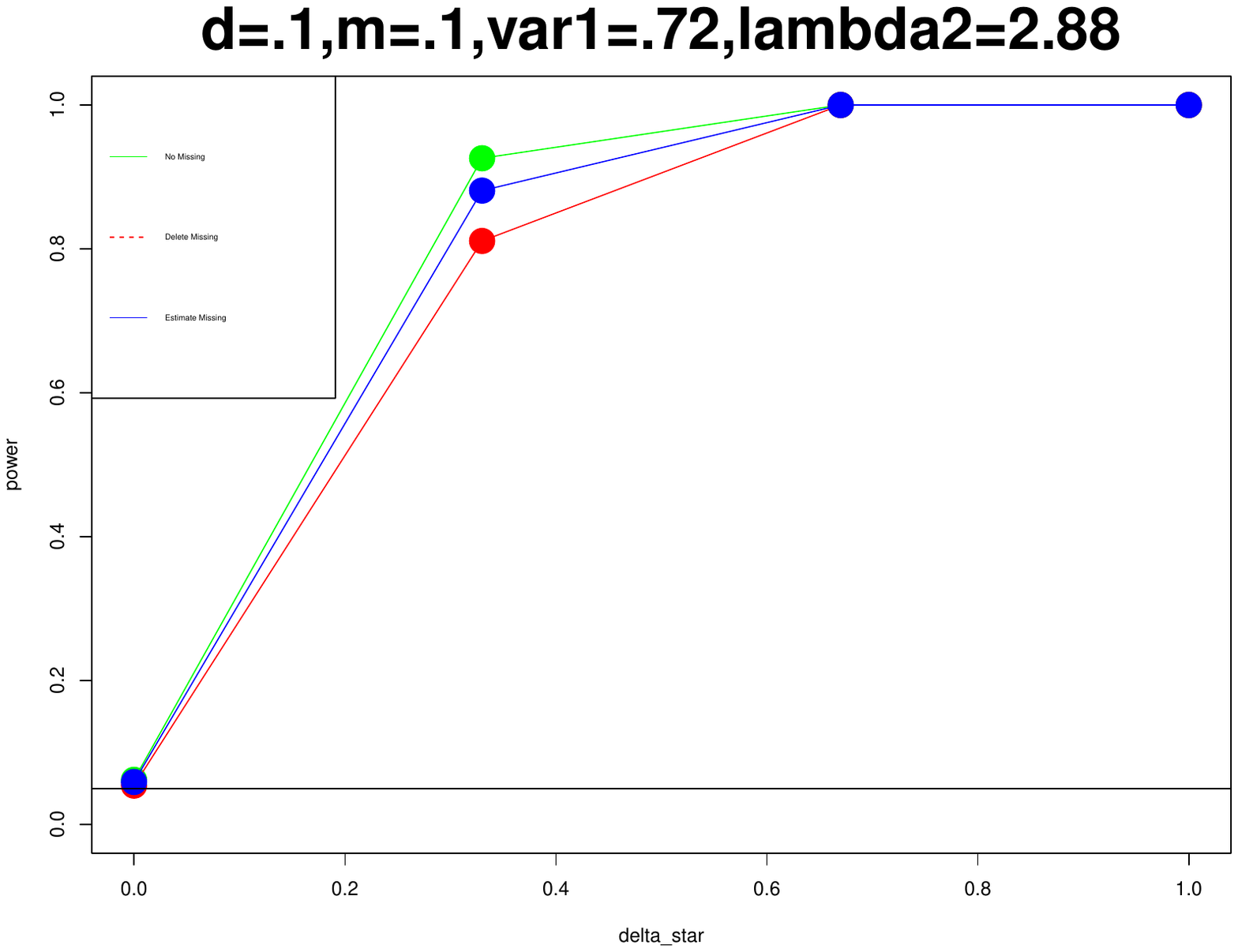}
\includegraphics[width = 2.3in, height = 1.3in]{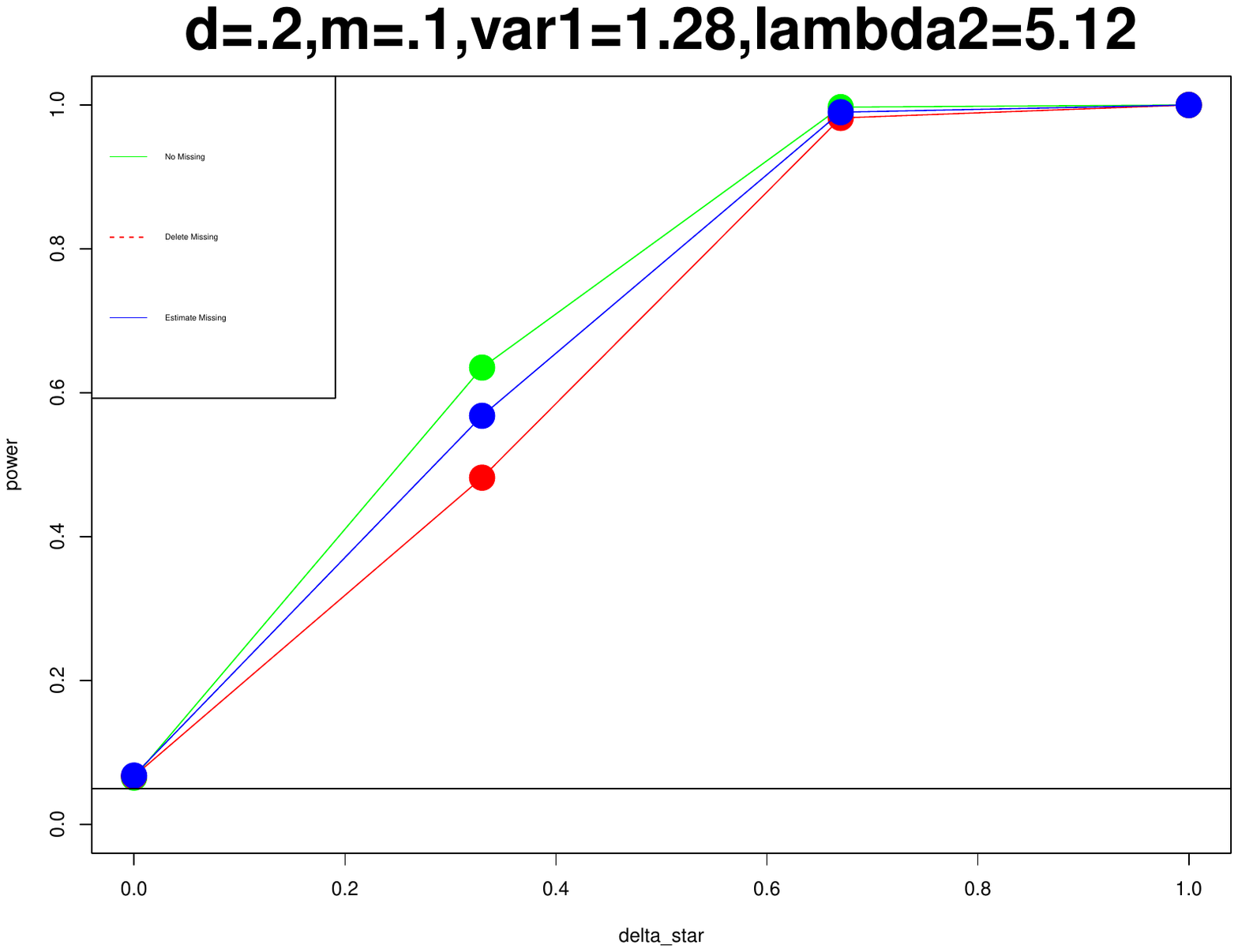}
\includegraphics[width = 2.3in, height = 1.3in]{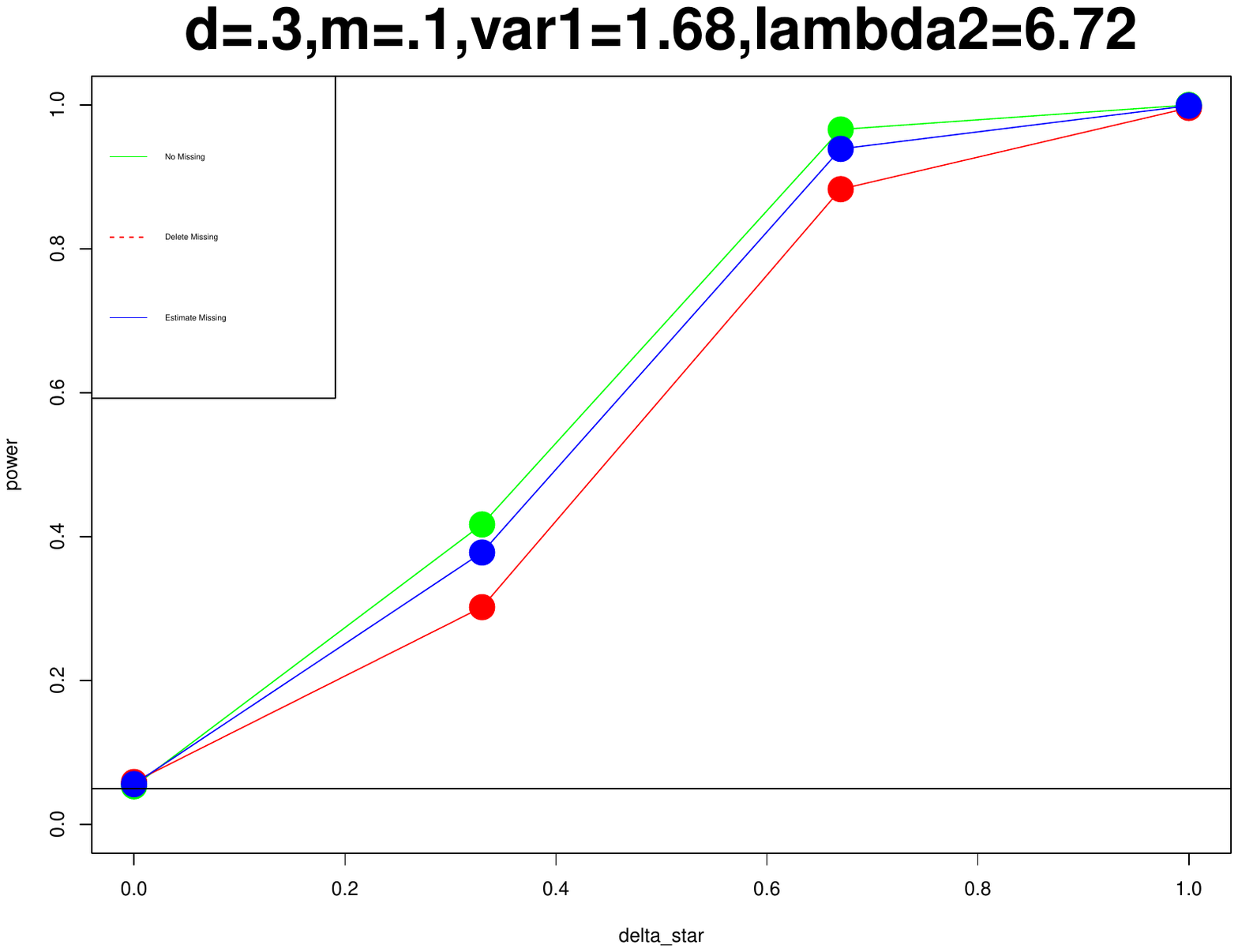}

\hspace{1.5cm}
$d=.1 \quad m=.1$
\hspace{3cm}
$d=.2 \quad m=.1$
\hspace{3cm}
$d=.3 \quad m=.1$

\includegraphics[width = 2.3in, height = 1.5in]{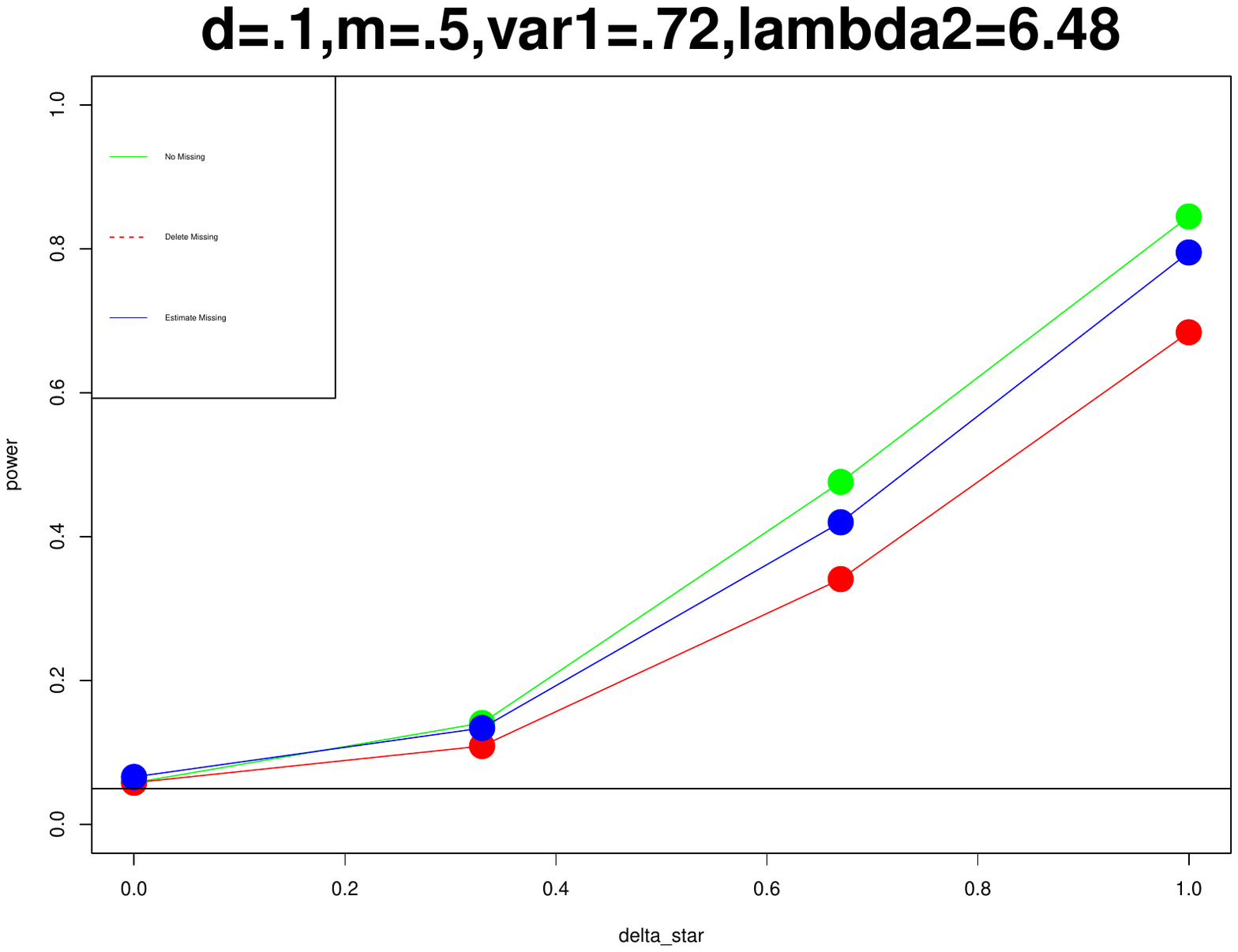}
\includegraphics[width = 2.3in, height = 1.5in]{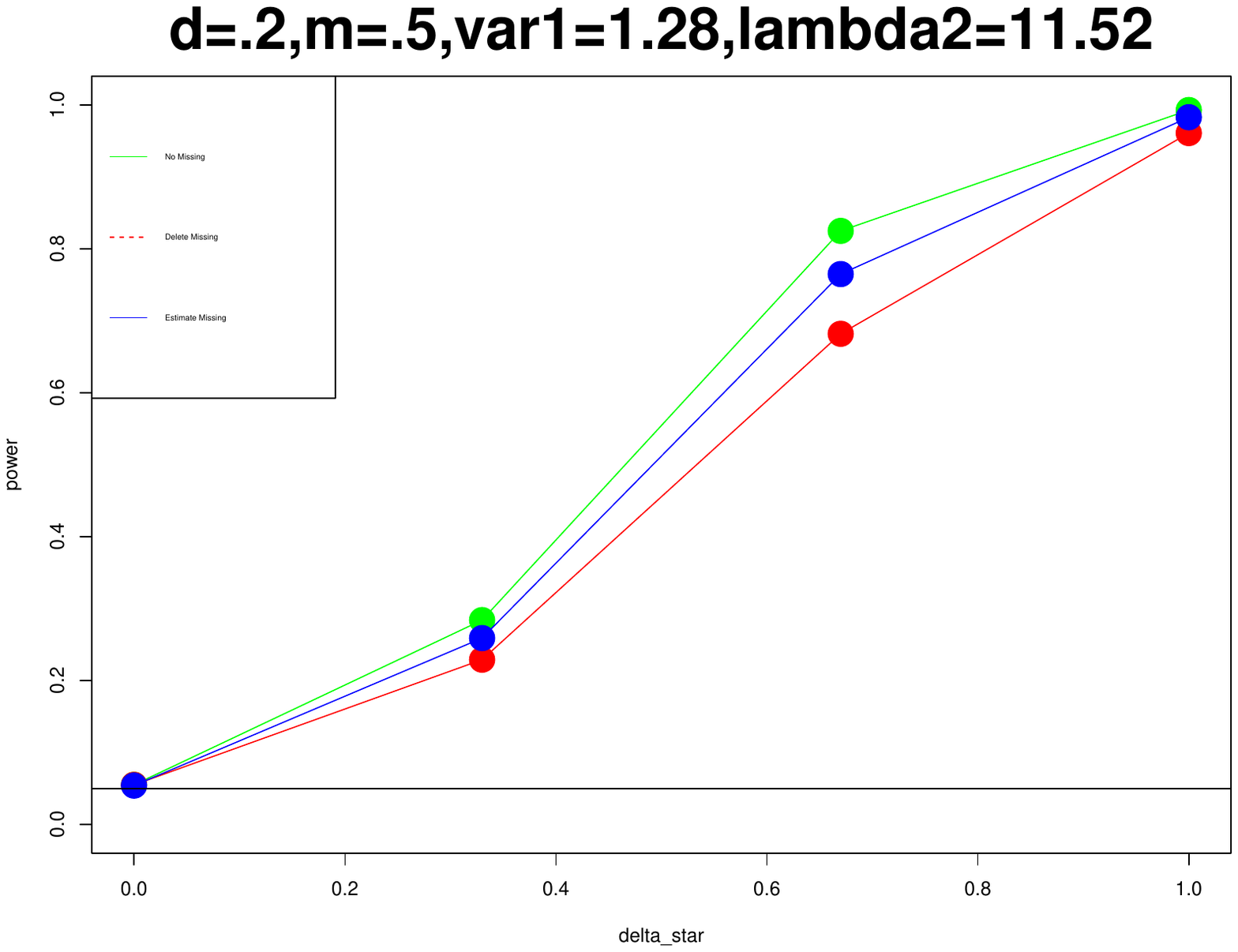}
\includegraphics[width = 2.3in, height = 1.5in]{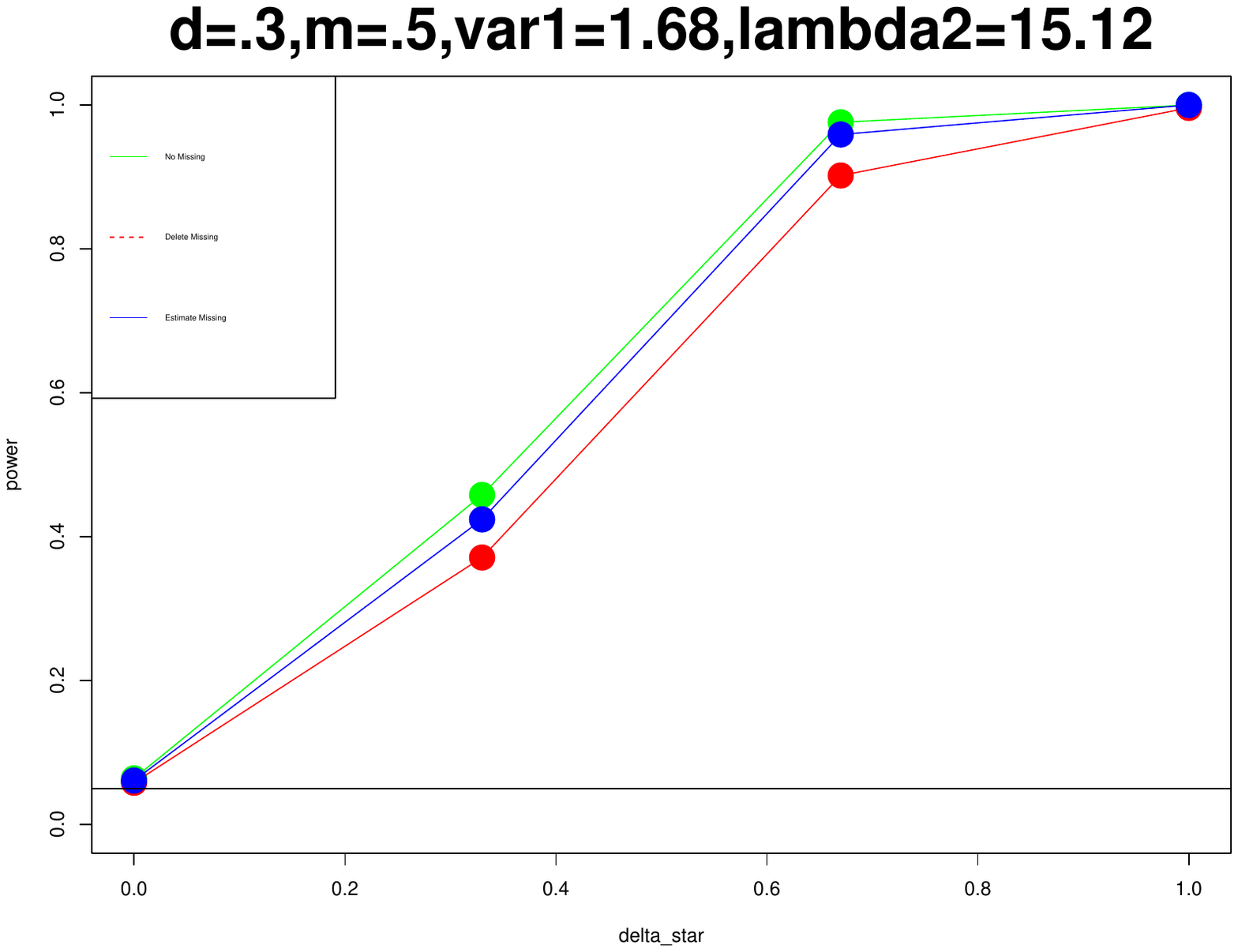}

\hspace{1.5cm}
$d=.1 \quad m=.5$
\hspace{3cm}
$d=.2 \quad m=.5$
\hspace{3cm}
$d=.3 \quad m=.5$

\includegraphics[width = 2.3in, height = 1.5in]{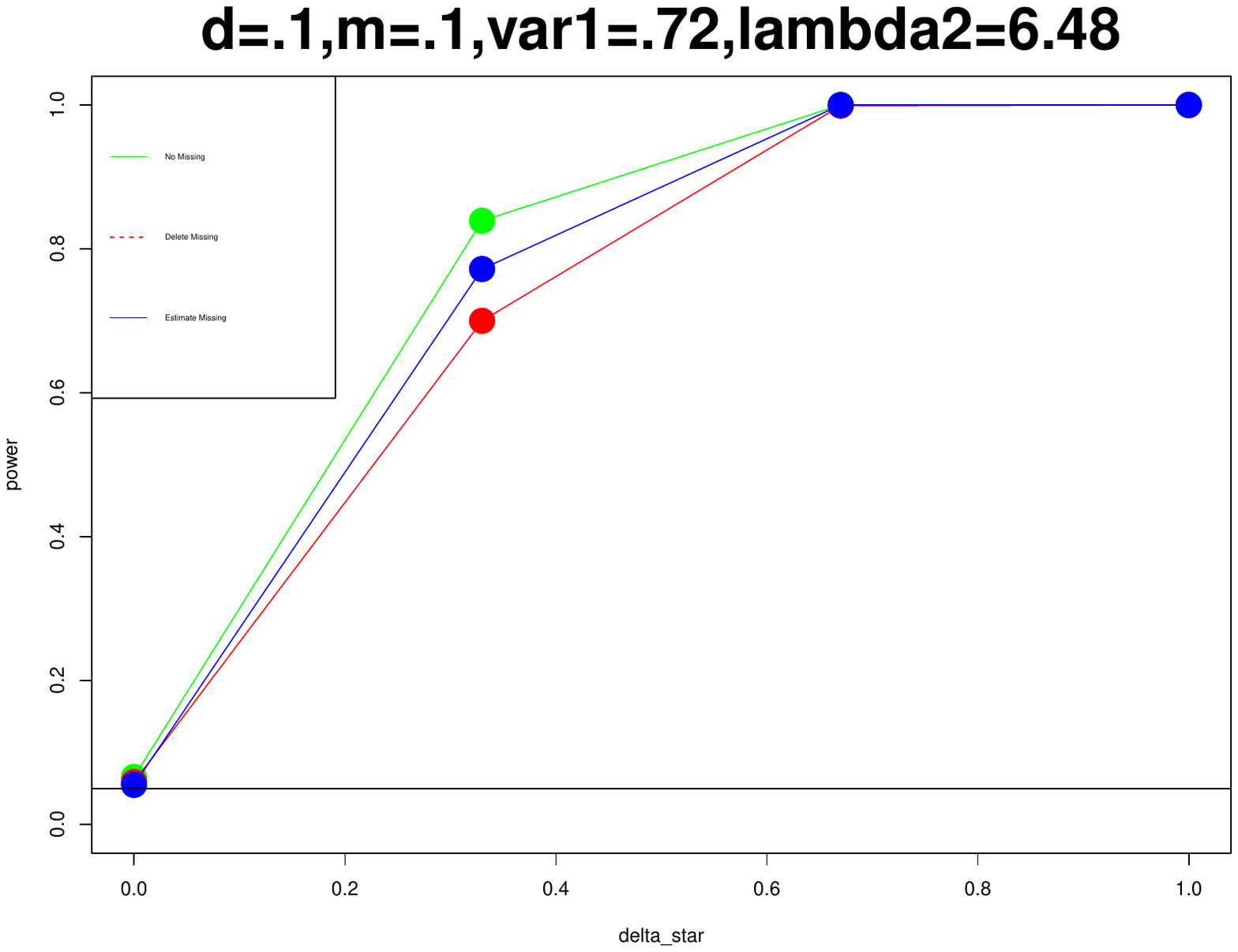}
\includegraphics[width = 2.3in, height = 1.5in]{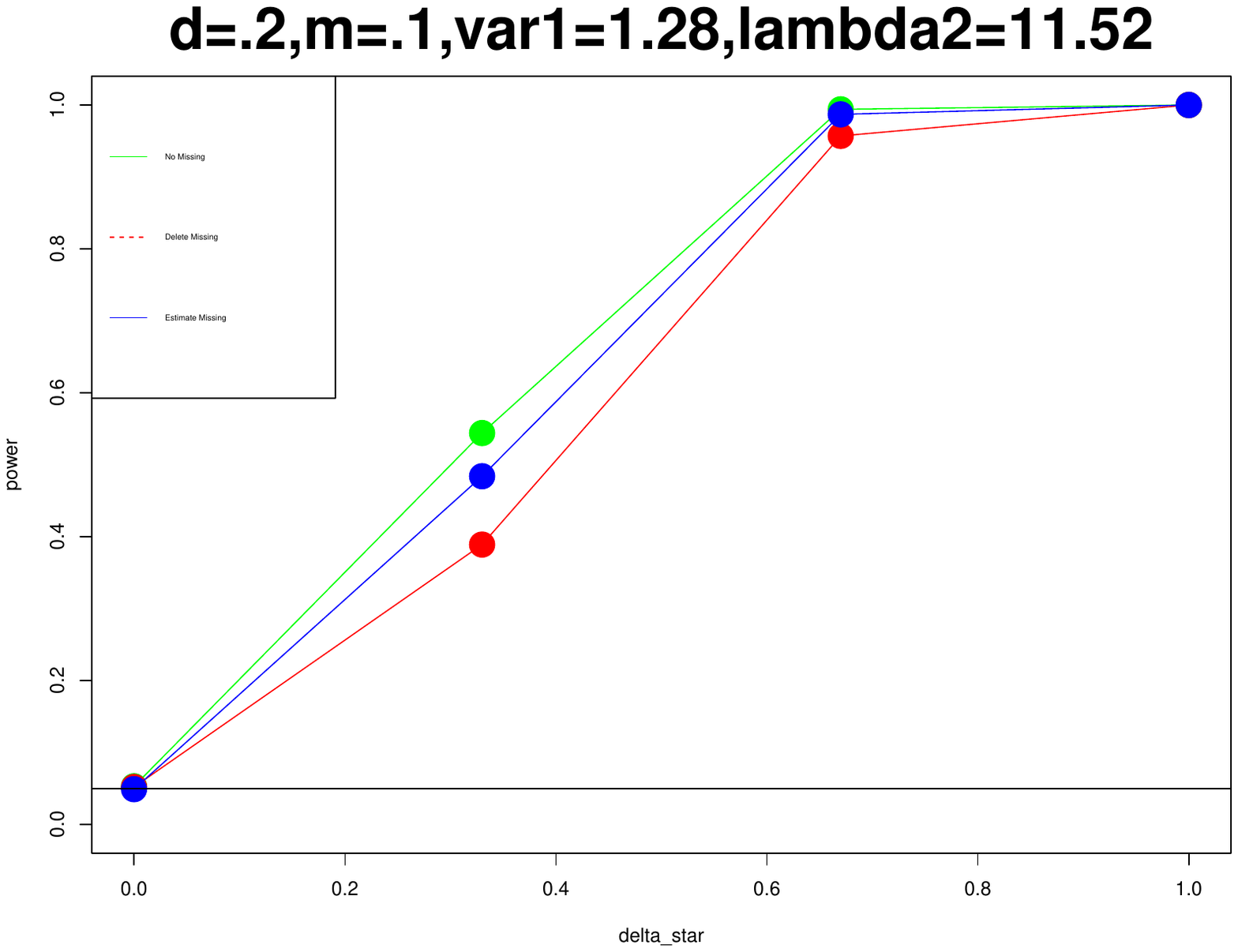}
\includegraphics[width = 2.3in, height = 1.5in]{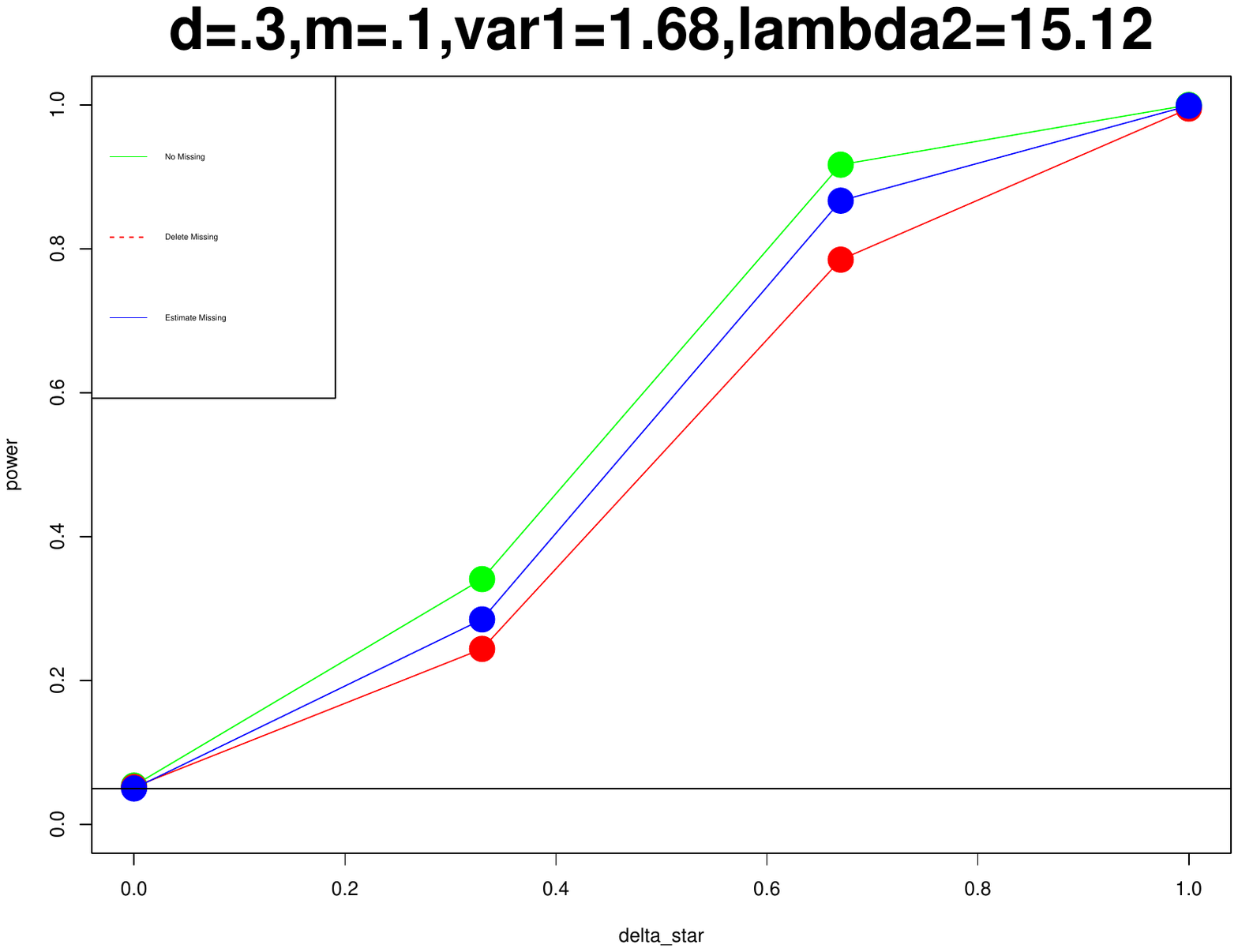}

\subsubsection{When both Traits have Poisson Distribution}

\hspace{1.5cm}
$d=.1 \quad m=.5$
\hspace{3cm}
$d=.2 \quad m=.5$
\hspace{3cm}
$d=.3 \quad m=.5$

\includegraphics[width = 2.3in, height = 1.3in]{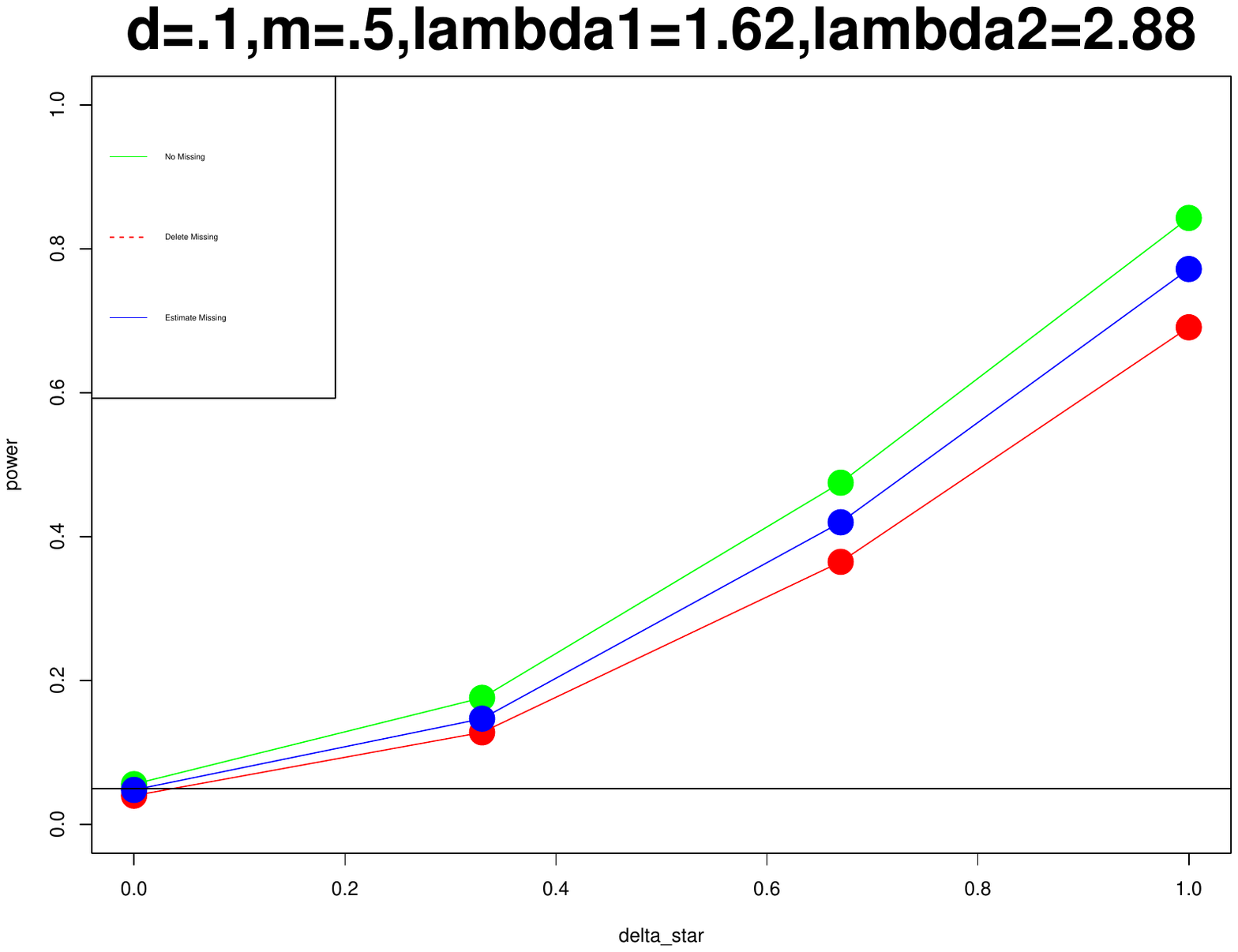}
\includegraphics[width = 2.3in, height = 1.3in]{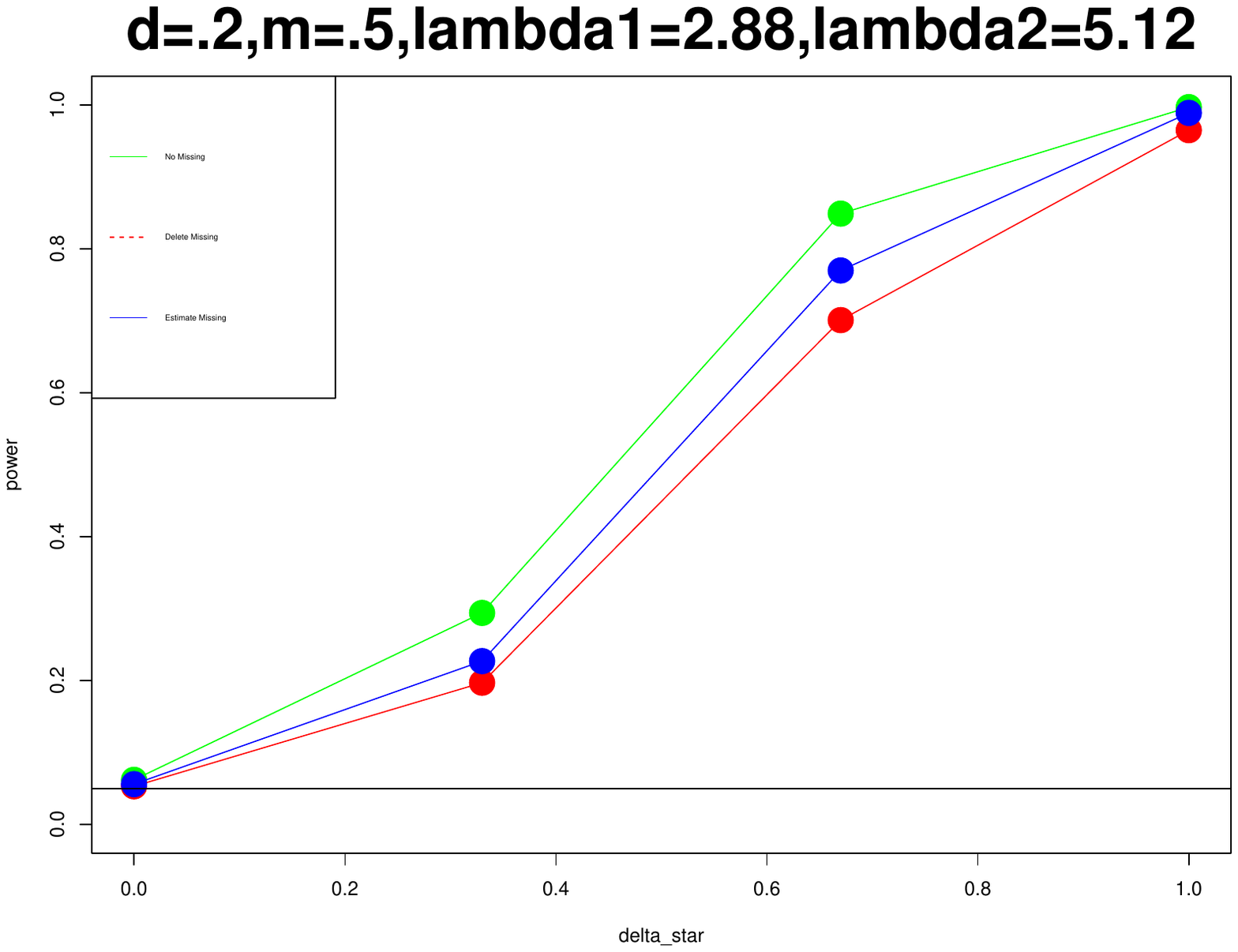}
\includegraphics[width = 2.3in, height = 1.3in]{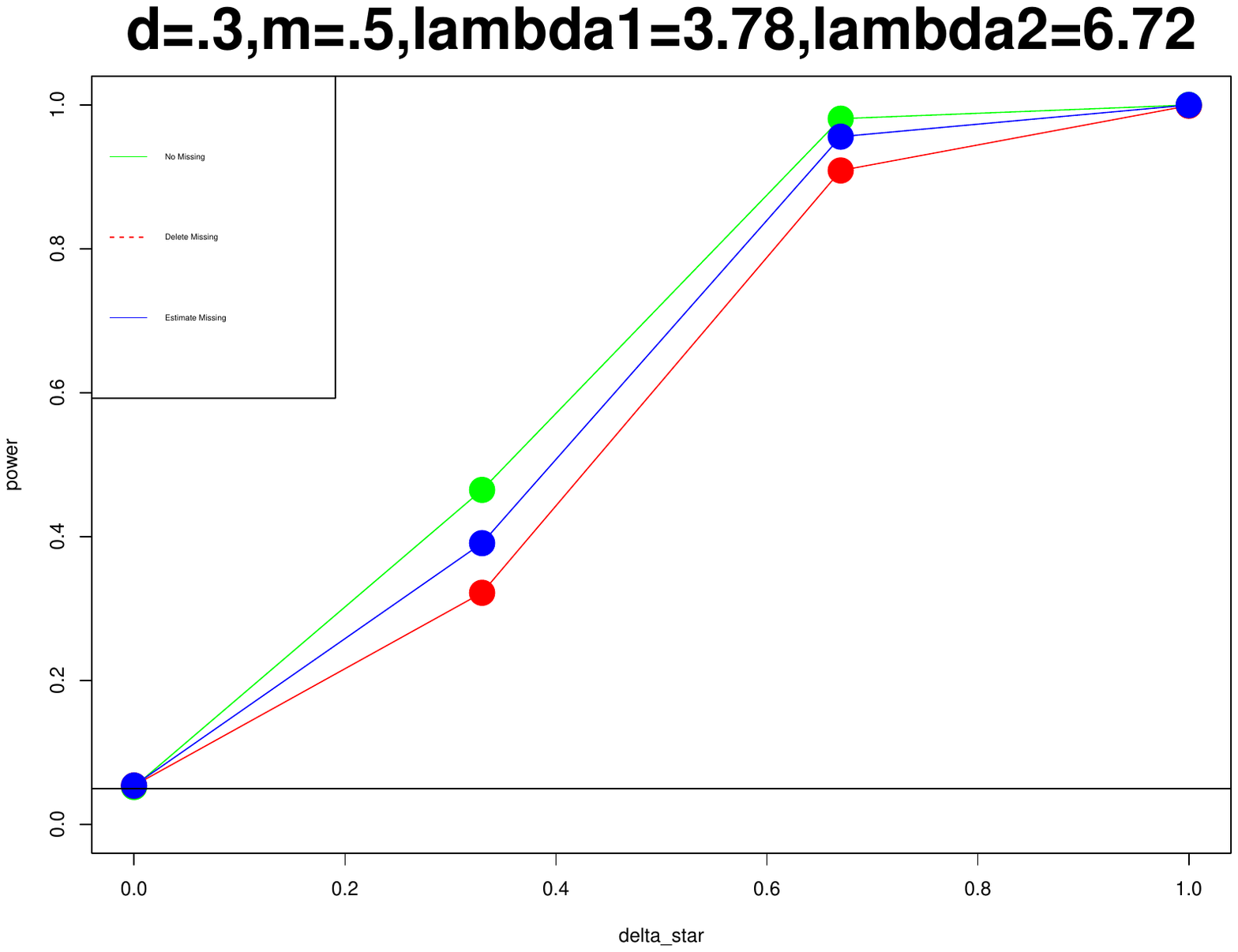}

\hspace{1.5cm}
$d=.1 \quad m=.1$
\hspace{3cm}
$d=.2 \quad m=.1$
\hspace{3cm}
$d=.3 \quad m=.1$

\includegraphics[width = 2.3in, height = 1.3in]{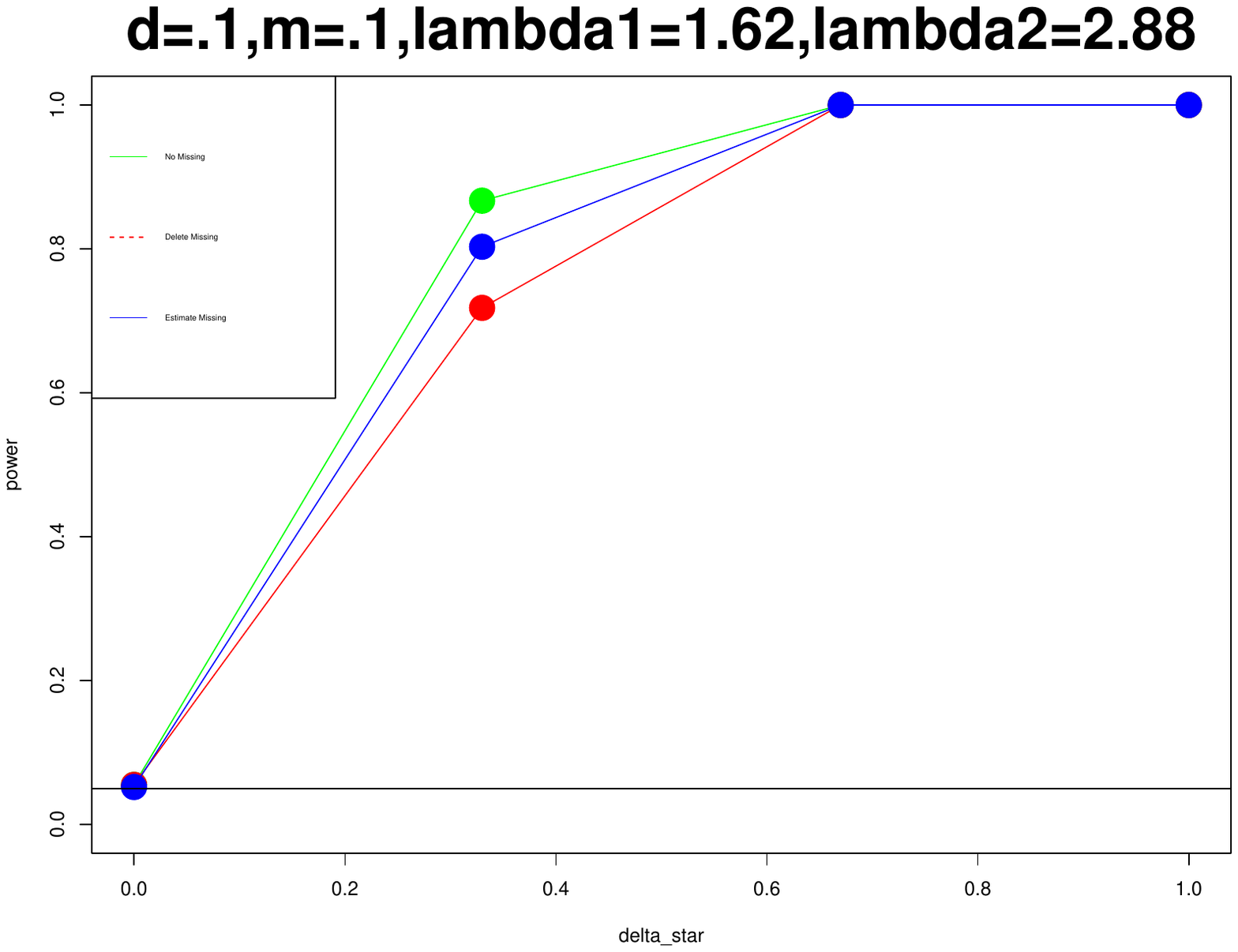}
\includegraphics[width = 2.3in, height = 1.3in]{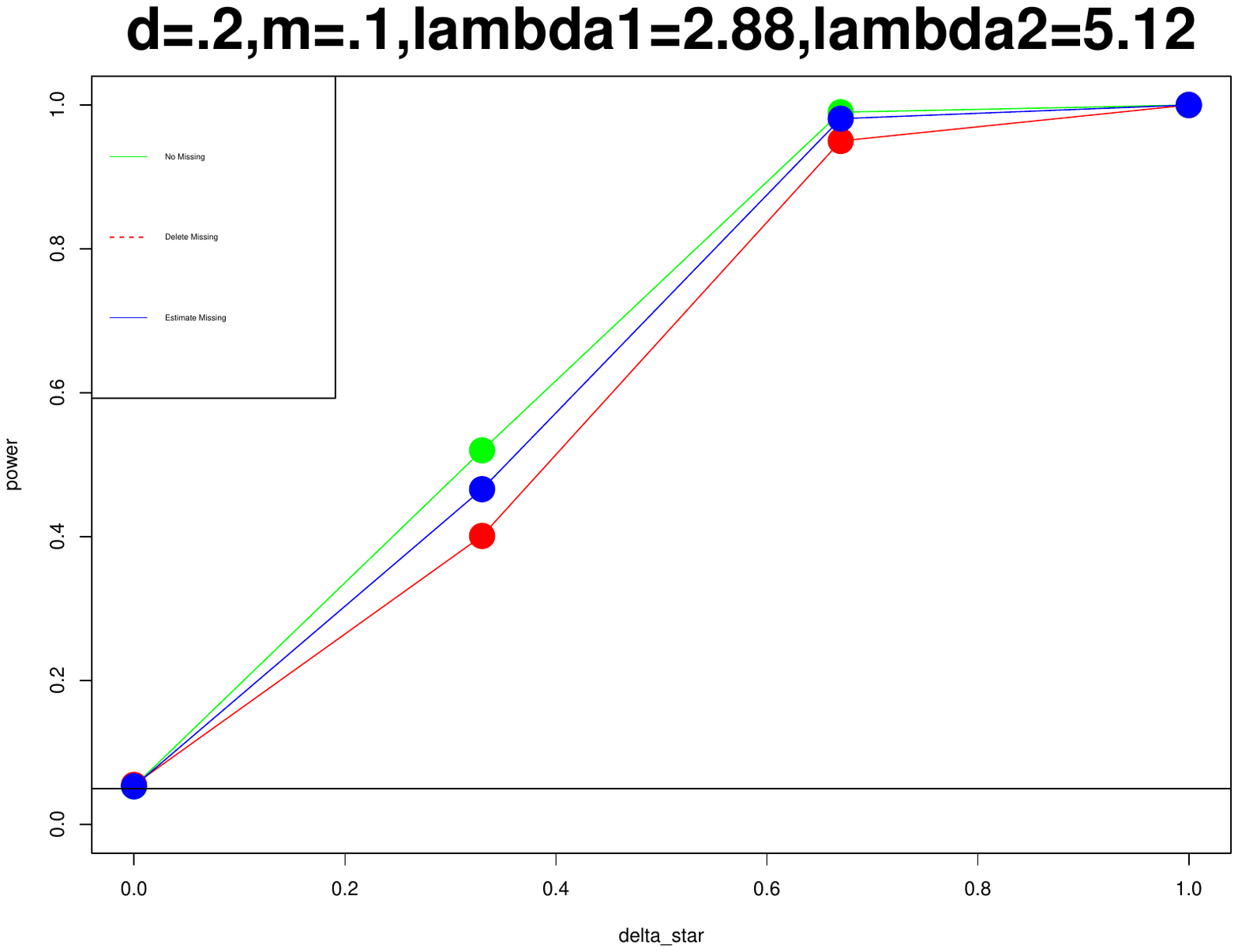}
\includegraphics[width = 2.3in, height = 1.3in]{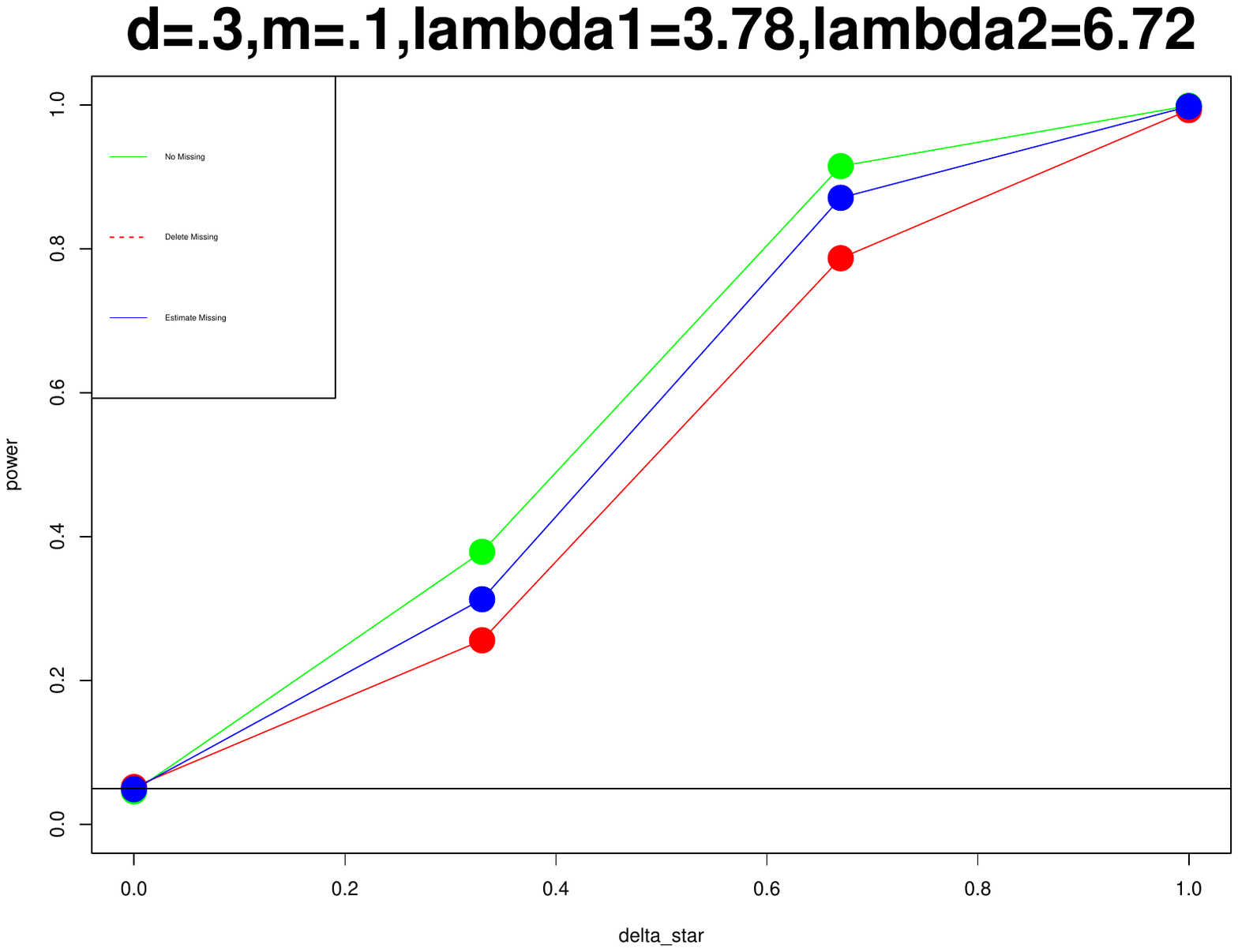}

\hspace{1.5cm}
$d=.1 \quad m=.1$
\hspace{3cm}
$d=.2 \quad m=.1$
\hspace{3cm}
$d=.3 \quad m=.1$

\includegraphics[width = 2.3in, height = 1.3in]{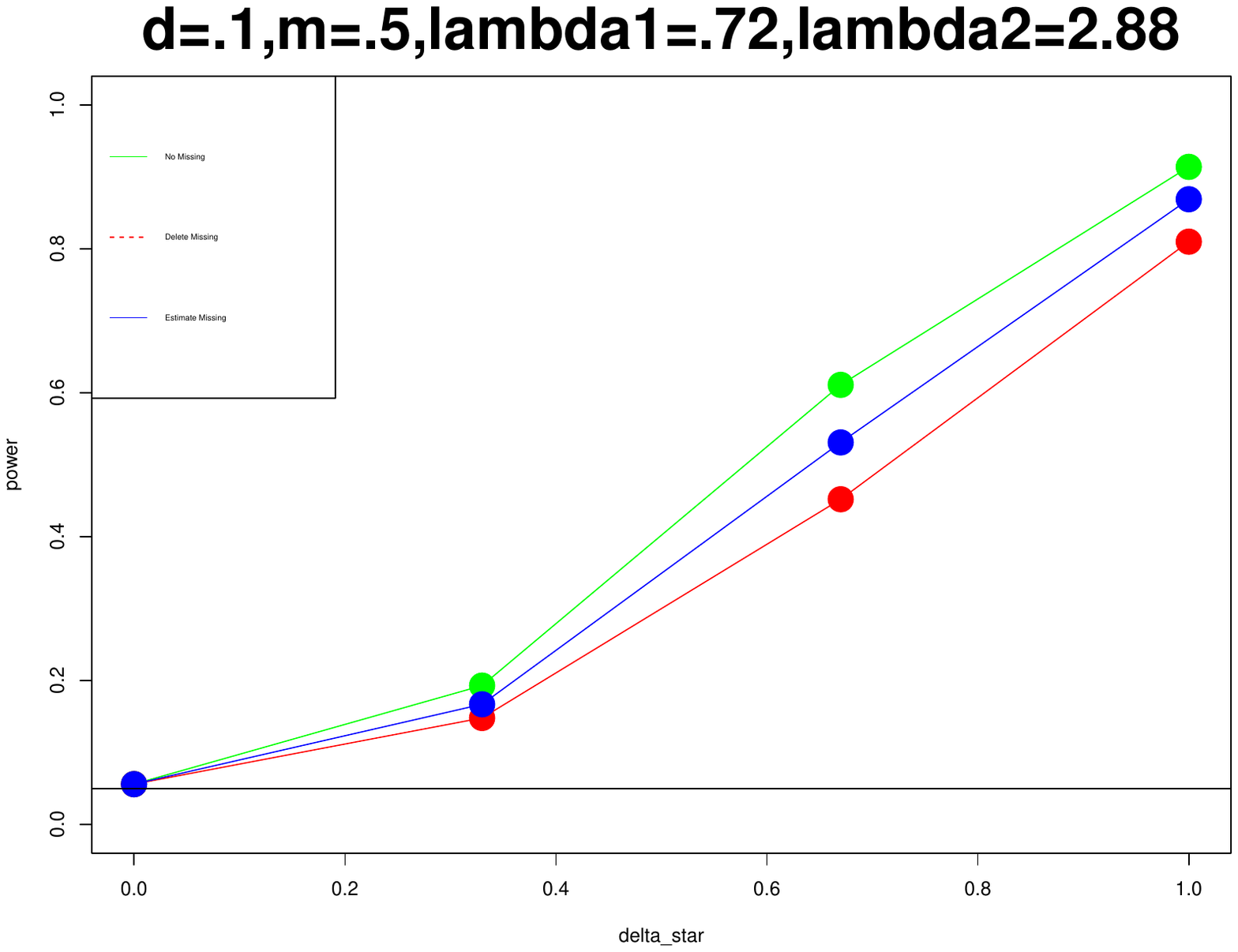}
\includegraphics[width = 2.3in, height = 1.3in]{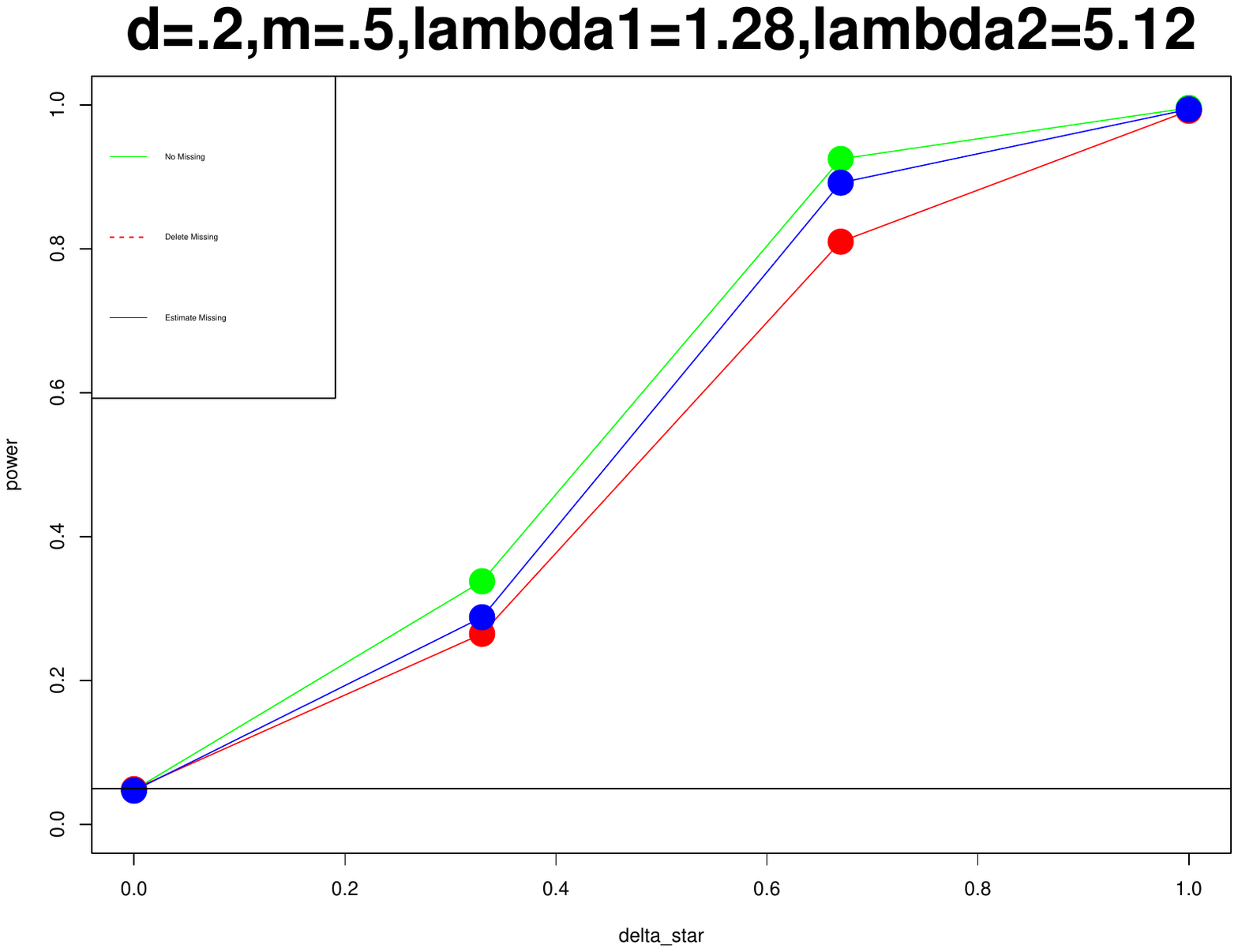}
\includegraphics[width = 2.3in, height = 1.3in]{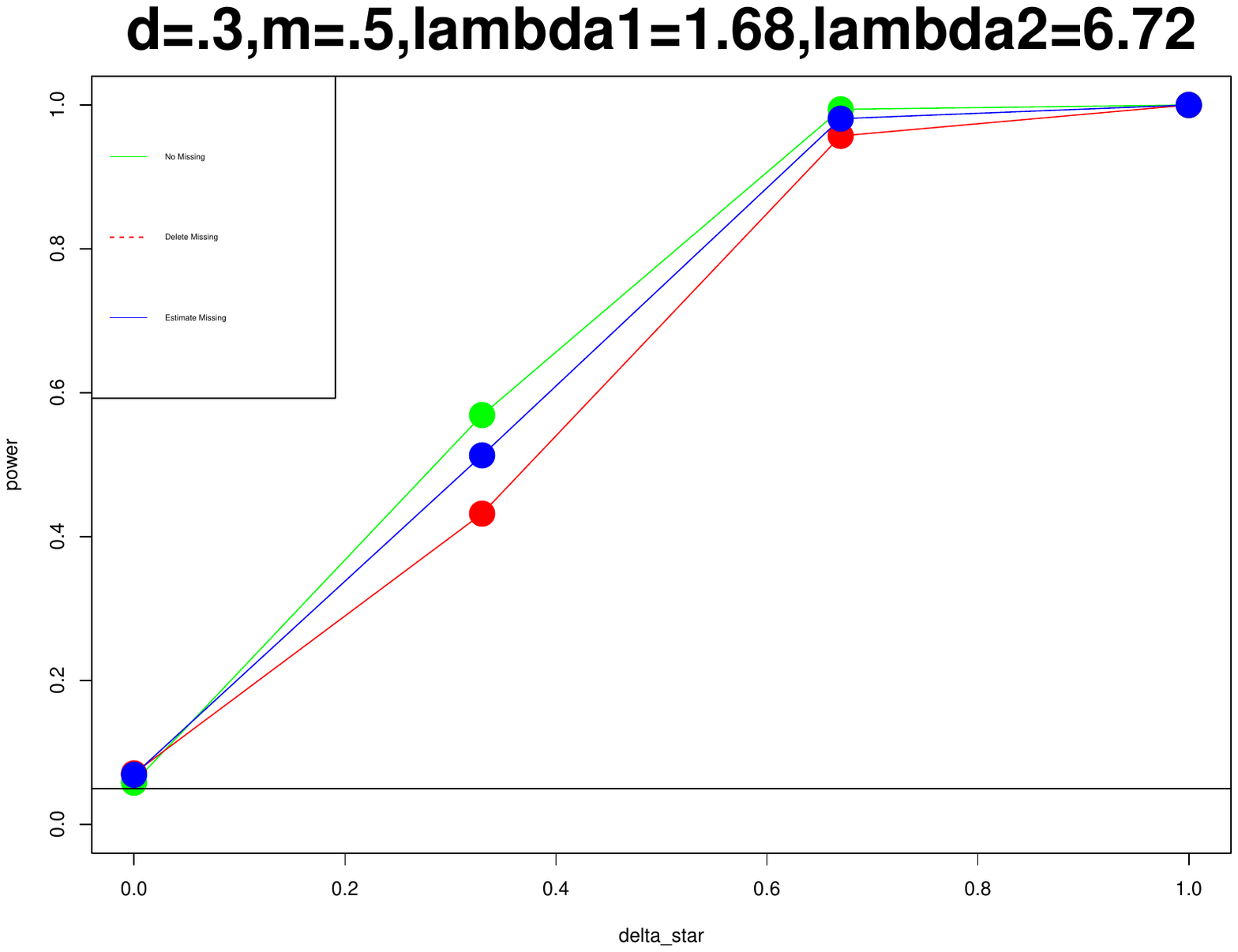}

\hspace{1.5cm}
$d=.1 \quad m=.5$
\hspace{3cm}
$d=.2 \quad m=.5$
\hspace{3cm}
$d=.3 \quad m=.5$

\includegraphics[width = 2.3in, height = 1.3in]{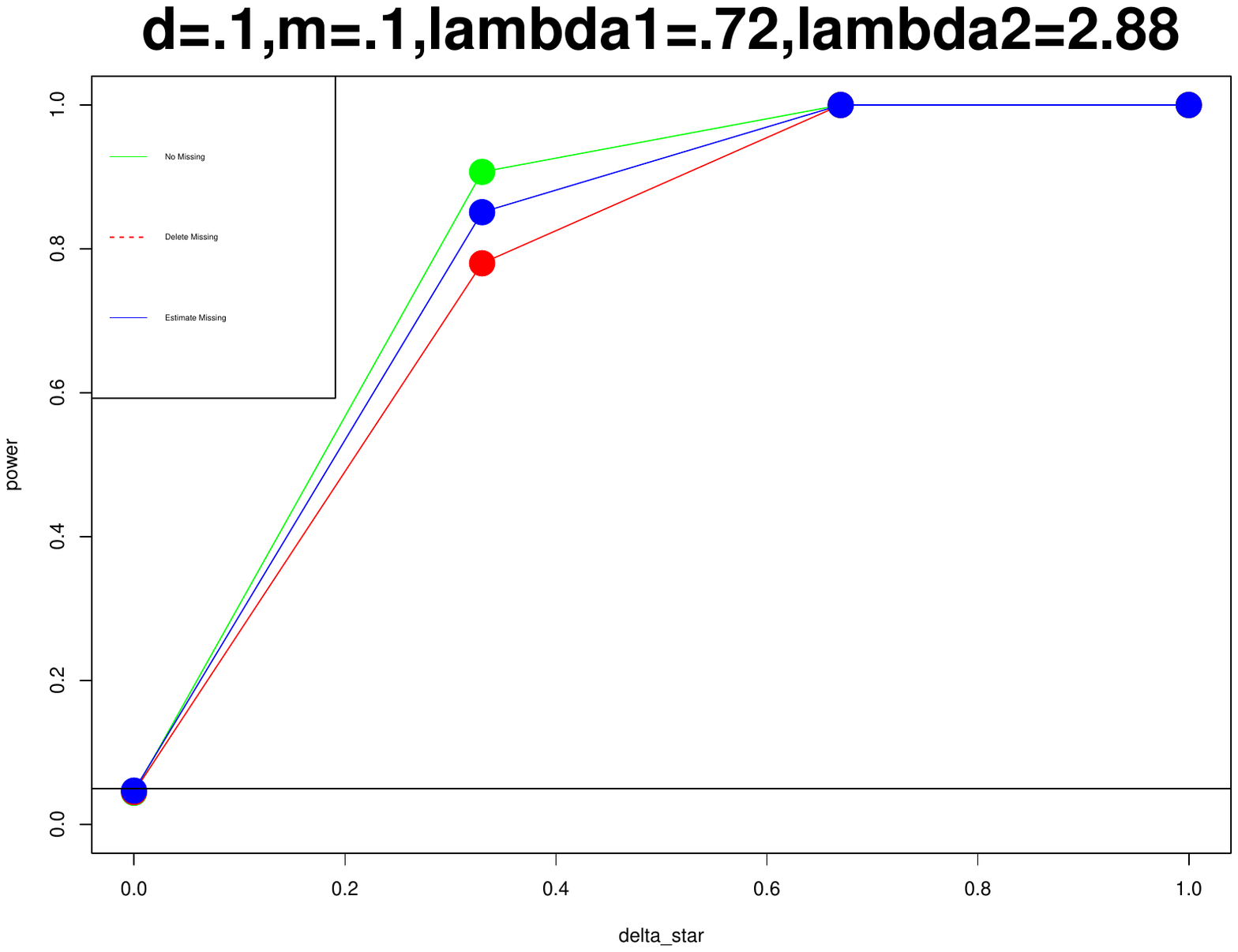}
\includegraphics[width = 2.3in, height = 1.3in]{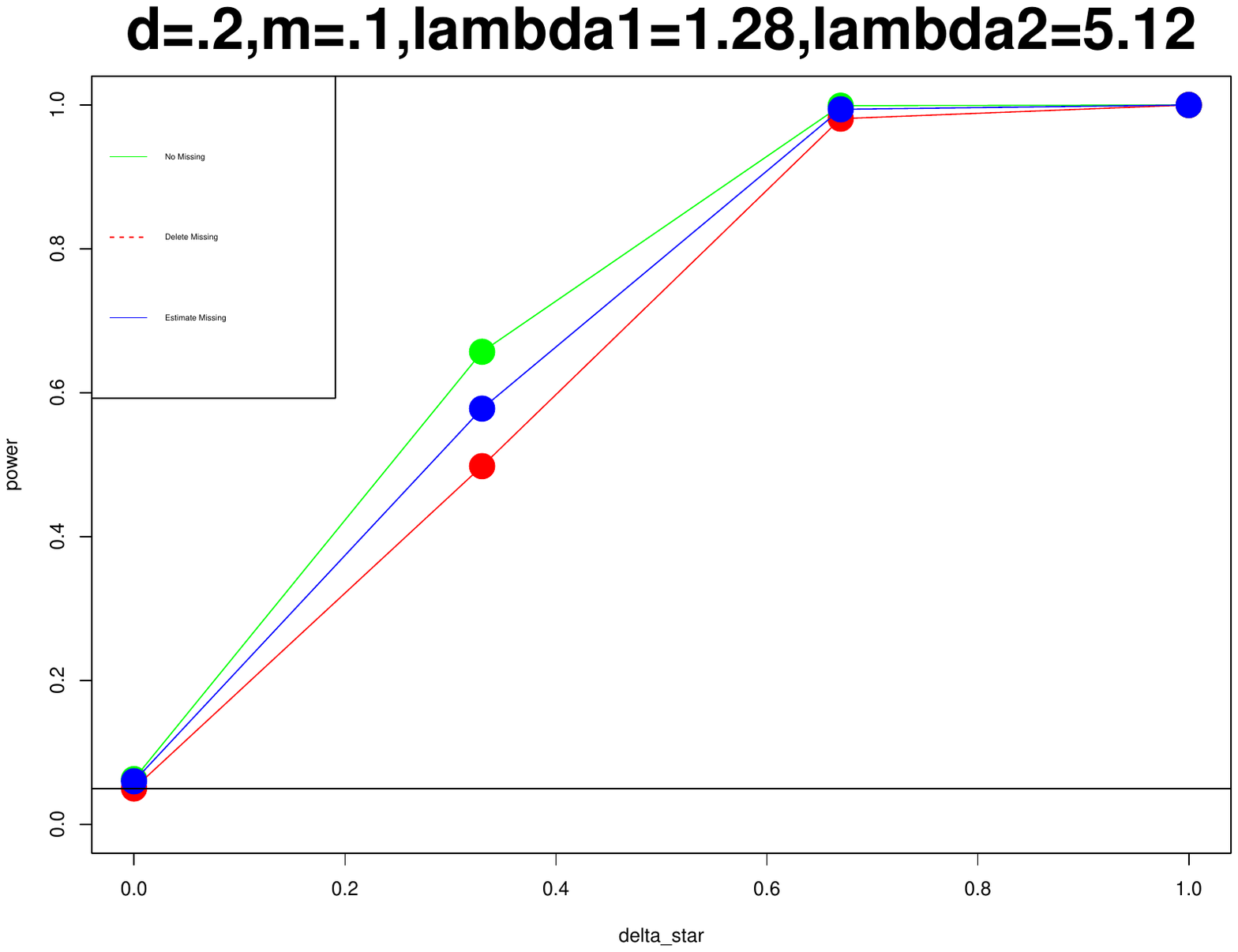}
\includegraphics[width = 2.3in, height = 1.3in]{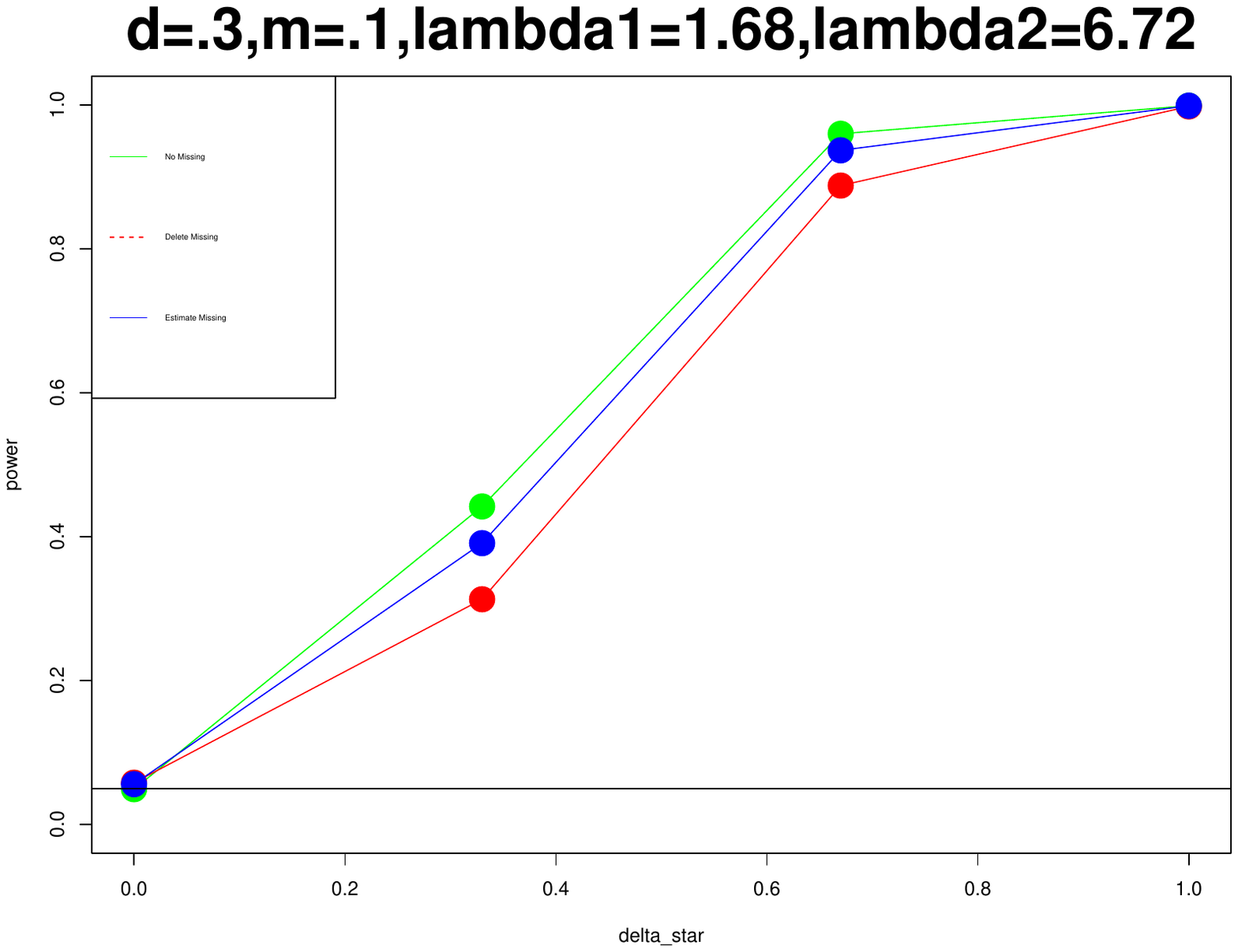}

\subsubsection{When one Trait has Normal Distribution and other Trait has Chi Squares Distribution}

\hspace{1.5cm}
$d=.1 \quad m=.5$
\hspace{3cm}
$d=.2 \quad m=.5$
\hspace{3cm}
$d=.3 \quad m=.5$

\includegraphics[width = 2.3in, height = 1.4in]{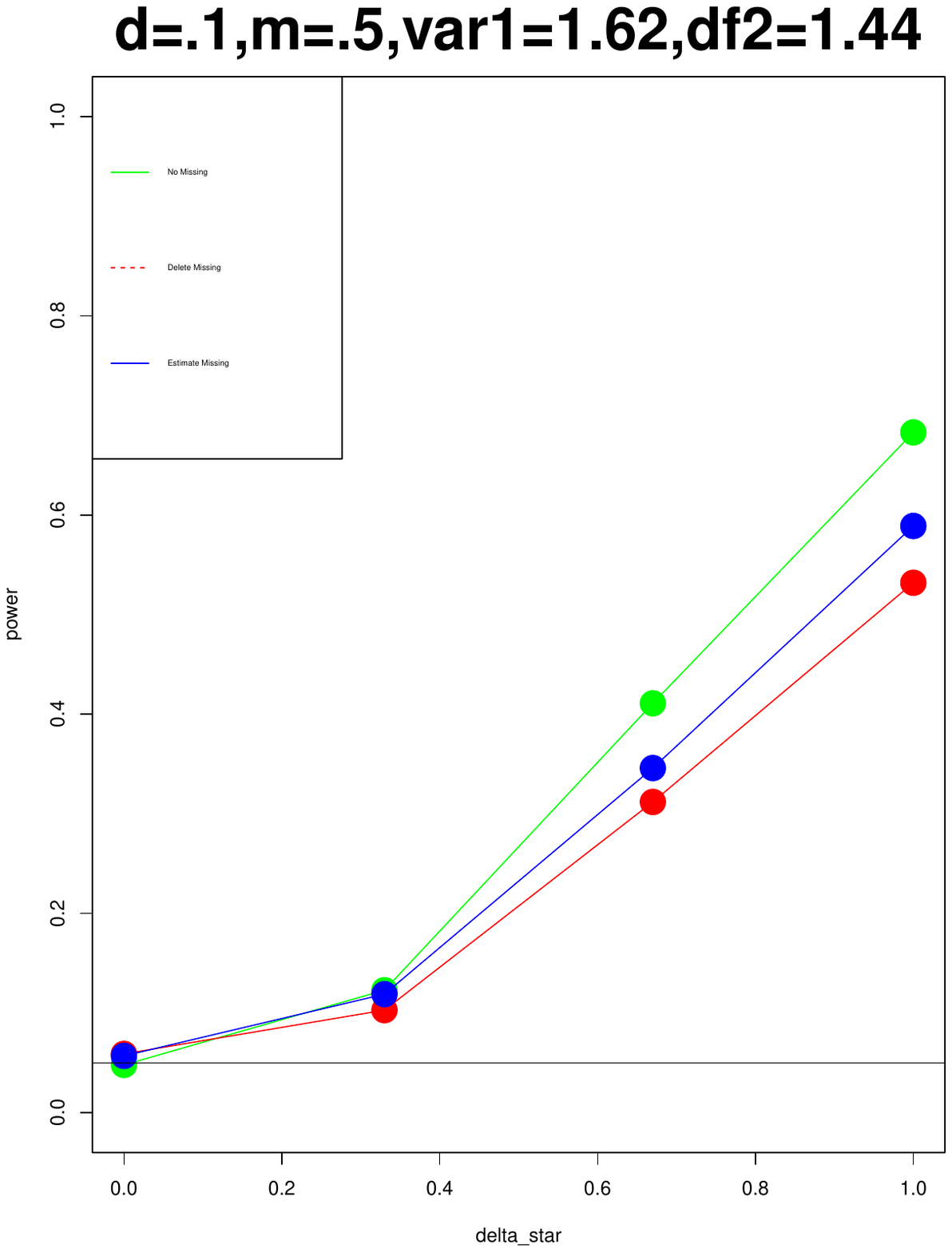}
\includegraphics[width = 2.3in, height = 1.4in]{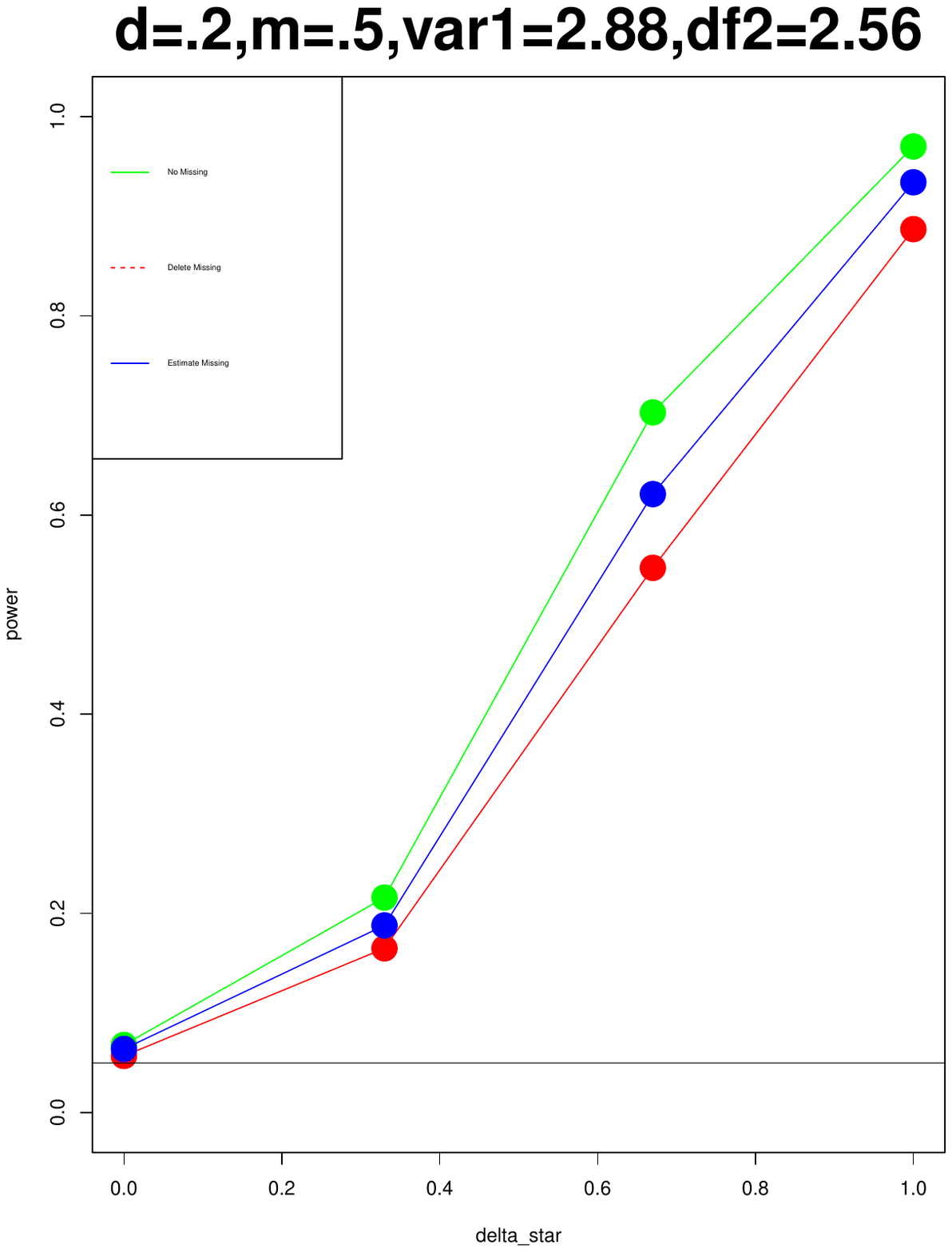}
\includegraphics[width = 2.3in, height = 1.4in]{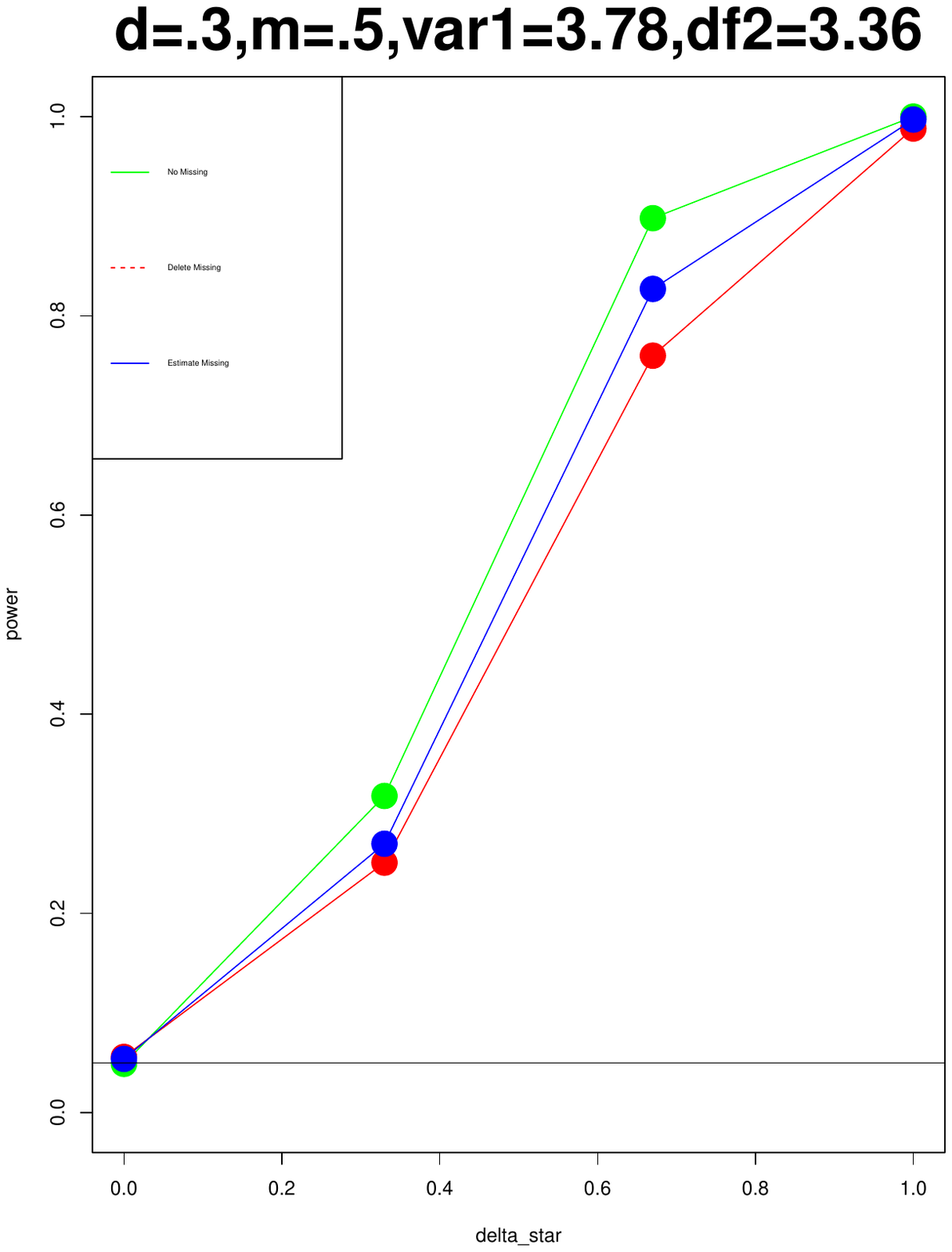}

\hspace{1.5cm}
$d=.1 \quad m=.1$
\hspace{3cm}
$d=.2 \quad m=.1$
\hspace{3cm}
$d=.3 \quad m=.1$

\includegraphics[width = 2.3in, height = 1.4in]{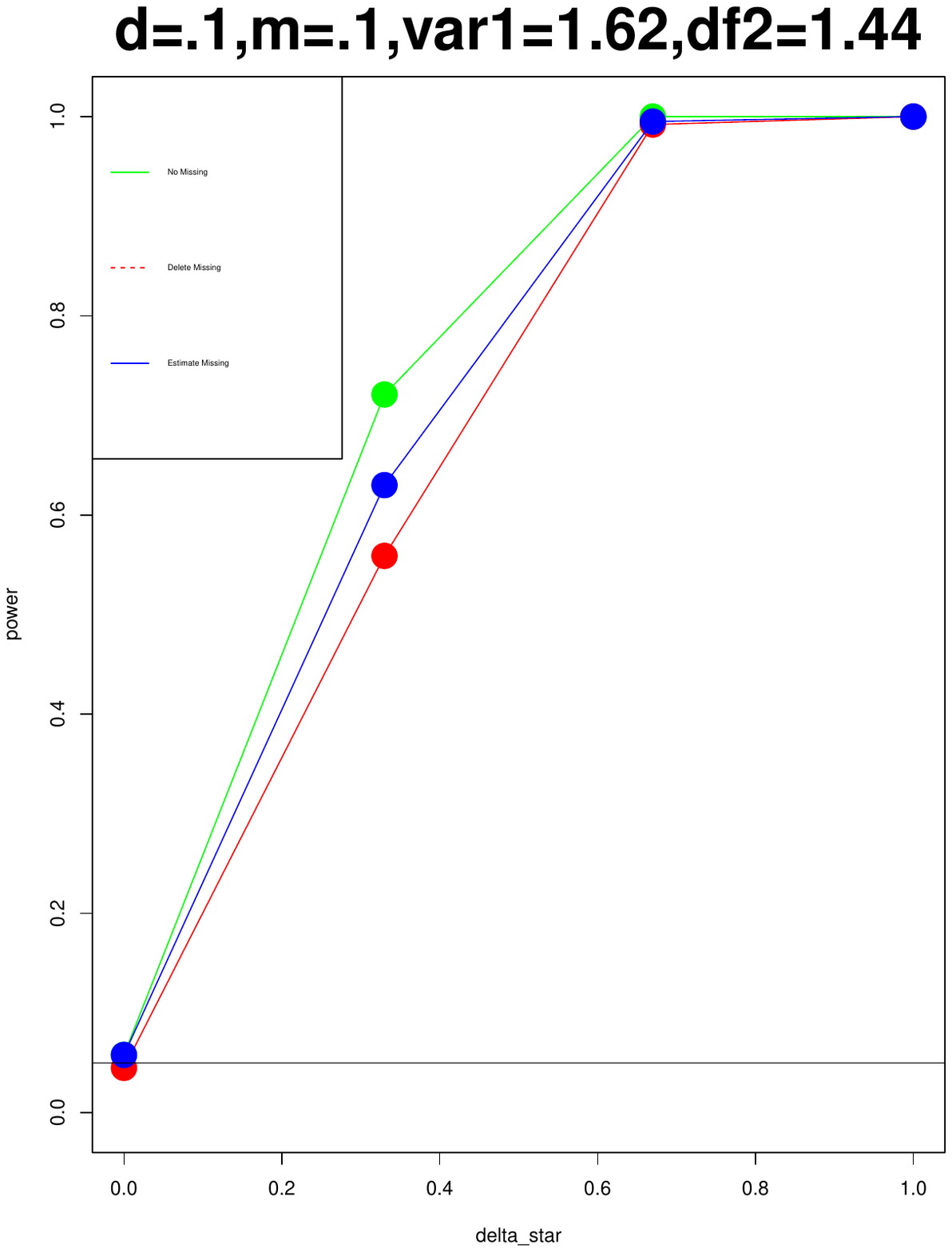}
\includegraphics[width = 2.3in, height = 1.4in]{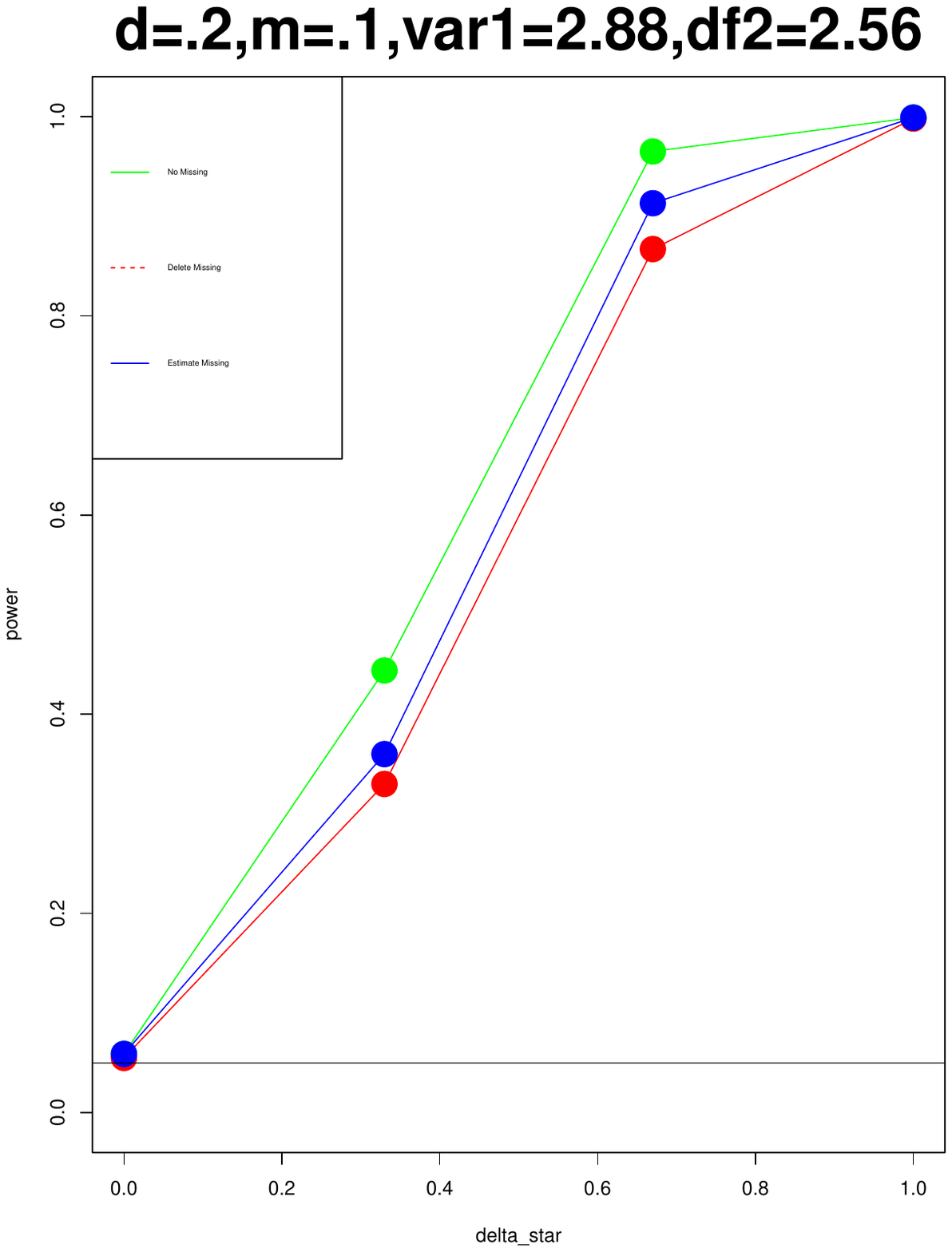}
\includegraphics[width = 2.3in, height = 1.4in]{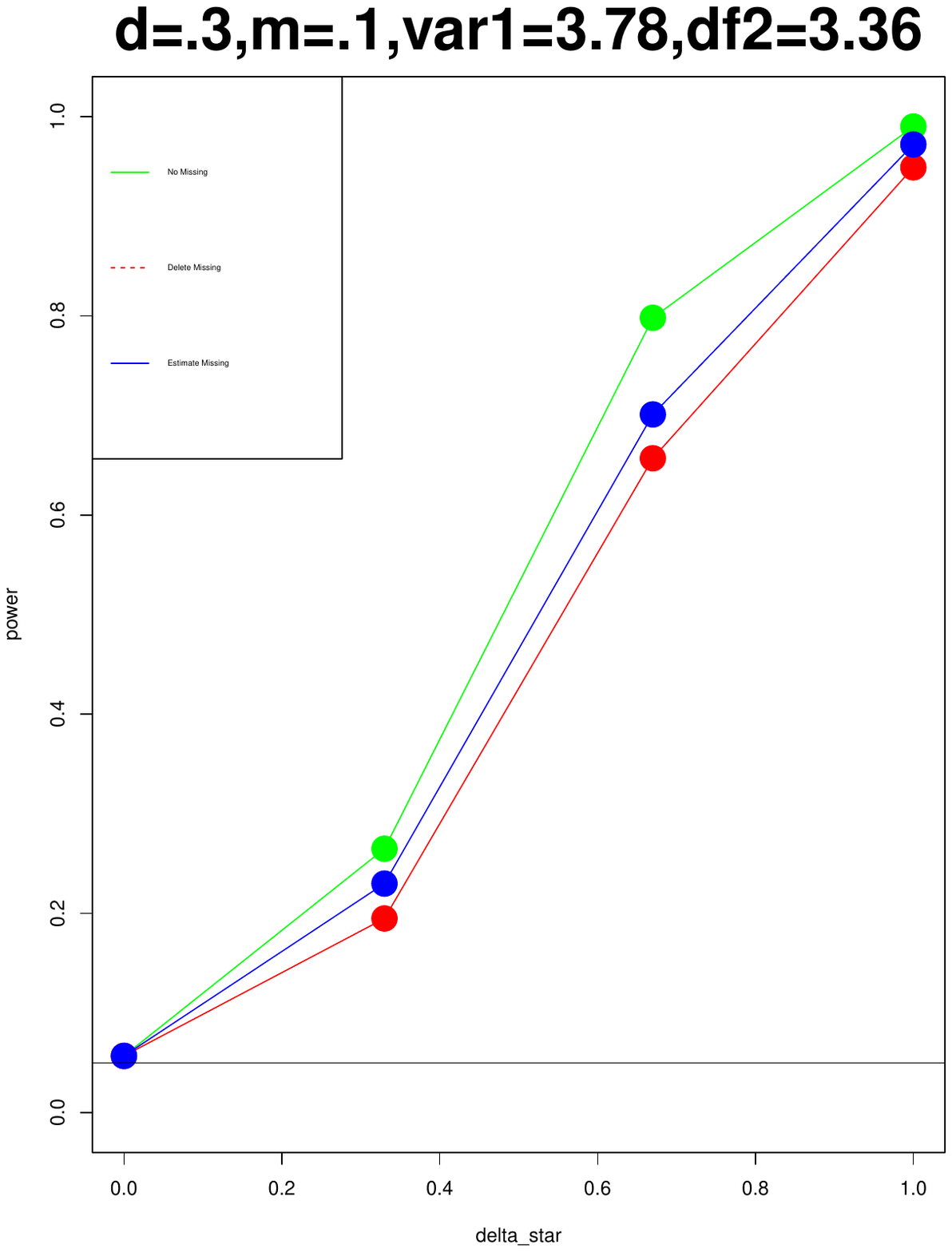}

\hspace{1.5cm}
$d=.1 \quad m=.1$
\hspace{3cm}
$d=.2 \quad m=.1$
\hspace{3cm}
$d=.3 \quad m=.1$

\includegraphics[width = 2.3in, height = 1.4in]{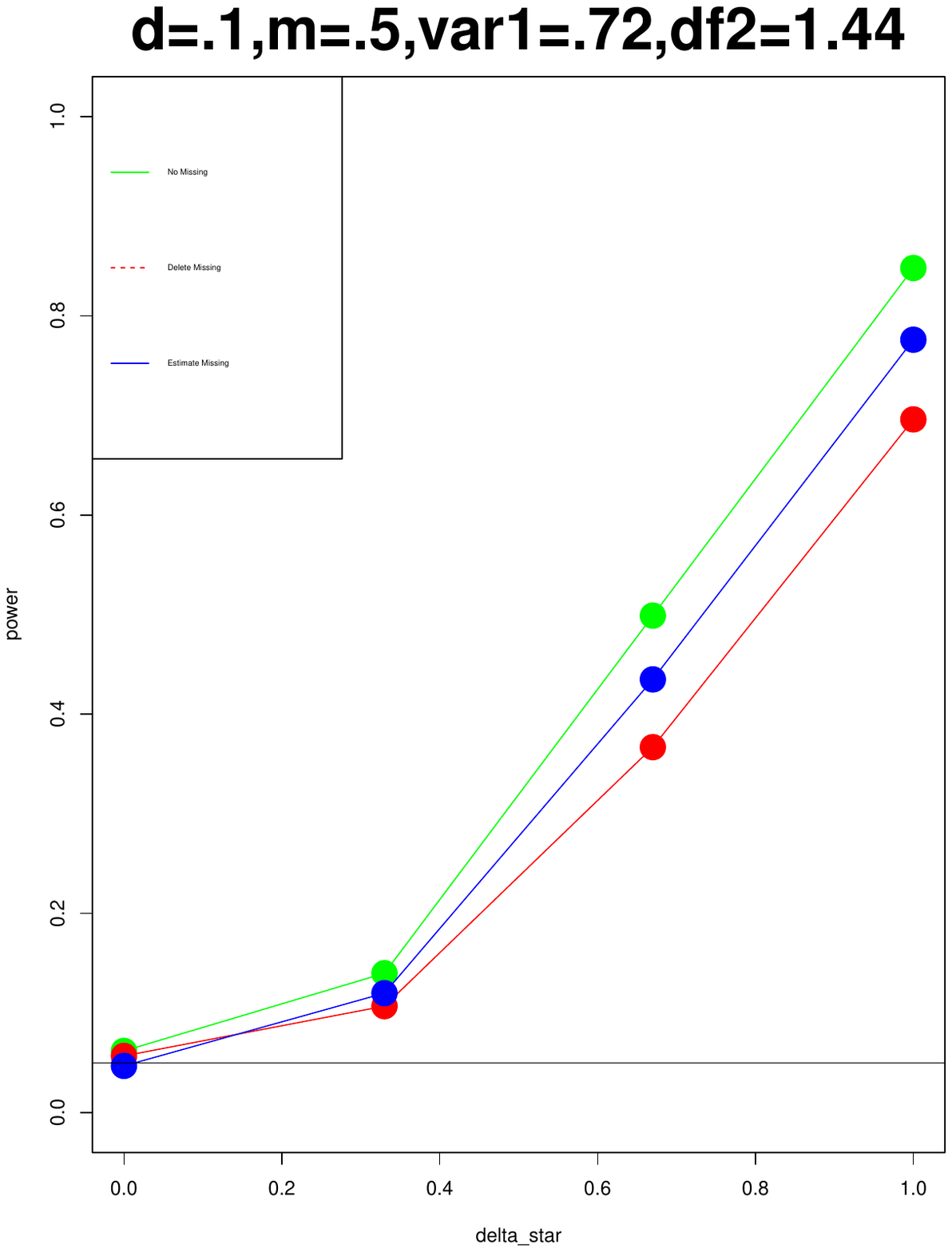}
\includegraphics[width = 2.3in, height = 1.4in]{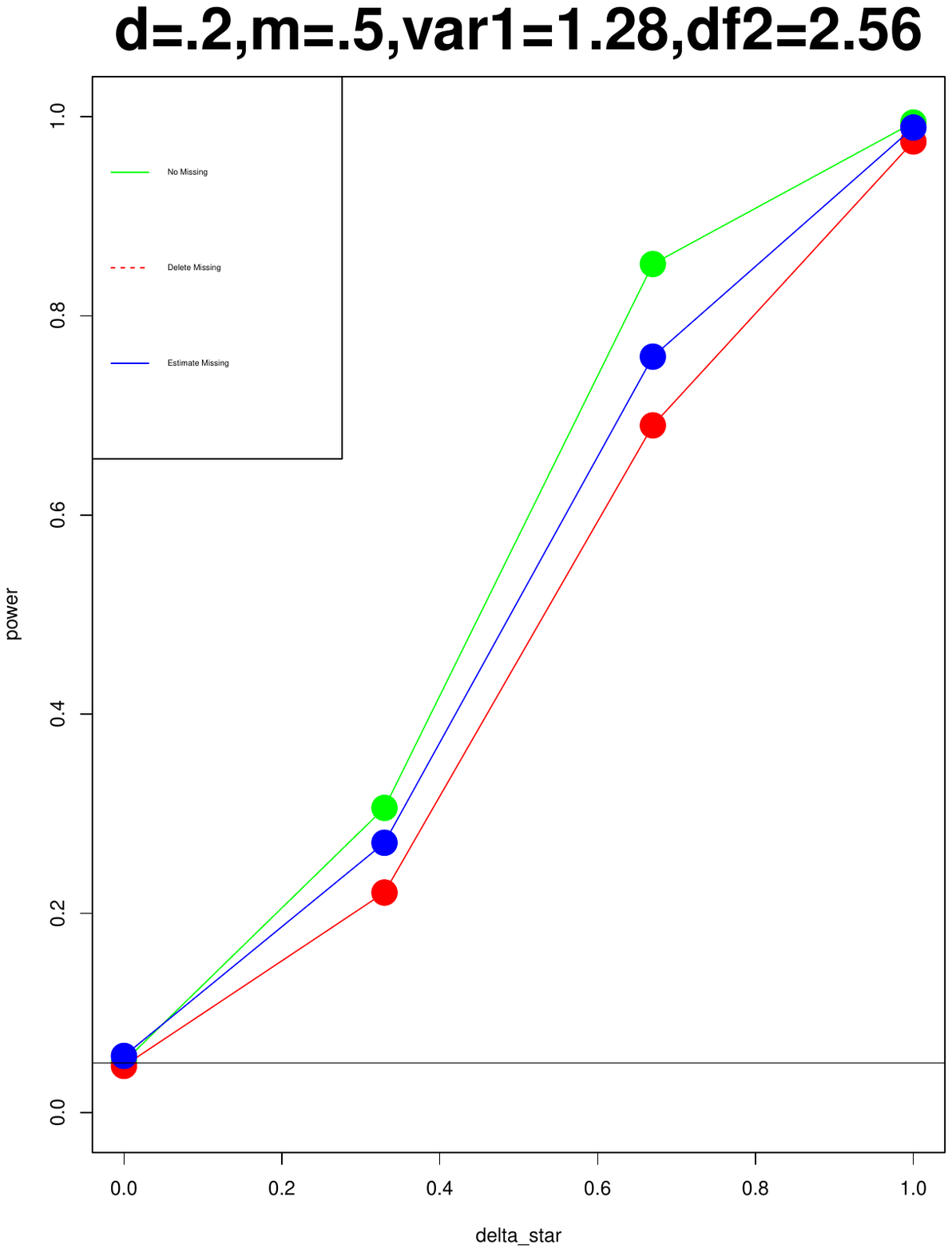}
\includegraphics[width = 2.3in, height = 1.4in]{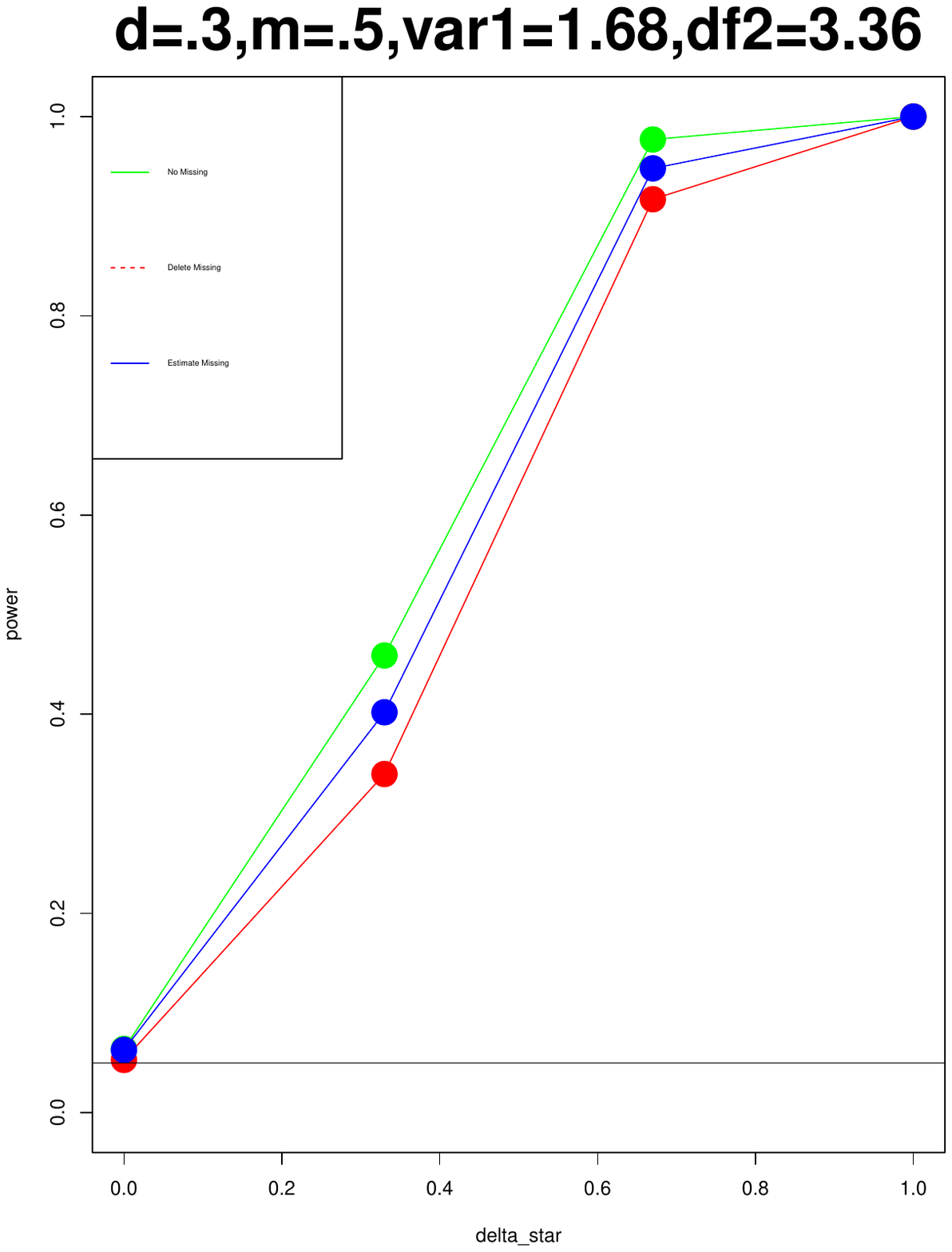}

\hspace{1.5cm}
$d=.1 \quad m=.5$
\hspace{3cm}
$d=.2 \quad m=.5$
\hspace{3cm}
$d=.3 \quad m=.5$

\includegraphics[width = 2.3in, height = 1.4in]{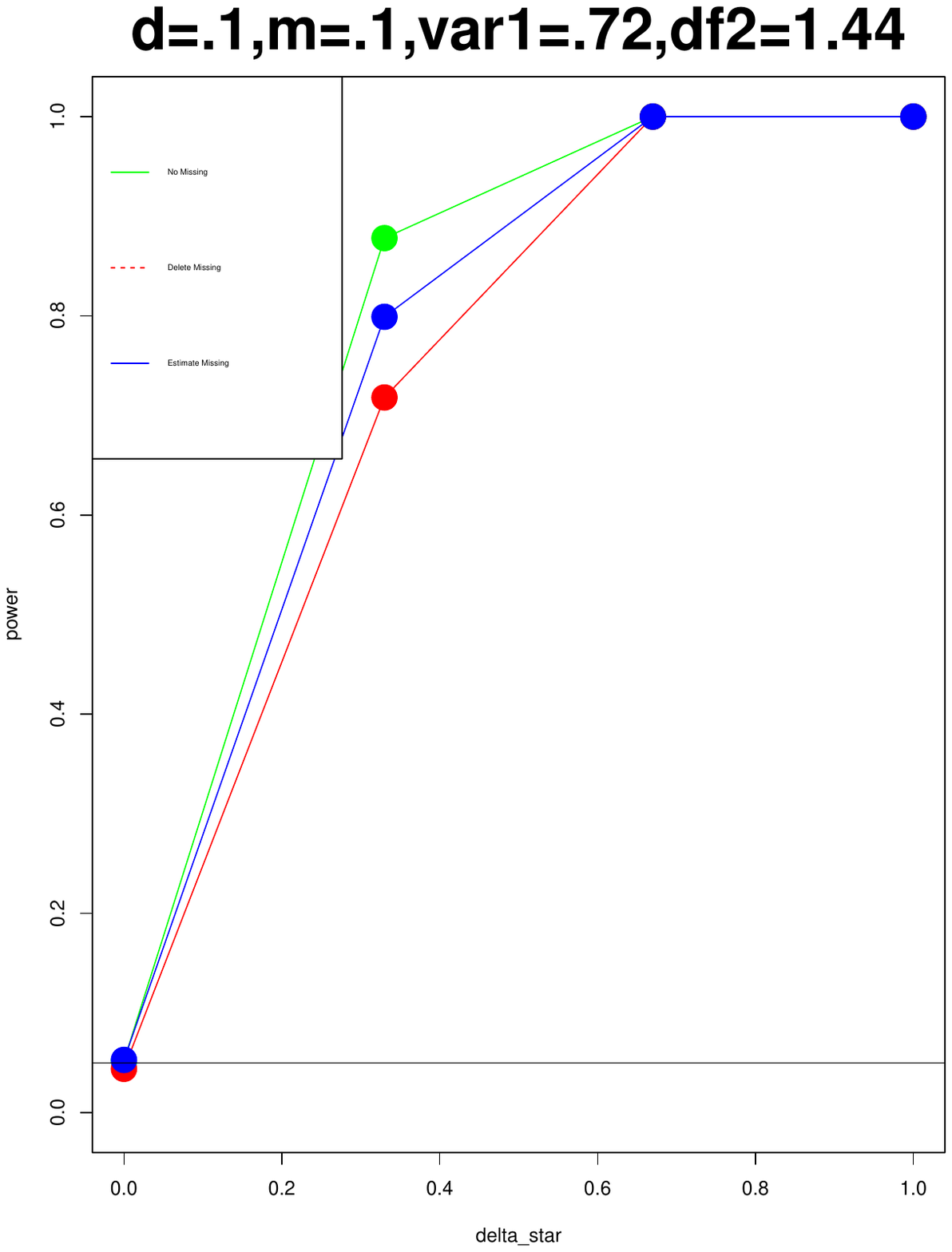}
\includegraphics[width = 2.3in, height = 1.4in]{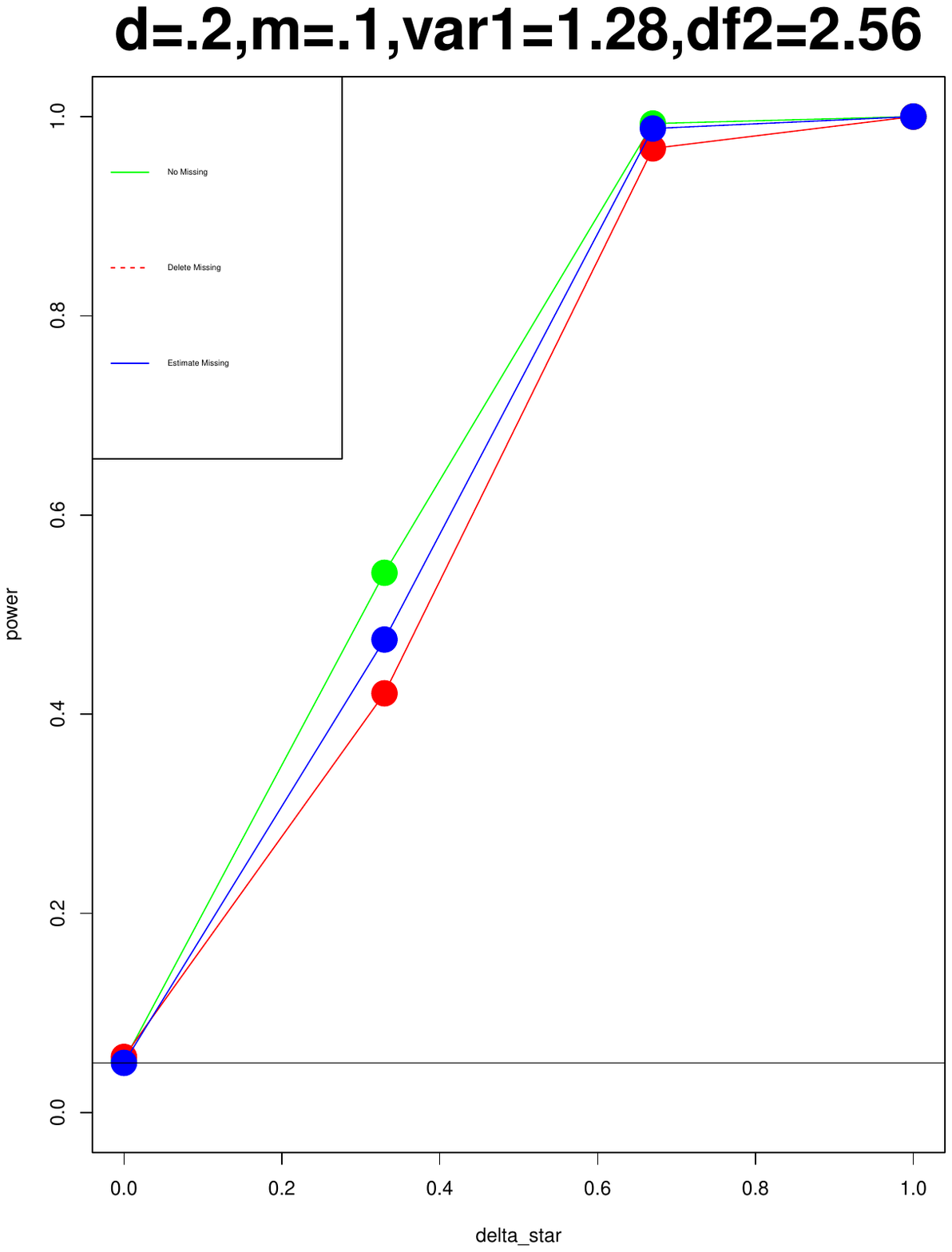}
\includegraphics[width = 2.3in, height = 1.4in]{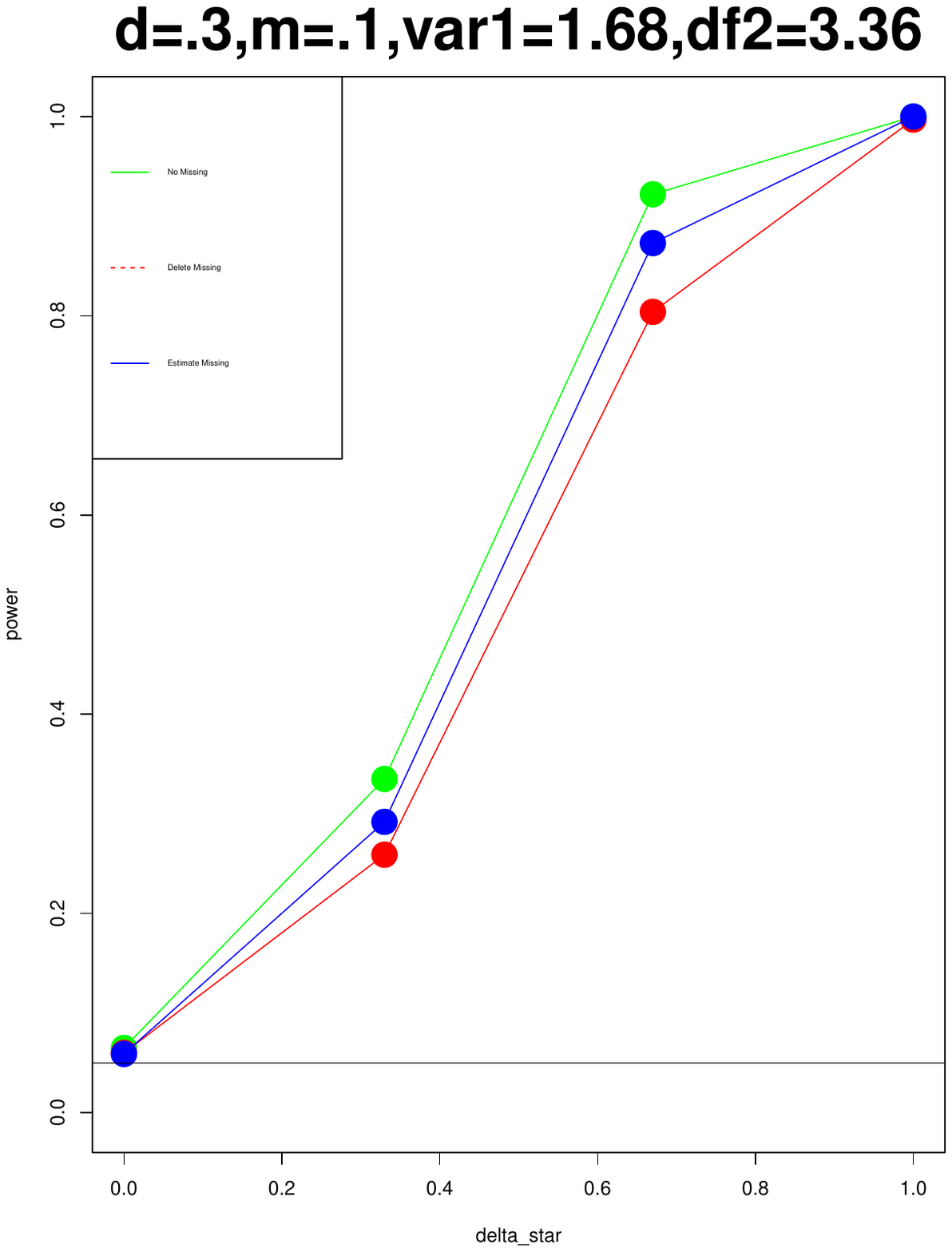}

\hspace{1.5cm}
$d=.1 \quad m=.1$
\hspace{3cm}
$d=.2 \quad m=.1$
\hspace{3cm}
$d=.3 \quad m=.1$

\includegraphics[width = 2.3in, height = 1.4in]{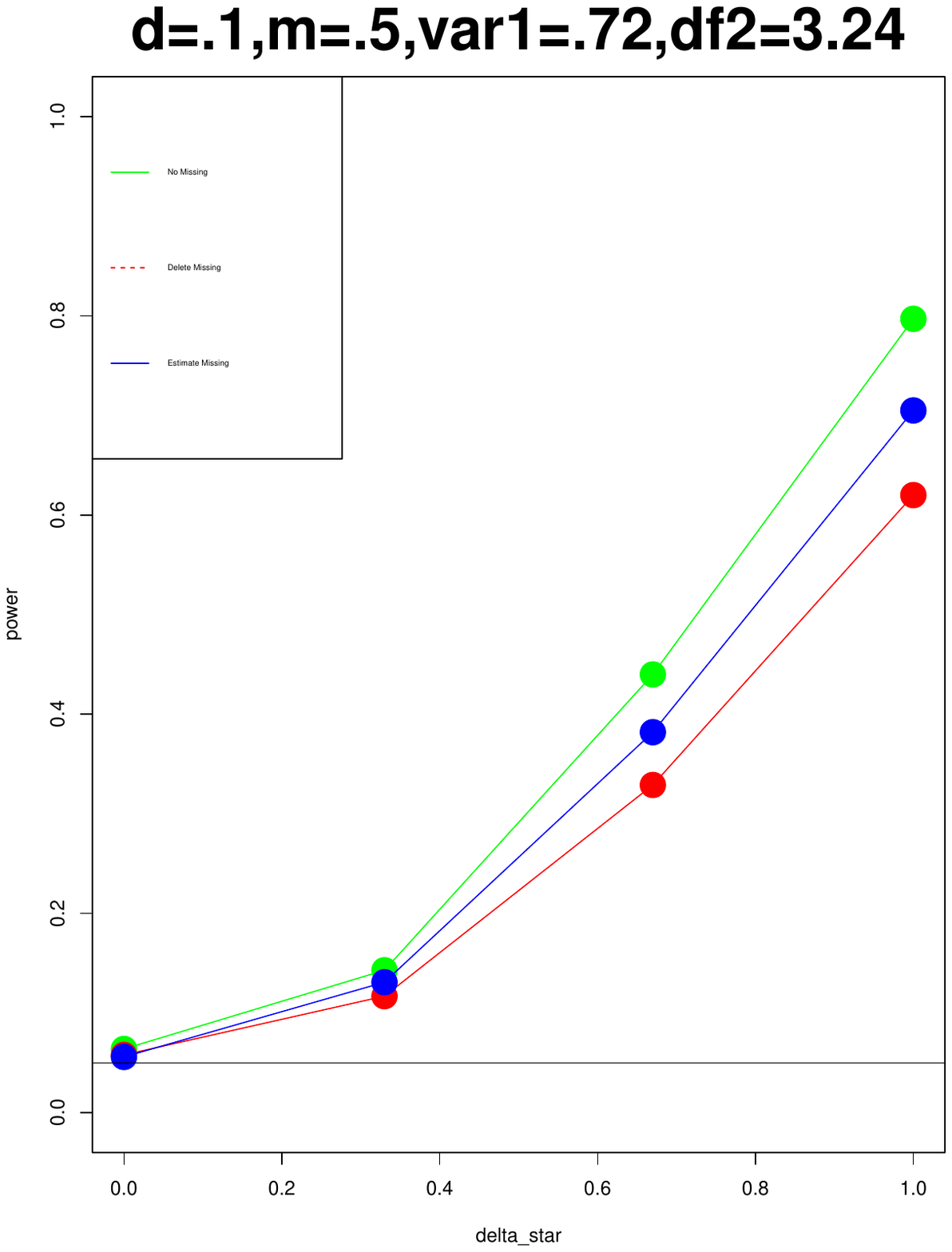}
\includegraphics[width = 2.3in, height = 1.4in]{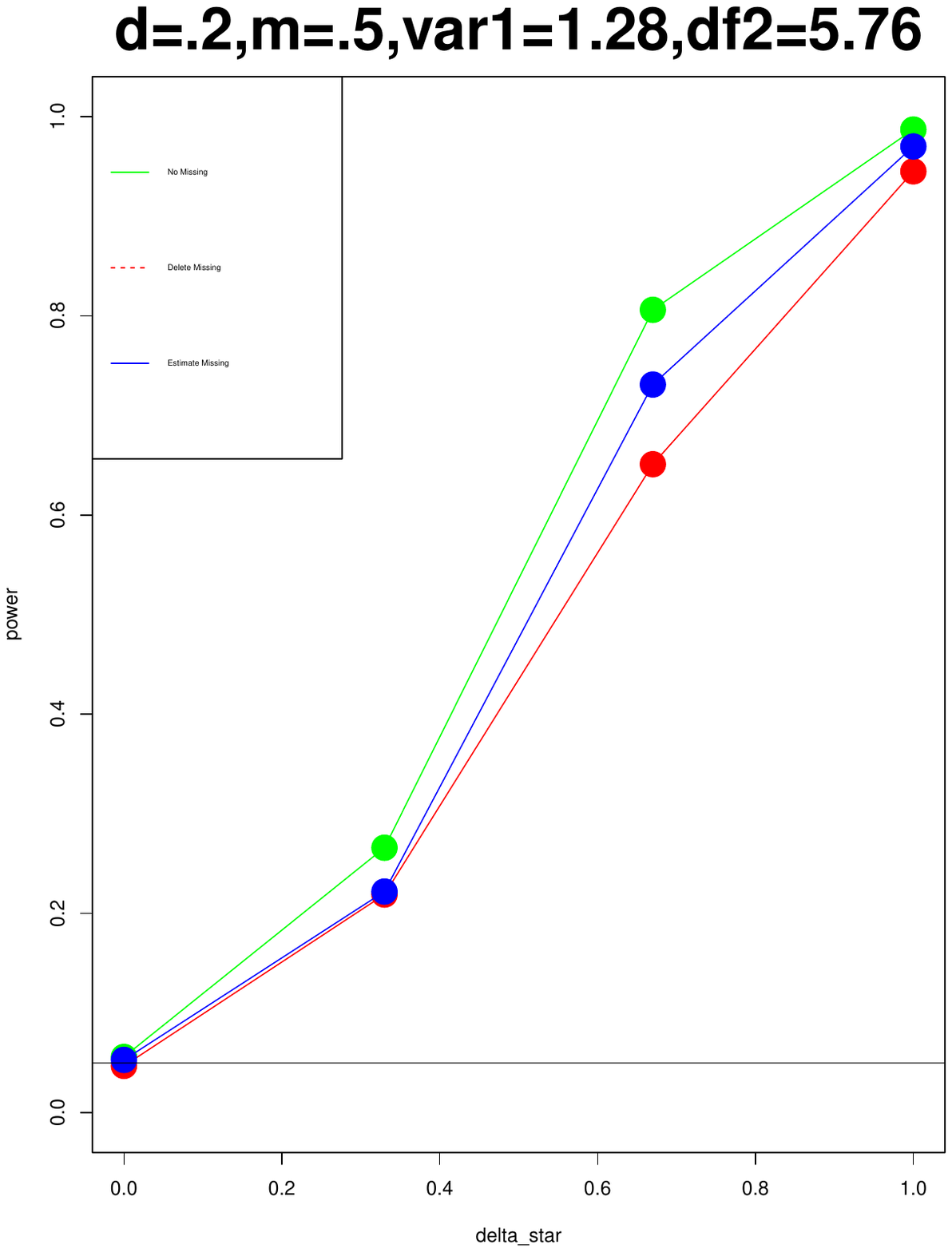}
\includegraphics[width = 2.3in, height = 1.4in]{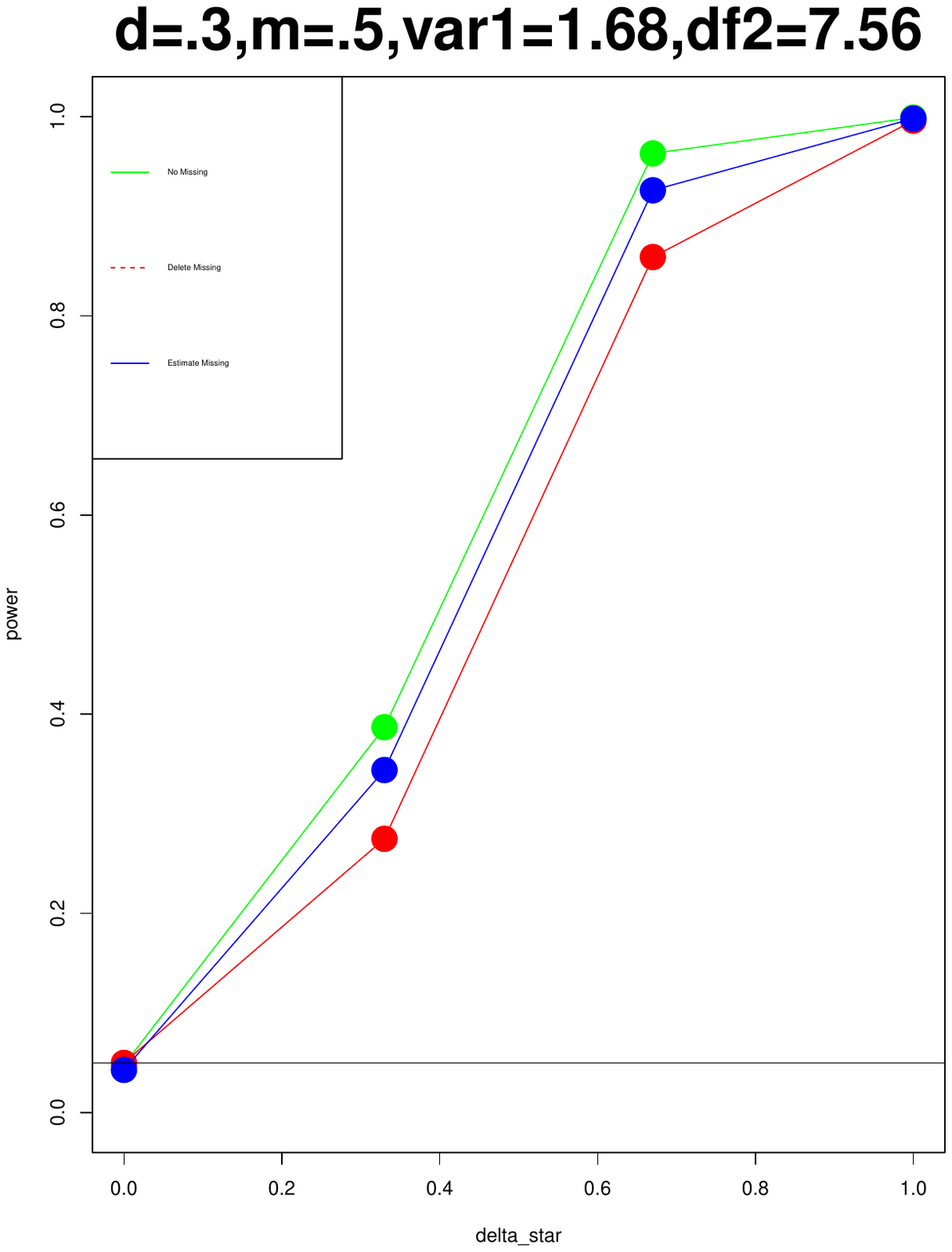}

\hspace{1.5cm}
$d=.1 \quad m=.5$
\hspace{3cm}
$d=.2 \quad m=.5$
\hspace{3cm}
$d=.3 \quad m=.5$

\includegraphics[width = 2.3in, height = 1.4in]{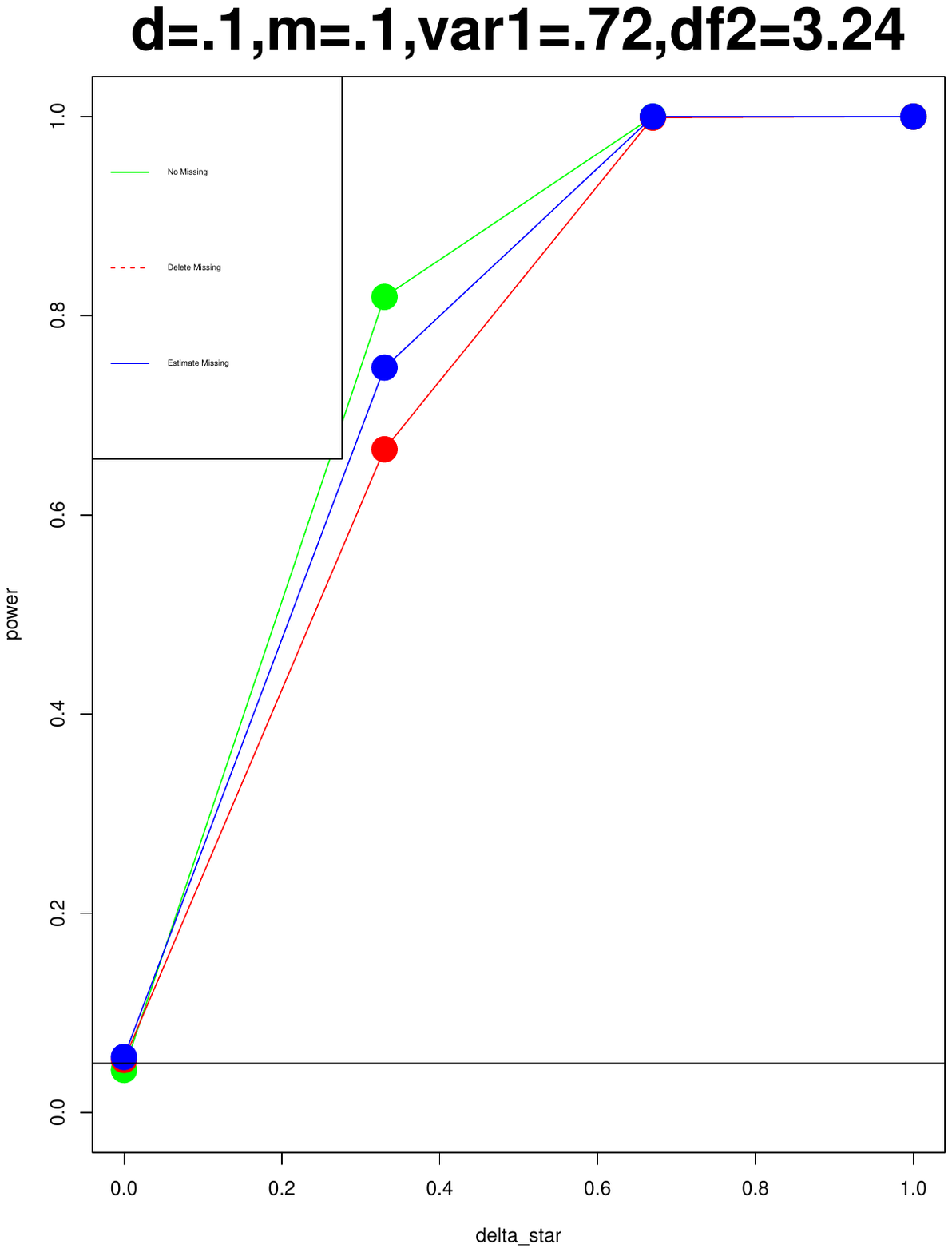}
\includegraphics[width = 2.3in, height = 1.4in]{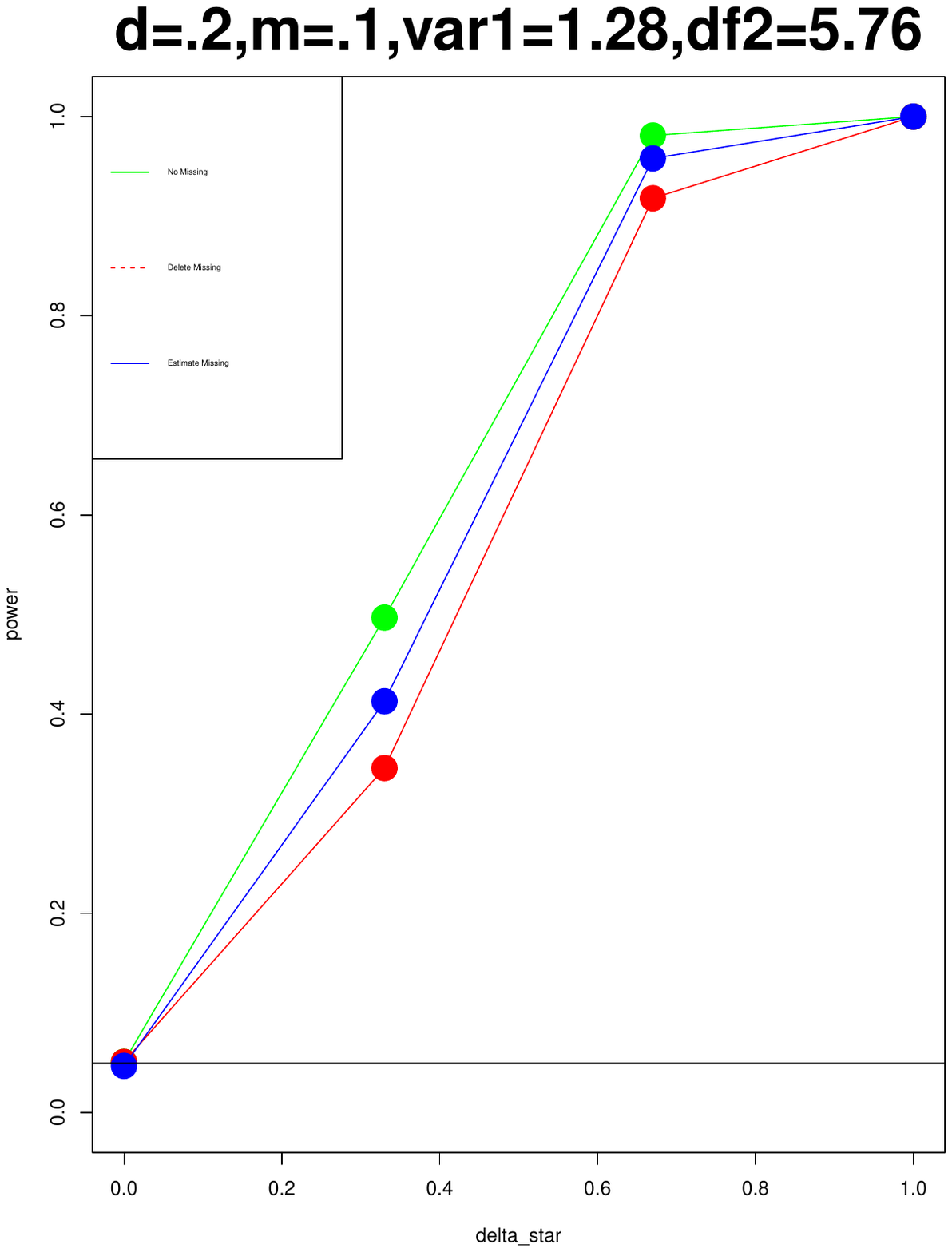}
\includegraphics[width = 2.3in, height = 1.4in]{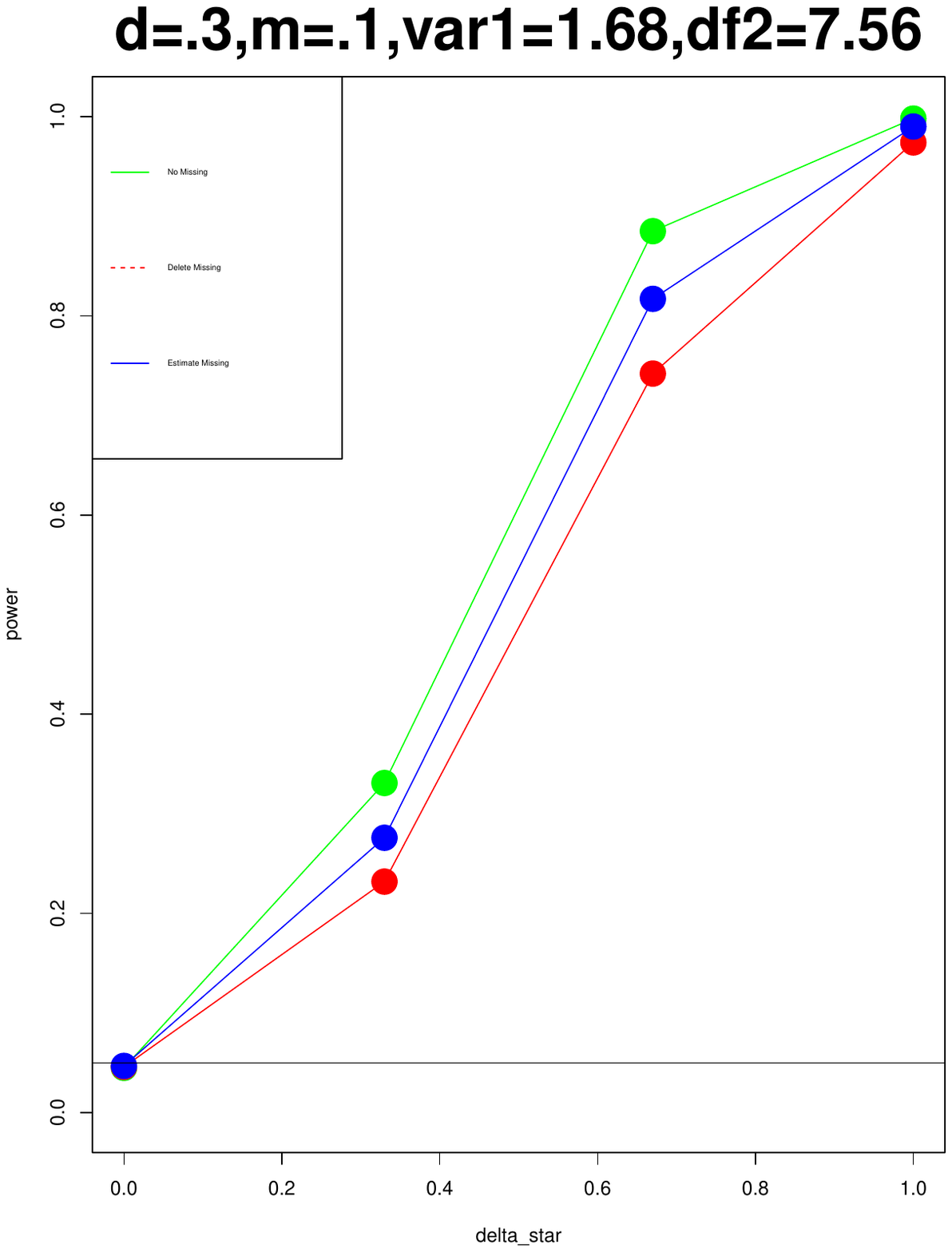}

\subsubsection{When one Trait has Poisson Distribution and other Trait has Chi Squares Distribution}

\hspace{1.5cm}
$d=.1 \quad m=.5$
\hspace{3cm}
$d=.2 \quad m=.5$
\hspace{3cm}
$d=.3 \quad m=.5$

\includegraphics[width = 2.3in, height = 1.4in]{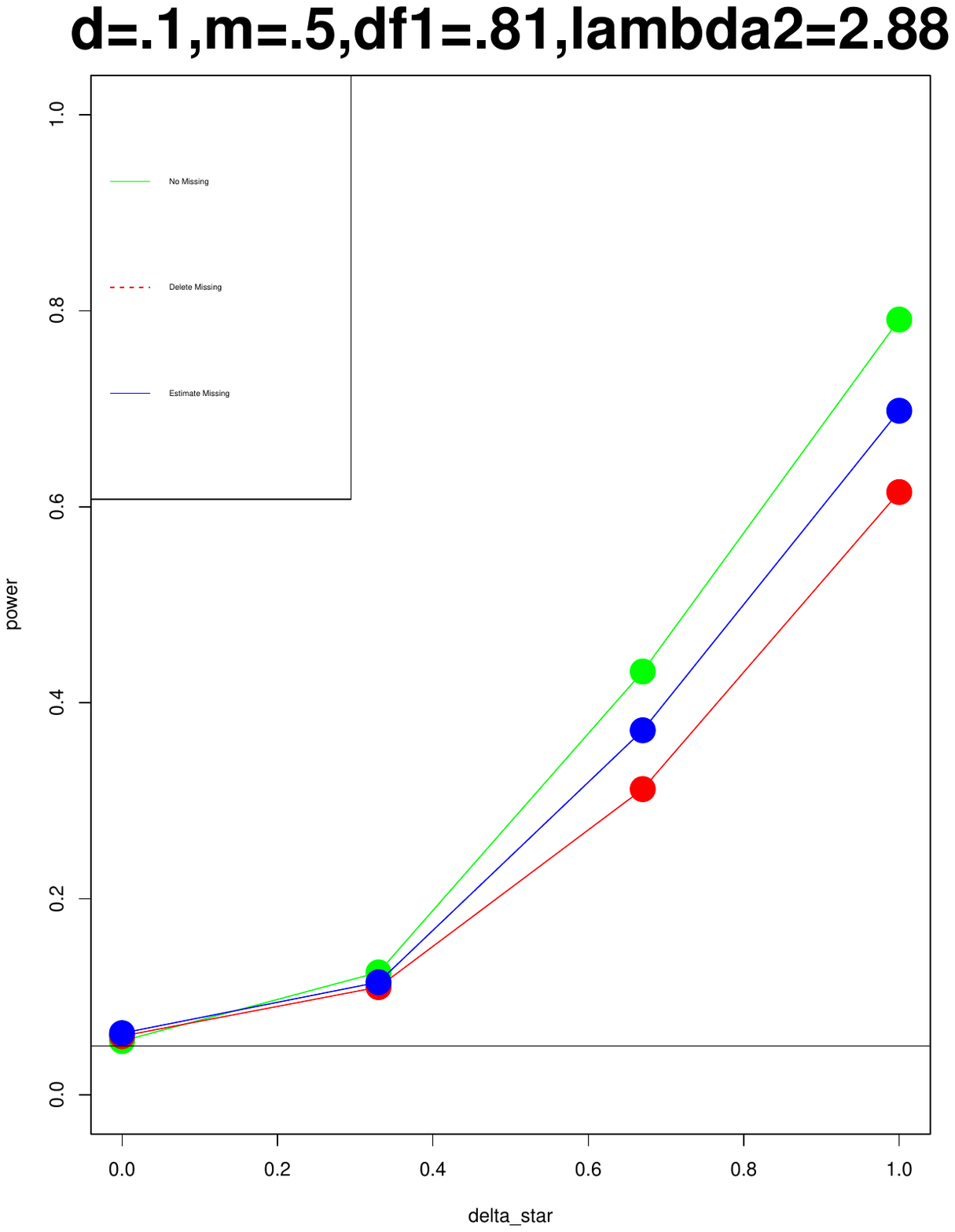}
\includegraphics[width = 2.3in, height = 1.4in]{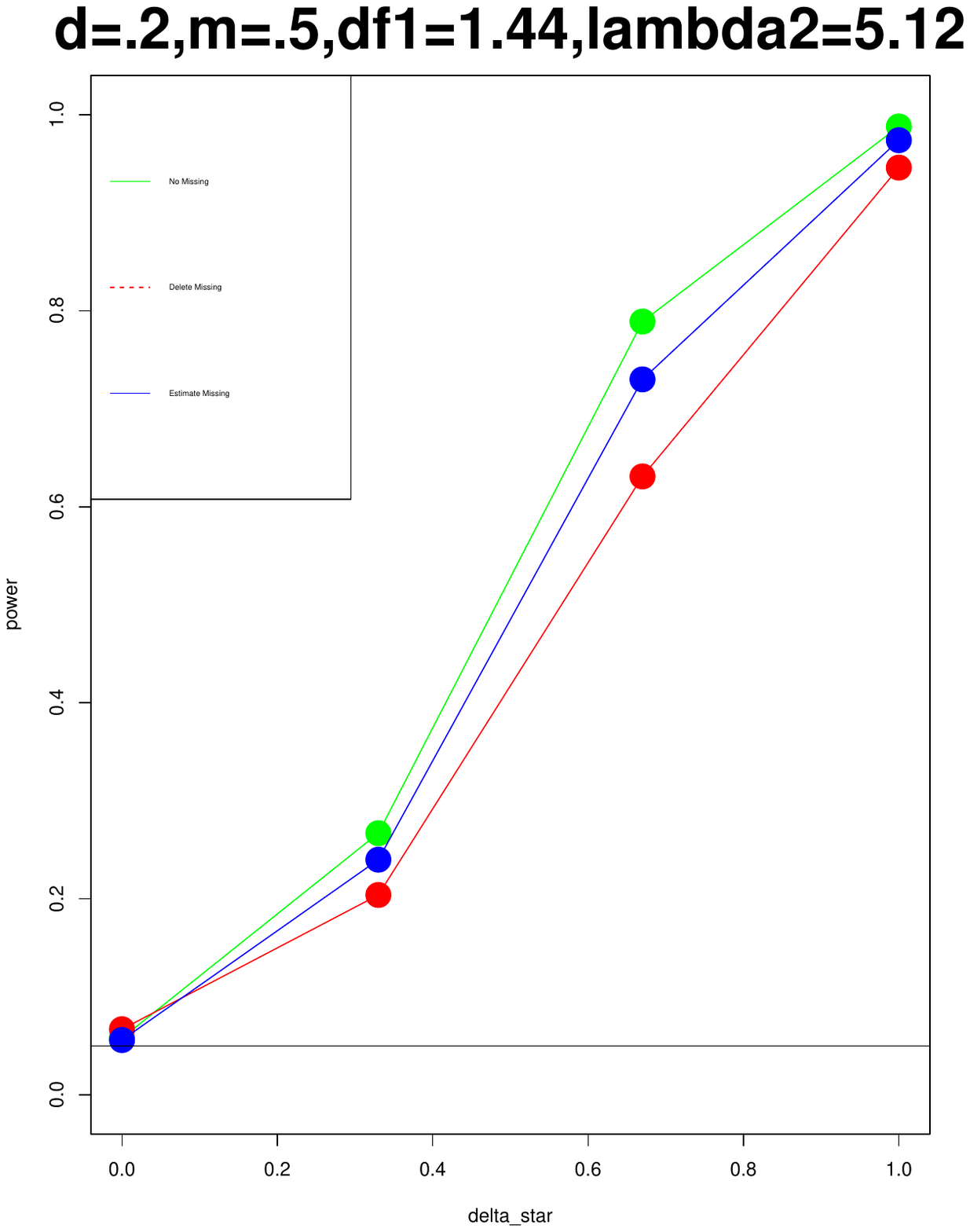}
\includegraphics[width = 2.3in, height = 1.4in]{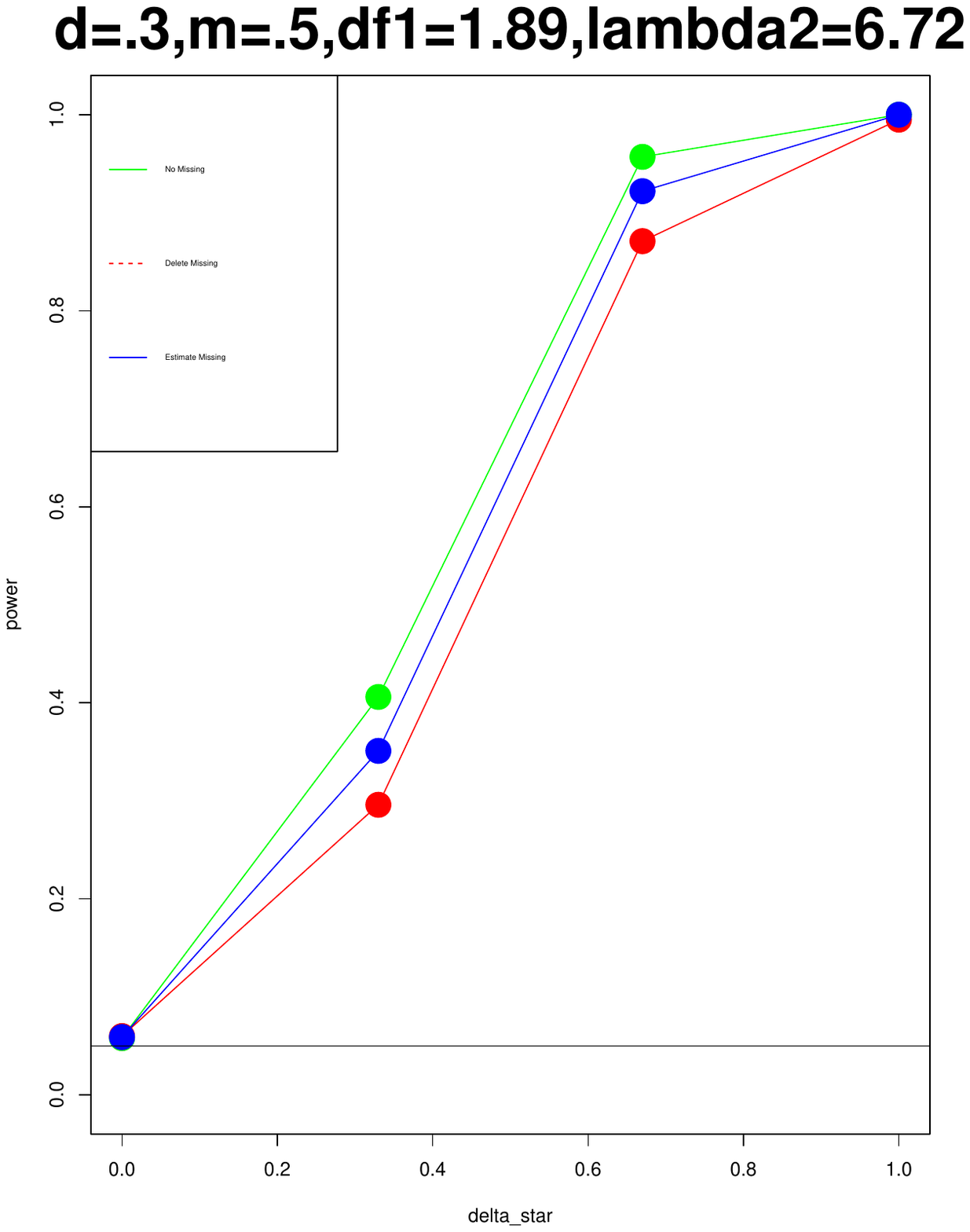}

\hspace{1.5cm}
$d=.1 \quad m=.1$
\hspace{3cm}
$d=.2 \quad m=.1$
\hspace{3cm}
$d=.3 \quad m=.1$

\includegraphics[width = 2.3in, height = 1.4in]{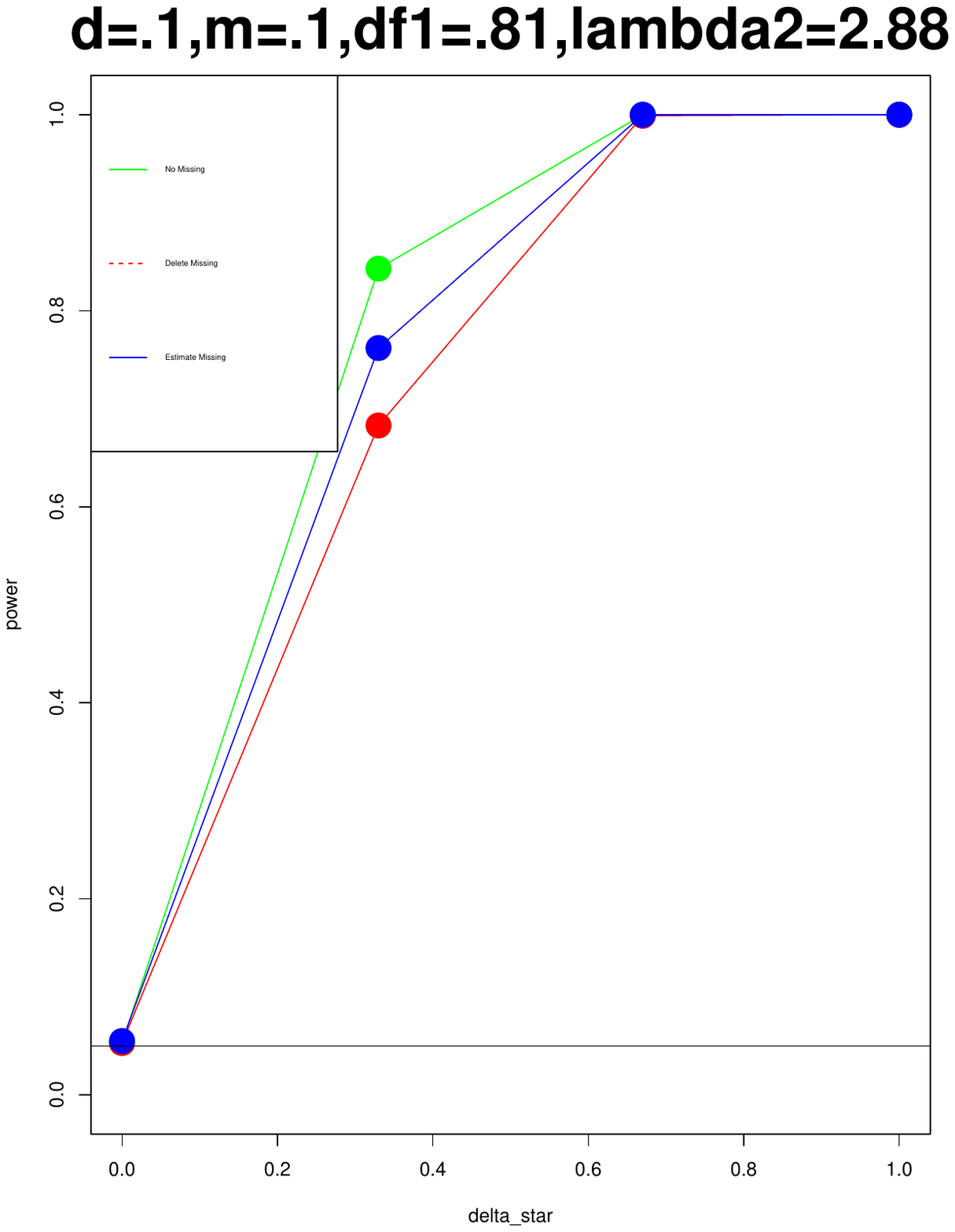}
\includegraphics[width = 2.3in, height = 1.4in]{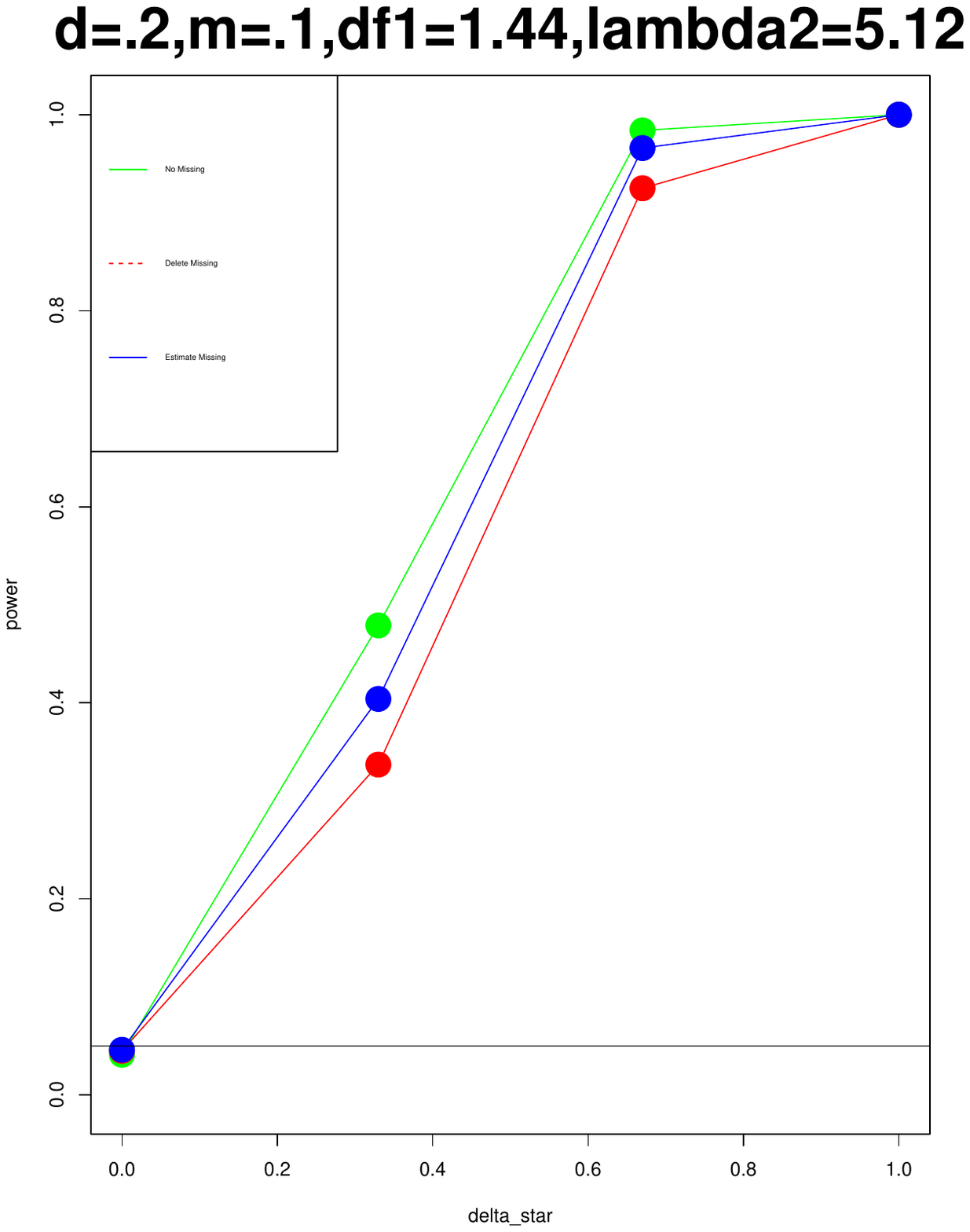}
\includegraphics[width = 2.3in, height = 1.4in]{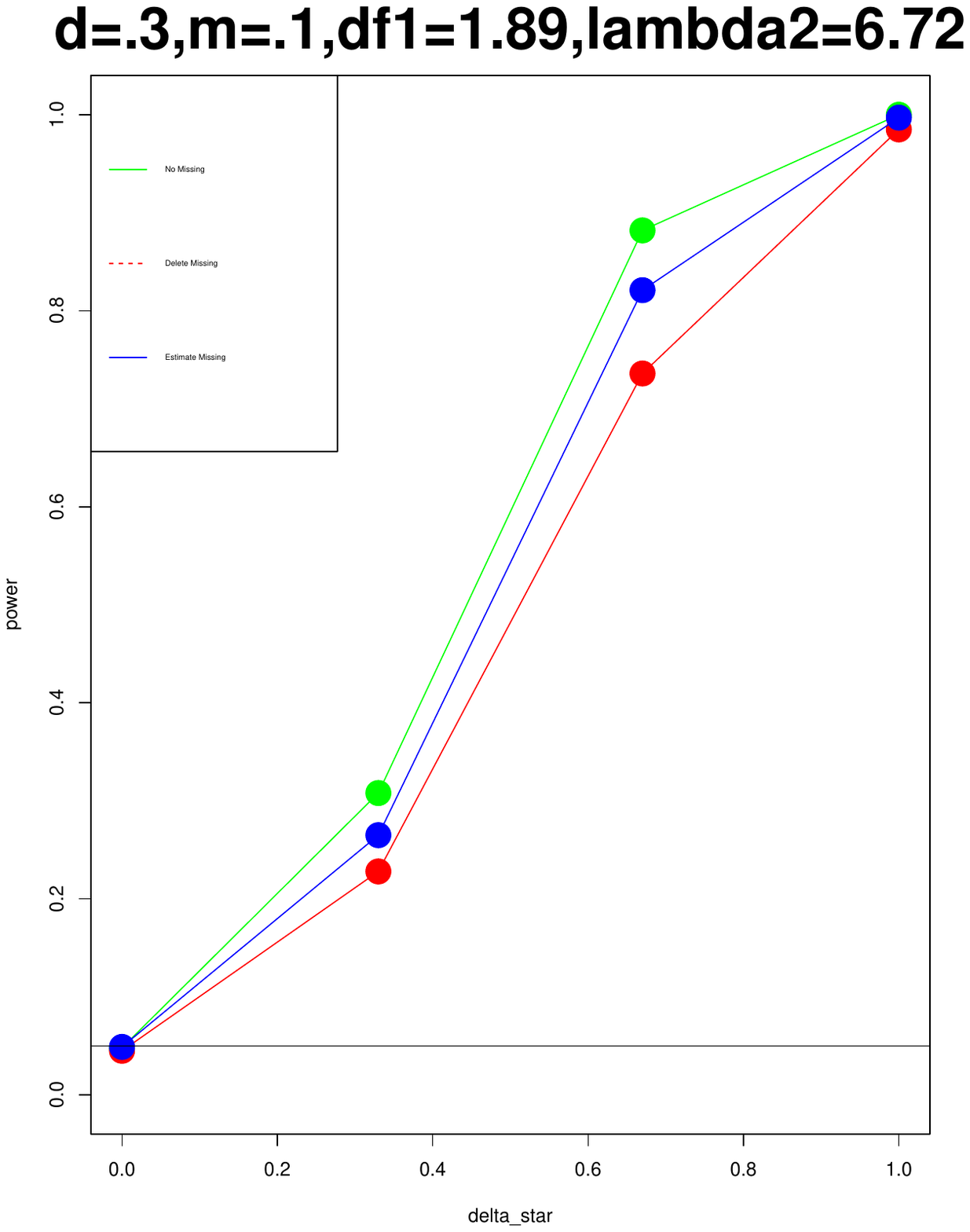}

\hspace{1.5cm}
$d=.1 \quad m=.1$
\hspace{3cm}
$d=.2 \quad m=.1$
\hspace{3cm}
$d=.3 \quad m=.1$

\includegraphics[width = 2.3in, height = 1.4in]{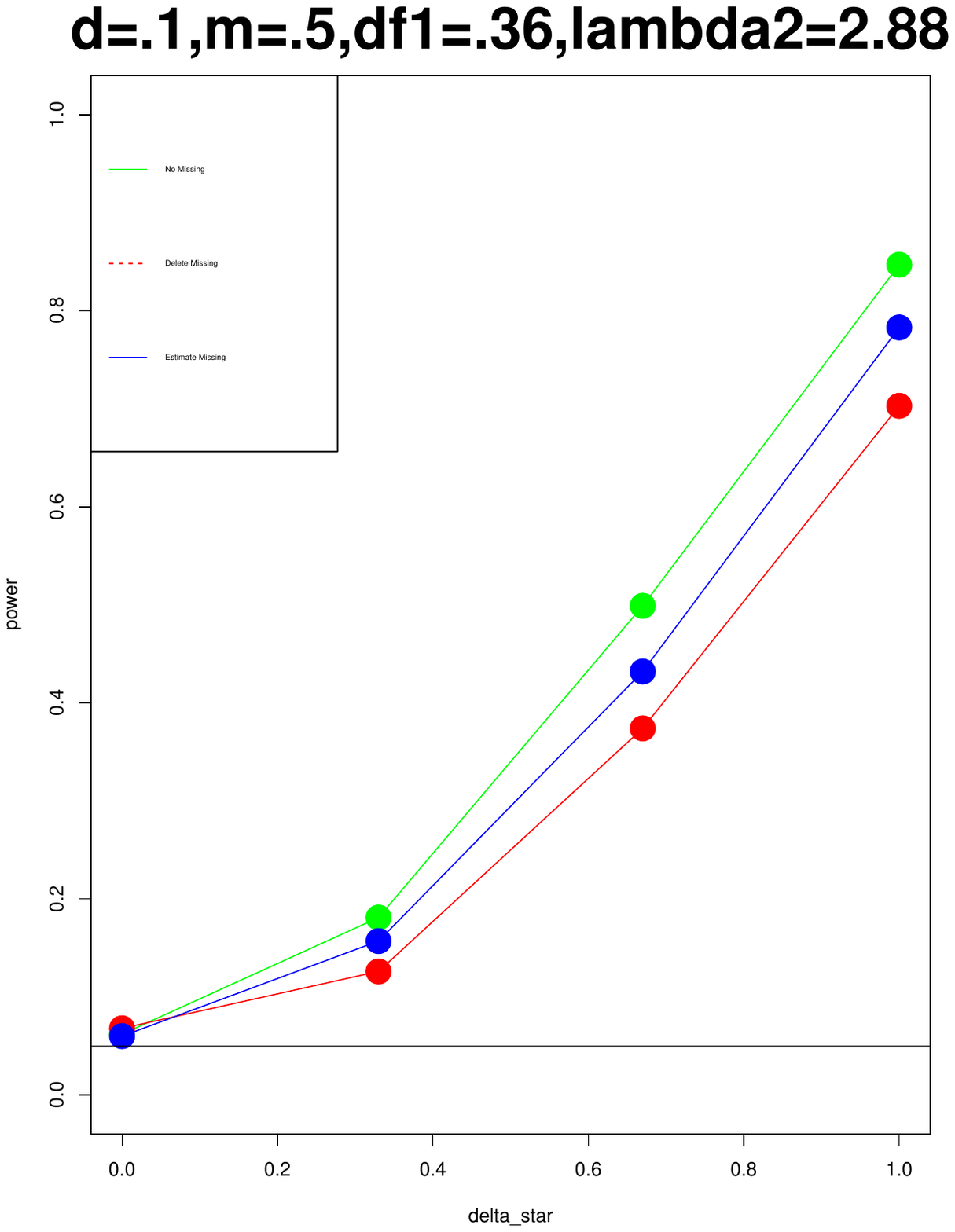}
\includegraphics[width = 2.3in, height = 1.4in]{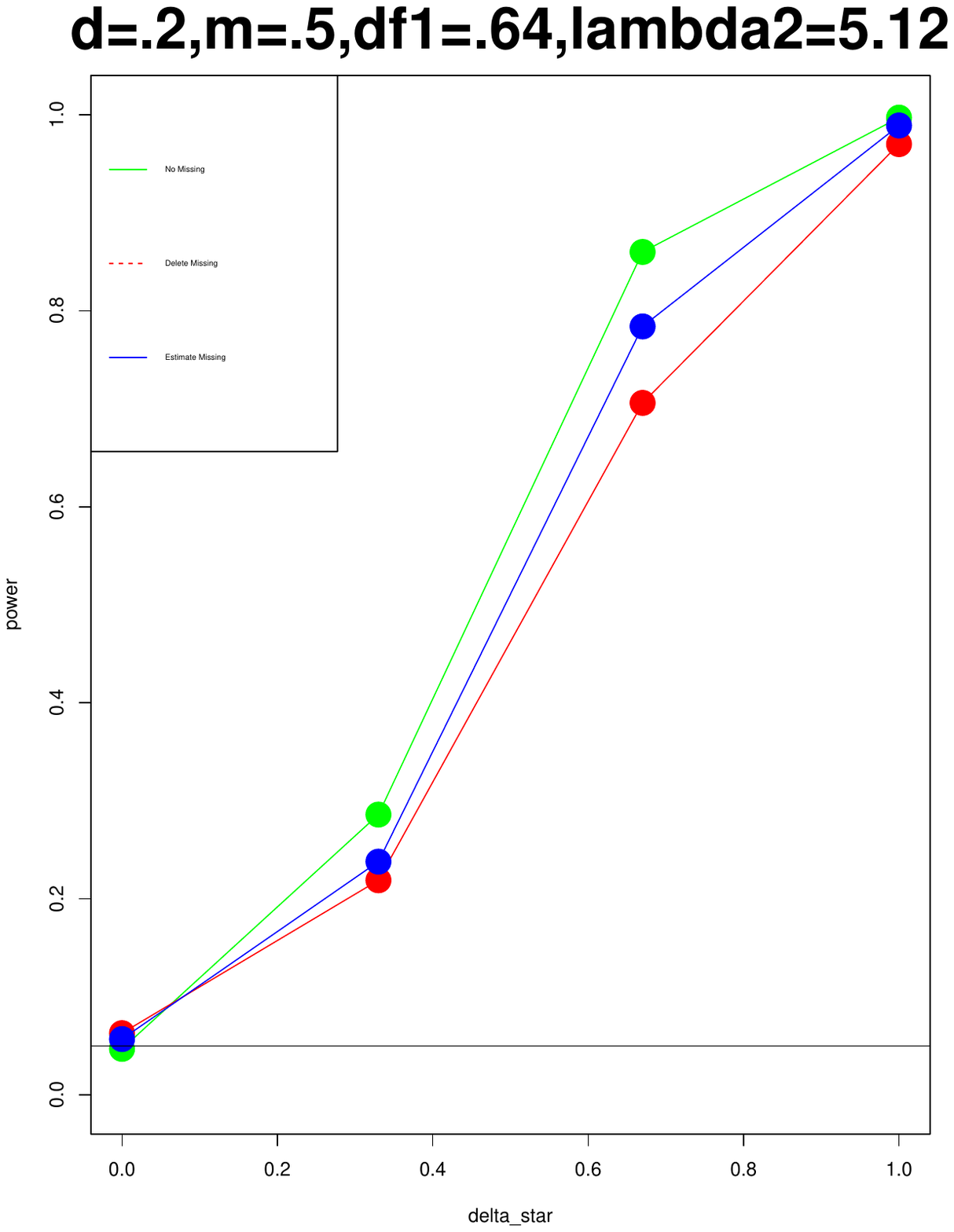}
\includegraphics[width = 2.3in, height = 1.4in]{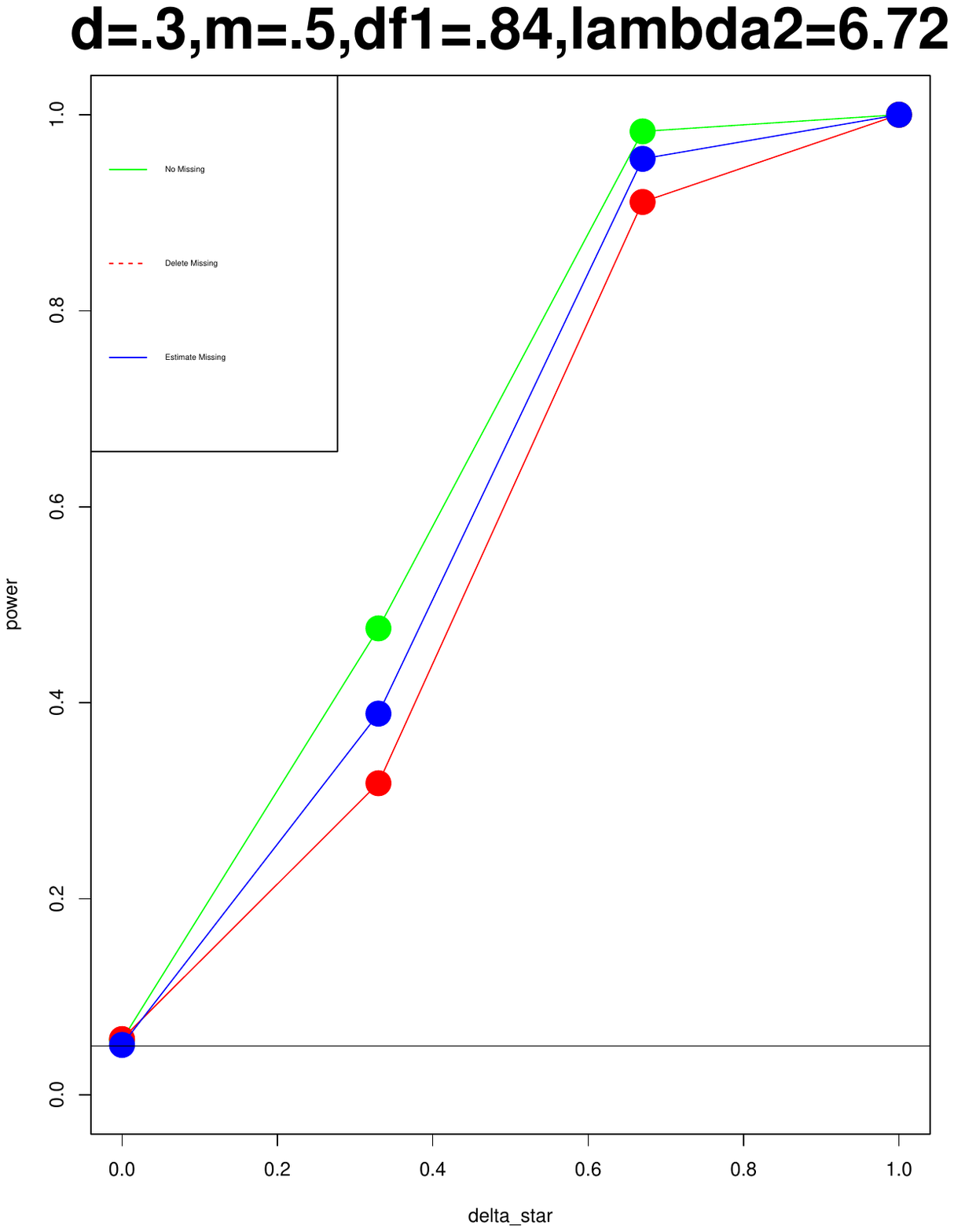}

\hspace{1.5cm}
$d=.1 \quad m=.5$
\hspace{3cm}
$d=.2 \quad m=.5$
\hspace{3cm}
$d=.3 \quad m=.5$

\includegraphics[width = 2.3in, height = 1.4in]{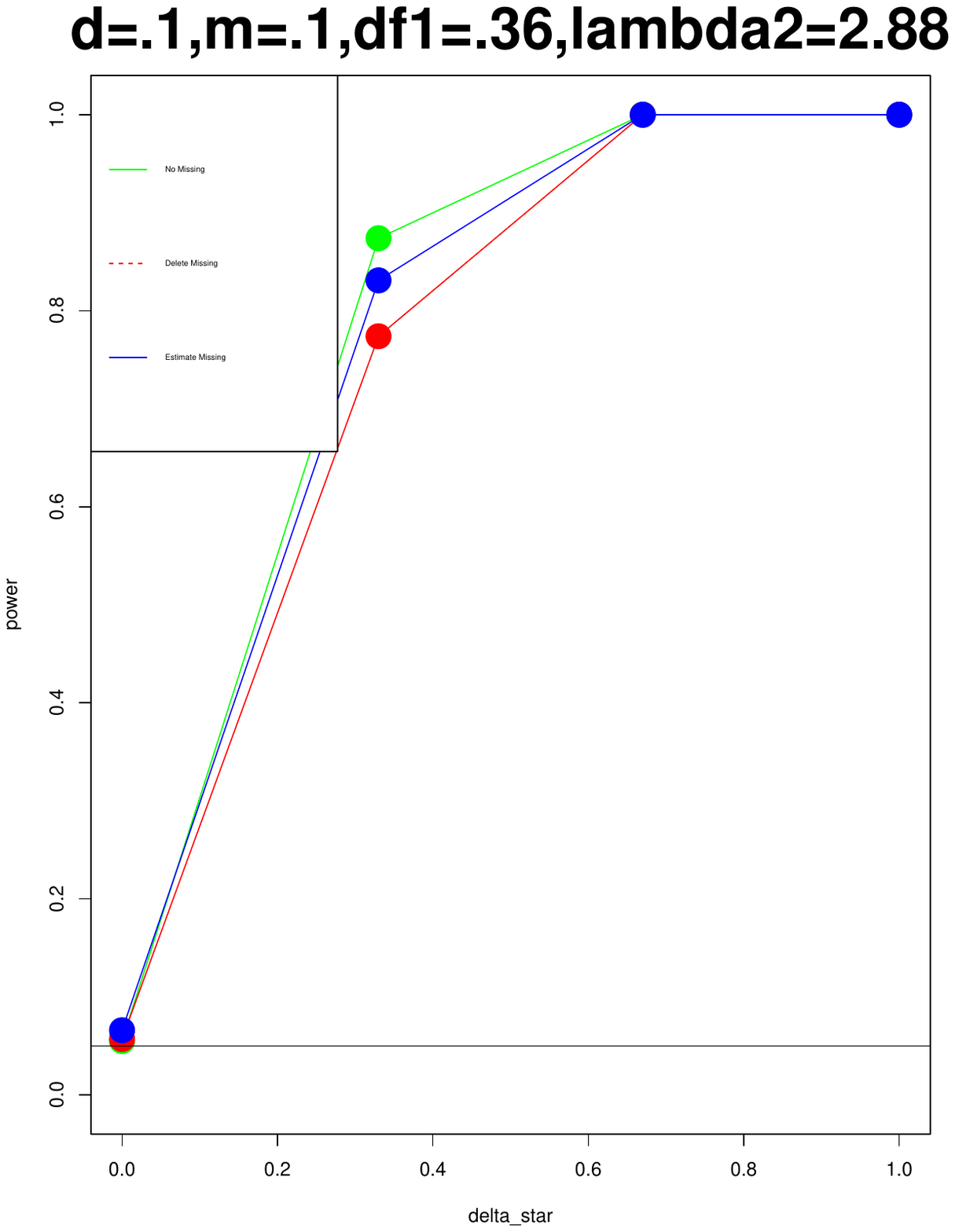}
\includegraphics[width = 2.3in, height = 1.4in]{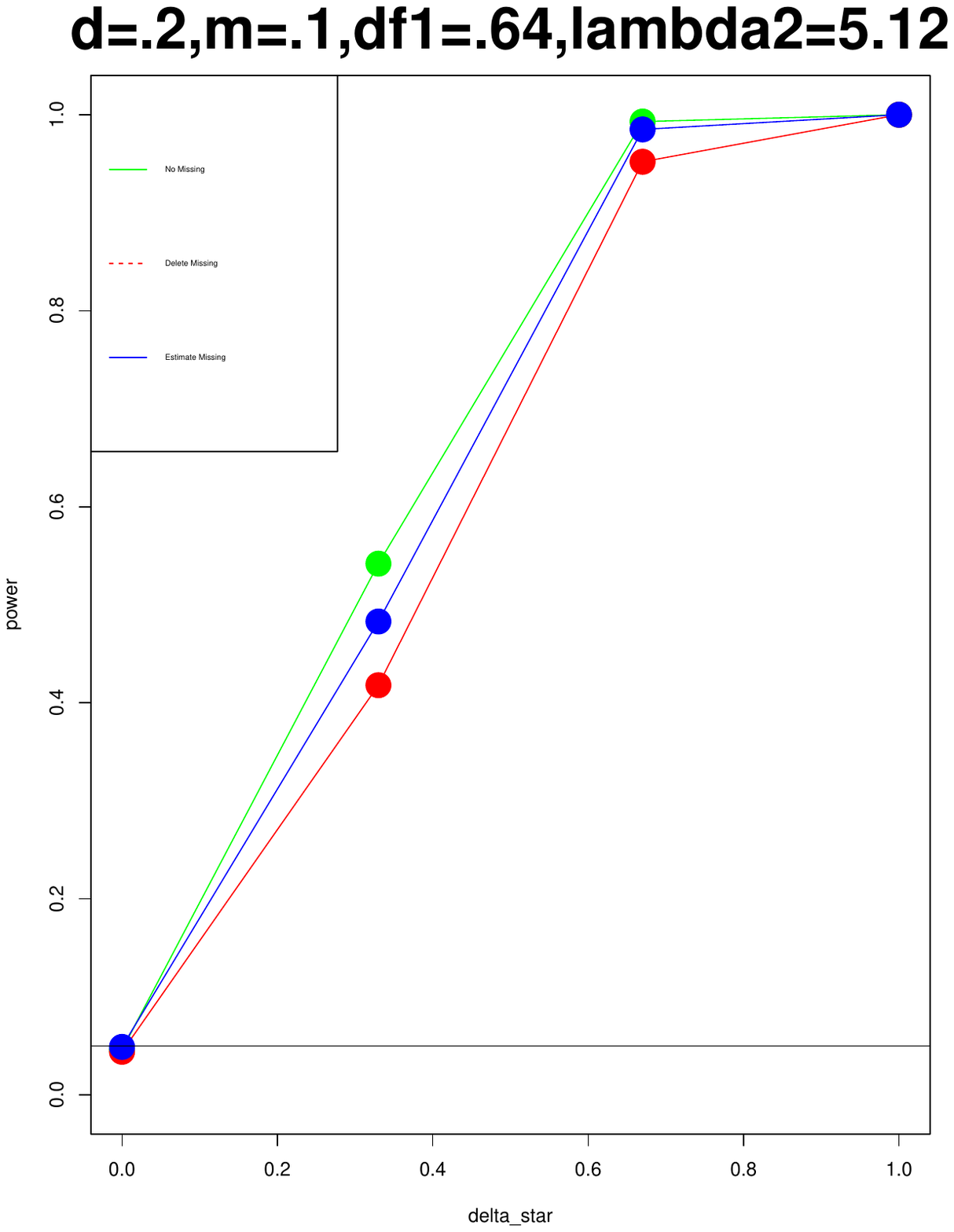}
\includegraphics[width = 2.3in, height = 1.4in]{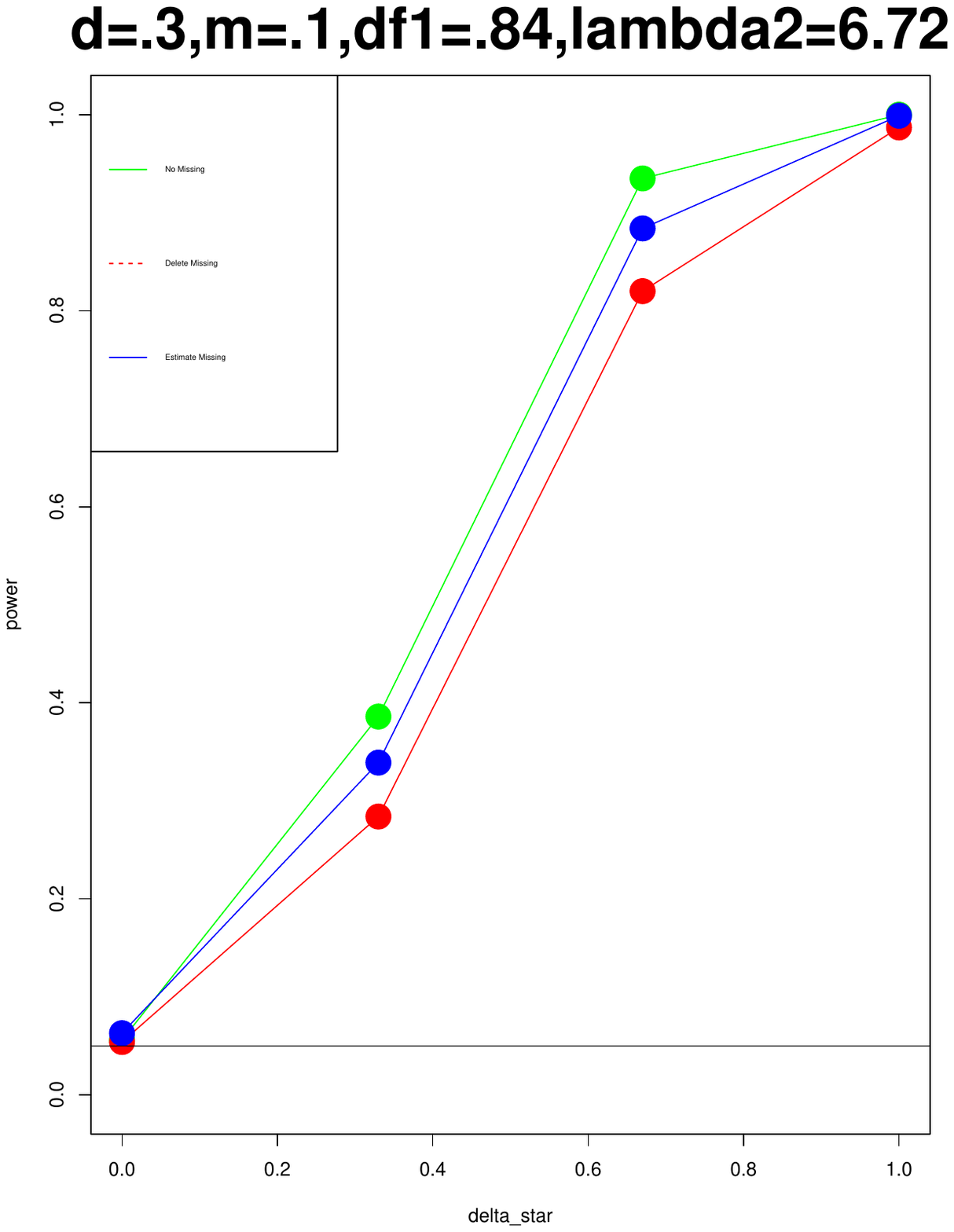}

\hspace{1.5cm}
$d=.1 \quad m=.1$
\hspace{3cm}
$d=.2 \quad m=.1$
\hspace{3cm}
$d=.3 \quad m=.1$

\includegraphics[width = 2.3in, height = 1.4in]{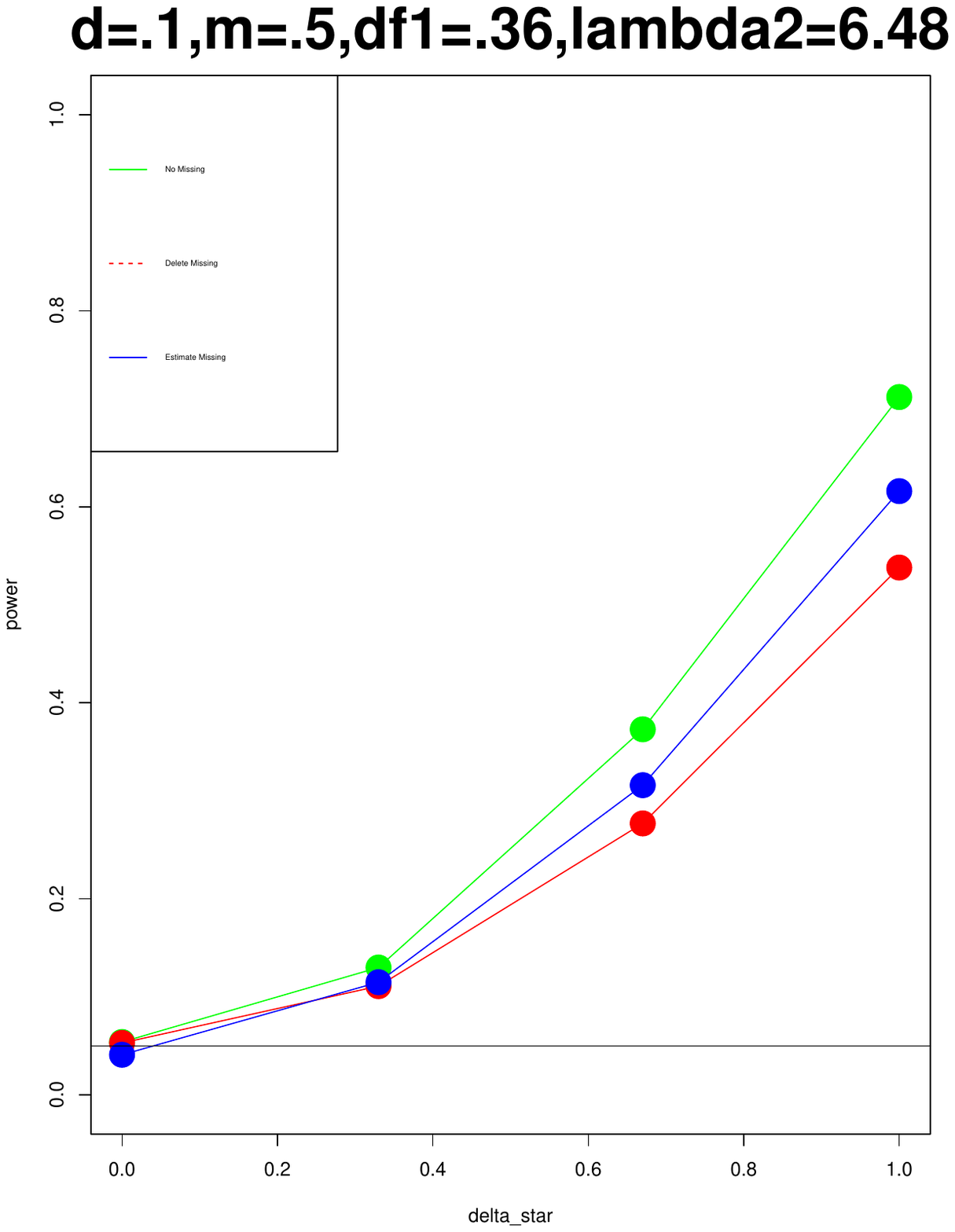}
\includegraphics[width = 2.3in, height = 1.4in]{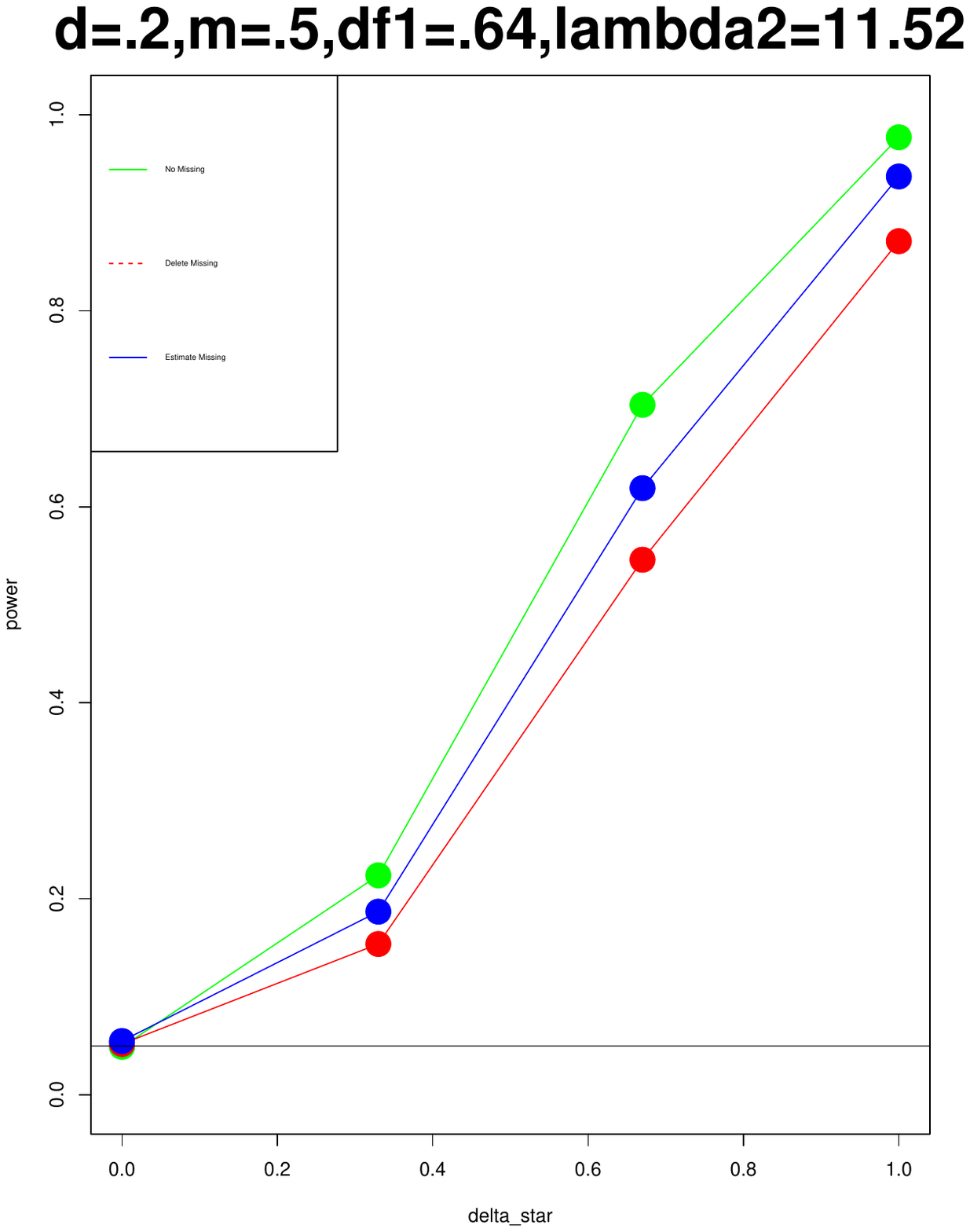}
\includegraphics[width = 2.3in, height = 1.4in]{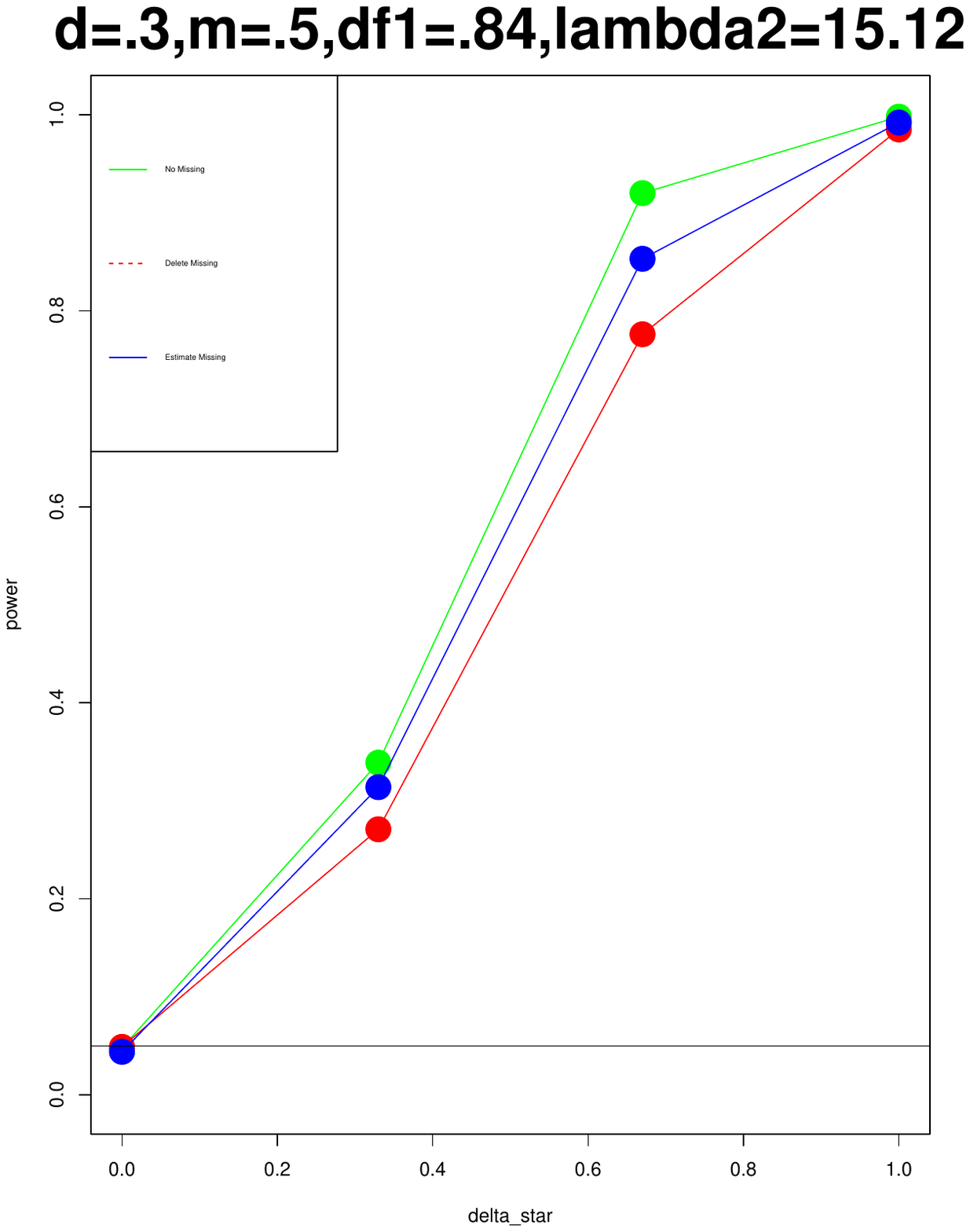}

\hspace{1.5cm}
$d=.1 \quad m=.5$
\hspace{3cm}
$d=.2 \quad m=.5$
\hspace{3cm}
$d=.3 \quad m=.5$

\includegraphics[width = 2.3in, height = 1.4in]{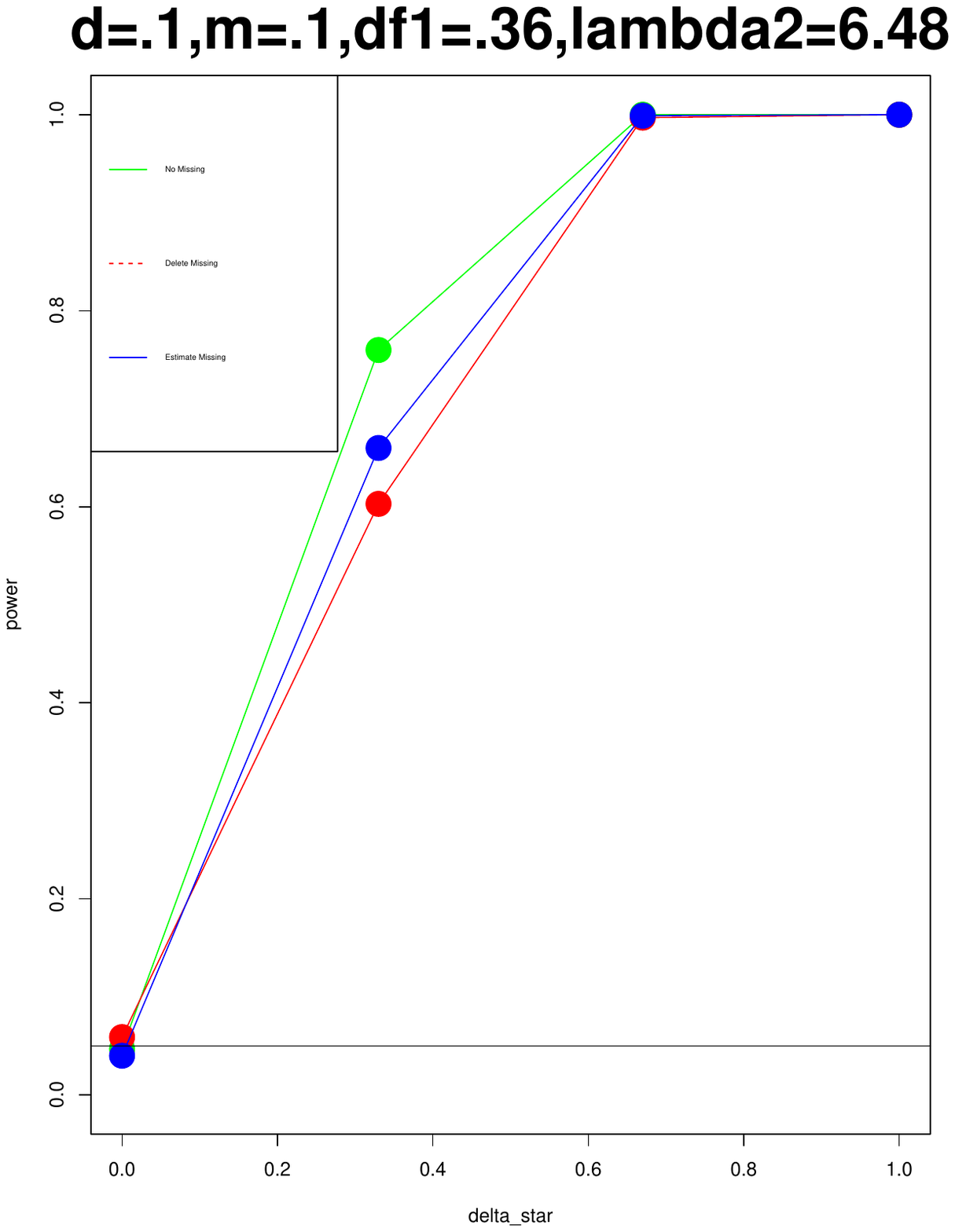}
\includegraphics[width = 2.3in, height = 1.4in]{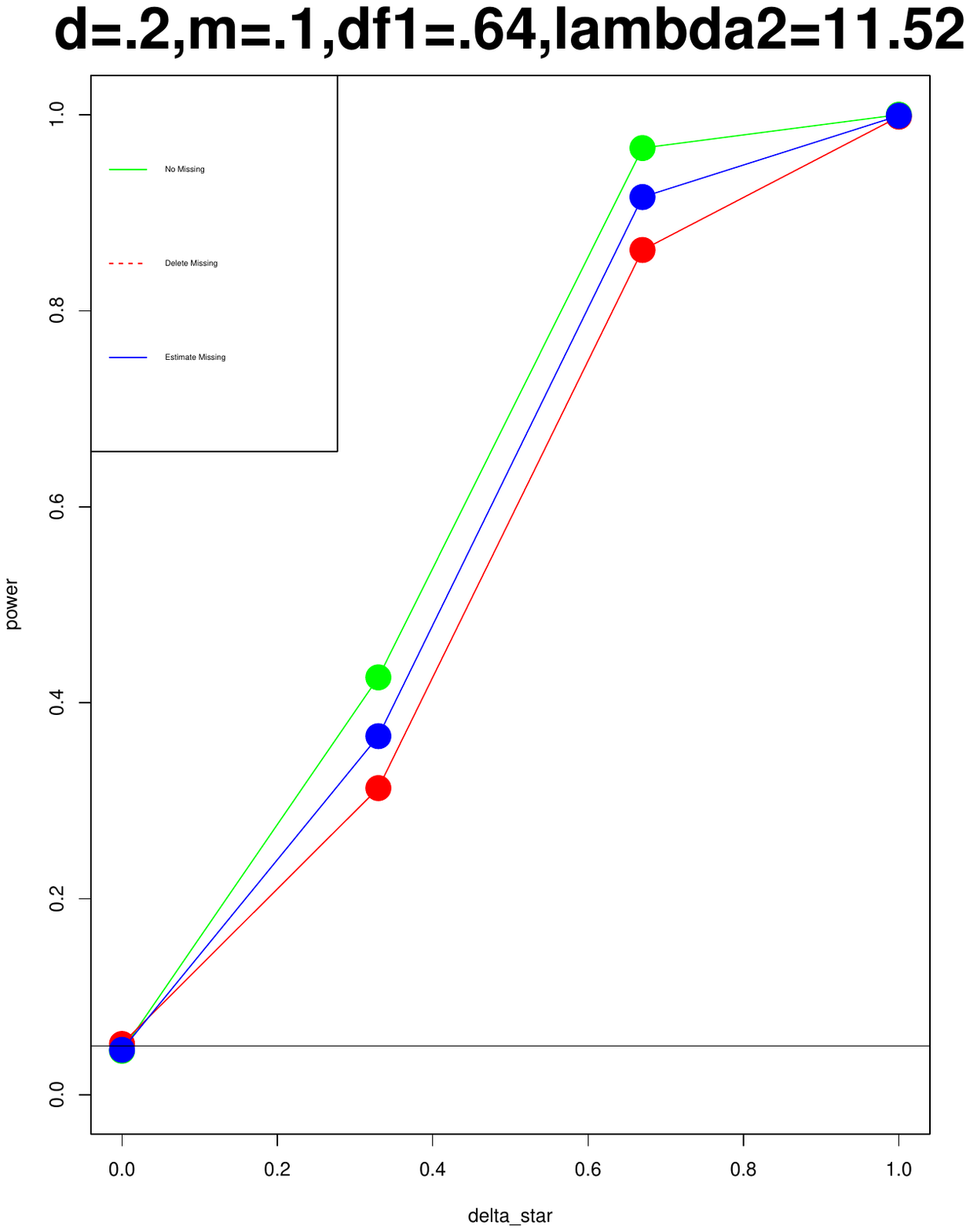}
\includegraphics[width = 2.3in, height = 1.4in]{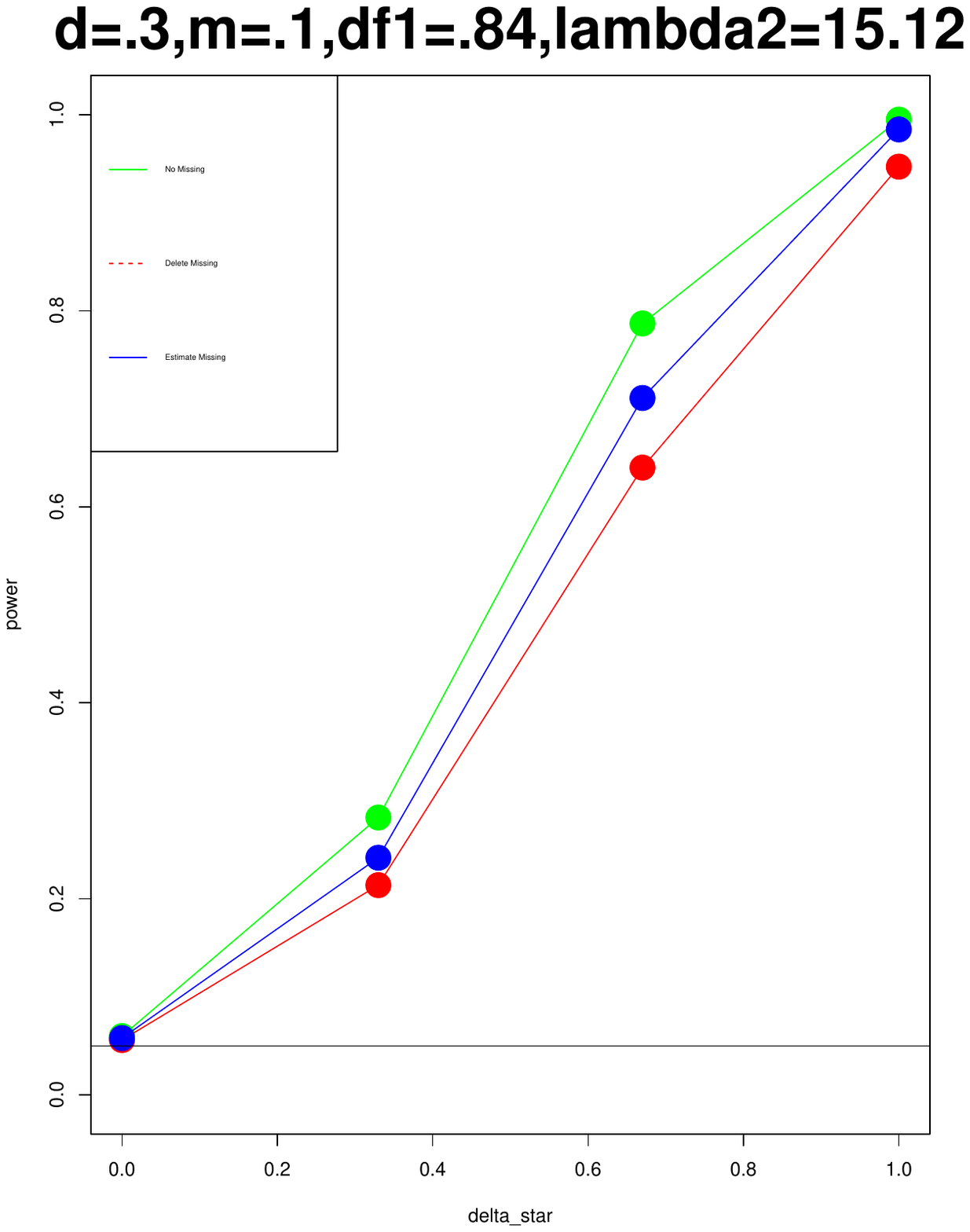}

\subsubsection{When both Traits have Chi Squares Distribution}

\hspace{1.5cm}
$d=.1 \quad m=.5$
\hspace{3cm}
$d=.2 \quad m=.5$
\hspace{3cm}
$d=.3 \quad m=.5$

\includegraphics[width = 2.3in, height = 1.4in]{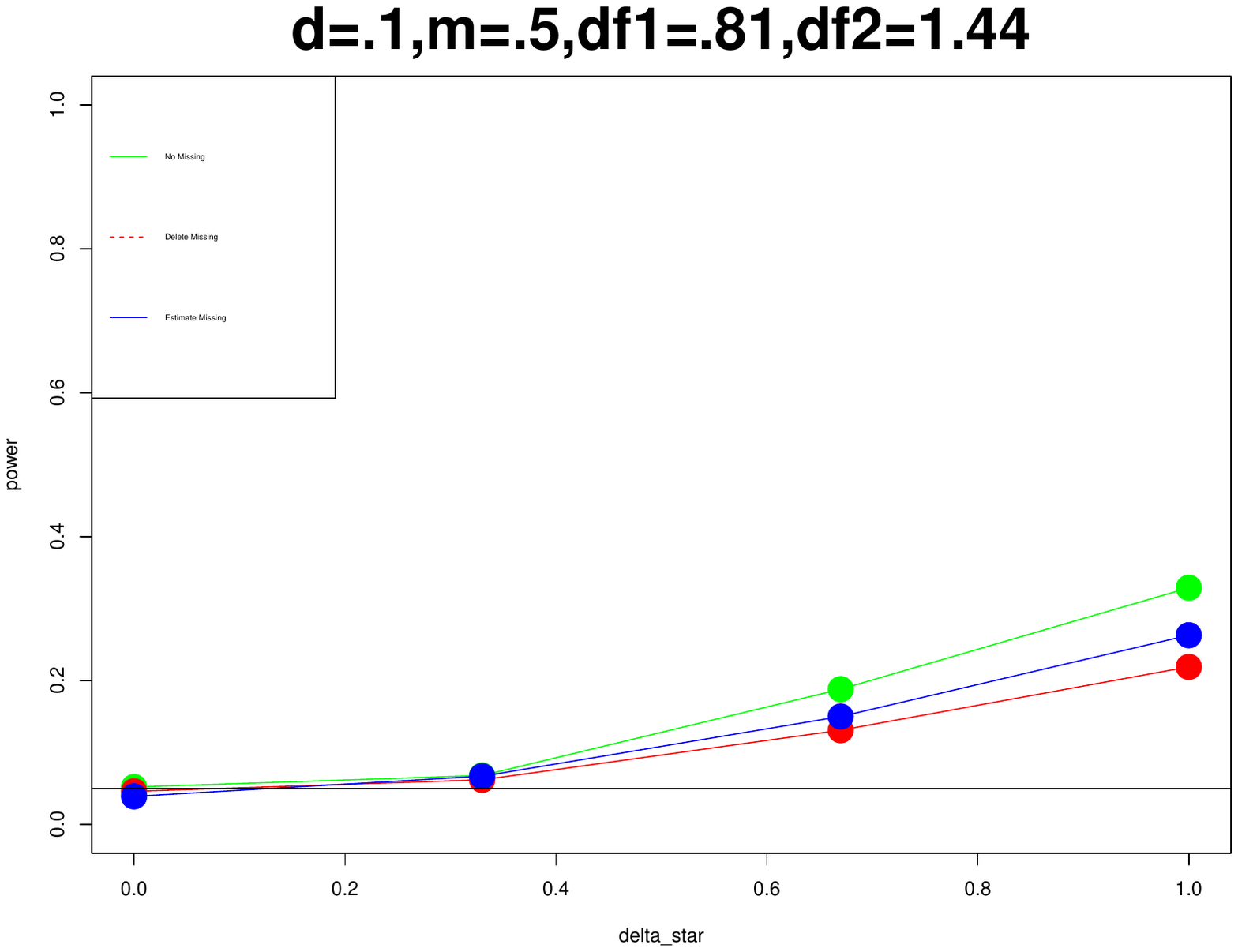}
\includegraphics[width = 2.3in, height = 1.4in]{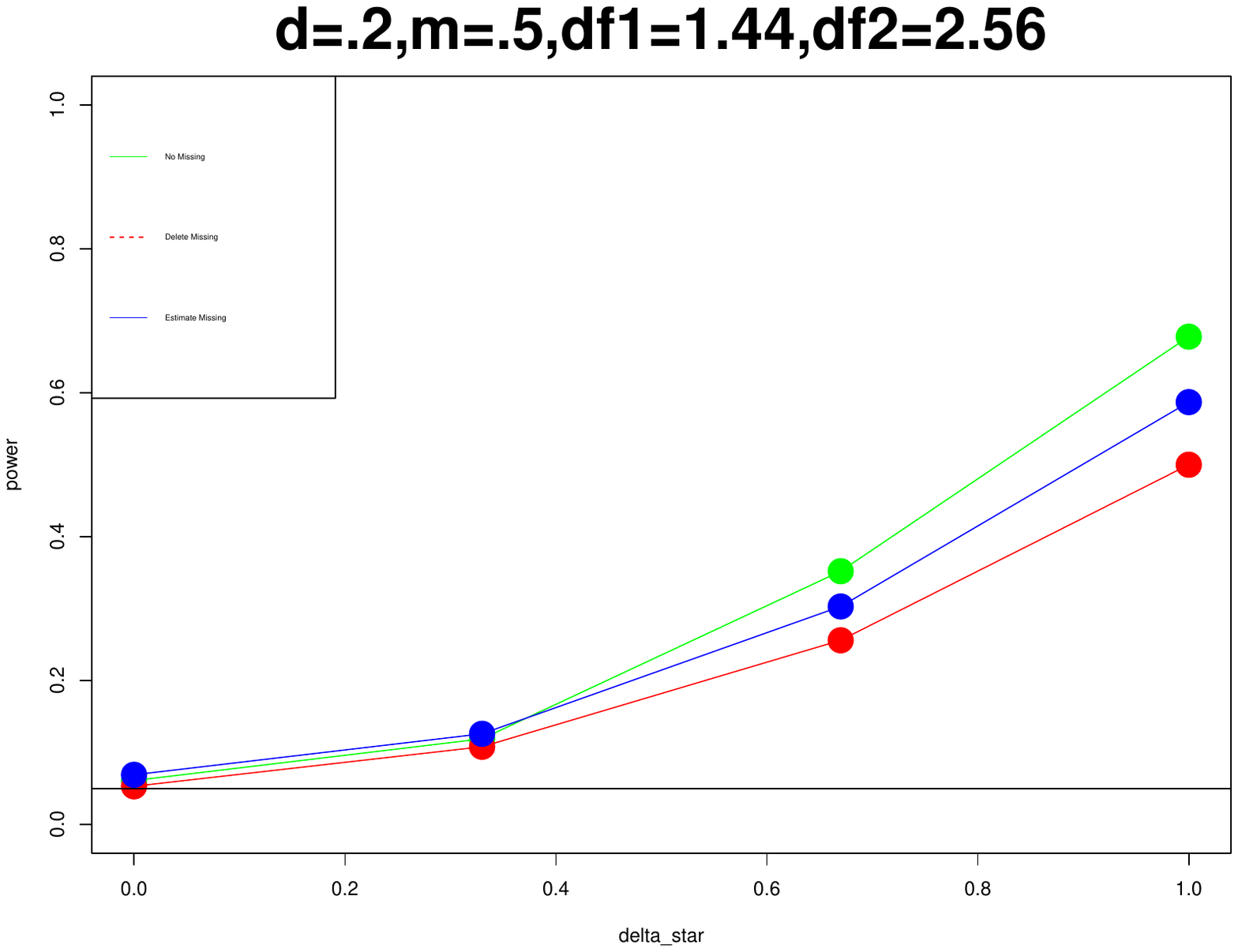}
\includegraphics[width = 2.3in, height = 1.4in]{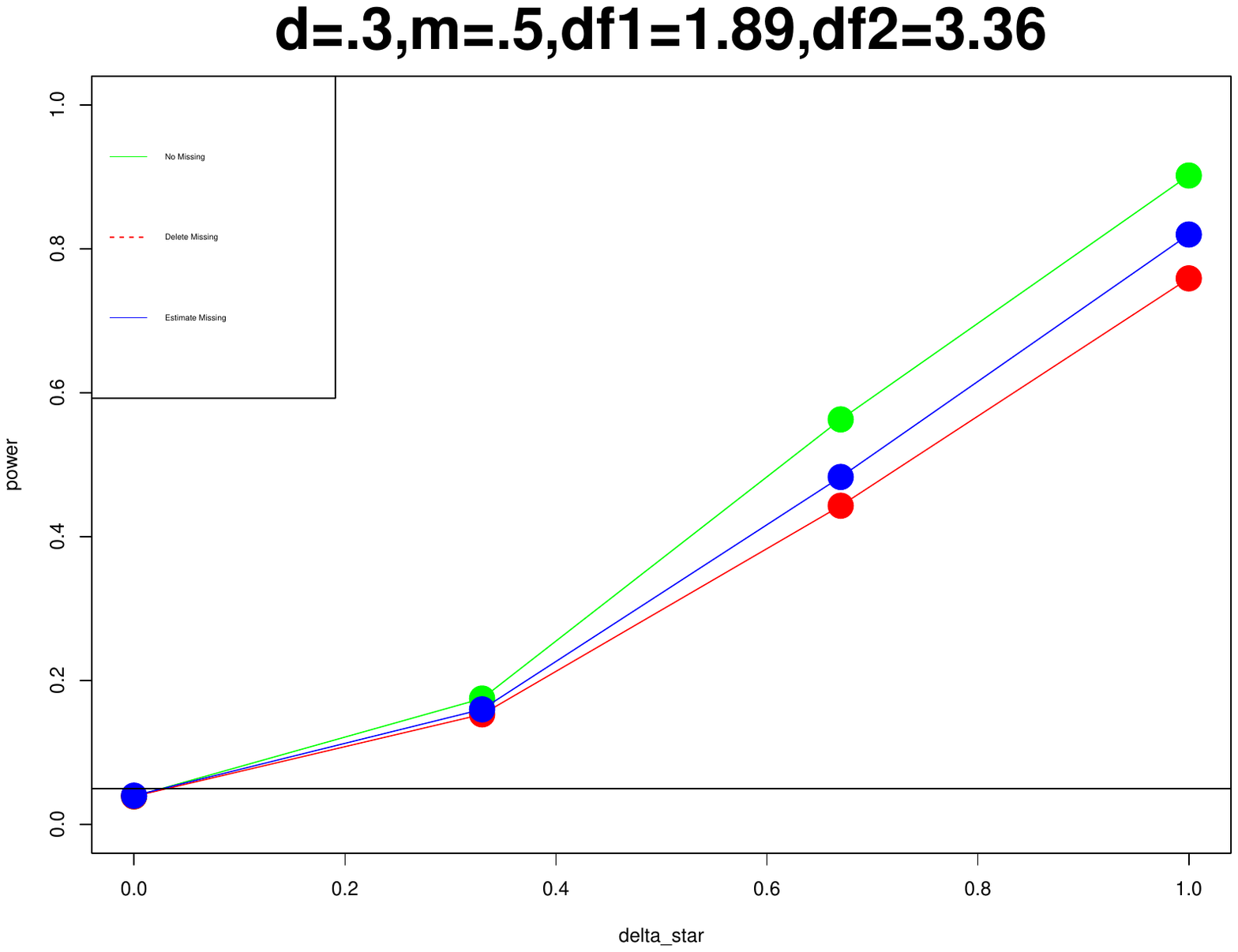}

\hspace{1.5cm}
$d=.1 \quad m=.1$
\hspace{3cm}
$d=.2 \quad m=.1$
\hspace{3cm}
$d=.3 \quad m=.1$

\includegraphics[width = 2.3in, height = 1.4in]{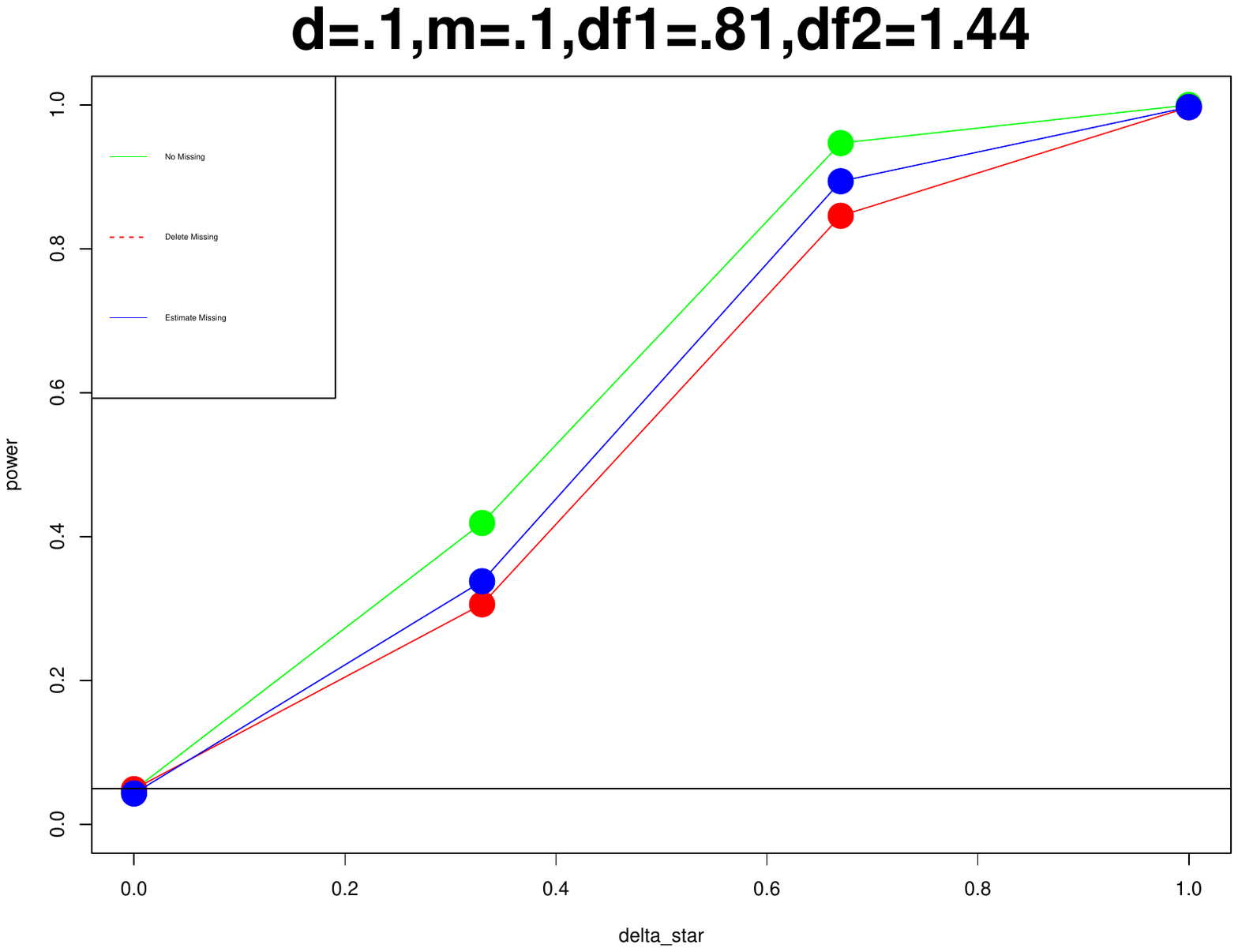}
\includegraphics[width = 2.3in, height = 1.4in]{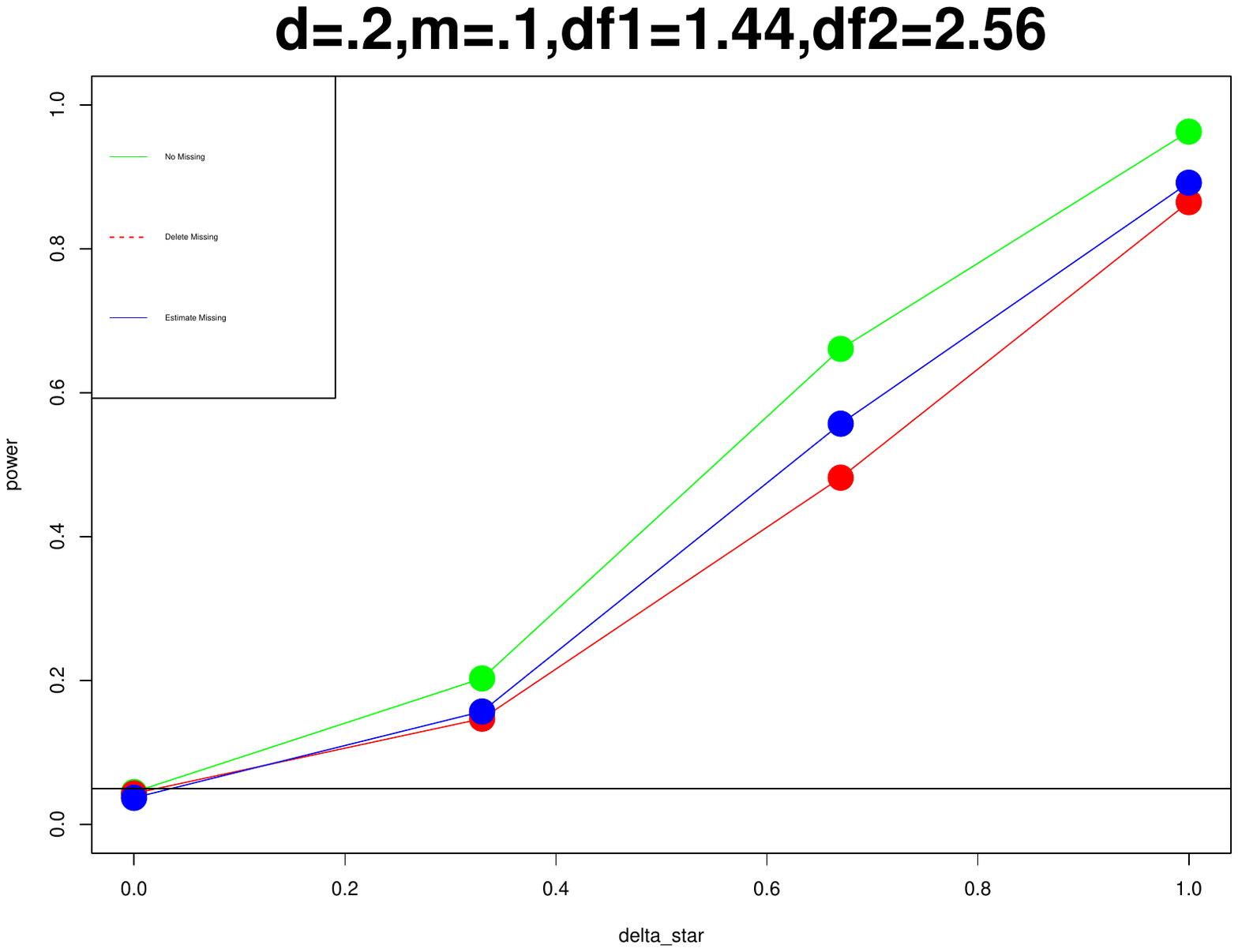}
\includegraphics[width = 2.3in, height = 1.4in]{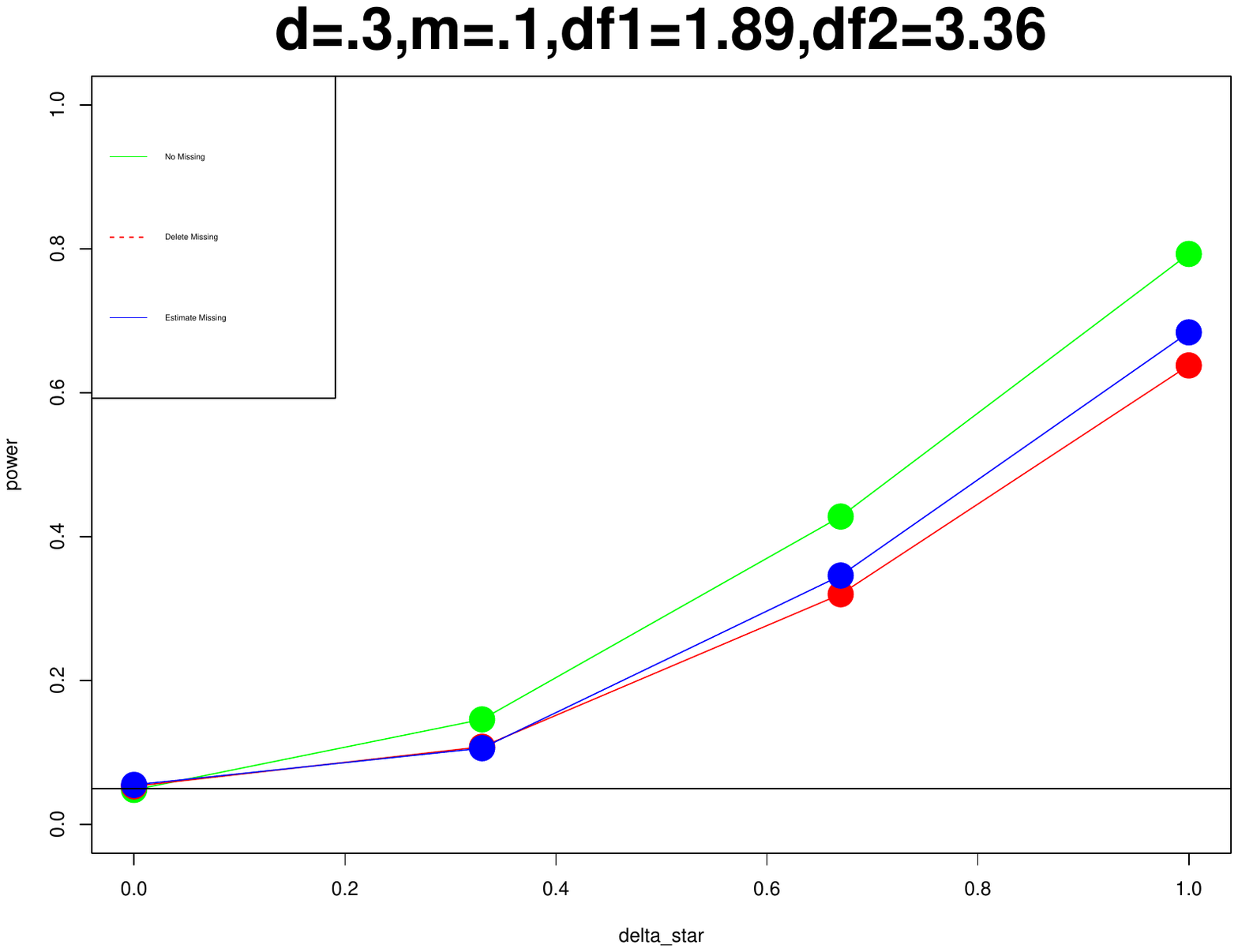}

\hspace{1.5cm}
$d=.1 \quad m=.1$
\hspace{3cm}
$d=.1 \quad m=.1$
\hspace{3cm}
$d=.1 \quad m=.1$

\includegraphics[width = 2.3in, height = 1.4in]{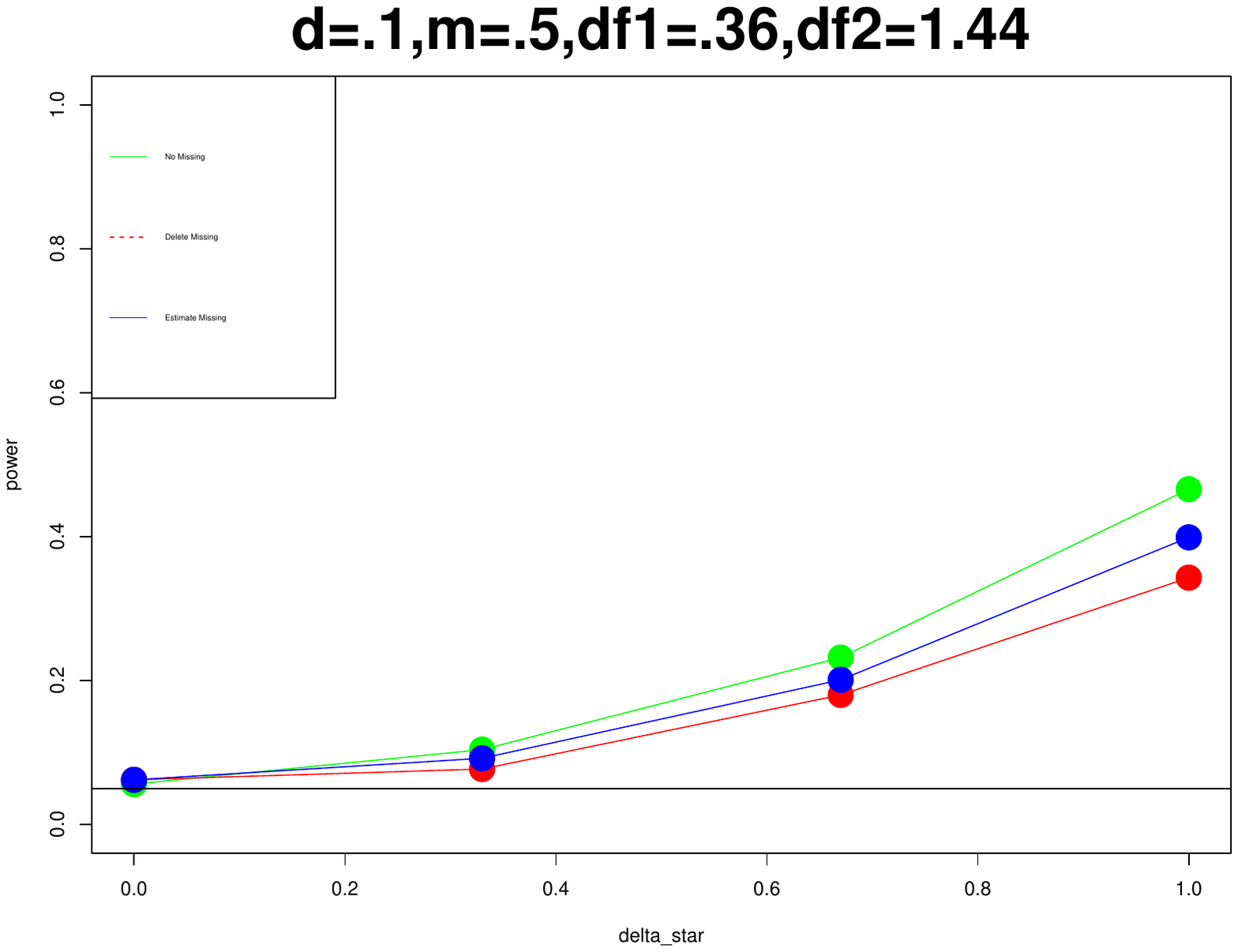}
\includegraphics[width = 2.3in, height = 1.4in]{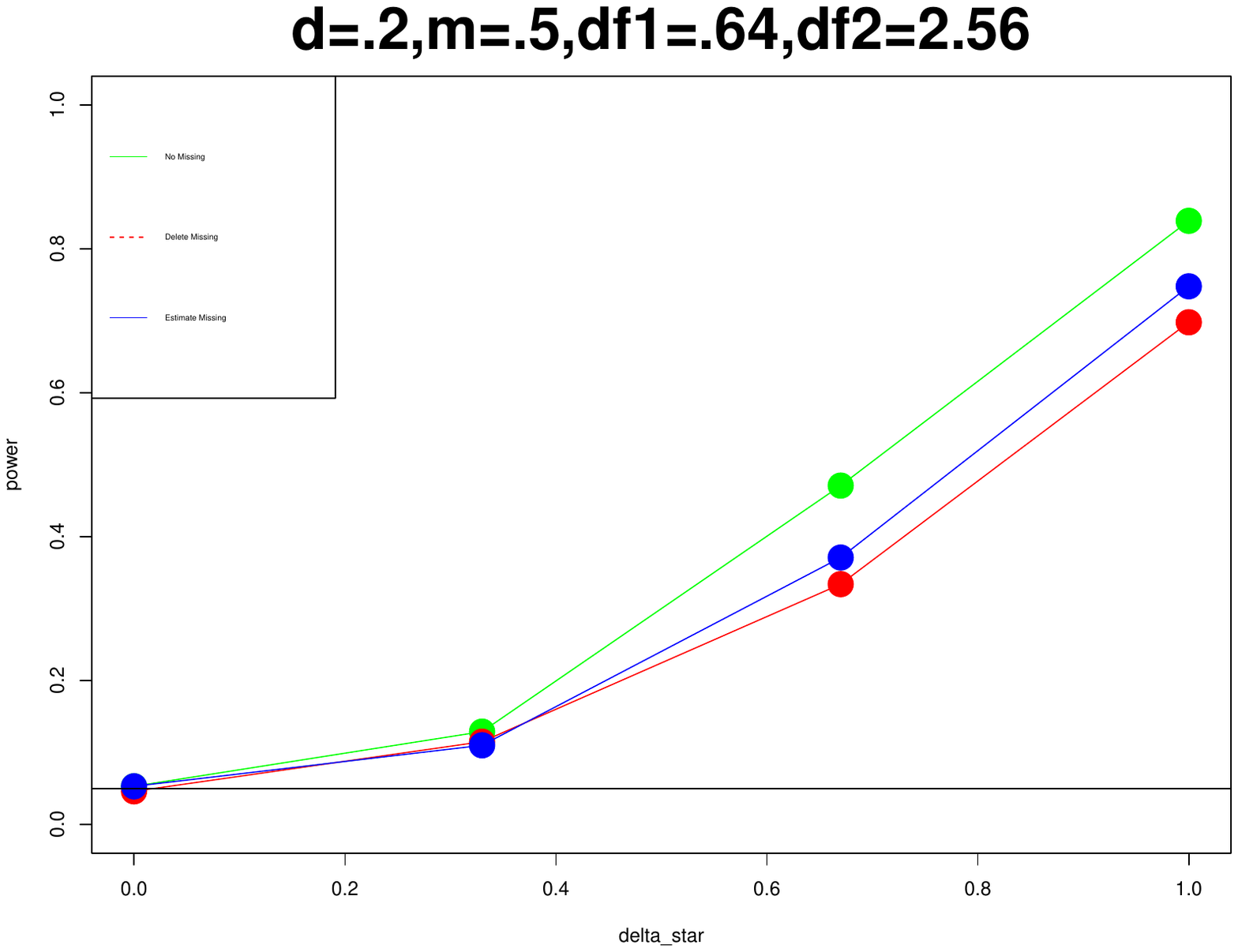}
\includegraphics[width = 2.3in, height = 1.4in]{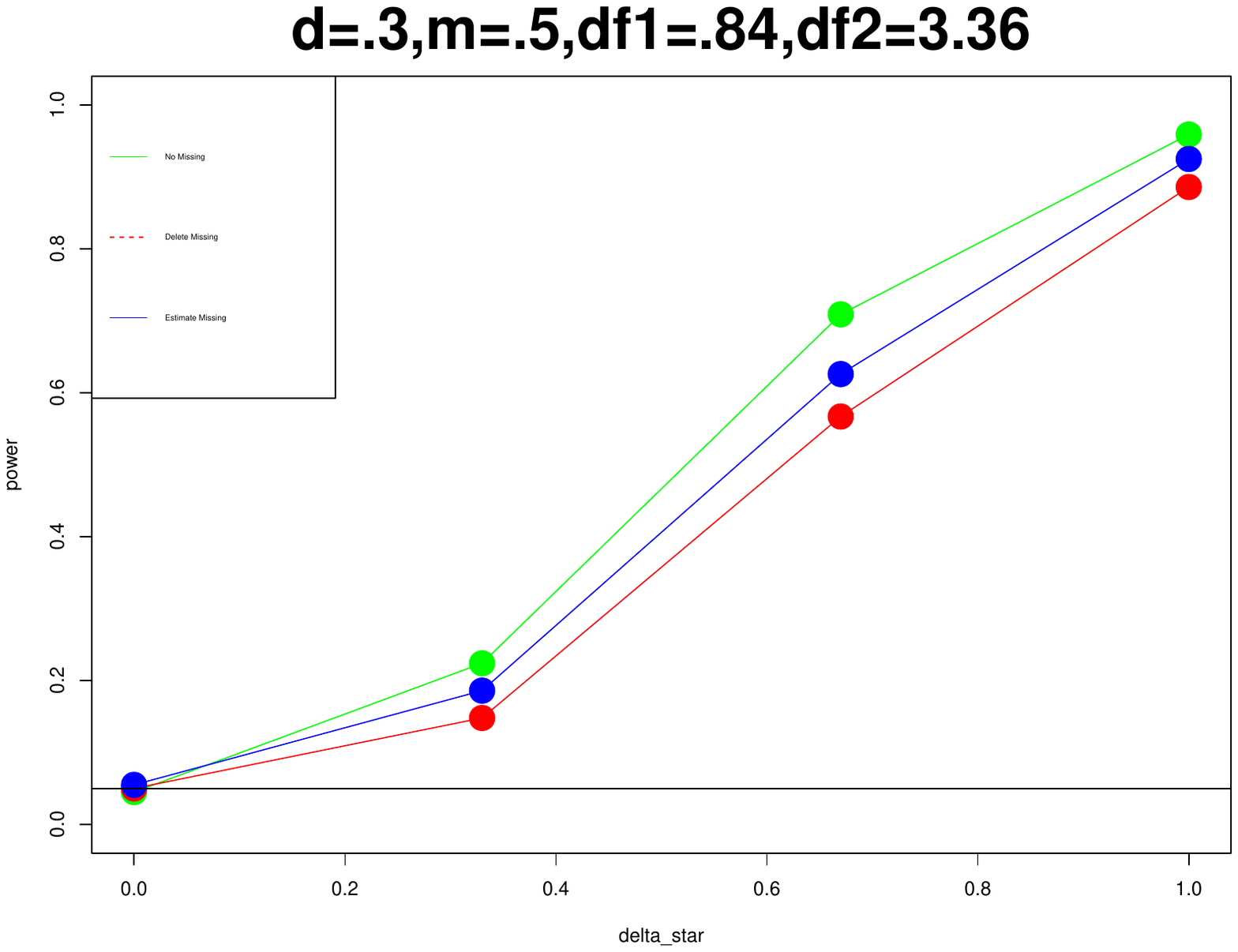}

\hspace{1.5cm}
$d=.1 \quad m=.5$
\hspace{3cm}
$d=.2 \quad m=.5$
\hspace{3cm}
$d=.3 \quad m=.5$

\includegraphics[width = 2.3in, height = 1.4in]{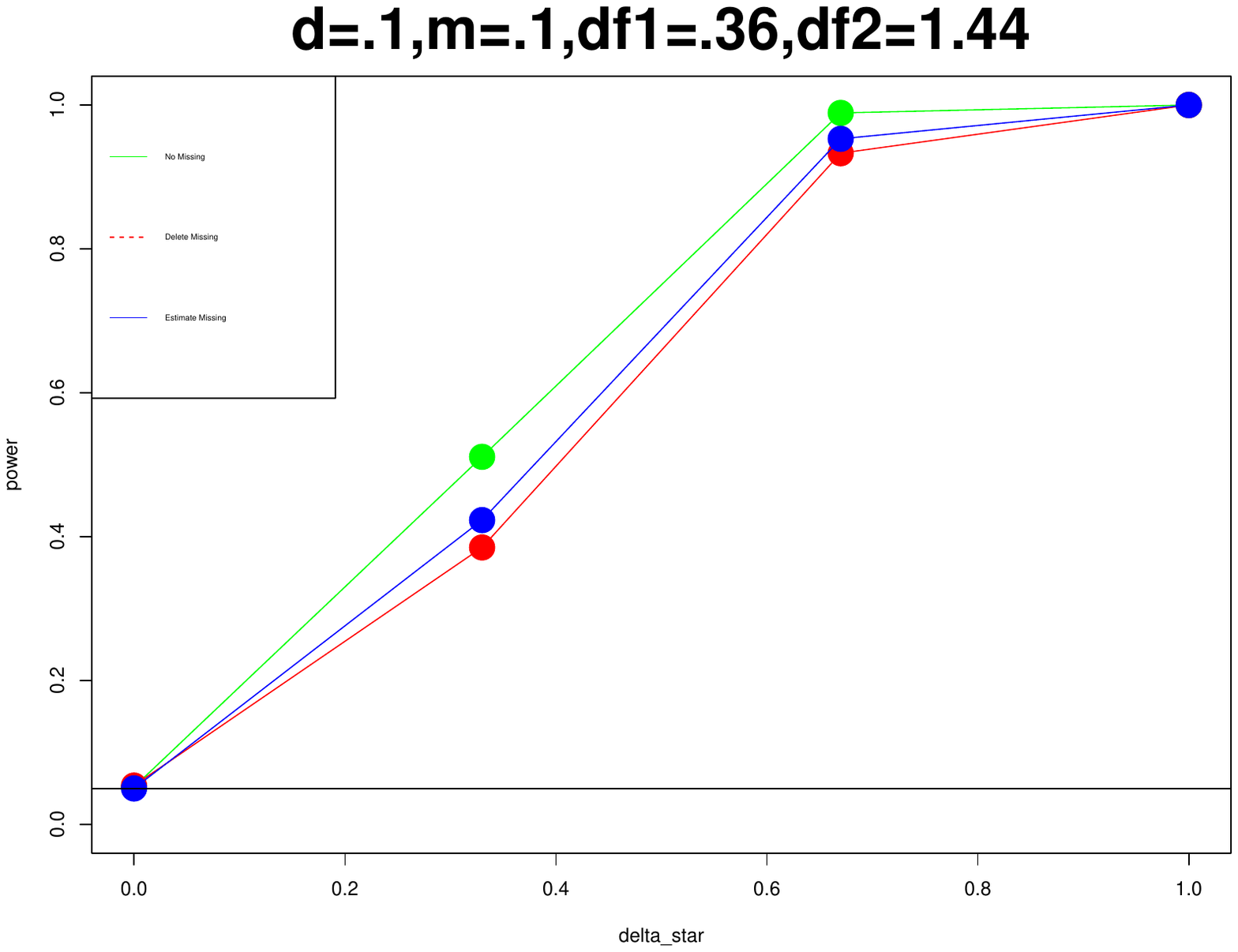}
\includegraphics[width = 2.3in, height = 1.4in]{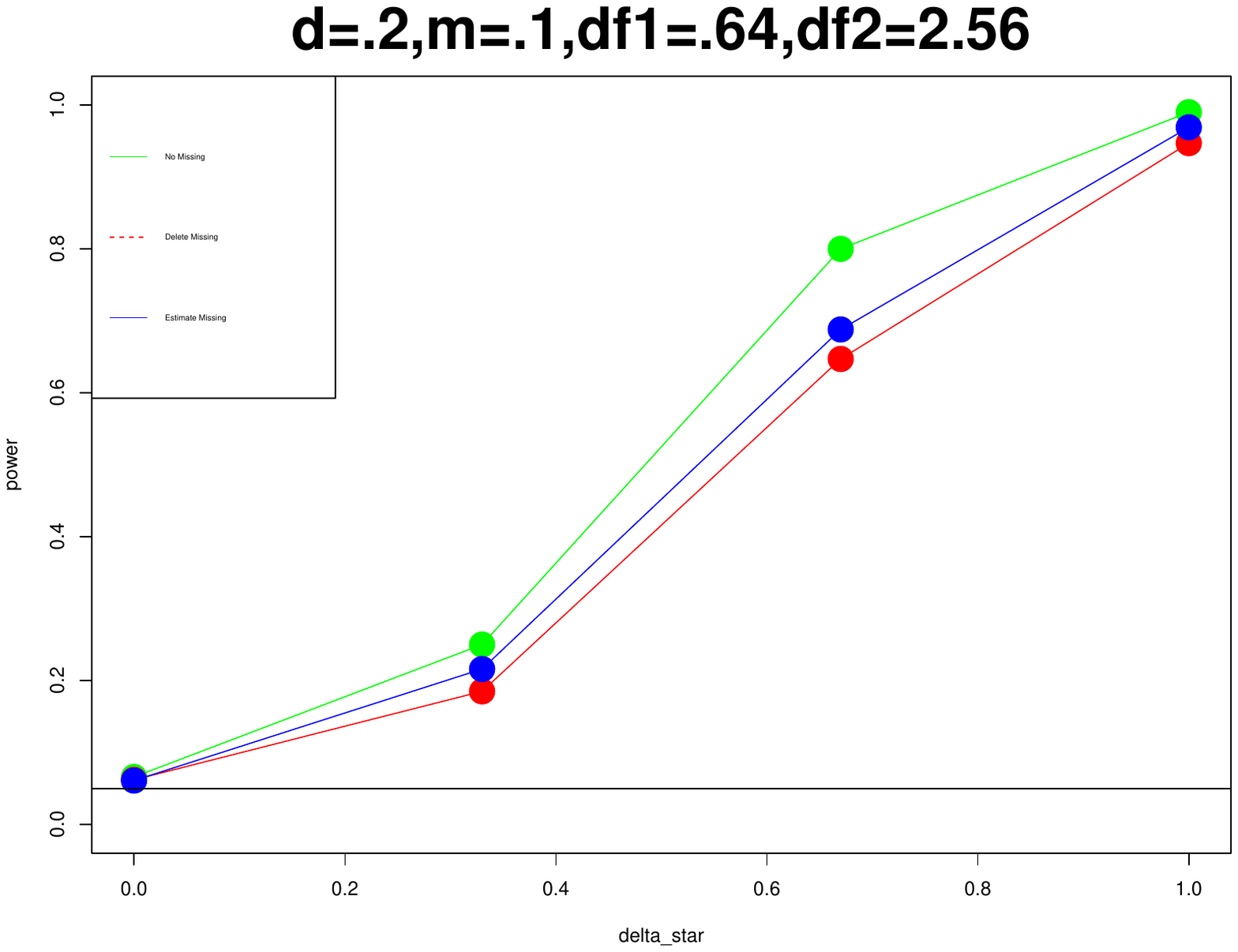}
\includegraphics[width = 2.3in, height = 1.4in]{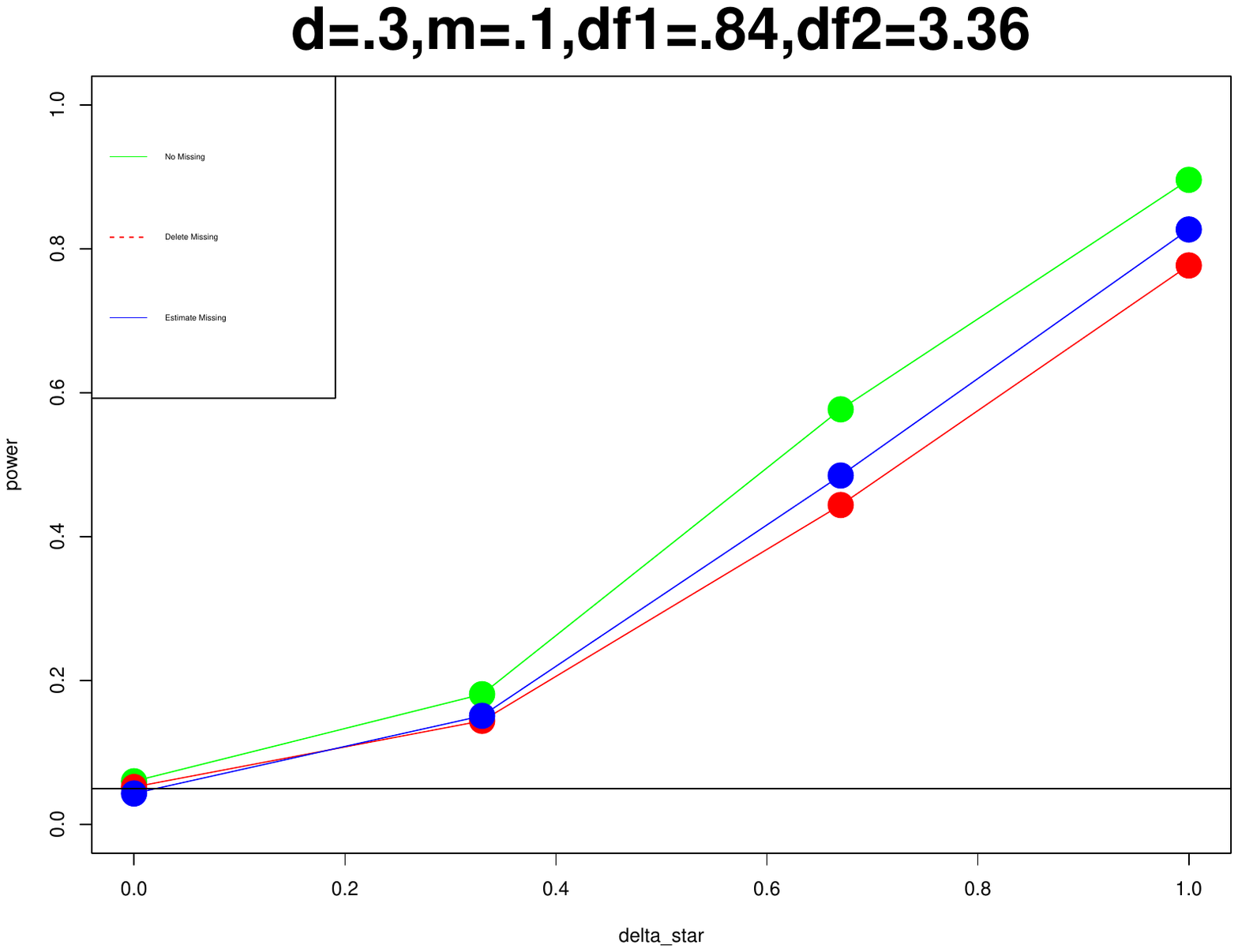}

\subsection{For Missing Type 2.4}

As stated in section 5, we have estimated the missing traits separately using the value of only trait present for those families.

In each of the cases simulation details are exactly same as the simulation details in the section 7.2.

\subsubsection{When both traits have Normal Distribution}

\hspace{1.5cm}
$d=.1 \quad m=.5$
\hspace{3cm}
$d=.2 \quad m=.5$
\hspace{3cm}
$d=.3 \quad m=.5$

\includegraphics[width = 2.3in, height = 1.4in]{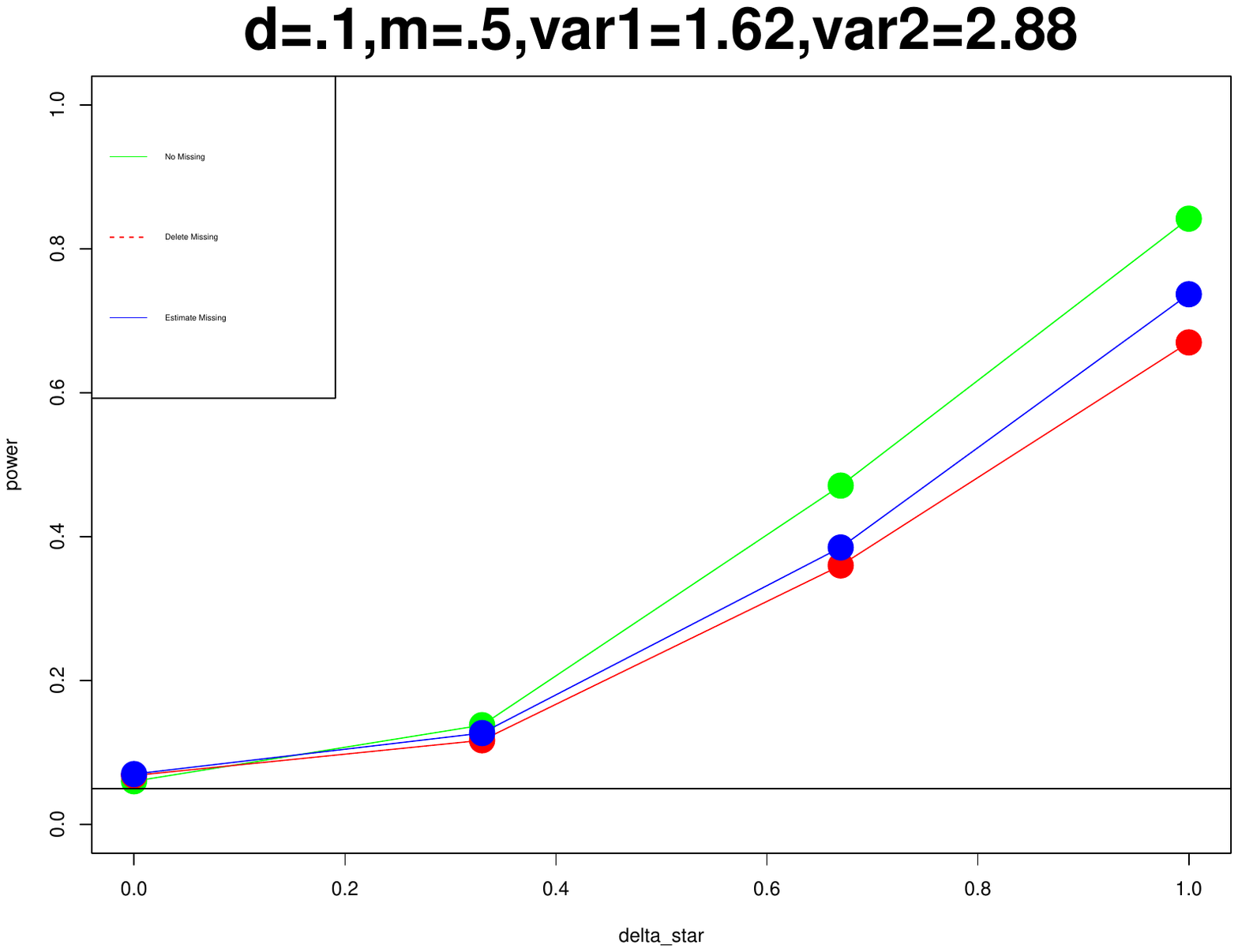}
\includegraphics[width = 2.3in, height = 1.4in]{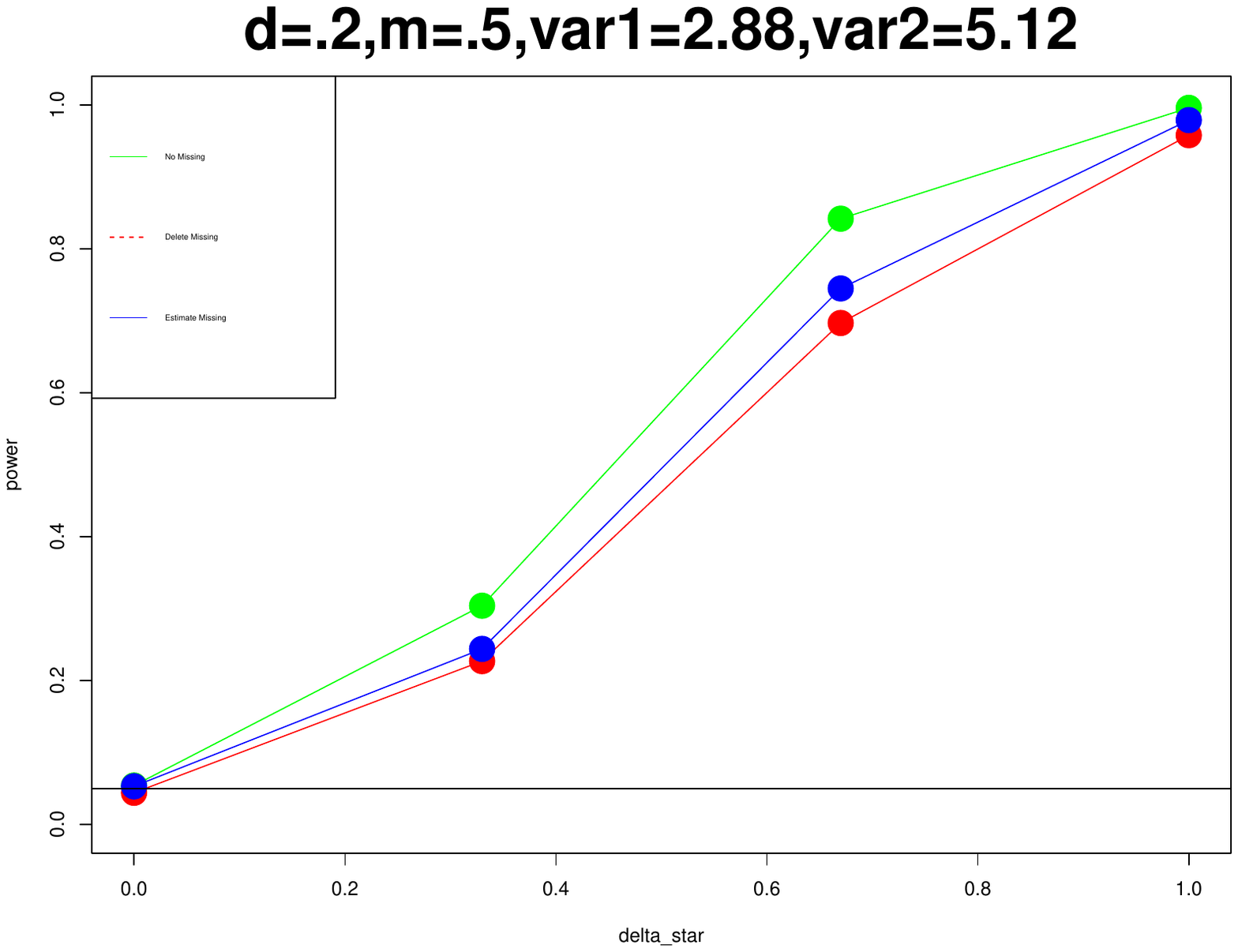}
\includegraphics[width = 2.3in, height = 1.4in]{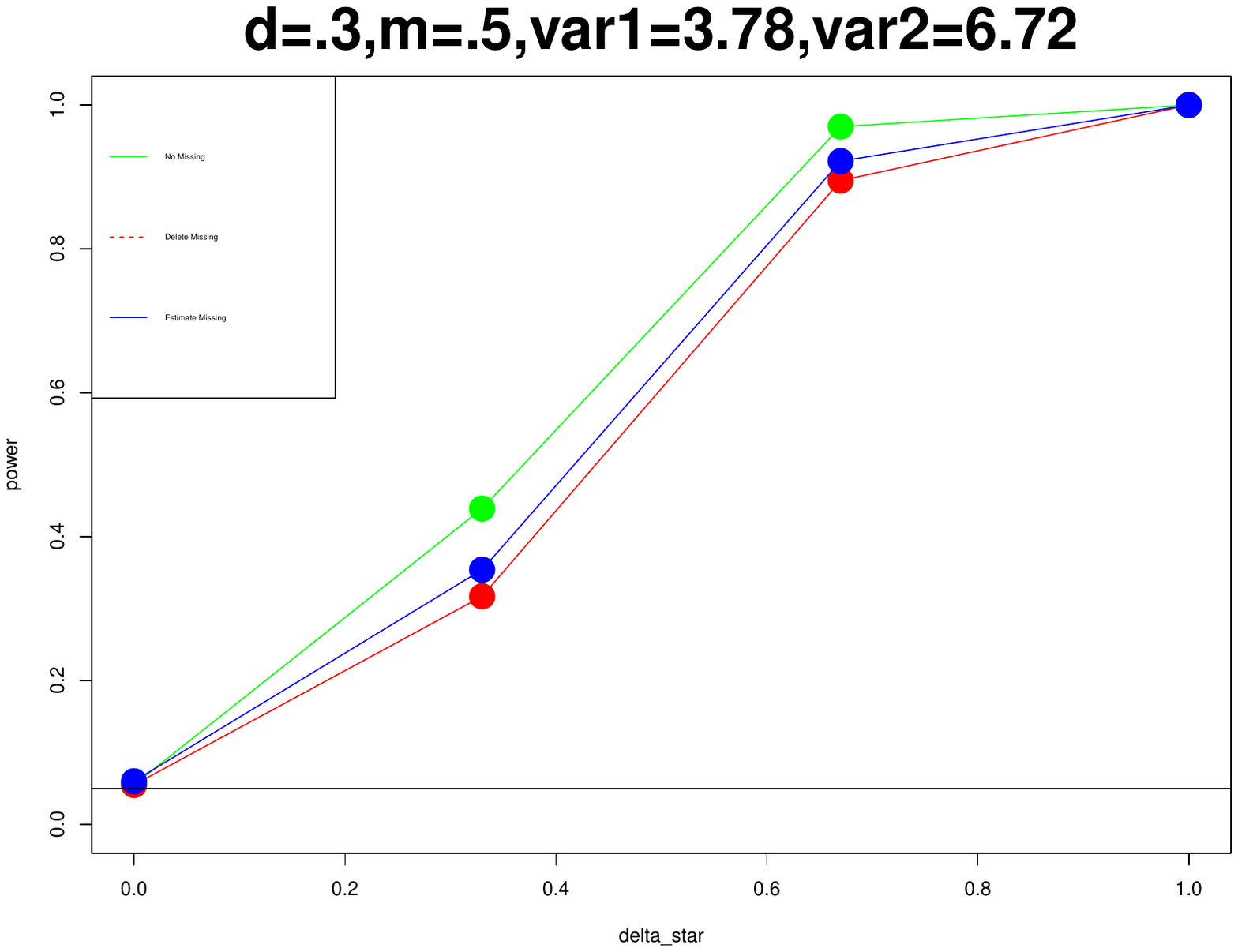}

\hspace{1.5cm}
$d=.1 \quad m=.1$
\hspace{3cm}
$d=.2 \quad m=.1$
\hspace{3cm}
$d=.3 \quad m=.1$

\includegraphics[width = 2.3in, height = 1.3in]{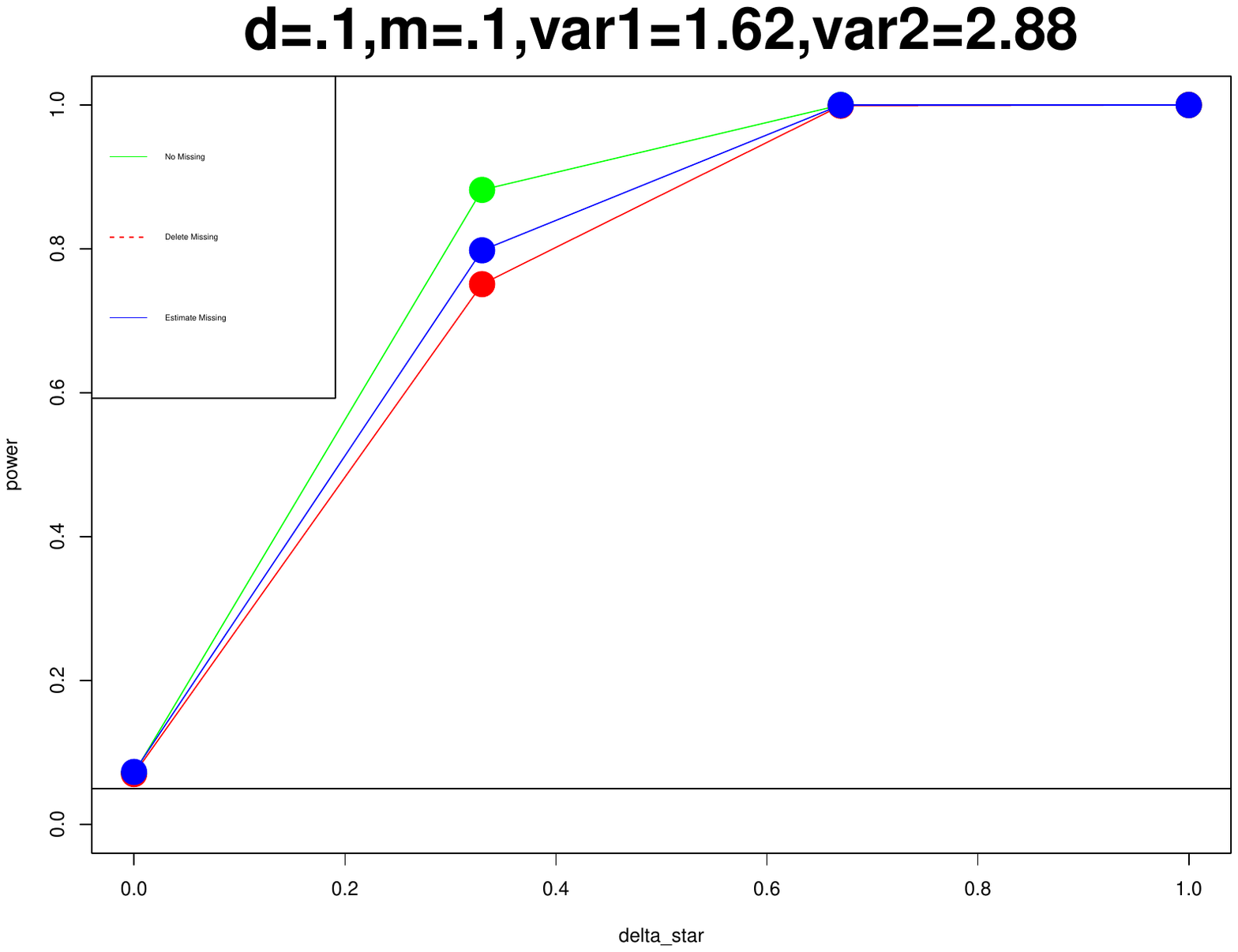}
\includegraphics[width = 2.3in, height = 1.3in]{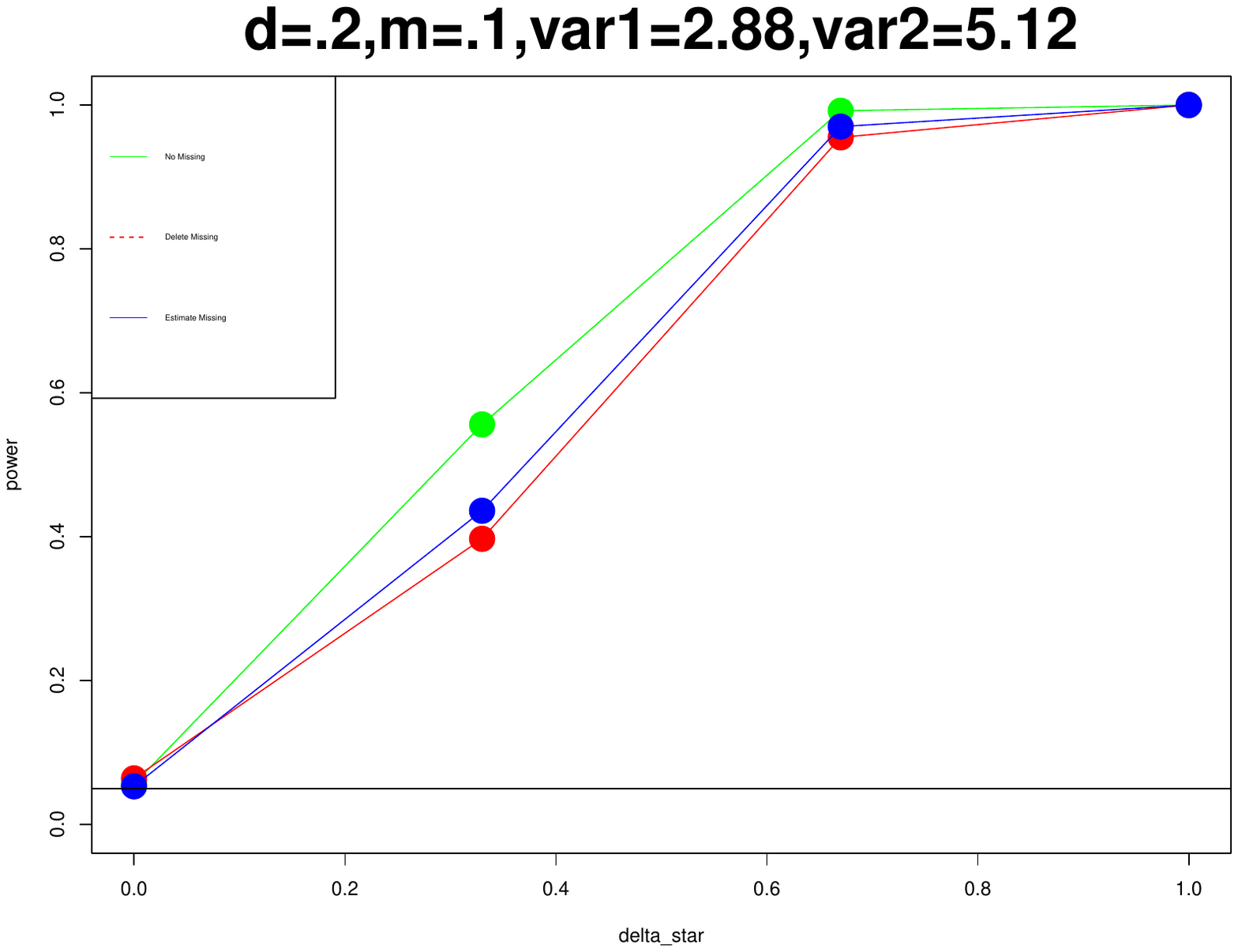}
\includegraphics[width = 2.3in, height = 1.3in]{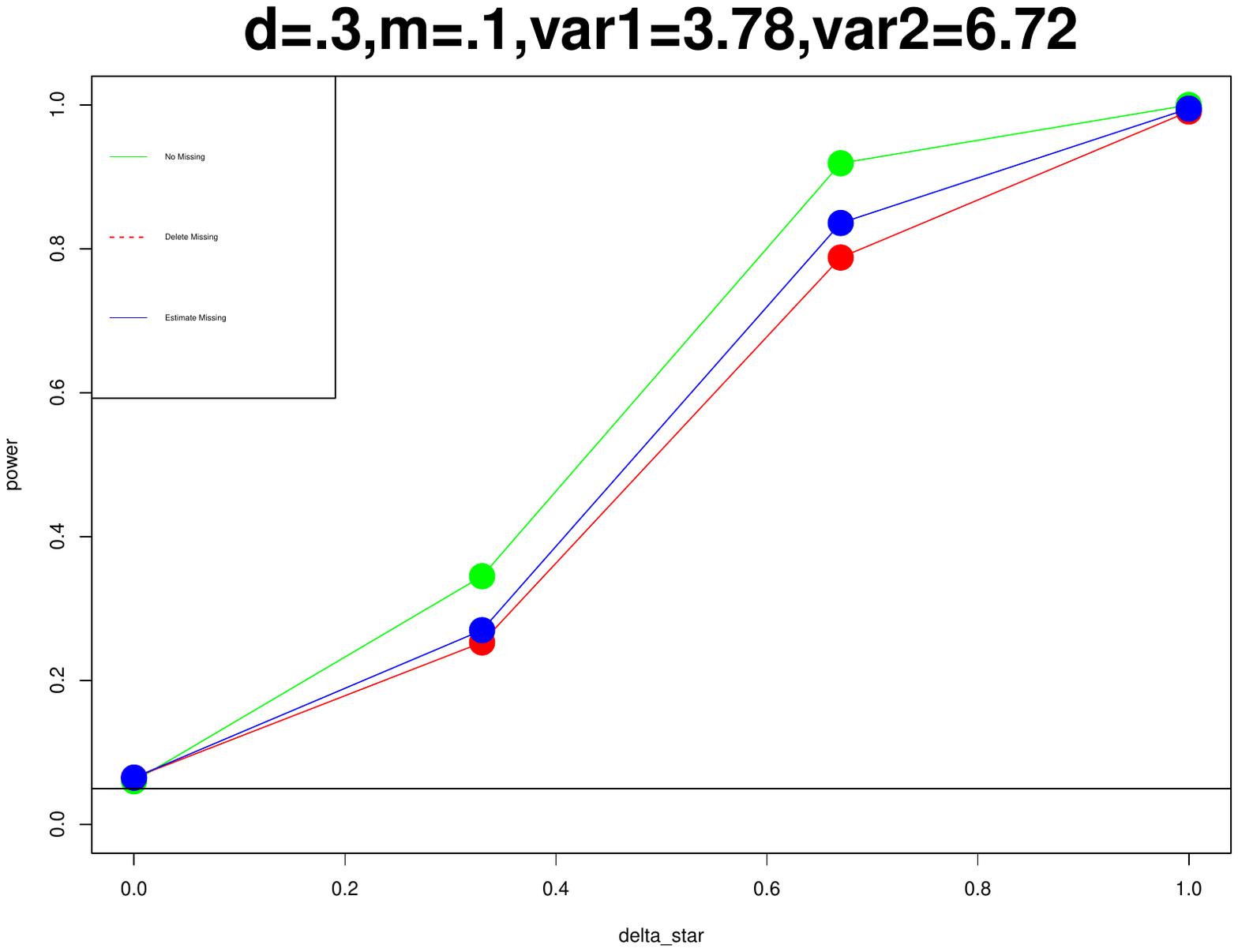}

\hspace{1.5cm}
$d=.1 \quad m=.1$
\hspace{3cm}
$d=.2 \quad m=.1$
\hspace{3cm}
$d=.3 \quad m=.1$

\includegraphics[width = 2.3in, height = 1.3in]{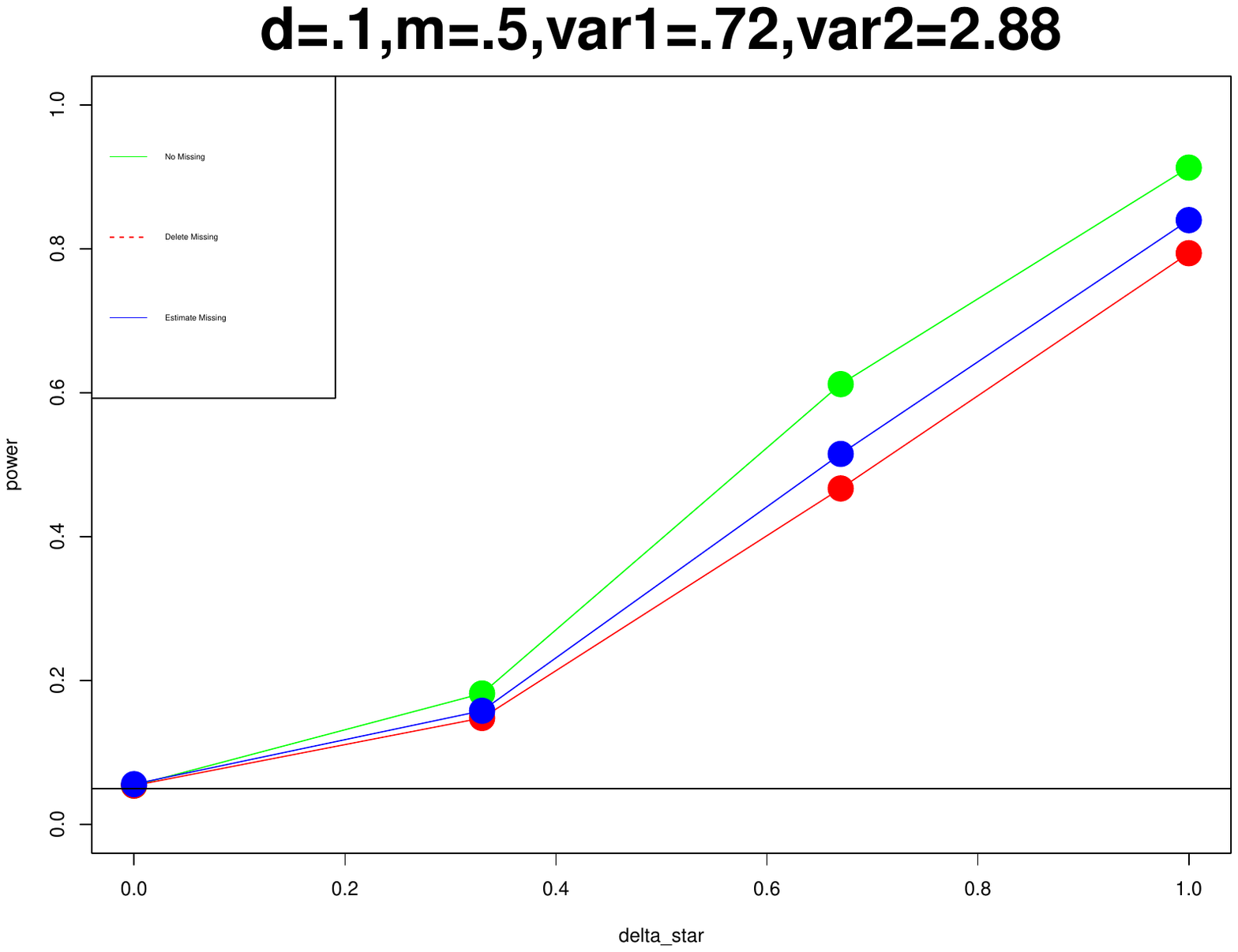}
\includegraphics[width = 2.3in, height = 1.3in]{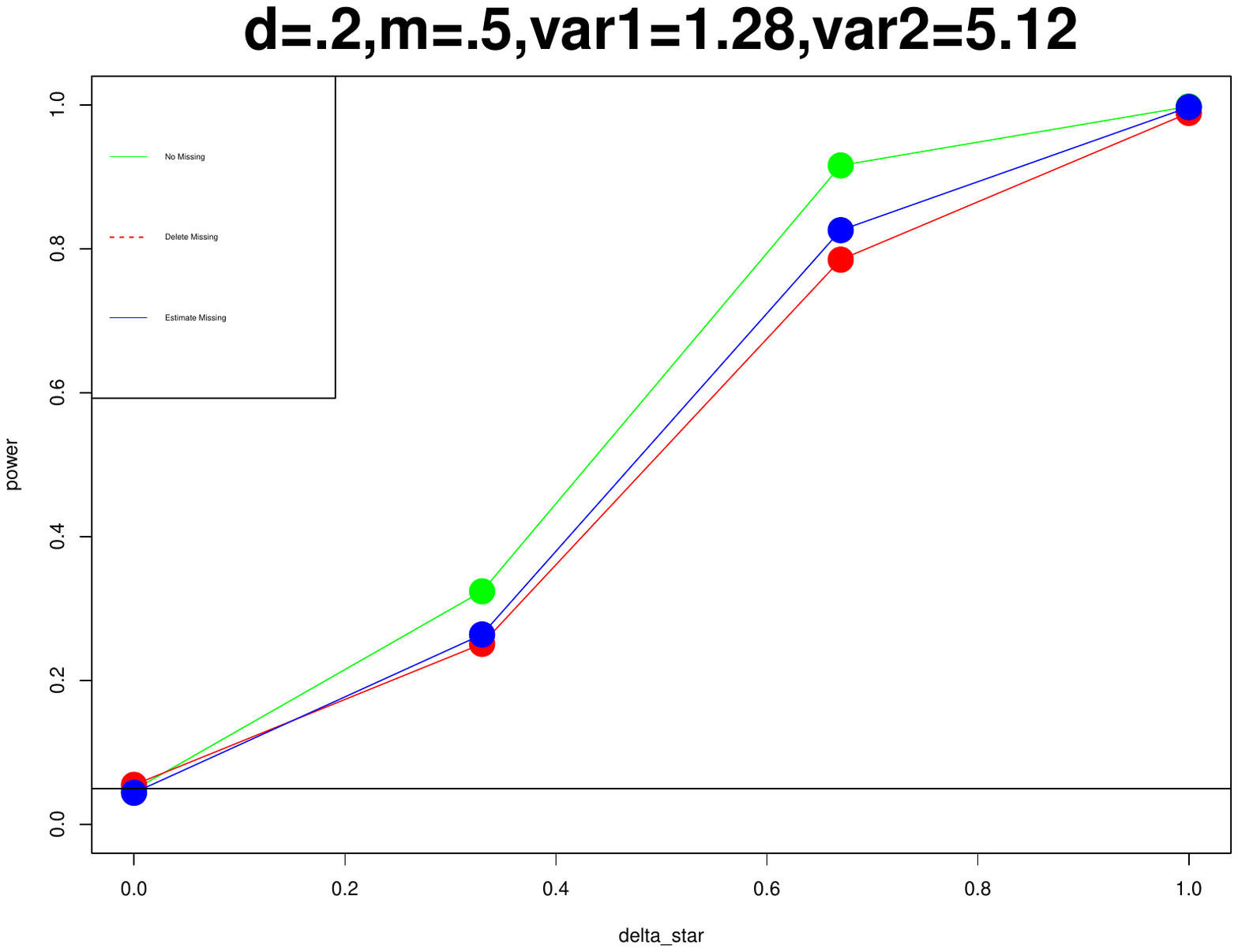}
\includegraphics[width = 2.3in, height = 1.3in]{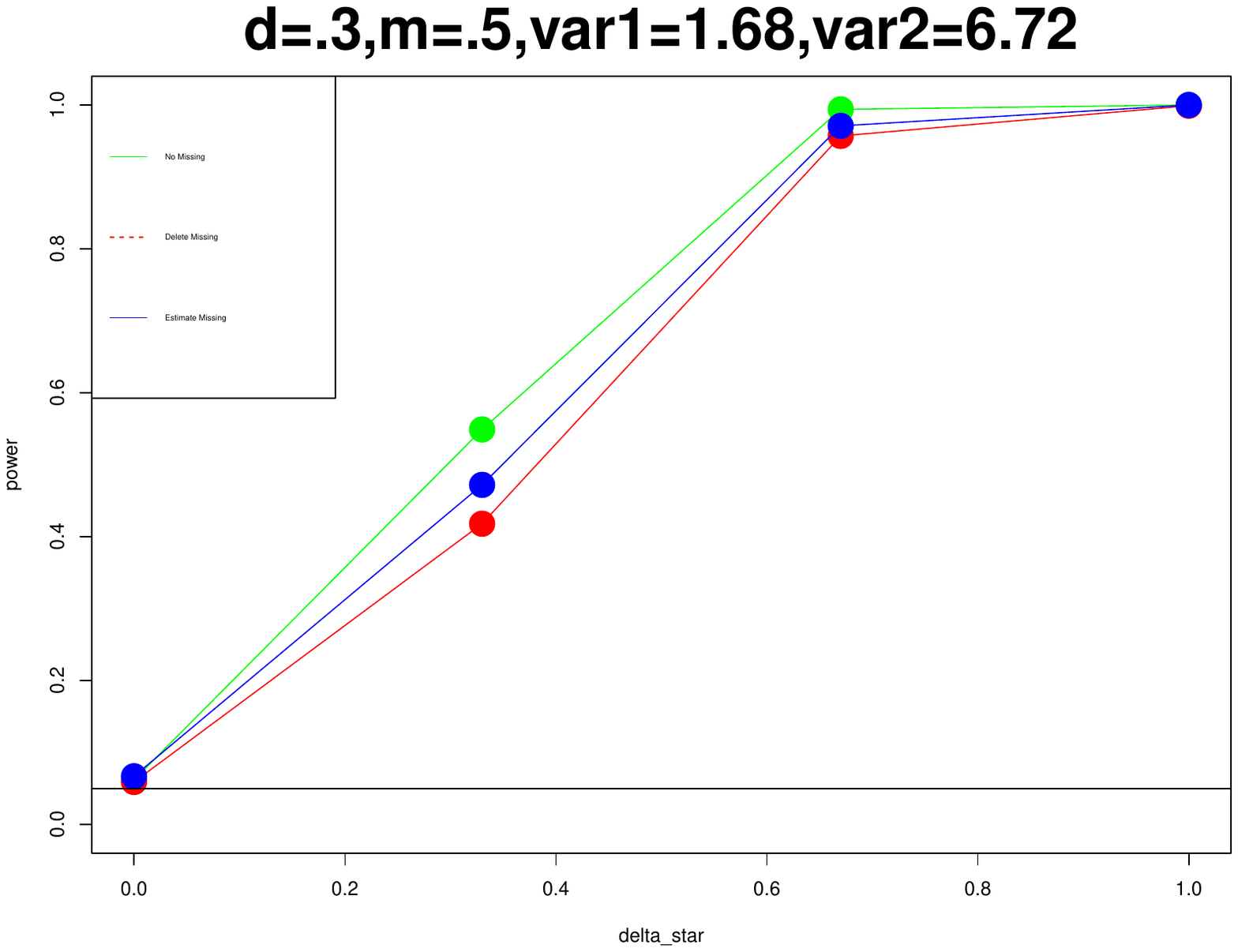}

\hspace{1.5cm}
$d=.1 \quad m=.5$
\hspace{3cm}
$d=.2 \quad m=.5$
\hspace{3cm}
$d=.3 \quad m=.5$

\includegraphics[width = 2.3in, height = 1.3in]{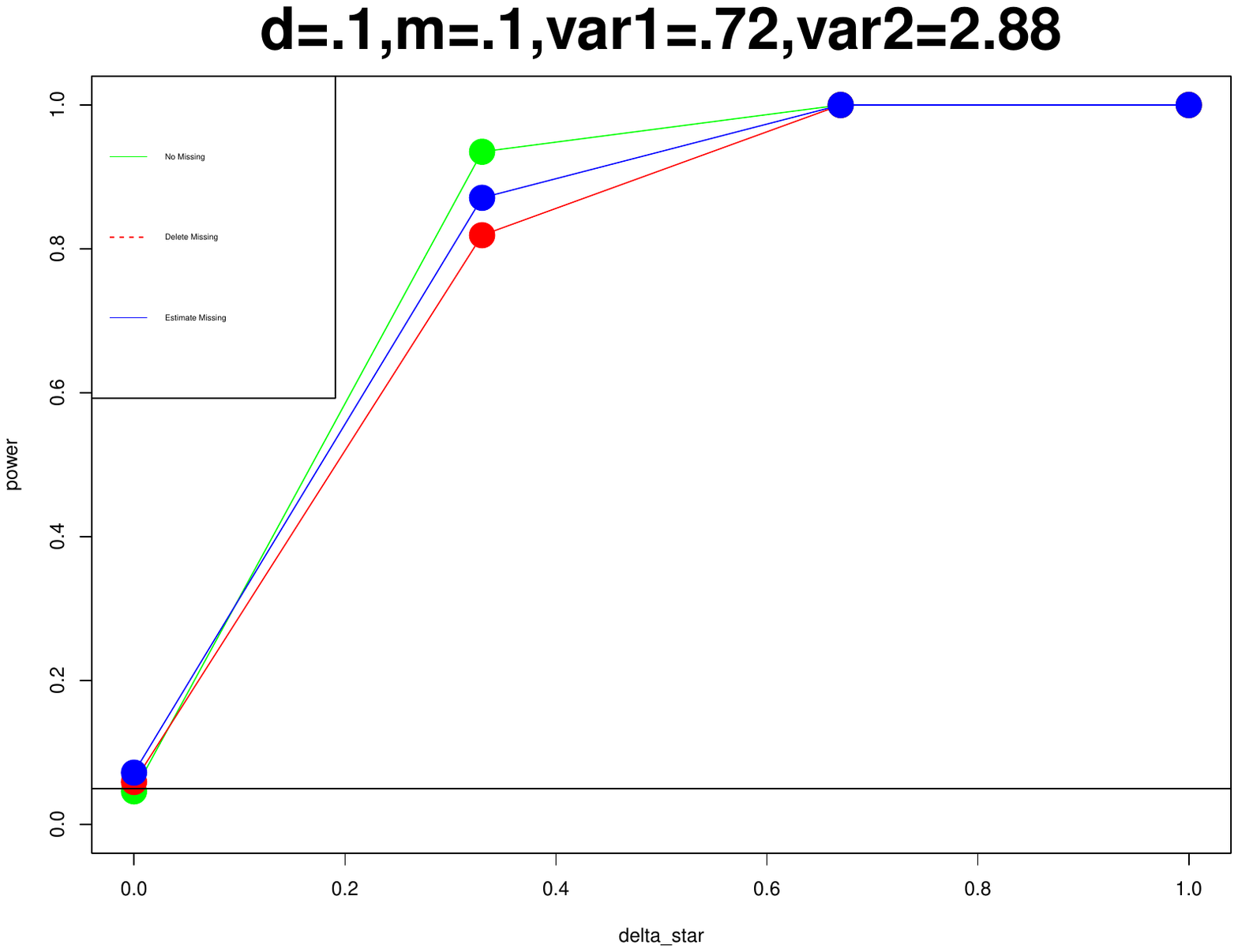}
\includegraphics[width = 2.3in, height = 1.3in]{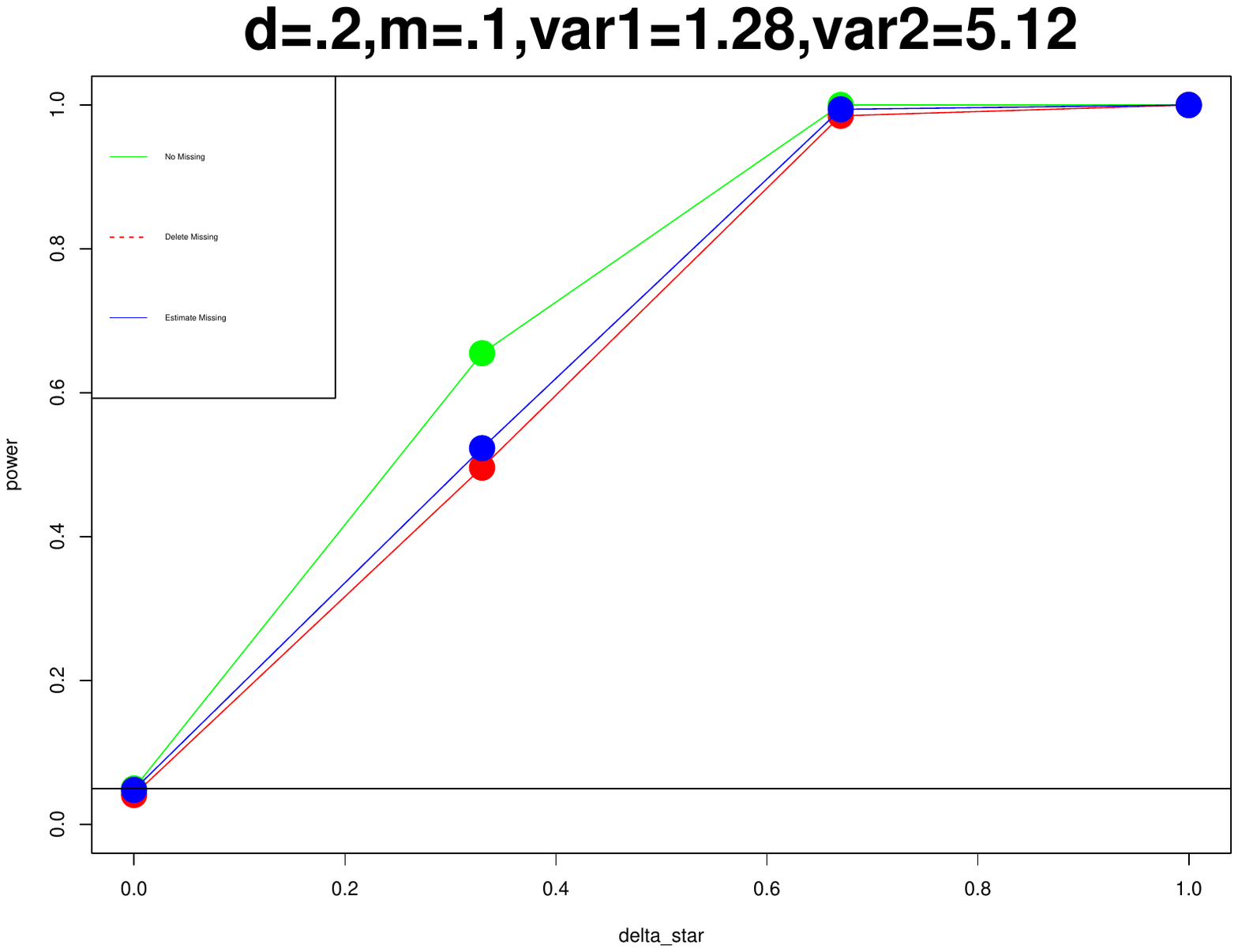}
\includegraphics[width = 2.3in, height = 1.3in]{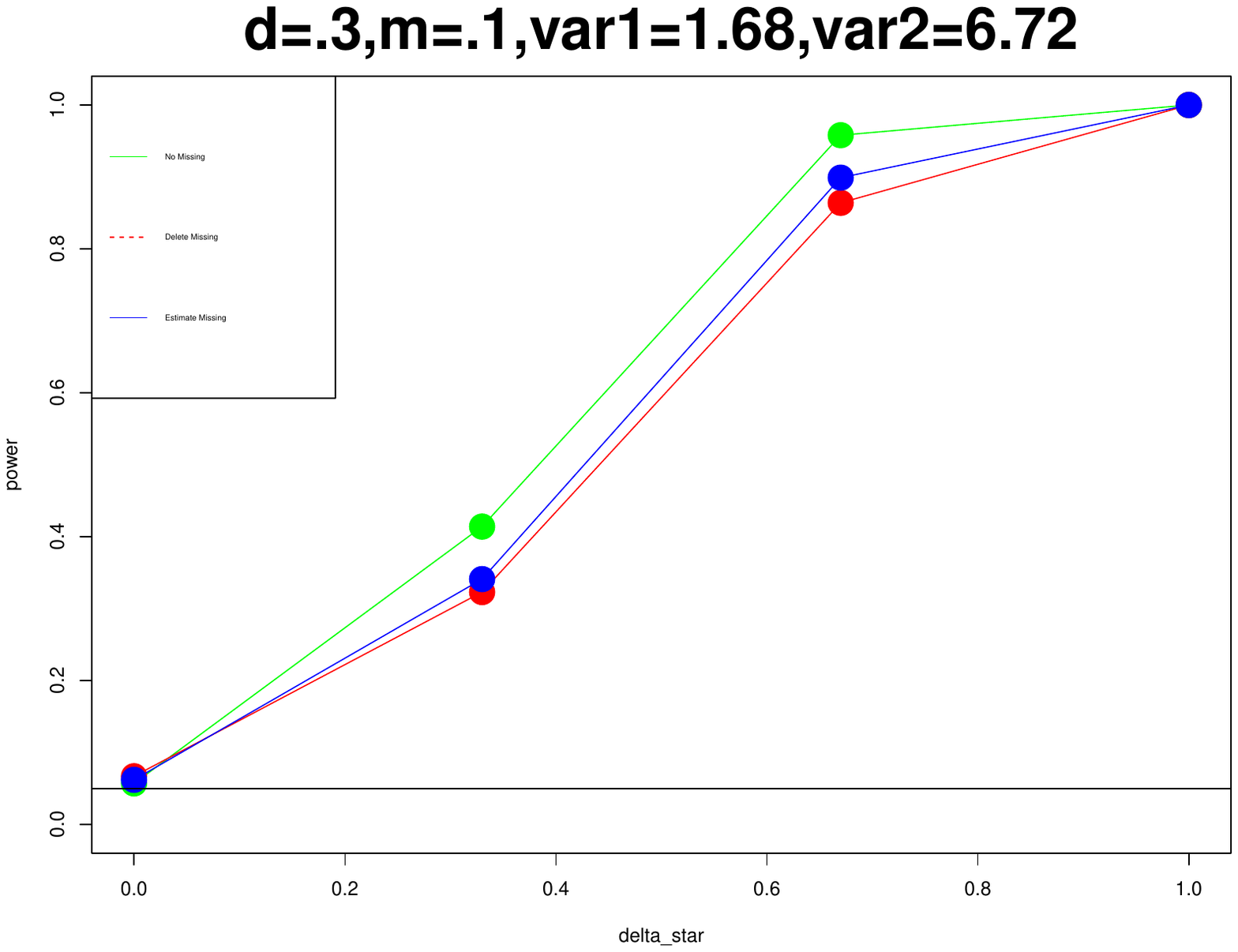}

\subsubsection{When one Trait has Normal Distribution and other Trait has Poisson Distribution}

\hspace{1.5cm}
$d=.1 \quad m=.5$
\hspace{3cm}
$d=.2 \quad m=.5$
\hspace{3cm}
$d=.3 \quad m=.5$

\includegraphics[width = 2.3in, height = 1.3in]{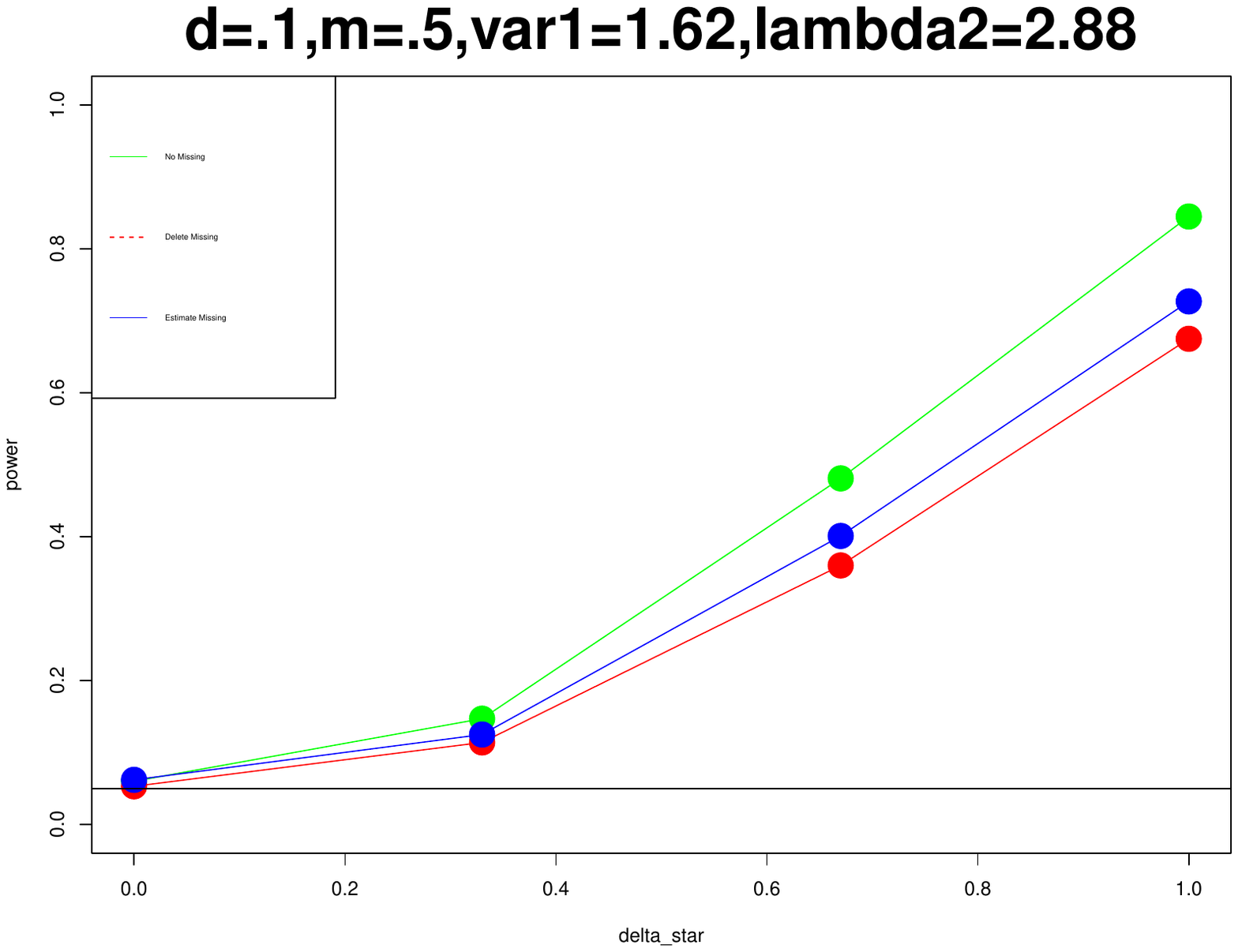}
\includegraphics[width = 2.3in, height = 1.3in]{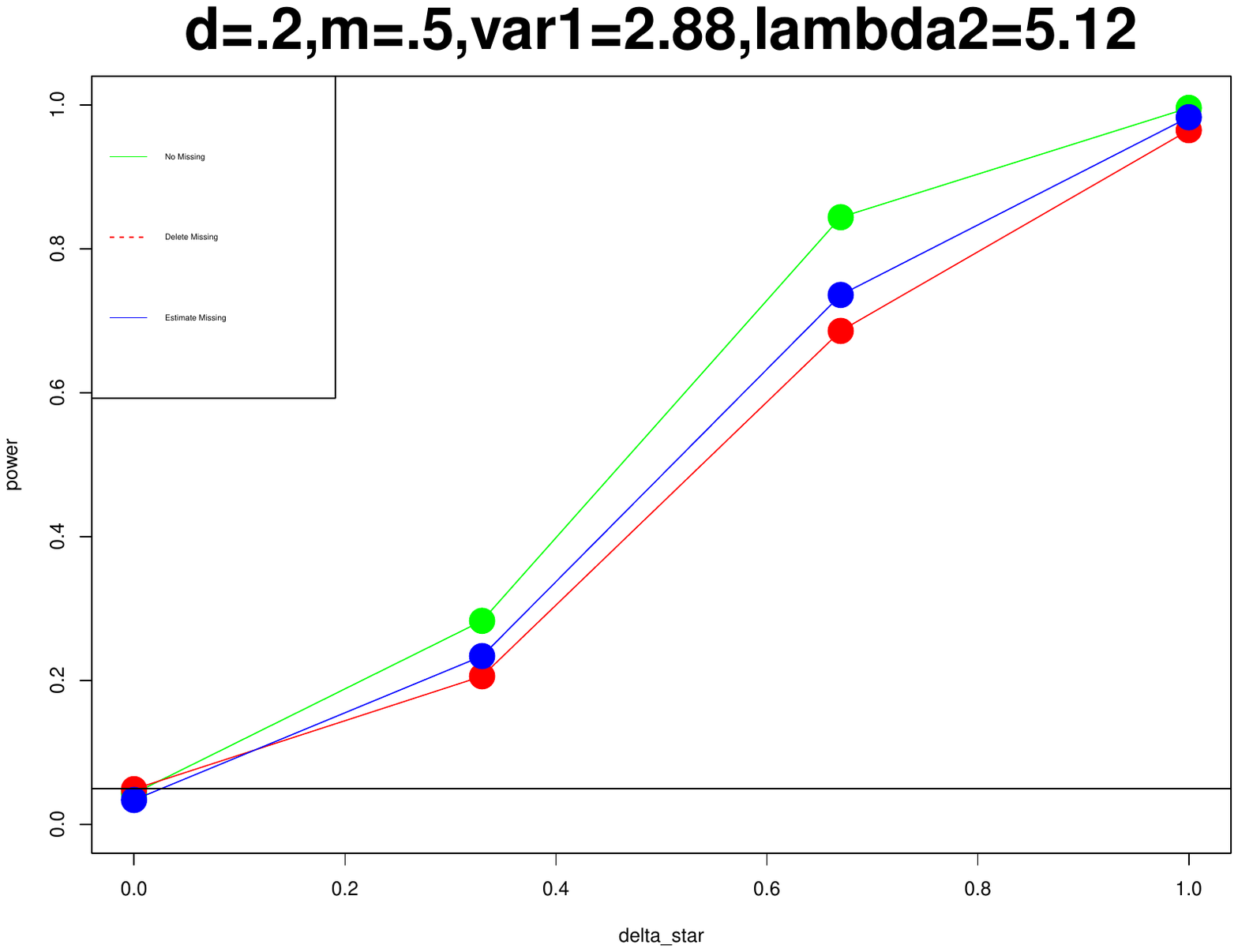}
\includegraphics[width = 2.3in, height = 1.3in]{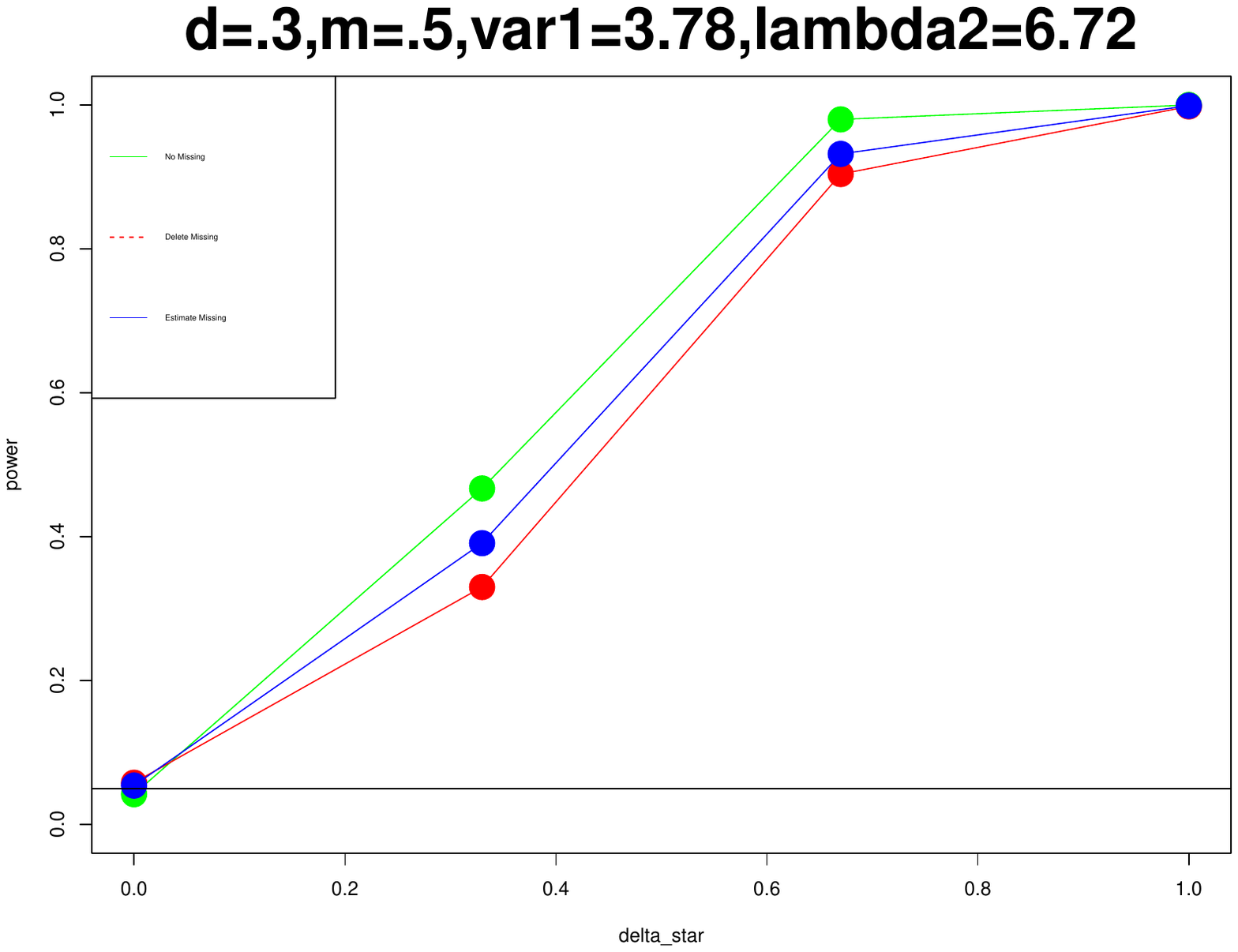}

\hspace{1.5cm}
$d=.1 \quad m=.1$
\hspace{3cm}
$d=.2 \quad m=.1$
\hspace{3cm}
$d=.3 \quad m=.1$

\includegraphics[width = 2.3in, height = 1.5in]{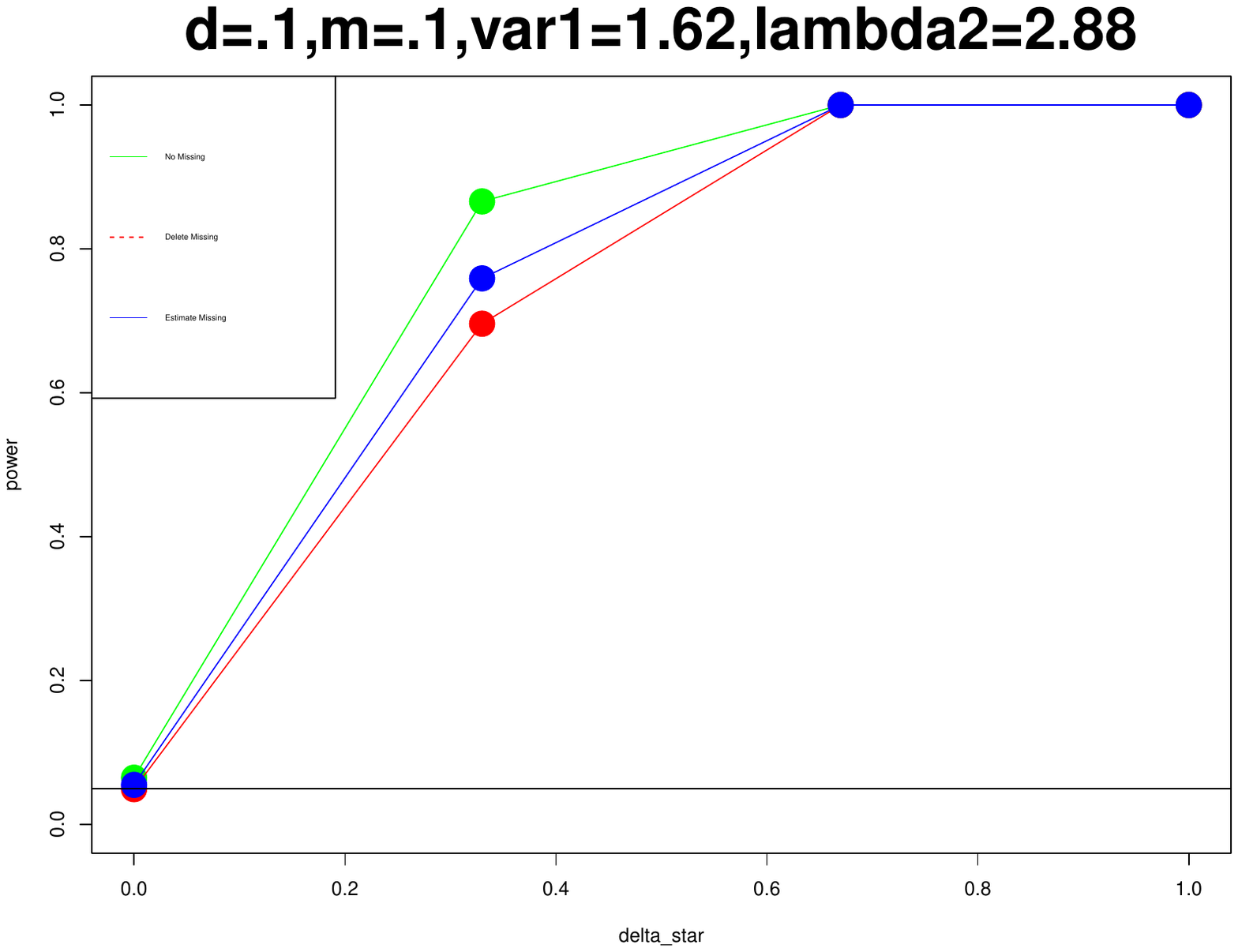}
\includegraphics[width = 2.3in, height = 1.5in]{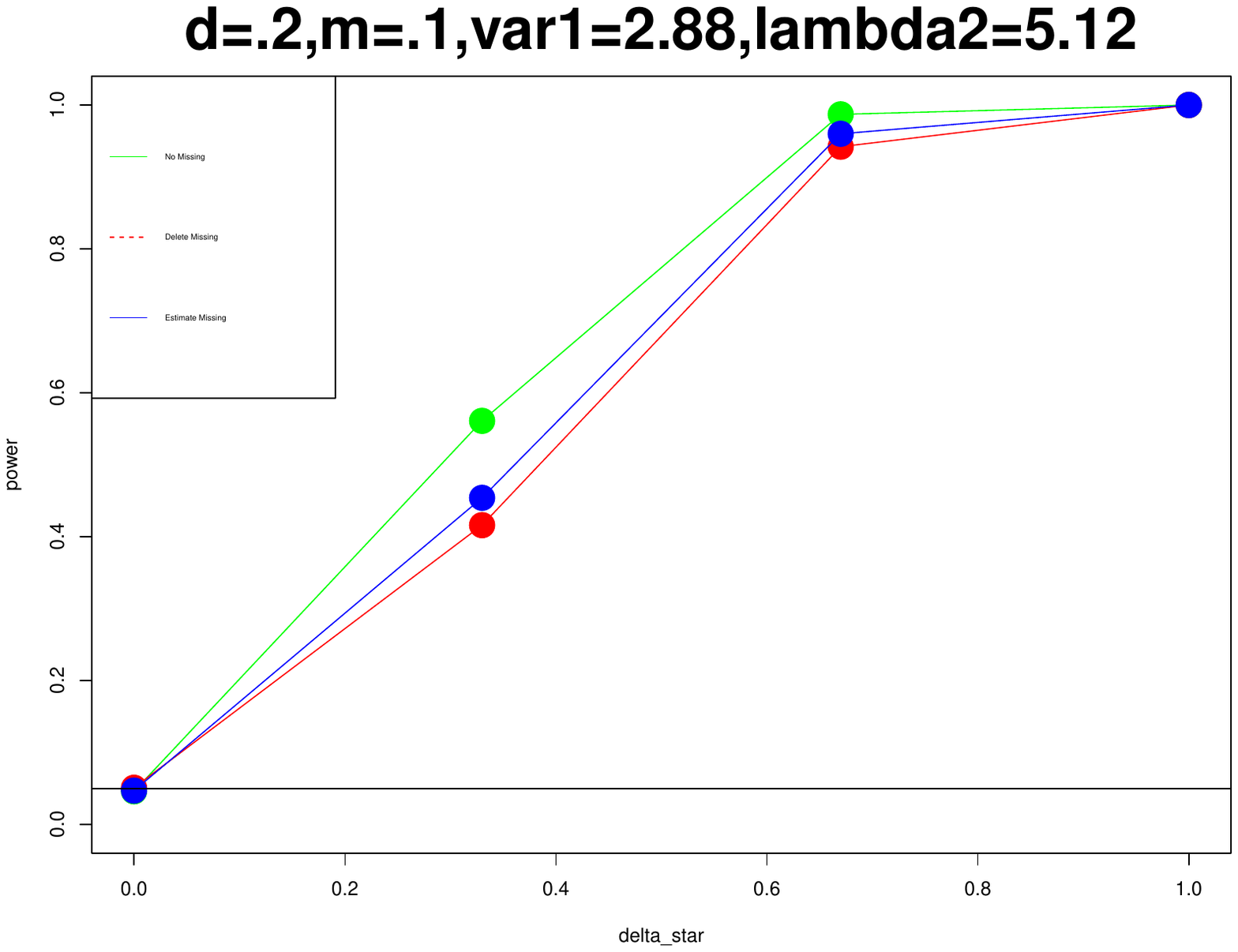}
\includegraphics[width = 2.3in, height = 1.5in]{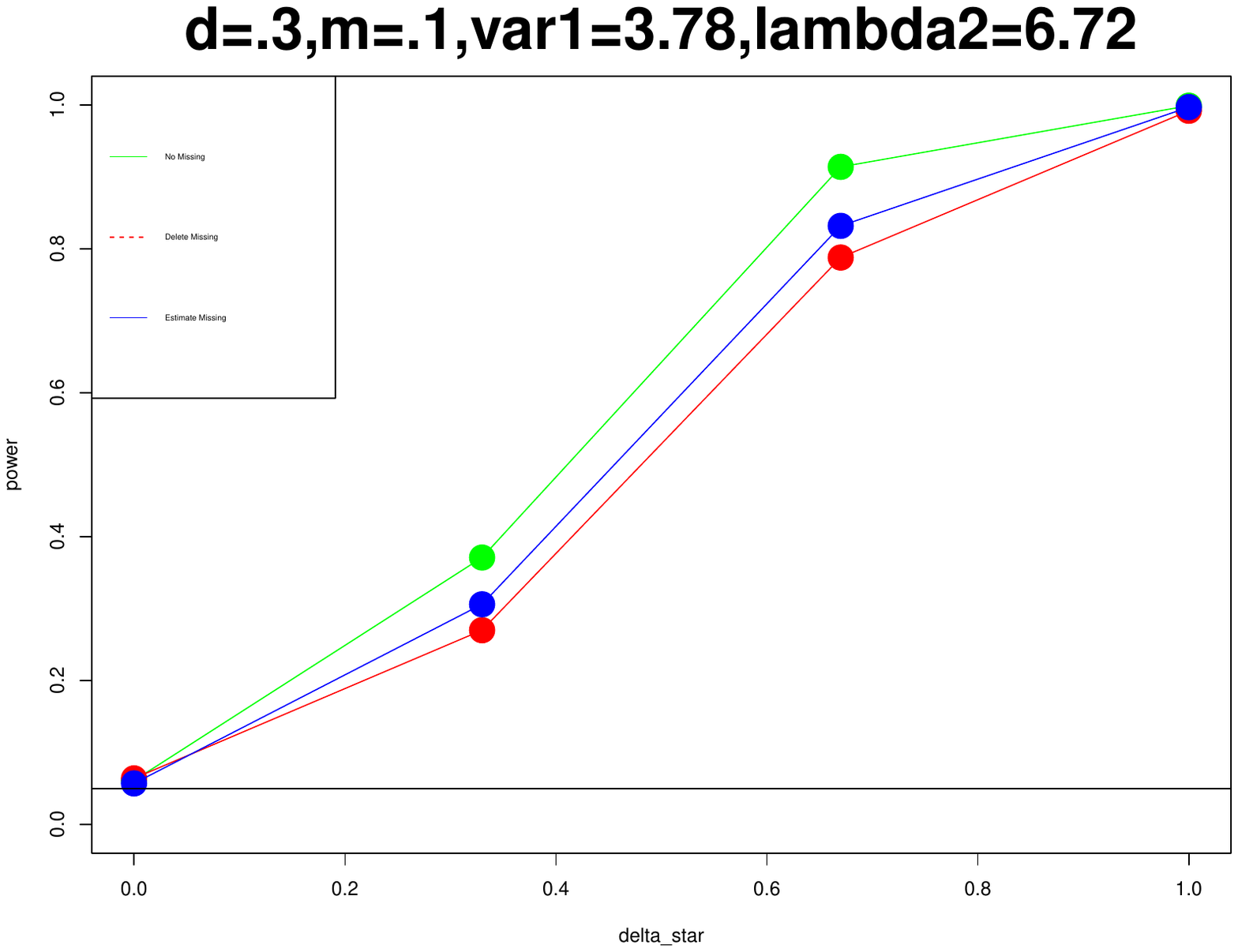}

\hspace{1.5cm}
$d=.1 \quad m=.1$
\hspace{3cm}
$d=.2 \quad m=.1$
\hspace{3cm}
$d=.3 \quad m=.1$

\includegraphics[width = 2.3in, height = 1.5in]{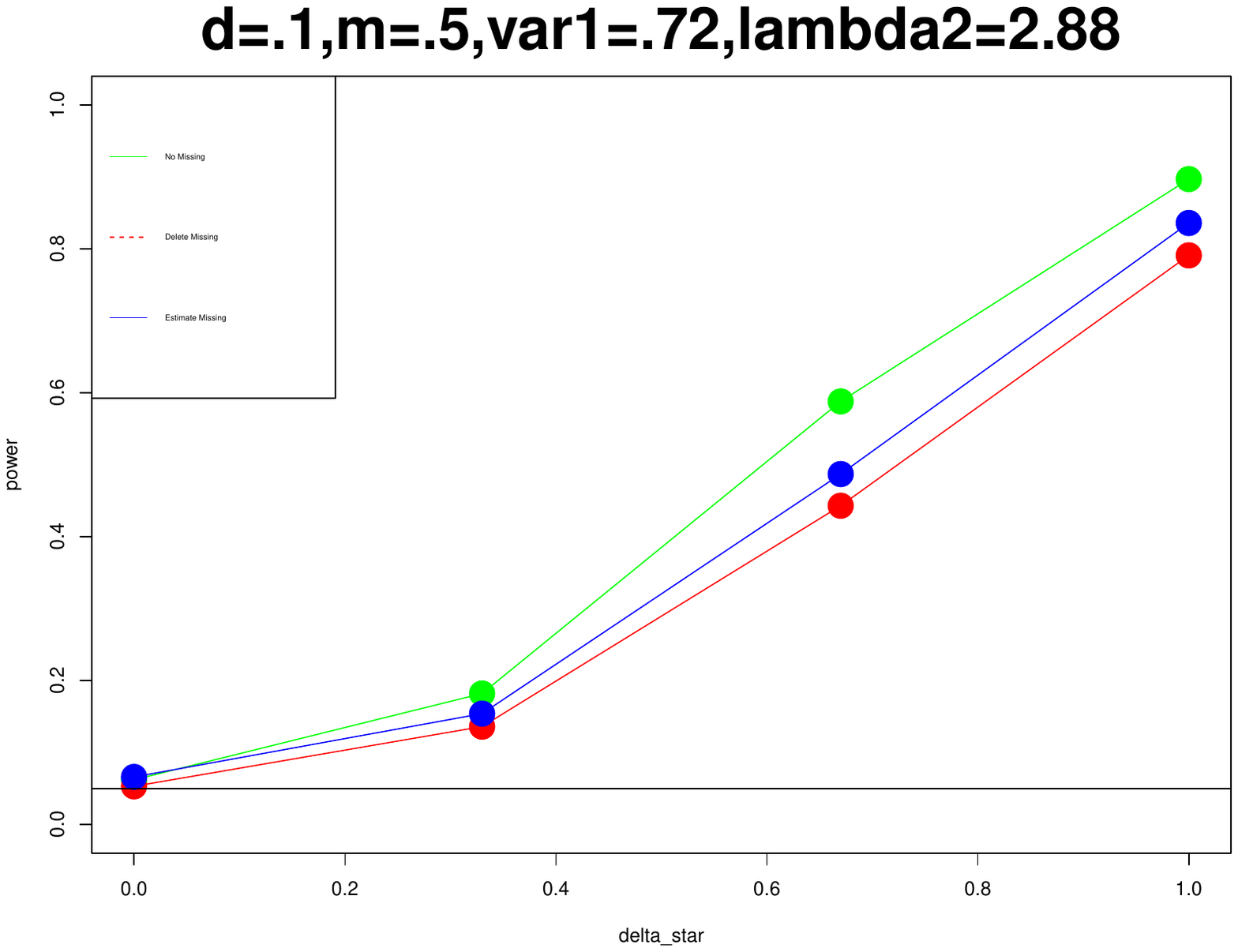}
\includegraphics[width = 2.3in, height = 1.5in]{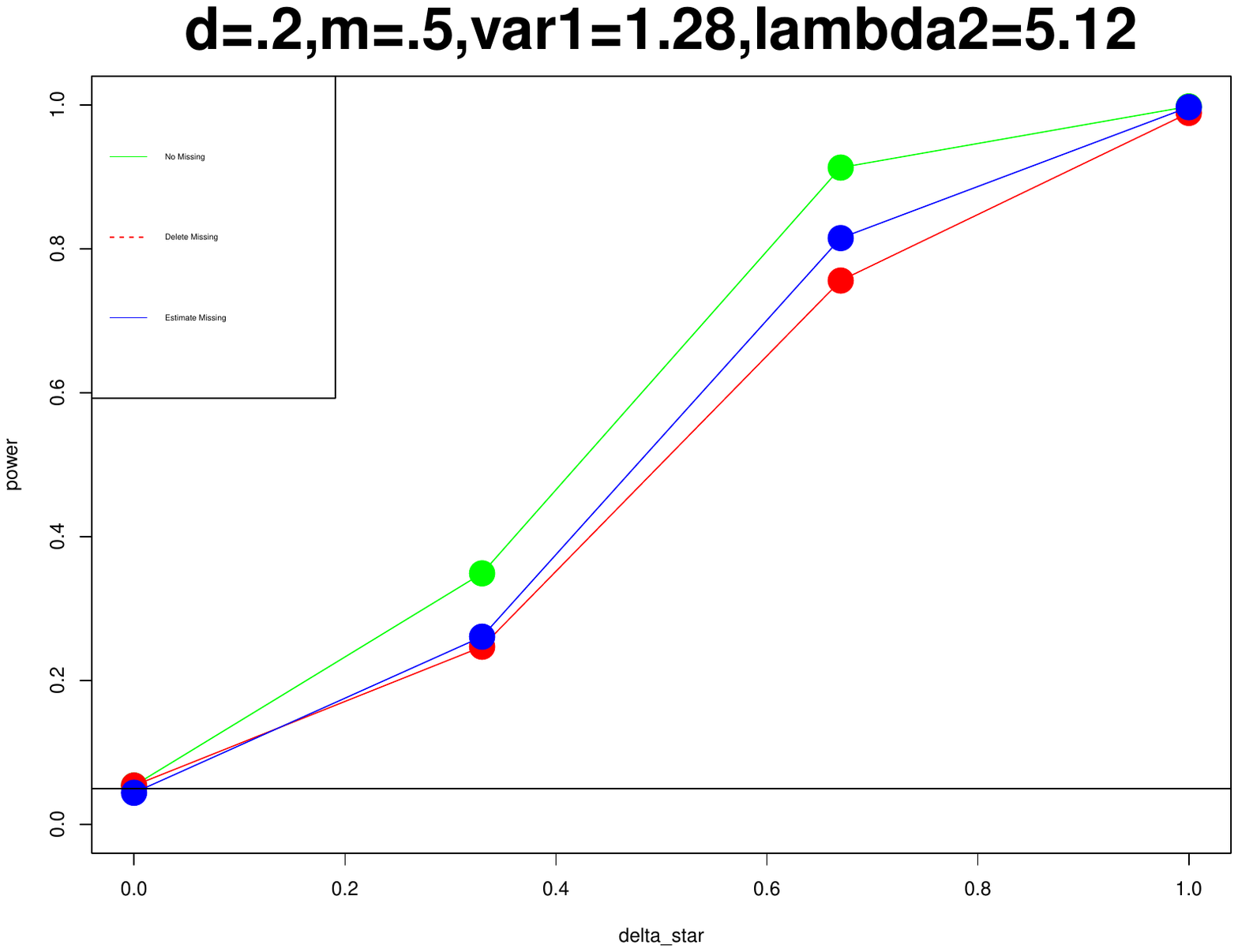}
\includegraphics[width = 2.3in, height = 1.5in]{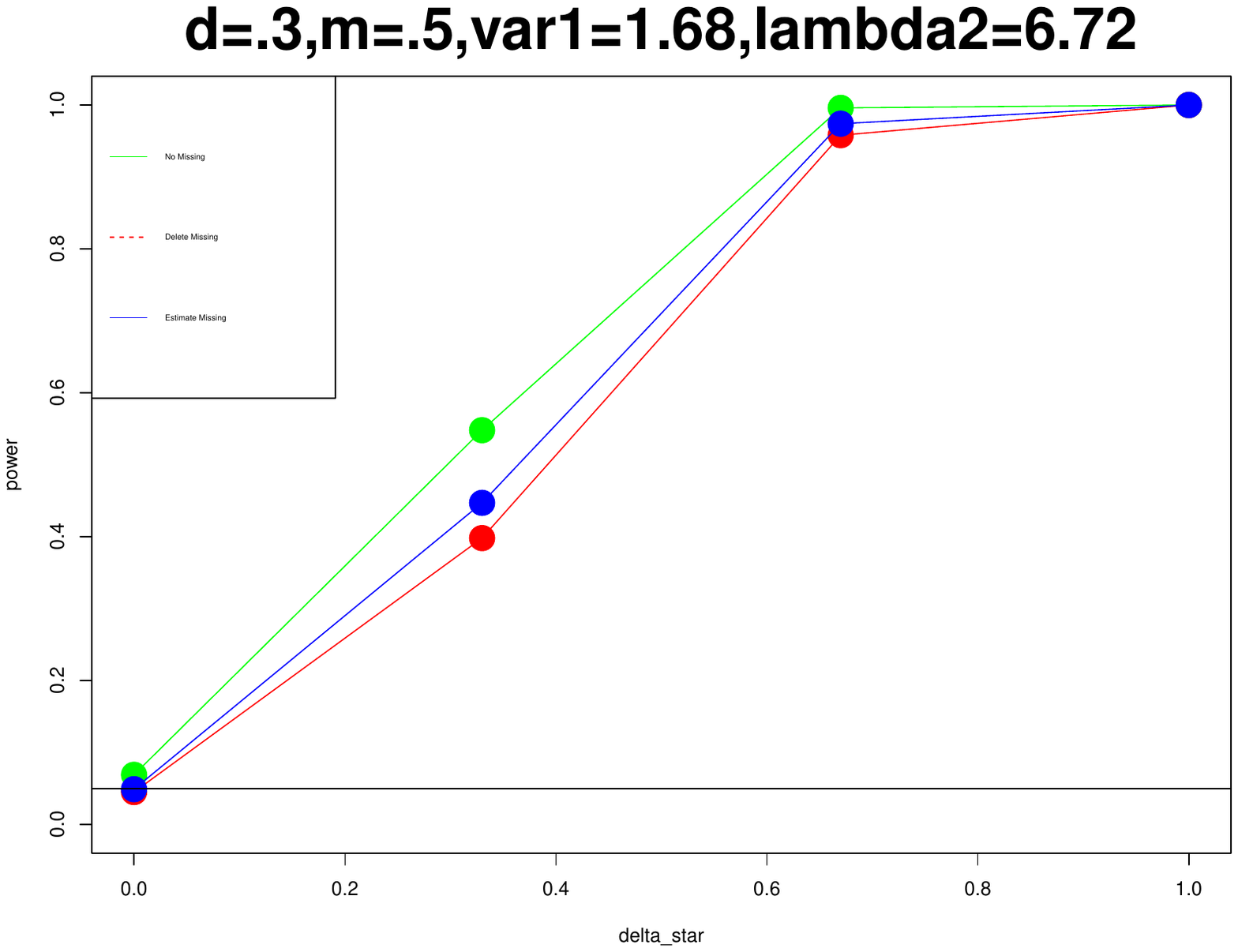}

\hspace{1.5cm}
$d=.1 \quad m=.5$
\hspace{3cm}
$d=.2 \quad m=.5$
\hspace{3cm}
$d=.3 \quad m=.5$

\includegraphics[width = 2.3in, height = 1.3in]{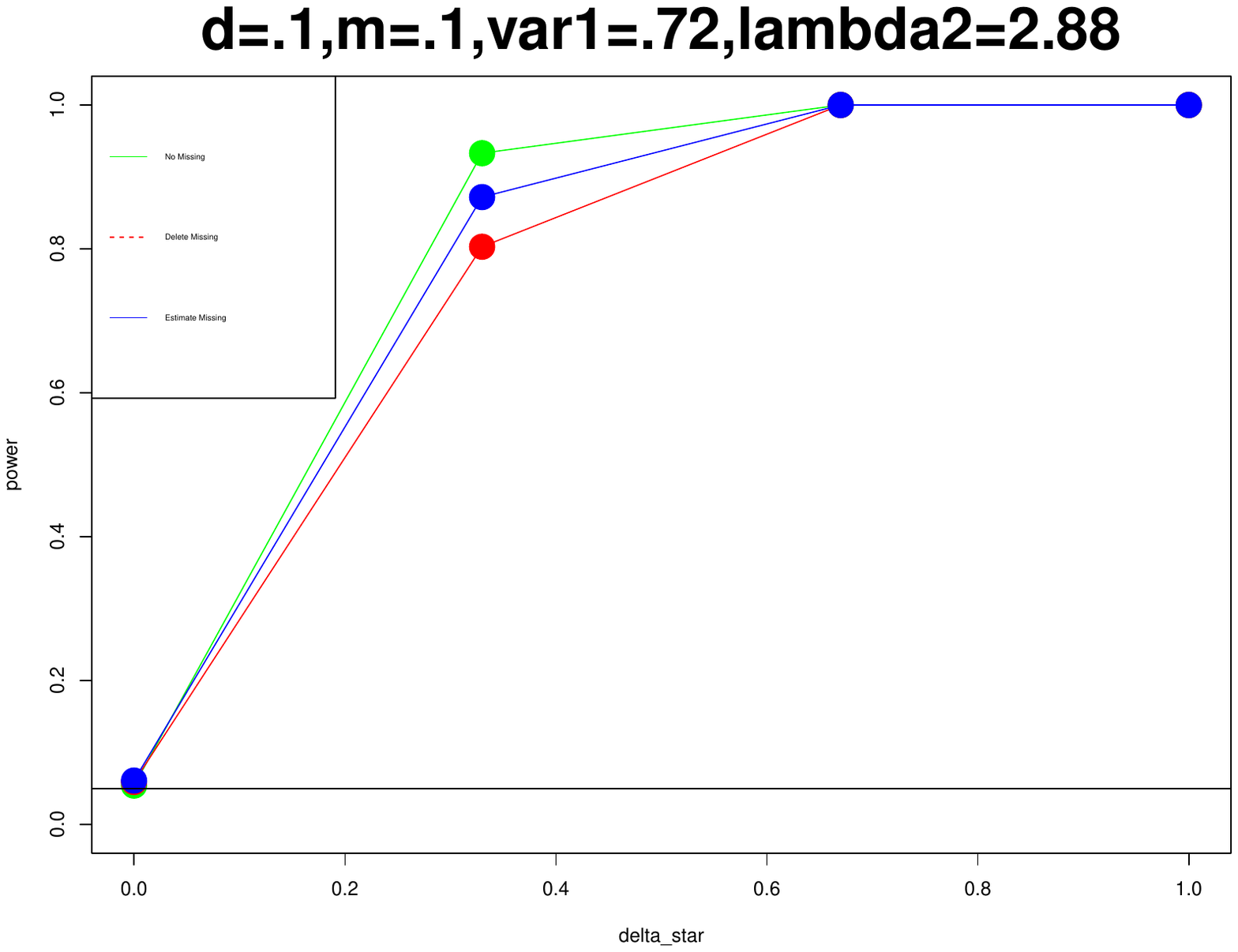}
\includegraphics[width = 2.3in, height = 1.3in]{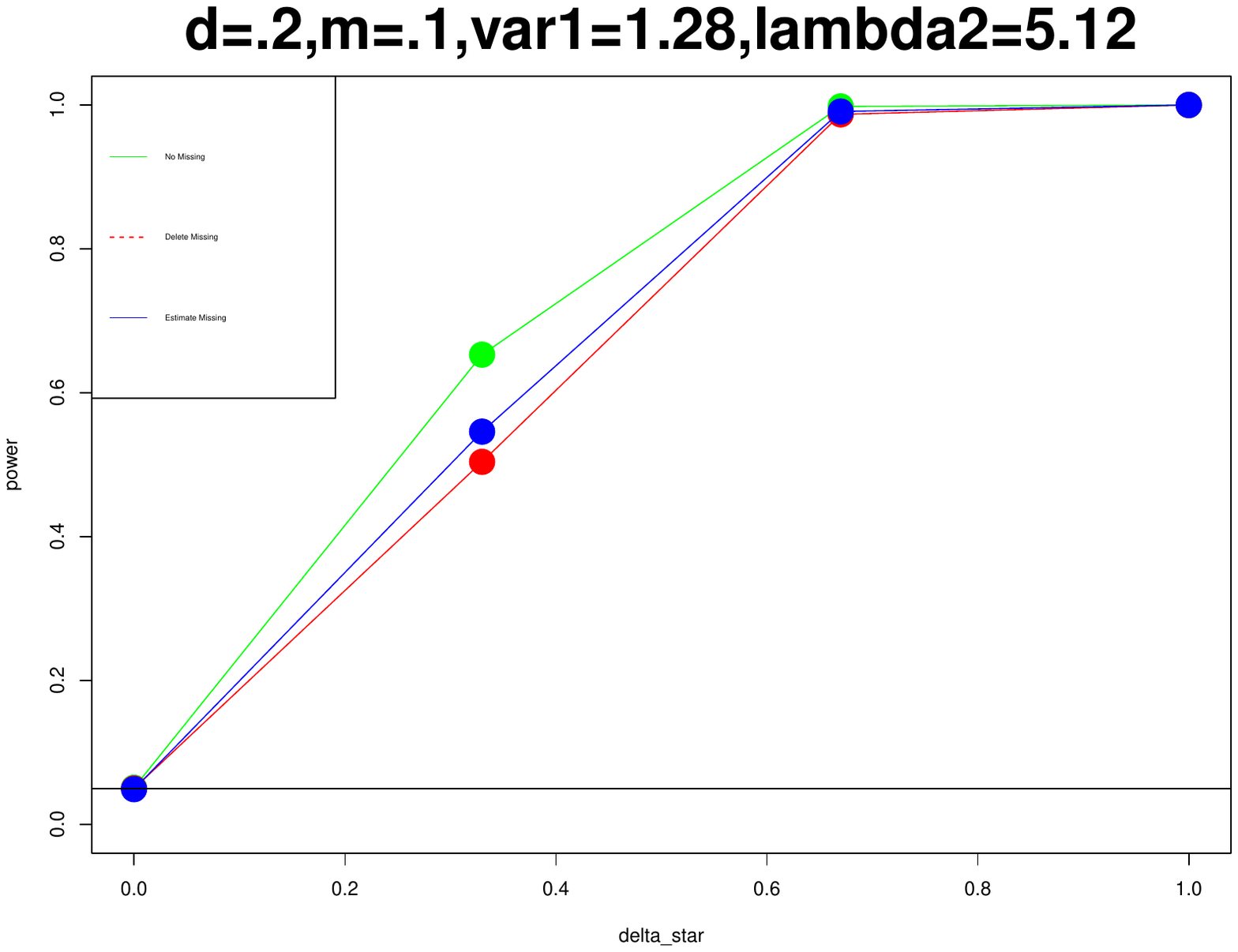}
\includegraphics[width = 2.3in, height = 1.3in]{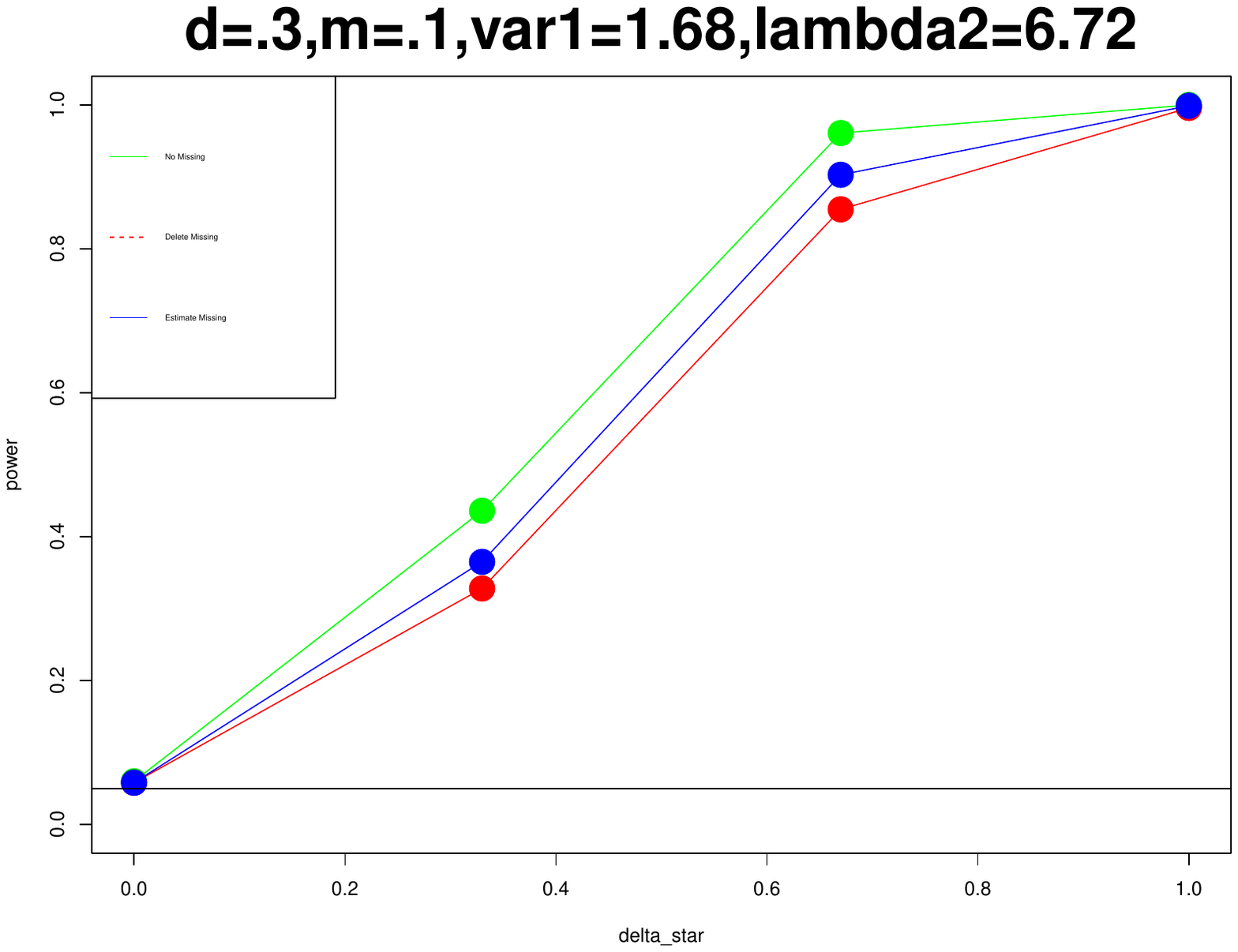}

\hspace{1.5cm}
$d=.1 \quad m=.1$
\hspace{3cm}
$d=.2 \quad m=.1$
\hspace{3cm}
$d=.3 \quad m=.1$

\includegraphics[width = 2.3in, height = 1.3in]{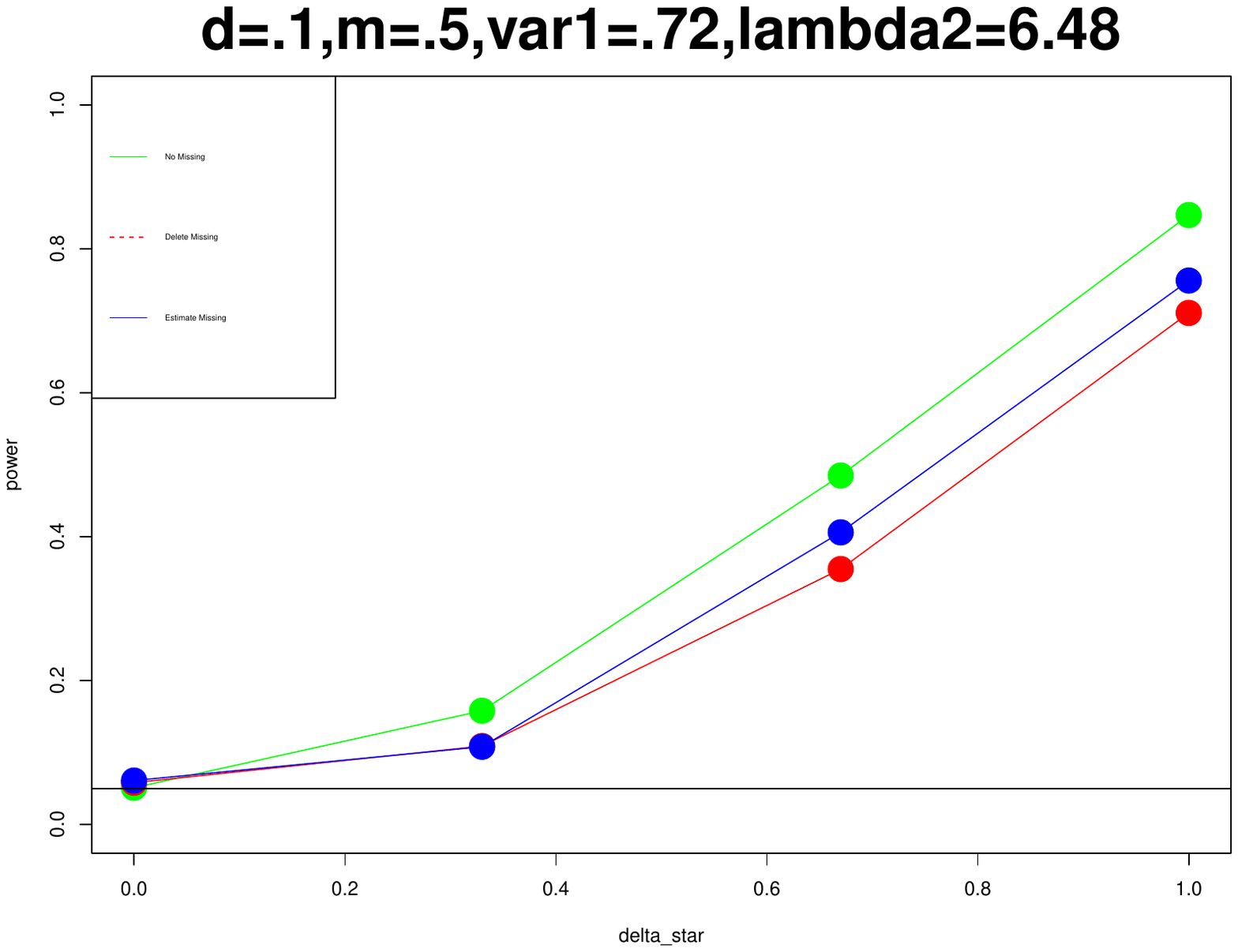}
\includegraphics[width = 2.3in, height = 1.3in]{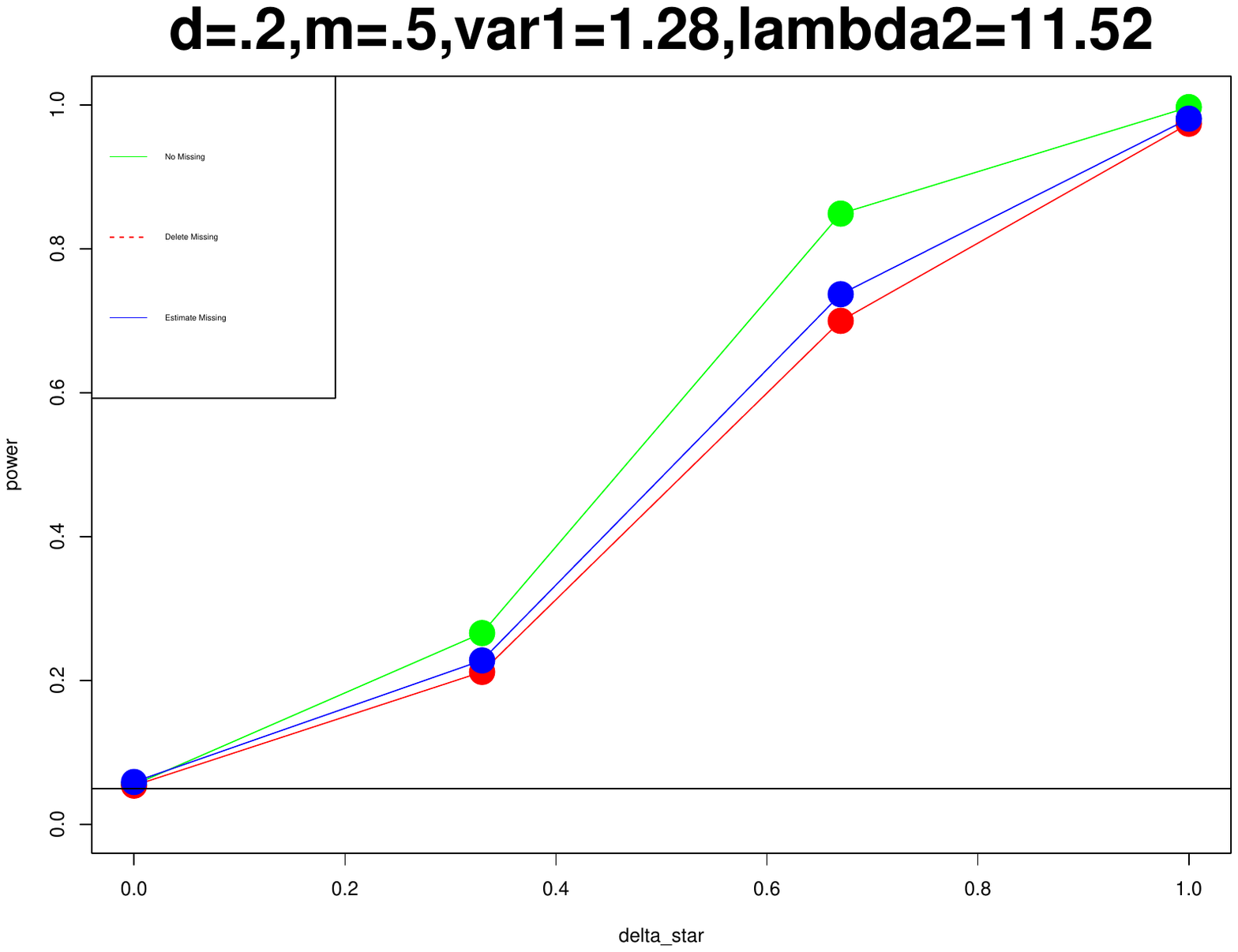}
\includegraphics[width = 2.3in, height = 1.3in]{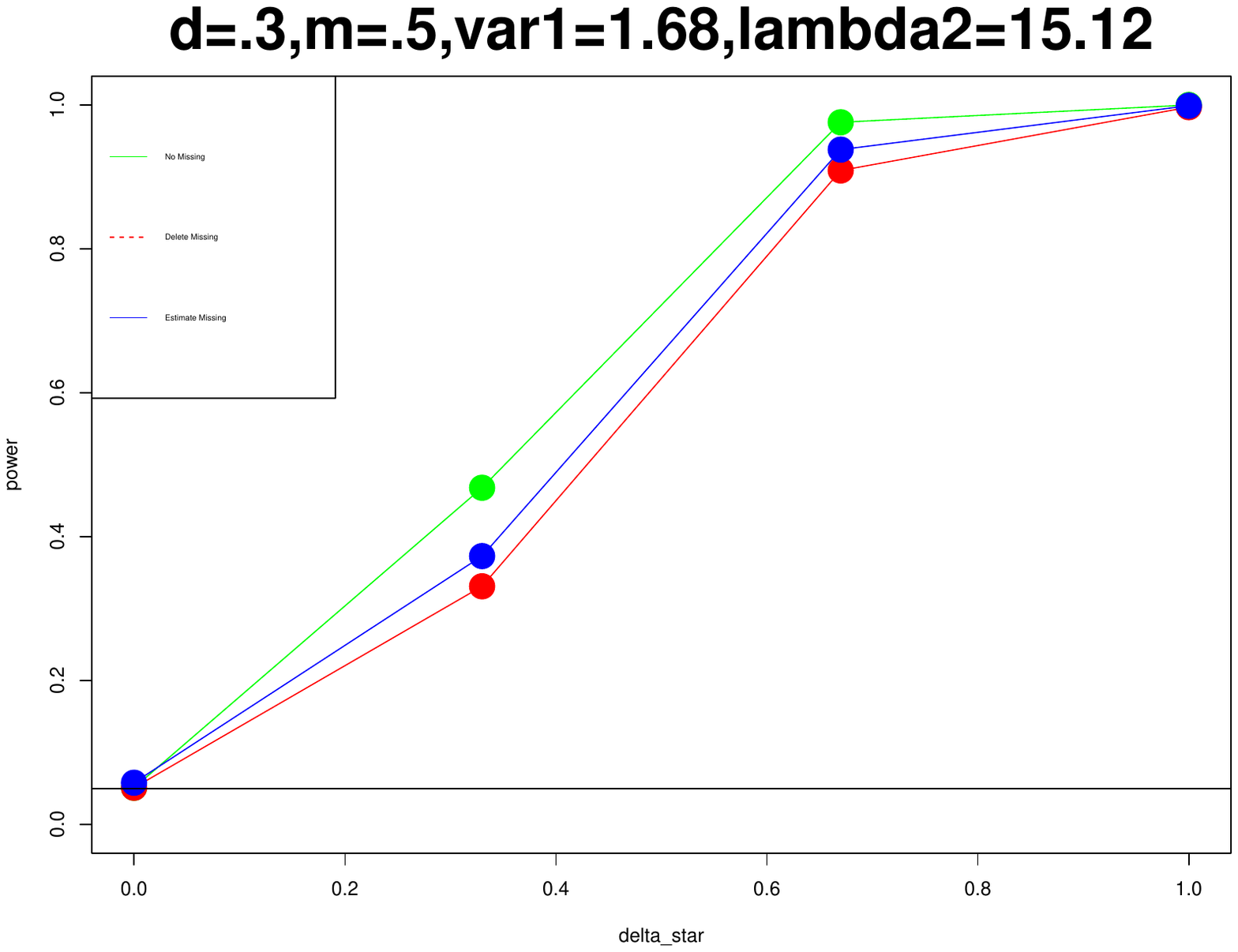}

\hspace{1.5cm}
$d=.1 \quad m=.5$
\hspace{3cm}
$d=.2 \quad m=.5$
\hspace{3cm}
$d=.3 \quad m=.5$

\includegraphics[width = 2.3in, height = 1.3in]{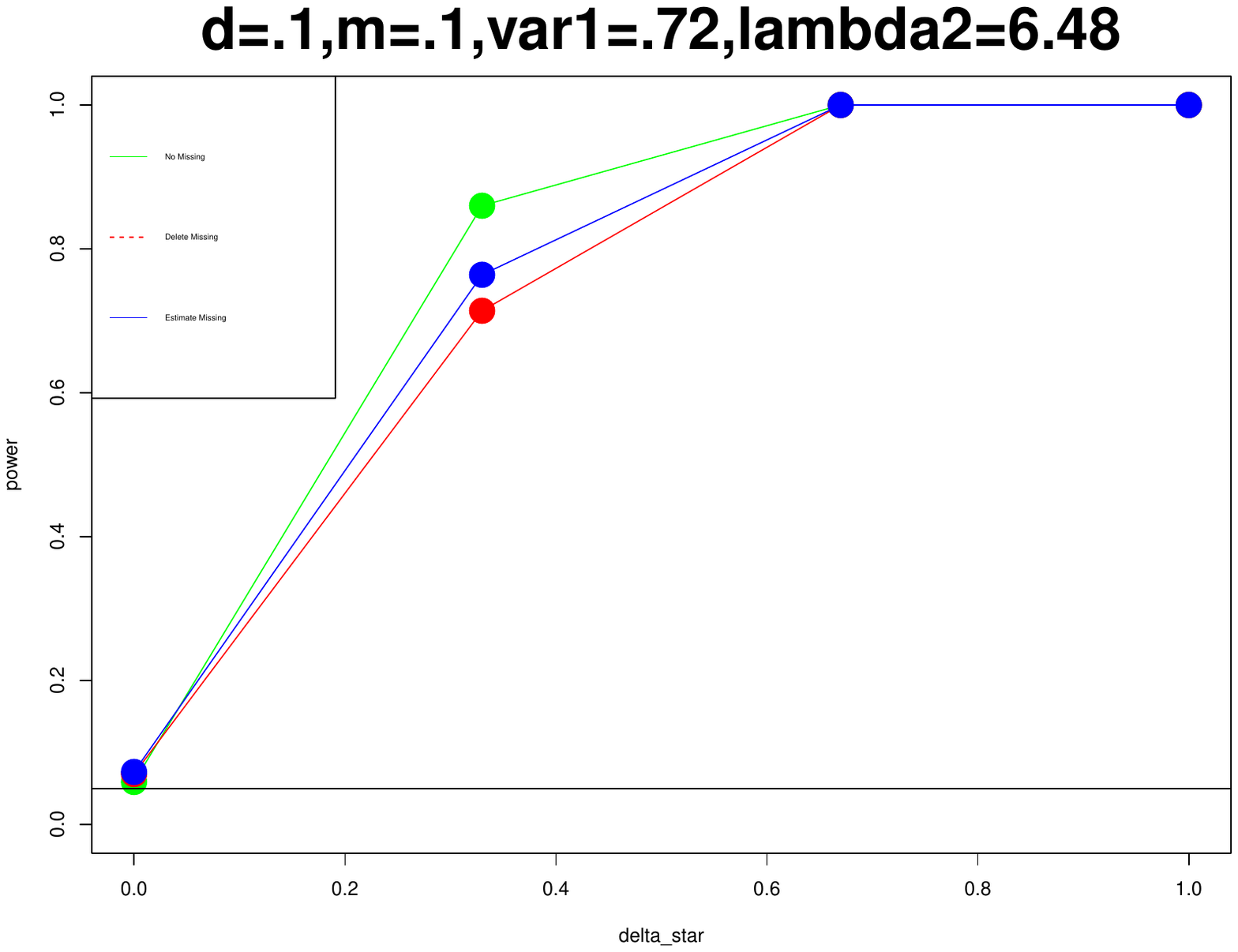}
\includegraphics[width = 2.3in, height = 1.3in]{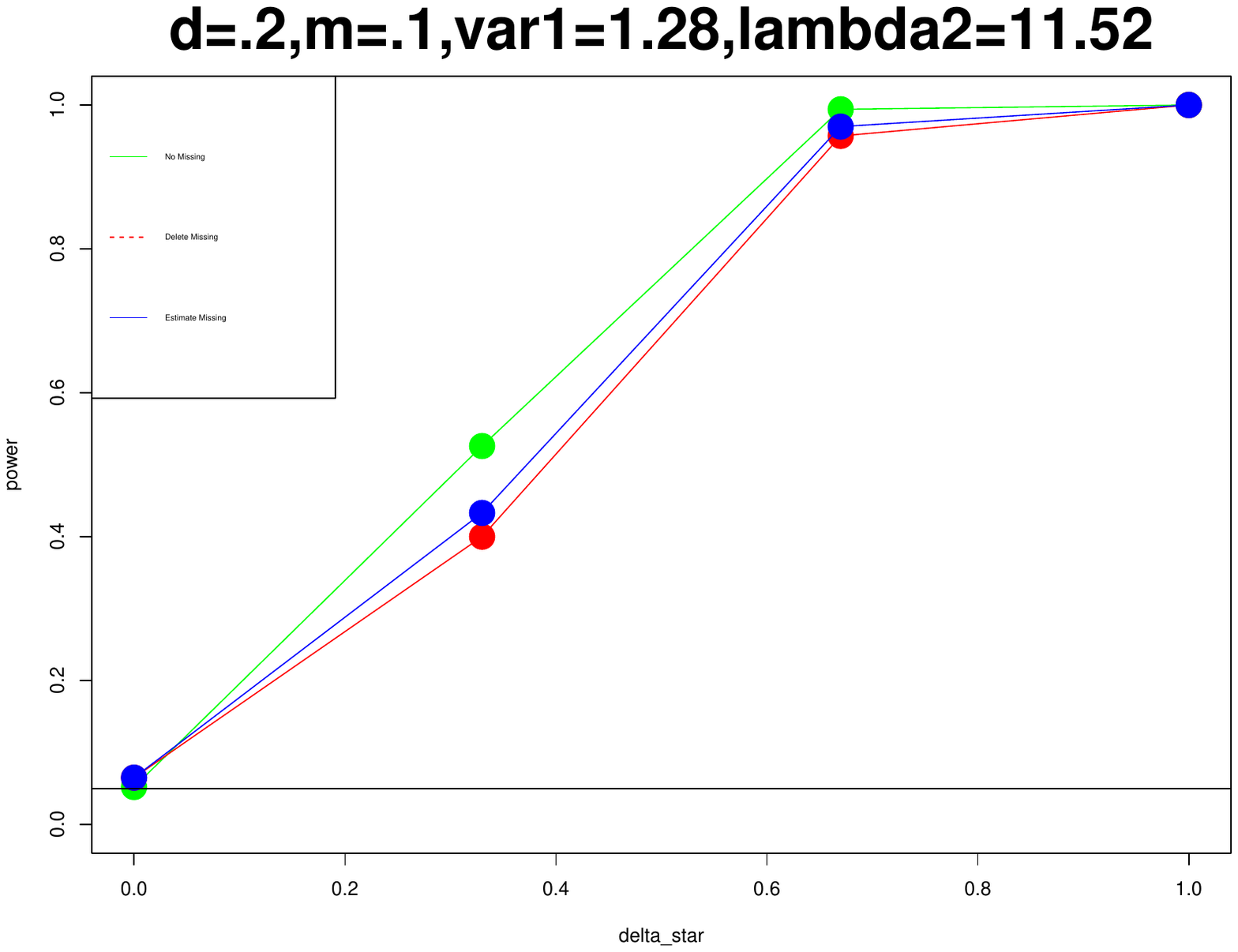}
\includegraphics[width = 2.3in, height = 1.3in]{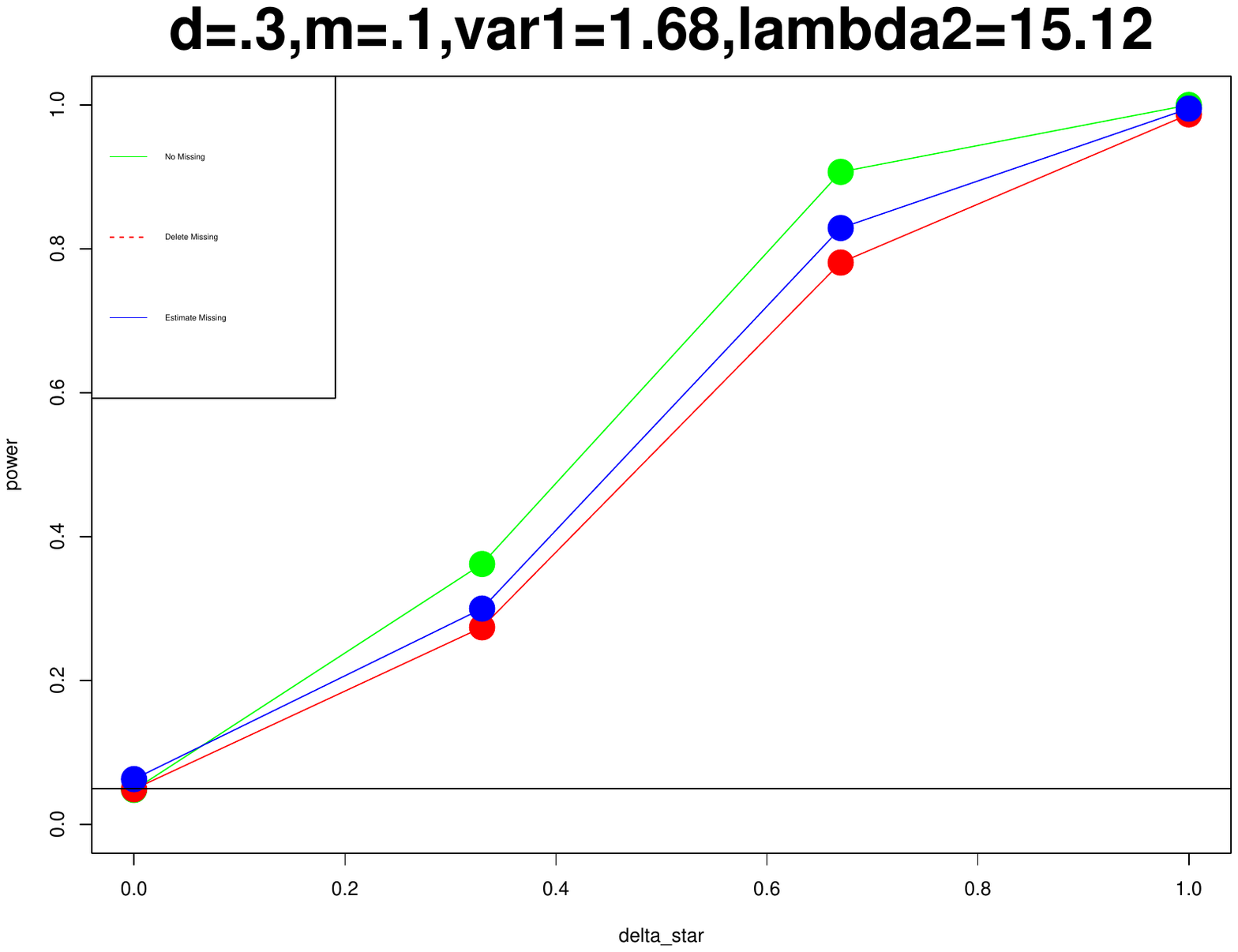}

\subsubsection{When both Traits have Poisson Distribution}

\hspace{1.5cm}
$d=.1 \quad m=.5$
\hspace{3cm}
$d=.2 \quad m=.5$
\hspace{3cm}
$d=.3 \quad m=.5$

\includegraphics[width = 2.3in, height = 1.4in]{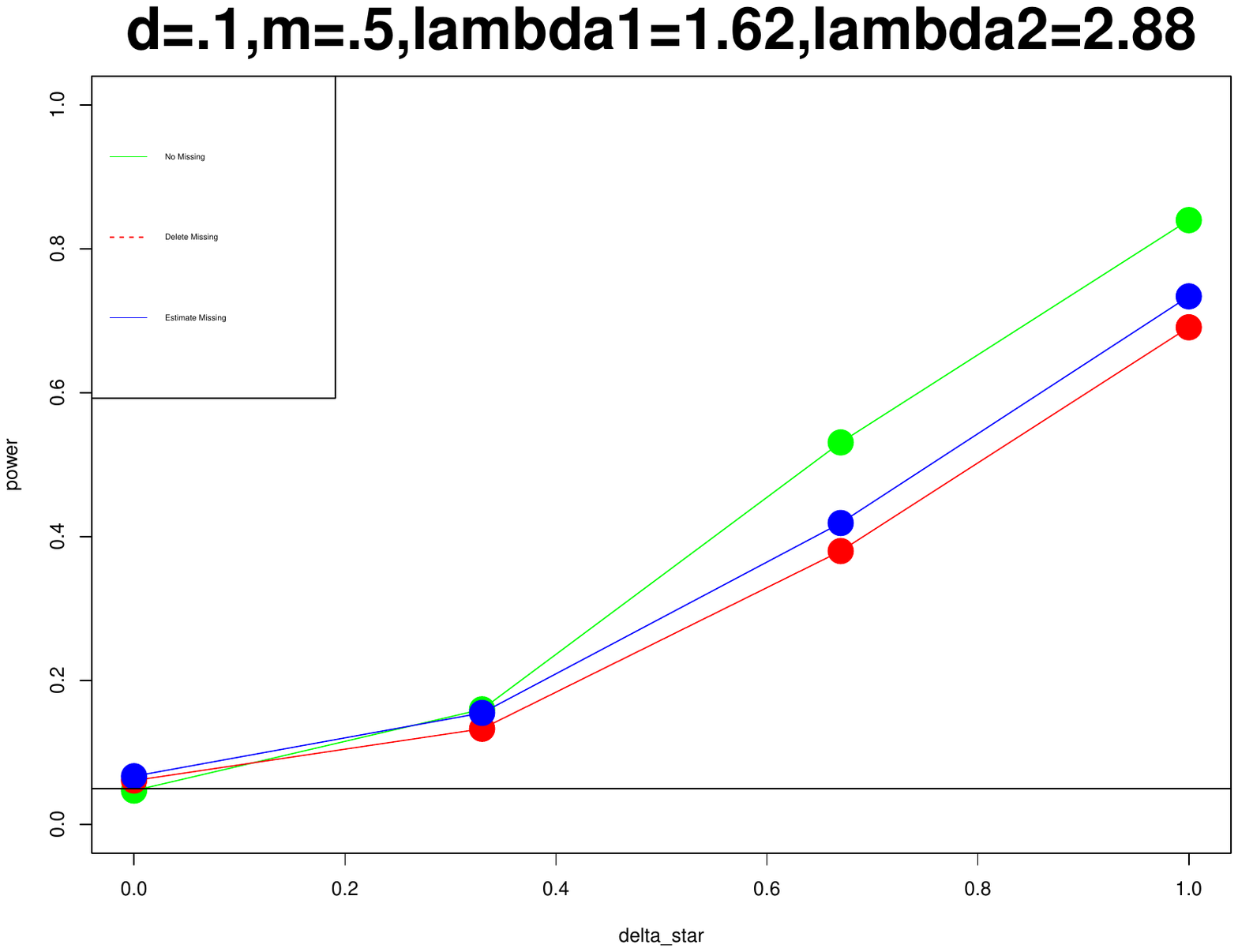}
\includegraphics[width = 2.3in, height = 1.4in]{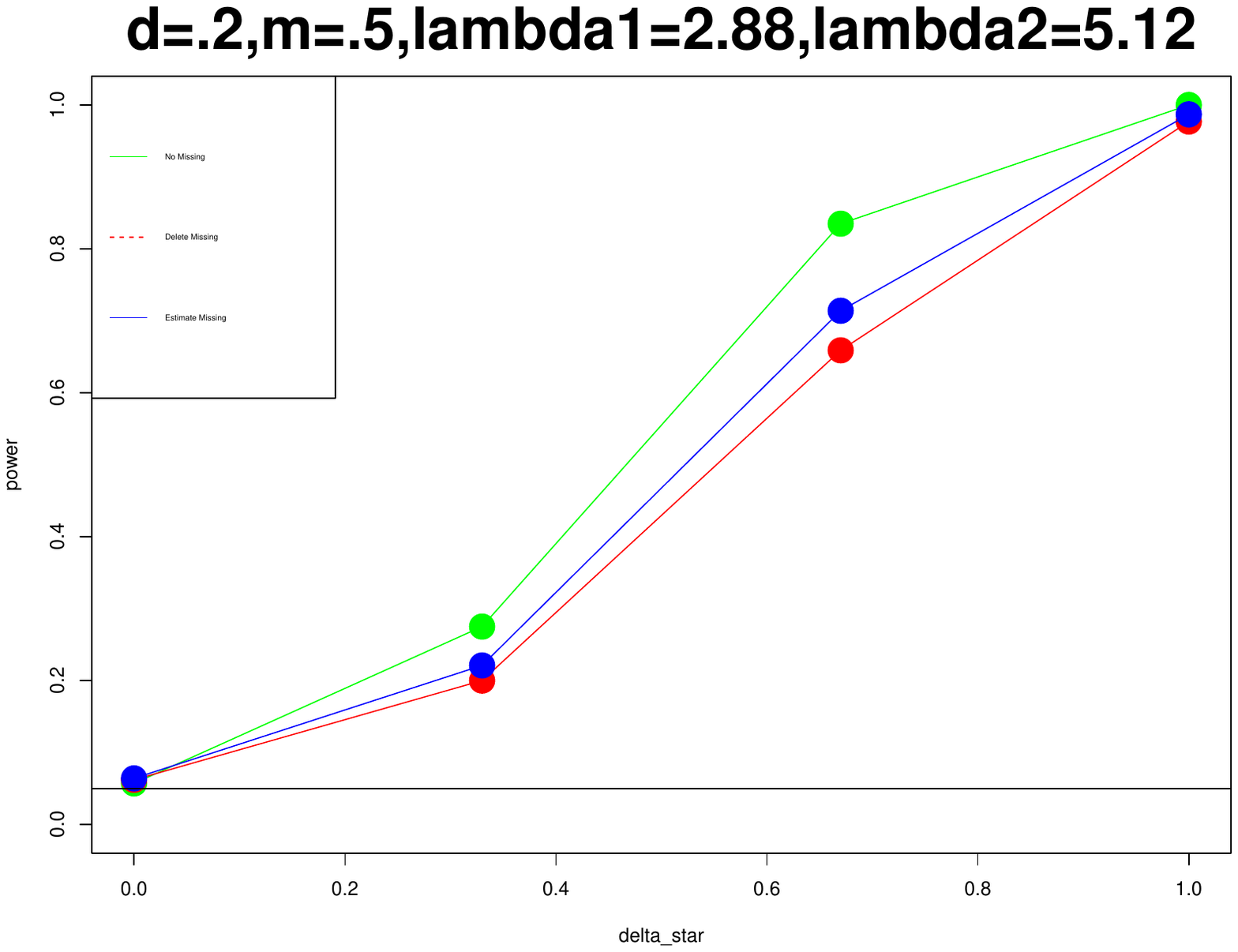}
\includegraphics[width = 2.3in, height = 1.4in]{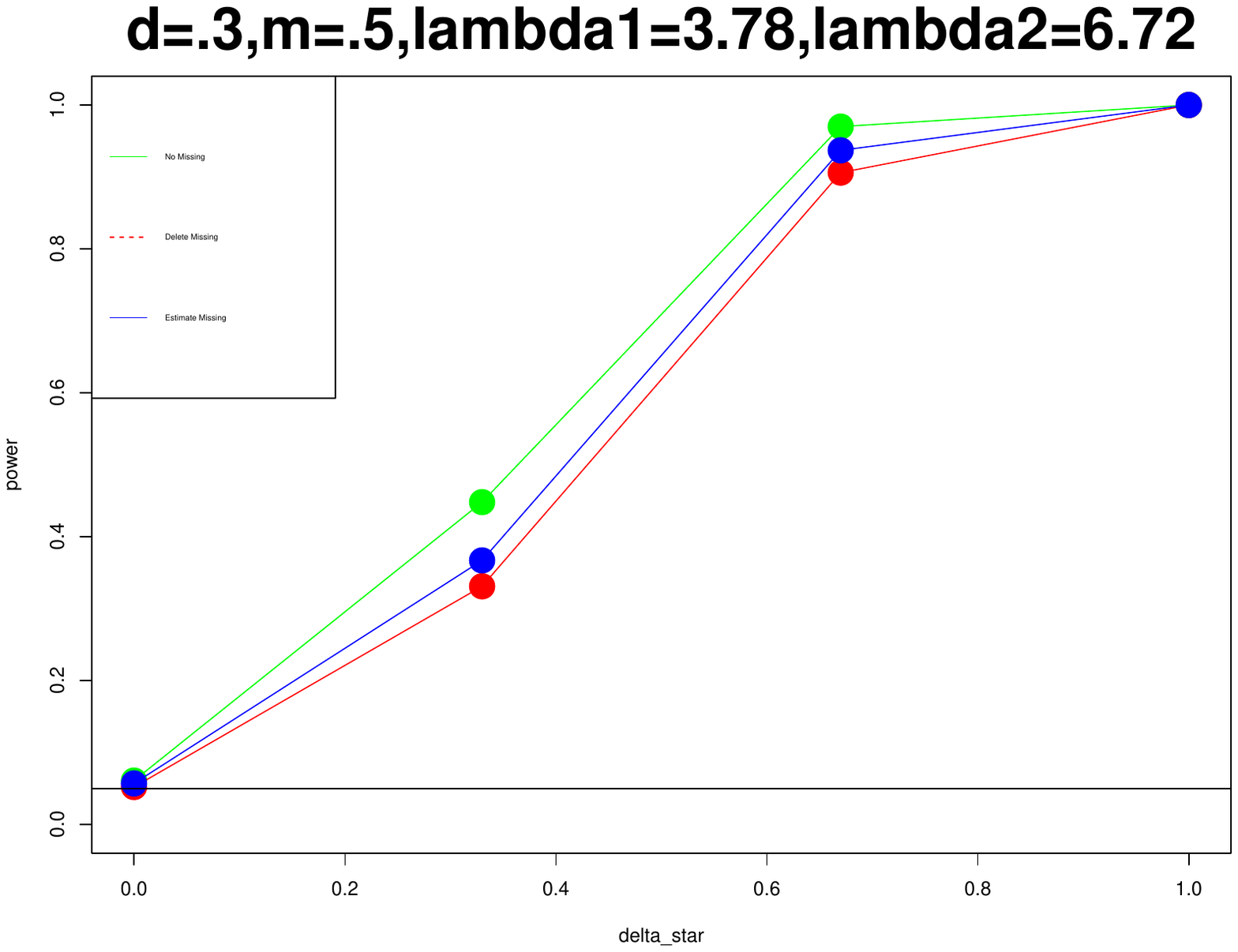}

\hspace{1.5cm}
$d=.1 \quad m=.1$
\hspace{3cm}
$d=.2 \quad m=.1$
\hspace{3cm}
$d=.3 \quad m=.1$

\includegraphics[width = 2.3in, height = 1.4in]{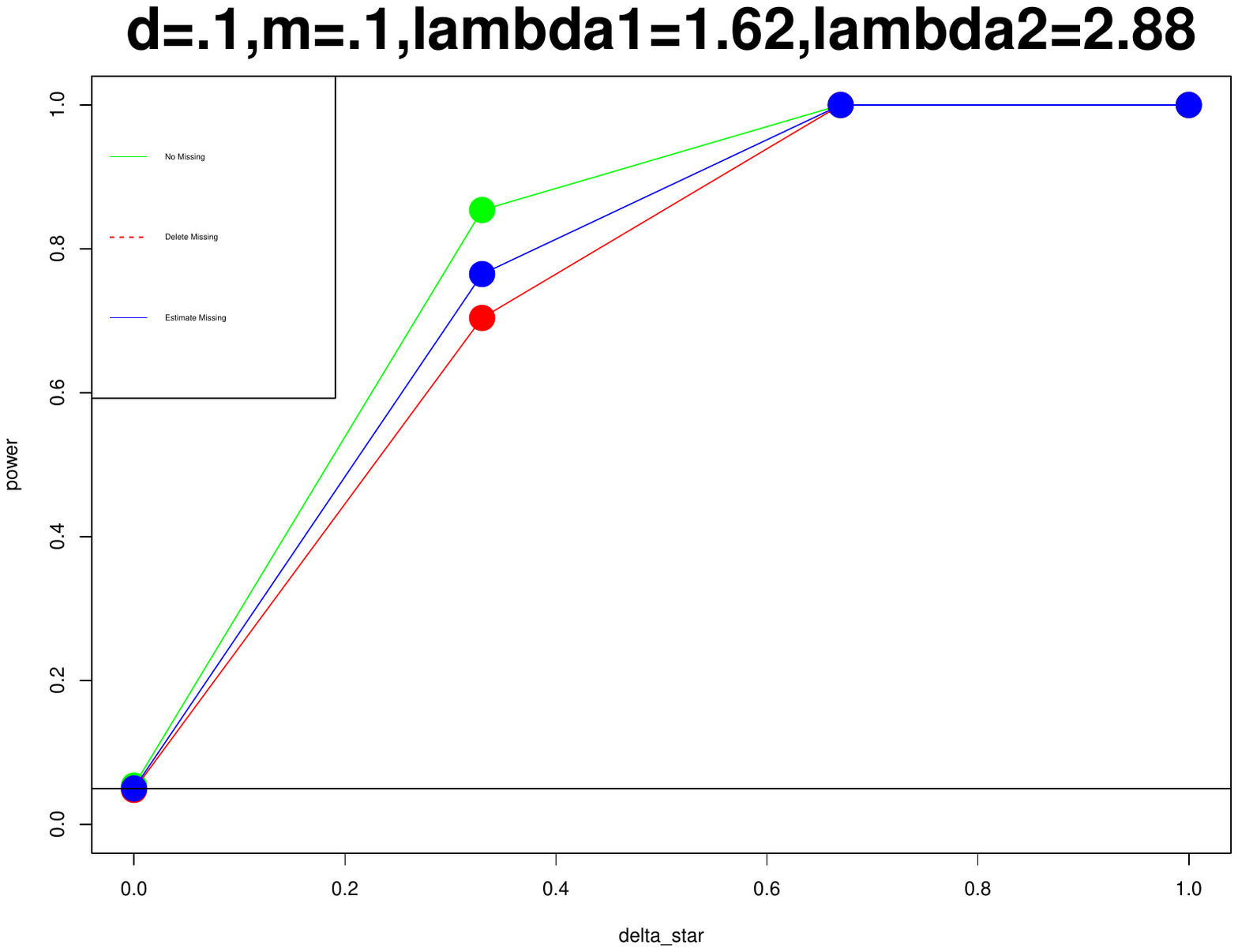}
\includegraphics[width = 2.3in, height = 1.4in]{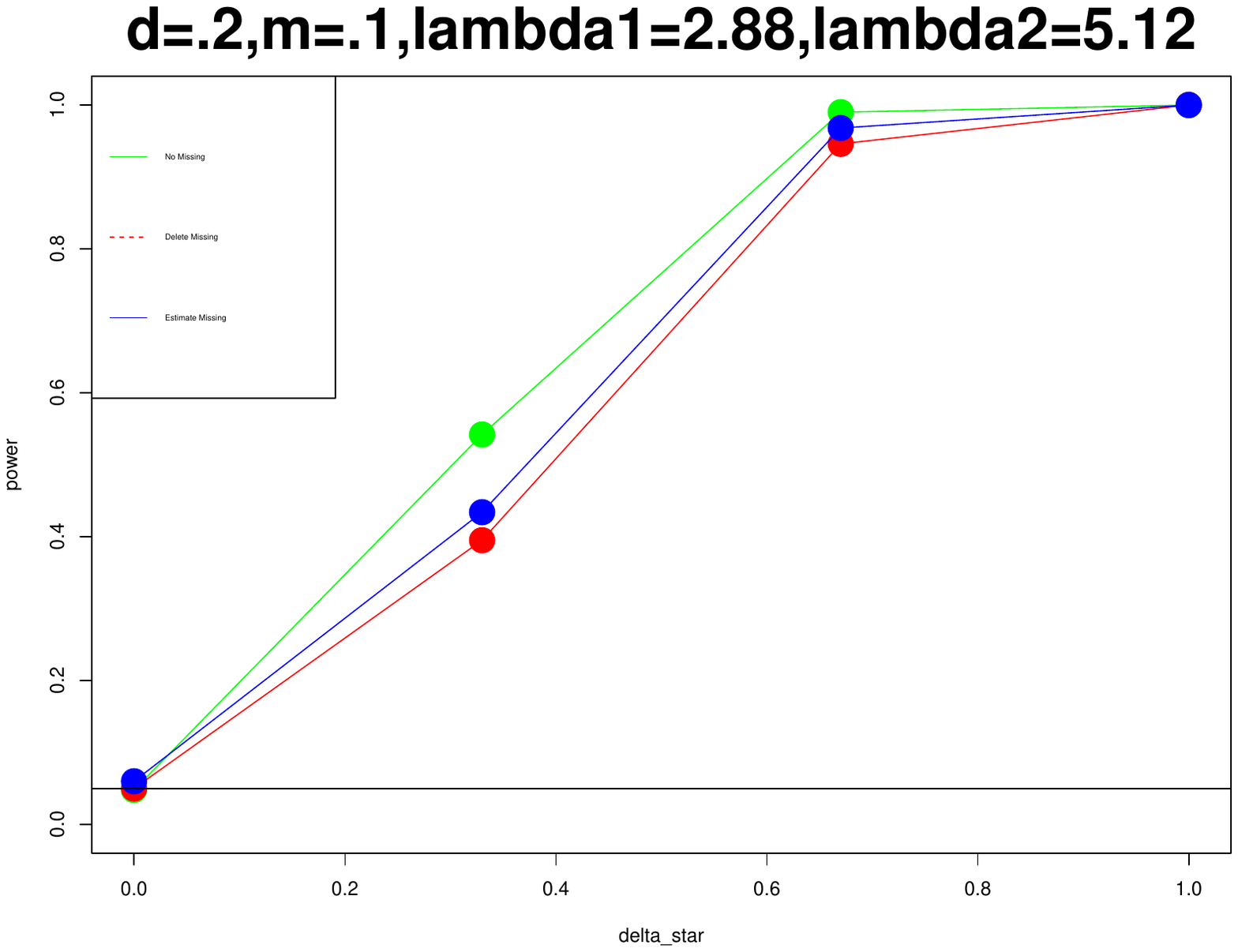}
\includegraphics[width = 2.3in, height = 1.4in]{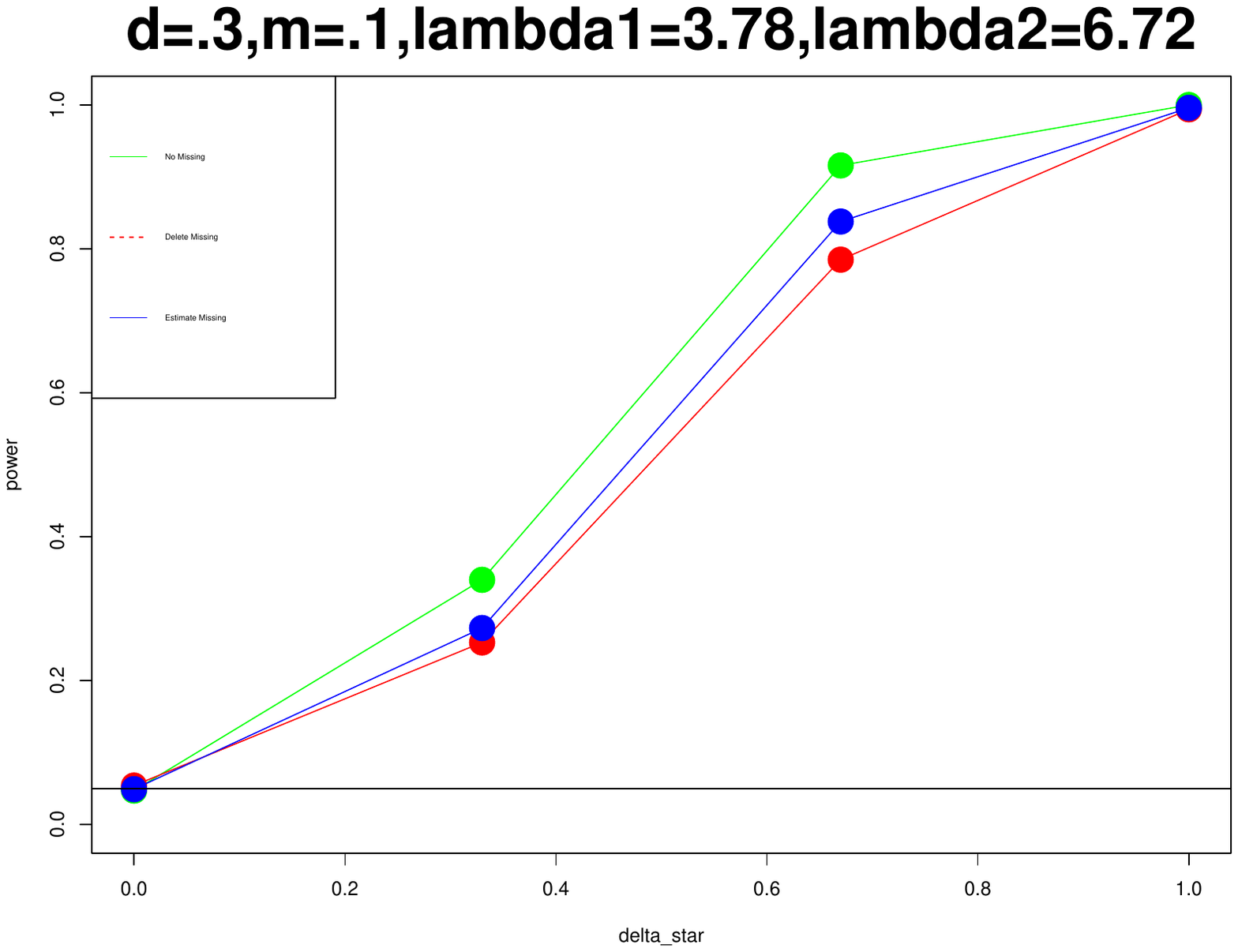}

\hspace{1.5cm}
$d=.1 \quad m=.1$
\hspace{3cm}
$d=.2 \quad m=.1$
\hspace{3cm}
$d=.3 \quad m=.1$

\includegraphics[width = 2.3in, height = 1.45in]{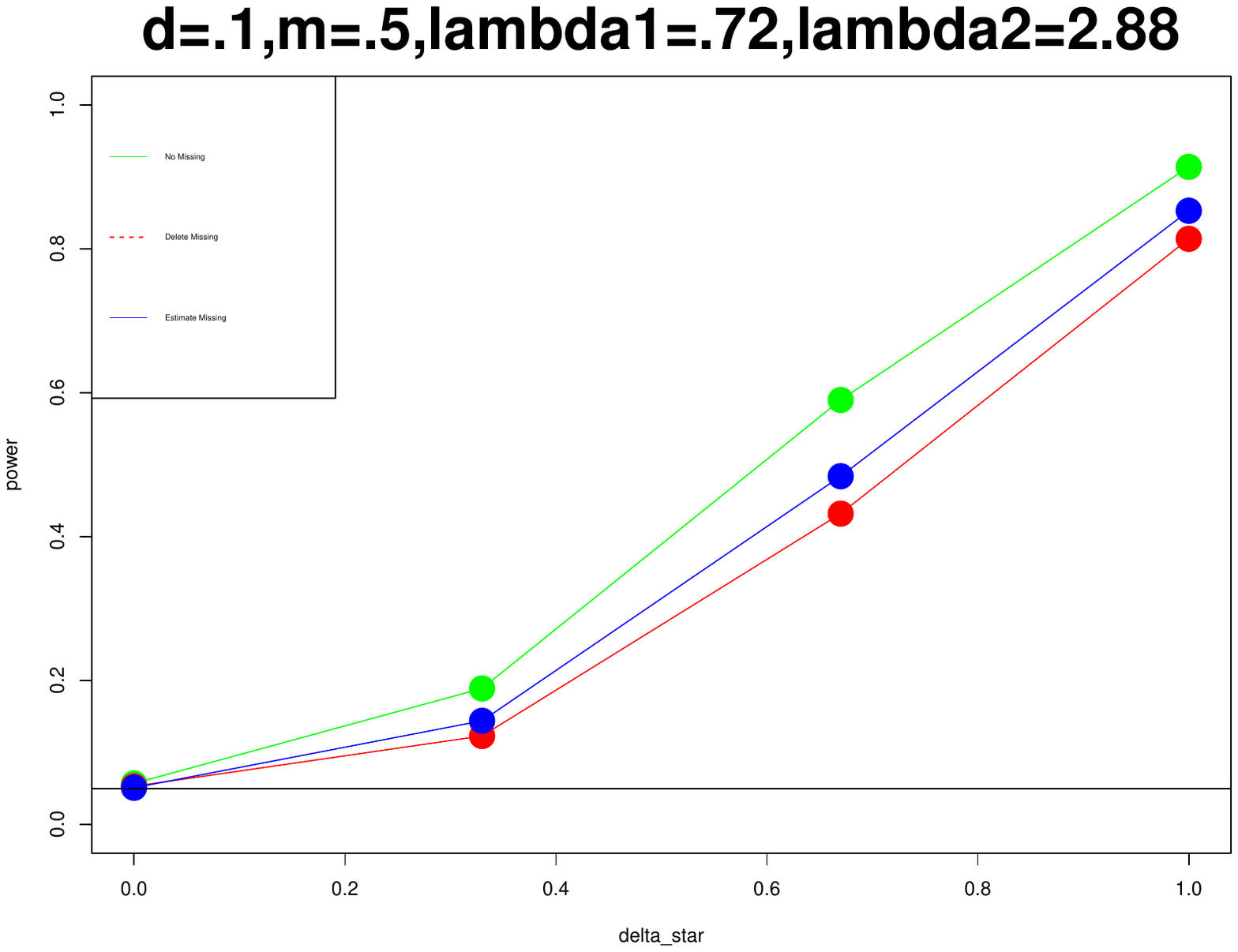}
\includegraphics[width = 2.3in, height = 1.45in]{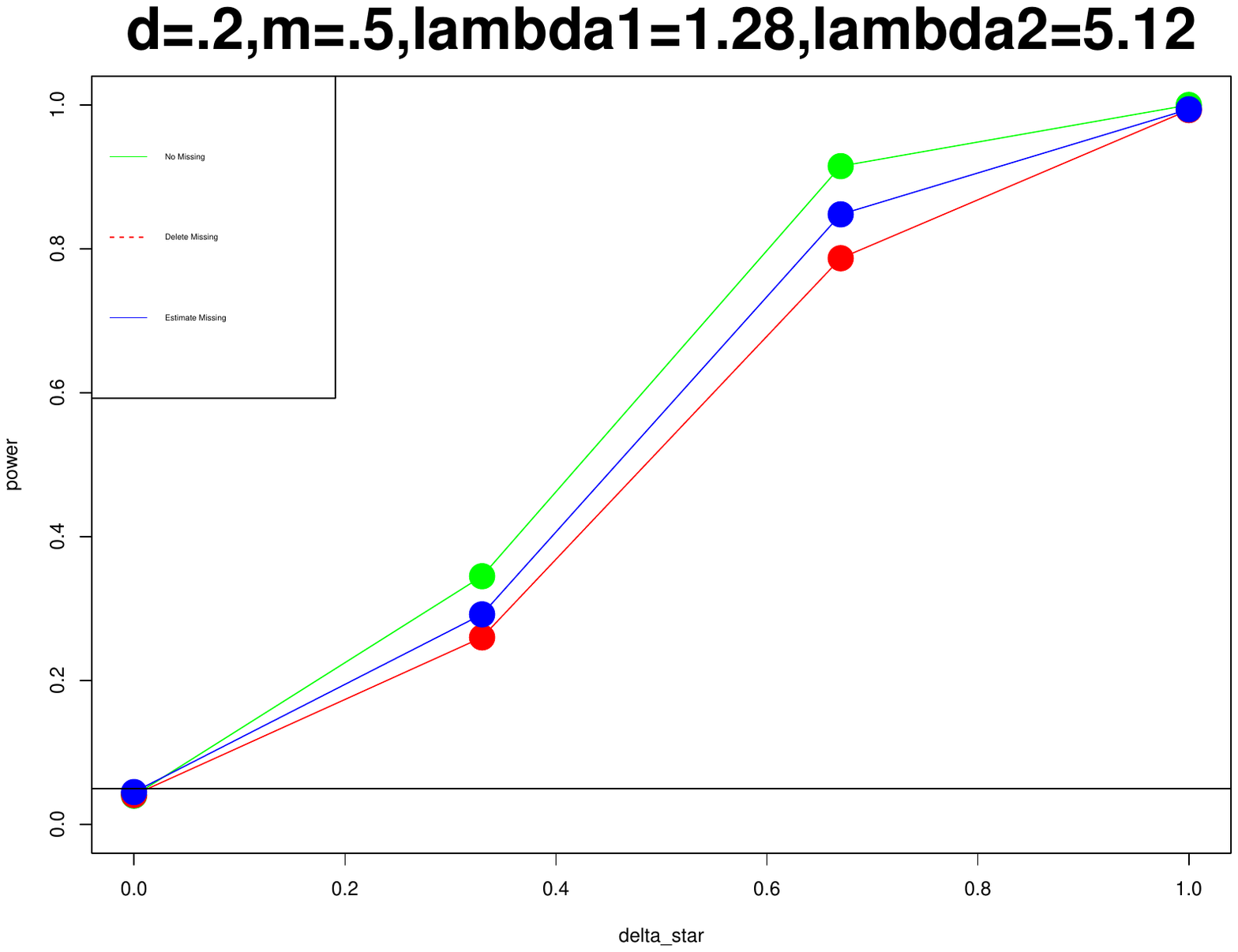}
\includegraphics[width = 2.3in, height = 1.45in]{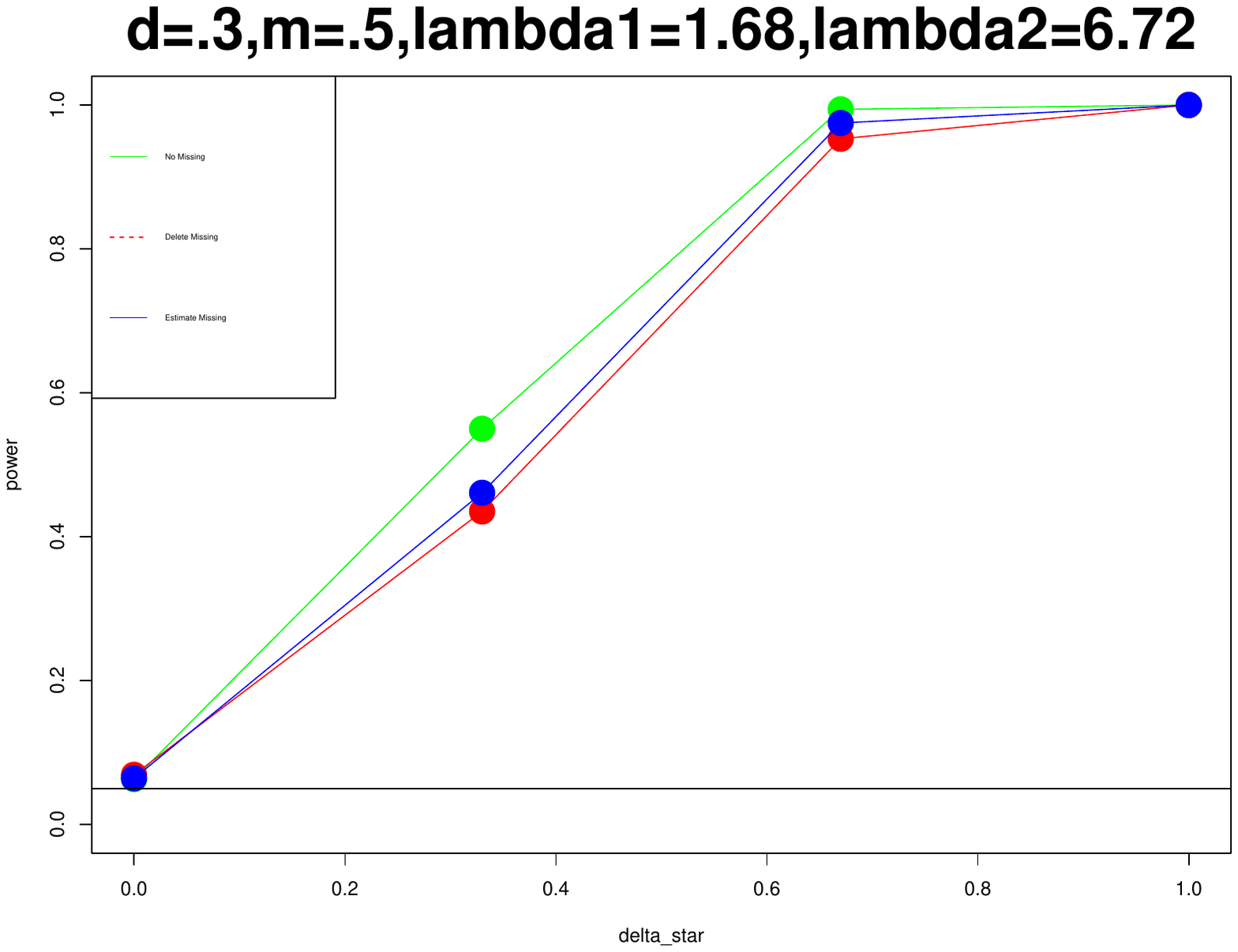}

\hspace{1.5cm}
$d=.1 \quad m=.5$
\hspace{3cm}
$d=.2 \quad m=.5$
\hspace{3cm}
$d=.3 \quad m=.5$

\includegraphics[width = 2.3in, height = 1.4in]{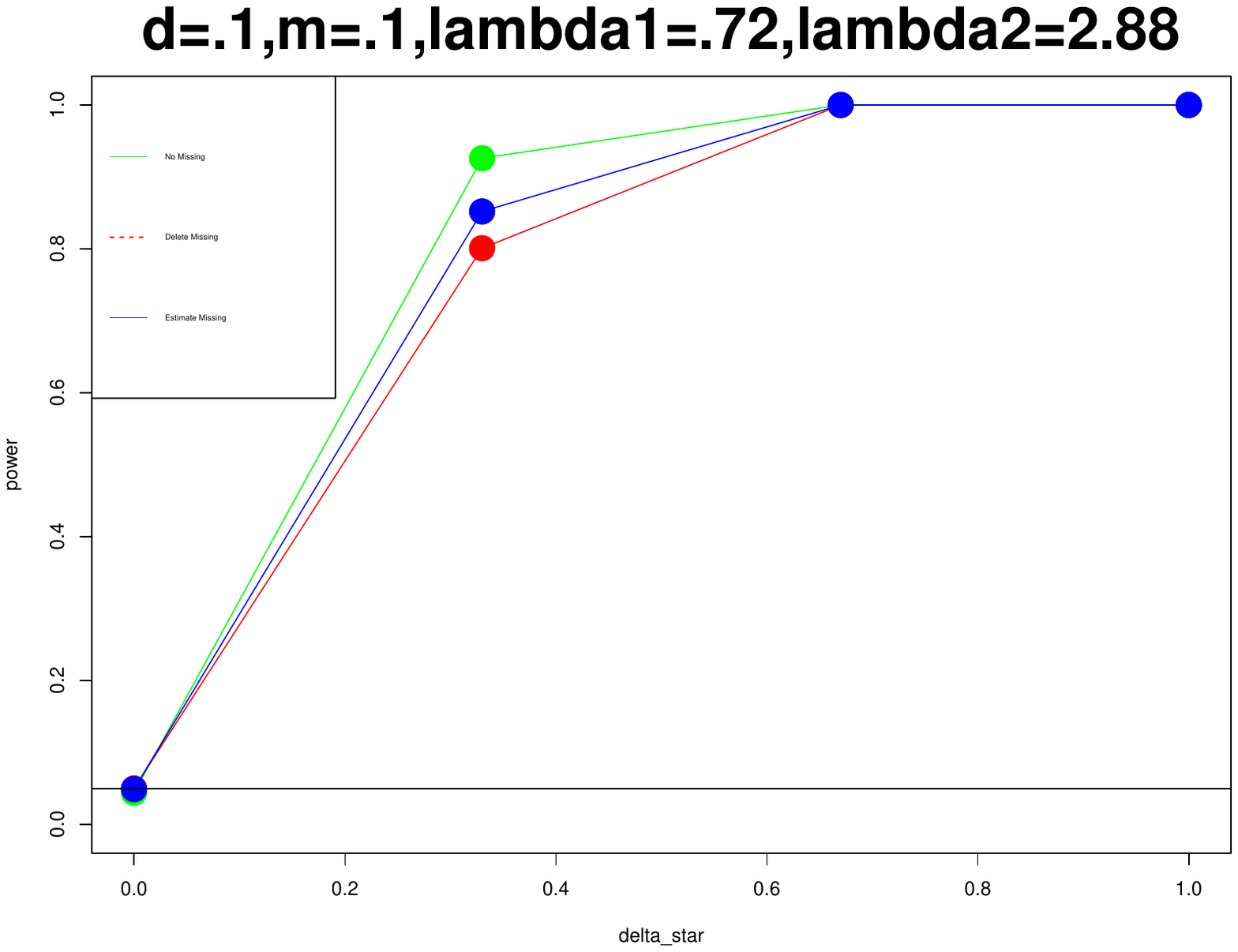}
\includegraphics[width = 2.3in, height = 1.4in]{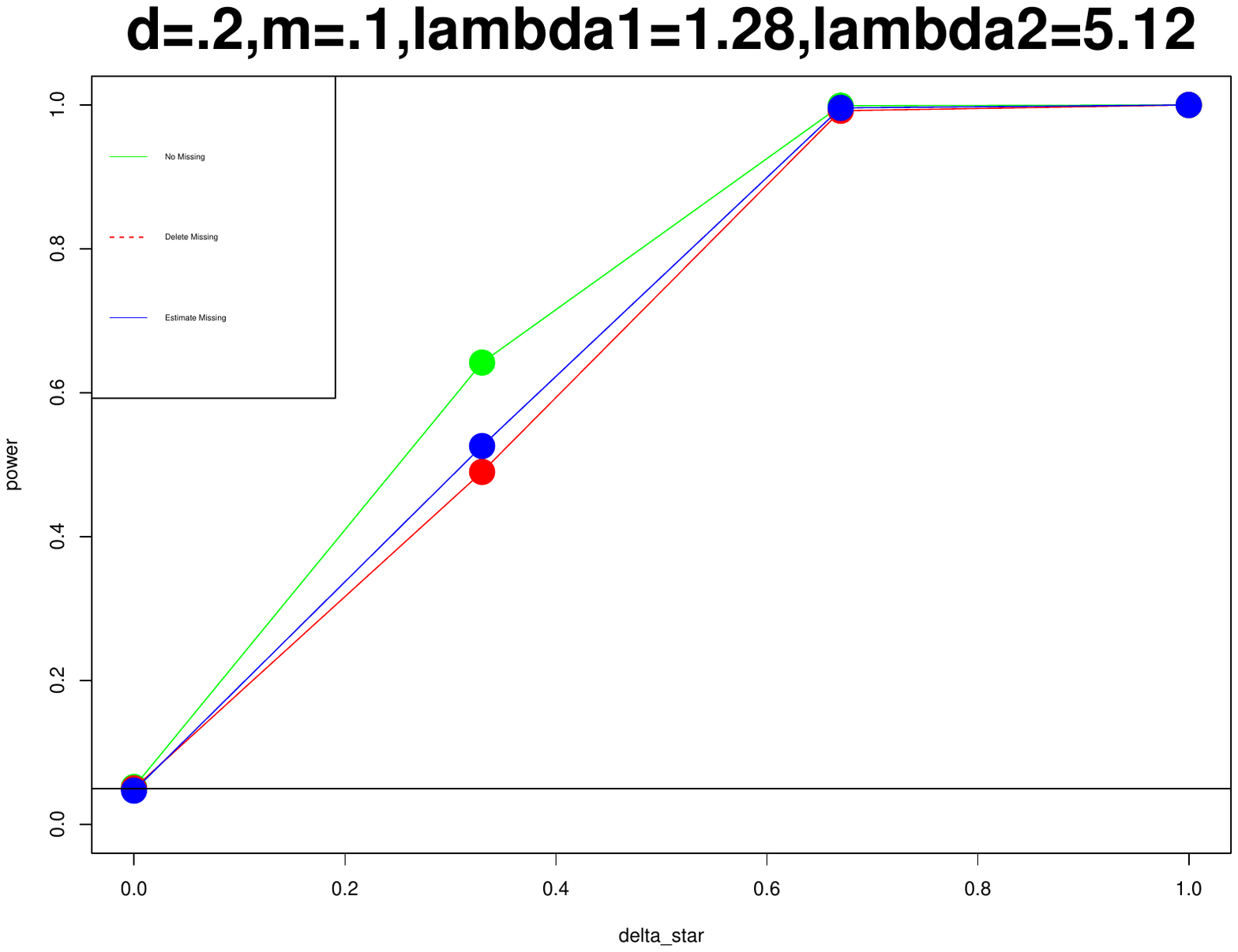}
\includegraphics[width = 2.3in, height = 1.4in]{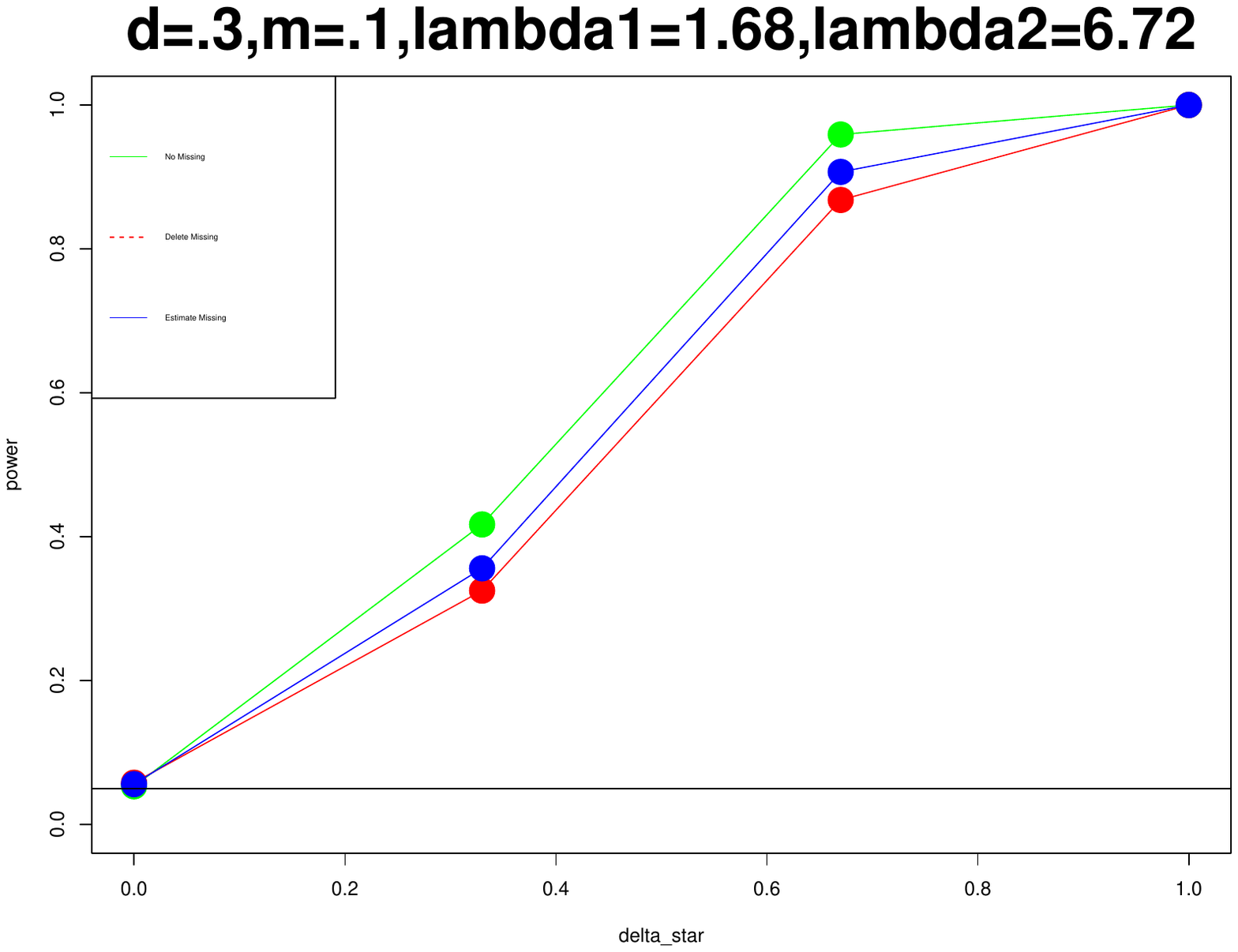}

\subsubsection{When one Trait has Normal Distribution and other Trait has Chi Squares Distribution}

\hspace{1.5cm}
$d=.1 \quad m=.5$
\hspace{3cm}
$d=.2 \quad m=.5$
\hspace{3cm}
$d=.3 \quad m=.5$

\includegraphics[width = 2.3in, height = 1.4in]{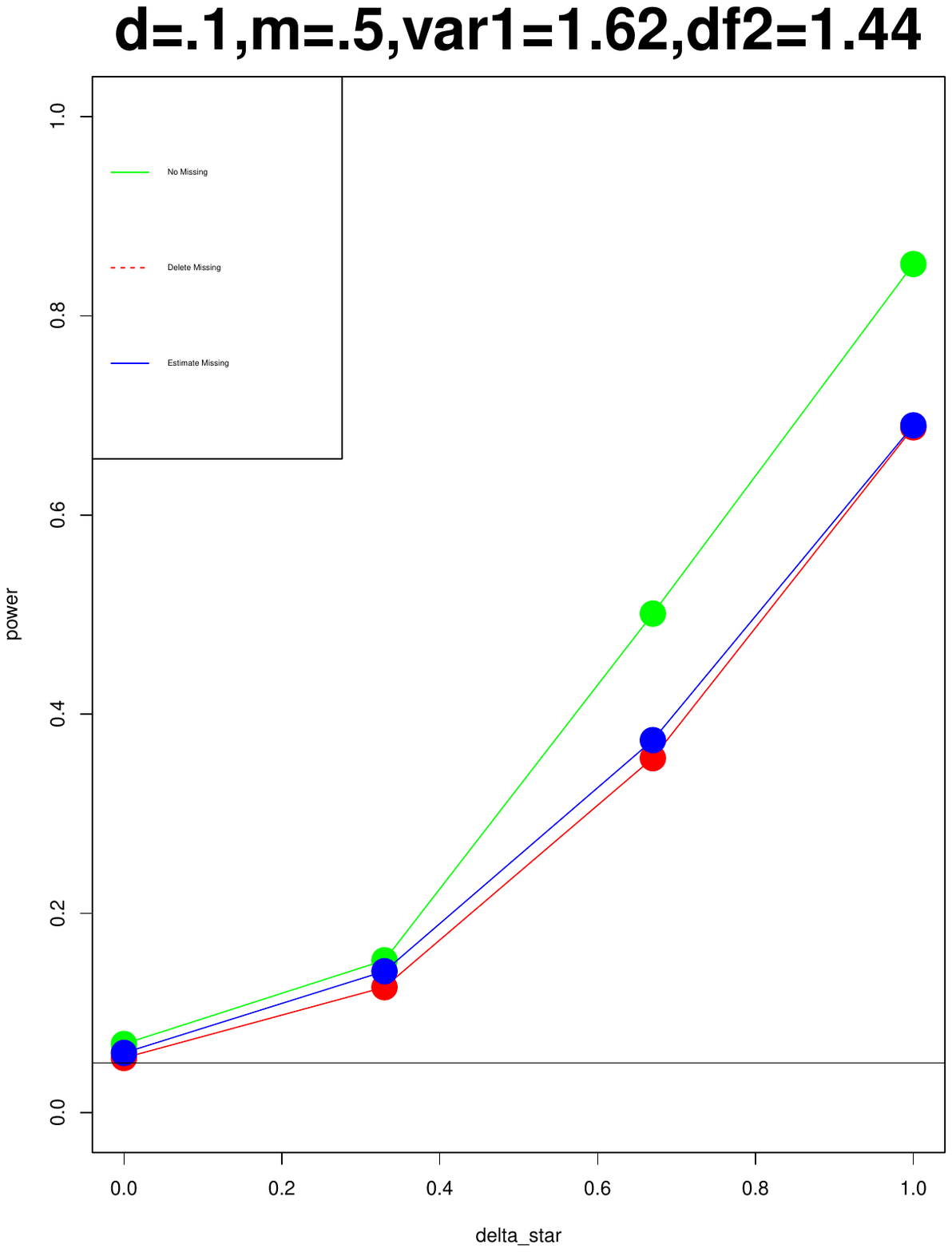}
\includegraphics[width = 2.3in, height = 1.4in]{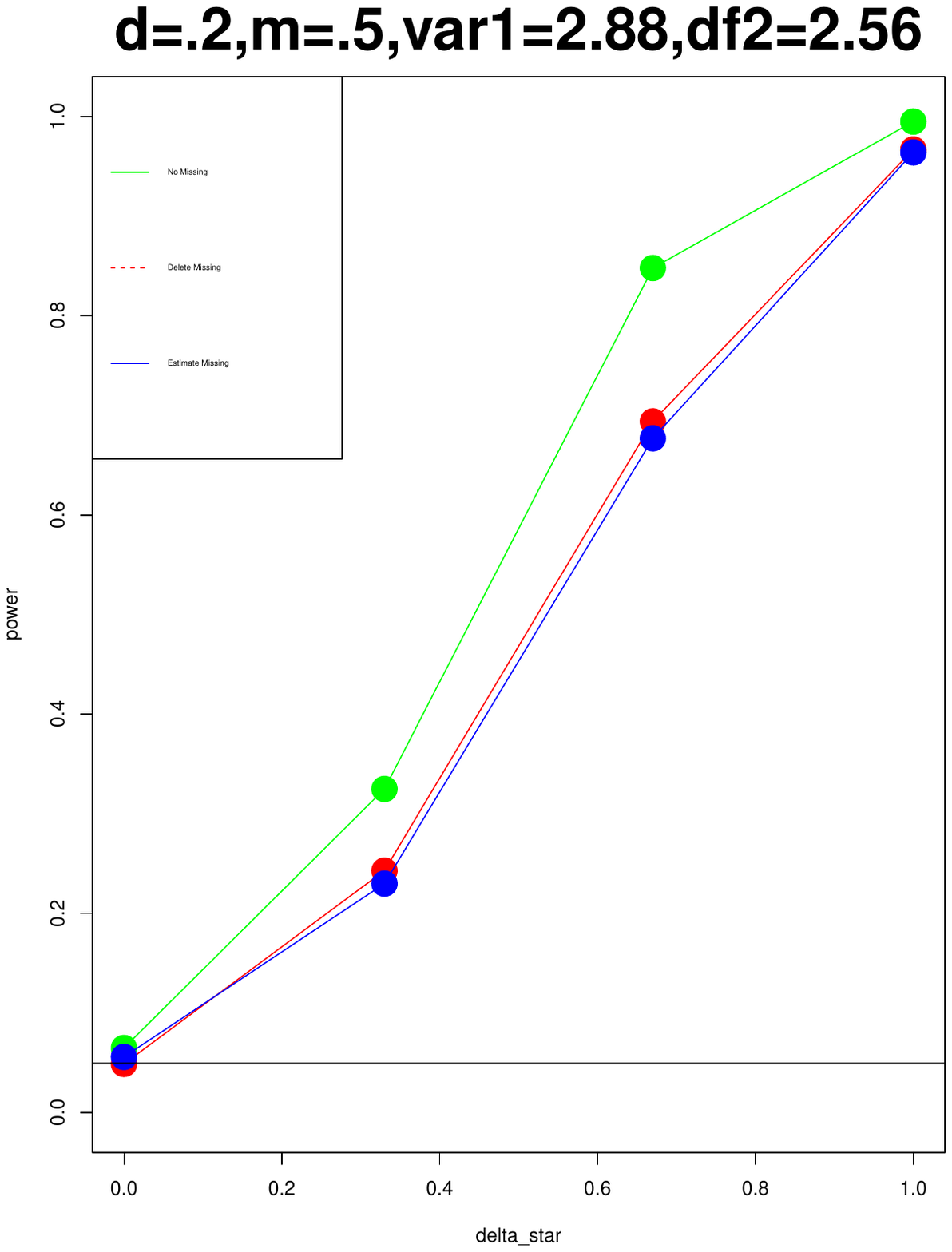}
\includegraphics[width = 2.3in, height = 1.4in]{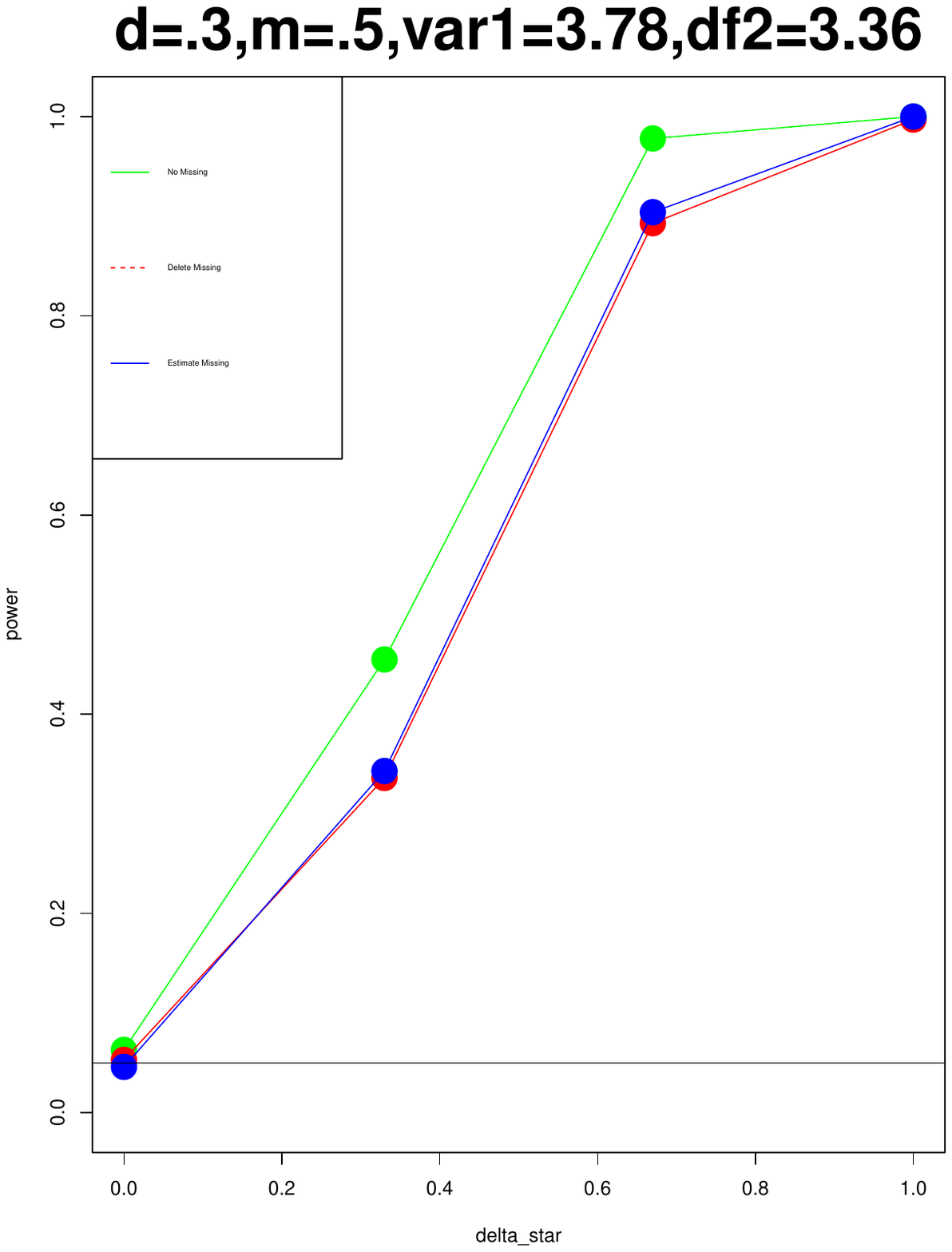}

\hspace{1.5cm}
$d=.1 \quad m=.1$
\hspace{3cm}
$d=.2 \quad m=.1$
\hspace{3cm}
$d=.3 \quad m=.1$

\includegraphics[width = 2.3in, height = 1.4in]{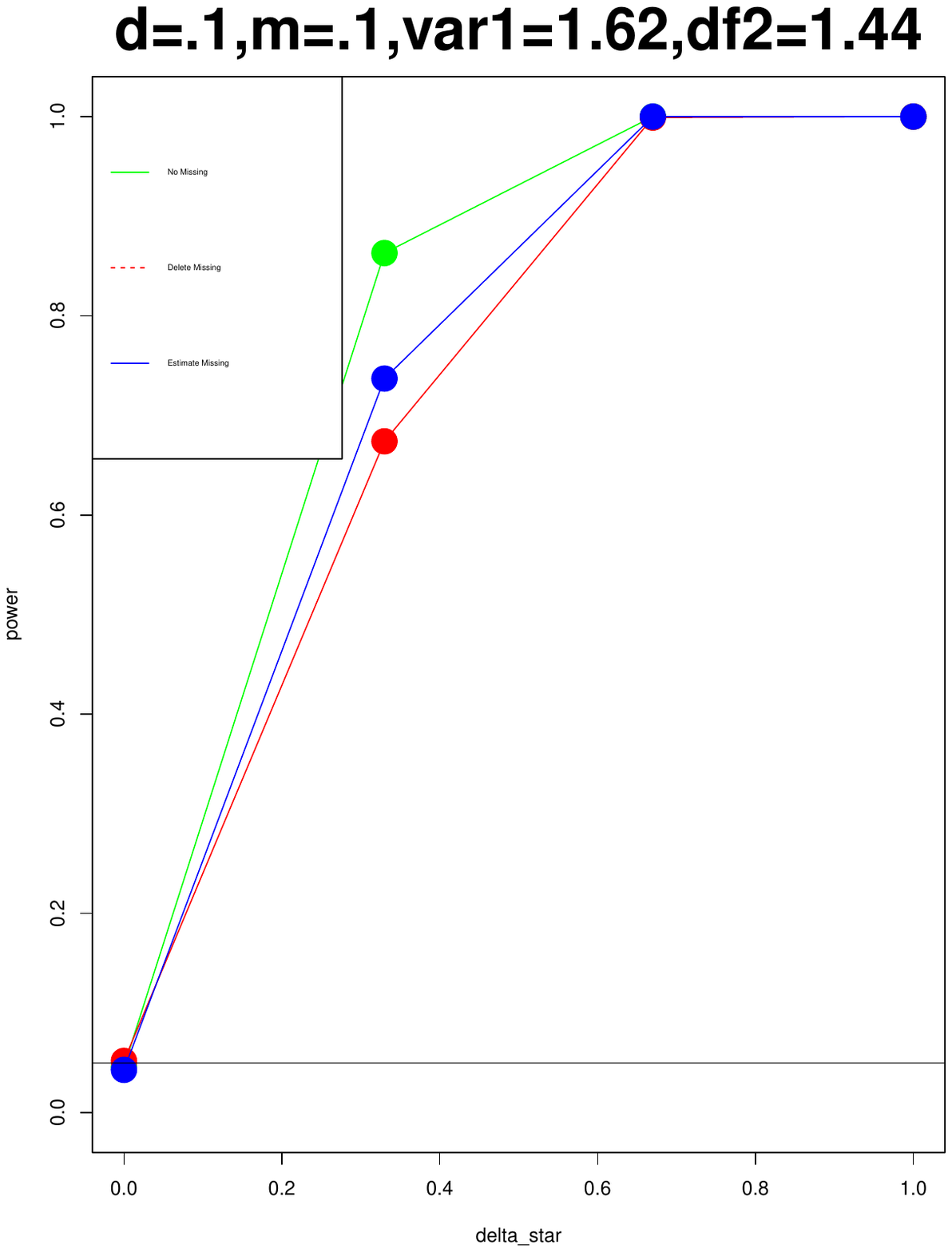}
\includegraphics[width = 2.3in, height = 1.4in]{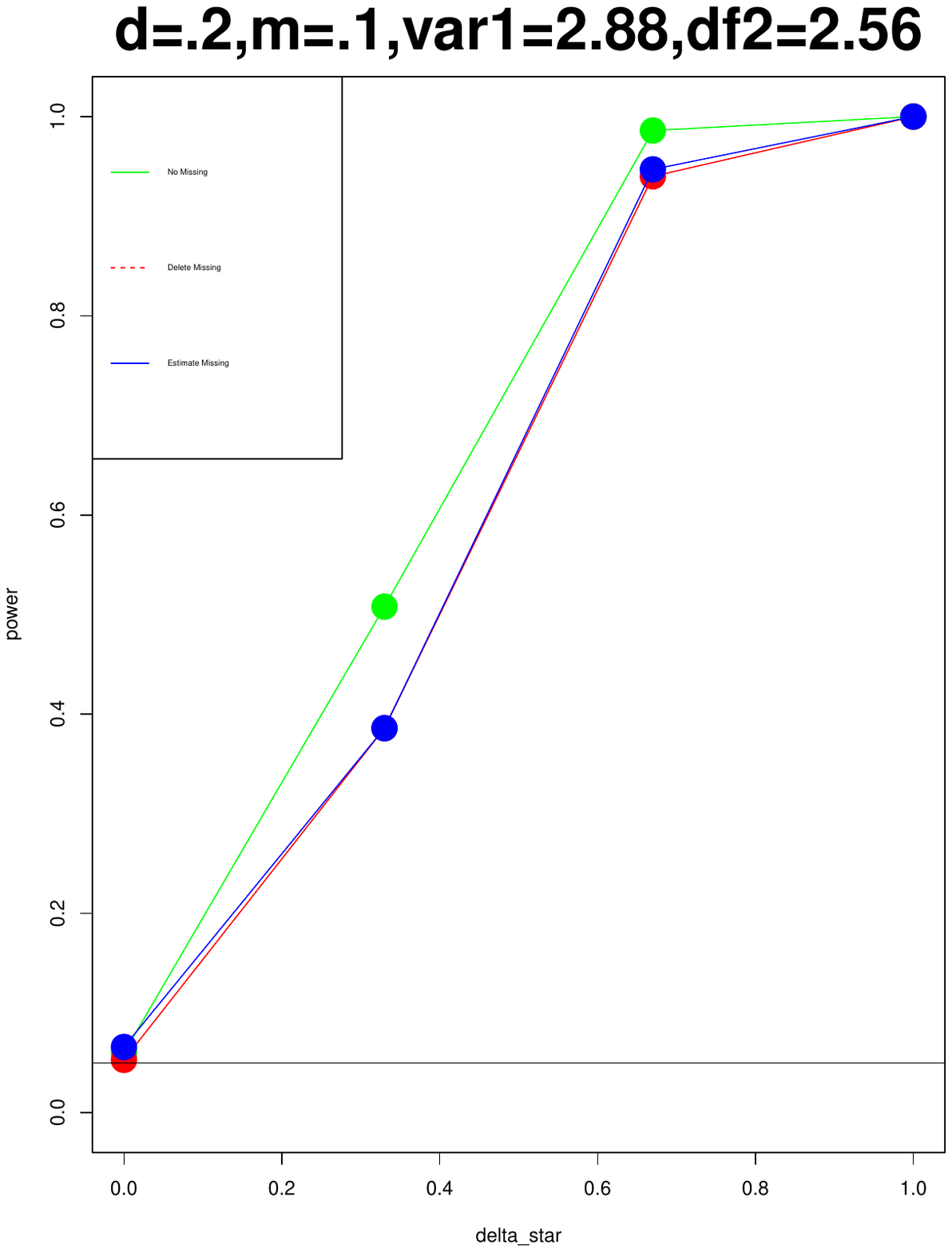}
\includegraphics[width = 2.3in, height = 1.4in]{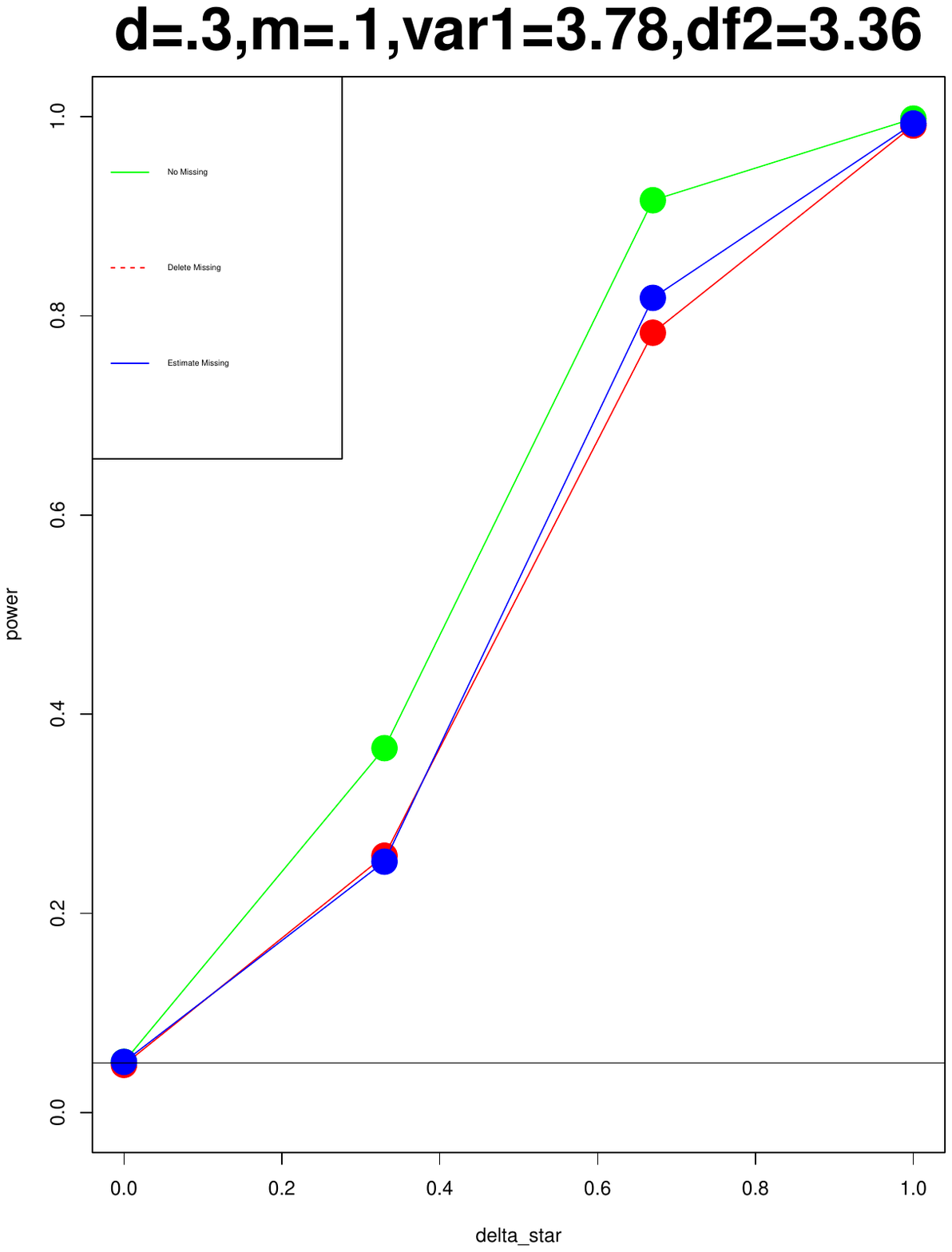}

\hspace{1.5cm}
$d=.1 \quad m=.1$
\hspace{3cm}
$d=.2 \quad m=.1$
\hspace{3cm}
$d=.3 \quad m=.1$

\includegraphics[width = 2.3in, height = 1.4in]{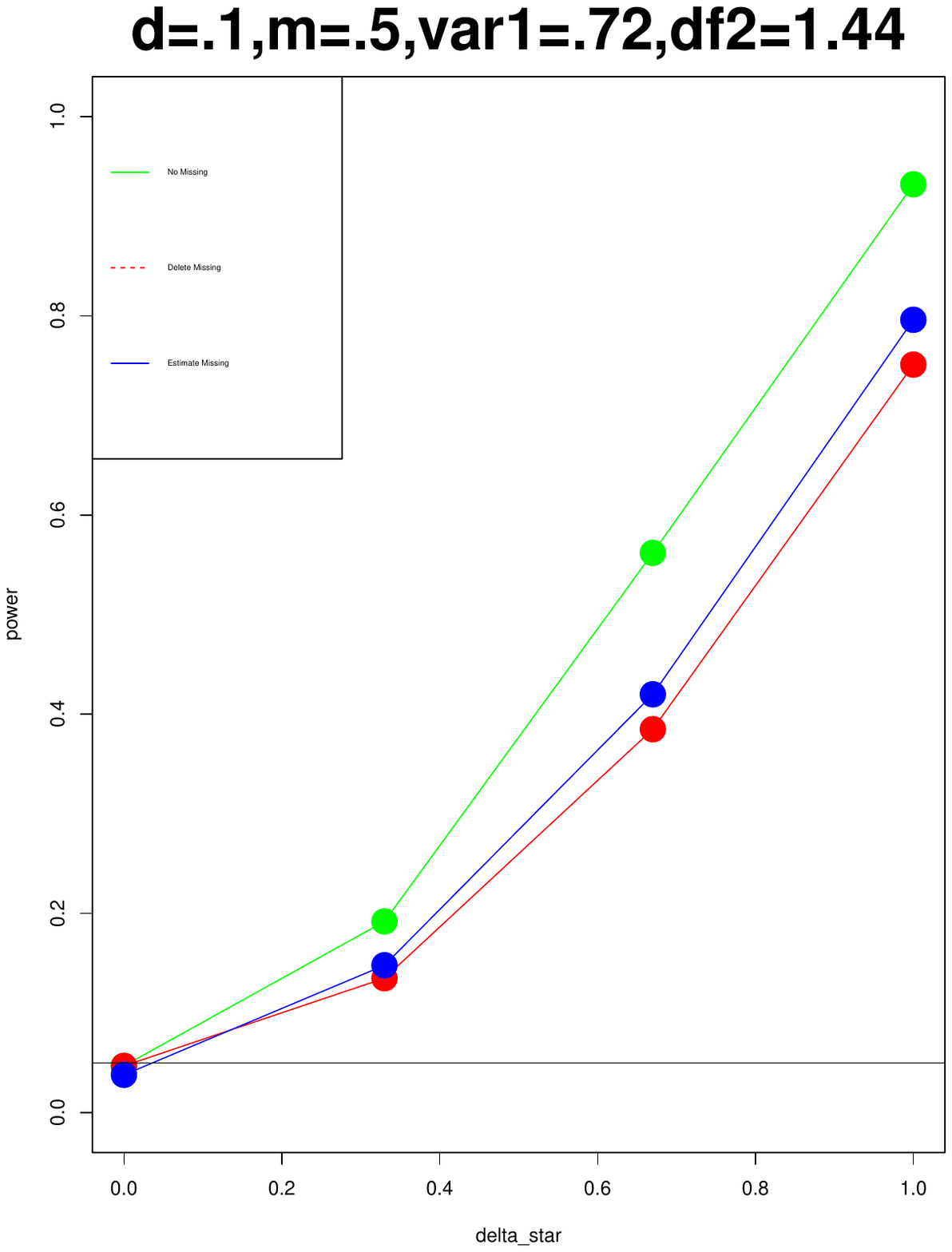}
\includegraphics[width = 2.3in, height = 1.4in]{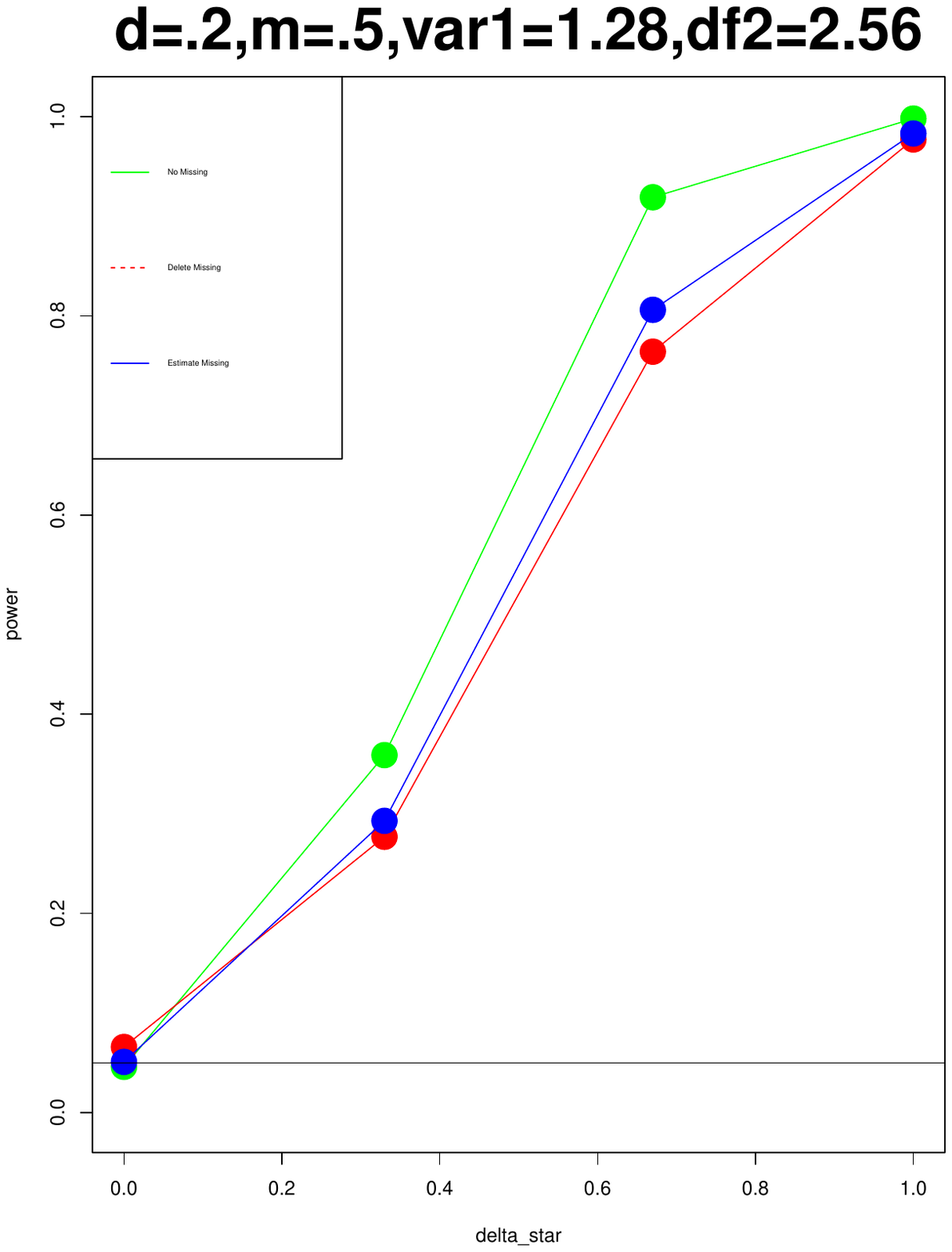}
\includegraphics[width = 2.3in, height = 1.4in]{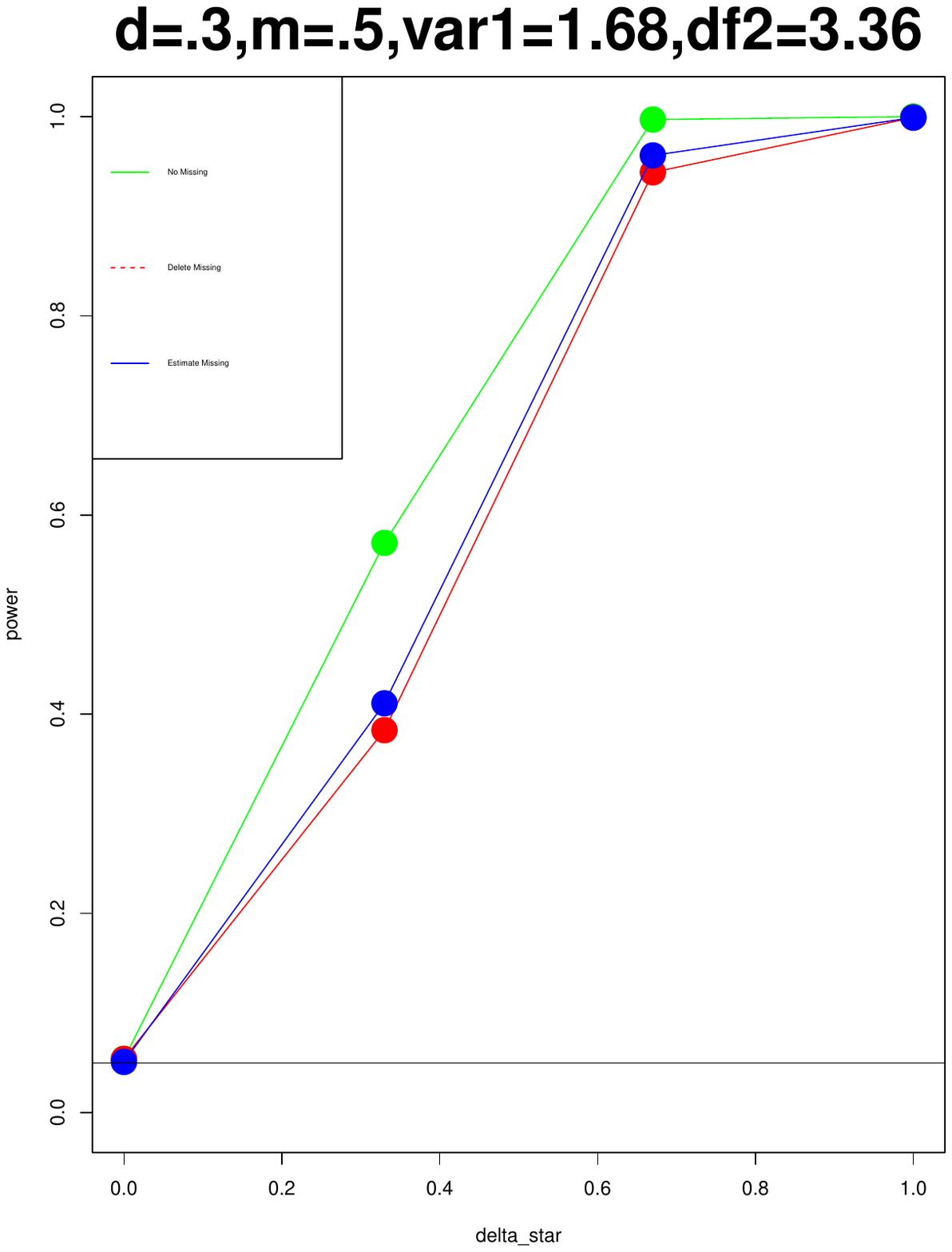}

\hspace{1.5cm}
$d=.1 \quad m=.5$
\hspace{3cm}
$d=.2 \quad m=.5$
\hspace{3cm}
$d=.3 \quad m=.5$

\includegraphics[width = 2.3in, height = 1.4in]{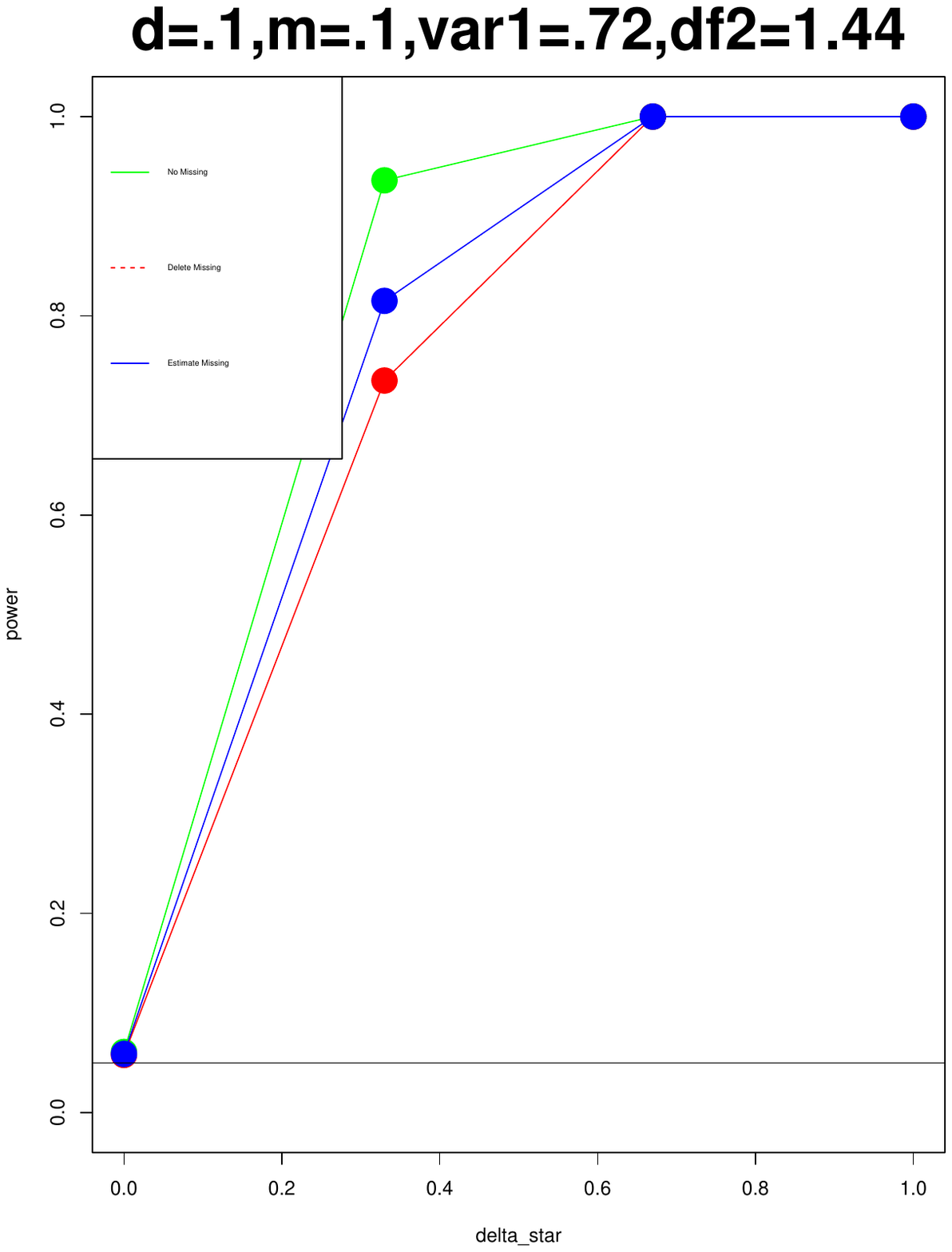}
\includegraphics[width = 2.3in, height = 1.4in]{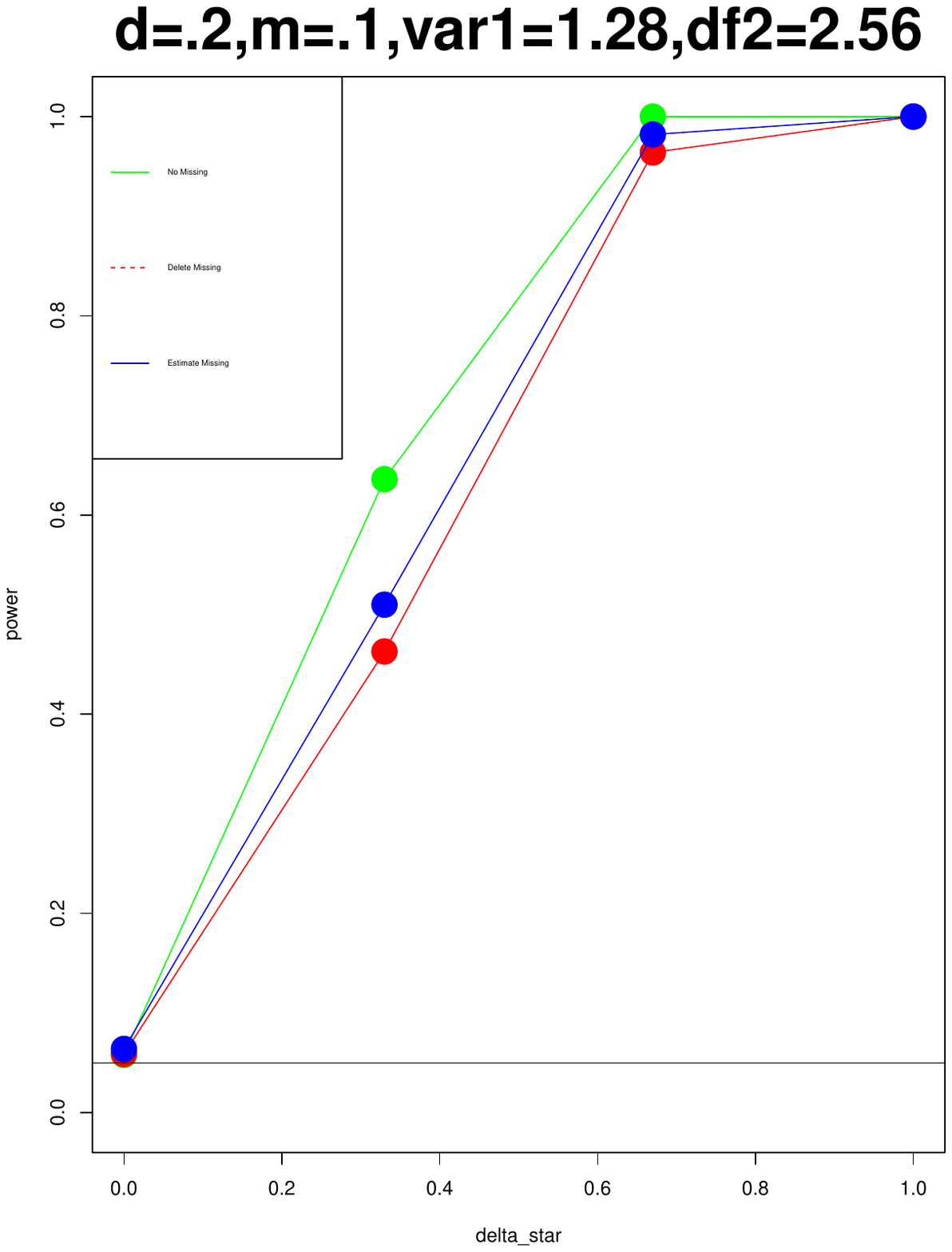}
\includegraphics[width = 2.3in, height = 1.4in]{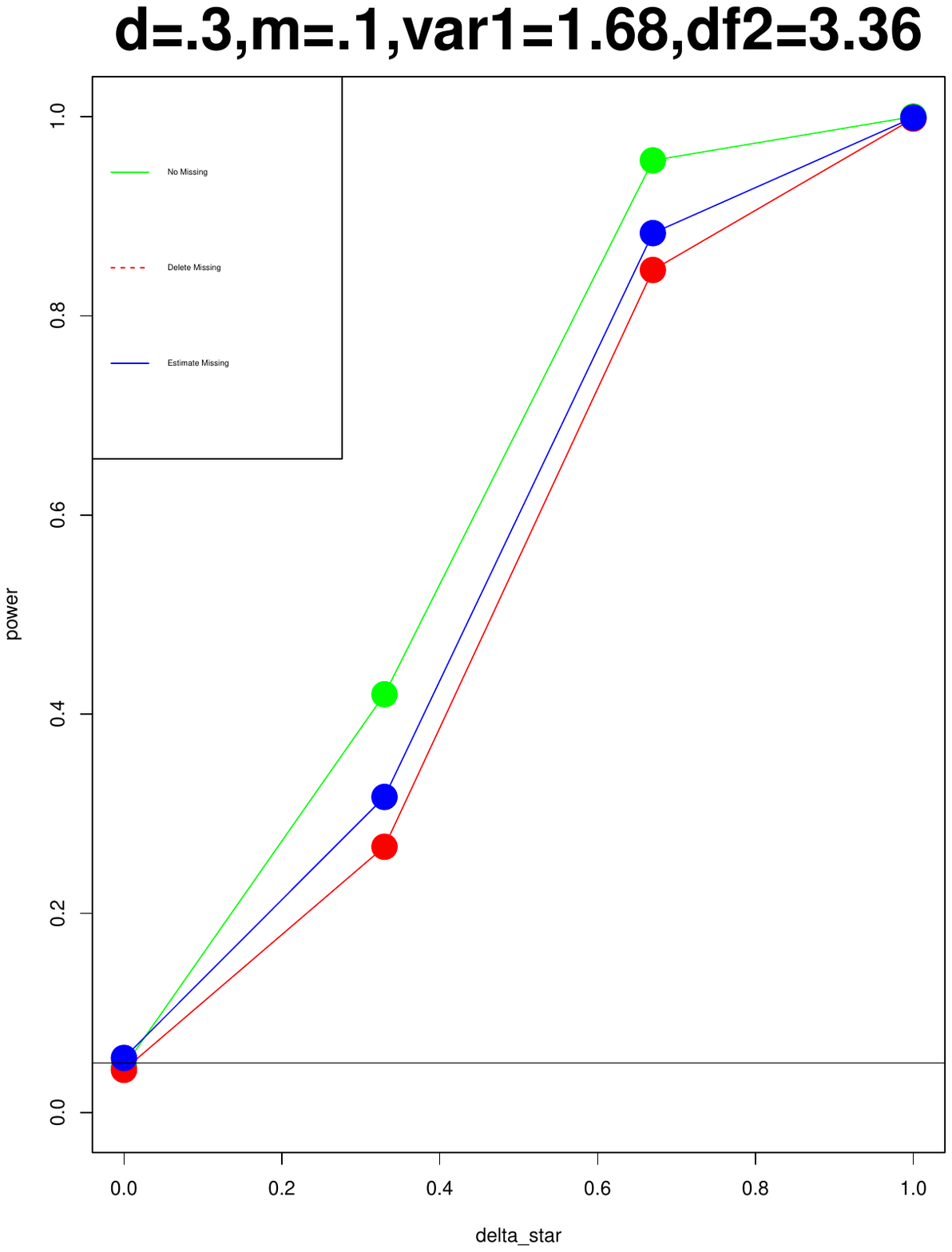}

\hspace{1.5cm}
$d=.1 \quad m=.1$
\hspace{3cm}
$d=.2 \quad m=.1$
\hspace{3cm}
$d=.3 \quad m=.1$

\includegraphics[width = 2.3in, height = 1.4in]{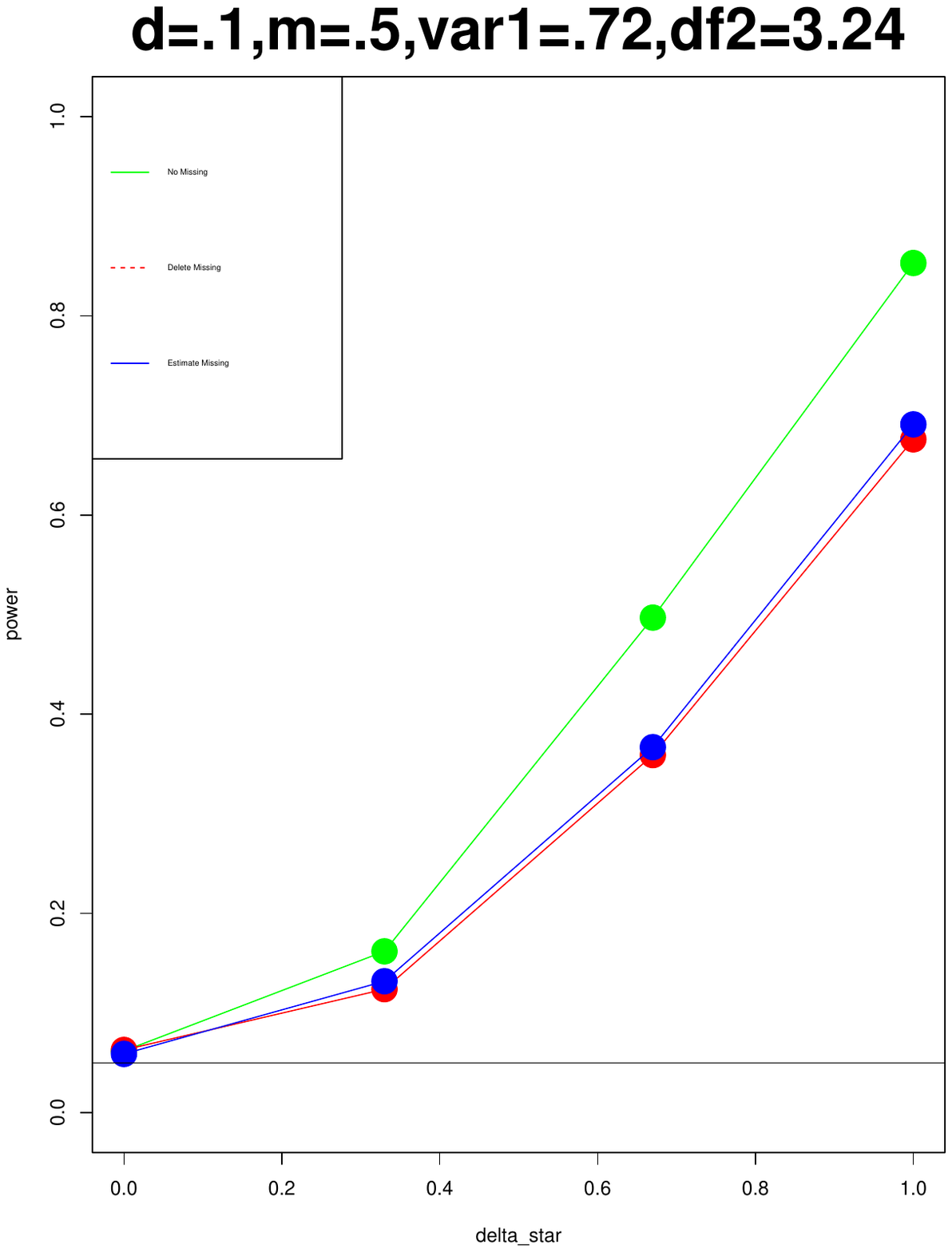}
\includegraphics[width = 2.3in, height = 1.4in]{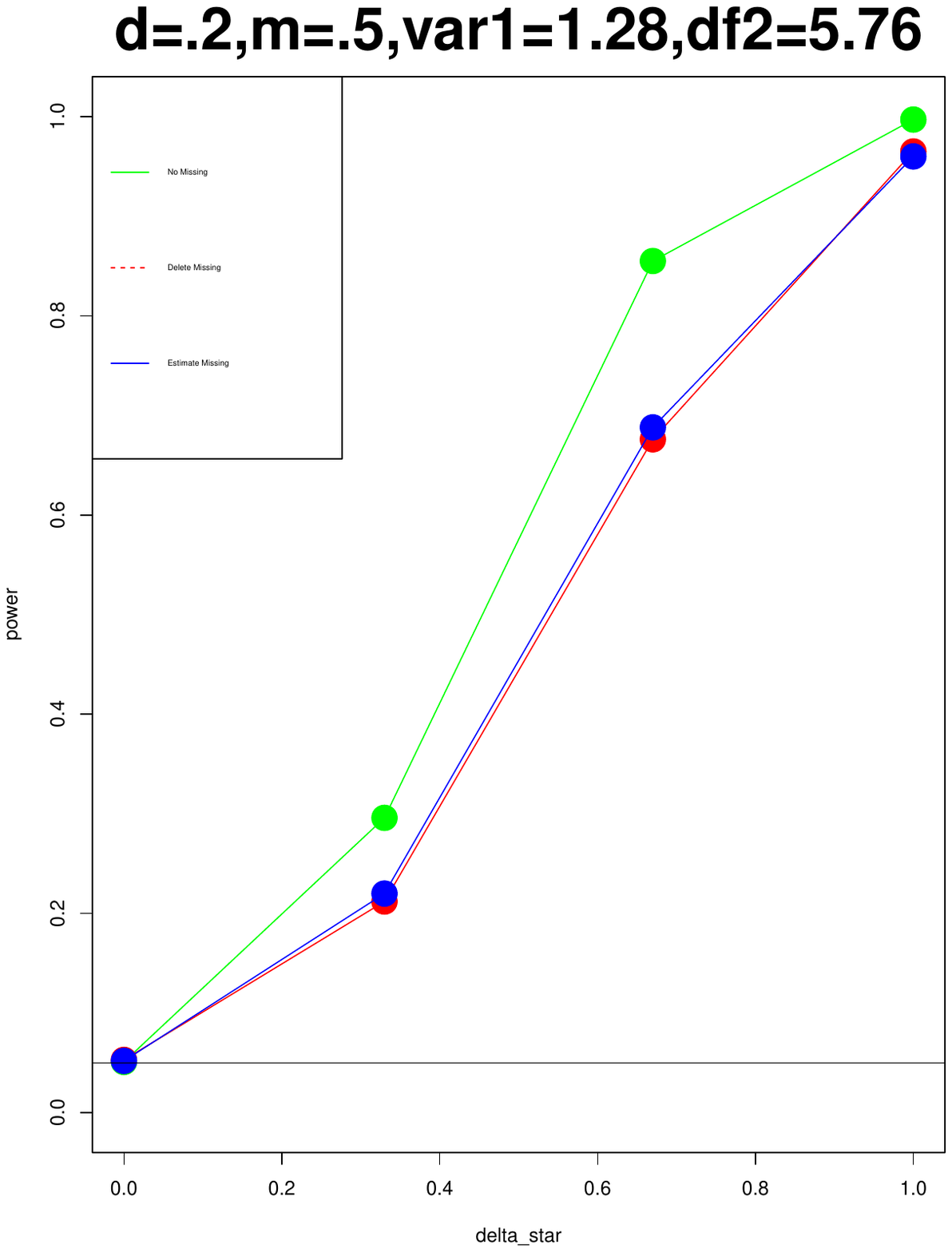}
\includegraphics[width = 2.3in, height = 1.4in]{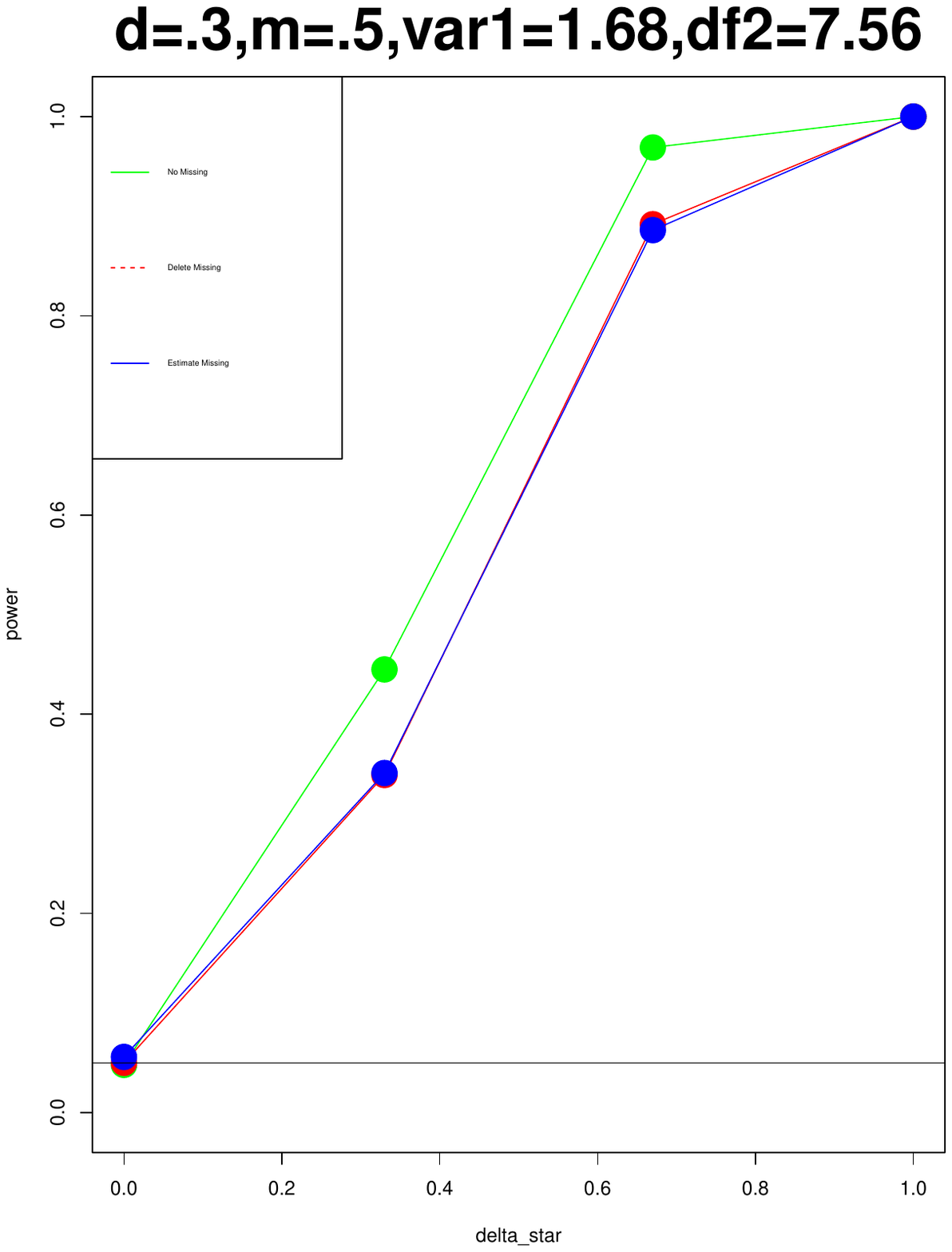}

\hspace{1.5cm}
$d=.1 \quad m=.5$
\hspace{3cm}
$d=.2 \quad m=.5$
\hspace{3cm}
$d=.3 \quad m=.5$

\includegraphics[width = 2.3in, height = 1.4in]{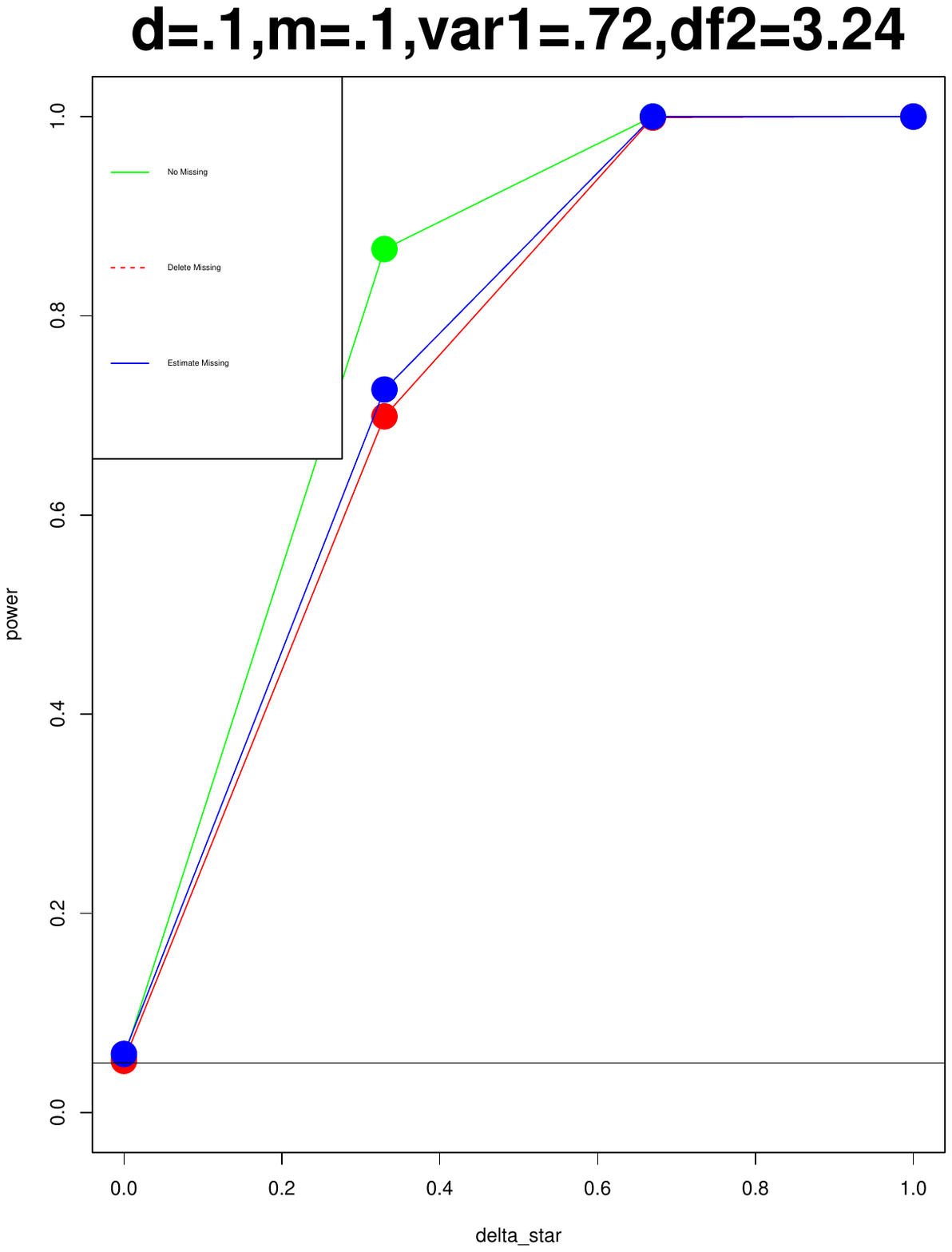}
\includegraphics[width = 2.3in, height = 1.4in]{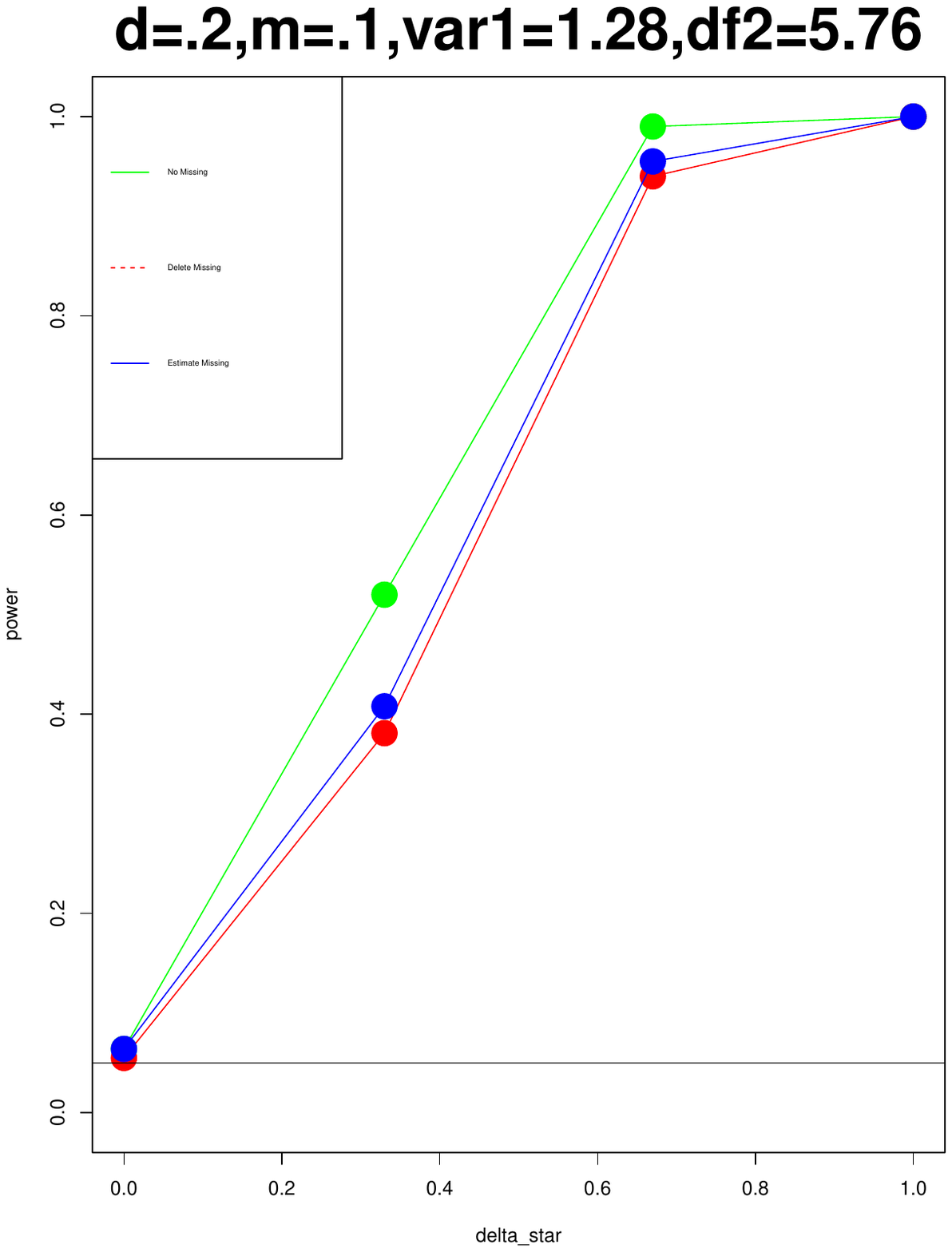}
\includegraphics[width = 2.3in, height = 1.4in]{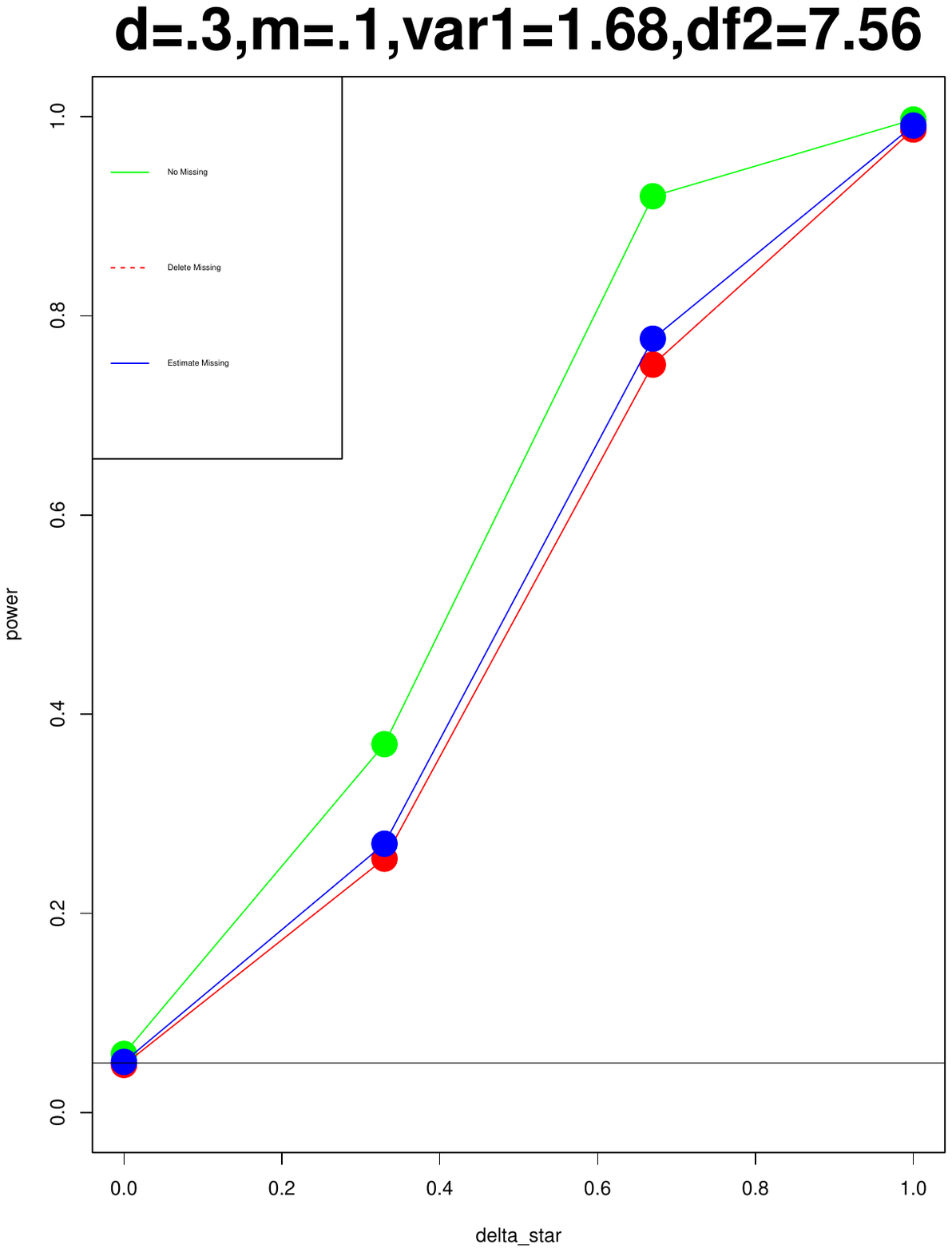}

\subsubsection{When one Trait has Poisson Distribution and other Trait has Chi Squares Distribution}

\hspace{1.5cm}
$d=.1 \quad m=.5$
\hspace{3cm}
$d=.2 \quad m=.5$
\hspace{3cm}
$d=.3 \quad m=.5$

\includegraphics[width = 2.3in, height = 1.4in]{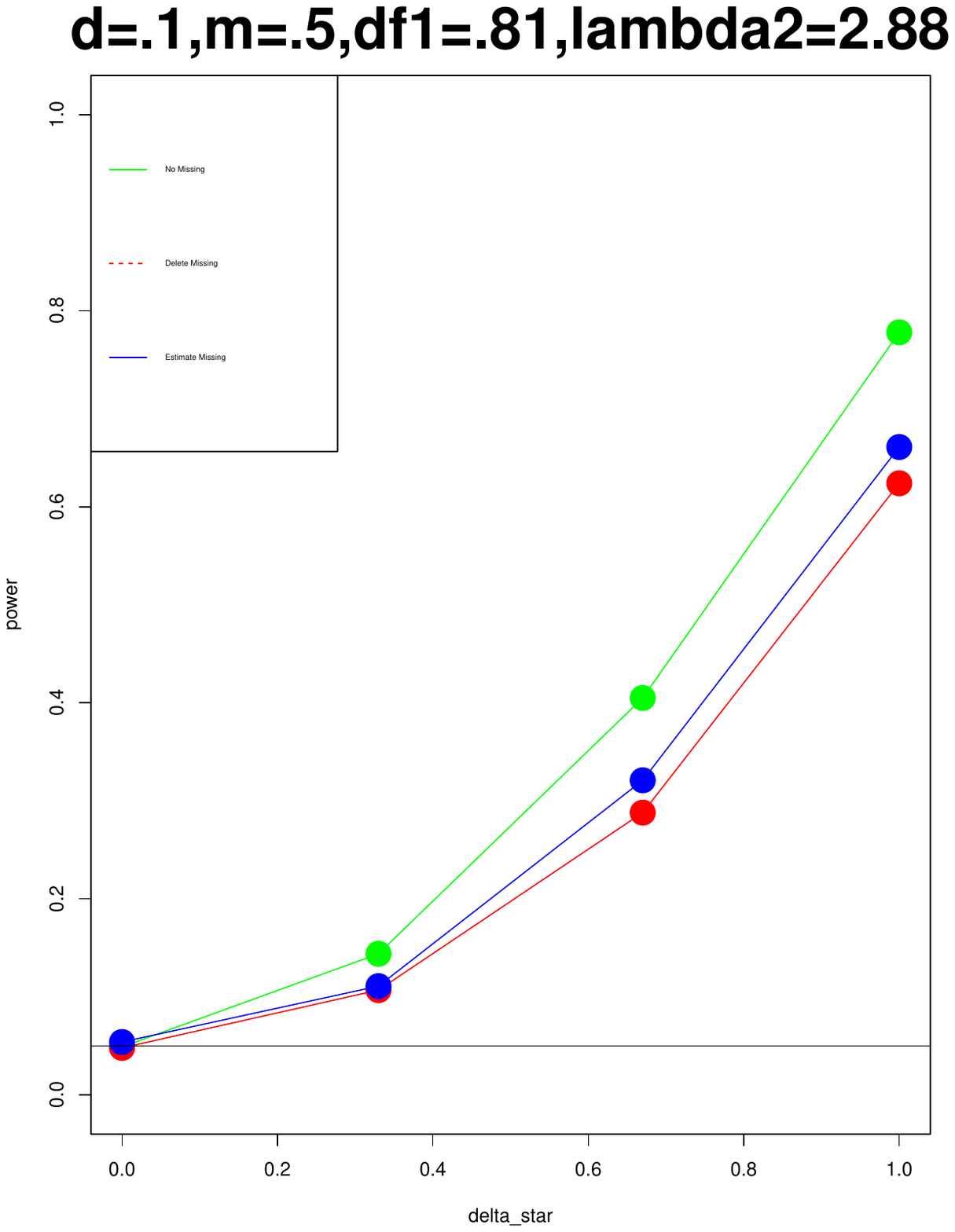}
\includegraphics[width = 2.3in, height = 1.4in]{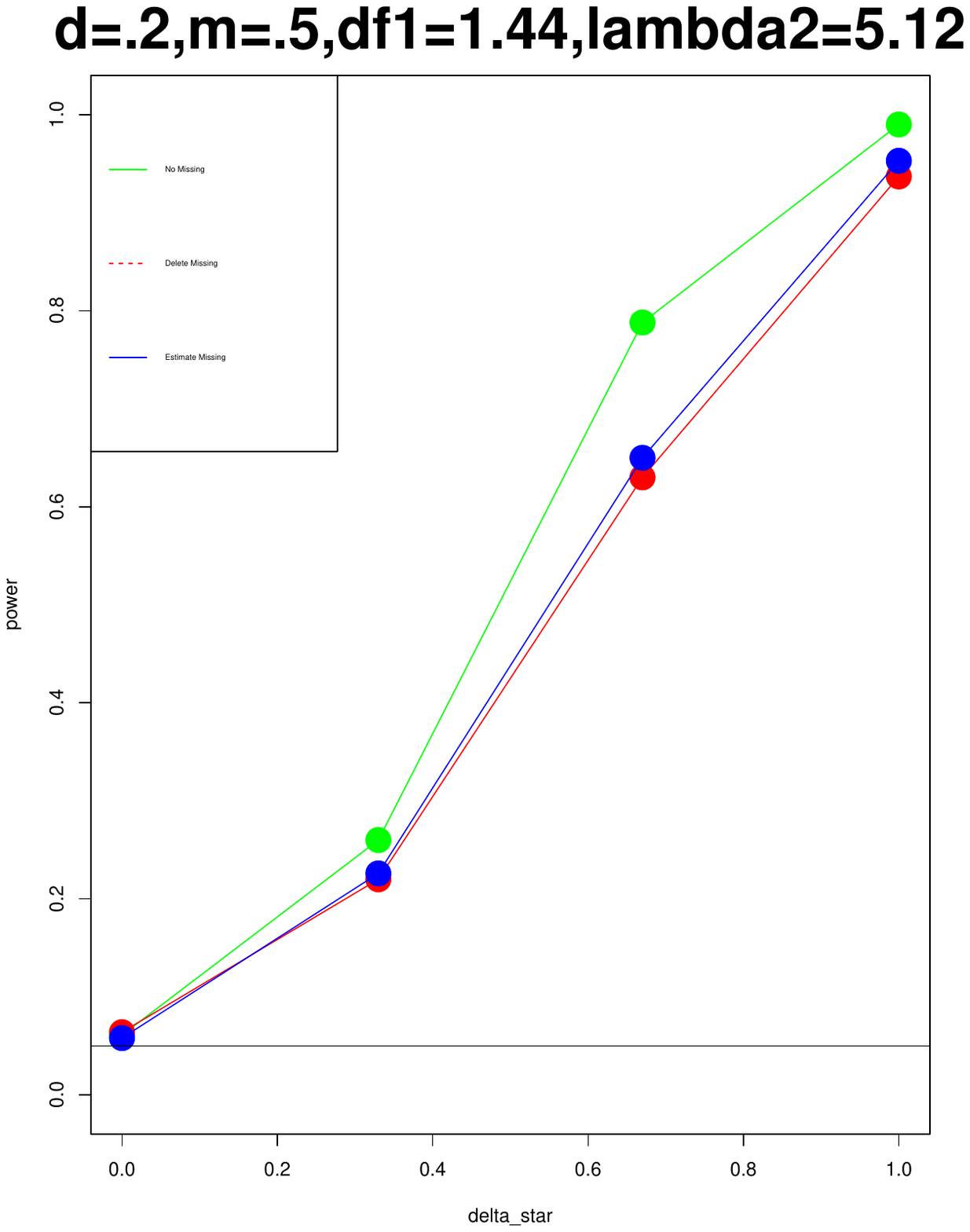}
\includegraphics[width = 2.3in, height = 1.4in]{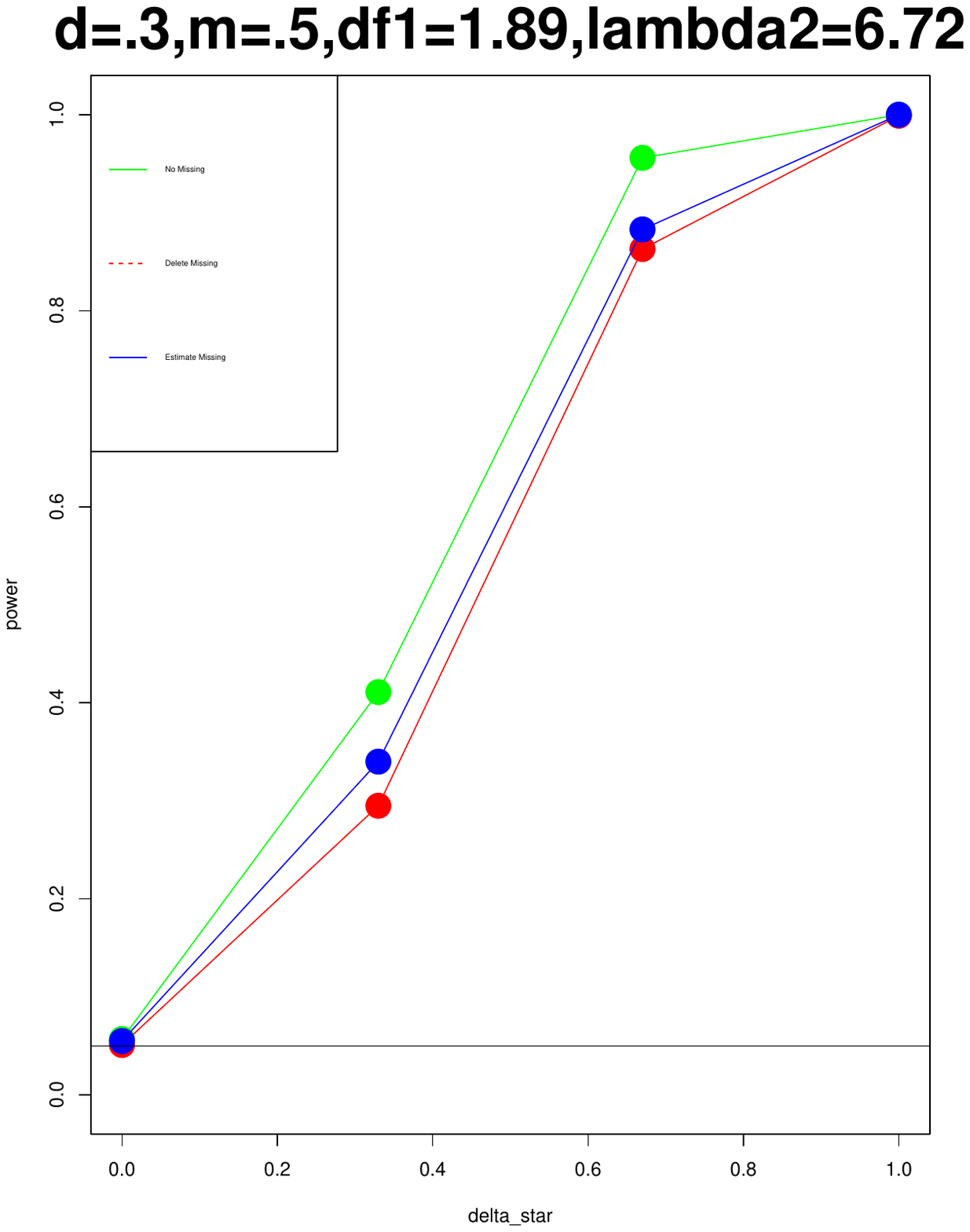}

\hspace{1.5cm}
$d=.1 \quad m=.1$
\hspace{3cm}
$d=.2 \quad m=.1$
\hspace{3cm}
$d=.3 \quad m=.1$

\includegraphics[width = 2.3in, height = 1.4in]{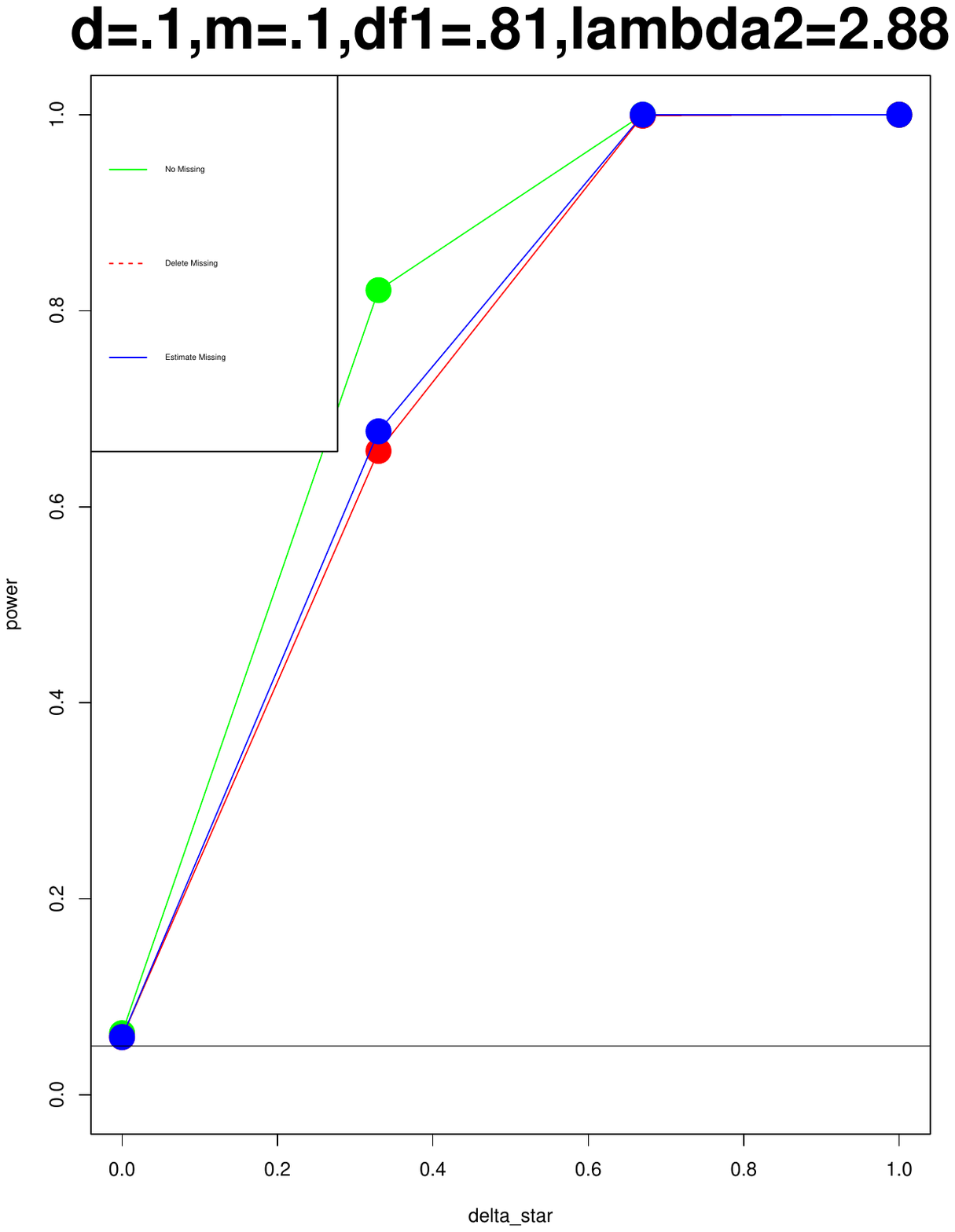}
\includegraphics[width = 2.3in, height = 1.4in]{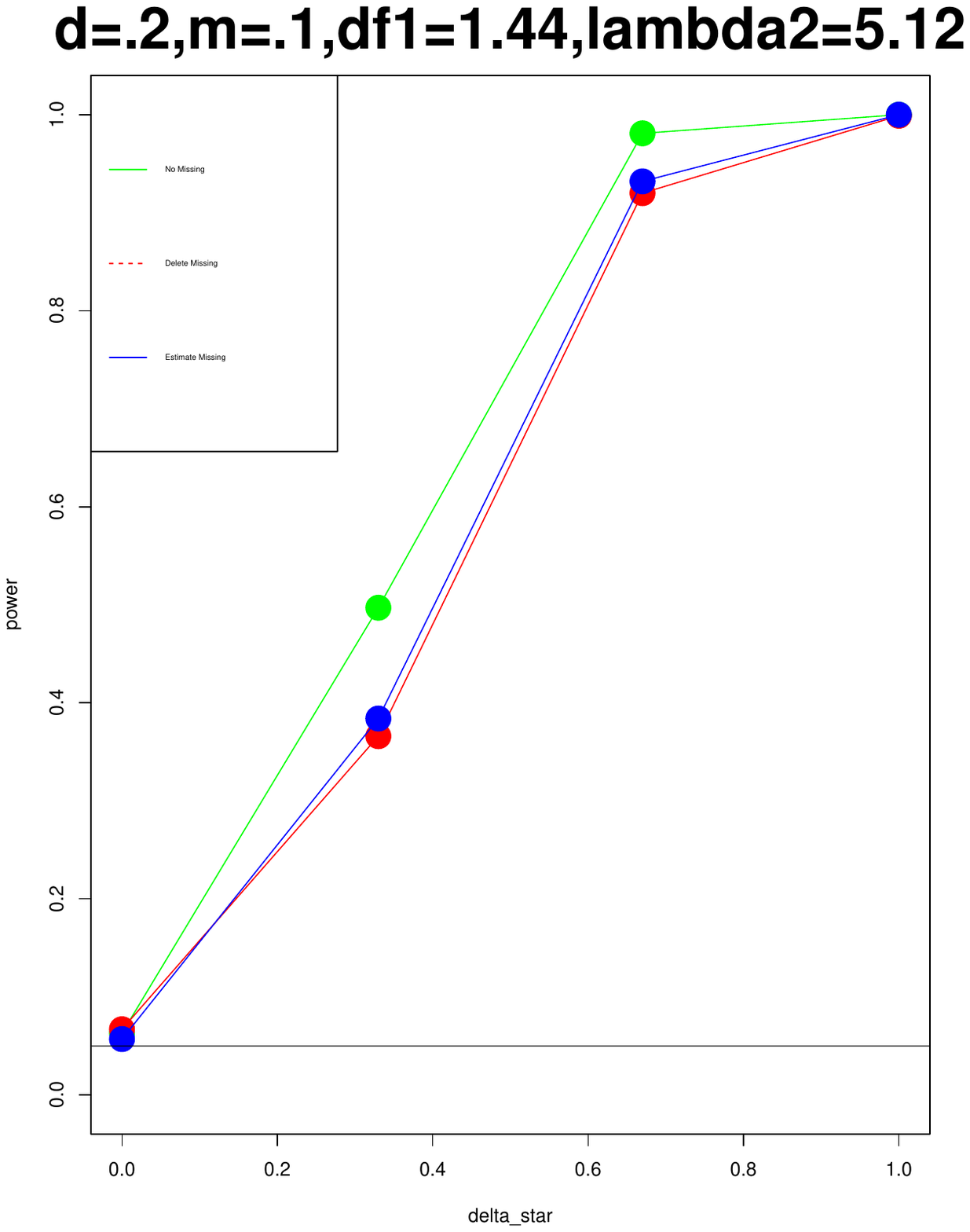}
\includegraphics[width = 2.3in, height = 1.4in]{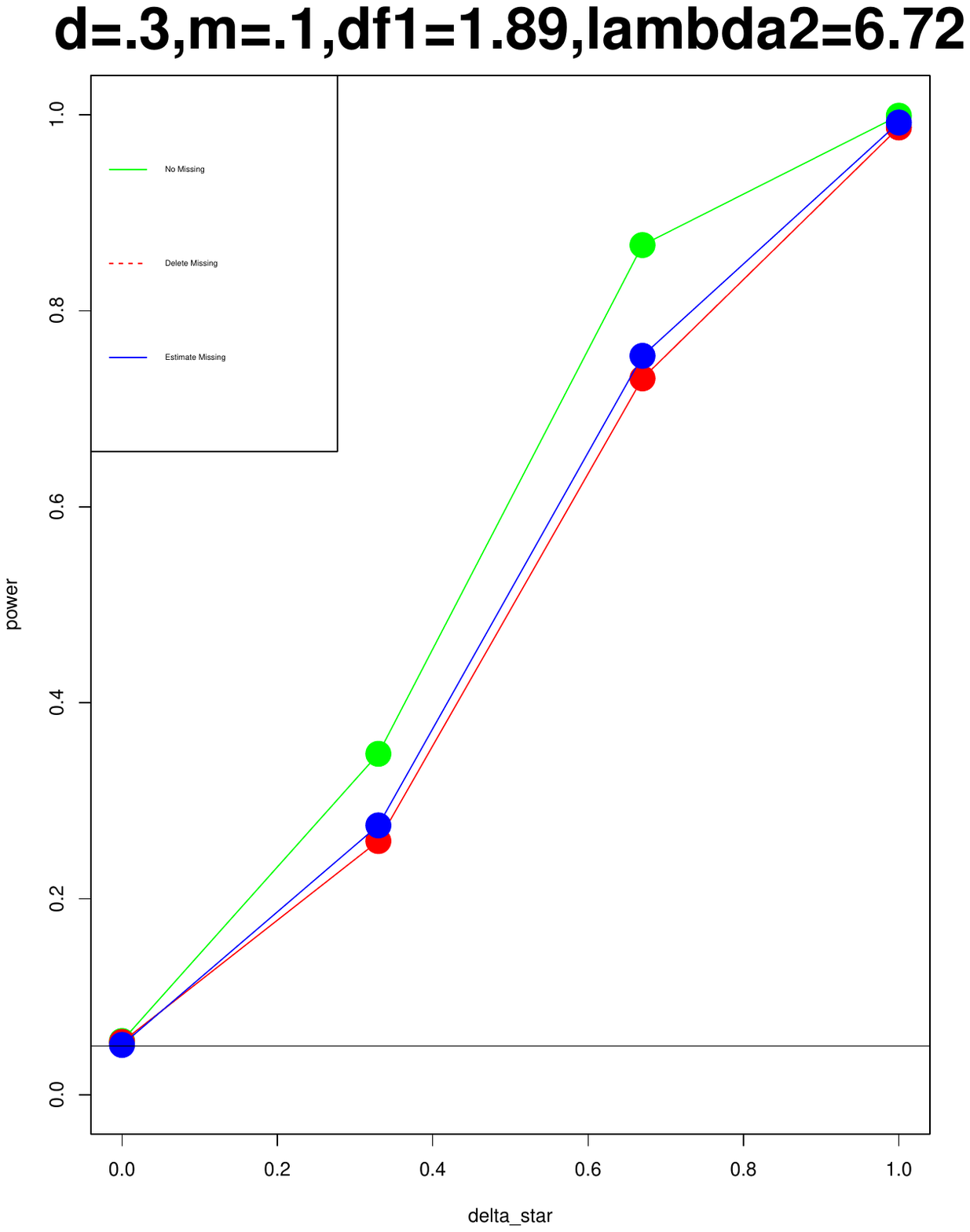}

\hspace{1.5cm}
$d=.1 \quad m=.1$
\hspace{3cm}
$d=.2 \quad m=.1$
\hspace{3cm}
$d=.3 \quad m=.1$

\includegraphics[width = 2.3in, height = 1.4in]{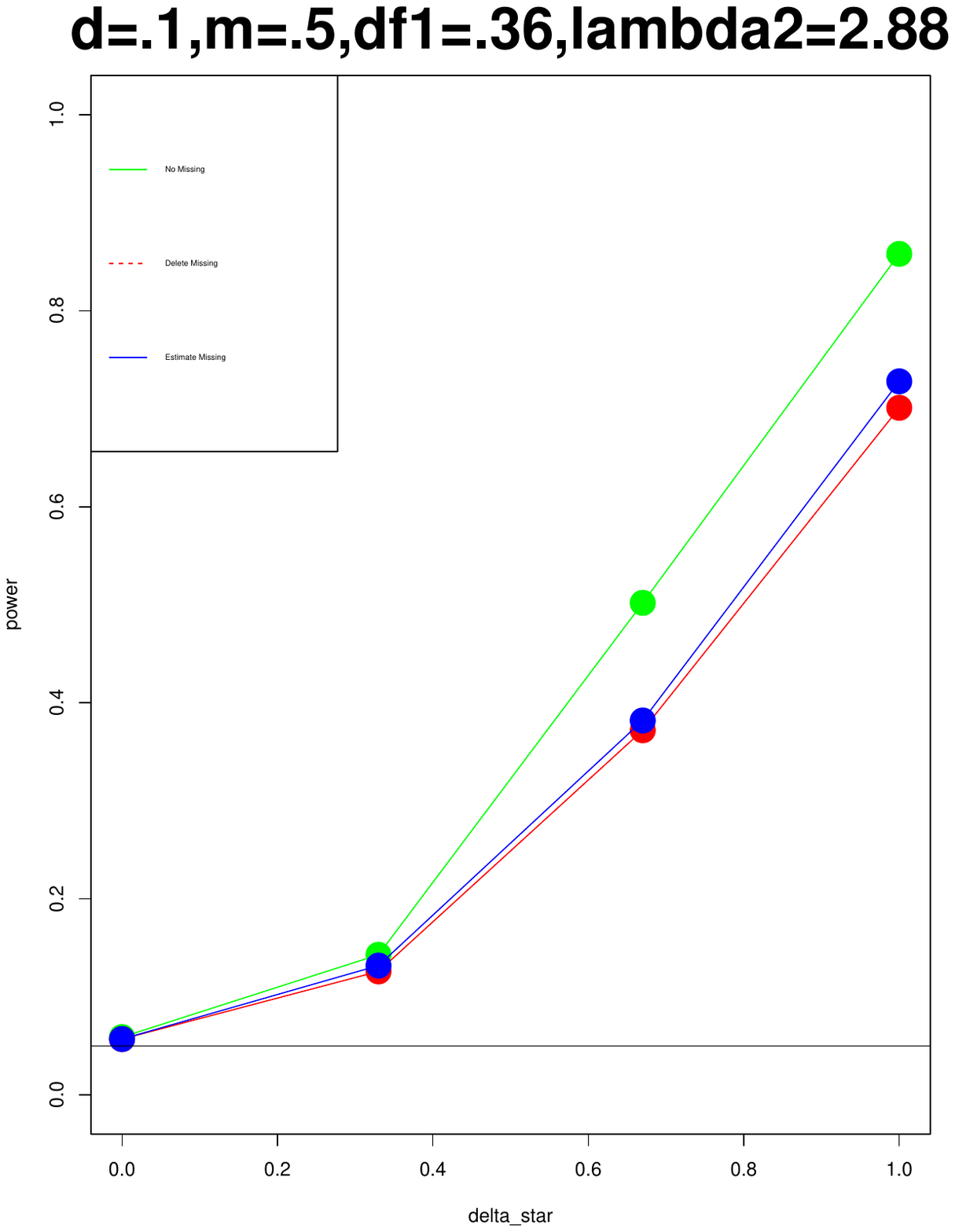}
\includegraphics[width = 2.3in, height = 1.4in]{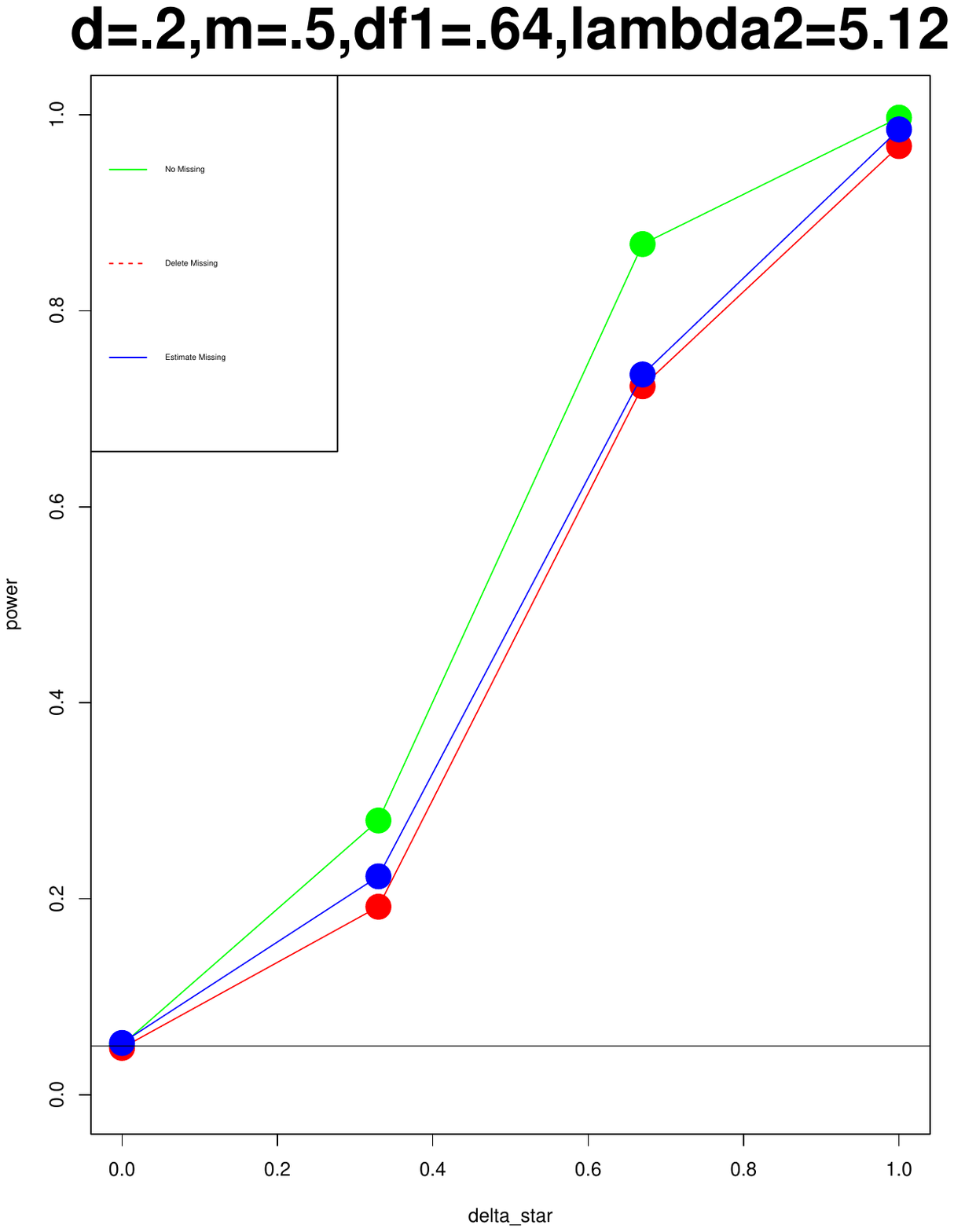}
\includegraphics[width = 2.3in, height = 1.4in]{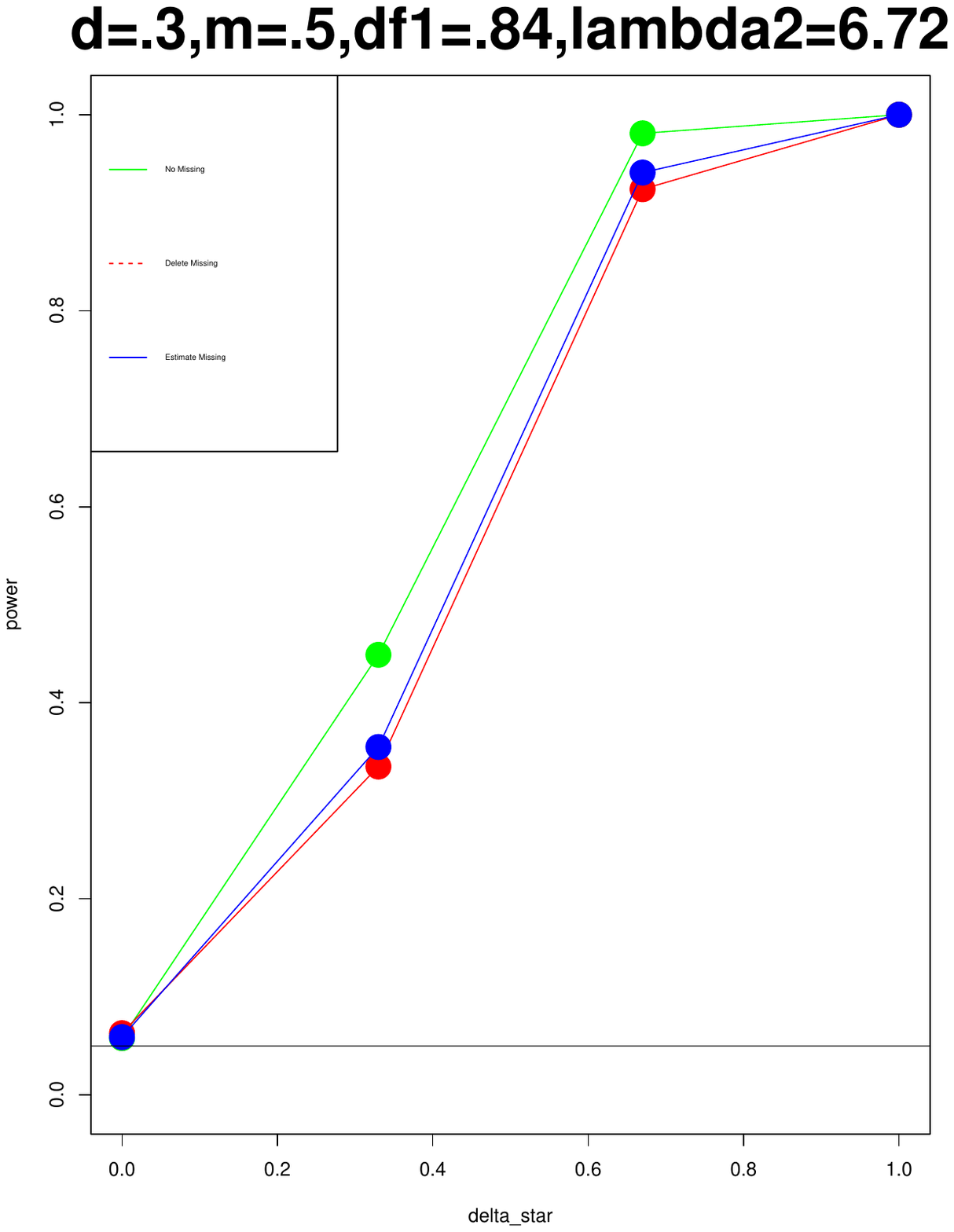}

\hspace{1.5cm}
$d=.1 \quad m=.5$
\hspace{3cm}
$d=.2 \quad m=.5$
\hspace{3cm}
$d=.3 \quad m=.5$

\includegraphics[width = 2.3in, height = 1.4in]{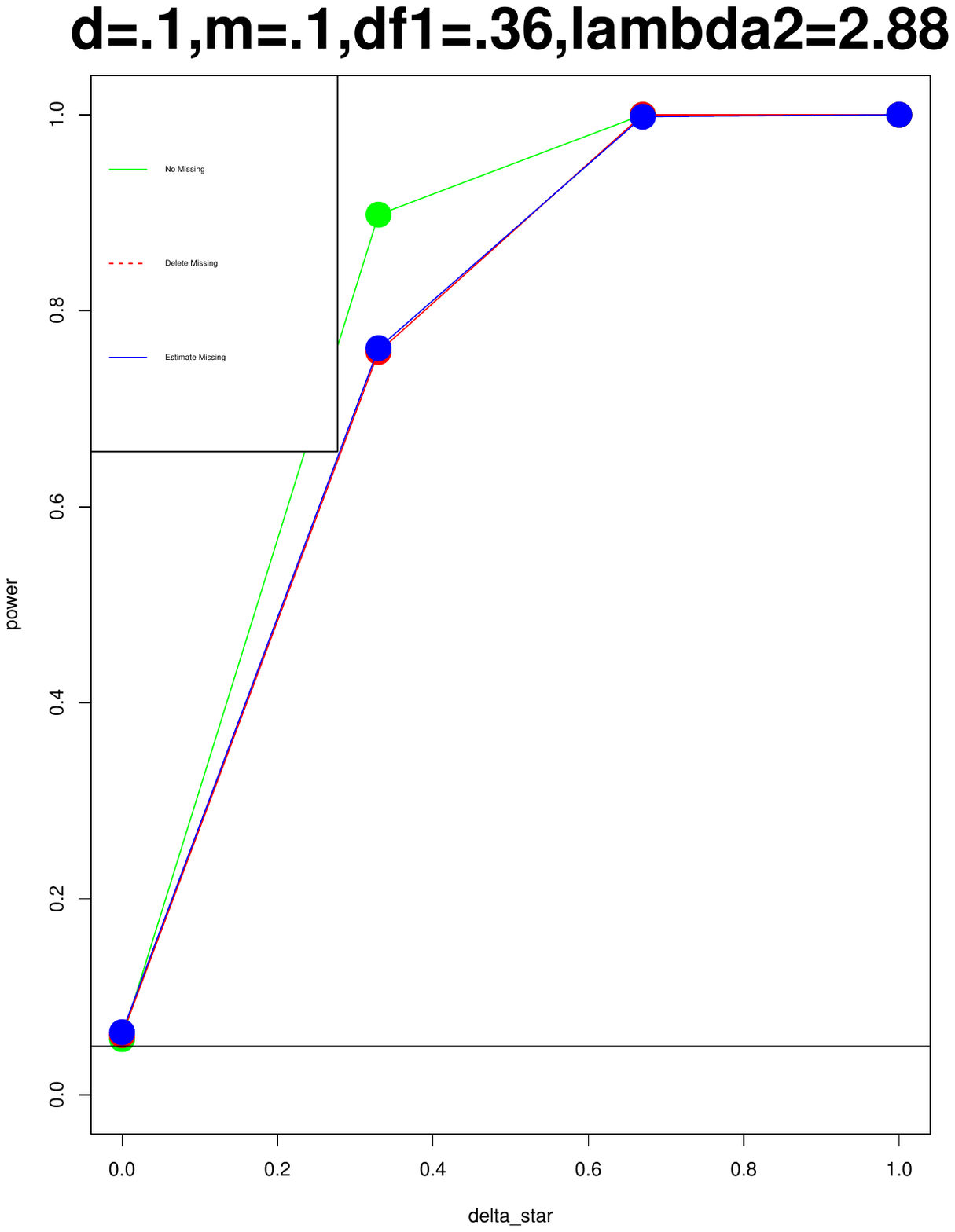}
\includegraphics[width = 2.3in, height = 1.4in]{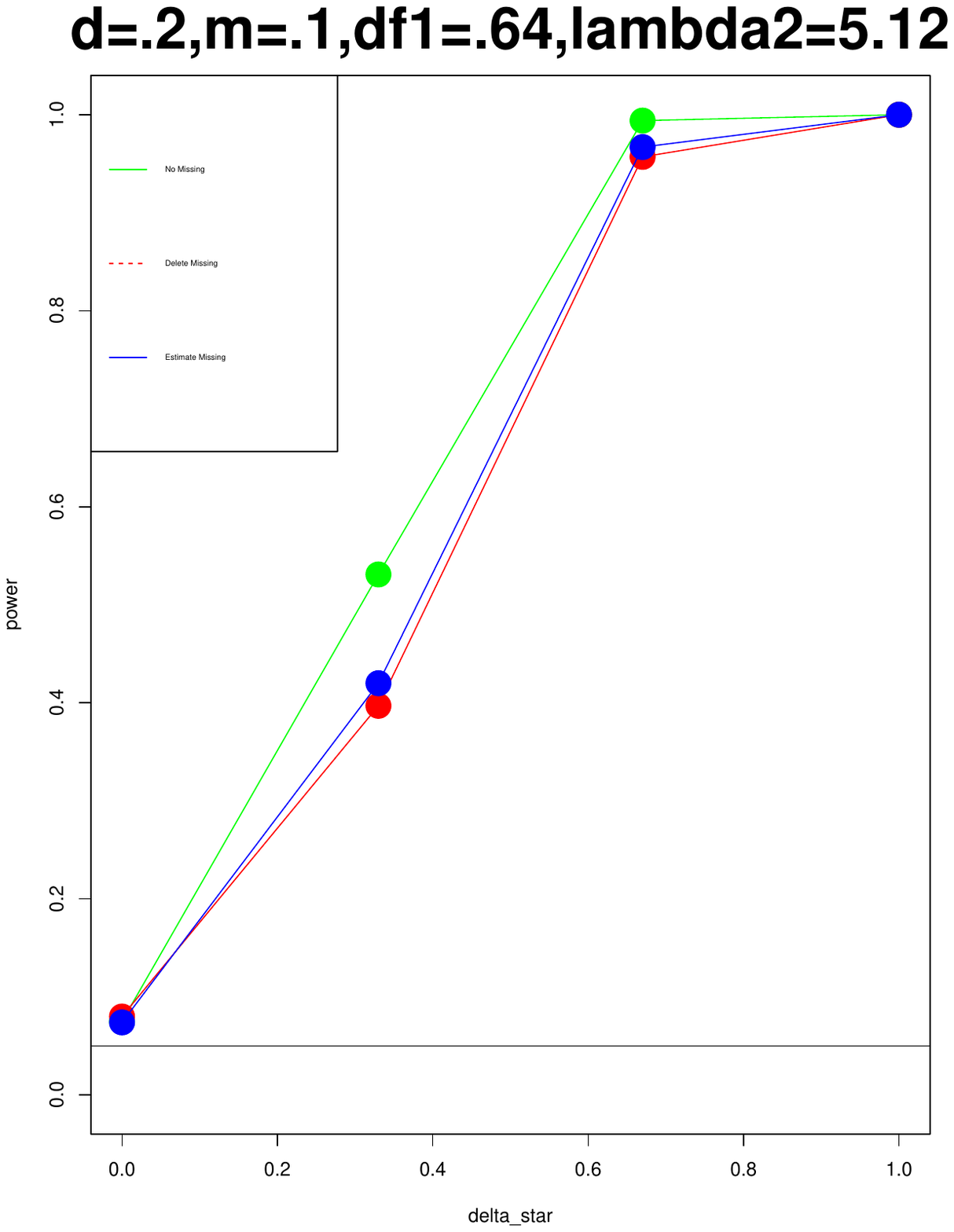}
\includegraphics[width = 2.3in, height = 1.4in]{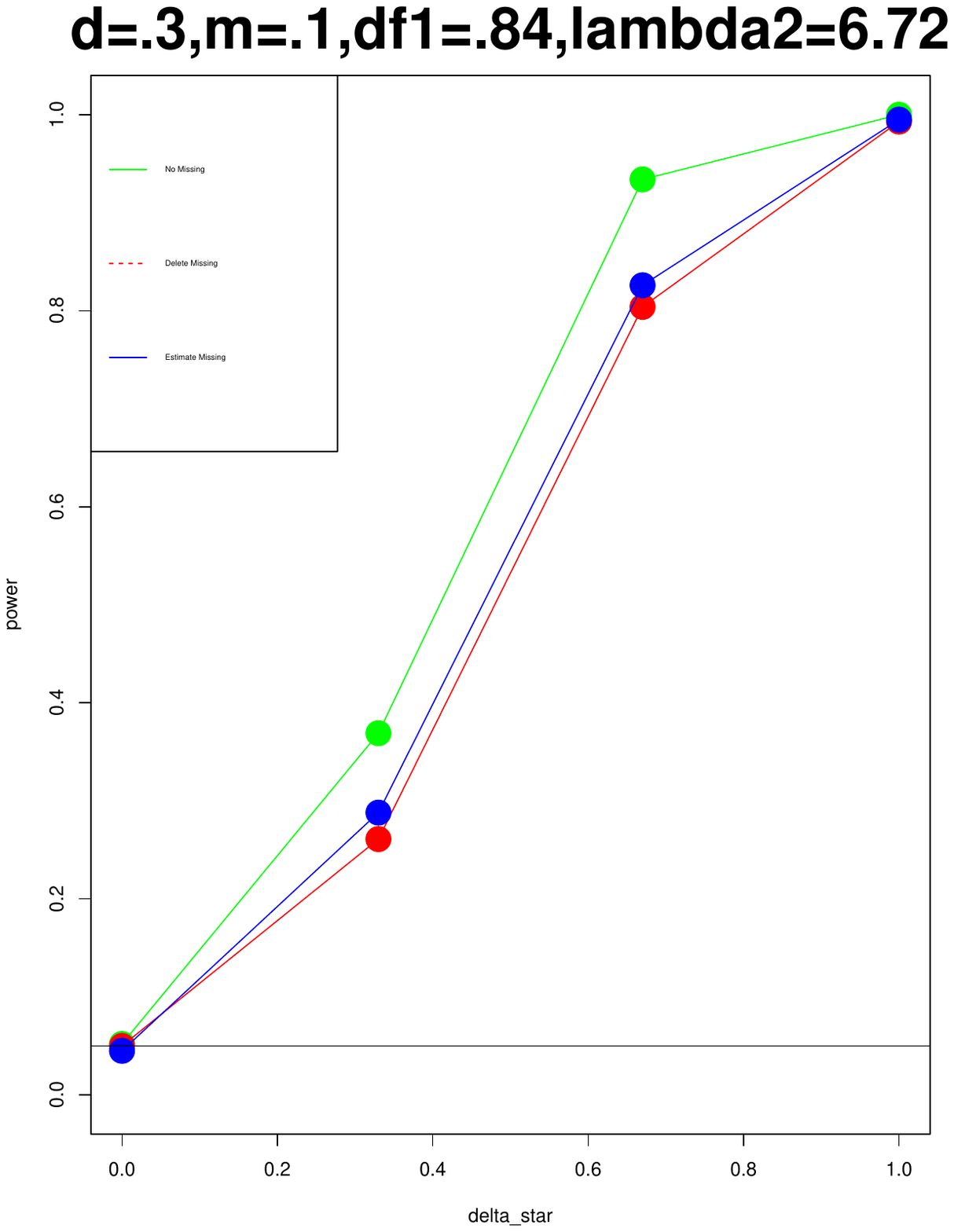}

\hspace{1.5cm}
$d=.1 \quad m=.1$
\hspace{3cm}
$d=.2 \quad m=.1$
\hspace{3cm}
$d=.3 \quad m=.1$

\includegraphics[width = 2.3in, height = 1.4in]{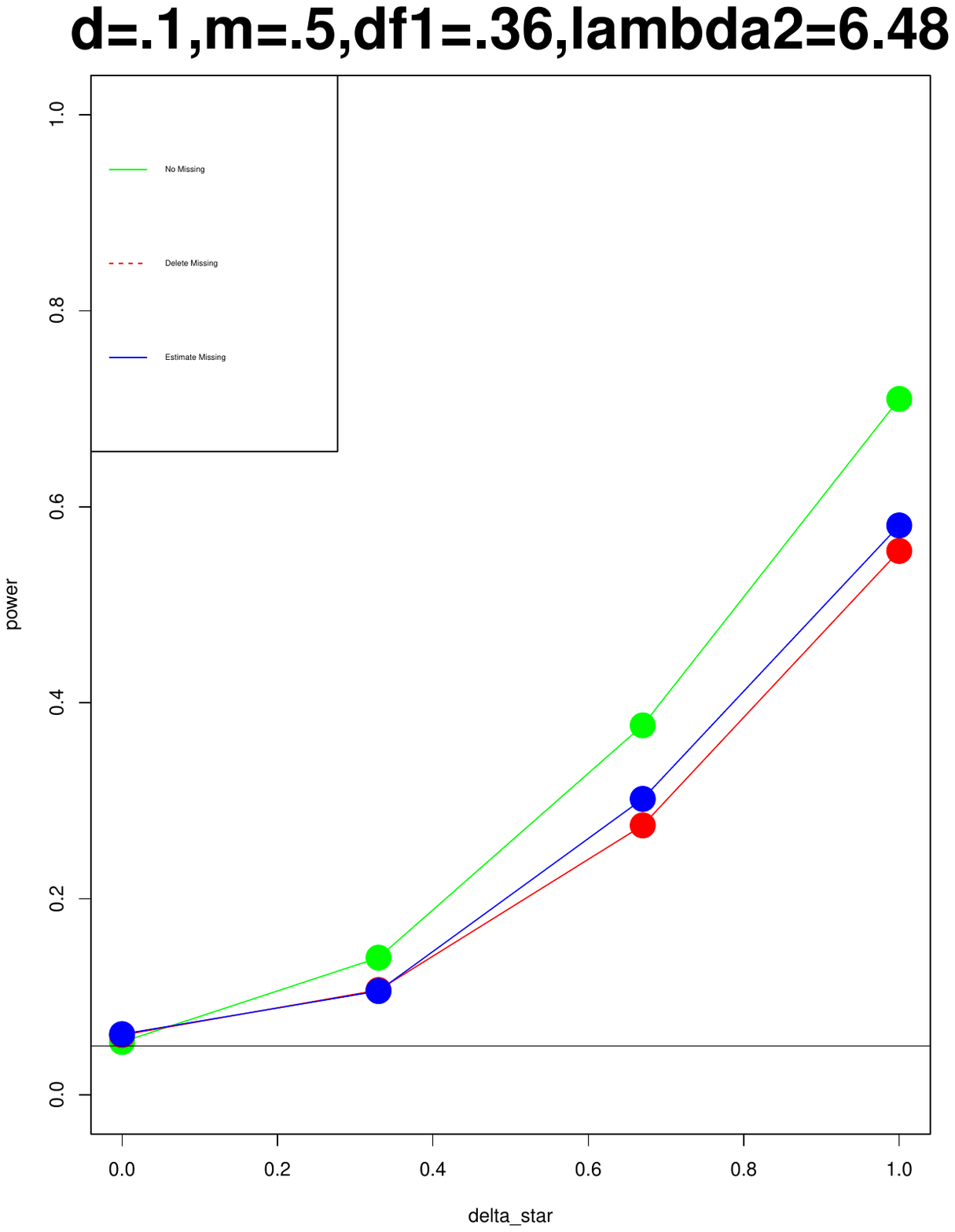}
\includegraphics[width = 2.3in, height = 1.4in]{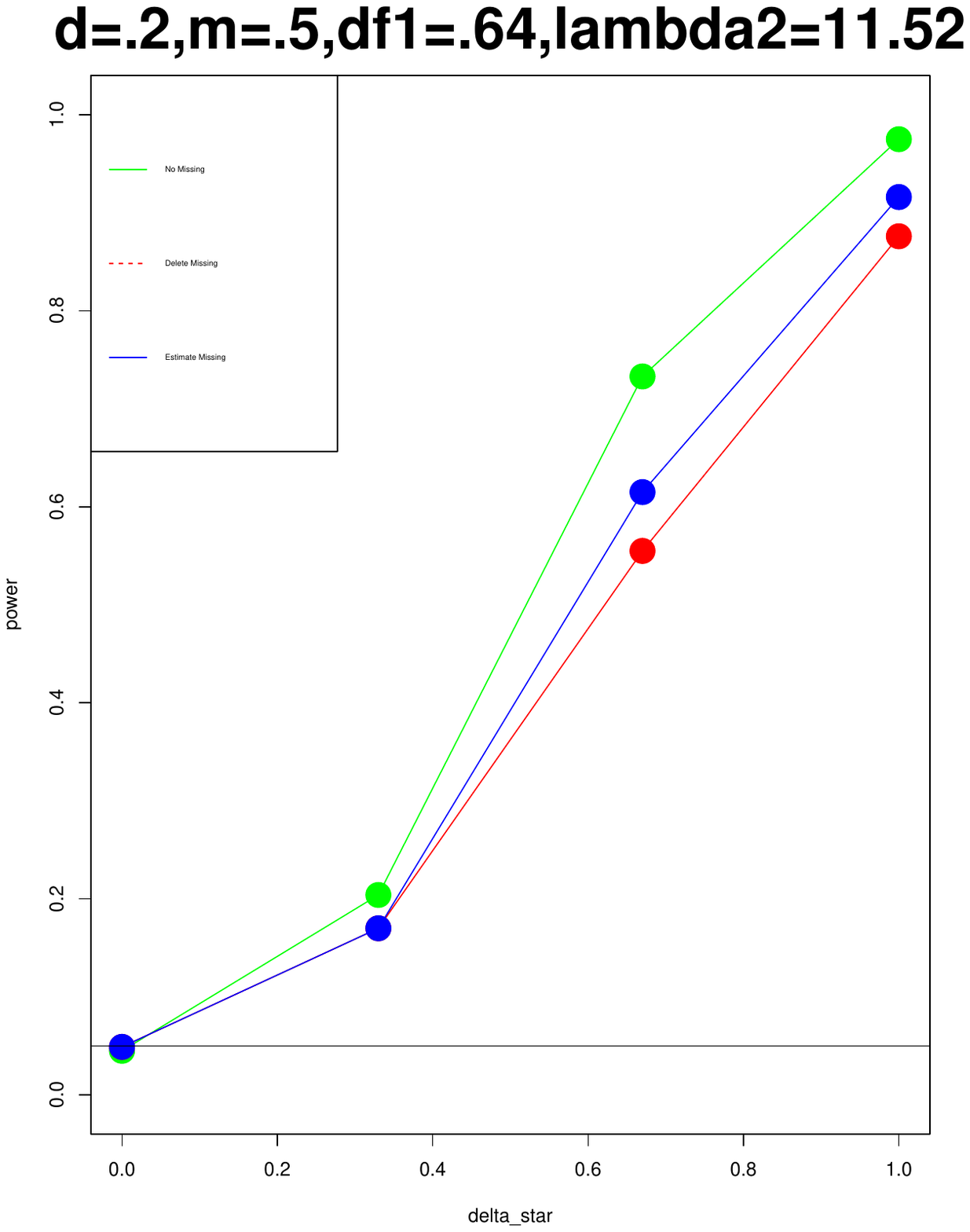}
\includegraphics[width = 2.3in, height = 1.4in]{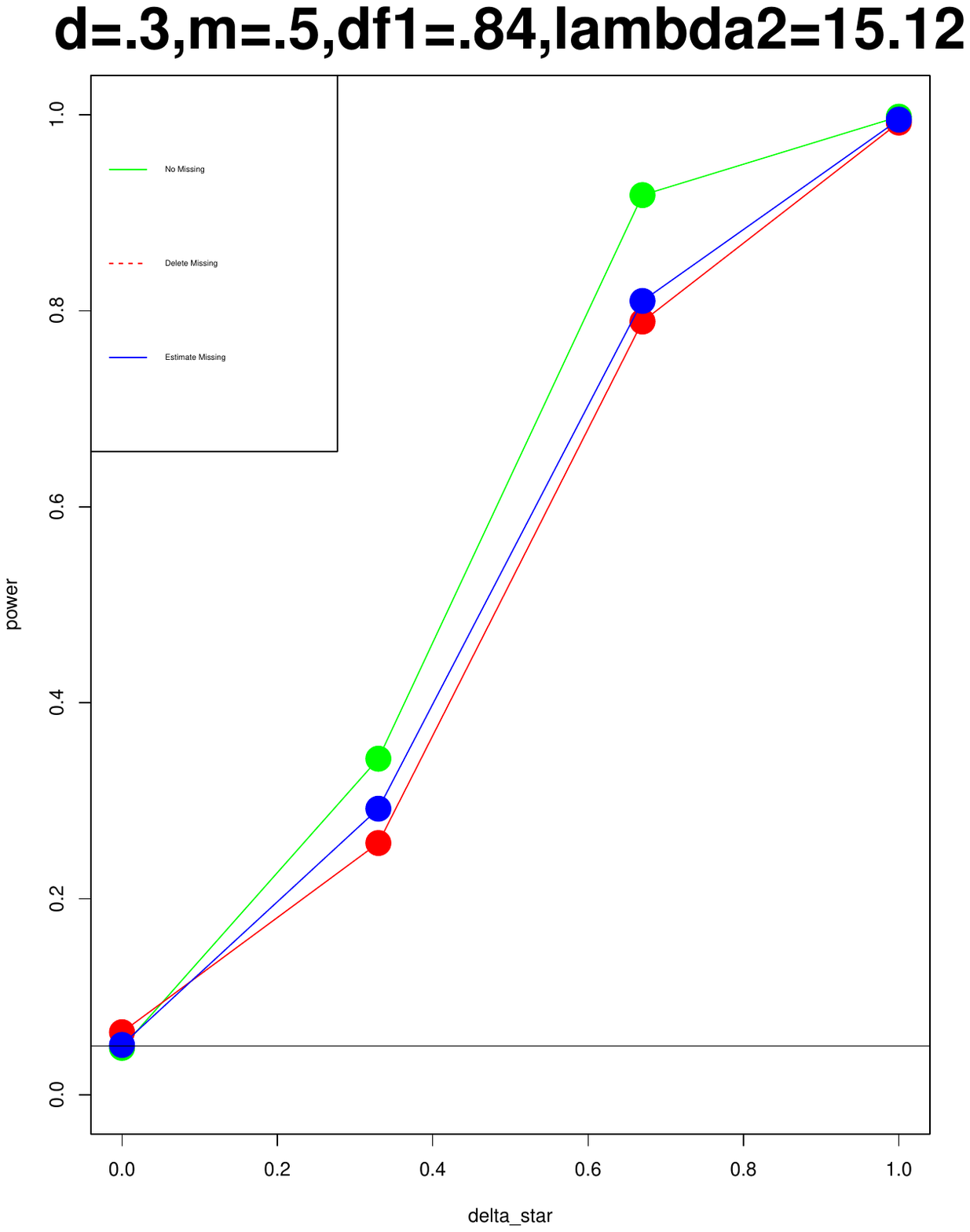}

\hspace{1.5cm}
$d=.1 \quad m=.5$
\hspace{3cm}
$d=.2 \quad m=.5$
\hspace{3cm}
$d=.3 \quad m=.5$

\includegraphics[width = 2.3in, height = 1.4in]{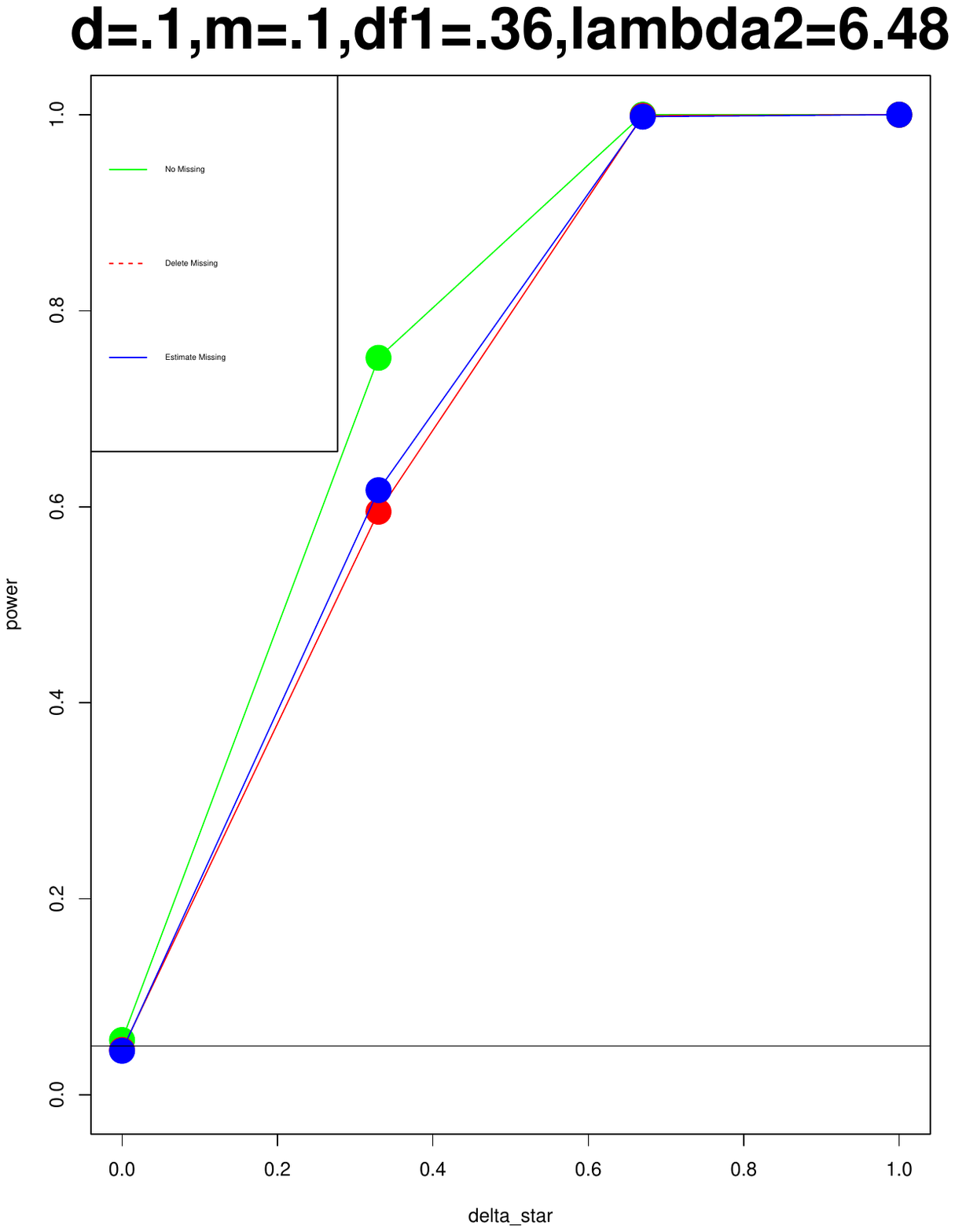}
\includegraphics[width = 2.3in, height = 1.4in]{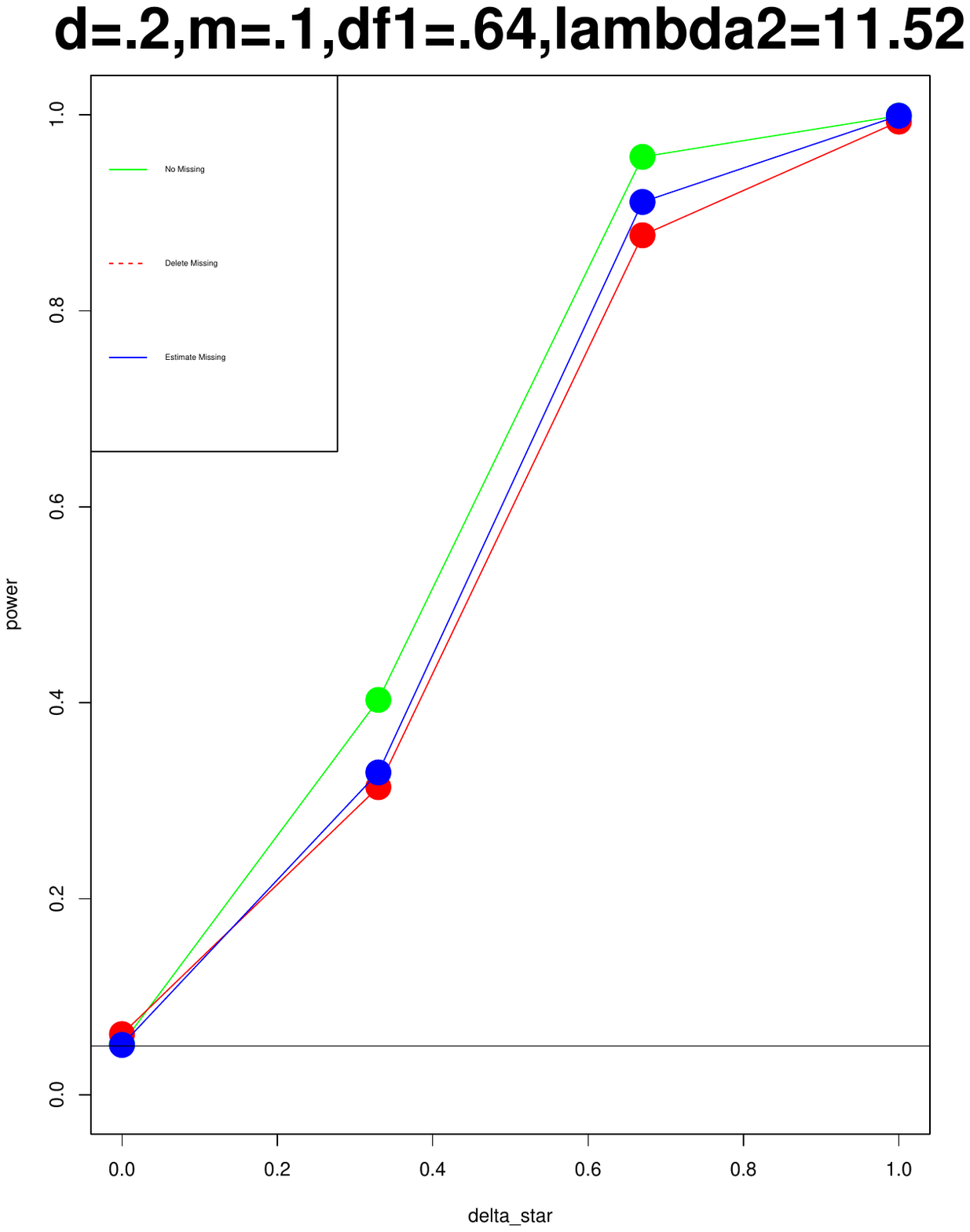}
\includegraphics[width = 2.3in, height = 1.4in]{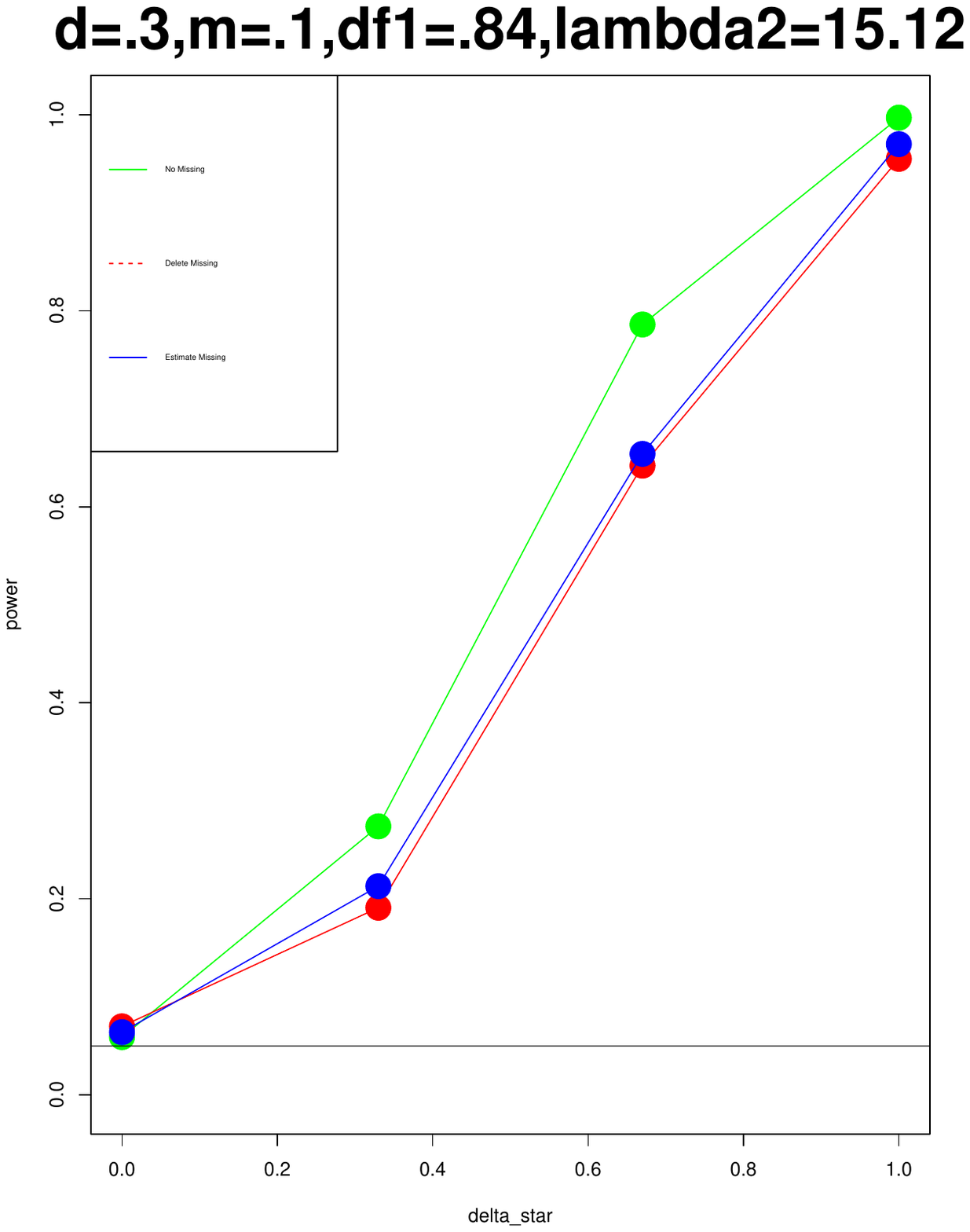}

\subsubsection{When both Traits have Chi Squares Distribution}

\hspace{1.5cm}
$d=.1 \quad m=.5$
\hspace{3cm}
$d=.2 \quad m=.5$
\hspace{3cm}
$d=.3 \quad m=.5$

\includegraphics[width = 2.3in, height = 1.4in]{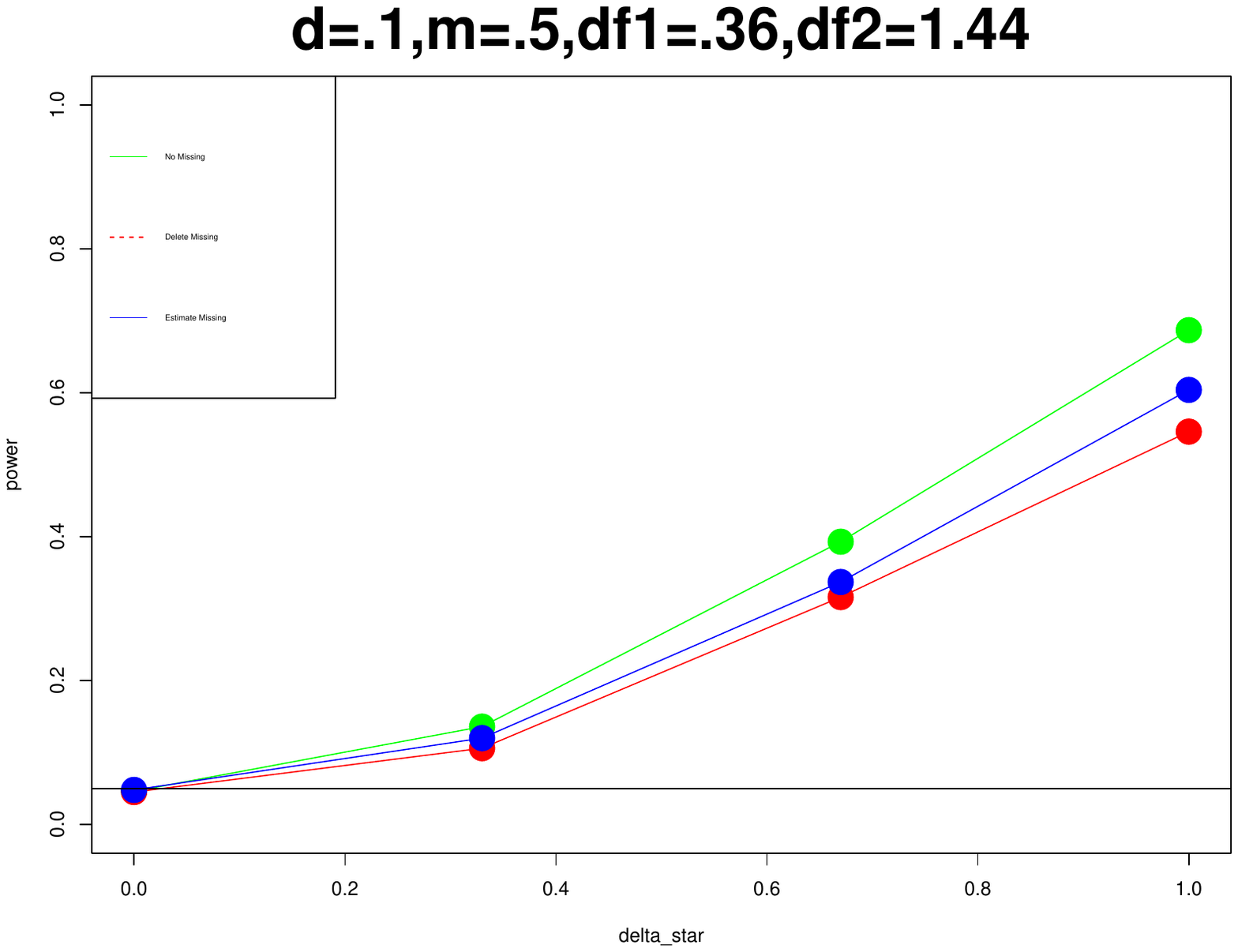}
\includegraphics[width = 2.3in, height = 1.4in]{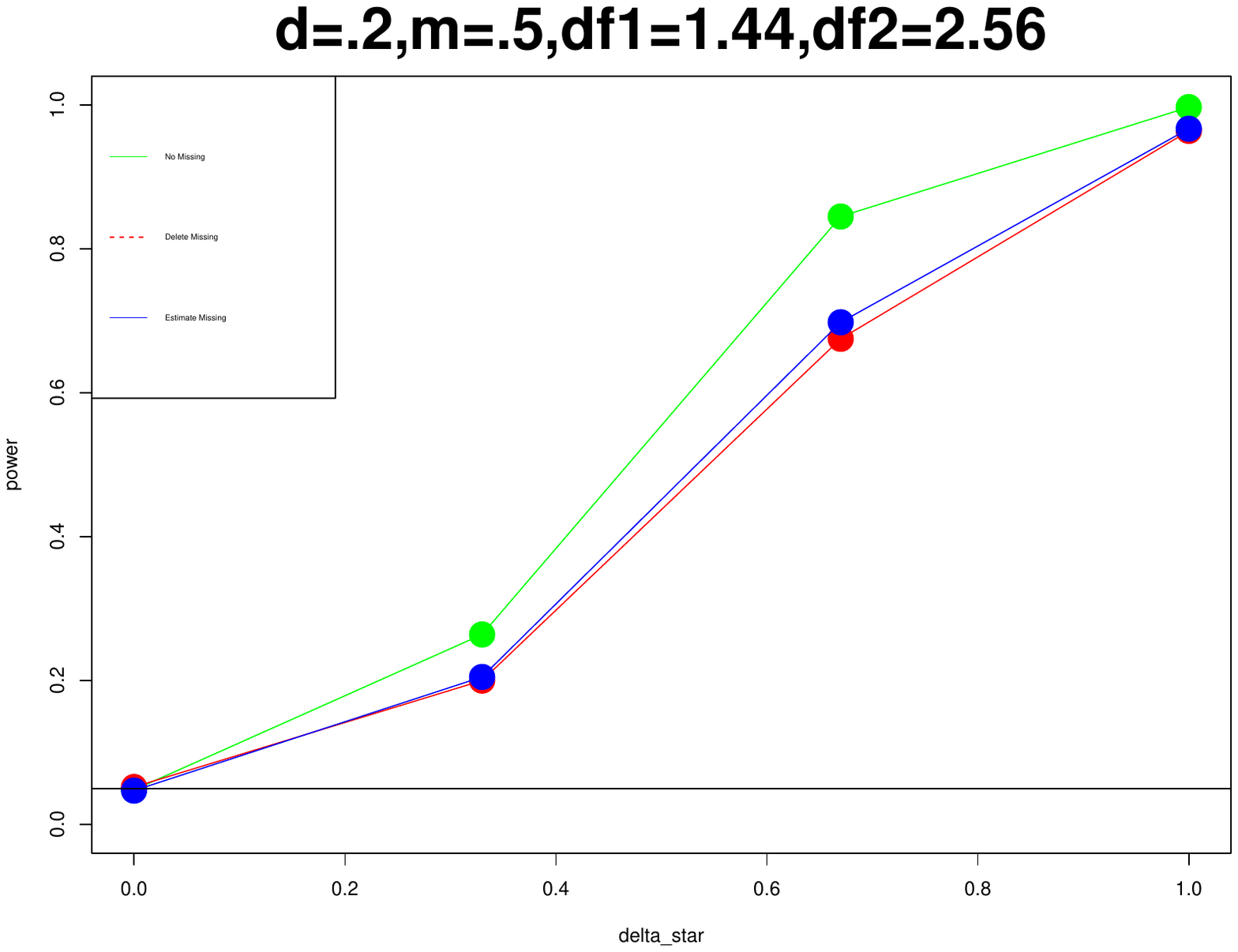}
\includegraphics[width = 2.3in, height = 1.4in]{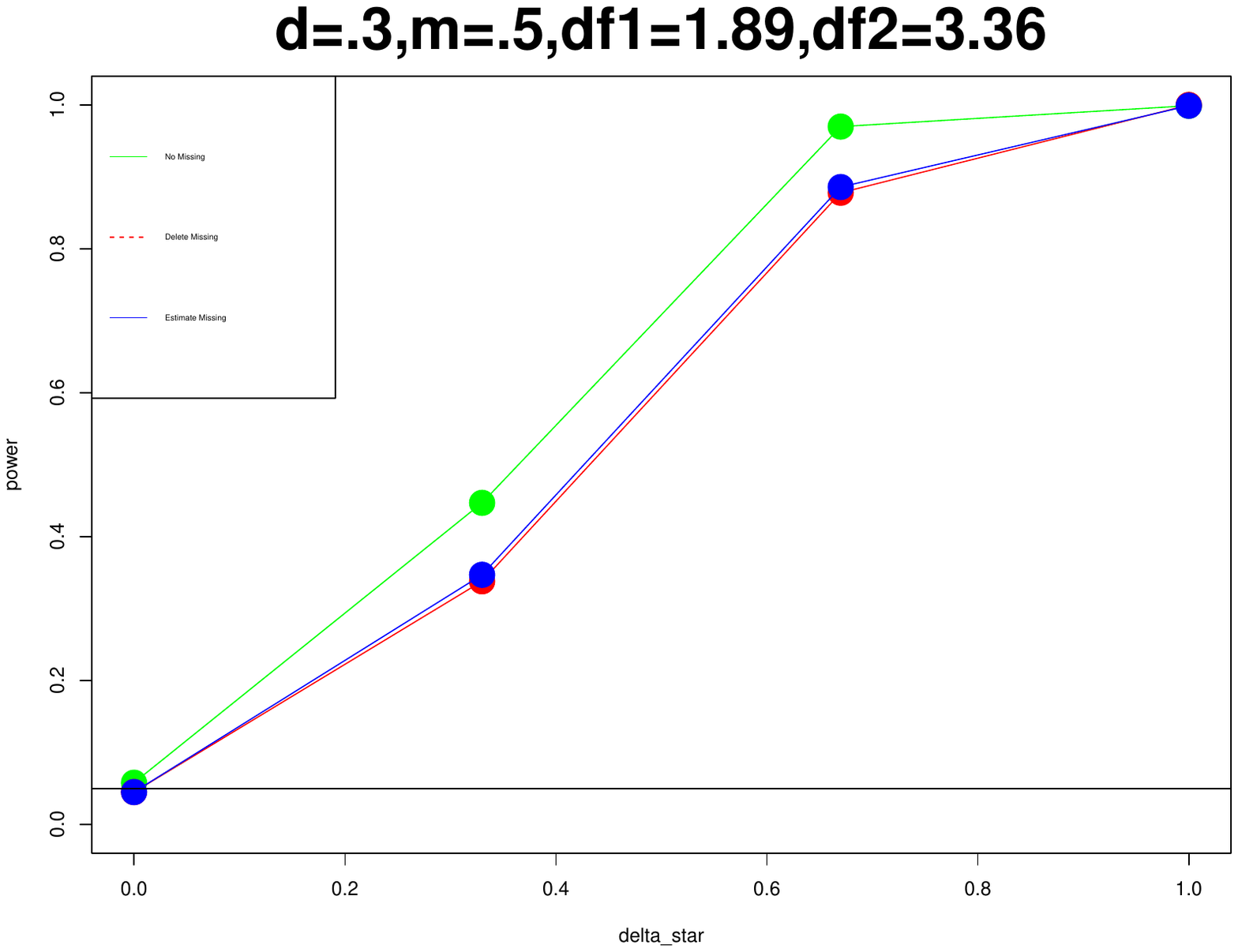}

\hspace{1.5cm}
$d=.1 \quad m=.1$
\hspace{3cm}
$d=.2 \quad m=.1$
\hspace{3cm}
$d=.3 \quad m=.1$

\includegraphics[width = 2.3in, height = 1.4in]{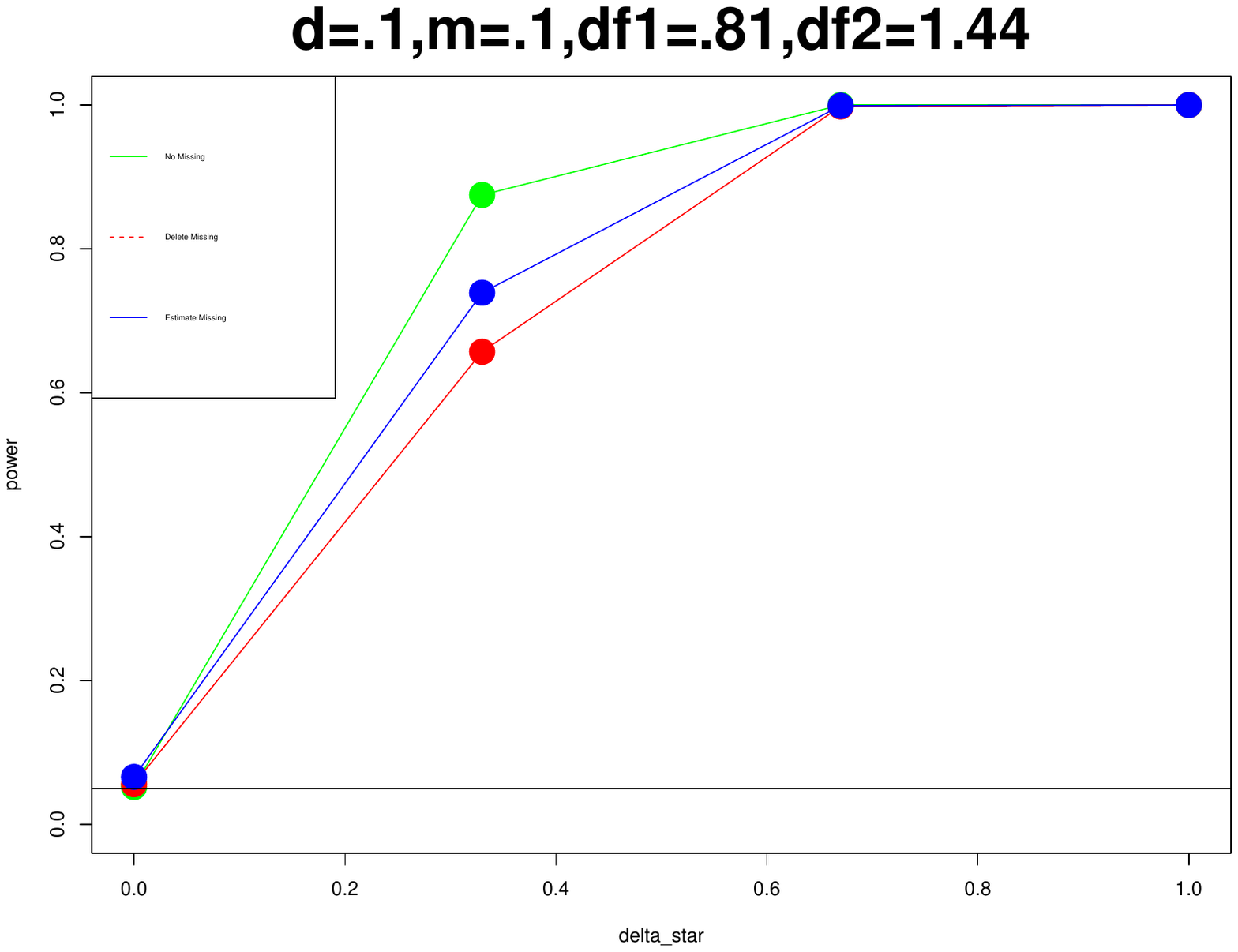}
\includegraphics[width = 2.3in, height = 1.4in]{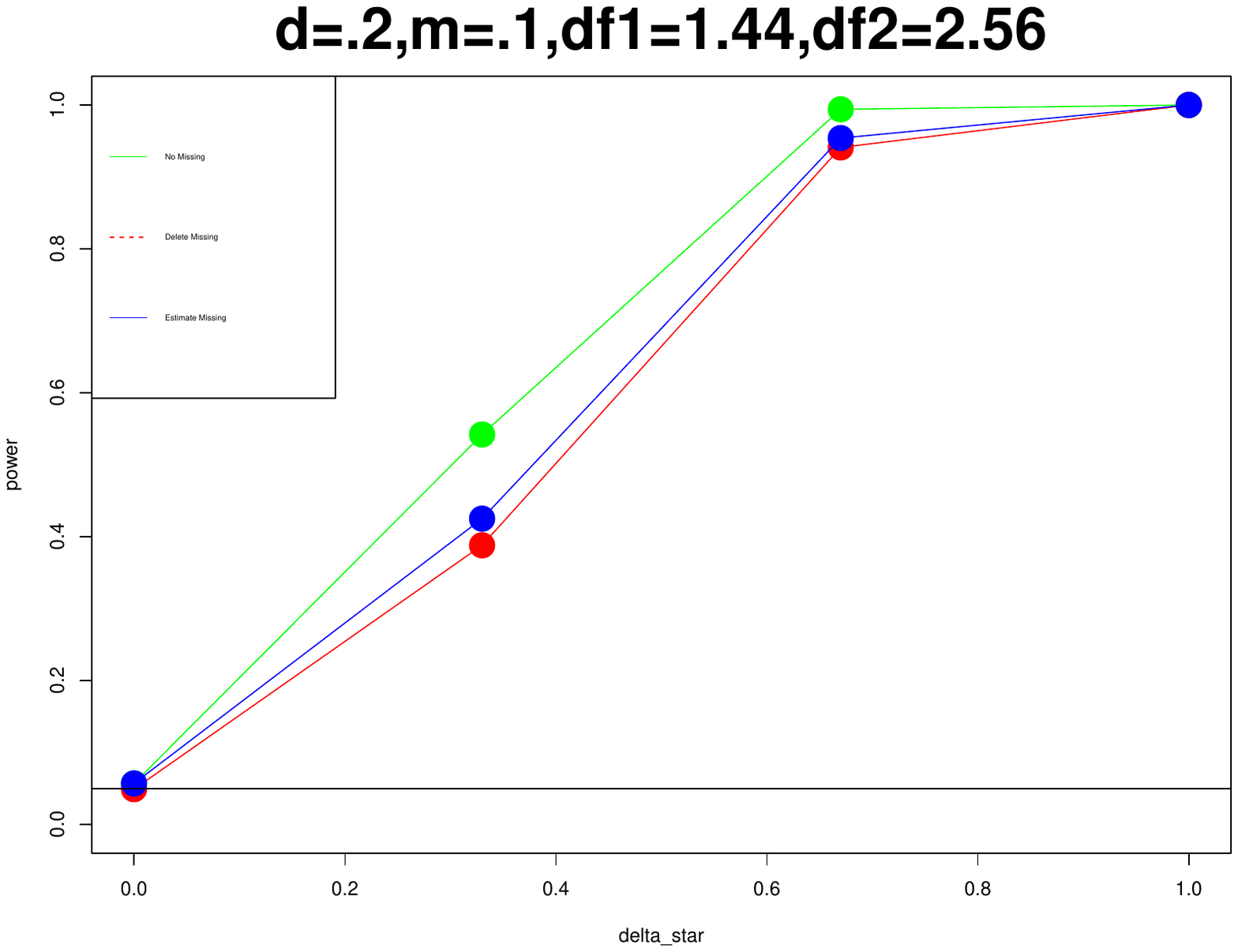}
\includegraphics[width = 2.3in, height = 1.4in]{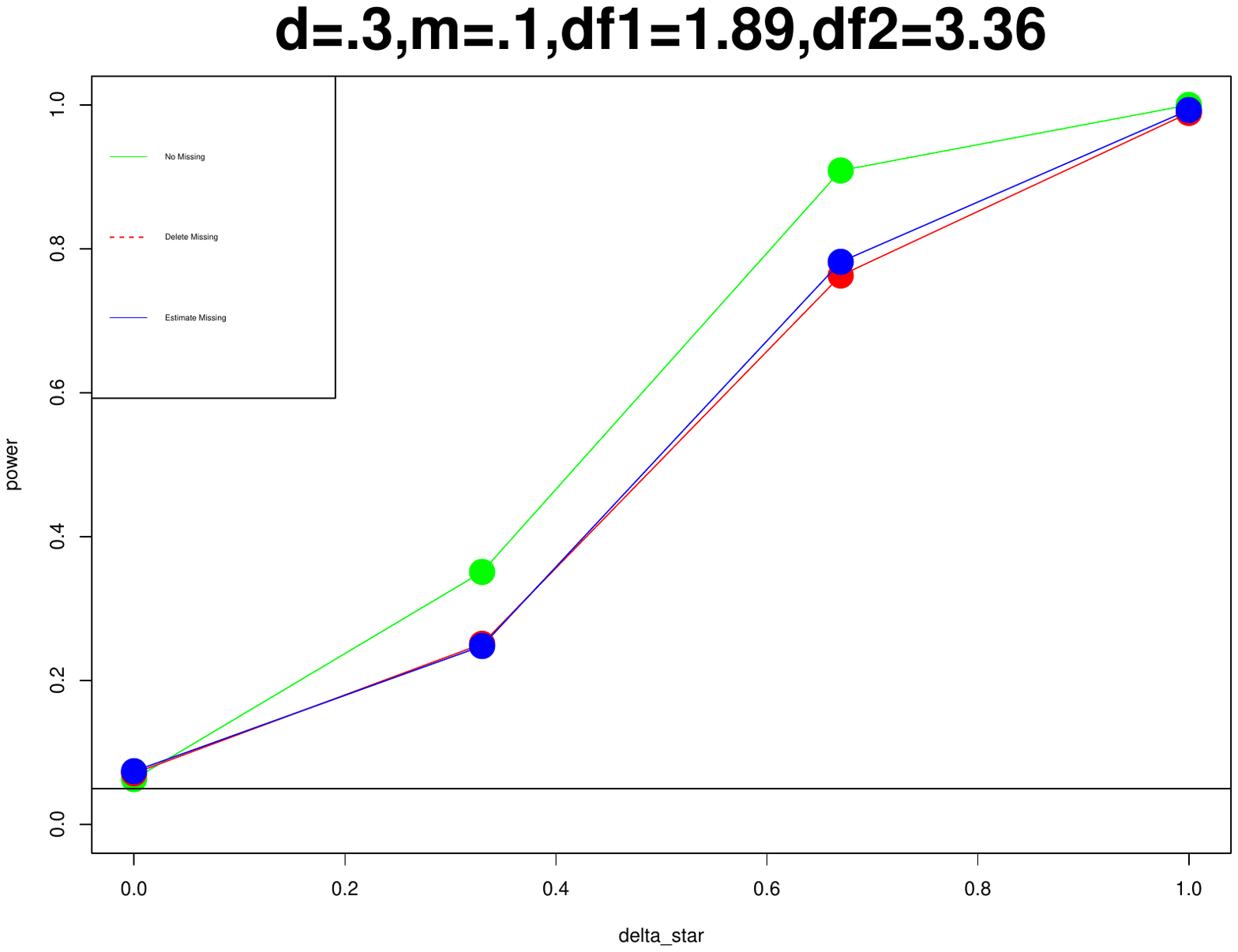}

\hspace{1.5cm}
$d=.1 \quad m=.1$
\hspace{3cm}
$d=.2 \quad m=.1$
\hspace{3cm}
$d=.3 \quad m=.1$

\includegraphics[width = 2.3in, height = 1.4in]{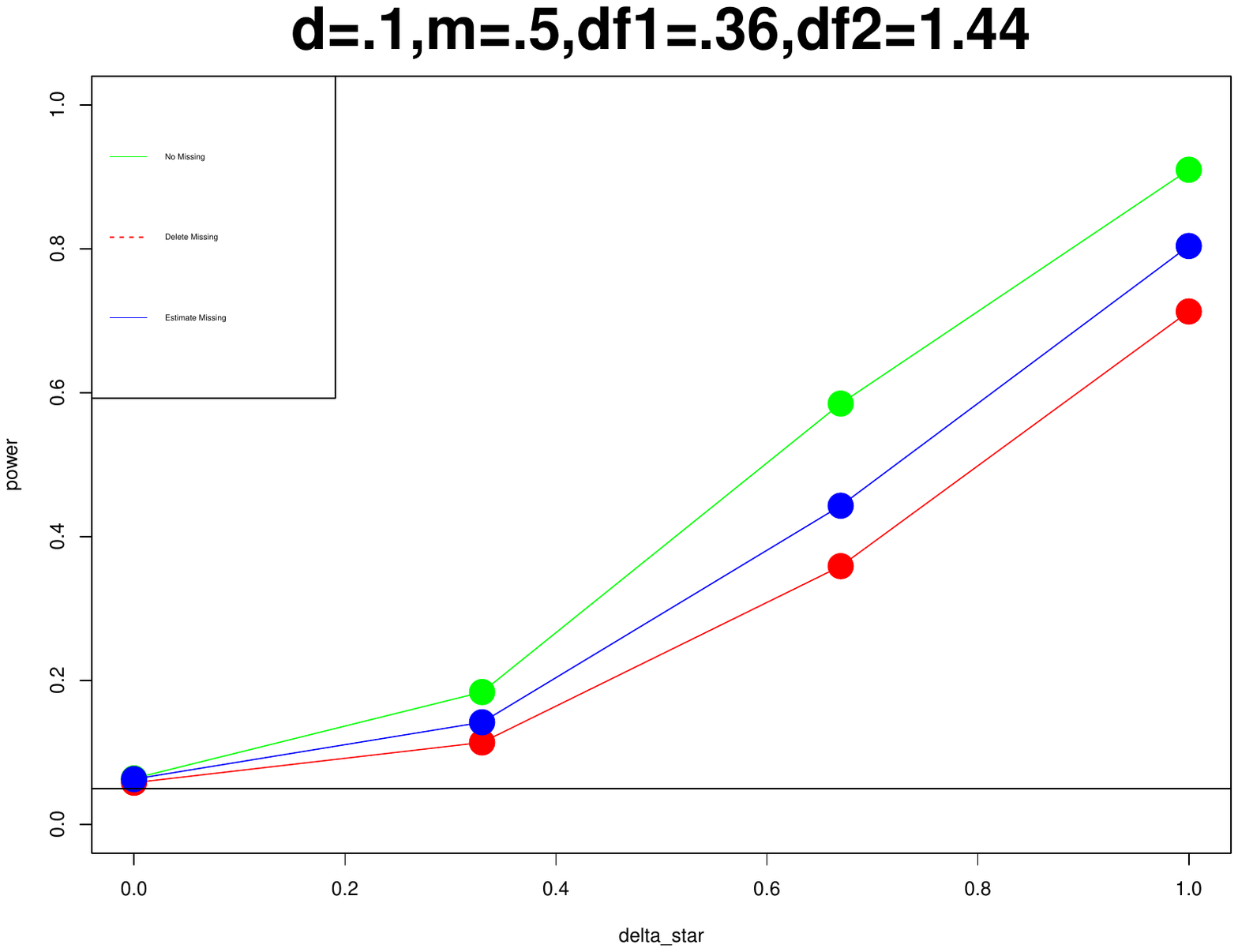}
\includegraphics[width = 2.3in, height = 1.4in]{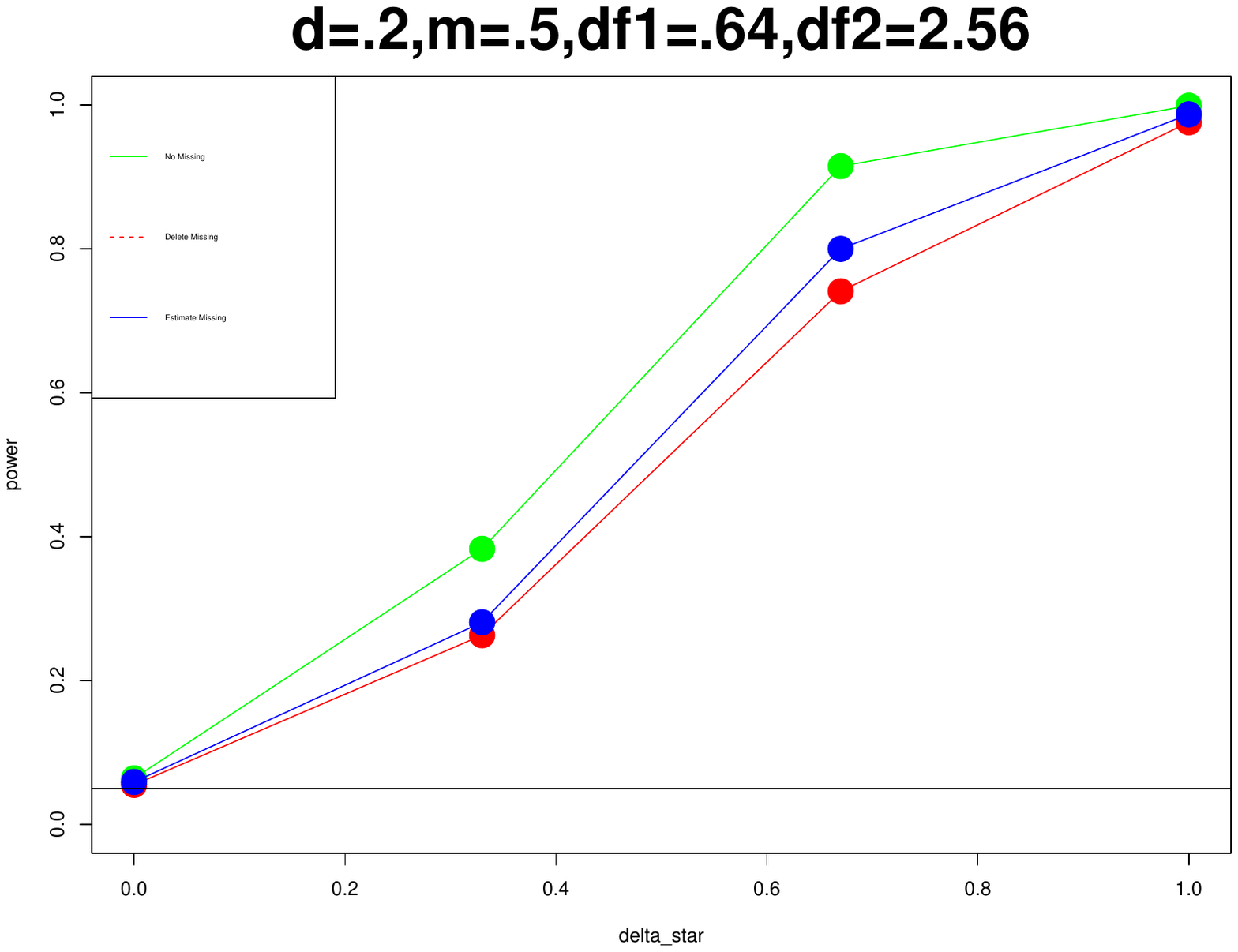}
\includegraphics[width = 2.3in, height = 1.4in]{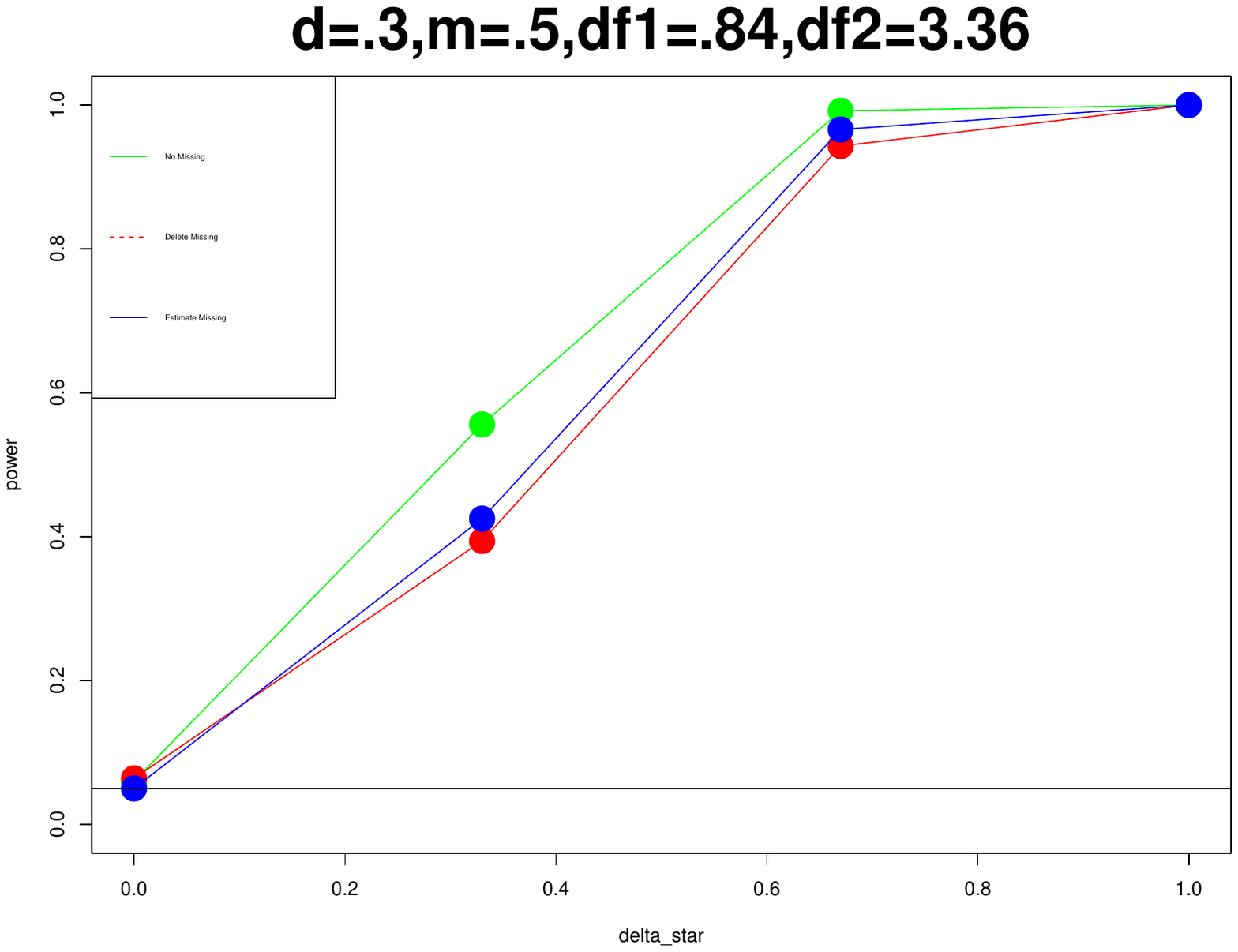}

\hspace{1.5cm}
$d=.1 \quad m=.5$
\hspace{3cm}
$d=.2 \quad m=.5$
\hspace{3cm}
$d=.3 \quad m=.5$

\includegraphics[width = 2.3in, height = 1.3in]{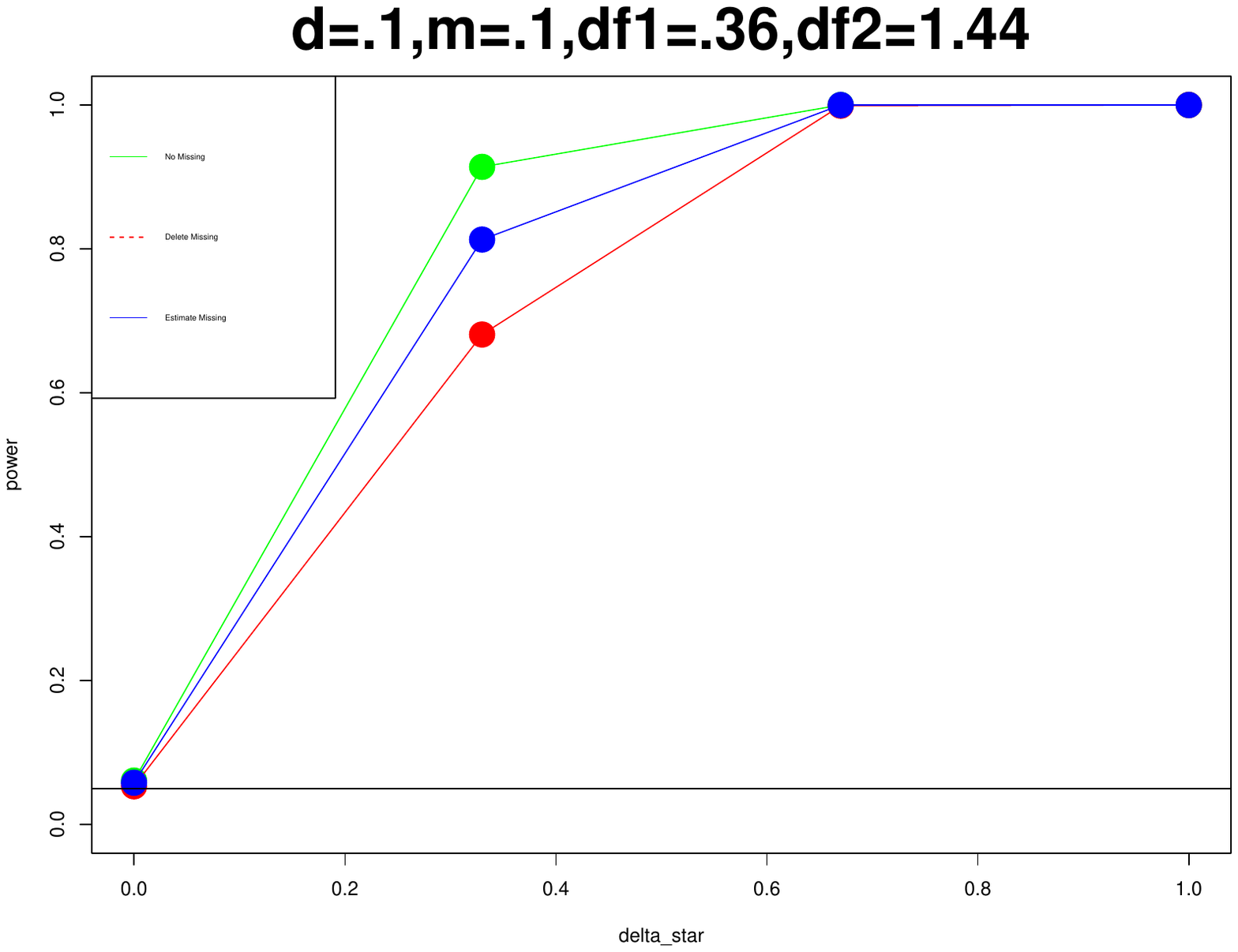}
\includegraphics[width = 2.3in, height = 1.3in]{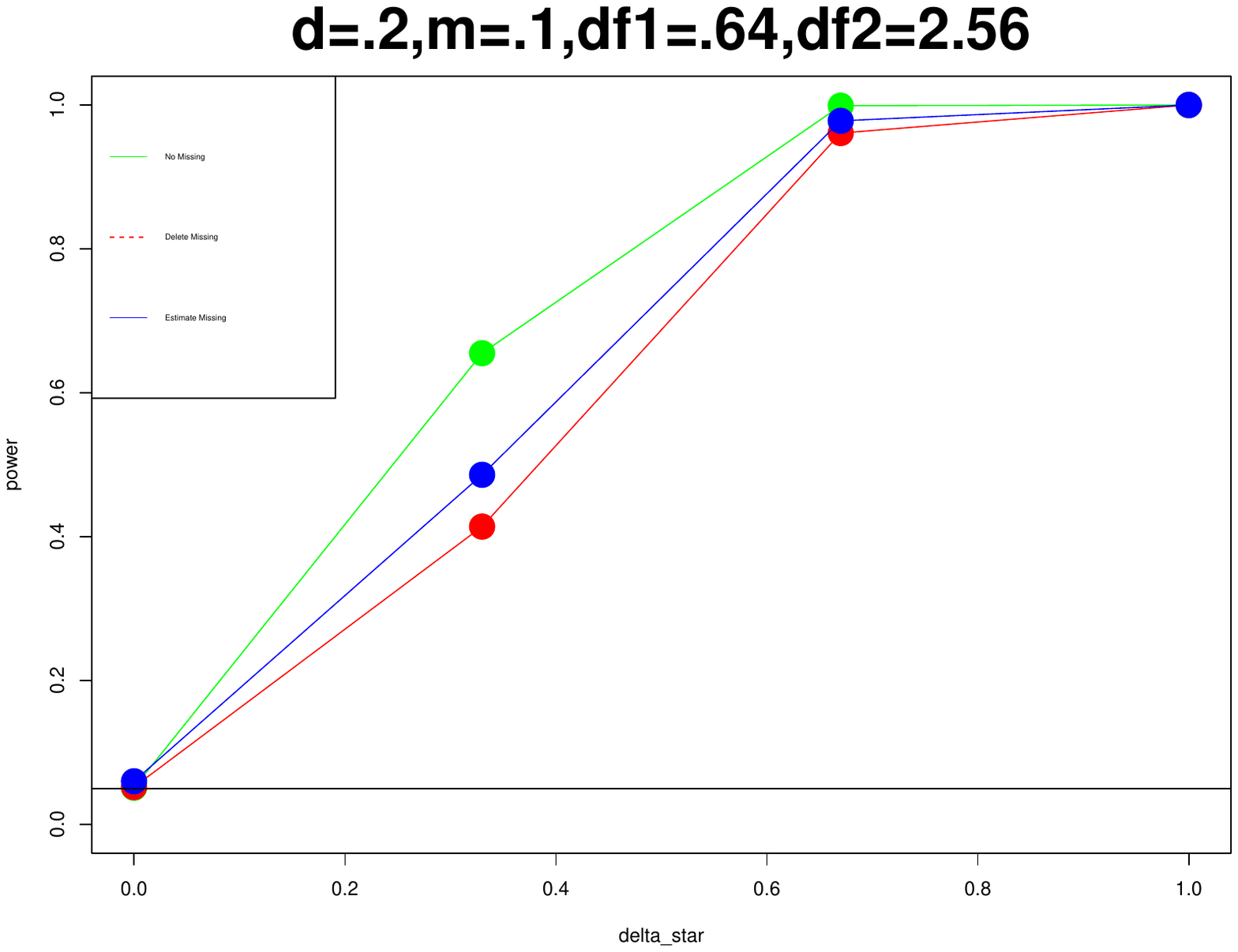}
\includegraphics[width = 2.3in, height = 1.3in]{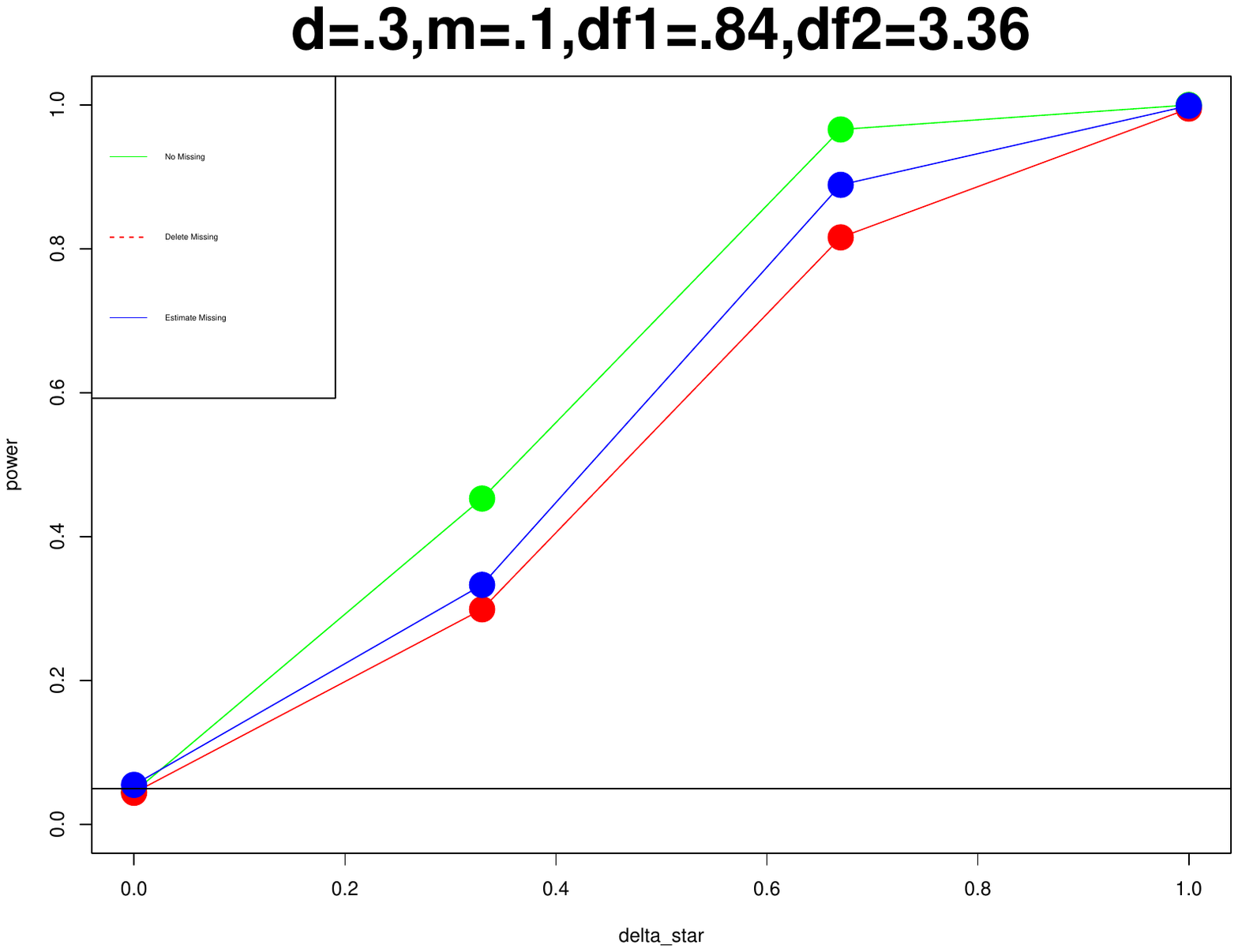}

\section{Conclusion and Remarks}
\begin{itemize}
\item \textbf{For Missing Type 1:}
The distinction between powers of estimated missing and deleted missing are more prominent when proportion of missing increases. For normal case we have taken two choices of $\beta$ and got similar results. We have done the same exercise by increasing $p^\star$ and observed same kind of results, only difference is that powers of those cases are higher and this is quite natural. So, we think the imputation method is equally good for different choices of $\beta$ and $p^\star.$

\item \textbf{For Missing Type 2.1:} 
For all the six combination of traits, we see that power curve for estimated missing data (blue line) is very close to the power curve for no missing (green line). So, we can conclude our imputation method is working very good for this type of missing.

\item \textbf{For Missing Type 2.3:}
Unlike missing type 2.1, here we did not compare among different strategies, rather we chose the strategy which makes more sense (as using same trait of other offspring makes more sense than using other trait of other offspring). In the plots we see that in most of the cases the blue line is well above the red line and this justifies the goodness of our imputation method.

\item \textbf{For Missing Type 2.4:}
In this type of missing we are estimating three trait values using only one trait, so the estimation is expected not be very good and as a result power of estimated missing curve is close to deleted missing one.
We observe that our imputation method in this type of missing working moderate for normal and poisson distribution but not that good for chi squares distribution.

\end{itemize}

We could not work for the missing type 2.2 and mixture of all these types of missing due to shortage of time. As we have seen that our imputation method is working good for most of the types of missing, we hope that it will work good for the mixture case. We are currently working on missing type 2.2 and hope that we shall be able to do that before the presentation.

\section{Acknowledgement}

The author would like to express most sincere gratitude and appreciation to the supervisor Prof. Saurabh Ghosh for his constant guidance and encouragement throughout the development of the project. The author also sincerely thanks Prof. Mausumi Bose , Prof. Deepayan Sarkar for their valuable suggestions.

\section{References}

\begin{itemize}

\item  Banerjee, K \& Ghosh, S. 2017. Mapping Multivariate Phenotypes in the Presence of Missing Data.\textit{ Project Report by MSTAT 2nd year Student.}

\item Galesloot TE, van Steen K, Kiemeney LA, Janss LL, Vermeulen SH. 2014. A comparison of multivariate genome-wide association methods. \textit{PloS One 9(4):e95923.}

\item Haldar, T. \& Ghosh, S. Statistical equivalent of the classical TDT for quantitative traits
and multivariate phenotypes. \textit{J. Genet. 94, 619–628 (2015)}.

\item Newman WP, Freedman DS, Voors AW, Gard PD, Srinivasan SR, Cresanta JL, Williamson GD, Webber LS, Berenson GS. 1986. Relation of serum lipoprotein levels and systolic blood pressure to early atherosclerosis. \textit{N Engl J Med 314(3): 138–144.}

\end{itemize}

\end{document}